\documentclass[twocolumn, traditabstract, longauth]{aa}
\usepackage[switch]{lineno}
\modulolinenumbers[5]
\usepackage[table,usenames,dvipsnames]{xcolor}

\usepackage{tikz}
\usetikzlibrary{calc,trees,positioning,arrows,chains,shapes.geometric,%
  decorations.pathreplacing,decorations.pathmorphing,shapes,%
  matrix,shapes.symbols
}

\tikzset{
  >=stealth',
  punktchain/.style={
    rectangle,
    rounded corners,
    draw=black, very thick,
    text width=12em,
    minimum height=3em,
    text centered,
    on chain},
  line/.style={draw, thick, <-},
  element/.style={
    tape,
    top color=white,
    bottom color=blue!50!black!60!,
    minimum width=8em,
    draw=blue!40!black!90, very thick,
    text width=10em,
    minimum height=3.5em,
    text centered,
    on chain},
  every join/.style={->, thick, shorten >=1pt},
  decoration={brace},
}
\usepackage{afterpage}
\usepackage[super]{nth}
\usepackage{graphicx}
\usepackage{epstopdf}
\usepackage{natbib}
\bibpunct{(}{)}{;}{a}{}{,}
\usepackage{amssymb}
\usepackage{amsmath}
\usepackage{longtable}
\usepackage{multirow}
\usepackage{txfonts}
\usepackage{ifthen}

\definecolor{linkcolor}{rgb}{0.6,0,0}
\definecolor{citecolor}{rgb}{0,0,0.75}
\definecolor{urlcolor}{rgb}{0.12,0.46,0.7}
\usepackage[breaklinks, colorlinks, urlcolor=urlcolor,
linkcolor=linkcolor, citecolor=citecolor, pdfencoding=auto]{hyperref}

\def\muKRJ{\ifmmode \,\mu$K$_{\rm RJ}\else \,$\mu$\hbox{K}$_{\rm RJ}$\fi}
\def\muKCMB{\ifmmode \,\mu$K$_{\rm CMB}\else \,$\mu$\hbox{K}$_{\rm CMB}$\fi}
\def\KCMB{\ifmmode \,$K$_{\rm CMB}\else \,\hbox{K}$_{\rm CMB}$\fi}
\def\Hz{\ifmmode $\,Hz$\else \,Hz\fi}
\def\nside#1{\ifmmode N_{\rm side}$\,=\,${#1}
                             \else $N_{\rm side}$\,=\,{#1}\fi}
\def\Herschel{\textit{Herschel}}

\newcommand{\mat}[1]{\tens{#1}}
\newcommand{\trans}[0]{{}^{\sf T}}
\newcommand{\inv}[0]{{}^{-1}}
\newcommand{\lfi}[0]{LFI}
\newcommand{\hfi}[0]{HFI}
\newcommand{\dacapo}[0]{\texttt{DaCapo}}
\newcommand{\sroll}[0]{\texttt{SRoll}}
\newcommand{\madam}[0]{\texttt{Madam}}
\newcommand{\libmadam}[0]{\texttt{libMadam}}
\newcommand{\quickpol}[0]{\texttt{QuickPol}}
\newcommand{\commander}[0]{\texttt{Commander}}
\newcommand{\sevem}[0]{\texttt{SEVEM}}

\newcommand{\npipe}[0]{\texttt{NPIPE}}
\newcommand{\healpix}[0]{\texttt{HEALPix}}
\newcommand{\grasp}[0]{\texttt{GRASP}}
\newcommand{\submm}[0]{submillimetre}
\newcommand{\wmap}[0]{\emph{WMAP}}

\renewcommand{\d}[0]{\vec{d}}
\newcommand{\x}[0]{\vec{x}}

\newcommand{\s}[0]{\vec{s}}
\renewcommand{\a}[0]{\vec{a}}
\newcommand{\m}[0]{\vec{m}}

\newcommand{\B}[0]{\tens{B}}

\renewcommand{\L}[0]{\tens{L}}

\newcommand{\N}[0]{\tens{N}}

\renewcommand{\S}[0]{\tens{S}}
\renewcommand{\r}[0]{\vec{r}}

\newcommand{\prtwo}{PR2}
\newcommand{\prthree}{PR3}
\newcommand{\srolltwo}[0]{\texttt{SRoll2}}

\def\setsymbol#1#2{\expandafter\def\csname #1\endcsname{#2}}
\def\getsymbol#1{\csname #1\endcsname}

\def\Planck{\textit{Planck}}

\newbox\tablebox    \newdimen\tablewidth
\def\leaderfil{\leaders\hbox to 5pt{\hss.\hss}\hfil}
\def\endPlancktable{\tablewidth=\columnwidth 
    $$\hss\copy\tablebox\hss$$
    \vskip-\lastskip\vskip -2pt}
\def\endPlancktablewide{\tablewidth=\textwidth 
    $$\hss\copy\tablebox\hss$$
    \vskip-\lastskip\vskip -2pt}
\def\tablenote#1 #2\par{\begingroup \parindent=0.8em
    \abovedisplayshortskip=0pt\belowdisplayshortskip=0pt
    \noindent
    $$\hss\vbox{\hsize\tablewidth \hangindent=\parindent \hangafter=1 \noindent
    \hbox to \parindent{$^#1$\hss}\strut#2\strut\par}\hss$$
    \endgroup}
\def\doubleline{\vskip 3pt\hrule \vskip 1.5pt \hrule \vskip 5pt}

\def\L2{\ifmmode L_2\else $L_2$\fi}

\def\DeltaT{\ifmmode \Delta T\else $\Delta T$\fi}
\def\deltat{\ifmmode \Delta t\else $\Delta t$\fi}
\def\fknee{\ifmmode f_{\rm knee}\else $f_{\rm knee}$\fi}
\def\Fmax{\ifmmode F_{\rm max}\else $F_{\rm max}$\fi}
\def\solar{\ifmmode{\rm M}_{\mathord\odot}\else${\rm M}_{\mathord\odot}$\fi}
\def\Msolar{\ifmmode{\rm M}_{\mathord\odot}\else${\rm M}_{\mathord\odot}$\fi}
\def\Lsolar{\ifmmode{\rm L}_{\mathord\odot}\else${\rm L}_{\mathord\odot}$\fi}
\def\inv{\ifmmode^{-1}\else$^{-1}$\fi}
\def\mo{\ifmmode^{-1}\else$^{-1}$\fi}
\def\sup#1{\ifmmode ^{\rm #1}\else $^{\rm #1}$\fi}
\def\expo#1{\ifmmode \times 10^{#1}\else $\times 10^{#1}$\fi}
\def\,{\thinspace}
\def\lsim{\mathrel{\raise .4ex\hbox{\rlap{$<$}\lower 1.2ex\hbox{$\sim$}}}}
\def\gsim{\mathrel{\raise .4ex\hbox{\rlap{$>$}\lower 1.2ex\hbox{$\sim$}}}}

\def\simprop{\mathrel{\raise .4ex\hbox{\rlap{$\propto$}\lower 1.2ex\hbox{$\sim$}}}}
\def\deg{\ifmmode^\circ\else$^\circ$\fi}
\def\pdeg{\ifmmode $\setbox0=\hbox{$^{\circ}$}\rlap{\hskip.11\wd0 .}$^{\circ}
          \else \setbox0=\hbox{$^{\circ}$}\rlap{\hskip.11\wd0 .}$^{\circ}$\fi}
\def\arcs{\ifmmode {^{\scriptstyle\prime\prime}}
          \else $^{\scriptstyle\prime\prime}$\fi}
\def\arcm{\ifmmode {^{\scriptstyle\prime}}
          \else $^{\scriptstyle\prime}$\fi}
\newdimen\sa  \newdimen\sb
\def\parcs{\sa=.07em \sb=.03em
     \ifmmode \hbox{\rlap{.}}^{\scriptstyle\prime\kern -\sb\prime}\hbox{\kern -\sa}
     \else \rlap{.}$^{\scriptstyle\prime\kern -\sb\prime}$\kern -\sa\fi}
\def\parcm{\sa=.08em \sb=.03em
     \ifmmode \hbox{\rlap{.}\kern\sa}^{\scriptstyle\prime}\hbox{\kern-\sb}
     \else \rlap{.}\kern\sa$^{\scriptstyle\prime}$\kern-\sb\fi}
\def\ra[#1 #2 #3.#4]{#1\sup{h}#2\sup{m}#3\sup{s}\llap.#4}
\def\dec[#1 #2 #3.#4]{#1\deg#2\arcm#3\arcs\llap.#4}
\def\deco[#1 #2 #3]{#1\deg#2\arcm#3\arcs}
\def\rra[#1 #2]{#1\sup{h}#2\sup{m}}

\def\dots{\relax\ifmmode \ldots\else $\ldots$\fi}
\def\WHzsr{\ifmmode $W\,Hz\mo\,sr\mo$\else W\,Hz\mo\,sr\mo\fi}
\def\mHz{\ifmmode $\,mHz$\else \,mHz\fi}
\def\GHz{\ifmmode $\,GHz$\else \,GHz\fi}
\def\mKs{\ifmmode $\,mK\,s$^{1/2}\else \,mK\,s$^{1/2}$\fi}
\def\muKs{\ifmmode \,\mu$K\,s$^{1/2}\else \,$\mu$K\,s$^{1/2}$\fi}
\def\muKRJs{\ifmmode \,\mu$K$_{\rm RJ}$\,s$^{1/2}\else \,$\mu$K$_{\rm RJ}$\,s$^{1/2}$\fi}
\def\muKHz{\ifmmode \,\mu$K\,Hz$^{-1/2}\else \,$\mu$K\,Hz$^{-1/2}$\fi}
\def\MJysr{\ifmmode \,$MJy\,sr\mo$\else \,MJy\,sr\mo\fi}
\def\MJysrmK{\ifmmode \,$MJy\,sr\mo$\,mK$_{\rm CMB}\mo\else \,MJy\,sr\mo\,mK$_{\rm CMB}\mo$\fi}
\def\microns{\ifmmode \,\mu$m$\else \,$\mu$m\fi}

\def\muK{\ifmmode \,\mu$K$\else \,$\mu$\hbox{K}\fi}
\def\microK{\ifmmode \,\mu$K$\else \,$\mu$\hbox{K}\fi}
\def\muW{\ifmmode \,\mu$W$\else \,$\mu$\hbox{W}\fi}
\def\kms{\ifmmode $\,km\,s$^{-1}\else \,km\,s$^{-1}$\fi}
\def\kmsMpc{\ifmmode $\,\kms\,Mpc\mo$\else \,\kms\,Mpc\mo\fi}
\providecommand{\sorthelp}[1]{}

\DeclareGraphicsExtensions{.eps, .pdf, .png, .jpg}

\graphicspath{ {./small_figures/} }

\begin{document}

\title{\Planck\ intermediate results.  LVII.  Joint \Planck\ LFI and HFI data processing }

\titlerunning{\npipe\ processing}
\authorrunning{Planck Collaboration}
\author{\small
Planck Collaboration: Y.~Akrami\inst{11, 49, 50}
\and
K. J.~Andersen\inst{50}
\and
M.~Ashdown\inst{56, 3}
\and
C.~Baccigalupi\inst{65}
\and
M.~Ballardini\inst{16, 33}
\and
A.~J.~Banday\inst{78, 6}
\and
R.~B.~Barreiro\inst{52}
\and
N.~Bartolo\inst{20, 53}
\and
S.~Basak\inst{71}
\and
K.~Benabed\inst{48, 73}
% ...
}
\institute{\small
APC, AstroParticule et Cosmologie, Universit\'{e} Paris Diderot, CNRS/IN2P3, CEA/lrfu, Observatoire de Paris, Sorbonne Paris Cit\'{e}, 10, rue Alice Domon et L\'{e}onie Duquet, 75205 Paris Cedex 13, France\goodbreak
\and
African Institute for Mathematical Sciences, 6-8 Melrose Road, Muizenberg, Cape Town, South Africa\goodbreak
\and
Astrophysics Group, Cavendish Laboratory, University of Cambridge, J J Thomson Avenue, Cambridge CB3 0HE, U.K.\goodbreak
\and
Astrophysics \& Cosmology Research Unit, School of Mathematics, Statistics \& Computer Science, University of KwaZulu-Natal, Westville Campus, Private Bag X54001, Durban 4000, South Africa\goodbreak
\and
CITA, University of Toronto, 60 St. George St., Toronto, ON M5S 3H8, Canada\goodbreak
\and
CNRS, IRAP, 9 Av. colonel Roche, BP 44346, F-31028 Toulouse cedex 4, France\goodbreak
\and
Cahill Center for Astronomy and Astrophysics, California Institute of Technology, Pasadena CA,  91125, USA\goodbreak
\and
California Institute of Technology, Pasadena, California, U.S.A.\goodbreak
\and
Computational Cosmology Center, Lawrence Berkeley National Laboratory, Berkeley, California, U.S.A.\goodbreak
\and
D\'{e}partement de Physique Th\'{e}orique, Universit\'{e} de Gen\`{e}ve, 24, Quai E. Ansermet,1211 Gen\`{e}ve 4, Switzerland\goodbreak
\and
D\'{e}partement de Physique, \'{E}cole normale sup\'{e}rieure, PSL Research University, CNRS, 24 rue Lhomond, 75005 Paris, France\goodbreak
\and
Departamento de Astrof\'{i}sica, Universidad de La Laguna (ULL), E-38206 La Laguna, Tenerife, Spain\goodbreak
\and
Departamento de F\'{\i}sica, Universidad de Oviedo, C/ Federico Garc\'{\i}a Lorca, 18 , Oviedo, Spain\goodbreak
\and
Department of Astrophysics/IMAPP, Radboud University, P.O. Box 9010, 6500 GL Nijmegen, The Netherlands\goodbreak
\and
Department of Physics \& Astronomy, University of British Columbia, 6224 Agricultural Road, Vancouver, British Columbia, Canada\goodbreak
\and
Department of Physics \& Astronomy, University of the Western Cape, Cape Town 7535, South Africa\goodbreak
\and
Department of Physics, Gustaf H\"{a}llstr\"{o}min katu 2a, University of Helsinki, Helsinki, Finland\goodbreak
\and
Department of Physics, Princeton University, Princeton, New Jersey, U.S.A.\goodbreak
\and
Department of Physics, University of California, Santa Barbara, California, U.S.A.\goodbreak
\and
Dipartimento di Fisica e Astronomia G. Galilei, Universit\`{a} degli Studi di Padova, via Marzolo 8, 35131 Padova, Italy\goodbreak
\and
Dipartimento di Fisica e Scienze della Terra, Universit\`{a} di Ferrara, Via Saragat 1, 44122 Ferrara, Italy\goodbreak
\and
Dipartimento di Fisica, Universit\`{a} La Sapienza, P. le A. Moro 2, Roma, Italy\goodbreak
\and
Dipartimento di Fisica, Universit\`{a} degli Studi di Milano, Via Celoria, 16, Milano, Italy\goodbreak
\and
Dipartimento di Fisica, Universit\`{a} degli Studi di Trieste, via A. Valerio 2, Trieste, Italy\goodbreak
\and
Dipartimento di Fisica, Universit\`{a} di Roma Tor Vergata, Via della Ricerca Scientifica, 1, Roma, Italy\goodbreak
\and
European Space Agency, ESAC, Planck Science Office, Camino bajo del Castillo, s/n, Urbanizaci\'{o}n Villafranca del Castillo, Villanueva de la Ca\~{n}ada, Madrid, Spain\goodbreak
\and
European Space Agency, ESTEC, Keplerlaan 1, 2201 AZ Noordwijk, The Netherlands\goodbreak
\and
Gran Sasso Science Institute, INFN, viale F. Crispi 7, 67100 L'Aquila, Italy\goodbreak
\and
HEP Division, Argonne National Laboratory, Lemont, IL 60439, USA\goodbreak
\and
Haverford College Astronomy Department, 370 Lancaster Avenue, Haverford, Pennsylvania, U.S.A.\goodbreak
\and
Helsinki Institute of Physics, Gustaf H\"{a}llstr\"{o}min katu 2, University of Helsinki, Helsinki, Finland\goodbreak
\and
IFPU - Institute for Fundamental Physics of the Universe, Via Beirut 2, 34014 Trieste, Italy\goodbreak
\and
INAF - OAS Bologna, Istituto Nazionale di Astrofisica - Osservatorio di Astrofisica e Scienza dello Spazio di Bologna, Area della Ricerca del CNR, Via Gobetti 101, 40129, Bologna, Italy\goodbreak
\and
INAF - Osservatorio Astronomico di Padova, Vicolo dell'Osservatorio 5, Padova, Italy\goodbreak
\and
INAF - Osservatorio Astronomico di Trieste, Via G.B. Tiepolo 11, Trieste, Italy\goodbreak
\and
INAF - Osservatorio Astronomico di Trieste, via G. B. Tiepolo 11, I-34143 Trieste, Italy\goodbreak
\and
INAF, Istituto di Radioastronomia, Via Piero Gobetti 101, I-40129 Bologna, Italy\goodbreak
\and
INAF/IASF Milano, Via E. Bassini 15, Milano, Italy\goodbreak
\and
INFN - CNAF, viale Berti Pichat 6/2, 40127 Bologna, Italy\goodbreak
\and
INFN, Sezione di Bologna, viale Berti Pichat 6/2, 40127 Bologna, Italy\goodbreak
\and
INFN, Sezione di Ferrara, Via Saragat 1, 44122 Ferrara, Italy\goodbreak
\and
INFN, Sezione di Milano, Via Celoria 16, Milano, Italy\goodbreak
\and
INFN, Sezione di Roma 2, Universit\`{a} di Roma Tor Vergata, Via della Ricerca Scientifica, 1, Roma, Italy\goodbreak
\and
IUCAA, Post Bag 4, Ganeshkhind, Pune University Campus, Pune 411 007, India\goodbreak
\and
Imperial College London, Astrophysics group, Blackett Laboratory, Prince Consort Road, London, SW7 2AZ, U.K.\goodbreak
\and
Infrared Processing and Analysis Center, California Institute of Technology, Pasadena, CA 91125, U.S.A.\goodbreak
\and
Institut d'Astrophysique Spatiale, CNRS, Univ. Paris-Sud, Universit\'{e} Paris-Saclay, B\^{a}t. 121, 91405 Orsay cedex, France\goodbreak
\and
Institut d'Astrophysique de Paris, CNRS (UMR7095), 98 bis Boulevard Arago, F-75014, Paris, France\goodbreak
\and
Institute Lorentz, Leiden University, PO Box 9506, Leiden 2300 RA, The Netherlands\goodbreak
\and
Institute of Theoretical Astrophysics, University of Oslo, Blindern, Oslo, Norway\goodbreak
\and
Instituto de Astrof\'{\i}sica de Canarias, C/V\'{\i}a L\'{a}ctea s/n, La Laguna, Tenerife, Spain\goodbreak
\and
Instituto de F\'{\i}sica de Cantabria (CSIC-Universidad de Cantabria), Avda. de los Castros s/n, Santander, Spain\goodbreak
\and
Istituto Nazionale di Fisica Nucleare, Sezione di Padova, via Marzolo 8, I-35131 Padova, Italy\goodbreak
\and
Jet Propulsion Laboratory, California Institute of Technology, 4800 Oak Grove Drive, Pasadena, California, U.S.A.\goodbreak
\and
Jodrell Bank Centre for Astrophysics, Alan Turing Building, School of Physics and Astronomy, The University of Manchester, Oxford Road, Manchester, M13 9PL, U.K.\goodbreak
\and
Kavli Institute for Cosmology Cambridge, Madingley Road, Cambridge, CB3 0HA, U.K.\goodbreak
\and
LAL, Universit\'{e} Paris-Sud, CNRS/IN2P3, Orsay, France\goodbreak
\and
Laboratoire de Physique Subatomique et Cosmologie, Universit\'{e} Grenoble-Alpes, CNRS/IN2P3, 53, rue des Martyrs, 38026 Grenoble Cedex, France\goodbreak
\and
Laboratoire de Physique Th\'{e}orique, Universit\'{e} Paris-Sud 11 \& CNRS, B\^{a}timent 210, 91405 Orsay, France\goodbreak
\and
Lawrence Berkeley National Laboratory, Berkeley, California, U.S.A.\goodbreak
\and
Low Temperature Laboratory, Department of Applied Physics, Aalto University, Espoo, FI-00076 AALTO, Finland\goodbreak
\and
Max-Planck-Institut f\"{u}r Astrophysik, Karl-Schwarzschild-Str. 1, 85741 Garching, Germany\goodbreak
\and
Mullard Space Science Laboratory, University College London, Surrey RH5 6NT, U.K.\goodbreak
\and
National Centre for Nuclear Research, ul. L. Pasteura 7, 02-093 Warsaw, Poland\goodbreak
\and
SISSA, Astrophysics Sector, via Bonomea 265, 34136, Trieste, Italy\goodbreak
\and
San Diego Supercomputer Center, University of California, San Diego, 9500 Gilman Drive, La Jolla, CA 92093, USA\goodbreak
\and
School of Physical Sciences, National Institute of Science Education and Research, HBNI, Jatni-752050, Odissa, India\goodbreak
\and
School of Physics and Astronomy, Cardiff University, Queens Buildings, The Parade, Cardiff, CF24 3AA, U.K.\goodbreak
\and
School of Physics and Astronomy, Sun Yat-sen University, 2 Daxue Rd, Tangjia, Zhuhai, China\goodbreak
\and
School of Physics and Astronomy, University of Nottingham, Nottingham NG7 2RD, U.K.\goodbreak
\and
School of Physics, Indian Institute of Science Education and Research Thiruvananthapuram, Maruthamala PO, Vithura, Thiruvananthapuram 695551, Kerala, India\goodbreak
\and
Simon Fraser University, Department of Physics, 8888 University Drive, Burnaby BC, Canada\goodbreak
\and
Sorbonne Universit\'{e}, CNRS, UMR 7095, Institut d'Astrophysique de Paris, 98 bis bd Arago, 75014 Paris, France\goodbreak
\and
Sorbonne Universit\'{e}, Observatoire de Paris, Universit\'{e} PSL, \'{E}cole normale sup\'{e}rieure, CNRS, LERMA, F-75005, Paris, France\goodbreak
\and
Space Science Data Center - Agenzia Spaziale Italiana, Via del Politecnico snc, 00133, Roma, Italy\goodbreak
\and
Space Sciences Laboratory, University of California, Berkeley, California, U.S.A.\goodbreak
\and
The Oskar Klein Centre for Cosmoparticle Physics, Department of Physics, Stockholm University, AlbaNova, SE-106 91 Stockholm, Sweden\goodbreak
\and
Universit\'{e} de Toulouse, UPS-OMP, IRAP, F-31028 Toulouse cedex 4, France\goodbreak
% ...
}

\abstract{
We present the \npipe\ processing pipeline, which produces calibrated frequency
maps in temperature and polarization from data from the \Planck\ Low Frequency
Instrument (\lfi) and High Frequency Instrument (\hfi) using high-performance
computers.  \npipe\ represents a natural evolution of previous \Planck\
analysis efforts, and combines some of the most powerful features of the
separate LFI and HFI analysis pipelines.  For example, following the \lfi\ 2018
processing procedure, \npipe\ uses foreground
polarization priors during the
calibration stage in order to break scanning-induced degeneracies.  Similarly,
\npipe\ employs the \hfi\ 2018 time-domain processing methodology to correct
for bandpass mismatch at all frequencies.  In addition, \npipe\ introduces
several improvements, including, but not limited to: inclusion of the 8\,\% of data
collected during repointing manoeuvres; smoothing of the \lfi\ reference load
data streams; in-flight estimation of detector polarization parameters; and
construction of maximally independent detector-set split maps.  For
component-separation purposes, important improvements include: maps that retain
the CMB Solar dipole, allowing for high-precision relative calibration in
higher-level analyses; well-defined single-detector maps, allowing for robust
CO extraction; and HFI temperature maps between 217 and 857\GHz\ that are
binned into 0\parcm9 pixels (\nside{4096}), ensuring that the full angular
information in the data is represented in the maps even at the highest
\Planck\ resolutions.  The net effect of these improvements is lower levels of
noise and systematics in both frequency and component maps at essentially all
angular scales, as well as notably improved internal consistency between the
various frequency channels.  Based on the \npipe\ maps, we present the first
estimate of the Solar dipole determined through component separation across all
nine \Planck\ frequencies.  The amplitude is ($3366.6\pm 2.7$)\muK, consistent
with, albeit slightly higher than, earlier estimates.  From the large-scale
polarization data, we derive an updated estimate of the optical depth of
reionization of $\tau=0.051\pm0.006$, which appears robust with respect to data
and sky cuts.   There are 600 complete signal, noise
and systematics simulations of the full-frequency and detector-set maps.  As a
\Planck\ first, these simulations include full time-domain processing of the
beam-convolved CMB anisotropies.  The release of \npipe\ maps and simulations
is accompanied with a complete suite of raw and processed time-ordered data
and the software, scripts, auxiliary data, and parameter files needed to
improve further on the analysis and to run matching simulations.}


\maketitle

\tableofcontents

\section{Introduction} \label{sec:introduction}

This paper, the last by the Planck Collaboration, describes a \Planck\footnote{\Planck\
  (\url{http://www.esa.int/Planck}) is a project of the
  European Space Agency (ESA) with instruments provided by two
  scientific consortia funded by ESA member states and led by
  Principal Investigators from France and Italy, telescope reflectors
  provided through a collaboration between ESA and a scientific
  consortium led and funded by Denmark, and additional contributions
  from NASA (USA).} data-processing pipeline called \npipe\, and the calibrated data, maps, simulations, and other data products that it produces. \npipe\ represents a natural evolution of the previous \Planck\ analysis efforts made within the \Planck\ data
processing centres (DPCs) in Paris and Trieste, but is uniquely designed to analyse data from both the Low Frequency Instrument (\lfi) and the High Frequency Instrument (\hfi) within the same framework. Furthermore, \npipe\ is implemented to execute efficiently on massively parallel high-performance computing (HPC) systems.  Indeed, it was both developed and run on the HPC facilities hosted by the National Energy Research Scientific Computing Center (NERSC), motivating the ``{\tt N}'' in \npipe. The main technical design criteria for \npipe\ are therefore:
\begin{itemize}
\item the ability to handle both LFI and HFI data;
\item{efficient execution on HPC systems;}
\item{adaptability to evolving HPC architectures;}
\item{\Planck\ data access based on Exchange File Format files rather than databases;}
\item{minimal I/O during processing.}
\end{itemize}

Running in a massively parallel HPC environment, \npipe\ overcomes some of the practical difficulties and limitations imposed on the pipelines operated by the \Planck\ DPCs.  With the above design requirements in mind, \npipe\ developed incrementally, with the first step being a prototyping platform for testing new implementations of existing \hfi\ preprocessing modules. Early success
in 4-K line removal and inclusion of the data taken during repointing manoeuvres (excluded in the DPC processing) presaged the development of a full data-processing pipeline from raw time-ordered data (TOD) all the way to maps.  Since most of the DPC pipeline modules were deeply integrated into the database-driven architecture at the DPC, this led to fresh implementations of a majority of the processing modules. The capability to handle \lfi\ data was added while studying reduction of the high-frequency noise in the \lfi\ TOD through improved decorrelation of the $1/f$ noise.

Having a single data-processing pipeline that could handle both \lfi\ and \hfi\ data led to significant improvements in the calibration
procedures, which combine the strengths of both \lfi\ and \hfi\ calibration. Specifically, the \npipe\ calibration procedure is
a synthesis of the high S/N approach that works well for the \hfi\ 353-GHz channel and the noise-limited approach that is critical to \lfi\ 70-GHz calibration.  Numerous smaller improvements to the data-processing modules were motivated by experiences with the nine \Planck\ frequencies and their diverse features.  The \lfi- and \hfi-DPC pipelines evolved in the post-launch period, often in response to instrument-specific effects that emerged as the calibration accuracy improved.   \npipe\ builds on knowledge accumulated through the years, and takes a global approach, aiming at coherent treatment of the full \Planck\ data set.

In what follows, we detail \npipe\ processing and a full data set of \Planck\ maps that result from it. These \npipe\ data maps are supported by a suite of simulated maps that capture the relevant noise and systematic errors present in the data. We also compare the \npipe\ results to the two previous public releases of \Planck\ temperature and polarization maps, namely the second data release in 2015 (``\prtwo''; \citealt{planck2014-a01}) and the third release in 2018 (``\prthree''; \citealt{planck2016-l01}).  Comparison with both earlier releases allows the reader to assess the magnitude of the differences we find.

To characterize the new maps, we consider a few important applications for which component separation is essential. Indeed, robustly supporting component-separation applications ranks among the primary motivations for the \npipe\ products, and several features have been added specifically to meet the requirements of future astrophysical analysis. Among these are: maps that retain the CMB Solar dipole, and thereby allow for joint component separation and high-precision relative calibration; robust single-detector temperature-only sky maps that allow for fine-grained CO extraction, and thereby weaker degeneracies with respect to CMB, thermal dust, and free-free emission; and \hfi\ maps between 217 and 857\GHz\ pixelized at a
{\tt HEALPix}\footnote{\url{https://healpix.sourceforge.io}. 
    The square roots of the uniform pixel areas at \nside{1024}, 2048, and
    4096 are 3\parcm4, 1\parcm7, and 0\parcm88, respectively.
} \citep{gorski2005}
resolution of \nside{4096}, corresponding to a pixel size of 0\parcm88, which ensures properly signal-bandwidth-limited (not limited by pixel size)
sampling of the highest resolution \Planck\ beams of 5\arcm.

Taking advantage of the new \npipe\ maps, we present the first \Planck-only astrophysical sky model that simultaneously constrains the CMB Solar dipole, higher-order CMB temperature fluctuations, low-frequency foregrounds, thermal dust, and CO line emission. This joint analysis has at least three major advantages compared to previous analyses.  First, it results in a global \lfi+\hfi\ CMB Solar dipole estimate with minimal foreground contamination. Second, it eliminates the need for estimating dipole residuals per frequency band during component separation, and thereby reduces large-scale degeneracies in the astrophysical component maps. Third, it allows for high-precision relative, inter-frequency gain calibration during component separation. The
resulting astrophysical components are compared to corresponding previous releases, and are found to be consistent with earlier maps, but exhibiting lower noise and systematics. Similarly, the derived CMB dipole parameters are also consistent with previous estimates, although the new result has a slightly higher amplitude. Properly taking into account uncertainties from higher-order CMB fluctuations \citep{thommesen:2019}, the \npipe\ results are consistent with the \Planck\ 2018 constraints at the 1--2$\,\sigma$ level. 

Finally, we constrain the optical depth of reionization $\tau$ with the large-scale polarization \npipe\ maps. Due to the low levels of
systematic residuals in these maps, we observe good consistency among results derived from different frequency maps, and the resulting large-scale angular CMB power spectrum and cosmological parameters appear robust with respect to both data and sky cuts.

The release of \npipe\ maps and simulations is accompanied with the full software suite and input data so that the results can be
reproduced, and even improved, given sufficient computational resources. The rest of the paper is organized as follows. In
Sect.~\ref{sec:processing} we summarize the low-level \npipe\ data processing, highlighting in particular where \npipe\ differs from the DPC processing. In Sect.~\ref{sec:destriping} we comment on the destriping mapmaking algorithm employed by \npipe. In
Sect.~\ref{sec:dataset} we present and characterize the \npipe\ frequency and detector maps, and in Sect.~\ref{sec:simulations} we describe the associated end-to-end simulations. In Sect.~\ref{sec:comparison} we compare \npipe\ results with those of previous data releases.  In Sect.~\ref{sec:compsep} we discuss the astrophysical component maps derived from the \npipe-only data set, before considering the CMB Solar dipole in Sect.~\ref{sec:dipole} and the optical depth of reionization in Sect.~\ref{sec:tau}. Conclusions are presented in Sect.~\ref{sec:conclusions}, and algorithmic details are provided in the appendices.

\section{Data processing}
\label{sec:processing}

\npipe\ re-implements most of the Level~2 data processing performed at the \Planck\ DPCs \citep{planck2016-l02,planck2016-l03}, while introducing a number of detailed changes that will be described later in this section. We divide our processing into two main steps.

Local \emph{preprocessing} covers all the steps that can take place at the single-detector%
\footnote{\label{fnote:detector}In this paper, a ``detector'' is understood to be an \lfi\ radiometer or an  \hfi\ bolometer.  Each LFI horn and each polarized \hfi\ horn feeds two linearly polarized detectors aligned in orthogonal directions.  Note that each \lfi\ radiometer actually comprises two detector diodes that are co-added to optimize rejection of coherent fluctuations.
}
level without projecting the signal onto maps.  Furthermore, every preprocessing step (except for the mission averaging of the \lfi\ low-pass filter and spike templates) can be performed at the single-pointing-period\footnote{A pointing period or ring consists of the time to orient the spacecraft spin axis with three precise thruster burns (the \emph{repointing manoeuvre}) and the continuous science scan that follows the repointing.  Each repointing manoeuvre lasted about four minutes, with the majority of time spent passively slewing the spacecraft spin axis into the new orientation.  The spin axis was adjusted by 2\arcm\ during a standard repointing.  The subsequent science scan lasted between $35$ and $75$ minutes.
}  
level (see Sect.~\ref{sec:deconv}) without reference to data beyond it.

Global \emph{reprocessing} encompasses single- and multi-detector operations that require much or all of the data to be in memory.  Examples of systematics handled during reprocessing are gain fluctuations and residuals from the low-frequency bolometer transfer function.

The data are written to disk after preprocessing and reprocessing to flexibly allow development of these two major steps.  Furthermore, mapmaking of the reprocessed TOD can be separately optimized after the data are processed.

The parts of \npipe\ data processing that continue to employ DPC-derived products are the following.
\begin{itemize}
\item \lfi\ analogue-to-digital converter (ADC) correction.  We apply the DPC-provided correction profiles.
\item \hfi\ Initial ADC correction measurement.  We start from the DPC-provided correction profiles.
\item \hfi\ bolometer transfer-function measurement.  We deconvolve the DPC-measured transfer function.
\item Beam and focal plane geometry measurements.  Such changes can be implemented in \npipe, but the data set described and released here retains the DPC values.  Neither are expected to differ from \prthree, since we have not updated the bolometer transfer functions.
\end{itemize}

\subsection{Differences between \npipe\ and the PR3 data processing}

We list here the most important differences between \npipe\ and \prthree\ data processing, with section numbers for further details.

\begin{itemize}
\item We use a hybrid calibration scheme to work around a degeneracy in the \Planck\ calibration process.  The 30- and 353-GHz data are first calibrated along with the polarized sky much like PR3 \hfi\ is calibrated.  We then mimic the PR3 \lfi\ calibration and calibrate the intermediate CMB frequencies (44--217\GHz), using an approximation that the sky polarization can be captured with templates derived from the foreground frequencies (Sect.~\ref{sec:pestimates}).
The calibration for the CMB frequencies is thus based on temperature-only sky and polarized foreground templates (see Sect.~\ref{sec:pestimates}).  The use of this polarization prior significantly reduces the uncertainty in gain and the large-scale polarization systematics, but also requires a measurable and correctable transfer function affecting large-scale CMB polarization (described in Sect.~\ref{sec:ee_tf}).
\item We include the \Planck\ repointing manoeuvre data in all of our processing, leading to a roughly 8\,\% increase in total integration.
  time (\ref{sec:repointing}).
\item We retain the Solar dipole in the frequency maps for component separation.
\item We include in our pointing solution the latest star-camera distortion corrections that were developed for {\it Herschel\/} data processing (Sect.~\ref{sec:pointing}).
\item We correct all \lfi\ radiometers for 1-Hz housekeeping spikes, rather than just the 44-GHz ones (Sect.~\ref{sec:spike}).
\item We fit for more 4-K cooler lines, and use the global signal estimate (Sect.~\ref{sec:gestimate}) for signal removal (Sect.~\ref{sec:spike}).
\item We modify the \lfi\ sky--load differencing to include a low-pass filter that reduces the injection of uncorrelated noise
  \emph{into} the differenced signal (Sect.~\ref{sec:diff}).
\item We use our own signal estimate during \hfi\ glitch detection and removal (Sect.~\ref{sec:glitch_removal}).
\item We extend the \hfi\ glitch flags during transfer function deconvolution, to avoid leaking constrained-realization power from
  the gaps into the surrounding data.  This leads to less small-scale noise and lower noise correlations between half-rings\footnote{Half-ring data sets are built from the first and second halves of the pointing periods.} (Sect.~\ref{sec:deconv}).
\item We fit for more \hfi\ transfer-function harmonics, to better address the odd-even survey differences (Sect.~\ref{sec:tf}).
\item We subtract only the seasonally varying part of zodiacal emission, in order not to bias the dust component in the maps (Sect.~\ref{sec:zodi}).
\item We provide polarization-corrected, single-detector, and single-horn temperature maps at all frequencies (Sects.~\ref{sec:dmaps} and \ref{sec:hmaps}).
\item We correct the \hfi\ polarization angles and efficiencies, to address polarized signal-like residuals in single-detector maps
  (Sect.~\ref{sec:ppar}).
\item We bin our single-detector maps from 217 to 857\GHz\ at \nside{4096}, to fully sample the narrow beam.
\item We make detector-set maps for cross correlation analysis and systematics-level studies that are fully independently processed 
  (Sect.~\ref{sec:abmaps}).
\item We provide a consistent, low-resolution data set, with estimates of pixel-pixel noise covariance across all \Planck\ frequencies (Sect.~\ref{sec:lowres}).
\item We fit for signal distortion to address second-order ADC nonlinearity in the \hfi\ CMB channels from 100 to 217\GHz\
  (Sect.~\ref{sec:adcnl}).
\item We extend to \lfi\ data the \hfi\ approach of correcting for bandpass mismatch in the time domain (Sect.~\ref{sec:bpm}).
\item We destripe all data with \madam, using extremely short 167-ms baselines (Sect.~\ref{sec:destriping}).
\item We calibrate with apodized masks that allow us to access most of the sky while down-weighting Galactic regions where even small issues in the signal model can cause problems with calibration (Sect.~\ref{sec:masks}).
\end{itemize}

\subsection{Inputs}

The inputs to \npipe\ are the raw, digitized data as they were transmitted from the spacecraft and decompressed by the DPC Level~1 processing \citep{planck2013-p02,planck2013-p03}.  The \hfi\ TOD are initially corrected with the same ADC nonlinearity profiles as \prthree\ data.  The data are stored in \Planck\ exchange file format FITS files that we have extended to accommodate raw, undifferenced \lfi\ timelines.

Each \lfi\ horn feeds two polarization-orthogonal radiometers (``detectors'' in this paper, see footnote~\ref{fnote:detector}). Each radiometer comprises two diodes that alternate between observing the sky and a 4-K reference load.  In total, each \lfi\ horn provides four sky timelines and four reference timelines. From the 11 \lfi\ horns there are then 88~discrete timelines that are preprocessed into 22~detector timelines.

For each polarized \hfi\ horn there are two polarization-sensitive bolometers (PSBs labelled ``a'' and ``b'') at orthogonal angles of polarization sensitivity.  Some of the \hfi\ horns instead house nearly unpolarized ``spiderweb'' bolometers (SWBs).  There are 36 \hfi\ horns in total, 16 polarized horns and 20 nearly unpolarized.  In addition, two dark bolometers measured the thermal baseline on the focal plane.  With two of the SWBs irrecoverably compromised by random telegraphic signal, we have 52~bolometer timelines that are preprocessed into 50~optical timelines.

\subsubsection{Time span} \label{sec:timespan}

\hfi\ science operations spanned the period between 12 August 2009 and 15 January 2012, comprising almost five sky surveys,\footnote{The scan strategy was such that \Planck\ observed about 95\,\% of the sky over the course of six months, but the entire sky exactly twice every 12 months.  The approximately six-month periods are called ``surveys.''  See section~4 of \citet{planck2013-p01}, and particularly table~1, for precise definitions.
} 885~operational days (29~months), and 27\,000~pointing periods.  We follow \citet{planck2016-l03} and exclude the last 22~days (956~pointing periods) of the \hfi\ science data due to large drifts in the thermal baseline.  This leaves the fifth \hfi\ sky survey only 80\,\% complete.  The drifts were considered harmful both due to the correlated noise fluctuations they induce and due to the challenges the changes in the baseline introduce to ADC nonlinearity correction.

\lfi\ science operations continued until 10 October 2013, comprising 8.3 sky surveys, 1\,513 operational days (49.6 months), and 46\,000 pointing periods.  The \lfi\ maps in \prthree\ are based on the eight completed sky surveys \citep{planck2016-l02}.  \npipe\ maps include the additional $60$ days of the ninth sky survey.

\subsubsection{Pointing} \label{sec:pointing}

The boresight pointing in \npipe\ is based on the same raw attitude information as \prthree.  We processed the raw attitude quaternions to include a star-camera field-of-view distortion correction that was developed for \Herschel\ and updates some guide-star positions necessitated by the delay in the \Planck\ launch \citep{PT-CMOC-OPS-RP-6435-HSO-GF}.  The corrections are only a few arc seconds at most, but are systematic rather than statistical.

Reprocessing the attitude history allowed us to extend the low-pass filtering used to suppress high-frequency noise in the reconstructed pointing.  In \prthree, the filtering only covered the stable science scans, leaving the repointing manoeuvres considerably noisier.  In \npipe\ the same low-pass filter is used on both science scans and repointing manoeuvres, better
supporting our use of the repointing manoeuvre data in mapmaking.
  
\npipe\ uses the same PTCOR correction \citep{planck2014-a01} as \prthree\ for the thermal deformation of the angle between the boresight and the star camera.  For convenience, we apply the correction directly to the attitude history files (AHFs), thus removing the need to apply the correction every time the pointing is read from the AHFs.

\subsection{Local preprocessing}

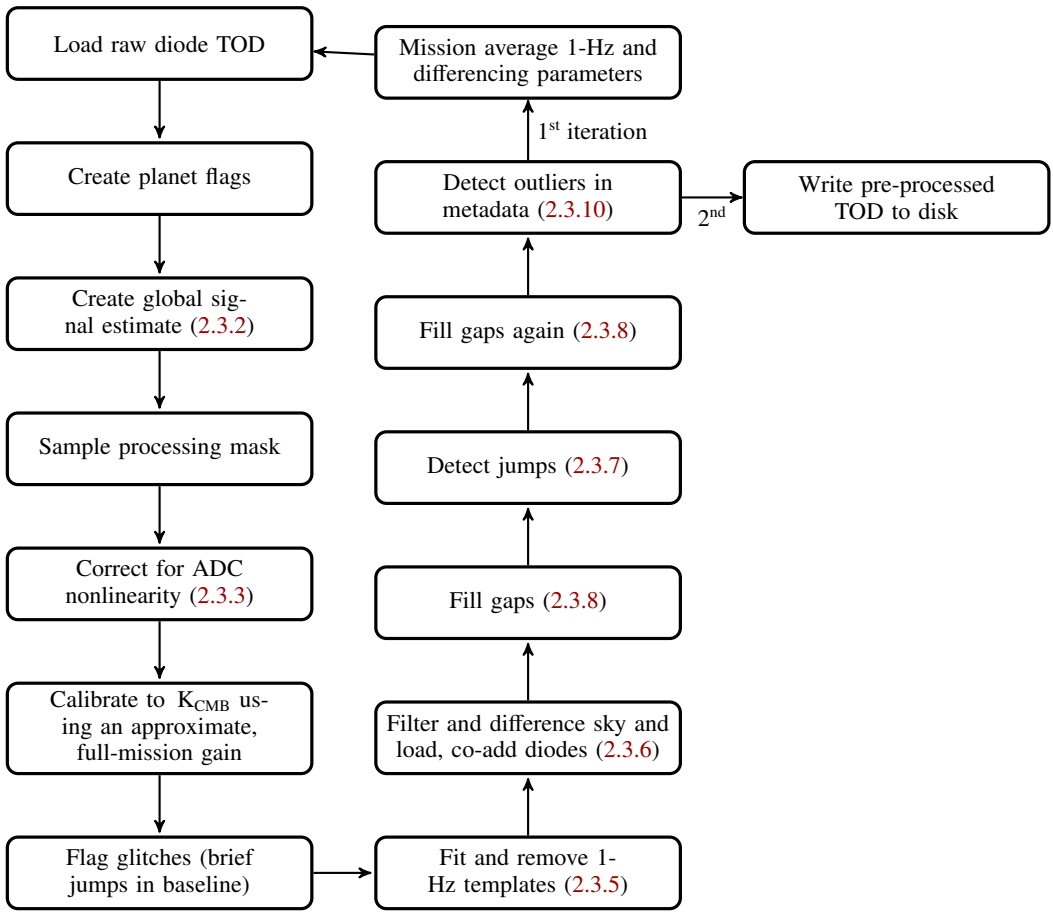
\begin{figure*}[htpb!]
  \centering
  \begin{tikzpicture} 
    [node distance=.8cm, start chain=going below]
    \node[punktchain, join, on chain=going below] (load) {Load raw diode TOD};
    \node[punktchain, join, on chain=going below] (pflags) {Create planet flags};
    \node[punktchain, join, on chain=going below] (gestimate) {Create global signal estimate (\ref{sec:gestimate})};
    \node[punktchain, join, on chain=going below] (mask) {Sample processing mask};
    \node[punktchain, join, on chain=going below] (adc) {Correct for ADC nonlinearity (\ref{sec:adc})};
    \node[punktchain, join, on chain=going below] (calib) {Calibrate to \KCMB\ using an approximate, full-mission gain};
    \node[punktchain, join, on chain=going below] (glitch) {Flag glitches (brief jumps in baseline)};
    \node[punktchain, join, on chain=going right] (lines) {Fit and remove 1-Hz templates (\ref{sec:spike})};
    \node[punktchain, join, on chain=going above] (diff) {Filter and difference sky and load, co-add diodes (\ref{sec:diff})};
    \node[punktchain, join, on chain=going above] (fill) {Fill gaps (\ref{sec:fill})};
    \node[punktchain, join, on chain=going above] (jumps) {Detect jumps (\ref{sec:jump})};
    \node[punktchain, join, on chain=going above] (fill2) {Fill gaps again (\ref{sec:fill})};
    \node[punktchain, join, on chain=going above] (outlier) {Detect outliers in metadata (\ref{sec:outlier})};
    \begin{scope}[start branch=out]
      \node[punktchain, on chain=going right] (write) {Write pre-processed TOD to disk};
    \end{scope}
    \node[punktchain, on chain=going above] (average) {Mission average 1-Hz and differencing parameters};
    \draw[->, thick] (average) -- (load);
    \draw[->, thick] (outlier) -- node[right] {\nth{1} iteration} (average);
    \draw[->, thick] (outlier) -- node[below] {\nth{2}} (write);
  \end{tikzpicture}
  \caption{\npipe\ preprocessing flow chart for \lfi, indicating the sections in which major steps are discussed.  The main loop runs twice, first fitting the 1-Hz spike templates and low-pass filter parameters independently on each pointing period, then mission-averaging the template amplitudes and filter parameters and running the entire processing pipeline again with these fixed values.}
  \label{fig:preproc_lfi}
\end{figure*}

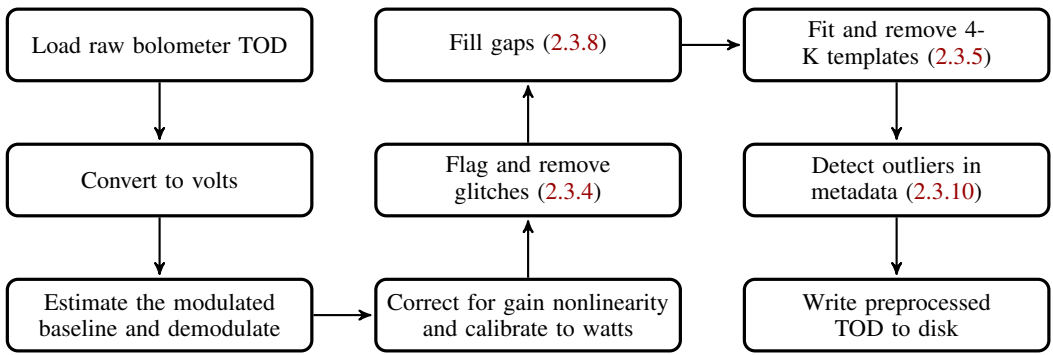
\begin{figure*}[htpb!]
  \centering
  \begin{tikzpicture}
    [node distance=.8cm, start chain=going below]
    \node[punktchain, join, on chain=going below] (load) {Load raw bolometer TOD};
    \node[punktchain, join, on chain=going below] (transf) {Convert to volts};
    \node[punktchain, join, on chain=going below] (demod) {Estimate the modulated baseline and demodulate};
    \node[punktchain, join, on chain=going right] (gaincorrect) {Correct for gain nonlinearity and calibrate to watts};
    \node[punktchain, join, on chain=going above] (glitch) {Flag and remove glitches (\ref{sec:glitch_removal})};
    \node[punktchain, join, on chain=going above] (fill) {Fill gaps (\ref{sec:fill})};
    \node[punktchain, join, on chain=going right] (lines) {Fit and remove 4-K templates (\ref{sec:spike})};
    \node[punktchain, join, on chain=going below] (outlier) {Detect outliers in metadata (\ref{sec:outlier})};
    \node[punktchain, join, on chain=going below] (write) {Write preprocessed\\TOD to disk};
  \end{tikzpicture}
  \caption{\npipe\ preprocessing flow chart for dark \hfi\ bolometers.}
  \label{fig:preproc_hfi_dark}
\end{figure*}

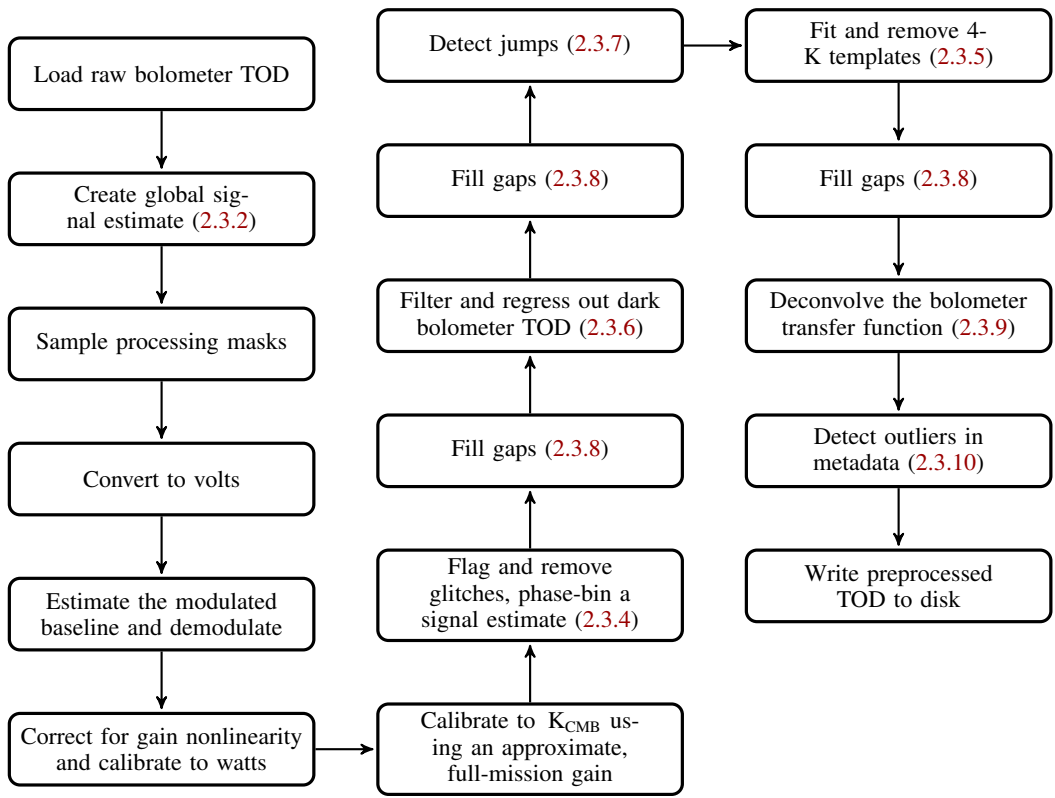
\begin{figure*}[htpb!]
  \centering
  \begin{tikzpicture}
    [node distance=.8cm, start chain=going below]
    \node[punktchain, join, on chain=going below] (load) {
      Load raw bolometer TOD};
    \node[punktchain, join, on chain=going below] (gestimate) {
      Create global signal estimate (\ref{sec:gestimate})};
    \node[punktchain, join, on chain=going below] (mask) {
      Sample processing masks};
    \node[punktchain, join, on chain=going below] (transf) {
      Convert to volts};
    \node[punktchain, join, on chain=going below] (demod) {
      Estimate the modulated baseline and demodulate};
    \node[punktchain, join, on chain=going below] (gaincorrect) {
      Correct for gain nonlinearity and calibrate to watts};
    \node[punktchain, join, on chain=going right] (calib) {
      Calibrate to \KCMB\ using an approximate, full-mission gain};
    \node[punktchain, join, on chain=going above] (glitch) {
      Flag and remove glitches, phase-bin a signal estimate
      (\ref{sec:glitch_removal})};
    \node[punktchain, join, on chain=going above] (fill) {
      Fill gaps (\ref{sec:fill})};
    \node[punktchain, join, on chain=going above] (decor) {
      Filter and regress out dark bolometer TOD (\ref{sec:diff})};
    \node[punktchain, join, on chain=going above] (fill2) {
      Fill gaps (\ref{sec:fill})};
    \node[punktchain, join, on chain=going above] (jumps) {
      Detect jumps (\ref{sec:jump})};
    \node[punktchain, join, on chain=going right] (lines) {
      Fit and remove 4-K templates (\ref{sec:spike})};
    \node[punktchain, join, on chain=going below] (fill3) {
      Fill gaps (\ref{sec:fill})};
    \node[punktchain, join, on chain=going below] (fill3) {
      Deconvolve the bolometer transfer function (\ref{sec:deconv})};
    \node[punktchain, join, on chain=going below] (outlier) {
      Detect outliers in metadata (\ref{sec:outlier})};
    \node[punktchain, join, on chain=going below] (write) {
      Write preprocessed\\TOD to disk};
  \end{tikzpicture}
  \caption{\npipe\ preprocessing flow chart for optical \hfi\ bolometers.}
  \label{fig:preproc_hfi}
\end{figure*}

We present flow charts of the preprocessing steps for \lfi\ in Fig.~\ref{fig:preproc_lfi}, dark \hfi\ bolometers in 
Fig.~\ref{fig:preproc_hfi_dark}, and optical \hfi\ bolometers in Fig.~\ref{fig:preproc_hfi}.  The flow charts are annotated with the
appropriate section numbers.

\subsubsection{Repointing manoeuvres} \label{sec:repointing}

\npipe\ was designed to retain the data from the times between fixed pointings of the spin axis that were previously neglected. While these repointing manoeuvres are marked by changes in the rotation of the spacecraft, these changes alter the angular momentum as little as possible to conserve propellant.  This means that the scanning rate on the sky does not change and the effective beam is unaffected.  The DPC pipelines omitted these data for a number of reasons.  Firstly, they lack the repetitive overlapping scans that allow effective compression of the data onto ring maps.  Secondly, the pointing solution is less accurate.  Lastly, there was some concern that the thermal environment onboard the spacecraft was less stable during and immediately after thruster burns.  With the benefit of hindsight, it is now clear that none of these concerns is very serious.

\npipe\ deals with the lack of repetitive scans by using a global signal estimate (Sect.~\ref{sec:gestimate}), and only using ring-compressed data for calibration and other template fitting.  Our pointing solution for the repointings is more stable than the earlier versions, due to our use of a low-pass filter to suppress small-timescale fluctuations (Sect.~\ref{sec:pointing}).  The concern for thermal fluctuations is greatly alleviated by our application of horn-symmetric detector weights (Sect.~\ref{sec:symmetrization}), leading to effective cancellation of thermal common modes within polarized horns. Tests show that the repointing-period data are perfectly usable (Sect.~\ref{sec:repointings}).

\subsubsection{Global signal estimate} \label{sec:gestimate}

Many of the preprocessing modules require an estimate of the detected sky signal.  In particular, the \hfi-DPC modules make extensive use of a phase-binned\footnote{Each pointing period consists of tens of repeating scanning circles. Individual detector data for one period form a one-dimensional data set that can be indexed and binned according to the spin phase of the spacecraft.
} signal estimate, where data from a single pointing period and a single detector are used to estimate the periodic signal present in the data.  This binning is made possible by the repetitive \Planck\ scanning strategy, which repeats the same scanning circle 39--65~times at a fixed rate of one revolution per minute.  Binning the data according to the spacecraft spin phase produces an unbiased estimate of the signal, but the phase-binned data contain a significant amount of noise that is made scan-synchronous by the binning.  Moreover, the phase-binned estimate is not applicable to the 4-min repointing manoeuvres at the beginnings of each science scan.

Except in glitch removal, \npipe\ uses an alternative that we call the ``global signal estimate.''  We sample a full-mission, single-detector temperature map from a previous iteration of \npipe, together with polarization from the full set of detectors within each frequency channel.  Using the last run of the pipeline as input to preprocessing is justified by the fact that the relevant, high S/N modes of the maps converged early in the development process\footnote{In fact, the large-scale temperature signal has remained essentially unchanged since Public Release 1 in 2013 \citep{planck2013-p01} except for small overall calibration adjustment.  One telling measure of the robustness of the temperature map is the small component separation residual discussed in Sect.~\ref{sec:compsep}.}.  This map already contains the Solar dipole.  We further add an estimate of the orbital dipole based on the known spacecraft velocity.  The time-dependent orbital dipole is not included in the map.  For \hfi, this global estimate of the sky signal is then convolved with the estimated bolometer transfer function in order to produce the required estimate of the sky signal.

\subsubsection{ADC nonlinearity correction} \label{sec:adc}

Both \Planck\ instruments experience detectable levels of nonlinearity in their analogue-to-digital conversion (ADC). The level of nonlinearity in \lfi\ data is much less than in \hfi, and is straightforward to correct.  \npipe\ applies the \lfi\ DPC correction profiles as part of our preprocessing.  Details of the \lfi\ nonlinearity and the measured correction can be found in \cite{planck2016-l02}.

For \hfi\, nonlinearity in the digitization was found to be the most important systematic error affecting large-scale polarization (see discussion in \citealt{planck2016-l03} and \citealt{Delouis:2019bub}).  The first-order manifestation of the nonlinearity is apparent gain changes that correlate with changes in the level of signal in the detector.  Secondarily we observe distortions of the signal itself.

After the end of HFI observations, a two-year campaign was carried out during the \lfi\ extended mission to gather statistical information about the ADC effects.  The problem is not fully tractable, since the digitized data transmitted to Earth are downsampled onboard the spacecraft.  Nevertheless, the derived ADC correction profiles greatly improve the effective gain stability in the data, as well as consistency between the odd- and even-parity data sets split from the square-wave-modulated data \citep{planck2014-a08}.  Even after applying these ADC correction profiles, however, residual ADC nonlinearity (ADCNL) remains the primary systematic error in the large-scale \hfi\ polarization data \citep{planck2014-a10}.  \npipe\ begins with the same initial ADCNL-corrected TOD as \prthree, but uses a different model to measure and remove the residual ADCNL (see Sect.~2.4.2).

\subsubsection{Glitch removal} 
\label{sec:glitch_removal}

Cosmic-ray glitches in the \hfi\ data (see \citealt{planck2013-p03e}) are detected, fitted, and removed in \npipe\ using the same \texttt{despike} software that runs in the \hfi-DPC pipeline \cite{planck2011-1.7,planck2013-p03}.  However, \npipe\ does its own signal removal and runs \texttt{despike} in the ``dark'' mode developed for processing the dark bolometer data.

For the signal estimate, we reorder the signal by the satellite spin phase and fit a sixth-order polynomial in a sliding window of three bin widths at a time, keeping the polynomial fit in the centre bin.  Each bin is as wide as the average FWHM of the beam at that frequency.  The DPC signal estimate uses 1\parcm5 bins regardless of the optical beam width.

We also experimented with using the global signal estimate in place of the phase-binned signal estimate.  This has the advantage that the signal estimate does not include noise from the pointing period that is being analysed.  However, since the signal estimate and the resulting glitch residuals touch virtually every datum in \hfi, it was safer to use the local signal estimate instead of potentially biasing the entire data set with errors in the global signal estimate.

Using a phase-binned signal estimate introduces a noise bias in the glitch fits as the fitted glitch templates attempt to accommodate  some of the noise in the signal estimate.  This bias was evident in the noise estimates derived from half-ring difference maps, which were about a percent short of the full-data noise due to correlated noise cancellation \citep{planck2014-a14}.  Our FWHM-wide binning scheme and the extra smoothing from the polynomial fits limits the amount of noise in the signal estimate and the amount of half-ring bias.  Nevertheless, users of \npipe\ products are advised to be cognizant of the fact that the small-scale noise in the half-ring maps is correlated at a low level due to correlated errors in the glitch residuals.

We have raised the glitch detection threshold to reflect improvements in the signal estimate and to reduce the rate of false positives.  Our short-baseline destriping (see Sect.~\ref{sec:destriping}) is also better equipped to remove residual glitch tails than the ring offset model applied in DPC processing. The net effect of all the changes in glitch detection is to reduce the overall glitch detection rates to an average of 8\,\% in each detector, about half that of the DPC pipeline. While this may seem like a large difference, the energy distribution of the glitches is such that the affected glitch population is mostly right at the detection limit.  In Sect.~\ref{sec:comparison} we show that the resulting \npipe\ maps have lower noise than their 2018 counterparts, despite allowing more glitches through, and even after correcting for the added integration time.

\subsubsection{Removal of frequency spikes} \label{sec:spike}

The drive electronics for the 4-K cooler cause interference in the \hfi\ data at many harmonics.  We follow the DPC approach in fitting time-domain sine and cosine templates at the harmonic frequencies.  The DPC pipeline fitted for nine harmonics, namely, 10, 16.7, 20, 30, 40, 50, 60, 70, and 80\Hz. We have added a further six harmonics that are transiently detected in some of the bolometers, namely 16.0, 25.0, 43.4, 46.2, 47.6, and 56.8\Hz.

\begin{figure*}[htpb!]
  \includegraphics[width=0.48\linewidth]{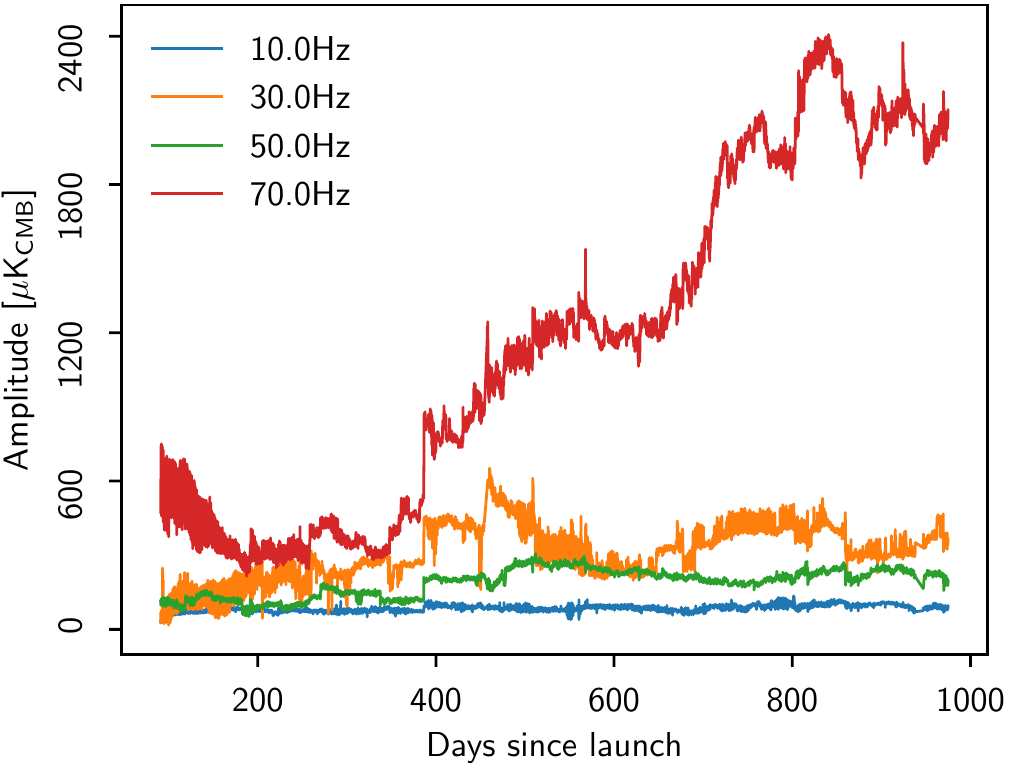}
  \includegraphics[width=0.48\linewidth]{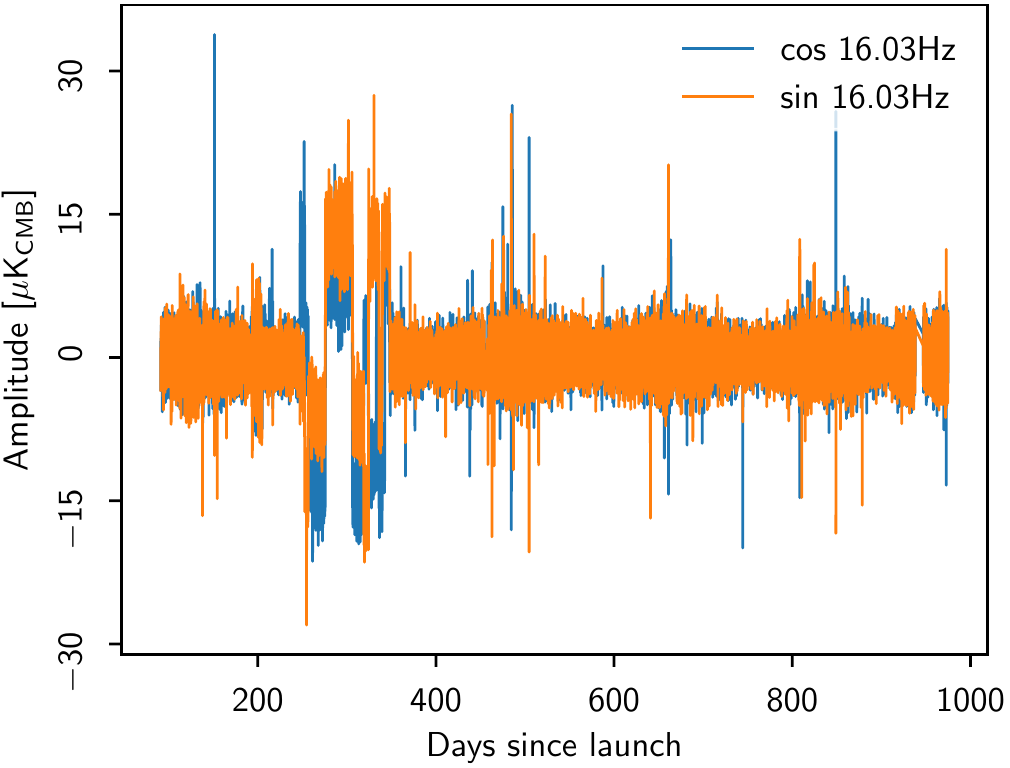}
  \caption{Interference from the drive electronics of the 4-K cooler in the \hfi\ detectors. \emph{Left:} Measured 4-K line amplitudes for the most intense lines in bolometer 143-3a.  \emph{Right:} New 16.03-Hz line template amplitudes for bolometer 100-1b.  For unknown reasons, the line is only detected intermittently around day 300.  For reference, the CMB temperature fluctuations typically fit within $\pm 250\muK$ and polarization fluctuations within $\pm 2\muK$; it is therefore clear that the cooler lines must be modelled and removed to high precision.  With the \Planck\ scan rate being $6\deg\,\mathrm{s}^{-1}$, the 10-Hz line corresponds to an angular scale of 36\arcm\ and multipole $\ell \approx 600$.  For higher-order harmonics of 10\,Hz, divide the angular scale and multiply the multipole with the appropriate factor.
}
\label{fig:4klines}
\end{figure*}

The line phases and amplitudes in the data are observed to change over timescales of minutes, although the source of the variation is not understood.  Fitting for the lines at the pointing period level leaves detectable residuals.  We have designated six harmonics as ``intense,'' meaning that their S/N allows for shorter fitting period ($5\,$min).  These intense lines are (in approximate order of intensity) at 70, 30, 10, 50, 16.7, and 20\Hz. The 5-min fitting period is a compromise between leaving line residuals and applying a notch filter to the instrumental noise. Figure~\ref{fig:4klines} shows the measured line amplitude for some of the lines in a particularly badly-affected bolometer, 143-3a.

Our pipeline uses the global signal estimate rather than a phase-binned signal estimate for spike removal.  The use of this approach avoids a resonant ring issue where the spin rate in some of the pointing periods causes the harmonic templates to be synchronous with the sky signal.

\begin{figure*}[htpb!]
  \center{
    \includegraphics[width=1.0\linewidth]{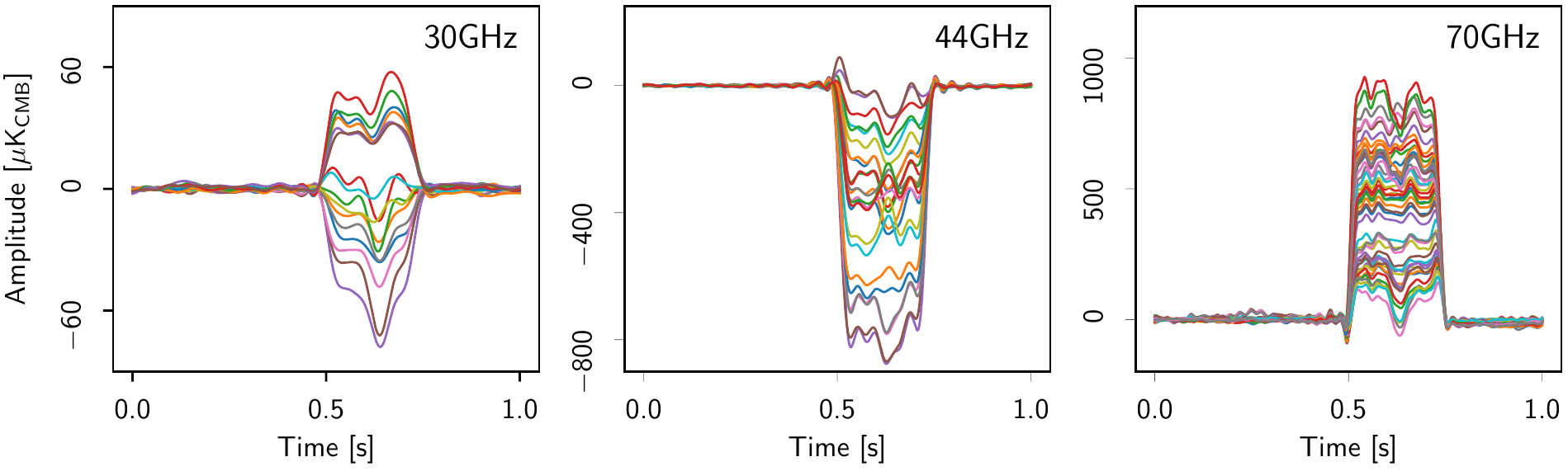}
  }
  \caption{
    \lfi\ spike templates built from the harmonic templates fitted to each of the 44~timelines.  Each of the coloured curves represents a different diode.  The CMB temperature fluctuations have a maximum amplitude of about $250\muKCMB$ and the polarization fluctuations about $2\muKCMB$.  The 1-Hz features would be seen at approximately $\ell=60$.
  }
  \label{fig:lfi_lines}
\end{figure*}

\lfi\ data show frequency spikes at 1\,Hz that are understood to originate in the housekeeping electronics.  To deal with these, we employ the same methods as with \hfi, constructing time-domain sine and cosine templates at $1\Hz$ and all of its harmonics (and their aliased counterparts) up to the Nyquist frequency.  With the \lfi\ sampling frequencies, this amounts to 47, 68, and 116 individual fitting frequencies at 30, 44, and 70\GHz\ respectively.  These template amplitudes were found to be stable (no model exists to predict their amplitude), so we averaged their amplitudes across the mission, then subtracted them from the TOD.  While the \lfi-DPC processing only corrected for the 1-Hz spikes in the most-affected 44-GHz data, \npipe\ fits and corrects for the electronic interference in all of the \lfi\ detectors.  Expanding the correction to all of the detectors is particularly important because the low-pass-filtered load signal does not cancel the interference as efficiently as an unfiltered signal.  We show co-added line templates for all \lfi\ diodes in Fig.~\ref{fig:lfi_lines}.

The \lfi-DPC approach to building the 1-Hz spike template differs from that of \npipe.  The DPC template was built by binning diode data into a 1-s-long template according to the phase within the 1-s period.   \npipe\ instead considers harmonic templates\footnote{``Harmonic'' in this context means sine and cosine templates needed to represent 1-s periodic signal, accounting for potential aliasing from downsampling the signal.  The frequencies of the templates are of the form $nf+kf_\mathrm{r}$, where $n$ is a non-negative integer, $f$ is the parasitic frequency ($1\Hz$), $k\in\{-1,0,1\}$ and $f_\mathrm r$ is the floating-point modulo of sampling frequency, $f_\mathrm{s}$, and $f$.} to model the parasitic oscillation.  In the limit of excellent S/N, the two approaches are equivalent, since the DPC template can be binned arbitrarily finely.  For shorter fitting periods and lower S/N, the harmonic templates are a more economical decomposition of the parasitic effect, and allow for a more robust fit.

\subsubsection{Sky--load differencing and thermal decorrelation} \label{sec:diff}

Both \lfi\ and \hfi\ perform decorrelation to reduce time-correlated ($1/f$) noise fluctuations in the data.  The \lfi\ $1/f$ templates are the reference-load timestreams, one for each diode.  The \hfi\ thermal templates come from the two dark bolometers.

Traditionally the \lfi\ sky--load differencing has relied on a scaling factor, $R$:
\begin{linenomath*}
\begin{equation}
  \label{eq:11}
  \mathrm{difference} = \mathrm{sky} - R \times \mathrm{load}.
\end{equation}
\end{linenomath*}
For \prthree, the \lfi\ DPC calculated the scaling factor for each operational day.  These least squares estimates of $R$ balance two opposing effects:
\begin{enumerate}
\item low-frequency $1/f$ fluctuations in the load timestreams are highly correlated with the fluctuations in the sky timestream;
\item high-frequency noise in the load timestream is essentially uncorrelated with the sky timestream.
\end{enumerate}
A value of $R$ calculated in this way suppresses the $1/f$ noise, while actually slightly \emph{increasing} the high-frequency noise in the differenced signal.

\begin{figure*}[htpb!]
  \includegraphics[width=0.48\linewidth]{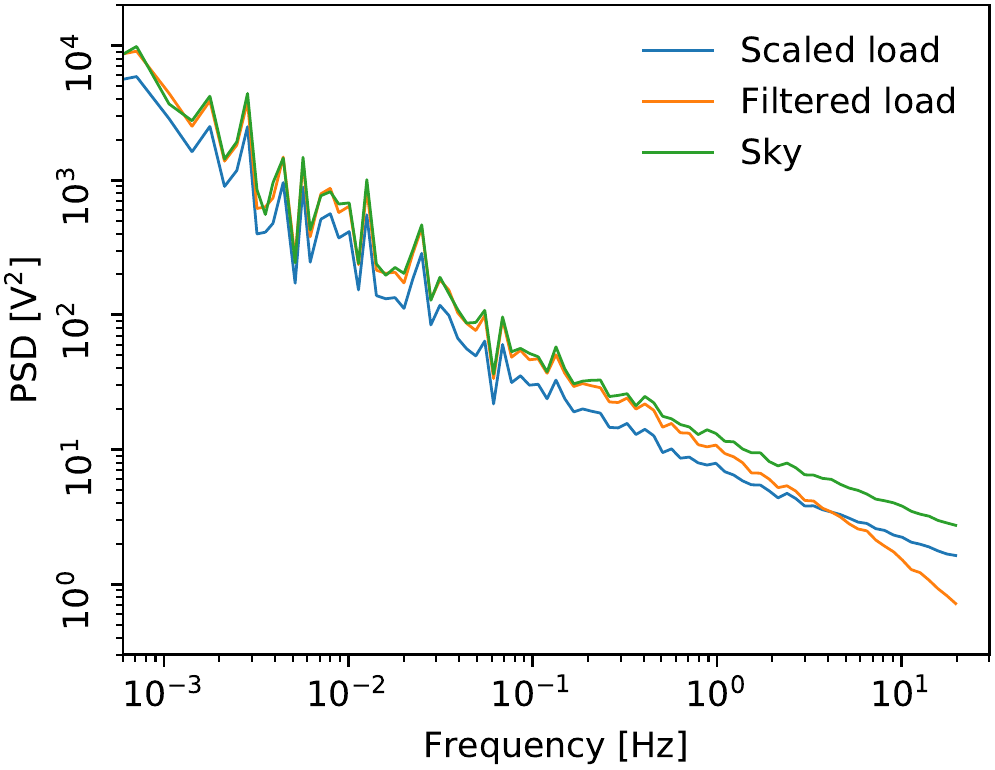}
  \includegraphics[width=0.48\linewidth]{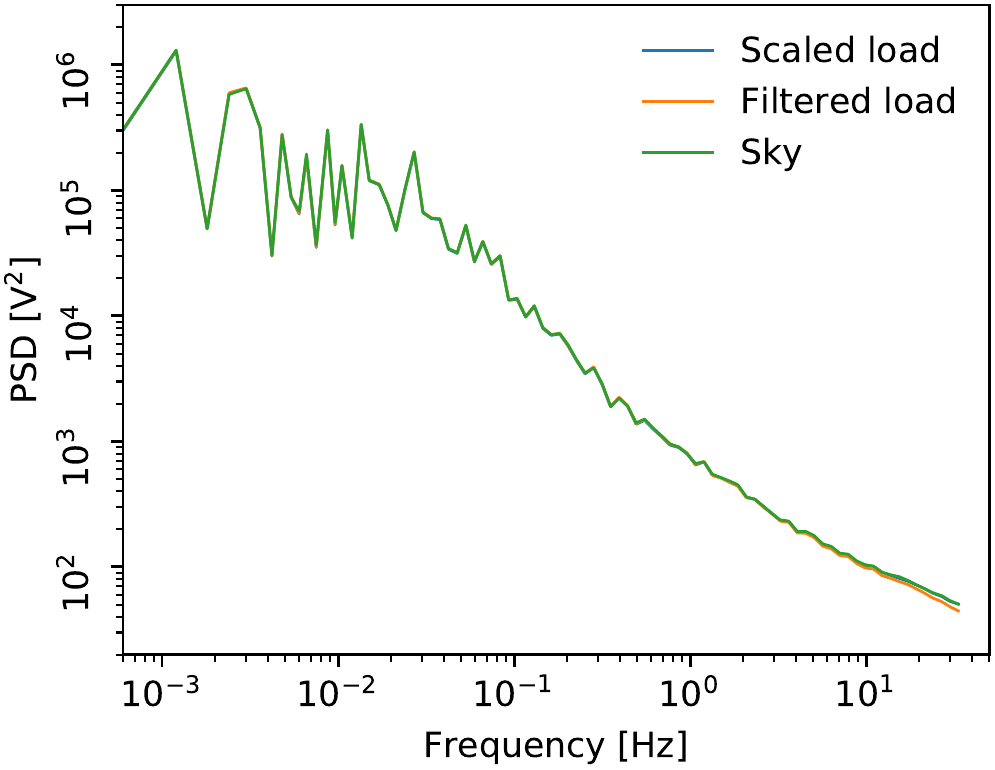}
  \\
  \includegraphics[width=0.48\linewidth]{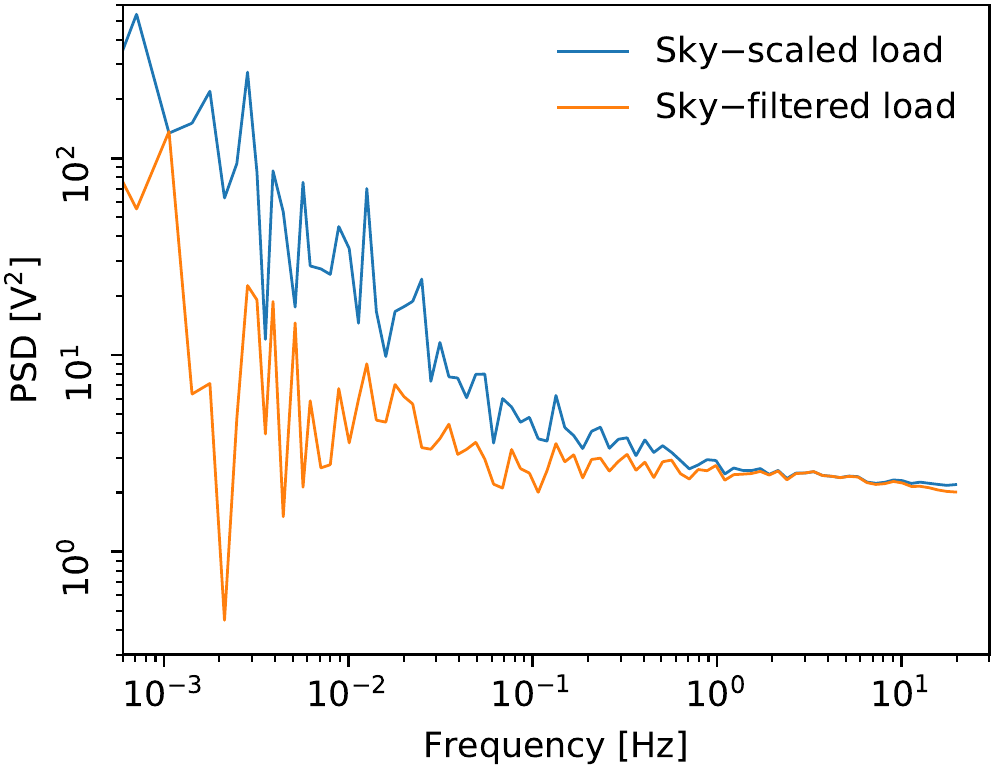}
  \includegraphics[width=0.48\linewidth]{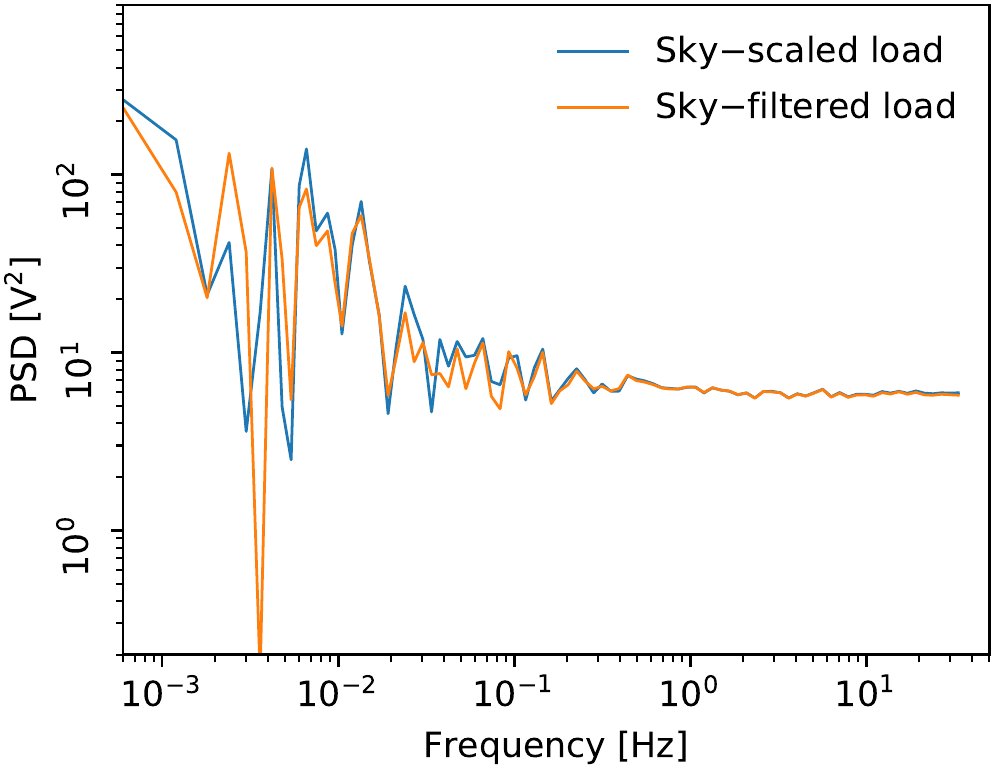}
  \caption{\npipe\ applies a low-pass filter to the \lfi\ load signal before decorrelation.  In \prthree, a single scaling coefficient was fitted each operational day.  We demonstrate the two approaches for one pointing period on two diodes: LFI2500 (left), with a relatively low amount of $1/f$ noise; and LFI1800 (right), which is dominated by $1/f$ at all frequencies. \emph{Top}: Power spectral densities of sky and best-fit load signals. \emph{Bottom}: Power spectral densities of sky$-$load differenced signals.  The PSDs were binned into 300 logarithmically-spaced bins for clarity. The scaled load template leaves more noise at both high and low frequencies than a filtered version, where the uncorrelated power is suppressed.  When the $1/f$ fluctuations dominate across the spectral domain, the scaling and filtering approaches achieve effectively the same result.}
\label{fig:differenced}
\end{figure*}

In \npipe\ we have attempted to suppress the injection of uncorrelated high frequency noise into the differenced signal.  In a somewhat simplified model, the sky and load each have an independent white noise component, $\sigma_\mathrm{sky}$ and $\sigma_\mathrm{load}$, and share a fully correlated $1/f$ component with power $P_{\rm corr}(f)$.  The corresponding power spectral densities (PSD) are
\begin{eqnarray}
  \label{eq:15}
  P_\mathrm{sky}(f) & = & \sigma_\mathrm{sky}^2
                          + R_0^2\, P_\mathrm{corr}(f), \\
  P_\mathrm{load}(f) & = & \sigma_\mathrm{load}^2 + P_\mathrm{corr}(f), \\
  P_\mathrm{corr}(f) & = & \sigma^2\left[
                           1 + \left(\frac{f}{f_\mathrm{knee}}\right)^\alpha
                           \right]\,,
\end{eqnarray}
where $R_0$ is a scaling factor that eliminates the correlated component in the sky--load difference in Eq.~(\ref{eq:11}).  Note that we are adopting a notation where $\sigma^2$ has PSD units.
Using an arbitrary scaling factor, the PSD of the differenced timestream gives
\begin{linenomath*}
\begin{equation}
  \label{eq:16}
  P_\mathrm{diff}(f) = \sigma_\mathrm{sky}^2 + R^2\sigma_\mathrm{load}^2
  + (R_0 - R)^2 P_\mathrm{corr}(f).
\end{equation}
\end{linenomath*}
We can thus derive a frequency-dependent scaling factor, $R(f)$, that minimizes the noise power in the differenced signal:
\begin{linenomath*}
\begin{equation}
  \label{eq:17}
  R(f) = R_0 \frac{
    P_\mathrm{corr}(f)
  }{
    P_\mathrm{corr}(f) + \sigma_\mathrm{load}^2
  }\,.
\end{equation}
\end{linenomath*}
In this simplified model, Eq.~(\ref{eq:17}) defines an optimal low-pass filter that minimizes noise power in the differenced signal.  We show examples of the sky and load PSDs and the differenced PSD in Fig.~\ref{fig:differenced}, and compare the low-pass-filtered approach to a direct scaling of the load signal.  The two examples show that when the uncorrelated instrumental noise is comparable to the correlated $1/f$ noise, low-pass filtering can substantially improve both the low and high frequency noise in the differenced signal.

As a refinement to the model, we consider that a radiometer comprises two diodes, each with sky and load, and the two diodes are also correlated with each other, complicating the situation.  We use Eq.~(\ref{eq:17}) to inform us of a parametric
representation of a low-pass filter with only three free parameters:
\begin{linenomath*}
\begin{equation}
  \label{eq:18}
  F^{\rm lp}(f;R,\sigma,\alpha) = R^2\frac{
    f^\alpha
  }{
    f^\alpha + \sigma^2
  }
\end{equation}
\end{linenomath*}
and we use nonlinear minimization to optimize the low-pass filter for both diodes and their relative weights that define the co-added radiometer signal, $\vec d$:
\begin{multline}
  \label{eq:19}
  \vec d =
  w\cdot\left(
    \vec s_0 - F^{\rm lp}(\vec l_0;R_0,\sigma_0,\alpha_0)
  \right) + \\ +(1-w)\cdot\left(
   \vec s_1 - F^{\rm lp}(\vec l_1;R_1,\sigma_1,\alpha_1)
  \right)\,.
\end{multline}
Here ``$\vec s$'' and ``$\vec l$'' represent the sky and the load, respectively, and indices ``0'' and ``1'' identify the two diodes in the radiometer.  The fit is performed while marginalizing over a sky-signal estimate.

The best-fit parameters ($w,R_0,\sigma_0,\alpha_0,R_1,\sigma_1,\alpha_1$) display a fair amount of scatter from pointing period to pointing period.  However, their sample medians are very stable, suggesting that the scatter is driven by noise.  We decided to run the main \lfi\ preprocessing loop twice.  The first iteration is performed to measure the best-fit, low-pass filters (Eq.~\ref{eq:18}), as well as to construct 1-Hz spike templates (Sect.~\ref{sec:spike}) for every pointing period.  Then the parameters are averaged across the entire mission and the processing is repeated, keeping the filter and spike parameters fixed.

\subsubsection{Jump correction} \label{sec:jump}

The \Planck\ data also contain ``jumps,'' which are sharp changes in the baseline level that persist (unlike spikes).  The \npipe\ jump-correction module uses a matched filter to detect baseline changes in the TOD.  We set the width of the filter to approximately 4~minutes of data. The width was chosen as a compromise between S/N (wider is better) and $1/f$ noise fluctuations (narrower is better).

We use the global signal estimate for signal removal, avoiding obvious issues that arise from building a signal estimate from data that contain jumps we are trying to detect.  Once a jump is detected, the filtered signal provides an estimate of the jump size, which is then corrected for.  Since both the exact position and size of the jump are subject to uncertainty, we flag one filter width (4~minutes) of data around the detected and corrected jump.  This approach differs from the implementation used in the \hfi-DPC pipeline, which identified pointing periods with jumps by examining the cumulative sum of samples.

The total number of detected jumps varies by detector, but on the average we identify, correct, and flag about a thousand jumps in each detector, once pointing periods with outlier statistics (see Sect.~\ref{sec:outlier}) have been discarded.

\subsubsection{Gap filling} 
\label{sec:fill}

As discussed in Sects.~\ref{sec:glitch_removal} and \ref{sec:jump}, the two instruments are subject to glitches and jumps of different origin, and both require removal of sections of data.  To facilitate the use of Fourier techniques (e.g., bolometer transfer function deconvolution, Sect.~\ref{sec:deconv}), the gaps need to be filled with a constrained realization. 
  
\npipe\ deals with gap filling using the same basic approach for all \Planck\ channels.  For \lfi, since it has relatively few gaps, the signal part of the constrained realization comes from the global signal estimate (Sect.~\ref{sec:gestimate}), while for \hfi\ it comes from the phase-binned signal estimate. The decision to use the (noisy) phase-binned signal estimate instead of the global signal estimate was made out of an abundance of caution:  even small but systematic errors in the signal estimate used to fill 10--20\,\% of the data before applying a filter could potentially lead to detectable bias.  The $1/f$ noise fluctuations are matched using a 5th order polynomial fit to the signal-subtracted data, and the gap-filling procedure is completed with a white-noise realization.

\subsubsection{Deconvolution of the bolometer time response} \label{sec:deconv}

The time response of the \hfi\ bolometers gives the relation between the optical signal incident on the bolometers and the output of the readout electronics, characterized by a gain, and a time shift, dependent on the temporal frequency of the incoming signal \citep{planck2014-a08}.  It is described by a linear complex transfer function in the frequency domain, called the ``time transfer function,'' which must be deconvolved from the data.  \npipe\ performs this deconvolution at the end of preprocessing.  The gap-filled data are Fourier transformed, divided by the measured complex transfer function, and transformed signal back into the real domain.  The complex transfer functions are identical to those used in \prthree, and described in \citet{planck2014-a08}.

\begin{figure}[hbtp!]
  \center{
    \includegraphics[width=0.48\textwidth]{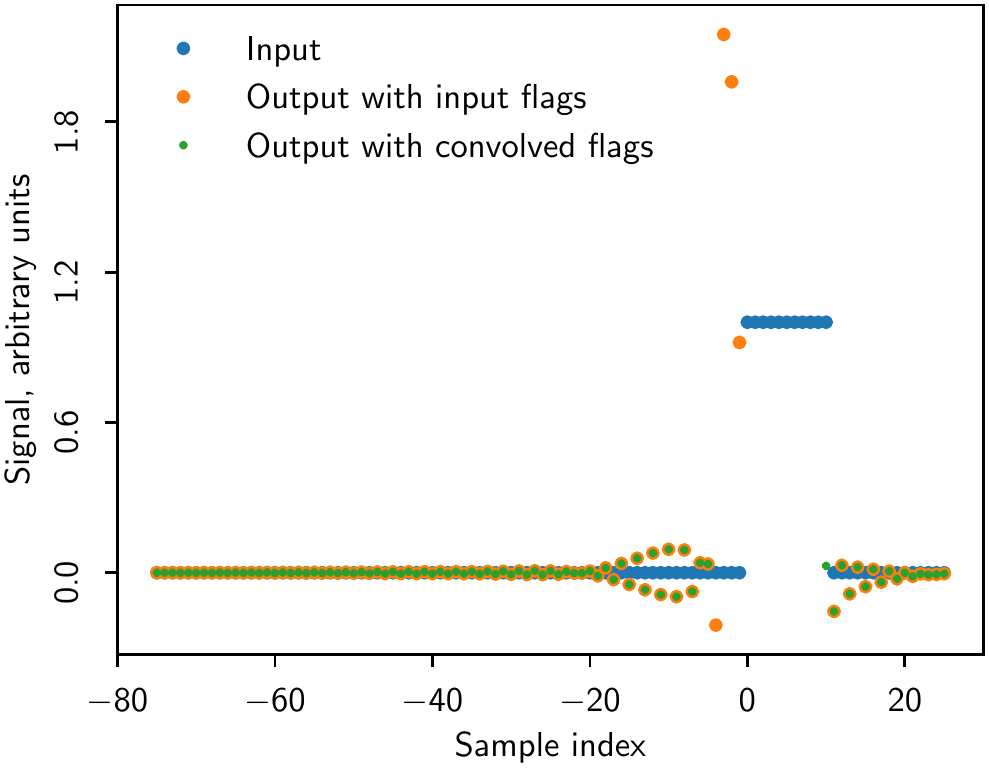}
  }
  \caption{Demonstration of the leakage of flagged signal onto science data. We created a zero (null) input signal (heavy blue line) with a gap between sample indices $0$ and $10$.  We then filled the gap with an assumed signal $= 1.0$, then deconvolved using the 100-2b transfer function.  Three samples immediately preceding the gap are significantly compromised by the signal in the gap.  Such samples are dropped in the \npipe\ analysis.
  }
  \label{fig:deconv_flags}
\end{figure}

We discovered that the application of the transfer function to glitch-removed and gap-filled data was picking up a considerable amount of power from the gap-filled samples and mixing it with the science data in the unflagged samples.  To suppress this effect, our deconvolution module transforms a delta function signal and measures a window over which more than 10\,\% of the deconvolved power can be attributed to the constrained realization present in the gaps.  These samples are flagged in addition to the usual quality flags.  The issue is illustrated in Fig.~\ref{fig:deconv_flags}.  The effect was not corrected for in \prthree, contributing a small amount of sky-synchronous noise (especially at $100\GHz$), responsible for several percent of the small-scale noise variance in the maps.

\begin{table*}[hbtb!]
  \begingroup
  \newdimen\tblskip \tblskip=5pt
  \caption{
    Discarded and quality-flagged data.
  }
  \label{tab:discarded}
  \nointerlineskip
  \vskip -3mm
  \footnotesize
  \setbox\tablebox=\vbox{
    \newdimen\digitwidth
    \setbox0=\hbox{\rm 0}
    \digitwidth=\wd0
    \catcode`*=\active
    \def*{\kern\digitwidth}
    \newdimen\signwidth
    \setbox0=\hbox{$-$}
    \signwidth=\wd0
    \catcode`!=\active
    \def!{\kern\signwidth}
 \halign{
      \hbox to 2.5cm{#\leaderfil}\tabskip 1em&
      \hfil#\hfil \tabskip 1em&
      \hfil#\hfil \tabskip 1em&
      \hfil#\hfil \tabskip 1em&
      \hfil#\hfil \tabskip 1em&
      \hfil#\hfil \tabskip 0pt\cr
    \noalign{\doubleline}
      \omit\hfil Frequency\hfil&
      \omit\hfil Outlier rings$^{\rm a}$\hfil&
      \omit\hfil Discarded rings$^{\rm b}$\hfil&
      \omit\hfil Discarded data$^{\rm c}$\hfil&Relative&Relative\cr
      \omit\hfil [GHz]\hfil&
      \omit\hfil  [\%] \hfil&
      \omit\hfil  [\%] \hfil&
      \omit\hfil  [\%] \hfil&
      \omit\hfil total samples$^{\rm d}$\hfil&
      \omit\hfil masked samples$^{\rm e}$\hfil\cr
      \noalign{\vskip 4pt\hrule\vskip 4pt}
      $*30$& $0.71$& $1.88$& *$2.62$& $1.119$& $1.095$\cr
      $*44$& $0.57$& $1.76$& *$2.19$& $1.120$& $1.098$\cr
      $*70$& $0.71$& $1.89$& *$2.84$& $1.114$& $1.091$\cr
      $100$& $0.97$& $2.17$& $20.00$& $1.167$& $1.163$\cr
      $143$& $1.16$& $2.36$& $21.53$& $1.167$& $1.160$\cr
      $217$& $0.95$& $2.16$& $20.98$& $1.221$& $1.209$\cr
      $353$& $0.90$& $2.08$& $21.78$& $1.179$& $1.174$\cr
      $545$& $0.89$& $2.06$& $15.93$& $1.068$& $1.066$\cr
      $857$& $0.93$& $2.11$& $17.76$& $1.019$& $1.015$\cr
      \noalign{\vskip 4pt\hrule\vskip 5pt}
    }
  }
\endPlancktablewide 
\tablenote {{\rm a}} Rings with anomalous glitch rate, noise level, gain, or numerous jumps.  A percentage of rings is not necessarily the same as a percentage of integration time, the ring length varies seasonally by almost a factor of 2.\par
\tablenote {{\rm b}} Total fraction of rings discarded because of outlier statistics and spacecraft events such as the sorption-cooler switchover, elevated spin-rate campaign and \lfi\ reboot on day $1085$ since launch.\par
\tablenote {{\rm c}} Total fraction of samples flagged because of pointing, jumps, cosmic-ray glitches, outlier pointing periods and spacecraft events.\par
\tablenote {{\rm d}} Ratio of total integration time in the full-frequency maps, in the sense (\npipe\ / 2018).\par
\tablenote {{\rm e}} Ratio of total integration time in the full-frequency maps, in the sense (\npipe\ / 2018), over the 50\,\% of the sky with the least foregrounds.  The distinction matters for \lfi, because the ninth survey --- included only in \npipe\ ---  lies almost entirely in the Galactic plane.\par
 \endgroup
\end{table*}

\subsubsection{Outlier pointing-period detection} \label{sec:outlier}

\npipe\ accumulates statistics of the data for each pointing period during the preprocessing.  Once all of the pointing periods are processed, the metadata are analysed for outlier periods, which are subsequently flagged completely.  We used glitch rate, apparent gain, and noise rms to test for outliers by looking for $>5 \sigma$ deviations in the statistics after removing a running average.  Rings with more than three such deviations were flagged as outliers.  Table~\ref{tab:discarded} shows the fraction of outlier pointing periods at each \Planck\ frequency.  It also shows the total fraction of pointing periods discarded after we also reject unusable data from the sorption-cooler switchover at the end of first year of operations, and data acquired during a $10$-day elevated spin-rate campaign.

\subsection{Global reprocessing} \label{sec:reprocessing}

\npipe\ applies a number of templates to correct the data using a generalized destriper, much like the \sroll\ method described in \cite{planck2014-a10} for \hfi\ large-scale polarization analysis.  Our data model can be cast in the standard destriping form:
\begin{linenomath*}
\begin{equation}
  \label{eq:destriper_model}
  \vec d = \mat P\,\vec m + \mat F\,\vec a + \vec n,
\end{equation}
\end{linenomath*}
where (multi-) detector time-ordered data $\vec d$ are a linear combination of the sought-for sky map $\vec m$, sampled into the time domain with the pointing matrix $\mat P$, and an arbitrary number of time-domain templates presented as columns of the sparse template matrix $\mat F$ and their amplitudes $\vec a$.   When $1/f$ noise offsets are included in $\mat F$, the noise term, $\vec n$, is approximated as white noise.

If we choose not to impose a prior on the distribution of the template amplitudes, $\vec a$, we can marginalize over the sky map, $\vec m$, and solve for the template amplitudes \citep{keihanen2004}:
\begin{linenomath*}
\begin{equation}
  \label{eq:13a}
  \vec a = \left(
    \mat F\trans \mat N\inv \mat Z \mat F
  \right)\inv \mat F\trans \mat N\inv \mat Z\,\vec d\,,
\end{equation}
\end{linenomath*}
where
\begin{linenomath*}
\begin{equation}
  \label{eq:13b}
  \mat Z = \tens{I} - \mat P \left(
    \mat P\trans \mat N^{-1} \mat P
  \right)^{-1} \mat P\trans \mat N^{-1}.
\end{equation}
\end{linenomath*}

Traditionally, $\mat F$ has comprised the baseline offset templates of a step function model of the $1/f$ noise\footnote{%
  We use the term $1/f$--noise for instrument noise that has a power spectral density (PSD) that can be approximated with a power law: $P(f) \propto f^\alpha$, where $f$ is the frequency and $\alpha$ is the slope of the spectrum.  The approximation is valid up to a so-called knee frequency, where $f^\alpha$ transitions into a high frequency noise plateau and other features.  See Fig.~\ref{fig:psds} for examples.}%
 -- one column for each baseline period (anywhere between 1\,s and 1\,hr) and every detector.  Here, we add additional columns to $\mat F$ for:
\begin{itemize}
\item linearized gain fluctuations (discussed in Sect.~\ref{sec:gain});
\item \hfi\ signal distortion (Sect.~\ref{sec:adcnl});
\item orbital dipole (Sect.~\ref{sec:diporb});
\item far-sidelobe pickup (Sect.~\ref{sec:fsl});
\item \hfi\ transfer-function residuals (Sect.~\ref{sec:tf});
\item 6-component zodiacal light model (Sect.~\ref{sec:zodi});
\item bandpass mismatch (Sect.~\ref{sec:bpm}); and 
\item foreground polarization (Sect.~\ref{sec:pestimates}).
\end{itemize}
Notice that there is no separate template for the Solar dipole; we consider it as an integral part of the sky, $\vec m$.  For this reason, \npipe\ calibration is not impacted by uncertainty in the Solar dipole estimate.

Polarization plays a key part in the degeneracy between the template amplitudes, $\vec a$, and the sky map $\vec m$ in Eq.~(\ref{eq:destriper_model}).  Especially at the CMB frequencies, where the foreground polarization is fainter, \Planck\ scanning strategy allows for apparent gain fluctuations that are compensated with scan-synchronous polarization residuals.  \lfi\ calibration in the 2015 and 2018 releases addressed this degeneracy by fixing the polarized part of $\vec m$ to a polarized sky model derived using \commander\ \citep{eriksen:2004}.  No such prior was used in the \hfi\ 2015 or 2018 processing.  We adopt the \lfi\ approach here and also use a polarization prior, but rather than requiring repeated iterations between \npipe\ and \commander, we sample smoothed \npipe\ 30-, 217-, and 353-GHz maps as polarization templates in $\mat F$.  When fitting for these polarization templates, the sky map $\vec m$ is unpolarized, effectively preventing the generation of spurious gain fluctuations that would be compensated by polarization errors.  This approach leads to significant improvement in the large-scale polarization systematics in the \hfi\ maps, particularly at $100\GHz$.   Once time-dependent gain fluctuations and other time-varying signals are corrected for, the last reprocessing iteration freezes the gains and then solves for bandpass mismatch over a polarized sky, without fitting for polarization templates.  This final fitting step also fits over a larger fraction of the sky than the previous iterations (Sect.~{\ref{sec:masks}}).

\npipe\ reprocessing solves for the template amplitudes, $\vec a$, in Eq.~(\ref{eq:destriper_model}) and produces a template-cleaned TOD:  $\vec d_\mathrm{clean} = \vec d - \mat F\,\vec a$.  Solving for the amplitudes is made nonlinear by the fact that the gain fluctuation template in $\mat F$ is derived from the sky map, $\vec m$.  Notionally, one could explicitly write out $\vec m$ in $\mat F$ in Eq.~(\ref{eq:destriper_model}) and solve for the amplitudes using nonlinear minimization techniques.  In practice, this is unfeasible owing to the dimensionality of the problem.  Instead, we iterate between solving for the amplitudes, $\vec a$, cleaning the TOD and updating $\vec m$ and $\mat F$.  The iterations also enable speeding up the template generation and fitting by compressing the data onto phase-binned \healpix\ rings,\footnote{Each pointing period is binned onto a subset of \healpix\ pixels at the destriping resolution.  We must exclude the repointing
    manoeuvres in doing so, since they break the repetitive scanning pattern we are leveraging.} 
while retaining the fully sampled data for cleaning and mapmaking.  \npipe\ reprocessing iterations follow these steps:

\begin{enumerate}
\item compress TOD onto \healpix\ rings;
\item solve for maximum likelihood template amplitudes $\vec a$ in Eq.~(\ref{eq:destriper_model}) \emph{using compressed data};
\item clean TOD using the amplitudes in $\vec a$; and
\item update the gain template by running \madam\ on the cleaned TOD.
\end{enumerate}
The recovered template amplitudes rapidly converge to zero as the cleaned data become consistent with the frequency estimate of the sky, $\vec m$.  In practice we only need three iterations for the residual errors to become negligible.

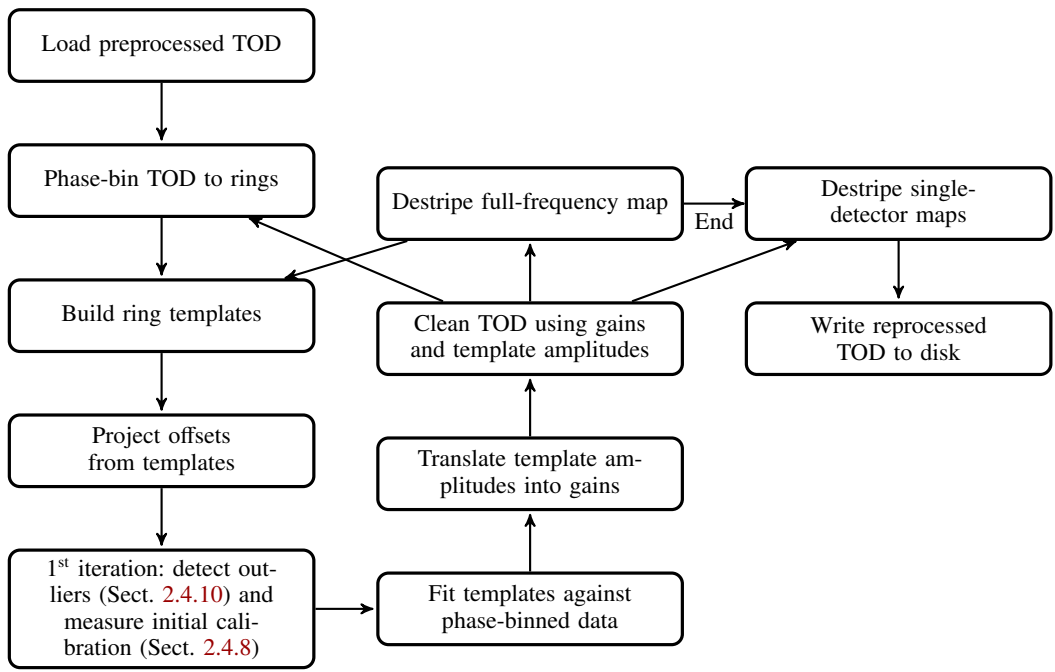
\begin{figure*}[htpb!]
  \centering
  \begin{tikzpicture}
    [node distance=.8cm, start chain=going below]
    \node[punktchain, join, on chain=going below] (load) {Load preprocessed TOD};
    \node[punktchain, join, on chain=going below] (compress) {Phase-bin TOD to rings};
    \node[punktchain, join, on chain=going below] (templates) {Build ring templates};
    \node[punktchain, join, on chain=going below] (project) {Project offsets from templates};
    \node[punktchain, join, on chain=going below] (first_iter) {\nth{1} iteration: detect outliers (Sect.~\ref{sec:rout}) and measure initial 
              calibration (Sect.~\ref{sec:incal})};
    \node[punktchain, join, on chain=going right] (destripe) {Fit templates against phase-binned data};
    \node[punktchain, join, on chain=going above] (gains) {Translate template amplitudes into gains};
    \node[punktchain, join, on chain=going above] (clean) {Clean TOD using gains and template amplitudes};
    \node[punktchain, join, on chain=going above] (full) {Destripe full-frequency map};
    \begin{scope}[start branch=out]
      \node[punktchain, on chain=going right] (single) {Destripe single-detector maps};
      \node[punktchain, join, on chain=going below] (write) {Write reprocessed TOD to disk};
    \end{scope}
    \draw[->, thick] (clean) -- (compress);
    \draw[->, thick] (clean) -- (single);
    \draw[->, thick] (full) -- (templates);
    \draw[->, thick] (full) -- node[below] {End} (single);
  \end{tikzpicture}
  \caption{\npipe\ reprocessing flow chart.  The first two columns contain the main iteration loop where the most recent frequency map is sampled into a new calibration template and the templates are iteratively fitted and subtracted from the TOD.  Reprocessing finishes with the third column where single-detector data are de-polarized using the full-frequency map.}
  \label{fig:reproc}
\end{figure*}

Figure~\ref{fig:reproc} shows the flow chart of \npipe\ reprocessing. We will now discuss how we build each of the time-domain templates in $\mat F$.

\subsubsection{Gain fluctuations} \label{sec:gain}

Calibrating the detector data is a fundamentally nonlinear problem.  We linearize the calibration problem by fitting a gain fluctuation template that measures deviations from the average gain for a given frequency.  The amplitude of the template, $\delta g$, provides us with a gain correction that we may apply to the detector in question:
\begin{linenomath*}
\begin{equation}
  \label{eq:calibration_factor}
  \vec d' = \frac{1}{1+\delta g}\vec d\,.
\end{equation}
\end{linenomath*}
Our gain template is the most recent iteration map downgraded to low resolution to improve S/N.  The sky map already includes the large Solar dipole.  Once the template is sampled, we add to it all of the time-domain templates that we have corrected for in the cleaned TOD.  That means adding the orbital dipole, far sidelobes, bandpass mismatch, \hfi\ transfer-function residuals, and zodiacal light.  It is true that a gain template sampled from the same sky map we are trying to estimate will have noise that is correlated with the detector TOD, but the effect is negligible because of the overwhelming number of samples accumulated into any sky pixel.

\begin{figure*}[htpb!]
  \includegraphics[width=\textwidth]{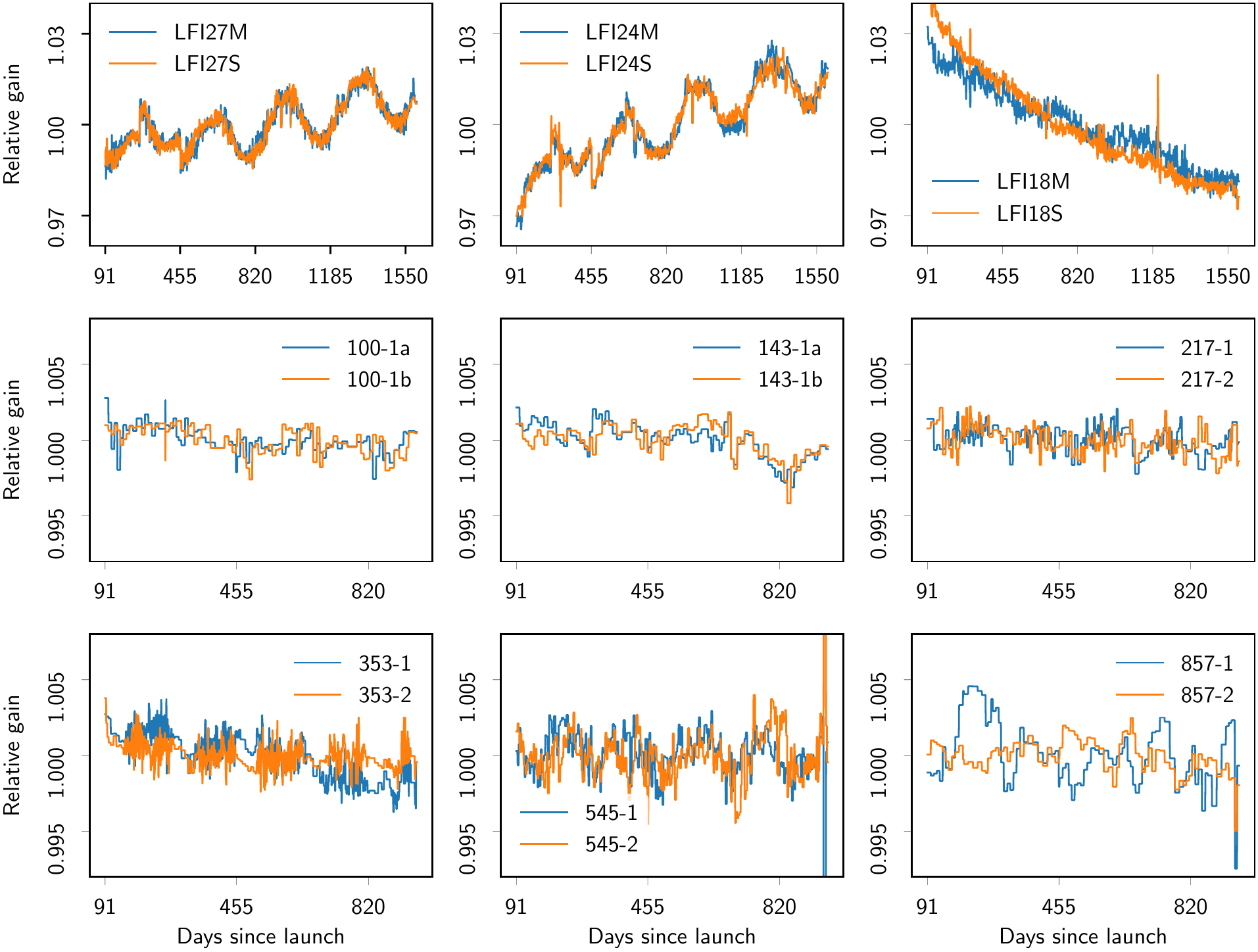}
  \caption{Sampling of measured gain fluctuations at all \Planck\ frequencies.  The top row shows 30, 44, and 70\GHz, from left to right, for representative detectors, with ``M'' and ``S'' referring to ``main'' and ``side'' (see \citealt{planck2016-l03}).  The second and third rows show \hfi\ frequencies, indicated in the detector names.  The vertical lines indicate complete observing years. Seasonal effects due to Solar distance are apparent in the 30- and 44-GHz gains, as well as discontinuities from switching to the redundant 20-K sorption-cooler system on day 455 and changing the transponder setting on day 270.
  }
  \label{fig:gains}
\end{figure*}

The gain template is split into disjoint chunks of time to trace changes in the gain.  Each gain step is long enough to amount to roughly equal S/N, except for the few steps that are broken by known discontinuities or run into a preset maximum step length of 300~pointing periods (about 10~days).  We have optimized the S/N threshold for each frequency to allow adequate tracking of gain fluctuations without excessively contributing to the map noise through noise-driven gain errors.  We show the measured gain fluctuations for a subset of \Planck\ detectors in Fig.~\ref{fig:gains}.  Unlike the \lfi-DPC gain solution, the \npipe\ gains are not filtered at all.  Details of the filtering required to render \lfi-DPC ring gains usable can be found in \cite{planck2014-a06}.

\subsubsection{Residual \hfi\ ADC nonlinearity} \label{sec:adcnl}

In previous \Planck\ releases, residual \hfi\  ADCNL was approximately addressed by correcting for apparent gain fluctuations \citep{planck2014-a09}.  Observed changes in the effective detector gain may be combined with the approximate level of the input signal to create an effective model of the residual ADC nonlinearity. This approach was shown to be effective in reducing large-scale polarization systematics in \cite{planck2016-l03}.  We refer to this model of the ADCNL causing multiplicative, gain-like fluctuations in the TOD as the ``linear gain model'' of ADCNL.

Despite the success of the linear gain model in suppressing the nonlinearity errors,  ADCNL was still the dominant source of large-scale polarization uncertainty in \cite{planck2016-l03}, prompting the use of inter-frequency cross-spectral methods for estimating the re-ionization optical depth from the maps \citep{planck2014-a10}.

\begin{figure}[htpb!]
  \center{
    \includegraphics[width=0.48\textwidth]{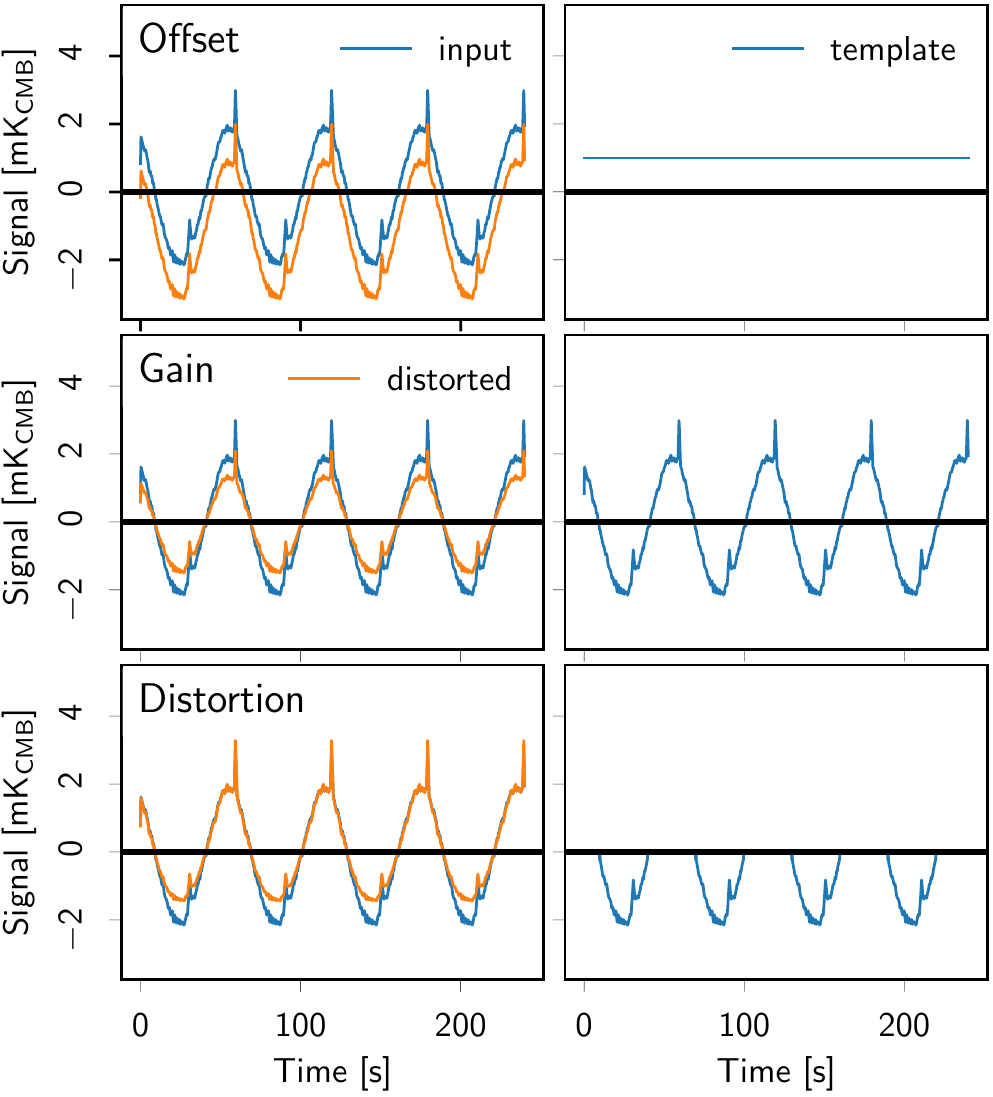}
  }
  \caption{Three degrees of signal distortion due to \hfi\ ADC nonlinearity (left panels) and the \npipe\ templates used to fit for them (right panels).  The blue curves give an interval of data from a specific bolometer.  The main variations seen come from the detector scanning across the dipole and the Galactic plane, including the ADCNL effects.  The orange curves show the steps taken to fit the distortions.
    \emph{Top:} the (harmless) signal offset is captured by destriping baselines (i.e., moving the orange curve upwards).
    \emph{Middle:} Linear gain fluctuations are addressed by calibration.
    \emph{Bottom:} Linear gain varies as a function of signal level and requires a separate gain template for the opposite ends of the
    signal range. This correction is unique to \npipe.
  }
  \label{fig:distortion}
\end{figure}

\begin{figure*}[htpb!]
  \includegraphics[width=\textwidth]{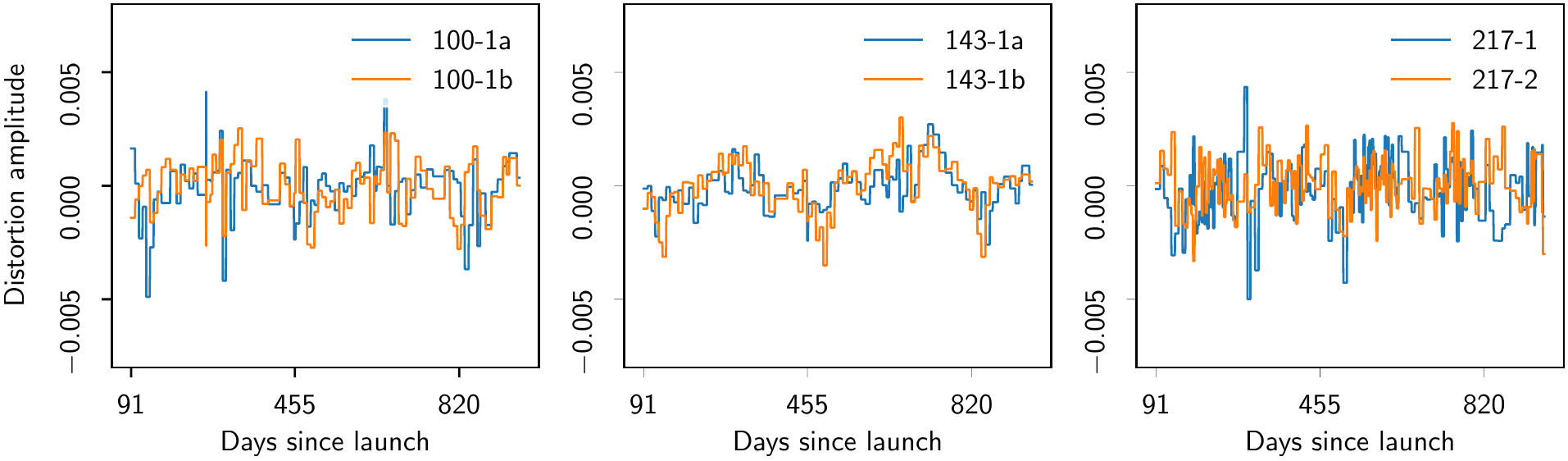}
  \caption{Sampling of measured signal distortions (cf.\ Sect.~\ref{sec:adcnl}) at 100, 143, and 217\GHz.  These are the only frequencies where the distortion template is fitted.  The distortion templates are fit at the same time steps as the gain templates in Fig.~\ref{fig:gains}.
  }
  \label{fig:distortions}
\end{figure*}

\npipe\ makes two important updates to the handling of residual ADCNL.  Firstly, we do not assume the apparent gain fluctuations to be exclusively sourced by the ADCNL.  In this way, our gain steps are free to address both apparent and real gain fluctuations.  Secondly, we add another calibration template that has all values above the median signal nulled.  The function of this second ``distortion'' template is to enable us to adapt to signal distortions that arise from a different effective gain being imposed at opposite ends of the signal range.  This is a natural extension of the linear gain model when the signal covers a wide dynamic range and the extremes of the signal probe different registers of the ADC.  We demonstrate the first three leading orders of signal error due to  ADCNL in Fig.~\ref{fig:distortion}.  Any fluctuation in the TOD that can be fitted with the gain and distortion templates, could also be fitted with two distortion templates having the opposite signal halves nulled.  Arranging the templates into gain- and distortion-only parts simplifies the interpretation.  The template corrections are applied to the TOD by multiplying the median-subtracted signal and half-signal (samples that have values below the median) with the appropriate factors.  We show examples of the fitted distortion template amplitudes in Fig.~\ref{fig:distortions}.

It is worth emphasizing that the \npipe\ model for the ADCNL is not a precise model that derives the correction from known inputs.  Such a model is, in fact, impossible to construct, due to the down-sampling that occurred onboard the spacecraft.  Instead, we simply fit piece-wise stationary time-domain templates to the mismatch between detectors, in an effort to reduce temperature-to-polarization leakage that would otherwise dominate the large-scale polarization.  The shape of the templates is well motivated by our understanding of the ADCNL.  The language of linear gain drifts is useful in describing the signal effects, but there is no way of disentangling actual changes in bolometer gain (expected to be of the order of $10^{-5}$) from apparent drifts sourced by ADCNL.

\subsubsection{Orbital dipole} \label{sec:diporb}

The motion of \Planck\ around the Sun causes a seasonal Doppler modulation of the CMB monopole.  The spacecraft position is known to an accuracy better than $1\,\mathrm{km}$ and the velocity better than $1\,\mathrm{cm}\,\mathrm{s}^{-1}$ \citep{godard2009}.  The CMB monopole is $2.72548\,\mathrm{K}\pm0.57\,\mathrm{mK}$ \citep{fixsen2009}.  Because both the spacecraft motion and the CMB monopole are extremely well known, this $\sim$300\muKCMB\ dipole\footnote{The range of the orbital dipole signature depends seasonally on the phase of the $7.5^\circ$ spin axis precession and the position of each detector on the focal plane.} signature is our best absolute calibrator.  We fit for an orbital dipole template in each detector TOD and find a per-frequency calibration coefficient that brings the average orbital dipole template to the expected level.  We do not require that all detectors observe an identical dipole, since far sidelobe effects alter the perceived amplitude at the individual horn level.  We do, however, subtract the same orbital dipole from both detectors in a horn. Far sidelobe effects are approximately corrected by convolving the orbital dipole template up to the second-order quadrupole (sometimes called the ``kinematic quadrupole'') with a \grasp\footnote{The \grasp\ software was de\-vel\-oped by TICRA (Co\-pen\-hagen, DK) for analysing gen\-eral re\-flec\-tor antennas (\url{http://www.ticra.it}).} estimate of the far sidelobes.

The orbital dipole fits provide for absolute calibration for all frequency channels except for 857\GHz, where \npipe\ uses the same planetary calibration as in \prthree.

\subsubsection{Far-sidelobe pickup} \label{sec:fsl}

We calculate timeline estimates of the far-sidelobe (FSL) pickup by convolving \prtwo\ full-frequency maps (with a Solar dipole added) with a model of the beam.  We considered updating the convolved timelines from \prthree\ or \npipe\ temperature maps, but the differences would have been negligible.

To perform this convolution, we used available \grasp\ estimates (which are monochromatic) of the $4\pi$ detector beams.  Full spectral treatment of the frequency-dependent FSL would have amounted to a percent-level correction to a $10^{-3}$ level systematic.  The estimates are only available for the 30- to 353-GHz channels, so we used 353-GHz beams as proxies for 545 and 857\GHz\ channels.  Except at the \submm\ frequencies, the FSL template is too faint and degenerate to include in the fitting.  We simply subtract it and correct the amplitude of the subtracted template as our estimate of the gain fluctuations improves.  For 545 and 857\GHz, the FSL template is successfully fitted with the individual SWBs, yielding roughly consistent best-fit amplitudes between the detectors.

\subsubsection{\hfi\ transfer-function residuals} \label{sec:tf}

The time response of the bolometers to sky signals is complicated. \hfi\ uses planet observations and stacked cosmic-ray glitches to infer the shape of the bolometer transfer function and deconvolve it.  These calibrators are excellent for probing the intermediate- and high-frequency components of the transfer function, but leave an uncertainty about the relative values of the low- and high-frequency components.  The most striking effect is a relative change in phase between the Solar dipole and signals on much smaller angular scales.  We construct templates that correspond to the real and imaginary parts of a binned residual transfer function, fit the templates, construct the transfer function from the fit amplitudes, and deconvolve it from the signal.  The real part of the transfer function corresponds to a small, scale-dependent change in calibration, and can be fit reliably only at 353\GHz\ and above.  The imaginary part introduces a phase shift between different scales, and is fit at all \hfi\ frequencies.

We quantize the transfer-function bins to contain one or more harmonics of the 16.7-mHz spin frequency, and exponentially increase the bin width according to
\begin{linenomath*}
\begin{equation}
  \label{eq:12}
  w_b = \mathrm{floor}(\mathrm e^{b / 4}) / 60\,\mathrm{s},
\end{equation}
\end{linenomath*}
where $b$ is the band index, $w_b$ is the width of the band in Hz, the first band begins at $0.5 / 60\,$s, and the ``floor'' function returns the largest integer value that is smaller than its argument.  The last band is extended all the way to the Nyquist frequency.  To avoid a degeneracy with overall calibration, we omit the first band of the real transfer-function templates.  We find that we
can reliably fit up to 16 bands at 353\GHz\ and above.

In the \hfi\ 2018 processing, the empirical transfer function was fitted in only four bins.  A pattern of ``zebra stripes'' was identified in the 353-GHz odd--even survey difference maps and shown to be caused by the inability of the coarse-grained transfer-function model to track the actual residual.  The odd--even survey differences for both methods are described further in Sect.~\ref{sss:internal_consistency}.

\begin{table}[htpb!]
  \begingroup
  \newdimen\tblskip \tblskip=5pt
  \caption{\hfi\ transfer-function bins.
  }
  \label{tab:bands}
  \nointerlineskip
  \vskip -3mm
  \footnotesize
  \setbox\tablebox=\vbox{
    \newdimen\digitwidth
    \setbox0=\hbox{\rm 0}
    \digitwidth=\wd0
    \catcode`*=\active
    \def*{\kern\digitwidth}
    \newdimen\signwidth
    \setbox0=\hbox{$-$}
    \signwidth=\wd0
    \catcode`!=\active
    \def!{\kern\signwidth}
    \halign{
      \hbox to 1.5cm{#\leaderfil}\tabskip 1em&
      \hfil#\hfil\tabskip 1em&
      \hfil#\hfil\tabskip 0pt\cr
      \noalign{\doubleline}
      \omit\hfil Band \hfil&
      \omit\hfil First harmonic$^{\rm a}$ \hfil&
      \omit\hfil $f_\mathrm{min}$ [Hz] \hfil\cr
      \noalign{\vskip 3pt\hrule\vskip 4pt}
      *1& **1& $0.025$\cr
      *2& **2& $0.042$\cr
      *3& **4& $0.075$\cr
      *4& **6& $0.108$\cr
      *5& **9& $0.158$\cr
      *6& *$13$& $0.225$\cr
      *7& *$18$& $0.308$\cr
      *8& *$25$& $0.425$\cr
      *9& *$34$& $0.575$\cr
      $10$& *$46$& $0.775$\cr
      $11$& *$61$& $1.025$\cr
      $12$& *$81$& $1.358$\cr
      $13$& $106$& $1.775$\cr
      $14$& $139$& $2.325$\cr
      $15$& $181$& $3.025$\cr
      $16$& $235$& $3.925$\cr
      \noalign{\vskip 3pt\hrule\vskip 5pt}
    }
  }
\endPlancktable                   
\tablenote {{\rm a}} of the satellite spin rate, $16.7\mHz$.\par
\endgroup
\end{table}

\subsubsection{Zodiacal emission} \label{sec:zodi}

Interplanetary dust in the Solar System forms a cloud through which we observe the CMB.  Thermal emission from zodiacal dust is visible at high \Planck\ frequencies \citep{planck2013-pip88}, and has characteristic seasonal variations as the satellite moves with respect to the cloud.  The seasonal variations are of concern for our template fitting, since they might bias the gains and noise offsets.  We use the six geometric components described by \cite{kelsall1998} to construct six independent templates that capture the seasonal variations.  To preserve the mean zodiacal emission in the signal, our templates are the differences of expected emission between the current spacecraft position and another position 180\deg\ along Earth's orbit.  The opposite position is found by inverting the heliocentric position vector.  Furthermore, to boost the S/N we use the same template amplitude for all detectors in a frequency channel.

Our recovered zodiacal template amplitudes are roughly in agreement with the 2018 results \citep{planck2016-l03}, but offer no new information, owing to the removal of the common mode and the emphasis on cleaning the data of seasonal effects.  We defer the actual zodiacal emission removal to the component-separation process, where multi-frequency information offers better means of cleaning the frequencies where the emission is faint.  The recovered template emissivities are listed in Table~\ref{tab:zodi}.

One may assess the success of the zodiacal emission removal by examining the difference between maps made from odd and even surveys.  Comparison of \npipe\ and \prthree\ odd--even survey differences (see Sect.~\ref{sss:internal_consistency}) suggests that \npipe\ processing achieves at least the same level of zodiacal residuals as the 2018 processing; however, it is hard to
quantify the performance because of other features in the same survey-difference maps.

\begin{table*}[htpb!]
  \begingroup
  \newdimen\tblskip \tblskip=5pt
  \caption{Zodiacal template emissivities.  The statistical uncertainties for these amplitudes are found to be negligible.  See \citet{planck2013-pip88} for an explanation of the components.  Note that \npipe\ only fits and corrects for the time-varying part of the Zodiacal emission model.
  }
  \label{tab:zodi}
  \nointerlineskip
  \vskip -3mm
  \footnotesize
  \setbox\tablebox=\vbox{
    \newdimen\digitwidth
    \setbox0=\hbox{\rm 0}
    \digitwidth=\wd0
    \catcode`*=\active
    \def*{\kern\digitwidth}
    \newdimen\signwidth
    \setbox0=\hbox{$-$}
    \signwidth=\wd0
    \catcode`!=\active
    \def!{\kern\signwidth}
    \halign{
      \hbox to 3.5cm{#\leaderfil}\tabskip 1em&
      \hfil#\hfil\tabskip 1em&
      \hfil#\hfil\tabskip 1em&
      \hfil#\hfil\tabskip 1em&
      \hfil#\hfil\tabskip 1em&
      \hfil#\hfil \tabskip 1em&
      \hfil$#$\hfil\tabskip 1em&
      \hfil$#$\hfil\tabskip 0pt\cr
      \noalign{\doubleline}
      Frequency [GHz]&$\lambda$ [$\mu$m]\hfil&Cloud\hfil&Band1\hfil&Band2\hfil&Band3\hfil&
           \omit\hfil Blob\hfil&\omit\hfil Ring\hfil\cr
      \noalign{\vskip 3pt\hrule\vskip 4pt}
      857& *350& 0.285*& 0.53*&   0.274*&  0.833& !0.317*&  !0.782*\cr
      545& *550& 0.182*& 0.973&  0.32**&   1.37*&  !0.214*&  !0.438*\cr
      353& *849& 0.105*& 0.954&  0.207*&  1.00*&  !0.0883& !0.449*\cr
      217& 1382& 0.0391& 0.823& 0.0999& 0.685& !0.0211& !0.332*\cr
      143& 2096& 0.0106& 0.777& 0.0693& 0.494& !0.0207& !0.243*\cr
      100& 2998& 0.0124& 0.68*&  0.0779& 0.523& -0.0327&-0.0282\cr
      \noalign{\vskip 3pt\hrule\vskip 5pt}
    }
  }
\endPlancktablewide
\endgroup
\end{table*}

\subsubsection{Bandpass mismatch} \label{sec:bpm}

\Planck\ detectors within a frequency band have differing bandpasses, causing them to see foreground components at different intensities.  In theory, one can use a frequency-dependent model of the foreground emission, coupled with a set of measured detector bandpasses, to predict the amount of bandpass mismatch and derive correction maps.  In practice, uncertainties in both the measured bandpasses and the sky model require fitting the corrections directly against the data.  For all continuum-emission foregrounds (synchrotron, free-free, anomalous microwave emission, and dust), we use a high-resolution \commander\ sky model to predict the frequency derivative of the foreground emission, pixel-by-pixel, at the centre of the frequency band.  The fitting amplitude directly gives us an estimate of the relative difference in centre frequency between the detectors.  We also consider narrow-line emission from CO regions, which are fitted with a separate template.  The bandpass templates are \healpix\ maps that we bilinearly interpolate to the detector sample positions.

For added flexibility, we choose to include some of the foreground components as separate templates.  This compromises any direct physical interpretation of the amplitude of the bandpass-mismatch template (i.e., difference in centre frequency), but leads to better agreement between full-frequency and single-detector maps.  For the 30 to 70\GHz\ channels, these extra templates are the frequency derivatives of the \commander\  anomalous microwave emission (AME) and dust frequency components, chosen for their similarity to maps of the full-frequency versus single-detector mismatch.  For 70\GHz\ we also fit for an HCN ($J\,{=}\,1\,{\rightarrow}\,0$, 88.63\GHz) emission-line template.  For 100 to 857\GHz, we use the frequency derivative of the free-free component instead of AME or dust.

\subsubsection{Initial calibration} \label{sec:incal}

In order to speed up convergence, we perform a linear regression of the raw TOD against the 20\,\% of the sky nearest to the dipole extrema.  The target is the 2015 \Planck\ dipole, which we use as a crude initial calibration.  The fit provides an initial, fixed, calibration coefficient for each detector; these are overridden by the orbital dipole fit and relative gain corrections during the first reprocessing iteration.

\subsubsection{Submillimetre processing}

\Planck\ 545- and 857-GHz frequency channels have unique processing requirements.  Foregrounds dominate across the sky, wide bandpasses cover a number of potential emission lines, and the detectors, although nominally polarization-independent, are in fact weakly sensitive to polarization, with polarization efficiencies measured to be around 6\,\%.  In contrast, the \hfi\ PSBs have values around 90\,\%.  For 545\GHz, we marginalize over the \npipe\ 30-, 217-, and 353-GHz polarization maps, smoothed to 60\arcm, to project out detector mismatch due to polarization. For 857\GHz, we found that this approach does not improve detector agreement, and chose to overlook the small polarization sensitivity in reprocessing.

\subsubsection{Outlier detection} \label{sec:rout}

\npipe\ reprocessing searches for outlier pointing periods in the data at two stages.  First, a submatrix of $\mat F$ is fitted against the TOD, one detector and one pointing period at a time.  Pointing periods showing anomalous fitting coefficients are flagged.  Second, the destriper estimates the initial residual for every detector and every pointing period by binning, resampling, and subtracting the input TOD.  Again, anomalous periods are flagged.  These steps result in fewer than ten pointing periods permanently flagged for each detector.  The total numbers of additional discarded pointing periods (across all detectors) at each frequency is 23, 25, 65, 9, 16, 18, 14, 0, and 11 for 30\GHz\ through 857\GHz, respectively.

\subsubsection{Horn symmetrization} 
\label{sec:symmetrization}

Each pair of polarization-orthogonal detectors that shares a horn forms a single polarization-sensitive unit, with nearly identical optical properties.  We enforce this by:
\begin{itemize}
\item applying a joint set of quality flags to both detectors and dropping the same outlier data for each;
\item using a common noise weight for both detectors;
\item using the same pointing;
\item using the same FSL template;
\item using the same zodiacal-light templates and template amplitudes.
\end{itemize}
The purpose of the symmetrization is to limit the temperature-to-polarization (``T-to-P'') leakage.  Since the detectors in a horn share the same optical feed, the difference in samples collected simultaneously from the two is free from most of the optical mismatch that would otherwise cause leakage.  In fact, our symmetrization scheme has the same effect as differencing the TOD from the two detectors, then using the resulting difference to solve for sky polarization.

One potential source of T-to-P leakage is the sub-pixel structure that includes sharp foreground features as well as the dipole gradient.  Samples drawn from various corners of the pixel may indicate different signal levels, and if those observations are associated with different observing angles, the difference can be interpreted as polarization.  To the extent that the polarized pairs of detectors are optically aligned, this source of leakage is blocked by our symmetrization procedures.

We have verified that the symmetrization reduces T-to-P leakage in the Galactic plane and in the vicinity of compact sources, where
beam effects are pronounced.  While the effect is small, it was nevertheless considered strong enough to merit the increase in noise from symmetrizing the flags and using slightly less optimal noise weights for the detectors.

\subsubsection{Last iteration and processing masks}
\label{sec:masks}

Each iteration of the reprocessing uses a processing mask to avoid issues arising from sub-pixel structure, variable unresolved (``point'') sources, and small incompatibilities in our bandpass-mismatch templates.  The same mask is not necessarily optimal for all of the templates.  In particular, the gains require only the cleanest and most reliable parts of the sky to be considered, while the bandpass corrections can greatly benefit from including as much as possible of the intense Galactic emission.  To best serve both applications, we mask more of the sky while we iterate over the calibration, then freeze the gain solution and run the last iteration with more of the Galaxy exposed.

Point sources, especially variable radio sources, and other steep signal gradients on the sky may cause an error in the fitted template amplitudes. We use processing masks to avoid regions of the sky where the signal model is inaccurate.

\npipe\ reprocessing uses apodized processing masks.  This means that it is possible to down-weight regions of the sky
without discarding them completely when solving for the template amplitudes.  This approach is particularly useful for measuring the gain fluctuations both at the dipole extrema and over intense Galactic emission.  The use of apodized processing masks is a new development in TOD processing, but has been used extensively in map domain analysis.

We build our apodized masks from an earlier iteration of \npipe\ temperature maps.  First, we subtract \prtwo\ dipole and CMB map from each frequency map, and then find the scaling coefficients that suppress the foreground amplitude in each pixel below a fixed threshold.  These thresholds are not rigorously optimized, but rather are chosen to visually suppress regions of the sky where single-detector maps show notable disagreement with the full-frequency map.  Further adjustment is performed when the gain solutions show sharp and recurring seasonal trends.  Finally, point sources are added to the masks from the second \Planck\ compact source catalogue \citep{planck2014-a35}.

\begin{figure*}[htpb!]
  \center{
    \includegraphics[width=0.8\linewidth]{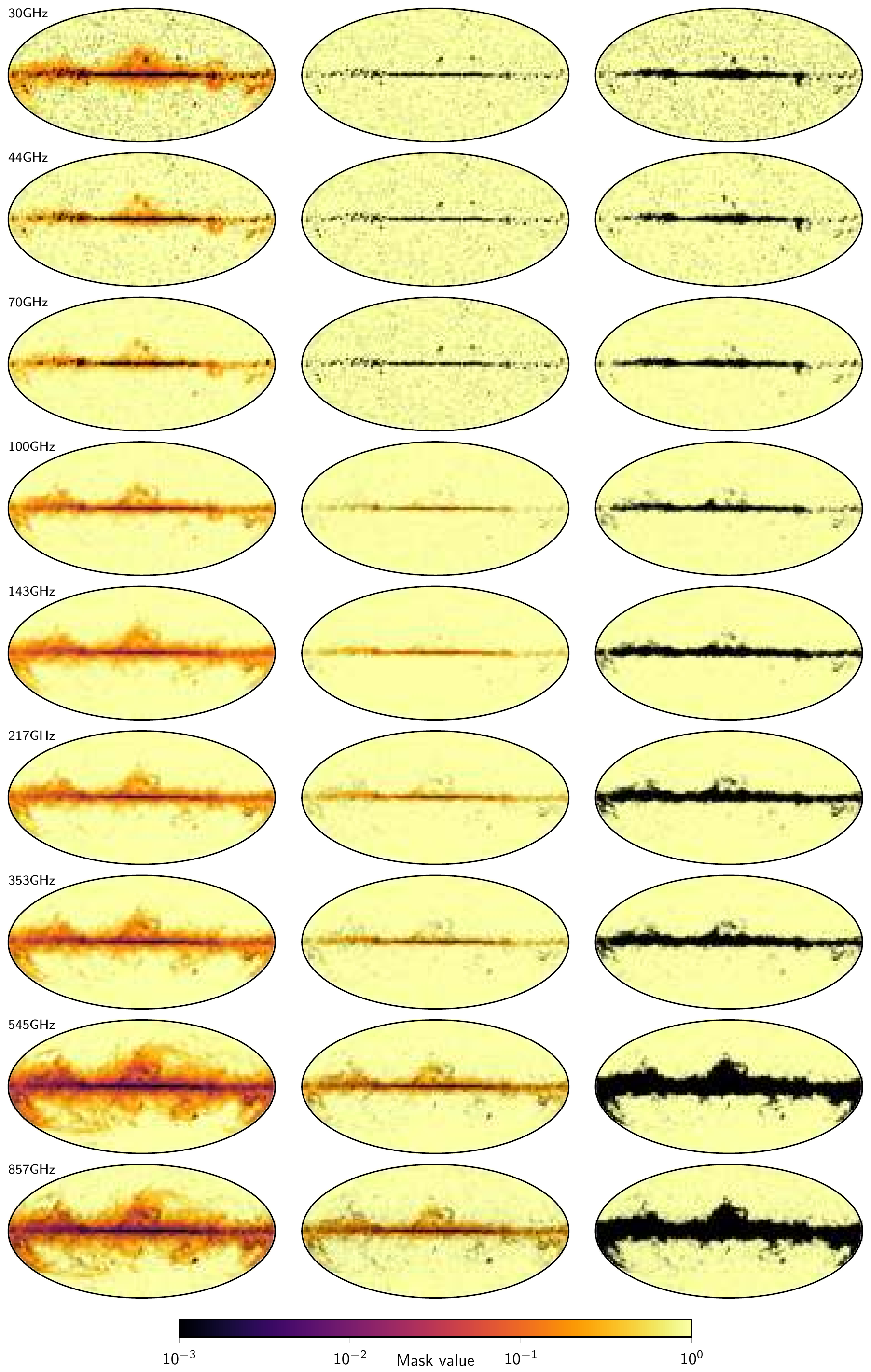}
  }
  \caption{The sharp and apodized processing masks used in \npipe\ processing. The three columns are for the calibration (left), bandpass correction (middle), and destriping (right) masks.  The destriping masks are explicitly binary because \madam\ (which we use for this step) does not support floating-point masks.
  }
  \label{fig:masks}
\end{figure*}

We present our processing masks in Fig.~\ref{fig:masks}.  The final short-baseline destriping uses binary masks, but we are able to use a large sky fraction due to the time-domain bandpass-mismatch corrections.

\subsubsection{Polarization estimates} \label{sec:pestimates}

The full polarized solution of the template amplitudes is degenerate at CMB frequencies and subject to large noise uncertainty.  Without external constraints, the template fitting converges to a self-consistent but sub-optimal solution, with large-scale polarization power leaking in from the dipole (see Sect.~\ref{sss:internal_consistency}).  These issues can be remedied by performing the template fits on a temperature-only sky.  Polarization cannot be overlooked without biasing the solution, but it is possible to marginalize over the sky polarization by adding additional templates that model the foreground polarization to the template matrix.  To this end, the 217-GHz reprocessing fits for 30- and 353-GHz polarization templates smoothed to 1\deg, while 44--143\GHz\ reprocessing fits for 30, 217, and 353\GHz.  For each template, we only allow for a single template amplitude across all detectors.  This improves the S/N at the cost of ignoring the polarized bandpass mismatch.  We do not explicitly fit for bandpass mismatch in polarization, but the sky model does predict a percent level correction based on the central frequencies fitted with temperature which is included when applying the bandpass correction.

To construct the time-domain templates, we smooth each foreground frequency map with a 1\deg-FWHM Gaussian beam and sample according to the detector polarization parameters and orientation.

\subsubsection{Degeneracies}

Including sampled sky maps as templates in Eq.~(\ref{eq:destriper_model}) creates a degeneracy.  For instance, the same bandpass-mismatch template is sampled for every detector, and we can shift power between the template amplitudes in $\vec a$ and the sky map, $\vec m$.  Such degeneracies prevents direct solution of Eq.~(\ref{eq:13a}); instead, we perform a conjugate
gradient (CG) iteration for $\vec a$ that minimizes the residual in
\begin{linenomath*}
\begin{equation}
  \label{eq:14}
  \left(
    \mat F\trans \mat N\inv \mat Z \mat F
  \right) \vec a =
  \mat F\trans \mat N\inv \mat Z\,\vec d.
\end{equation}
\end{linenomath*}
Once the CG iteration converges, we adjust the template amplitudes, $\vec a$, as follows:
\begin{itemize}
\item the bandpass-mismatch correction for each frequency must average to zero; 
\item relative gain corrections for each frequency must average to zero; and 
\item the average of all noise offsets must equal zero.
\end{itemize}
These adjusted template amplitudes are then used to correct the TOD.  An alternative (but equivalent) approach would have been to modify the destriping equation itself to include these priors. 

The matter of degenerate modes is a common topic in destriping.  They arise due to the projection matrix, $\mat Z$, which removes the sky-synchronous part of the signal.  Even in the case of traditional destriping that only fits for noise offsets, an overall offset of the solved map is set by the initial guess rather than being solved from the data.  Beyond that, any sky-synchronous mode in the templates is invisible to the destriper.

The unadjusted average of all of the gain fluctuations tends to be large (of order unity rather than zero) and matched by an opposite
average value in the orbital dipole template.  This simply reflects a degeneracy between the two templates, because the gain template includes the orbital dipole.  To deal with this, we subtract the frequency average of the gain fluctuations and adjust the orbital dipole amplitudes accordingly to keep the level of total orbital dipole unchanged.

There is a degeneracy between the templates used to model foreground polarization and the templates used to capture bandpass mismatch.  This degeneracy requires no active mitigation due to the way bandpass-mismatch correction is only applied in the final template fitting step that does not involve polarization templates.  For more discussion, see Appendix~\ref{app:degeneracy}.

\subsubsection{Polarization parameters} 
\label{sec:ppar}

During the \npipe\ development a number of diagnostics were produced to assess the performance of the template corrections.  One very useful diagnostic is a set of polarization-corrected, single-detector maps reflecting the final template amplitudes.  Any mismatch between the full-frequency temperature map and these single detector maps can be considered a residual and an indication of a deficiency in the data model or the fitting procedure.  Once other sources of residuals have been addressed, the remainder, especially for some of the 353-GHz bolometers, exhibits a residual signal that closely resembles maps of polarized sky
emission.  We suggest that these residuals arise from small errors in the polarization efficiencies or angles that we adopted from the DPCs.

We derive two time-domain polarization templates by scanning the full-frequency polarization map, one with the usual $Q$ and $U$ weights, and another using the polarization angle derivative of the $Q$ and $U$ weights.  We then bin these two templates using the same intensity weights as with the polarization-corrected, single-detector TOD.  Fitting the resulting two polarization templates against the single-detector residual map allows us to measure a correction to the polarization efficiency and polarization angle that we incorporate into a refined version of the instrument model.  The template fits are limited by other residuals in the single-detector maps, so their uncertainties are systematic, not statistical, and are difficult to assess without simulations.

We have been able to reduce the single-detector/full-frequency residuals only from 100 to 353\GHz\ (as shown in Table~\ref{table:polparams}).  The \lfi\ residual maps do not show a residual consistent with an error in the polarization parameters, either due to lack of such error or due to higher noise content.

One might argue that the binned full-mission maps are not appropriate for fitting the polarization parameters because they do not capture the fact that the detectors are rotated with respect to the pixels between different surveys.  We tried fitting for the polarized templates during destriping (similar to other templates), but found that the change in position angle over most of the sky is insufficient and the template fits are unreliable.  This is why the polarization parameter corrections are based on map domain fits.

\begin{table}[htbp!] 
  \begingroup
  \newdimen\tblskip \tblskip=5pt
  \caption{\npipe\ polarization efficiencies and angles.  We report corrections with respect to the ground-measured values used in \prthree.  In \cite{rosset2010} the statistical uncertainties for polarization efficiency are estimated to be 0.1--0.6\,\% for PSBs and SWBs operating between 100 and 353\GHz.  Polarization angle systematic uncertainties for the PSBs were estimated by \citet{rosset2010} to be 0\pdeg9 (not including a 0\pdeg3 common error); uncertainties for the SWBs were estimated to be up to 5\pdeg5.  For NPIPE, statistical uncertainties in the PSB polarization angles are small, while those for the SWBs are up to 2\pdeg3.  The \npipe\ uncertainties include statistical and systematic errors estimated using 60 simulations.
  }
  \label{table:polparams}
  \nointerlineskip
  \vskip -3mm
  \footnotesize
  \setbox\tablebox=\vbox{
    \newdimen\digitwidth
    \setbox0=\hbox{\rm 0}
    \digitwidth=\wd0
    \catcode`*=\active
    \def*{\kern\digitwidth}
    \newdimen\signwidth
    \setbox0=\hbox{$-$}
    \signwidth=\wd0
    \catcode`!=\active
    \def!{\kern\signwidth}
    \halign{
      \hbox to 2.2cm{#\leaderfil}\tabskip 1em&
      \hfil#\hfil \tabskip 1em&
      \hfil#\hfil \tabskip 1em&
      \hfil#\hfil \tabskip 1em&
      \hfil#\hfil \tabskip 0pt\cr
      \noalign{\doubleline}
      \noalign{\vskip -2pt}
      \omit\hfil Bolometer\hfil&
      \omit\hfil Pol. efficiency\hfil&
      \omit\hfil $\Delta$\hfil&
      \omit\hfil Pol. angle\hfil&
      \omit\hfil $\Delta$\hfil\cr
      \omit\hfil \hfil&
      \omit\hfil [\%]\hfil&
      \omit\hfil [\%]\hfil&
      \omit\hfil [deg]\hfil&
      \omit\hfil [deg]\hfil\cr
      \noalign{\vskip 3pt\hrule\vskip 4pt}
      100-1a& $94.35 \pm 1.03$& $-0.36$& *$23.32 \pm 0.28$ & $+0.22$\cr
      100-1b& $96.98 \pm 0.87$& $+2.67$& $111.93 \pm 0.24$& $+0.13$\cr
      100-2a& $96.03 \pm 0.45$& $-0.15$& *$44.26 \pm 0.18$& $-0.84$\cr
      100-2b& $89.74 \pm 0.45$& $-0.50$& $134.57 \pm 0.21$& $+0.27$\cr
      100-3a& $89.56 \pm 0.71$& $-0.53$& $179.33 \pm 0.14$& $-0.87$\cr
      100-3b& $93.23 \pm 1.06$& $-0.22$& *$90.29 \pm 0.20$& $+0.19$\cr
      100-4a& $93.88 \pm 0.63$& $-1.83$& $156.96 \pm 0.16$& $+0.46$\cr
      100-4b& $93.02 \pm 0.69$& $+0.75$& *$67.95 \pm 0.17$& $-0.06$\cr
      \noalign{\vskip 3pt\hrule\vskip 4pt}
      143-1a& $85.55 \pm 0.20$& $+2.32$& *$45.74 \pm 0.09$& $+0.24$\cr
      143-1b& $86.43 \pm 0.34$& $+1.84$& $137.66 \pm 0.12$& $-0.04$\cr
      143-2a& $85.35 \pm 0.28$& $-2.09$& *$45.19 \pm 0.11$& $-0.21$\cr
      143-2b& $88.97 \pm 0.20$& $-0.21$& $135.49 \pm 0.09$& $+0.29$\cr
      143-3a& $86.03 \pm 0.45$& $+2.13$& **$0.08 \pm 0.09$& $+0.38$\cr
      143-3b& $88.69 \pm 0.35$& $-1.25$& *$92.74 \pm 0.09$& $-0.36$\cr
      143-4a& $91.75 \pm 0.40$& $-1.36$& **$0.61 \pm 0.08$& $-0.29$\cr
      143-4b& $93.72 \pm 0.39$& $+0.88$& *$89.61 \pm 0.09$& $+0.31$\cr
      143-5&  *$7.52 \pm 0.30$& $+1.25$& *$68.95 \pm 1.03$& $+3.25$\cr
      143-6&  *$3.87 \pm 0.38$& $-0.30$& *$69.71 \pm 2.06$& $-0.89$\cr
      143-7&  *$3.06 \pm 0.32$& $+1.74$& *$95.24 \pm 2.96$& $-7.56$\cr
      \noalign{\vskip 3pt\hrule\vskip 4pt}
      217-1&  *$2.34 \pm 0.20$& $-1.50$& *$97.11 \pm 2.18$& $-1.29$\cr
      217-2&  *$3.33 \pm 0.20$& $+1.34$& *$95.87 \pm 1.57$&$+13.37$\cr
      217-3&  *$1.23 \pm 0.15$& $-2.72$& $163.02 \pm 2.27$& $-7.88$\cr
      217-4&  *$3.78 \pm 0.14$& $-0.61$& $121.20 \pm 1.56$& $+1.20$\cr
      217-5a& $96.14 \pm 0.12$& $+1.13$& *$47.08 \pm 0.04$& $+0.08$\cr
      217-5b& $97.35 \pm 0.15$& $+2.15$& $135.88 \pm 0.08$& $-0.22$\cr
      217-6a& $90.12 \pm 0.09$& $-4.82$& *$46.01 \pm 0.06$& $-0.09$\cr
      217-6b& $93.91 \pm 0.09$& $-1.48$& $135.97 \pm 0.04$& $+0.07$\cr
      217-7a& $94.09 \pm 0.15$& $+0.04$& $179.43 \pm 0.02$& $-0.17$\cr
      217-7b& $95.15 \pm 0.14$& $+1.48$& *$90.53 \pm 0.03$& $+0.03$\cr
      217-8a& $96.76 \pm 0.20$& $+2.55$& **$1.18 \pm 0.04$& $+0.88$\cr
      217-8b& $95.80 \pm 0.20$& $+1.68$& *$89.90 \pm 0.05$& $-0.80$\cr
      \noalign{\vskip 3pt\hrule\vskip 4pt}
      353-1&  *$0.23 \pm 0.13$& $-2.97$& $101.74 \pm 18.81$& $-1.36$\cr
      353-2&  *$7.49 \pm 0.05$& $+2.78$& $120.68 \pm 0.63$& $+6.08$\cr
      353-3a& $89.93 \pm 0.09$& $+0.95$& *$45.70 \pm 0.06$& $+0.20$\cr
      353-3b& $90.90 \pm 0.11$& $-1.17$& $133.64 \pm 0.07$& $-0.26$\cr
      353-4a& $87.35 \pm 0.13$& $+0.14$& *$45.67 \pm 0.05$& $-0.03$\cr
      353-4b& $89.85 \pm 0.08$& $-1.74$& $135.52 \pm 0.06$& $-0.08$\cr
      353-5a& $85.44 \pm 0.11$& $+0.75$& $177.87 \pm 0.02$& $+0.17$\cr
      353-5b& $90.95 \pm 0.11$& $+3.35$& *$89.43 \pm 0.04$& $-0.17$\cr
      353-6a& $84.99 \pm 0.58$& $-2.55$& **$0.22 \pm 0.12$& $+0.52$\cr
      353-6b& $92.56 \pm 0.66$& $+3.79$& *$88.86 \pm 0.09$& $-0.64$\cr
      353-7&  *$5.31 \pm 0.12$& $-2.68$& $126.81 \pm 0.70$& $+5.31$\cr
      353-8&  *$9.88 \pm 0.08$& $+2.06$& $139.30 \pm 0.35$& $+6.30$\cr
      \noalign{\vskip 3pt\hrule\vskip 5pt}
    }
  }
  \endPlancktable 
  \endgroup
\end{table}

\section{Destriping} 
\label{sec:destriping}

If the relevant systematics are sufficiently suppressed, low-frequency noise fluctuations in each detector begin to dominate large-scale uncertainties in the final maps.  In an effort to extract maximal information from the data, we have adopted an aggressive approach to destriping, fitting each detector with 167-ms baseline offsets that correspond to 1\deg\ steps on the sky.  These short baselines are used to capture as much of the low-frequency instrumental noise as is statistically possible, but they also adapt to other large-scale residuals such as ADC nonlinearities and gain fluctuations.  The baseline-offset solution is regularized by a prior on the baseline distribution \citep{keihanen2005,keihanen2010}, which is derived from the detector-noise power spectral densities (PSDs; see Sect.~\ref{sec:instrument_noise}).

\begin{figure*}[htpb!]
  \includegraphics[width=1.0\linewidth]{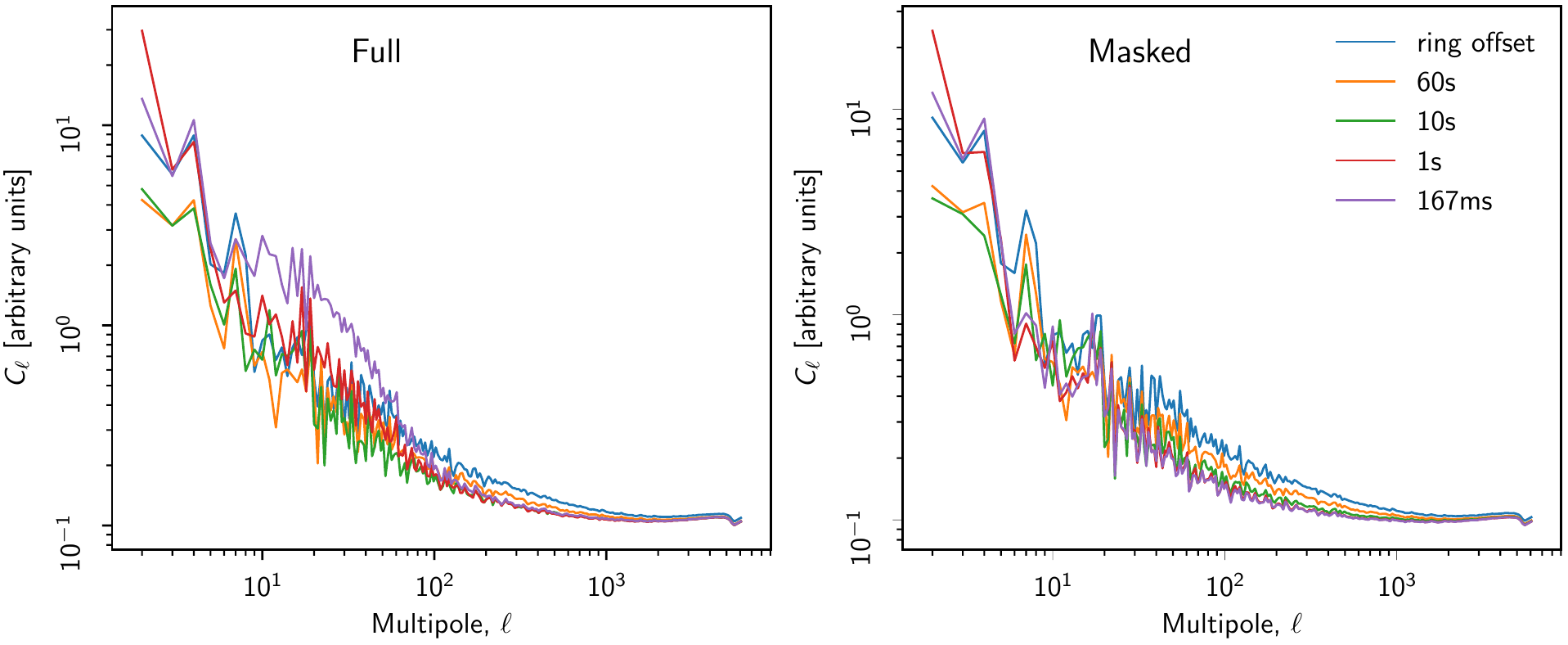}
  \caption{Power spectra of destriped dark-bolometer maps using various baseline lengths.  The ``full'' case on the left panel shows the pseudo-$C_\ell$ power spectrum over the entire map, including the Galactic plane that was masked out while solving for the baseline offsets.  The right panel shows the power spectrum taken over the sky pixels that were considered when solving for the baselines.  Shorter baselines clearly suppress noise above $\ell = 20$.  The 167-ms case shows an elevated residual at $\ell < 100$ inside the destriping mask, likely due to the filter's inability to constrain the solution.  The residual at $\ell < 20$ is dominated by the irreducible uncertainty set by the \Planck\ scan strategy, and cannot be improved without external information.
 }
  \label{fig:cl_dark}
\end{figure*}

We tested the efficacy of the short baselines by destriping and projecting dark-bolometer data from the two dark bolometers as if they were bolometers 857-1 and 857-3.  Figure~\ref{fig:cl_dark} shows the power spectra of maps destriped with various baseline lengths.  Using the dark bolometer data instead of synthetic TOD has the advantage that they include realistic glitch and 4-K line residuals.

In choosing the destriping resolution, we have balanced S/N, which improves with lower destriping resolution, with signal error, which is best when there is no detectable sub-pixel structure.  Our destriping resolution is \nside{512} for all \lfi\ frequencies, \nside{1024} for 100--353\GHz, and \nside{2048} for 545 and 857\GHz.  These resolutions allow for modest sub-pixel signal gradients outside the destriping mask, but do not lead to temperature-to-polarization leakage due to the horn symmetrization (Sect.~\ref{sec:symmetrization}).

Solving for short baselines also serves a secondary purpose in allowing us to apply the \madam\ approach for estimating the residual noise in the pixel-to-pixel covariance in a low-resolution version of the \npipe\ data set.  In \madam, the noise is broken down into baseline offsets and white noise.  When the baselines are too long to capture the $1/f$ noise fluctuations, \madam\ can only accommodate the extra power in the white noise component of the signal model, leading to mismatch between the noise matrix
and the actual noise in the maps.

\section{\npipe\ data set}
\label{sec:dataset}

We now describe the contents of the \npipe\ data release, highlighting differences between \npipe\ and other \Planck\ public releases that may be relevant to the user.  For information regarding the dissemination of the maps and the related software, see Appendix~\ref{app:release}.

\subsection{Maps}

All \npipe\ maps are calibrated to thermodynamic temperature units in kelvins (\KCMB).  The 857-GHz calibration is the same as in \prthree, but is converted into \KCMB\ temperature units using the measured bandpasses.  Note that \npipe\ maps include the Solar dipole; it may be removed using the parameters supplied in Sect.~\ref{sec:dipole}.

The conversion factor from thermodynamic temperature to flux density units is $58.04\MJysr/\KCMB$ for the 545-GHz maps and $2.27\MJysr/\KCMB$ for the 857-GHz maps (see \citealt{planck2013-p03d}).

\subsubsection{Frequency maps} \label{sec:fmaps}

\begin{figure*}[htpb!]
  \includegraphics[width=0.33\linewidth]{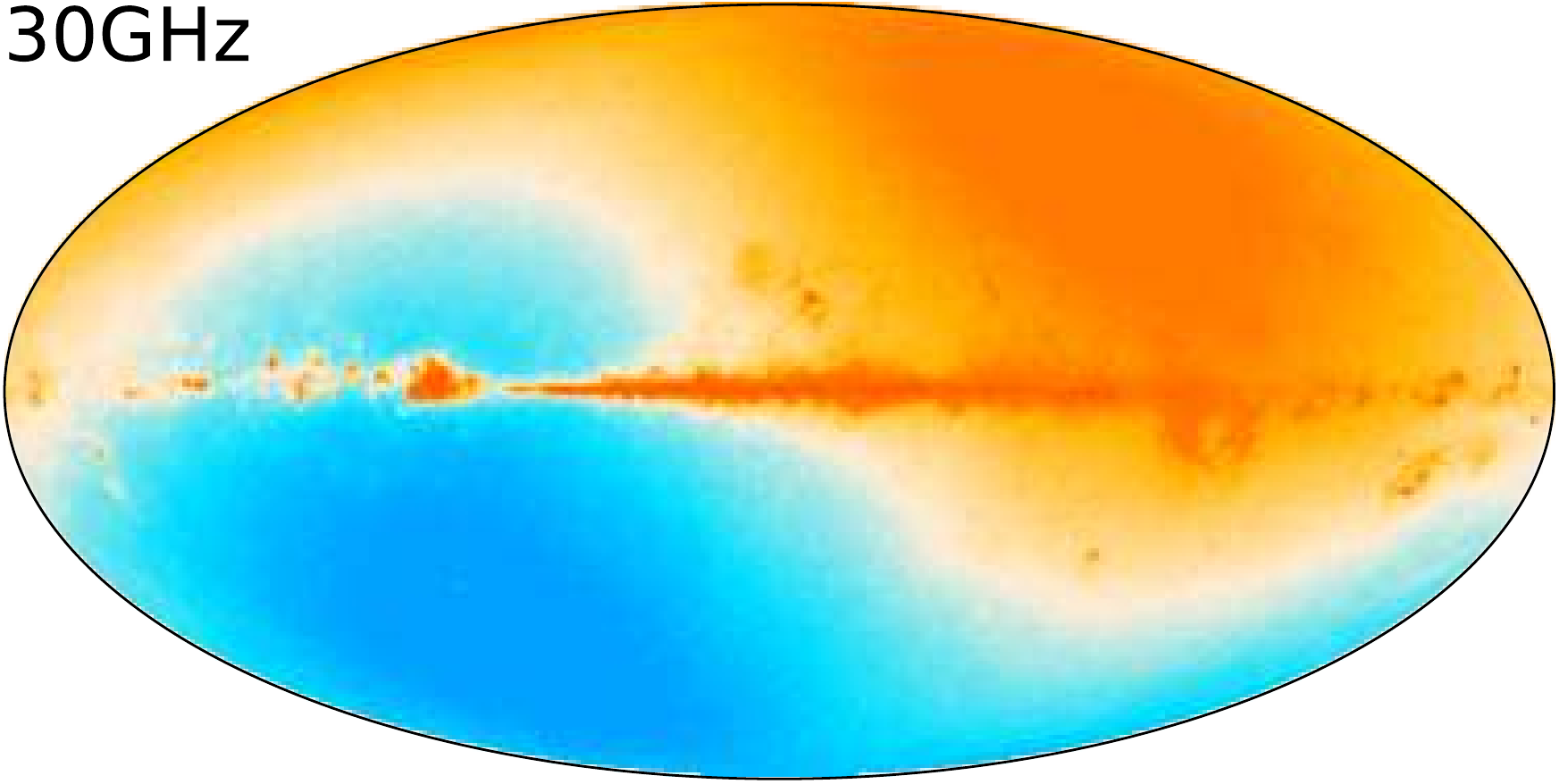}
  \includegraphics[width=0.33\linewidth]{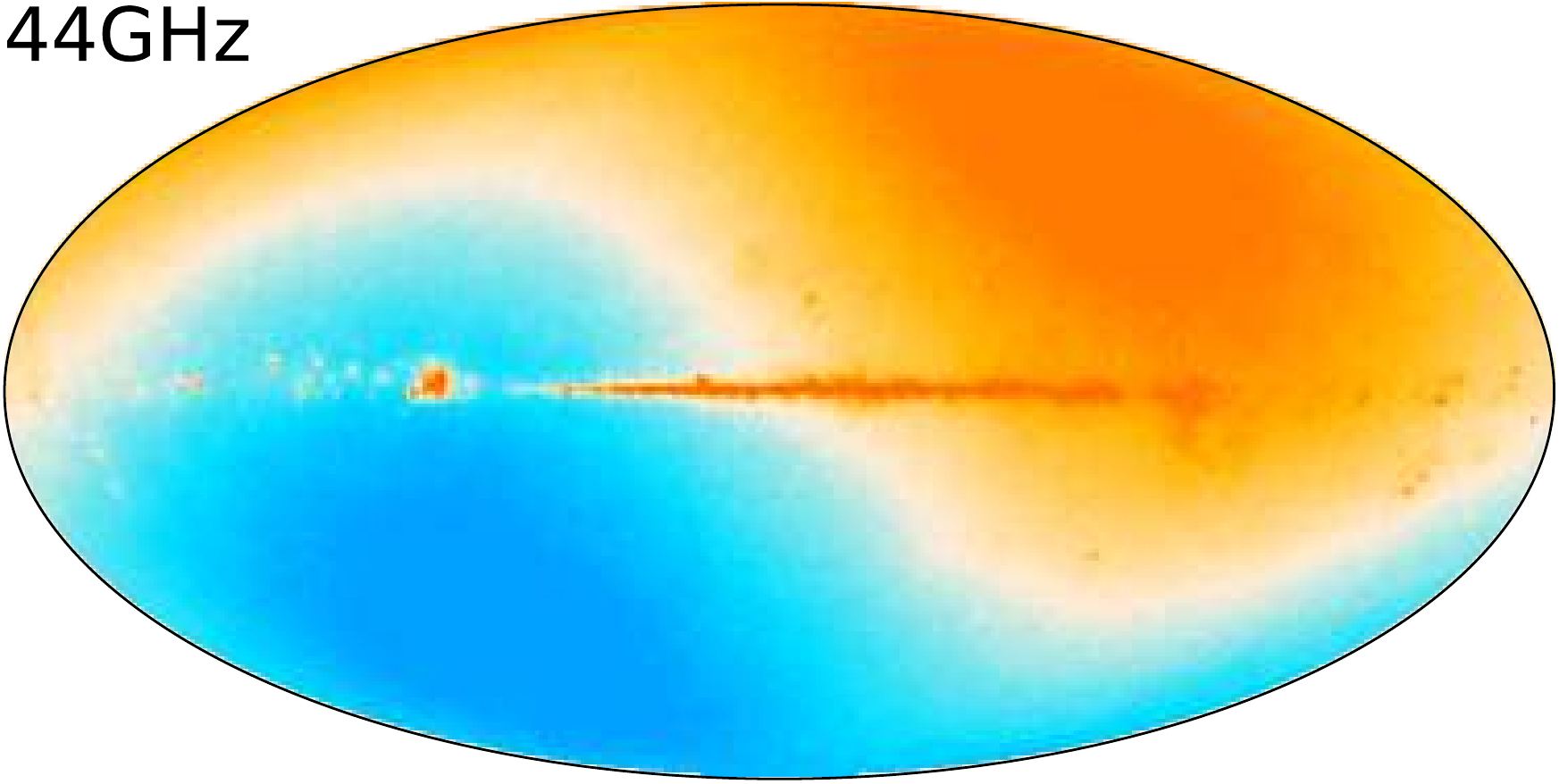}
  \includegraphics[width=0.33\linewidth]{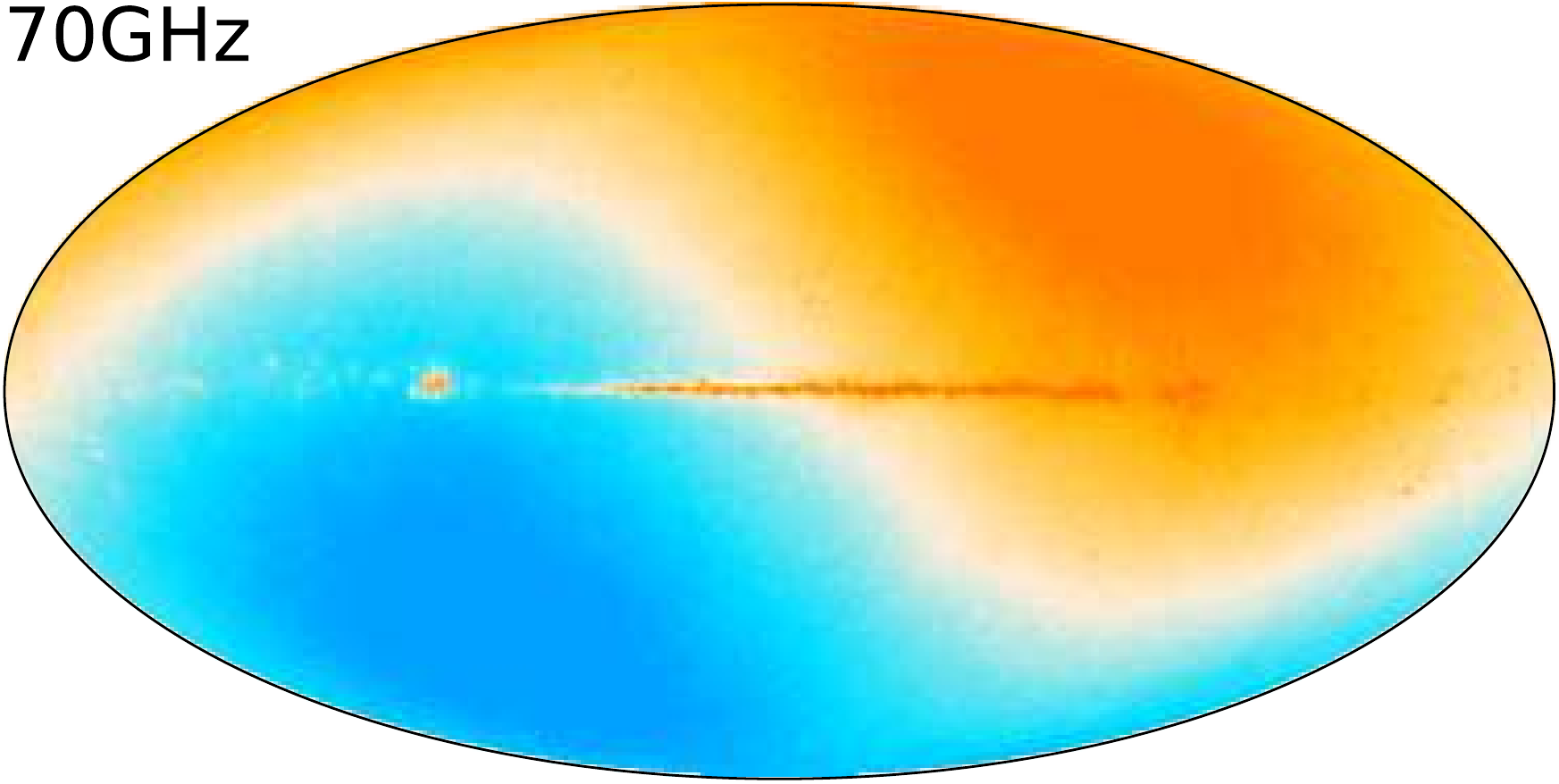}
  \\
  \includegraphics[width=0.33\linewidth]{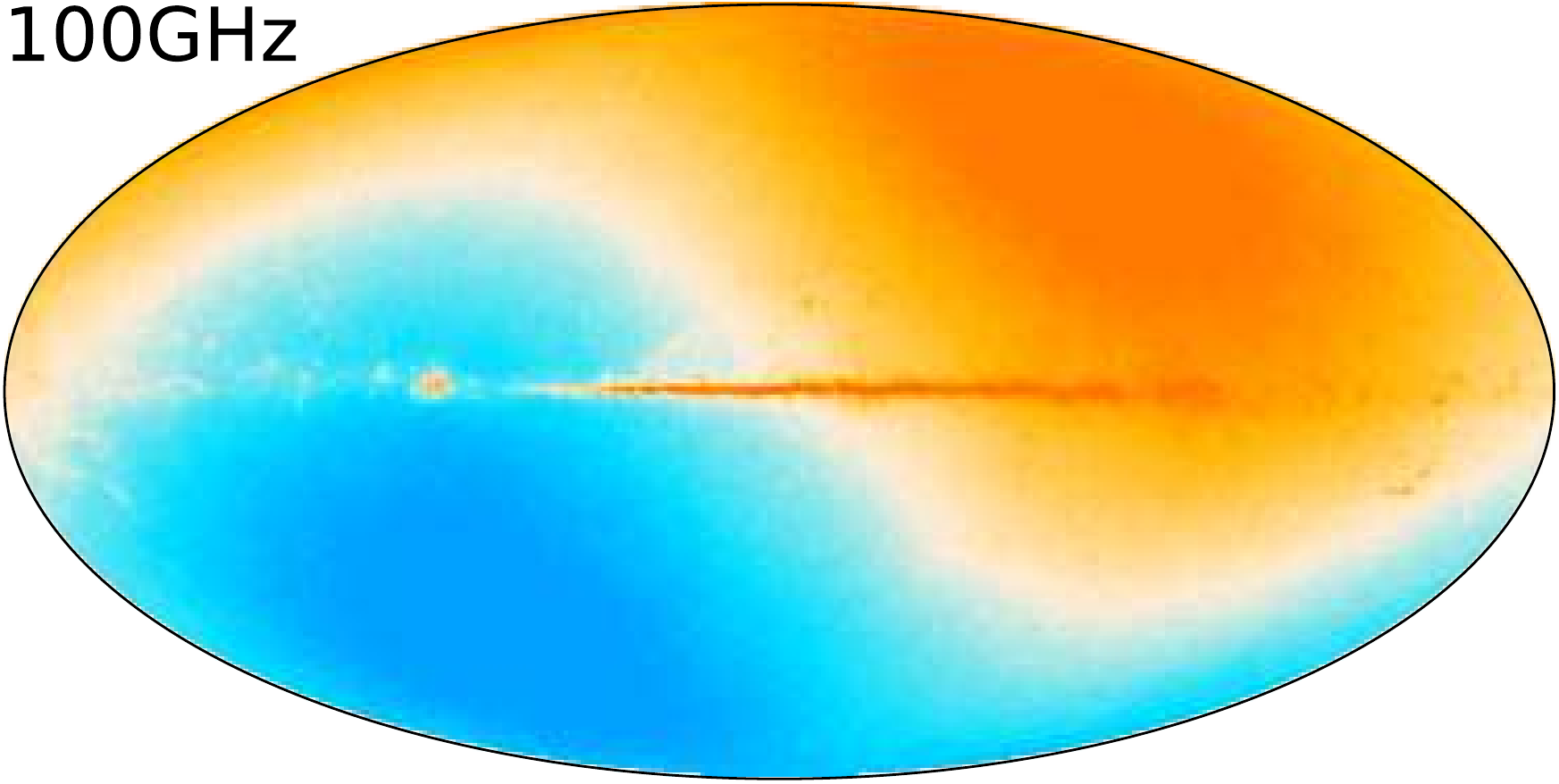}
  \includegraphics[width=0.33\linewidth]{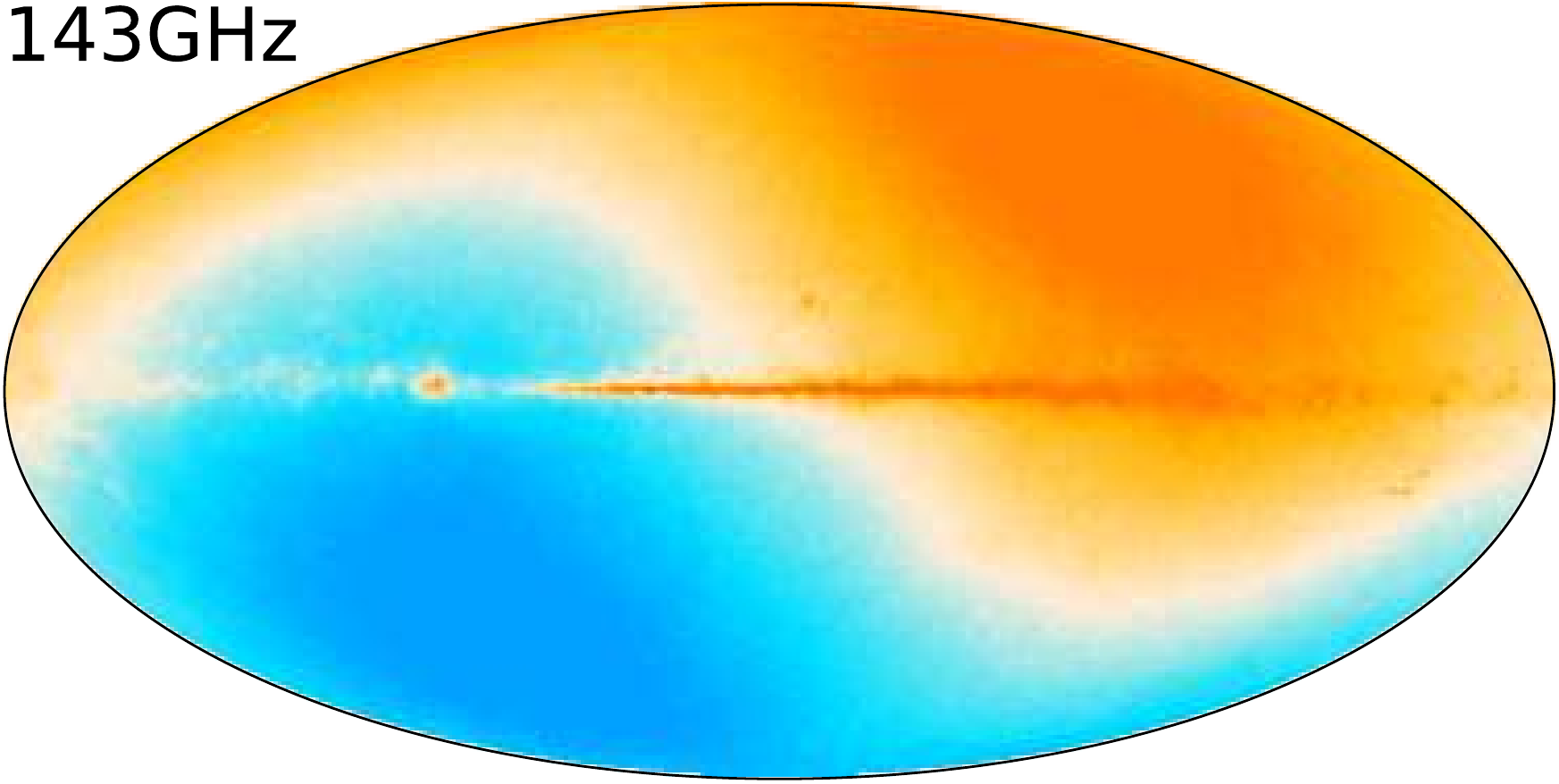}
  \includegraphics[width=0.33\linewidth]{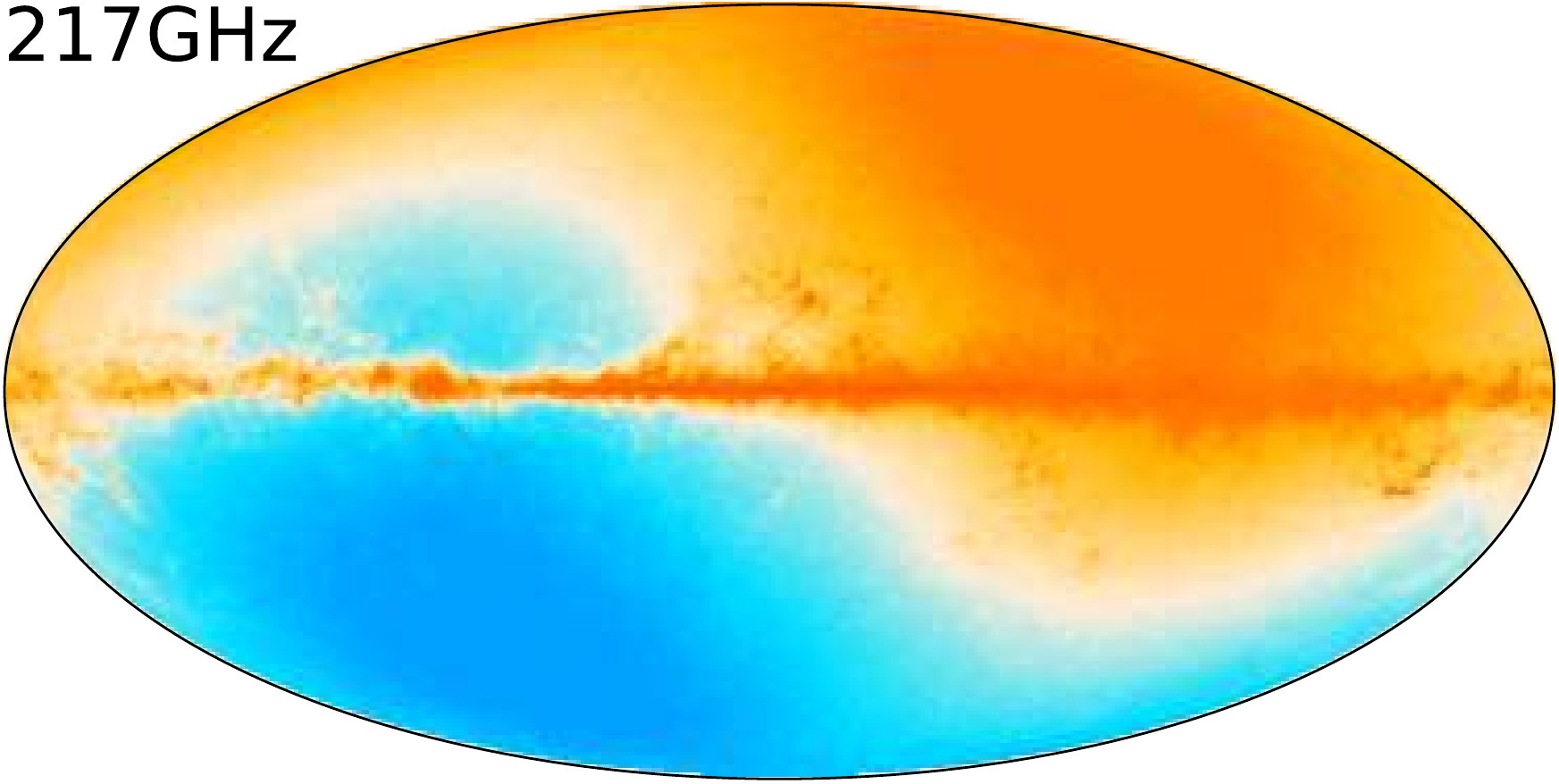}
  \\
  \includegraphics[width=0.33\linewidth]{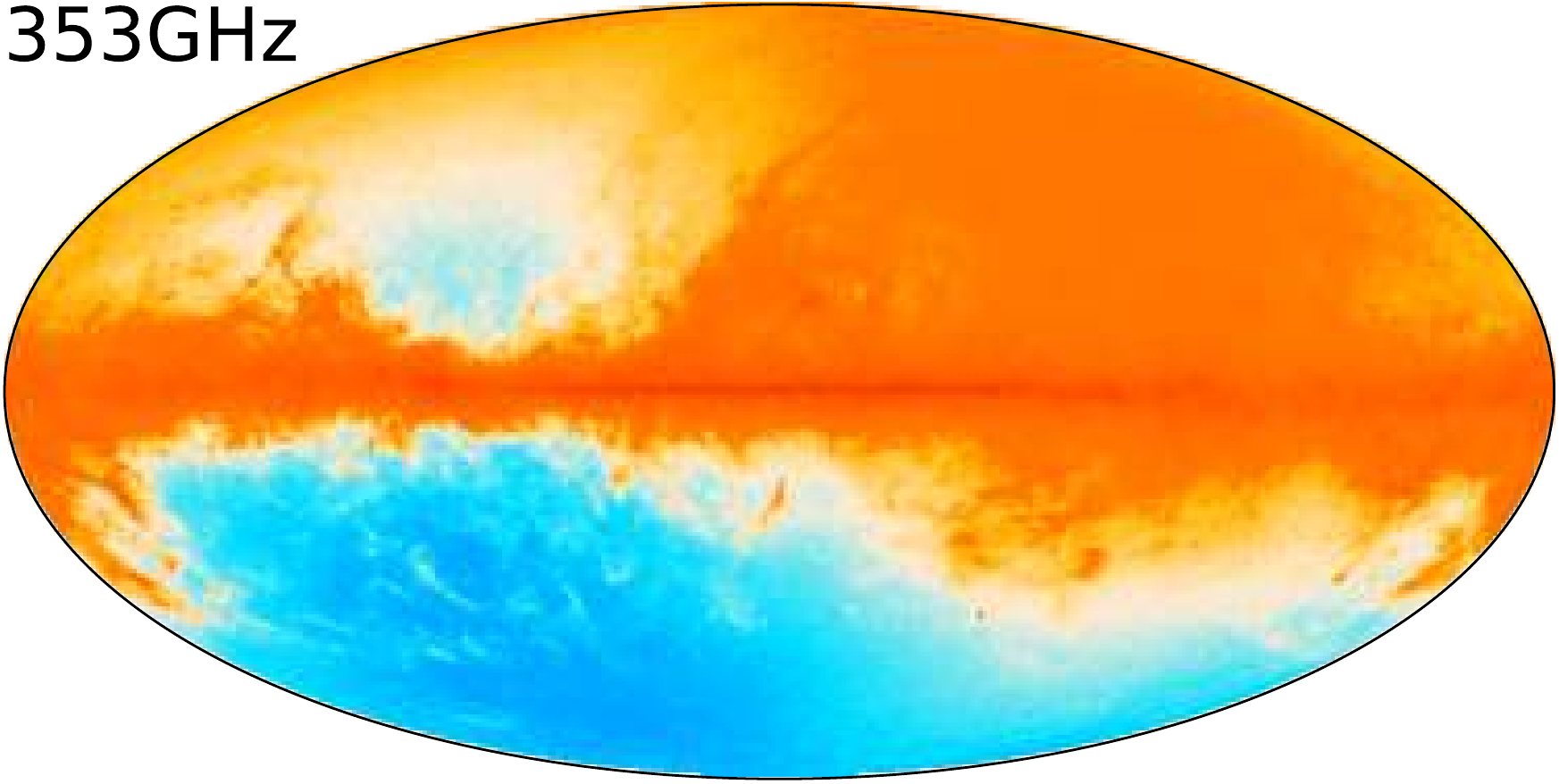}
  \includegraphics[width=0.33\linewidth]{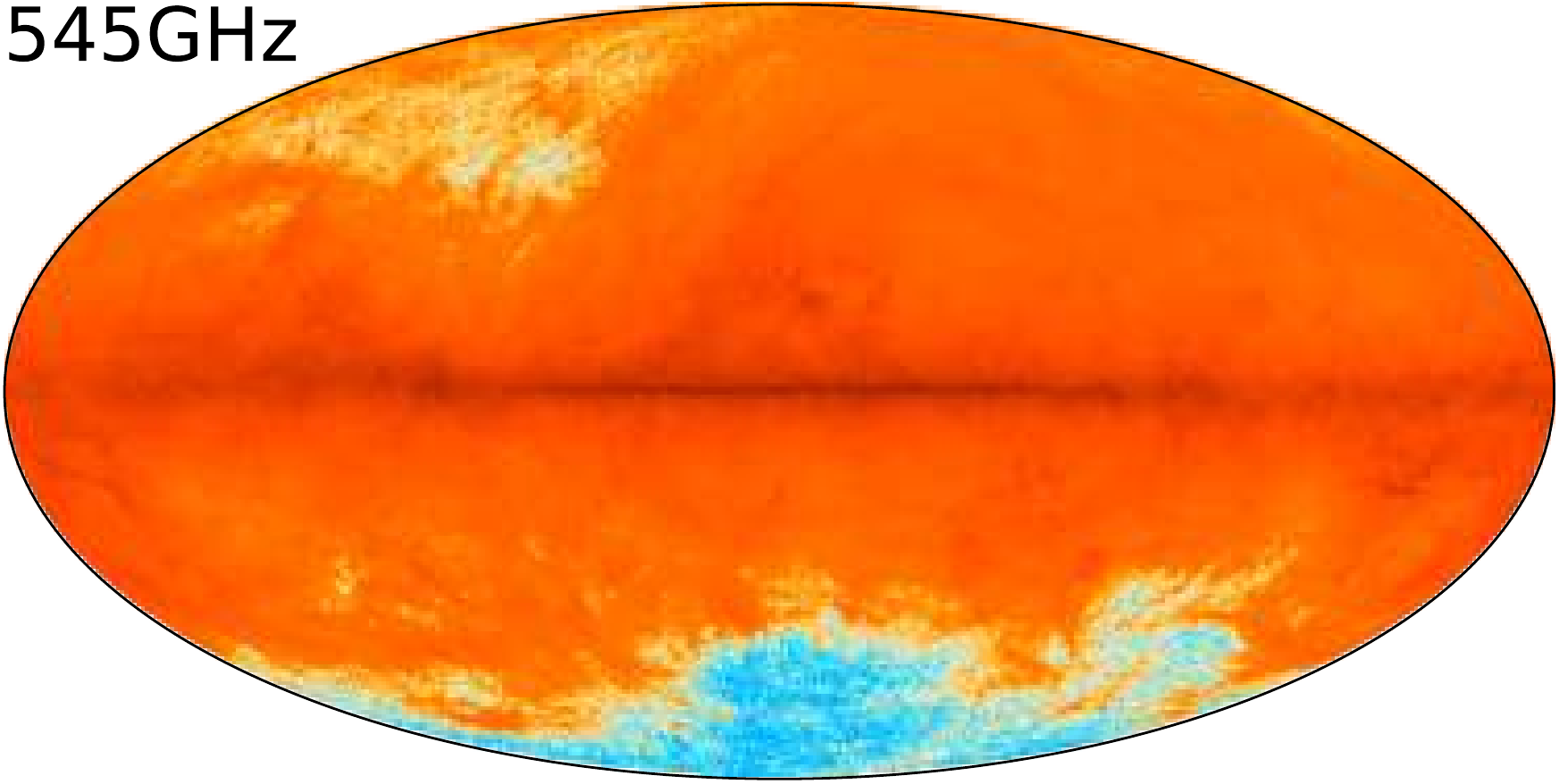}
  \includegraphics[width=0.33\linewidth]{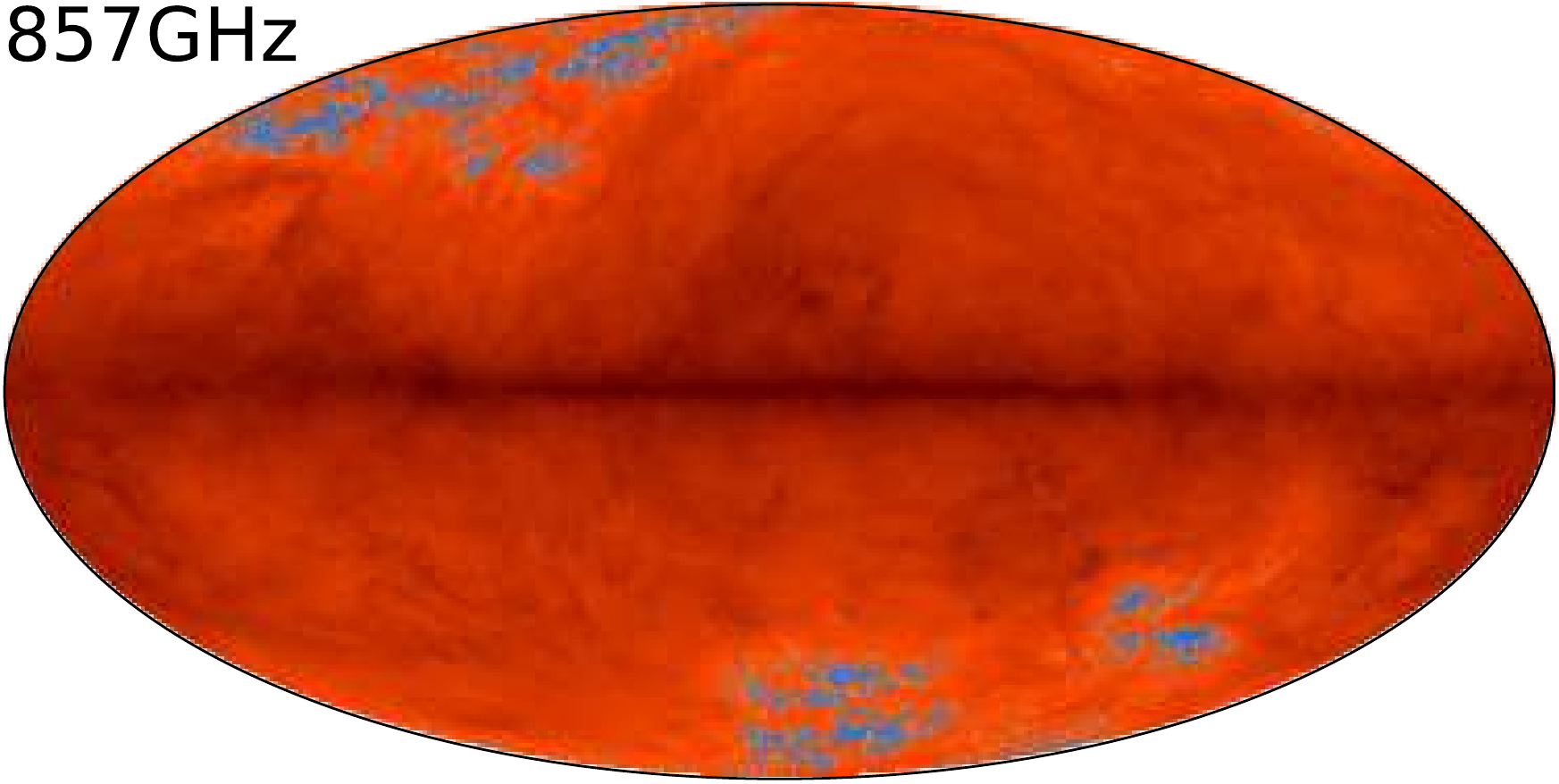}
  \\
  \includegraphics[width=1.0\linewidth]{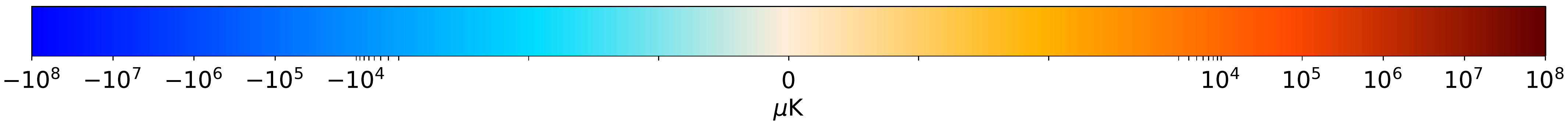}
  \caption{\npipe\ temperature maps, including the Solar dipole. The scaling is linear between $-3$ and $3\,$mK.
  }
  \label{fig:freqmaps_Tw}
\end{figure*}

\begin{figure*}[htpb!]
  \includegraphics[width=0.33\linewidth]{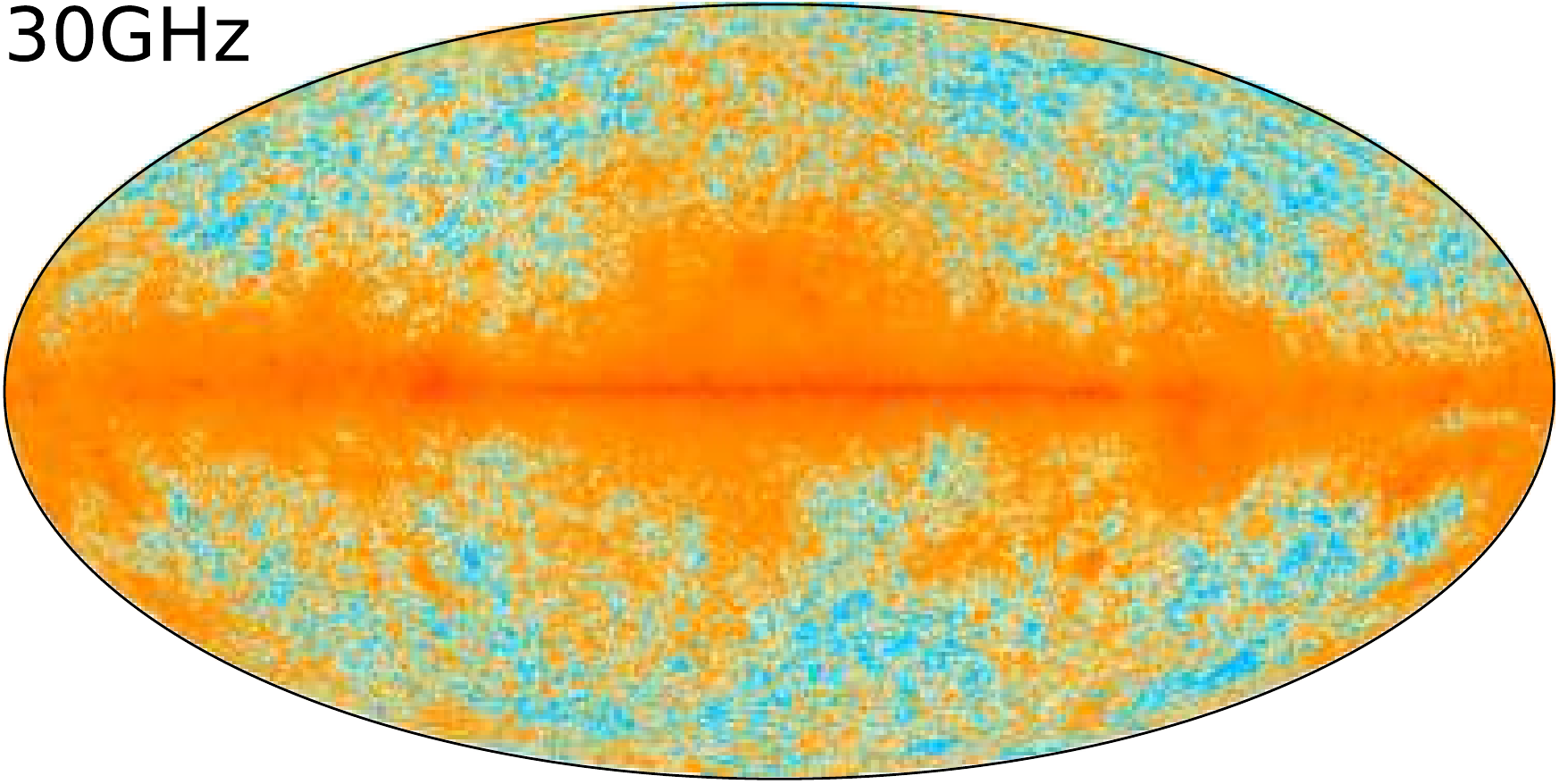}
  \includegraphics[width=0.33\linewidth]{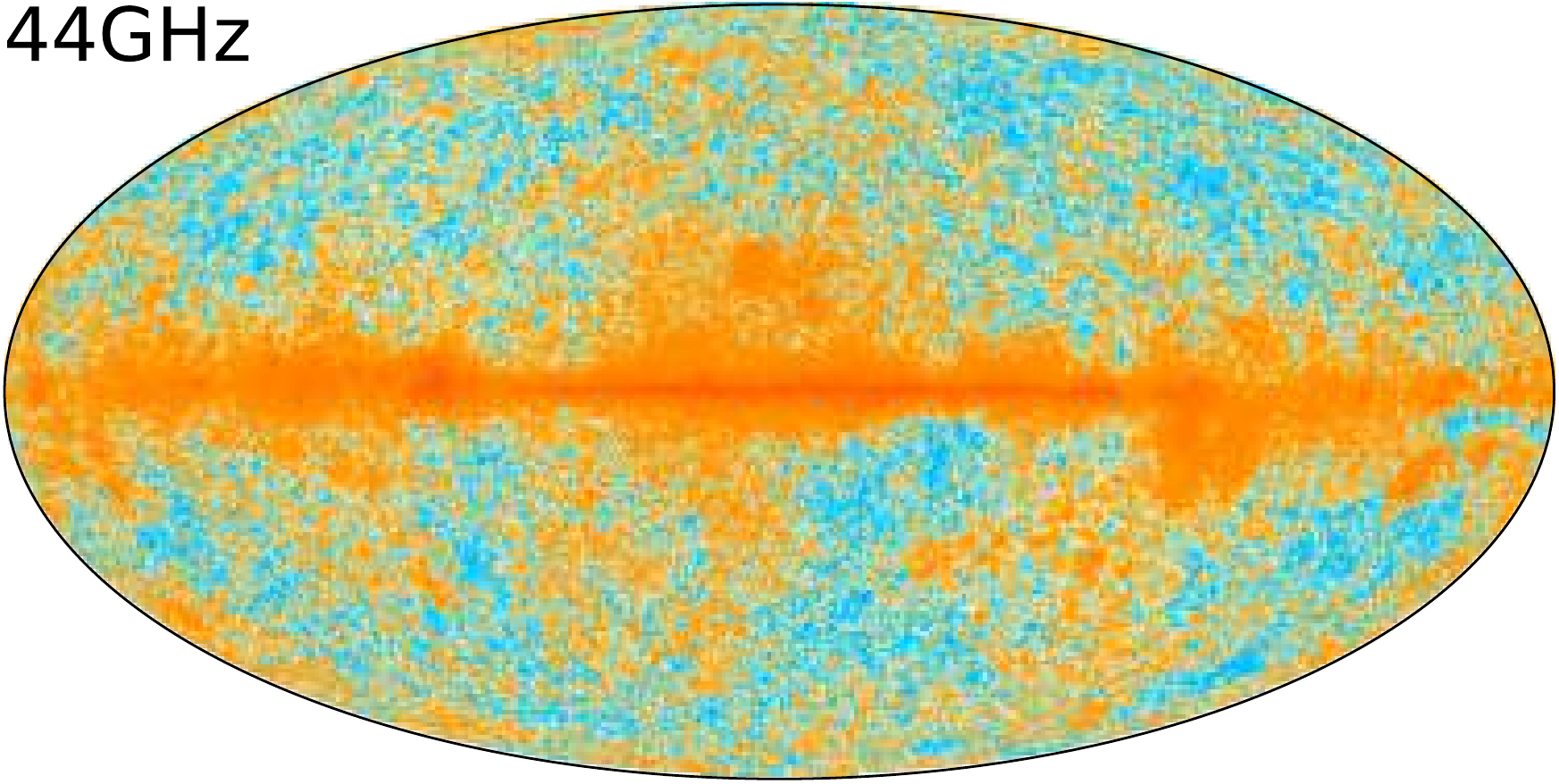}
  \includegraphics[width=0.33\linewidth]{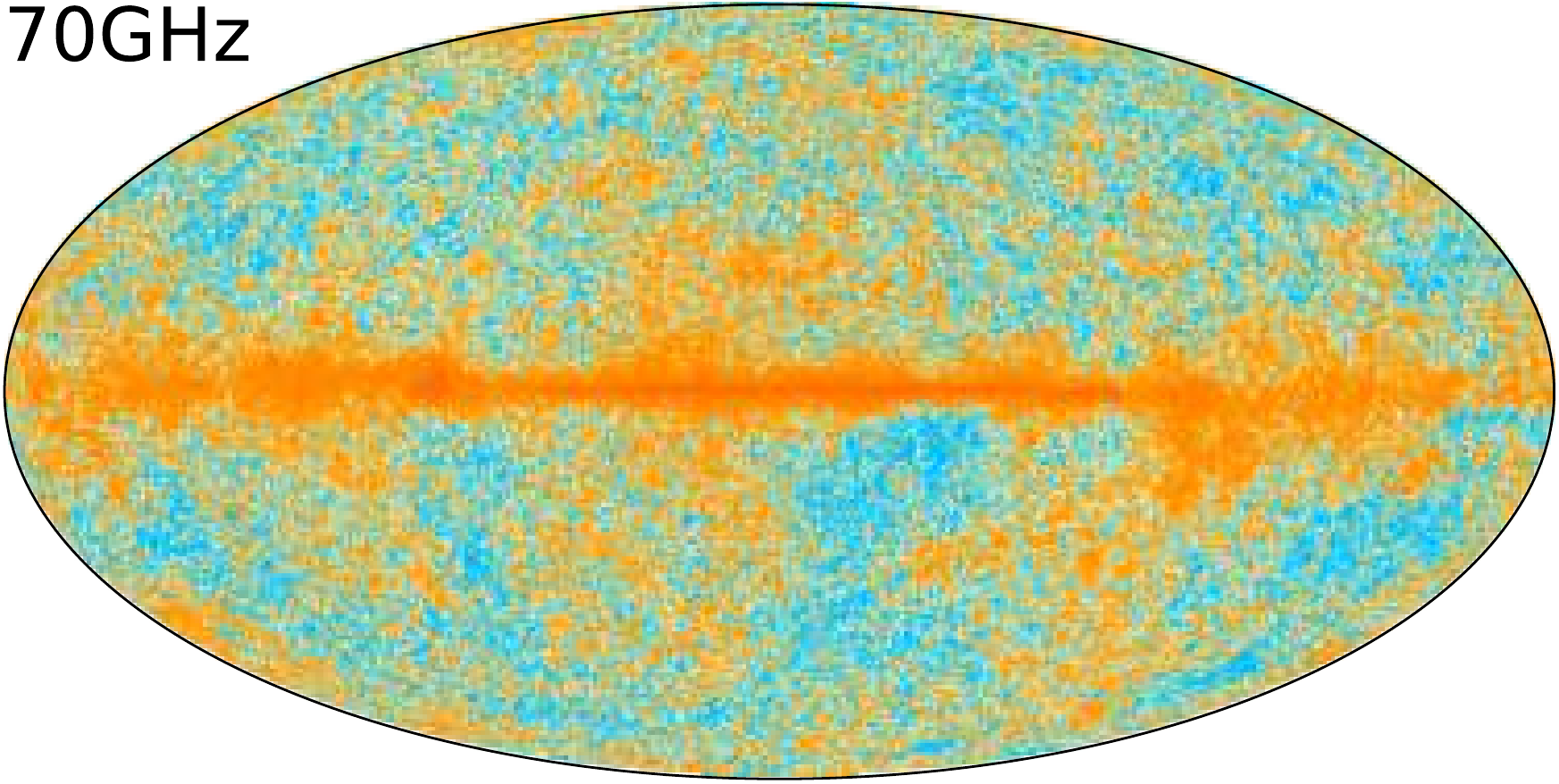}
  \\
  \includegraphics[width=0.33\linewidth]{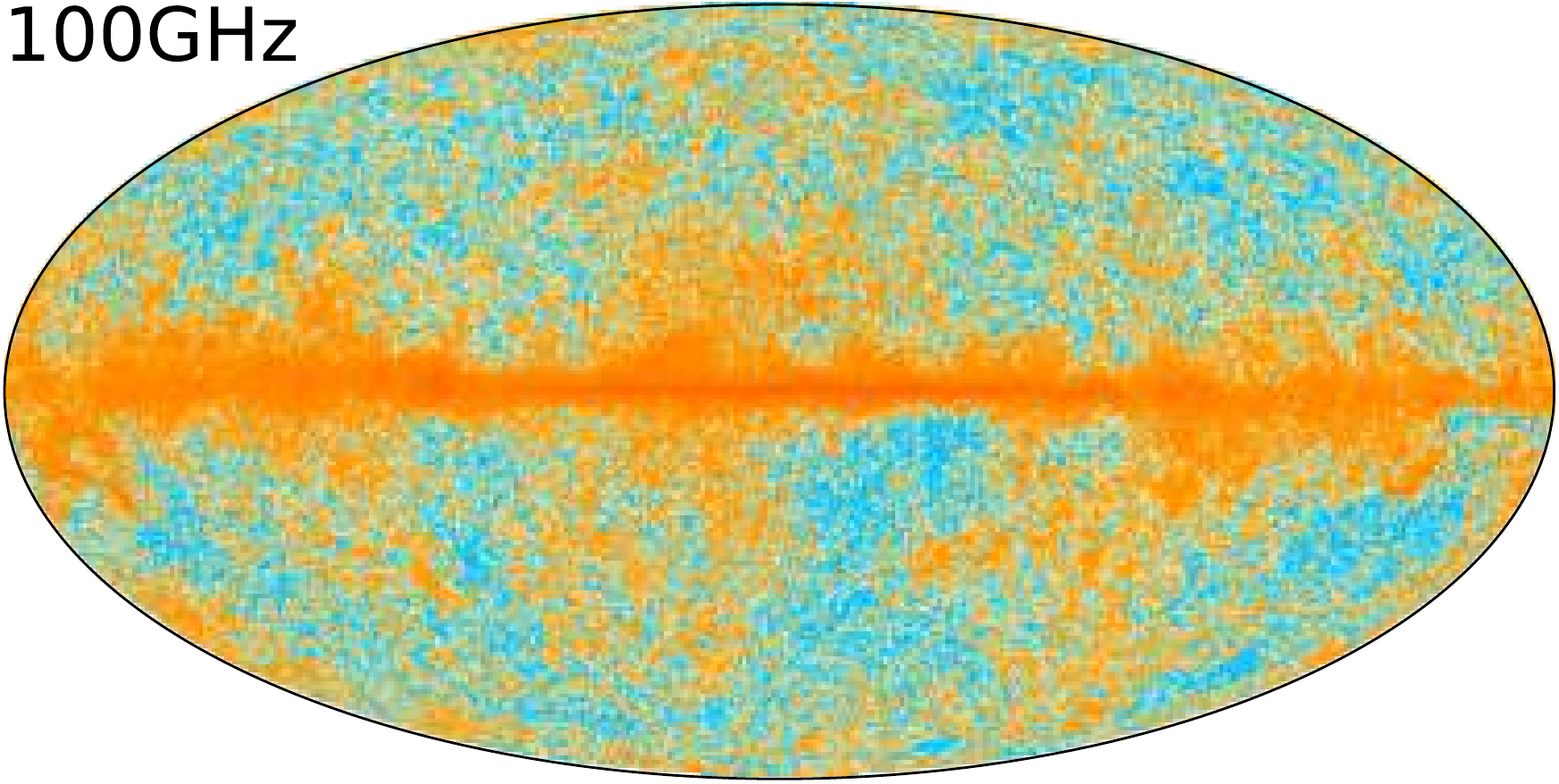}
  \includegraphics[width=0.33\linewidth]{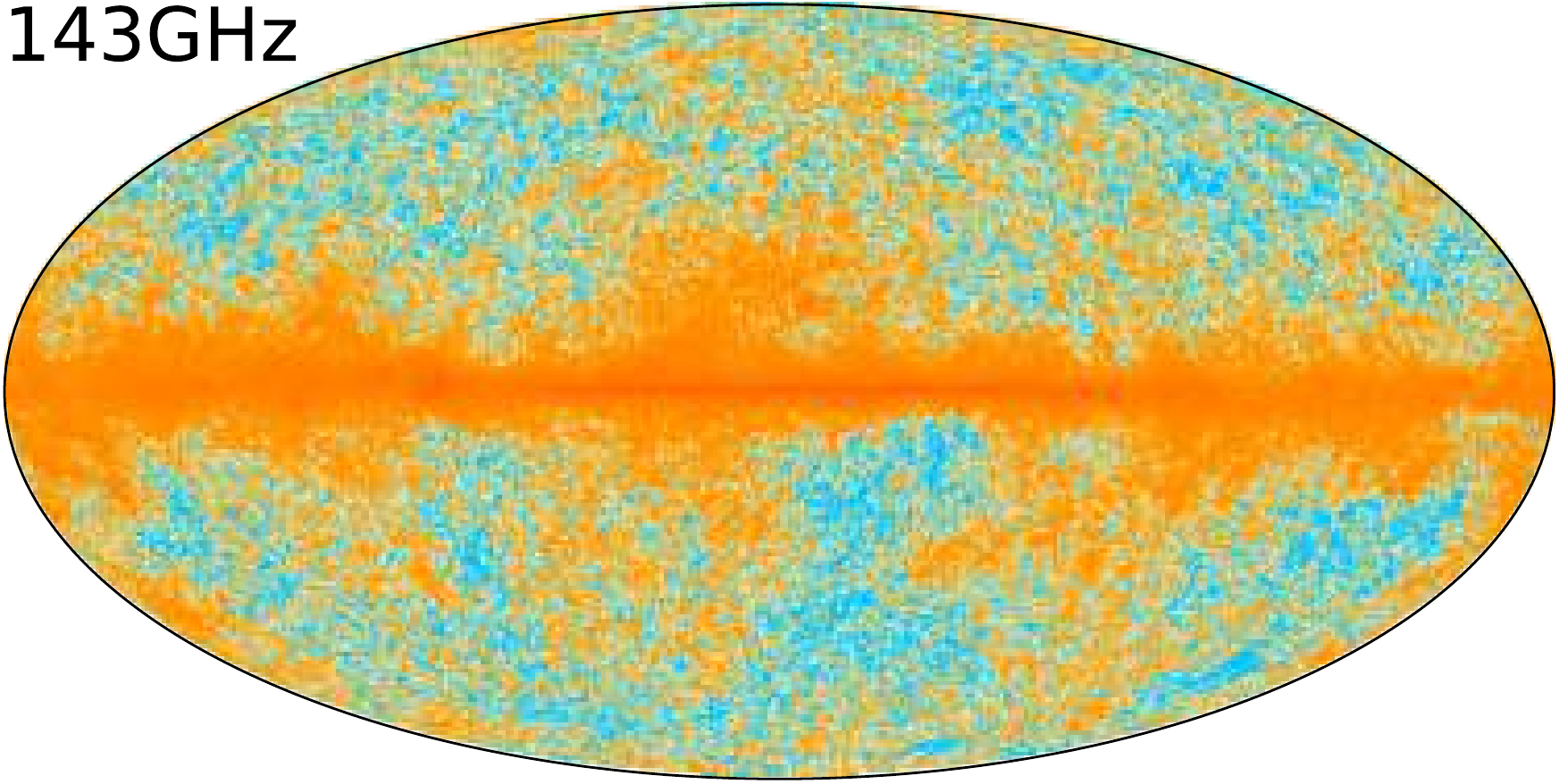}
  \includegraphics[width=0.33\linewidth]{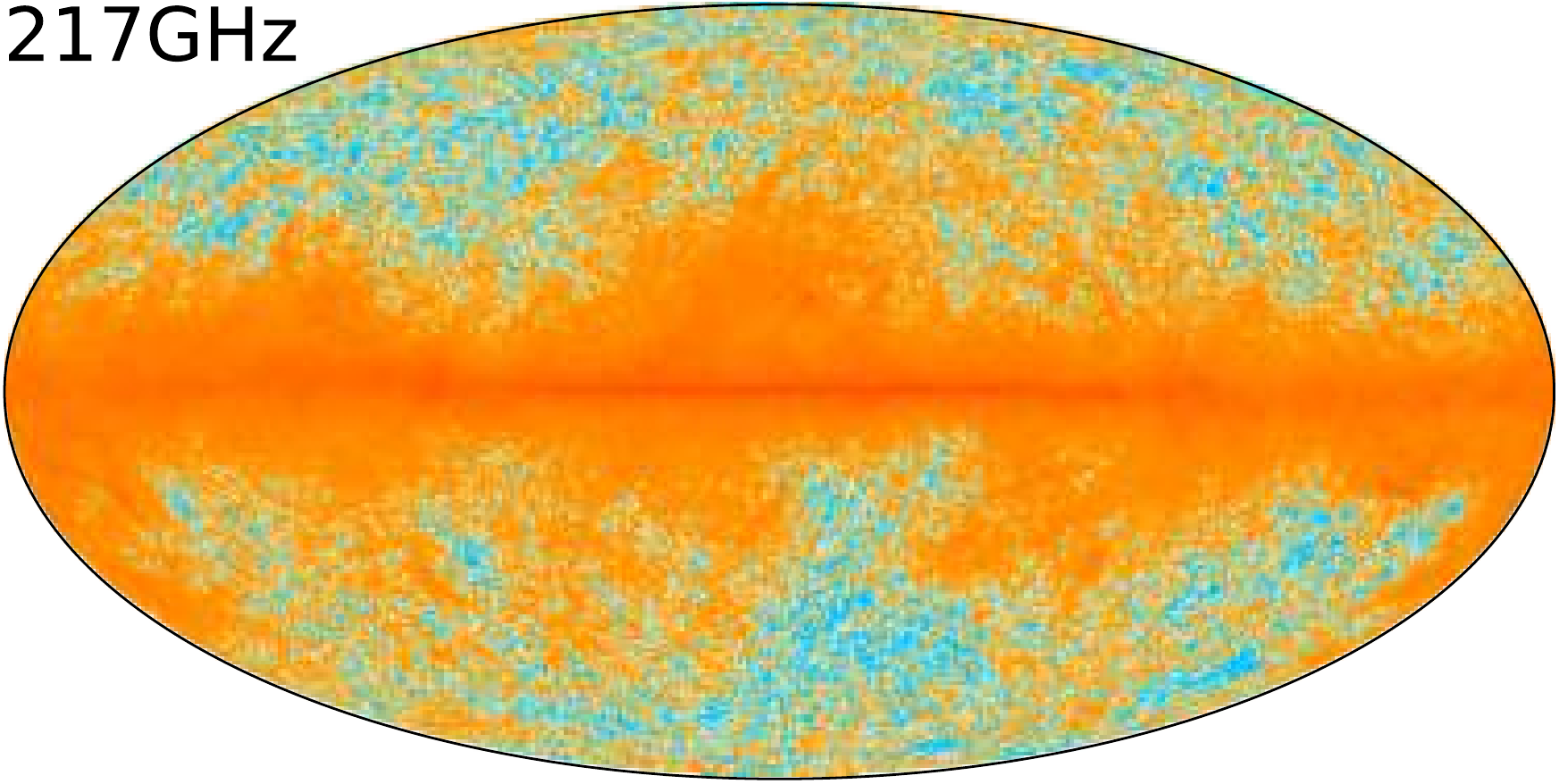}
  \\
  \includegraphics[width=0.33\linewidth]{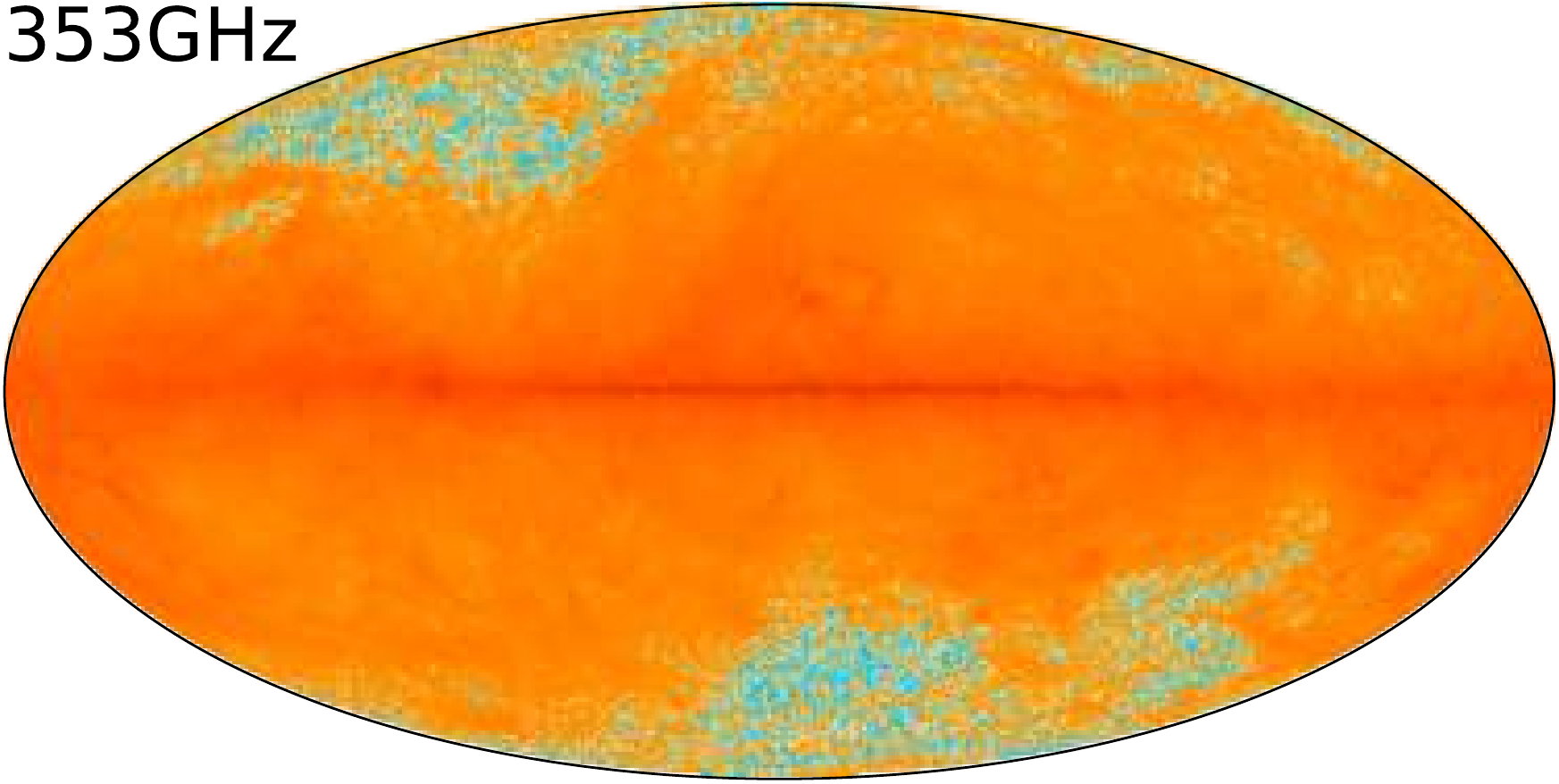}
  \includegraphics[width=0.33\linewidth]{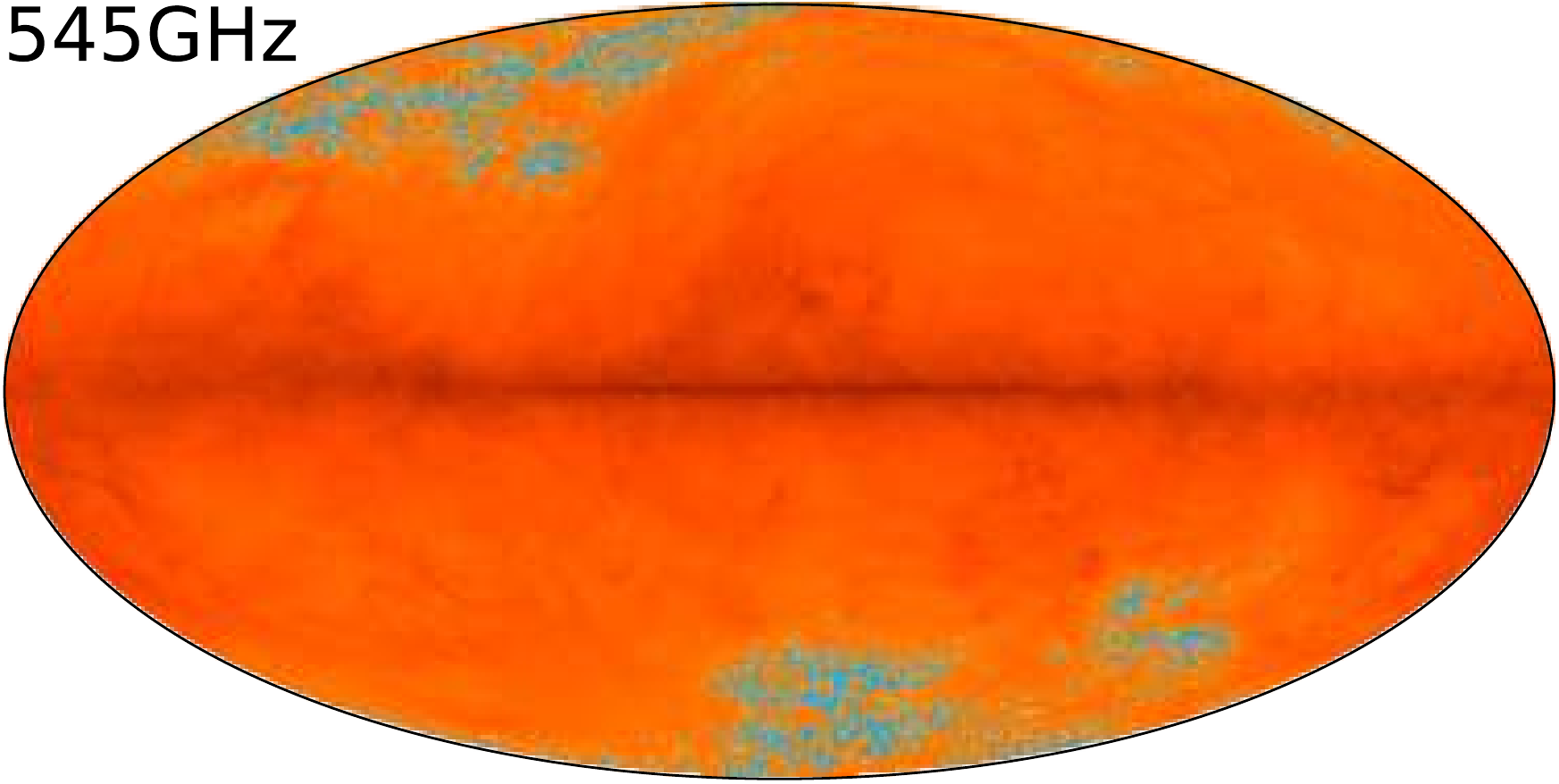}
  \includegraphics[width=0.33\linewidth]{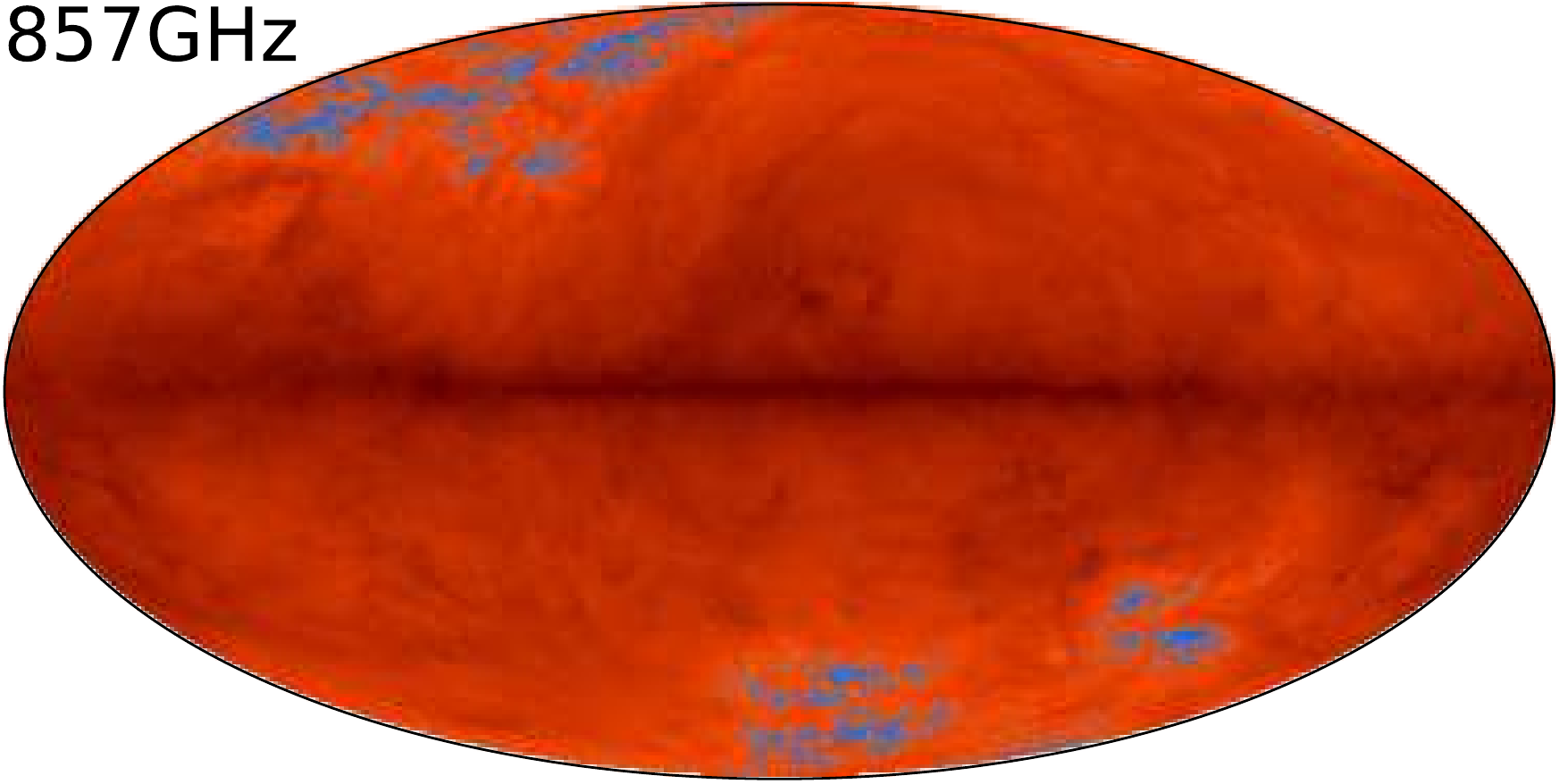}
  \\
  \includegraphics[width=1.0\linewidth]{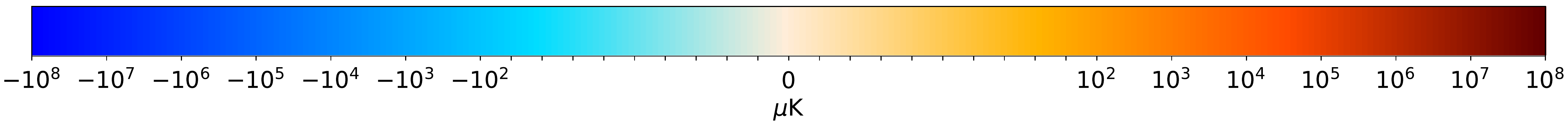}
  \caption{\npipe\ temperature maps, with the \Planck\ 2015 dipole removed.  The scaling is linear between $-100$ and $100\,\mu$K.
  }
  \label{fig:freqmaps_T}
\end{figure*}

\begin{figure*}[htpb!]
  \includegraphics[width=0.33\linewidth]{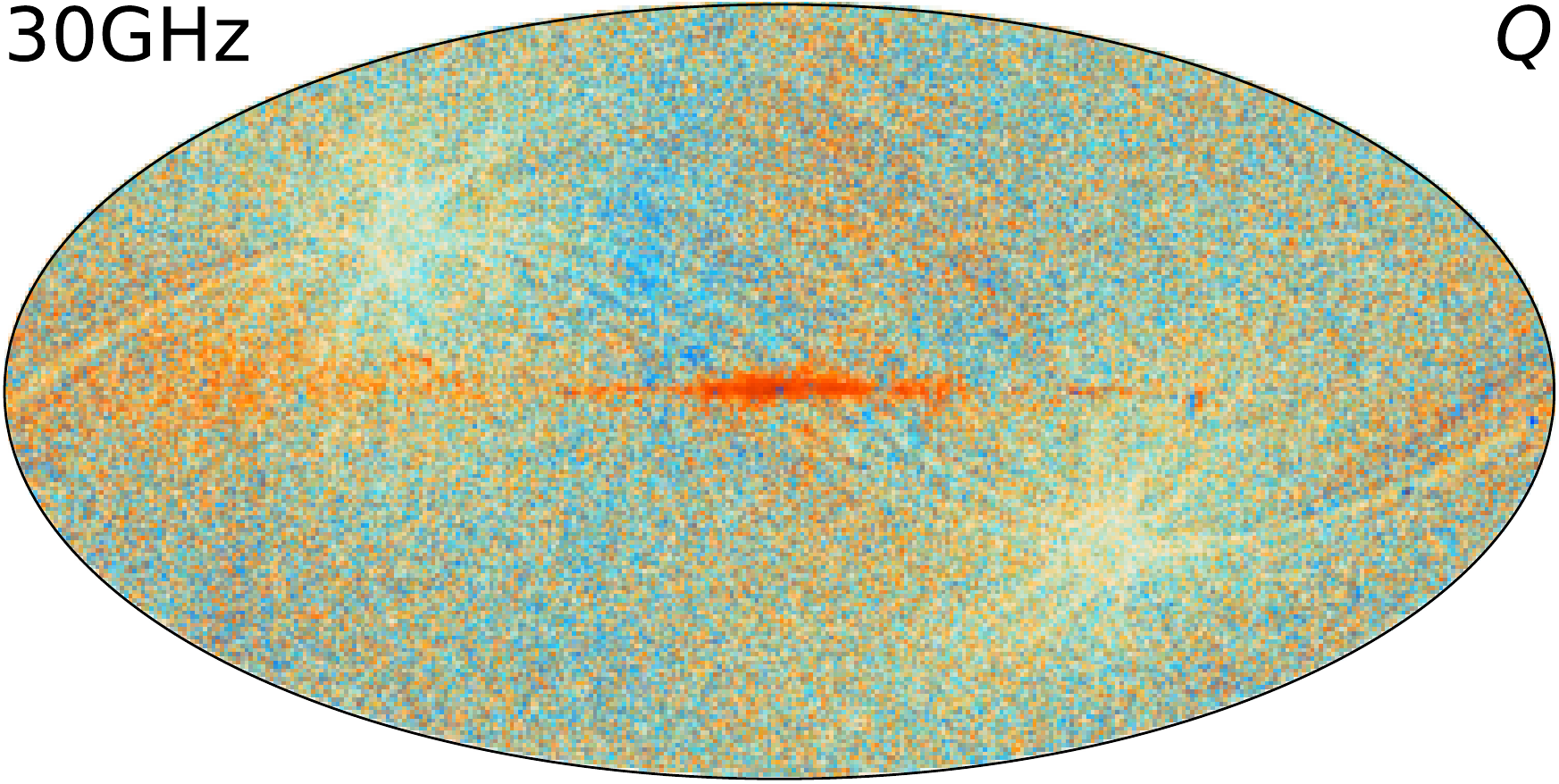}
  \includegraphics[width=0.33\linewidth]{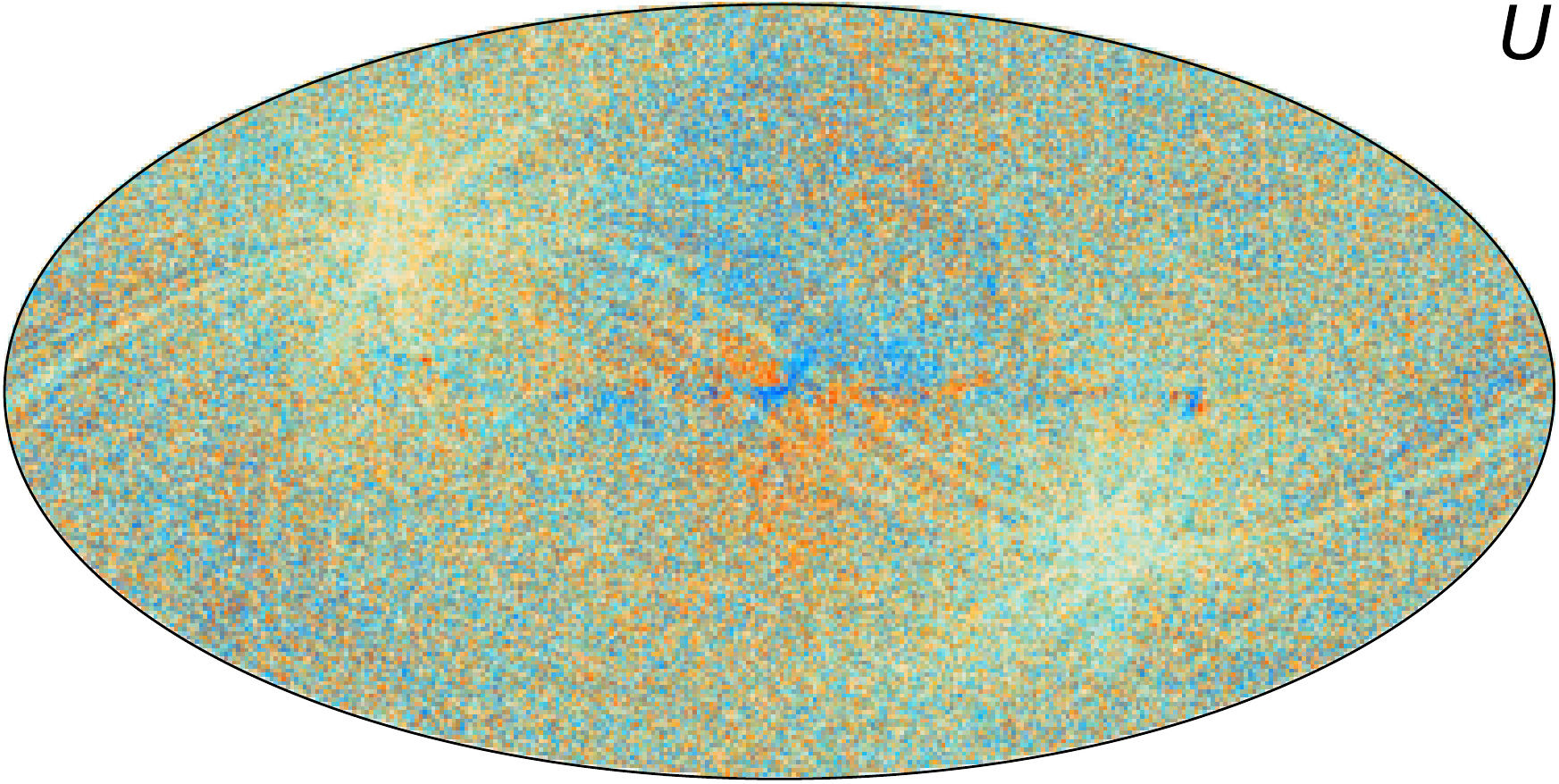}
  \includegraphics[width=0.33\linewidth]{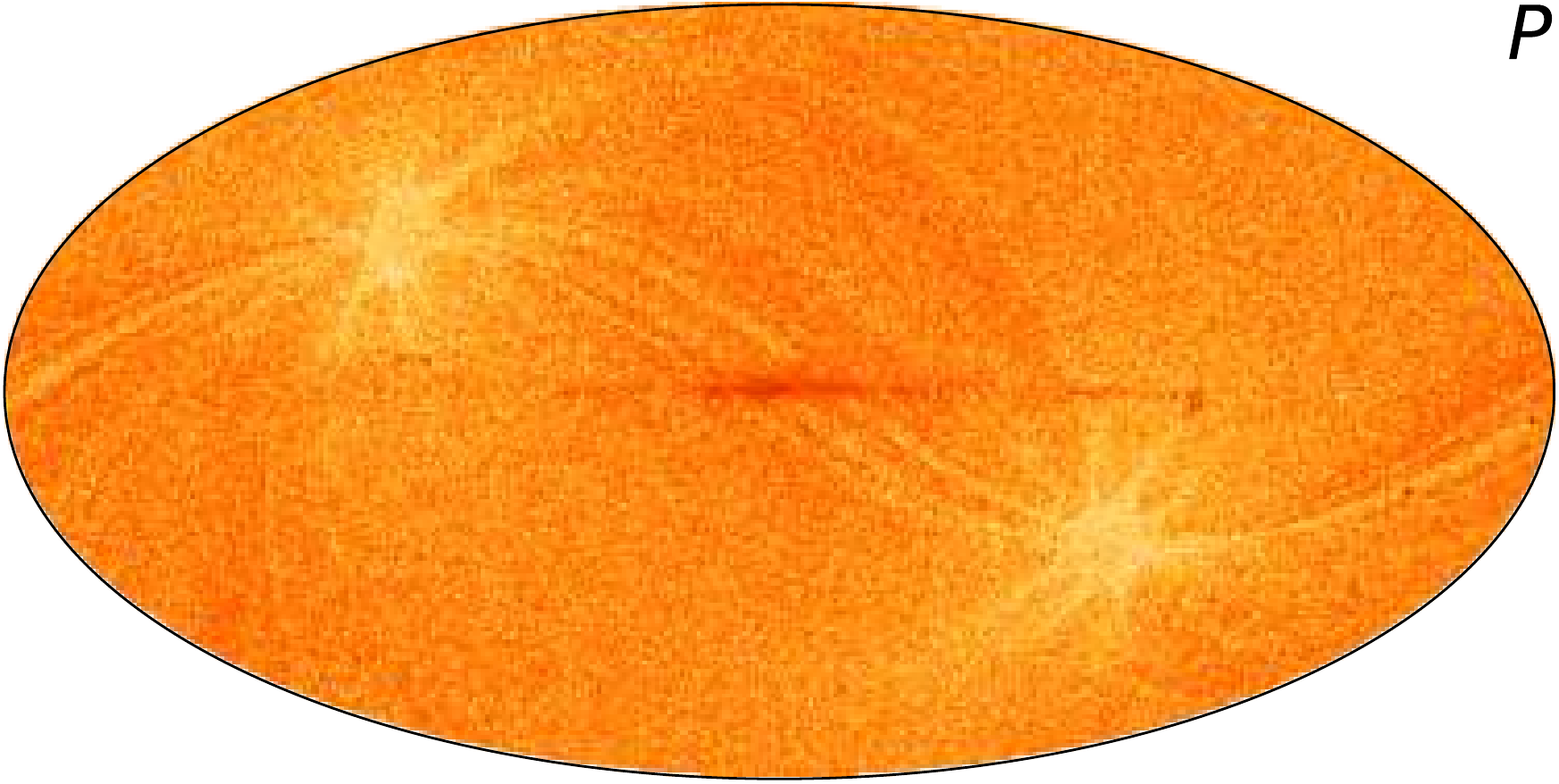}
  \\
  \includegraphics[width=0.33\linewidth]{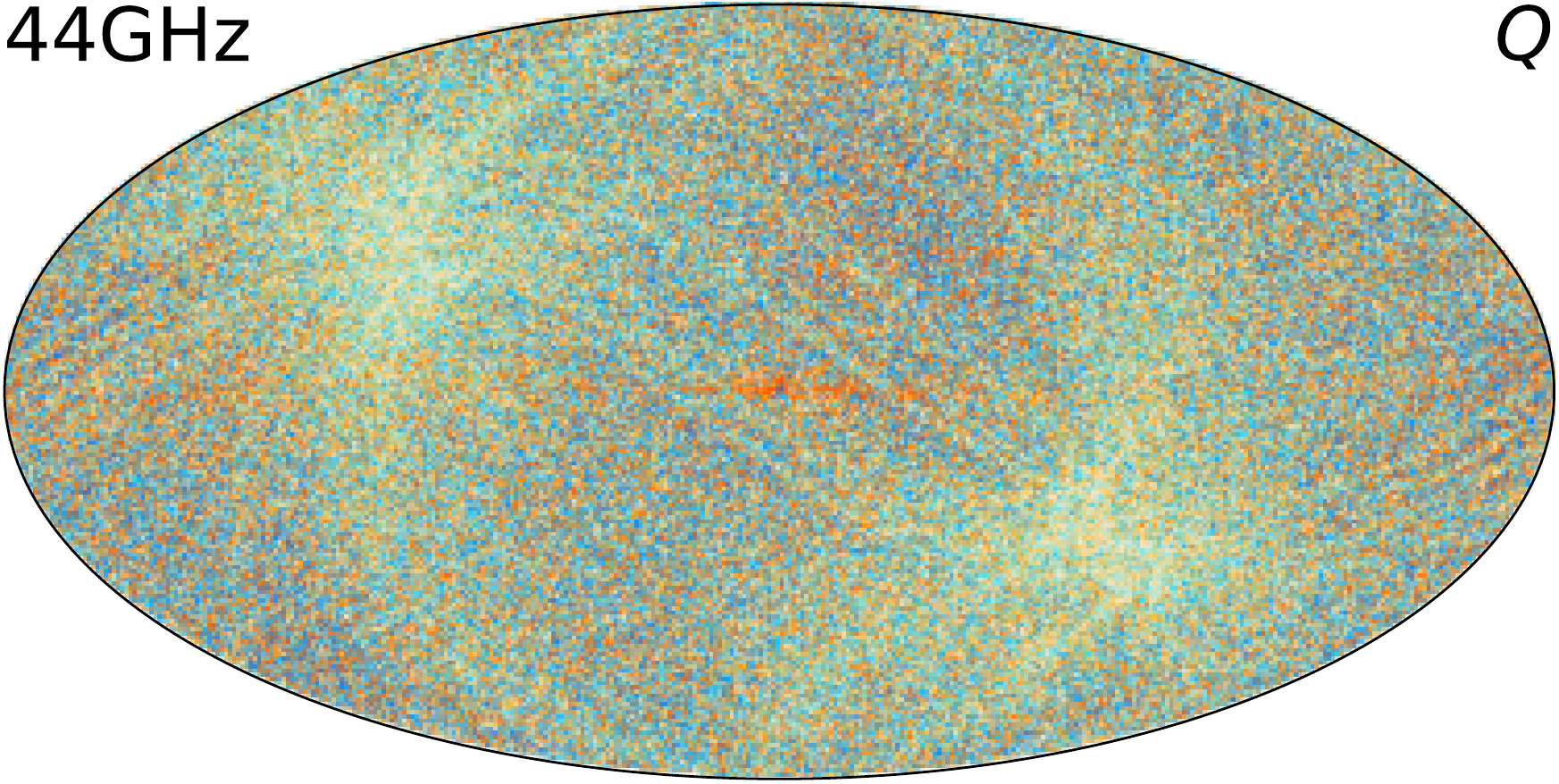}
  \includegraphics[width=0.33\linewidth]{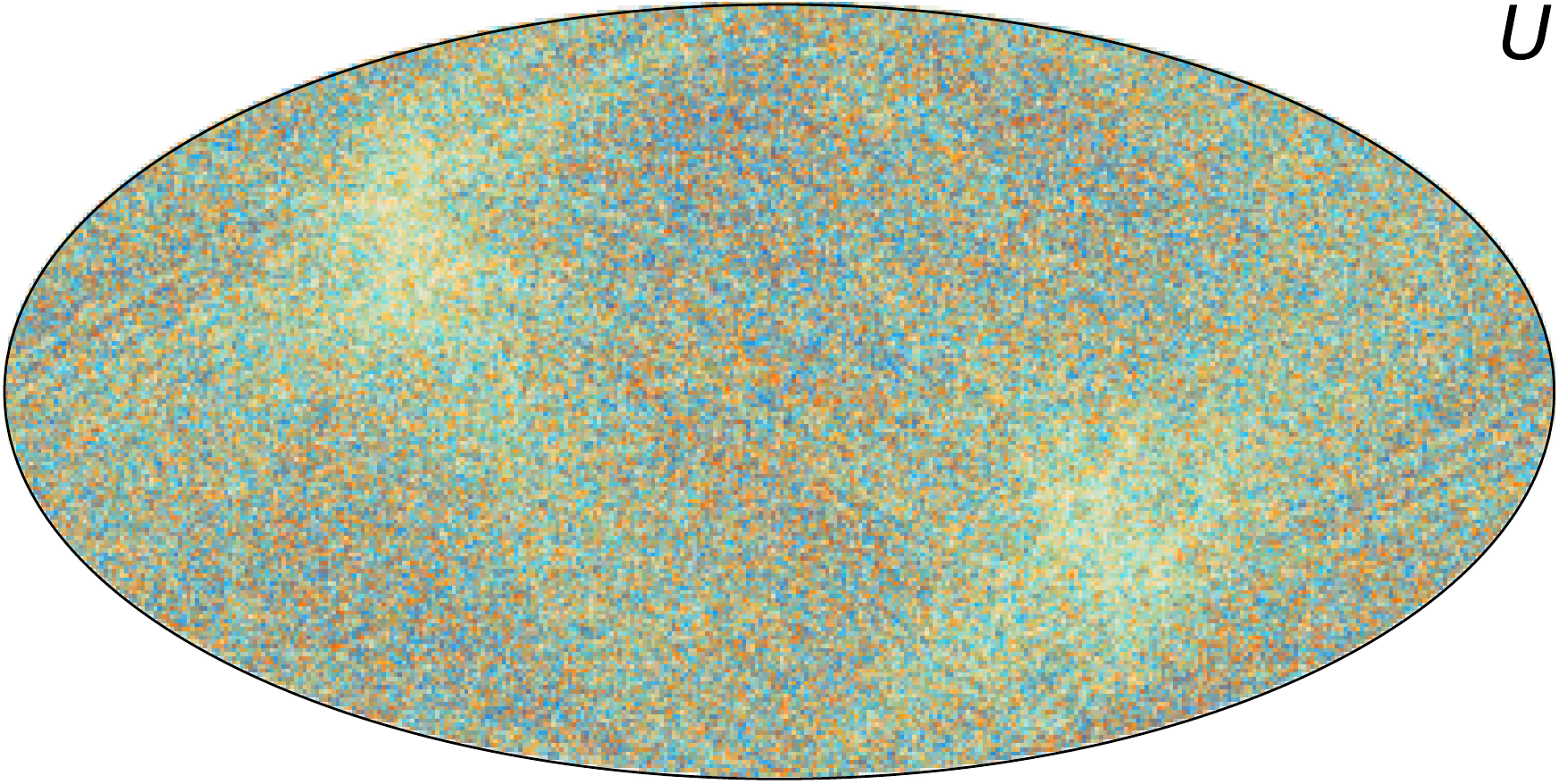}
  \includegraphics[width=0.33\linewidth]{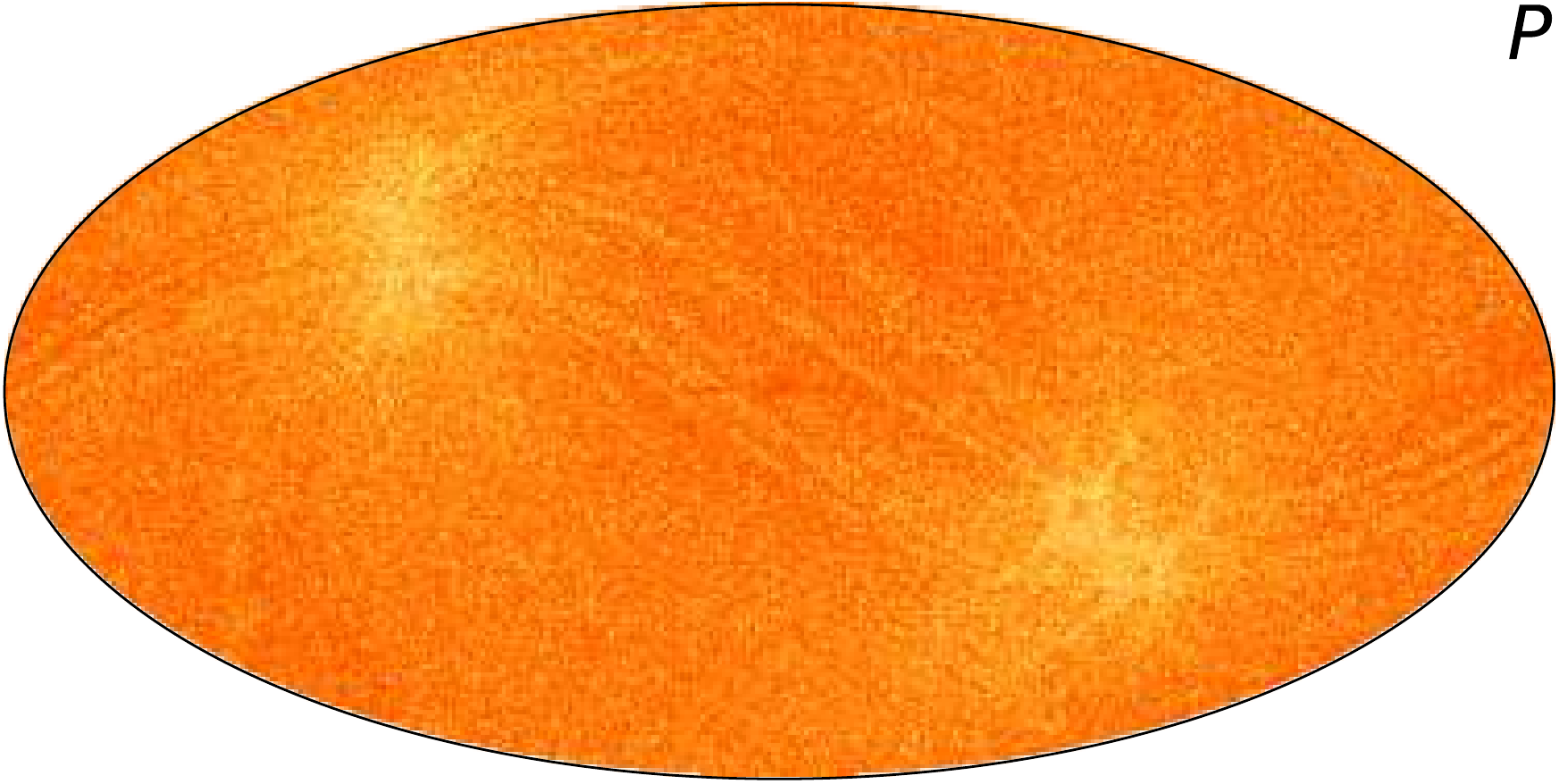}
  \\
  \includegraphics[width=0.33\linewidth]{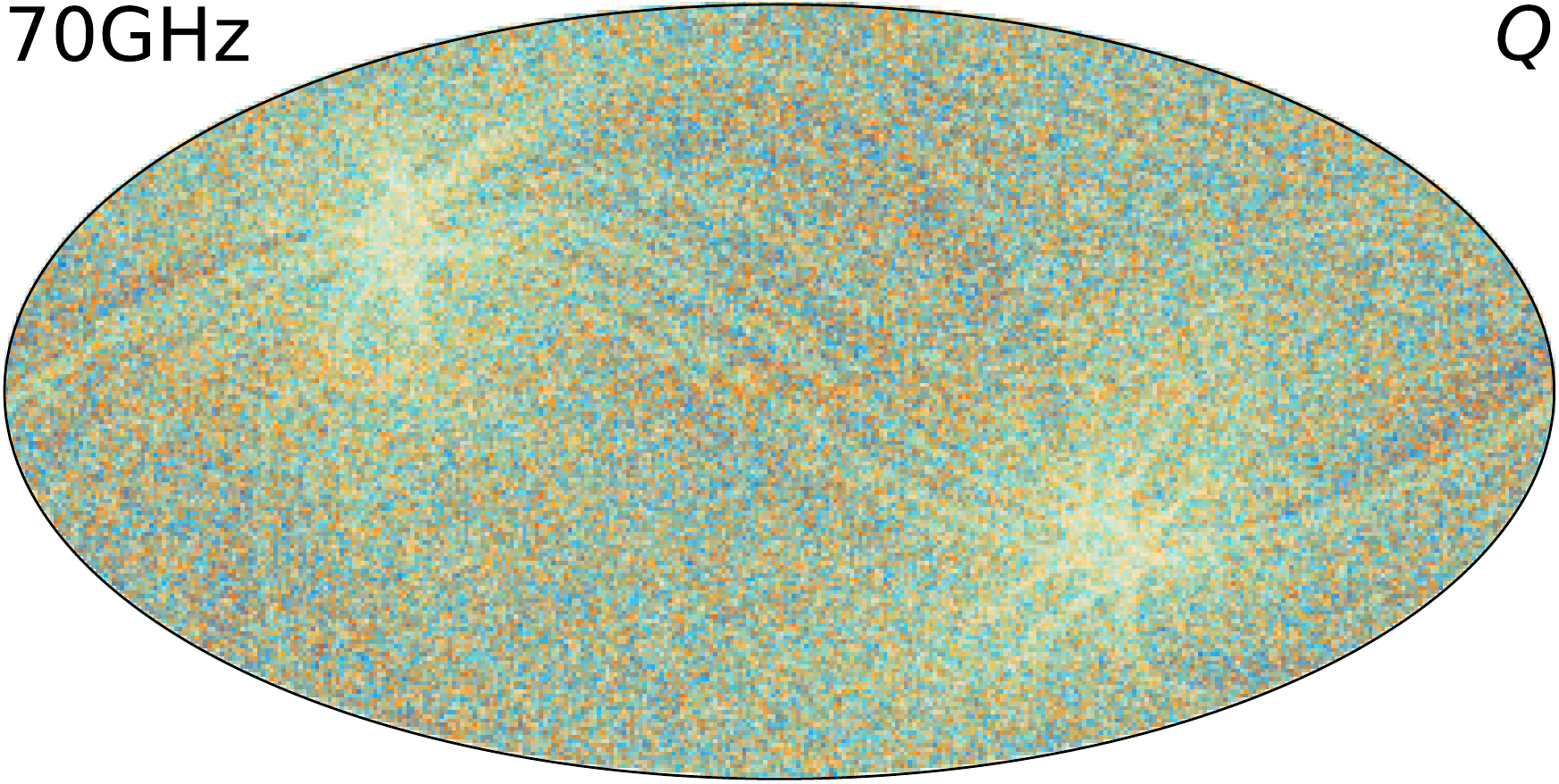}
  \includegraphics[width=0.33\linewidth]{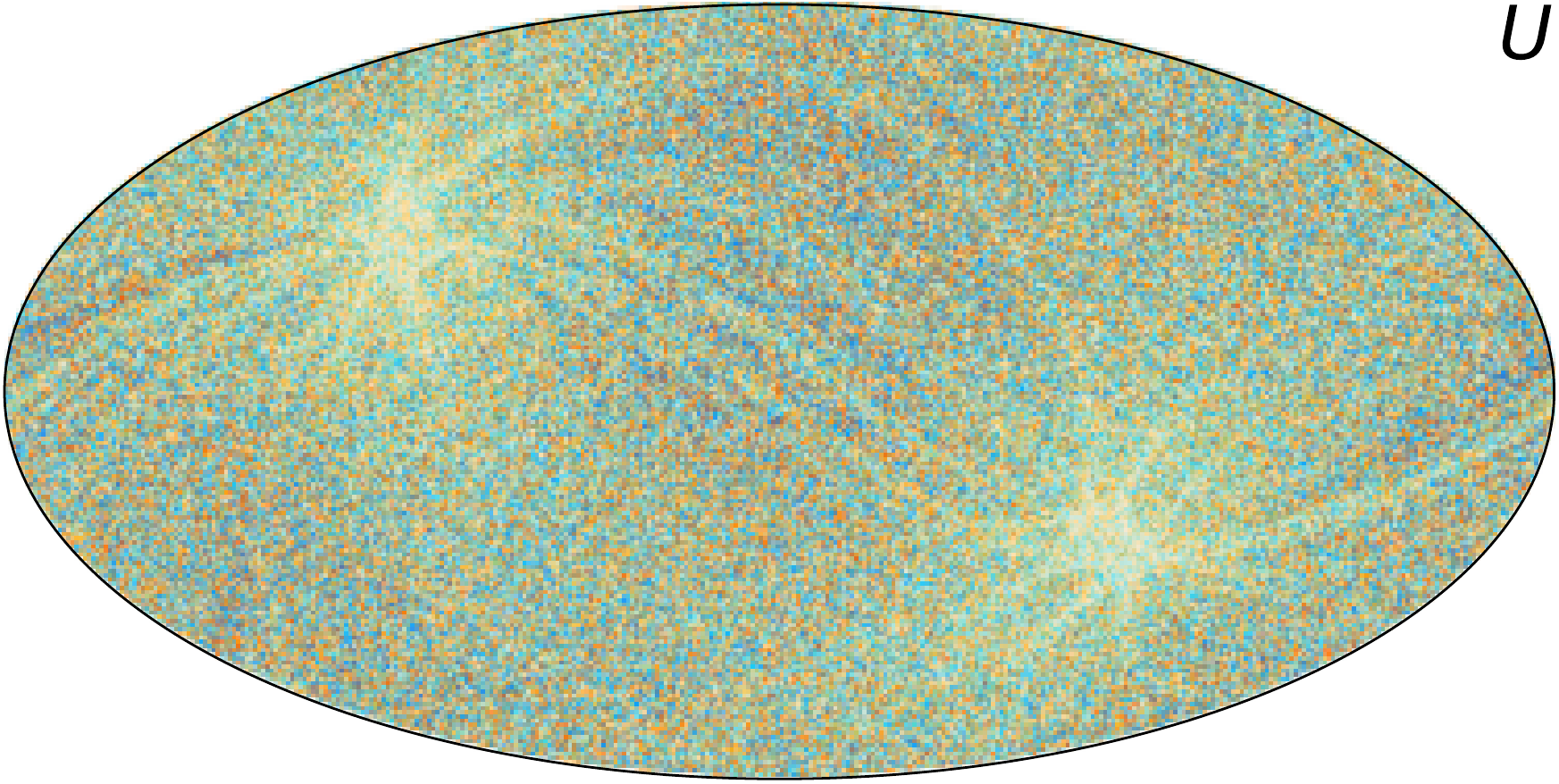}
  \includegraphics[width=0.33\linewidth]{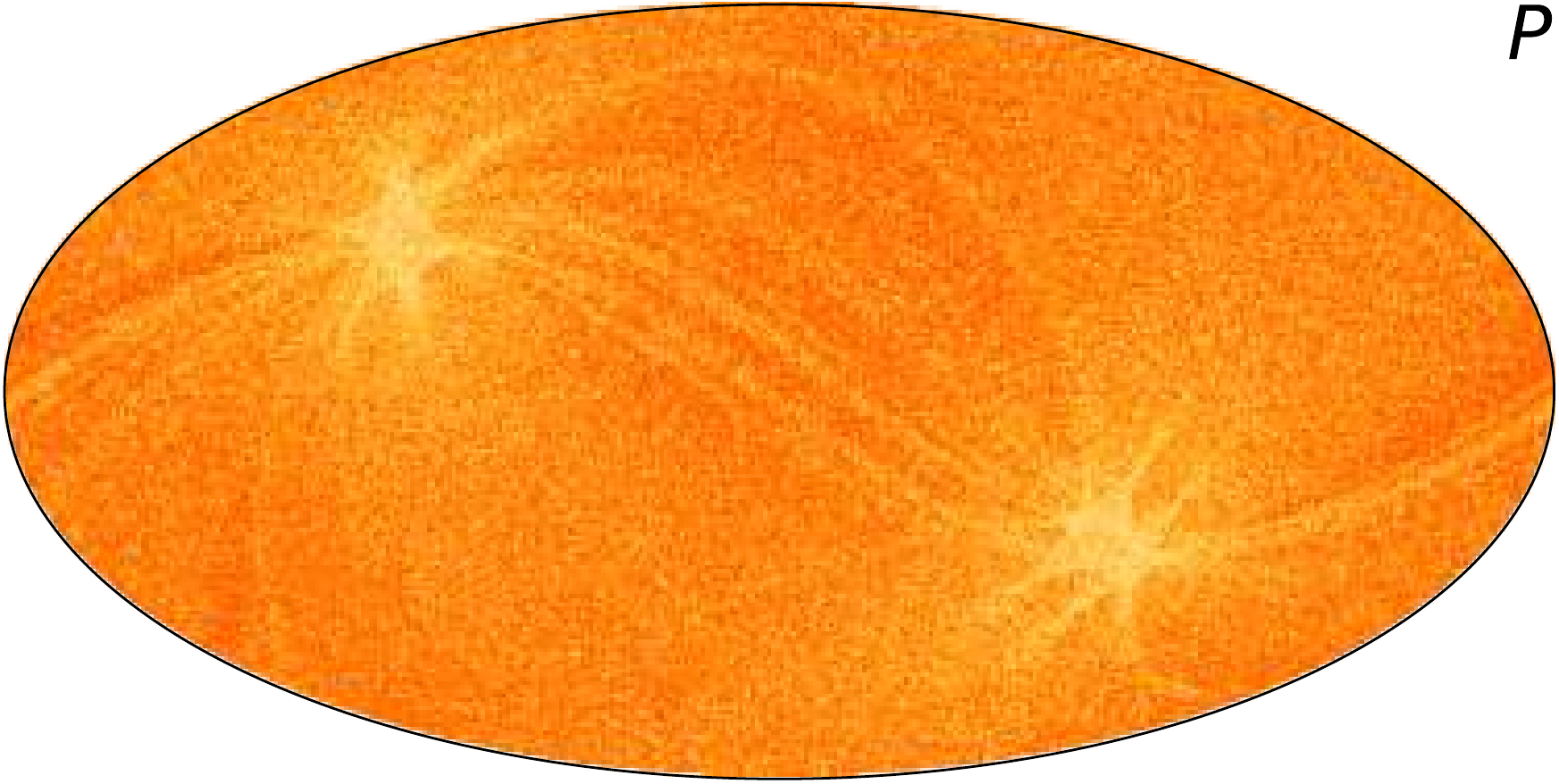}
  \\
  \includegraphics[width=0.33\linewidth]{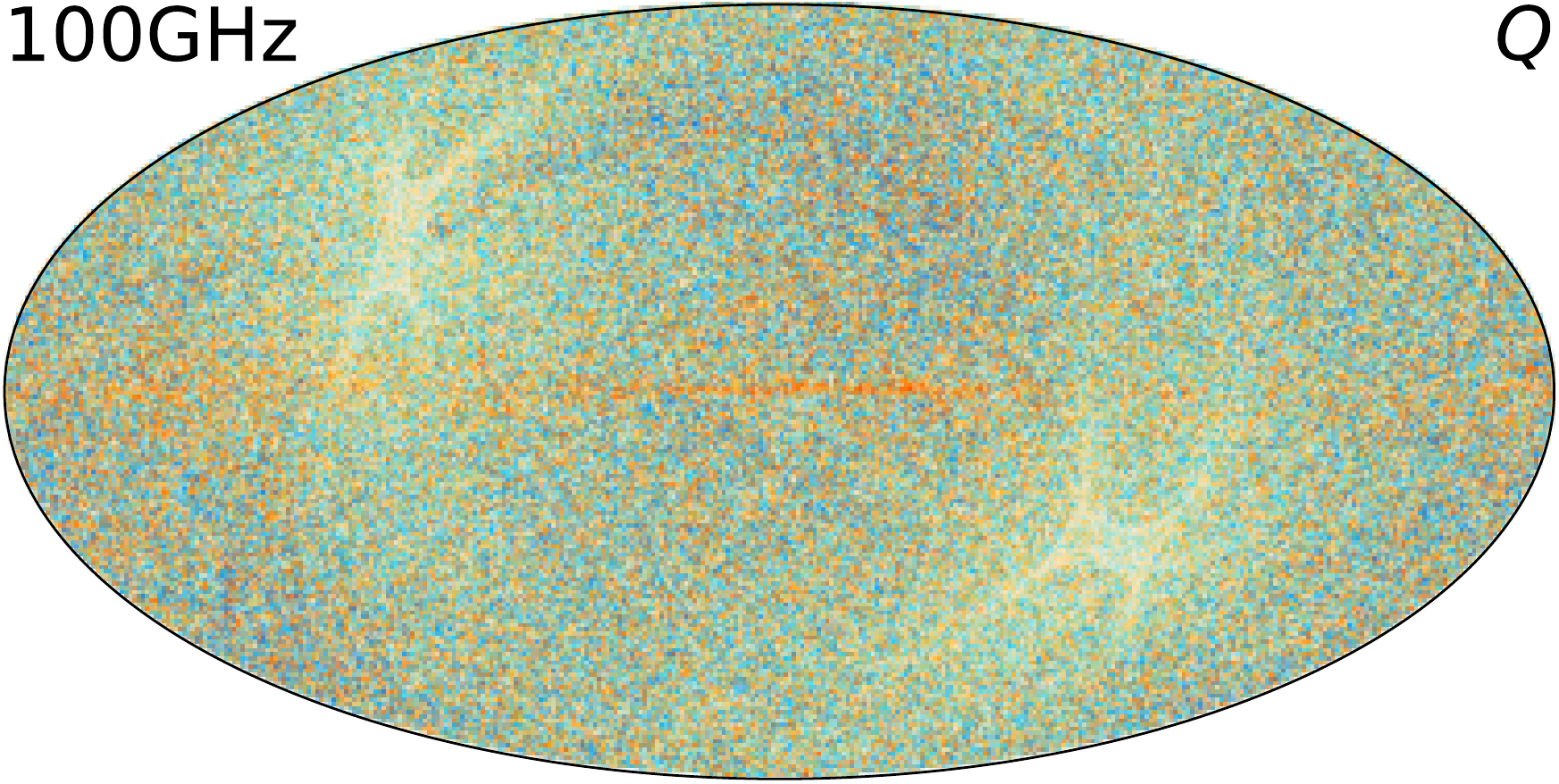}
  \includegraphics[width=0.33\linewidth]{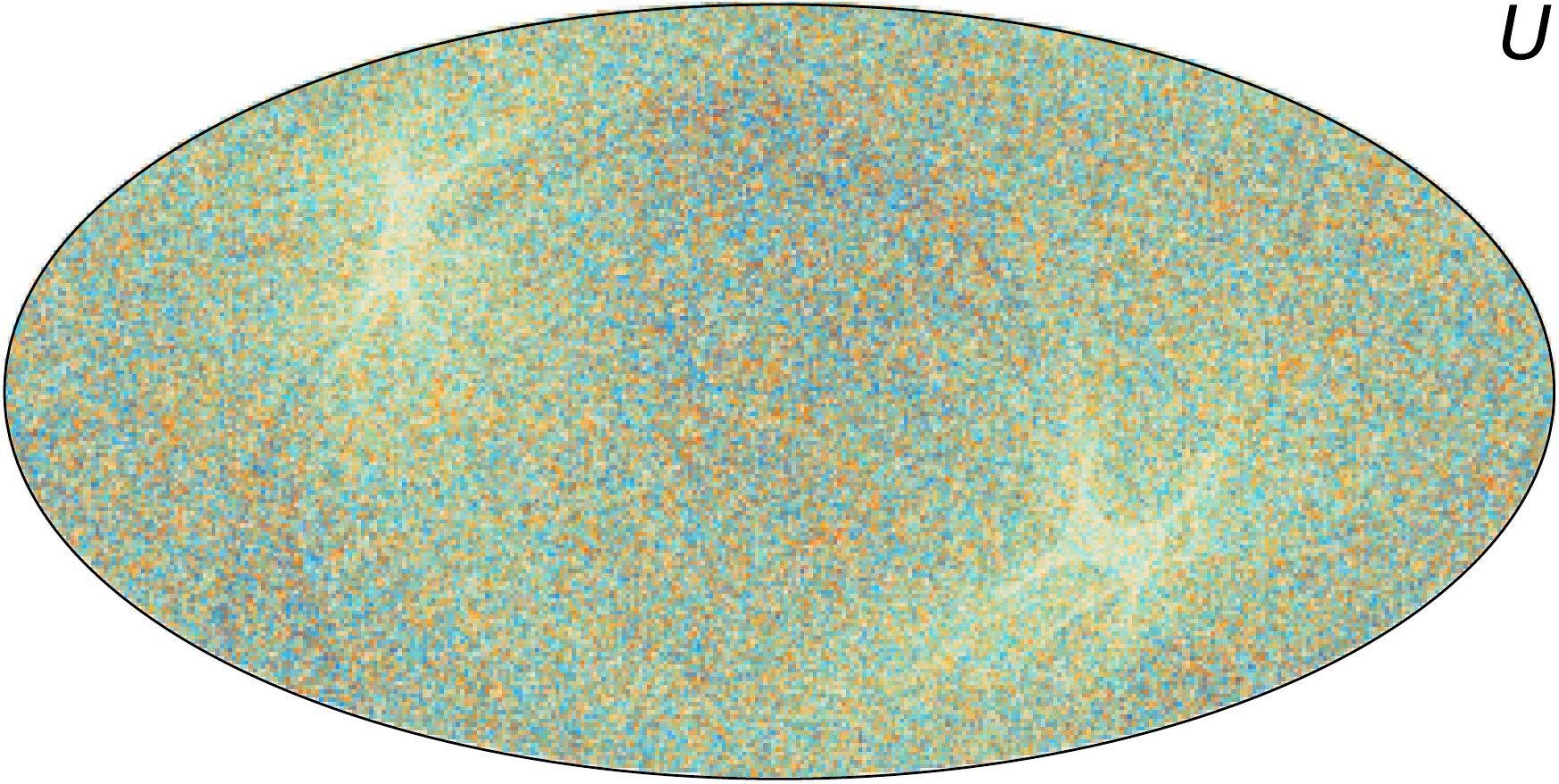}
  \includegraphics[width=0.33\linewidth]{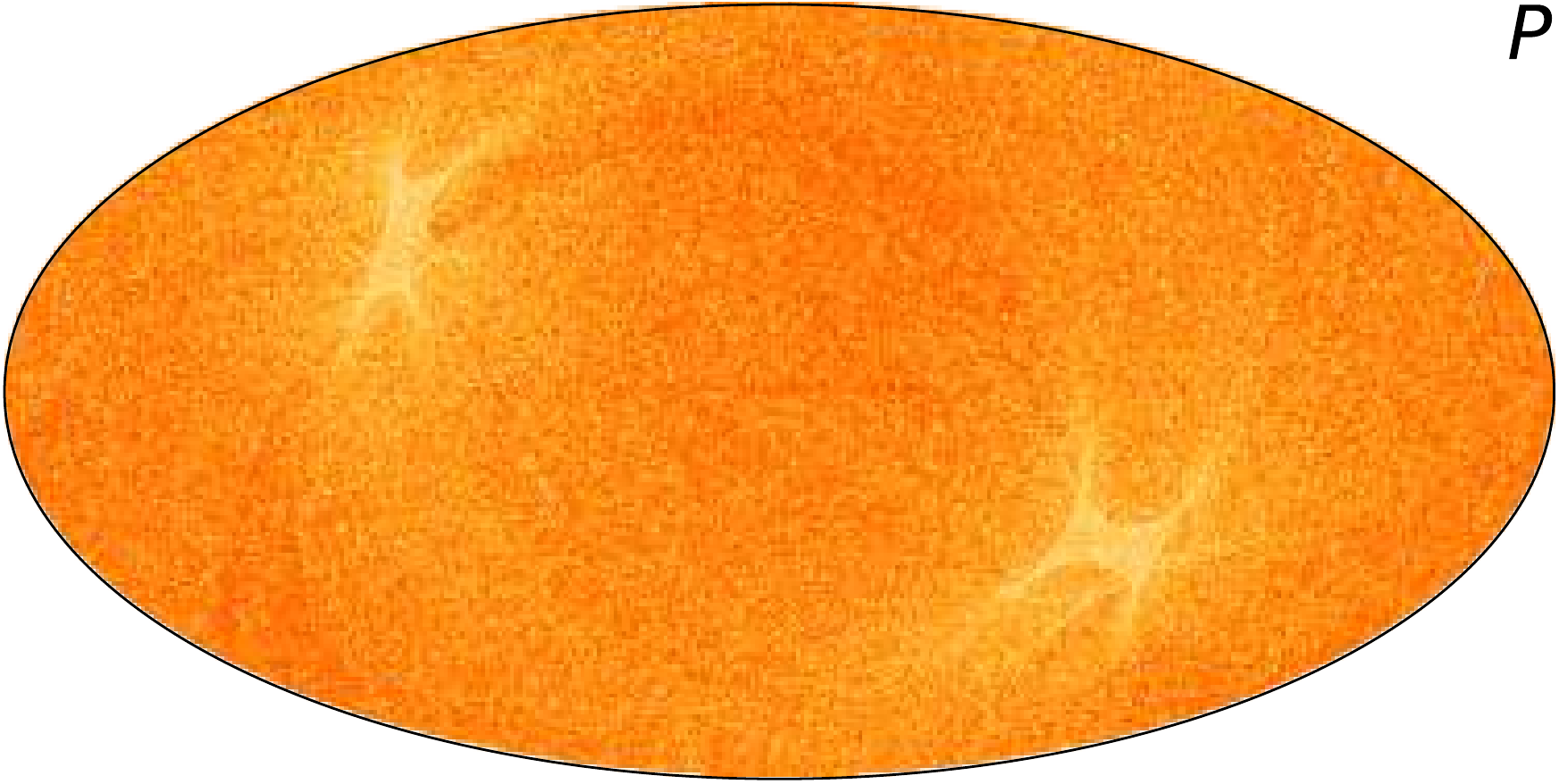}
  \\
  \includegraphics[width=0.33\linewidth]{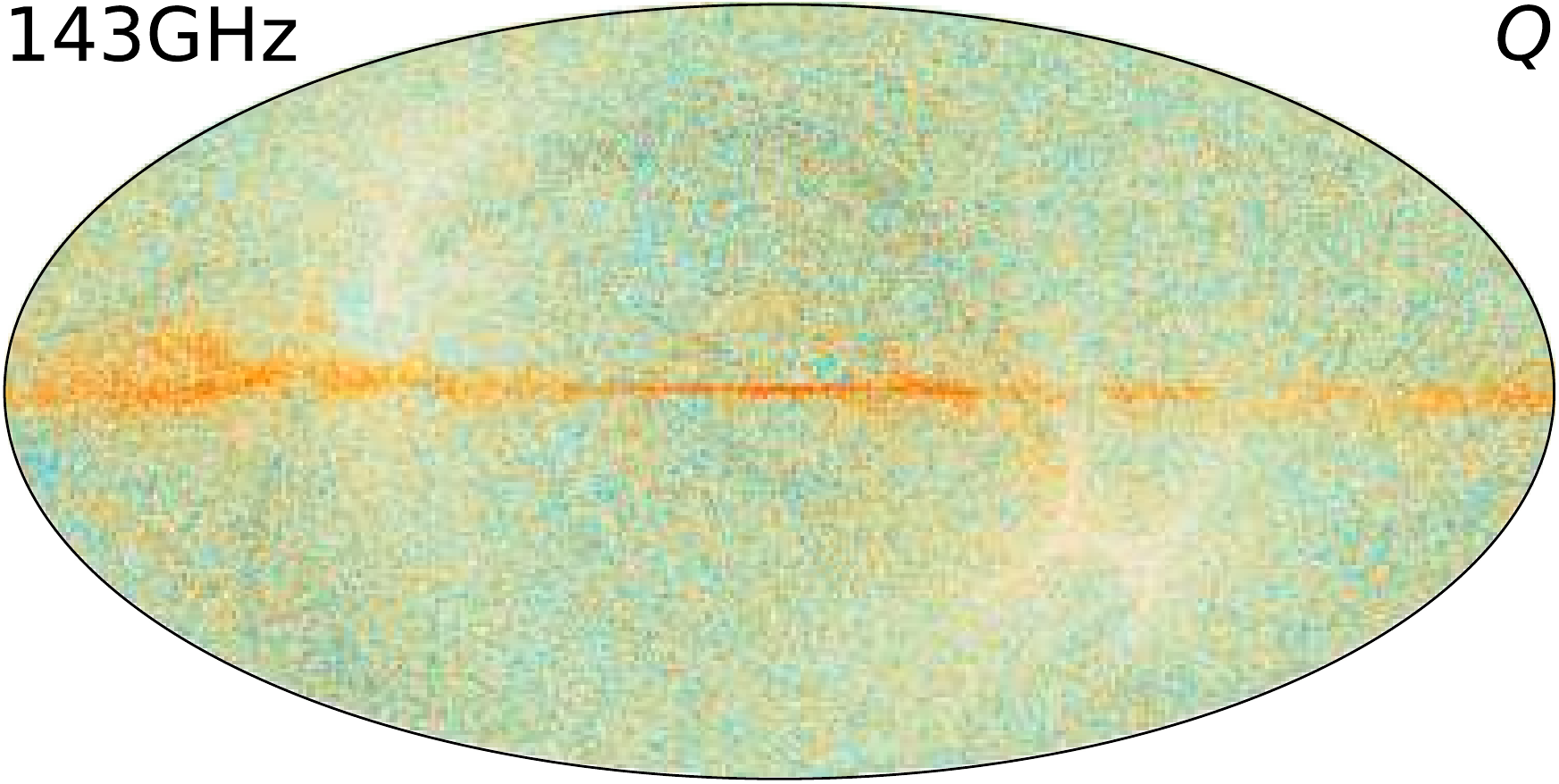}
  \includegraphics[width=0.33\linewidth]{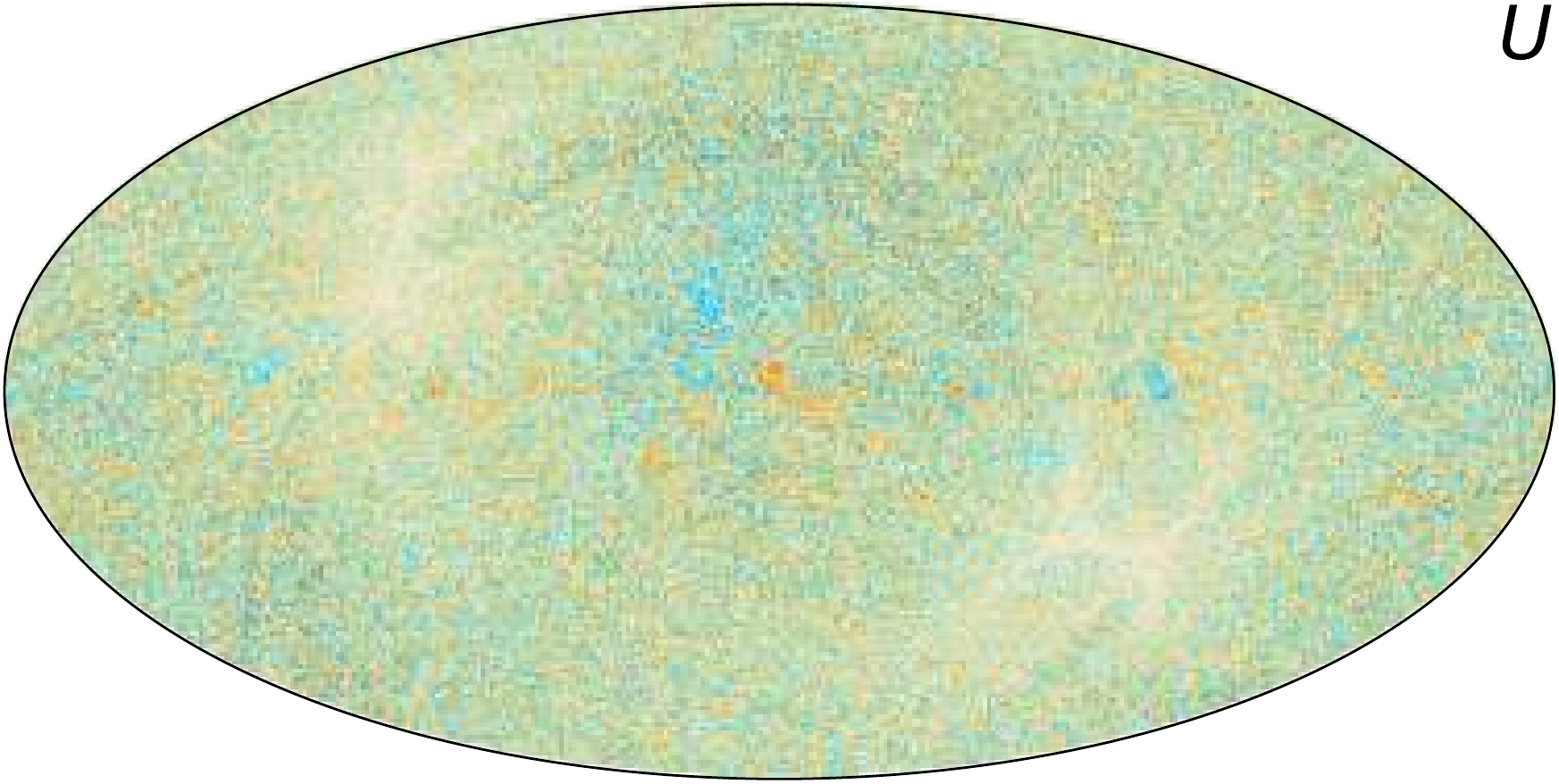}
  \includegraphics[width=0.33\linewidth]{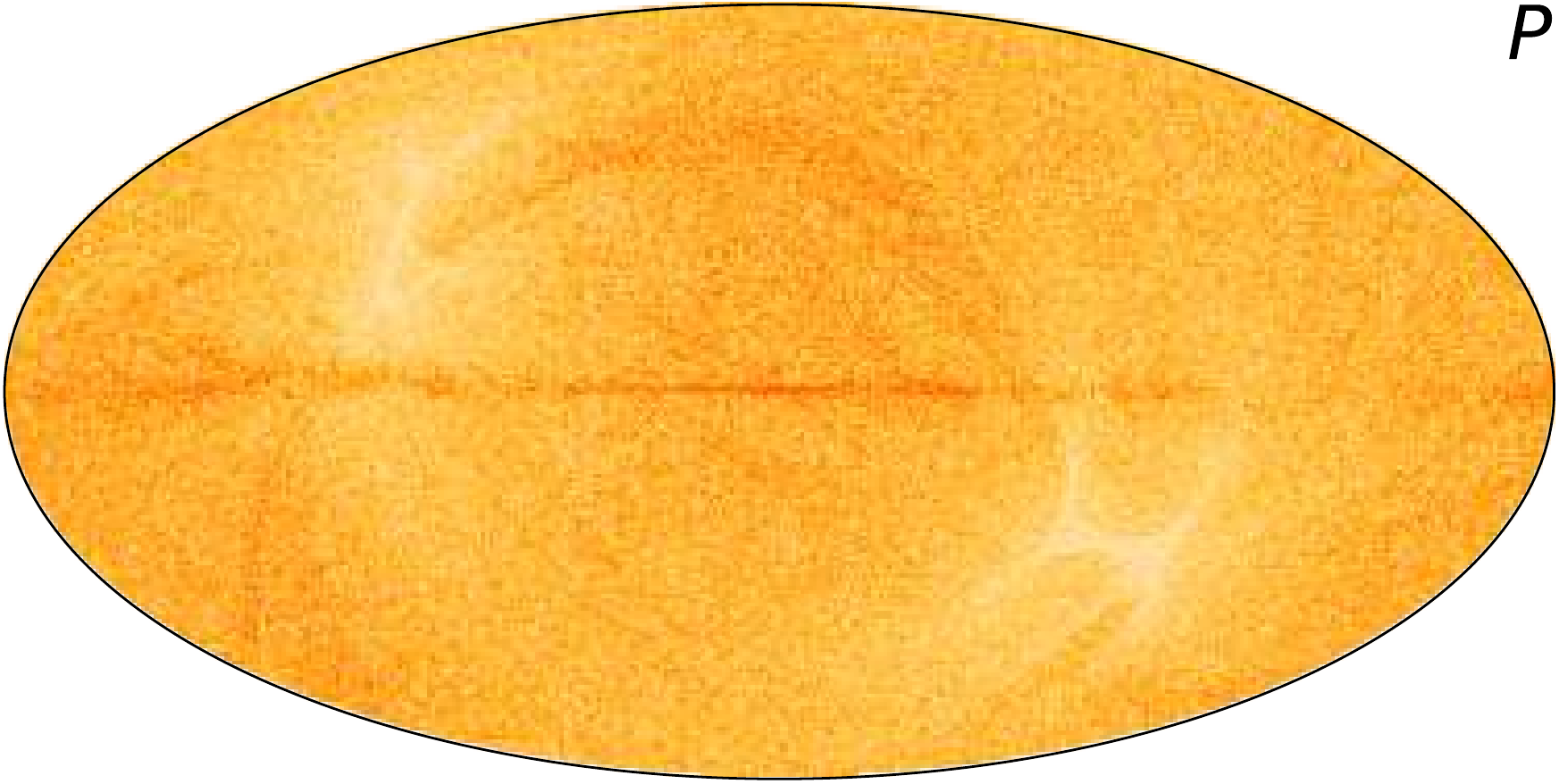}
  \\
  \includegraphics[width=0.33\linewidth]{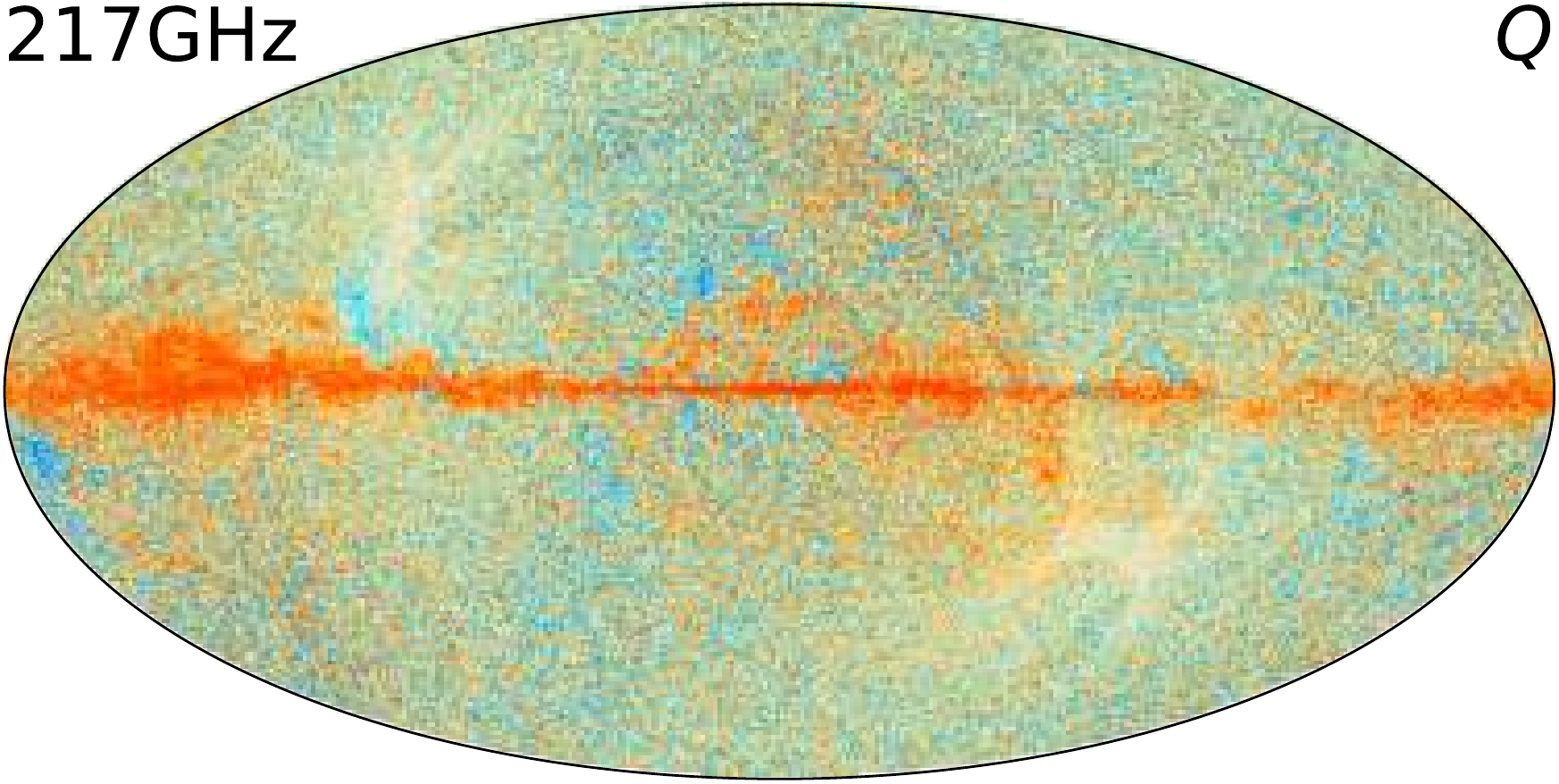}
  \includegraphics[width=0.33\linewidth]{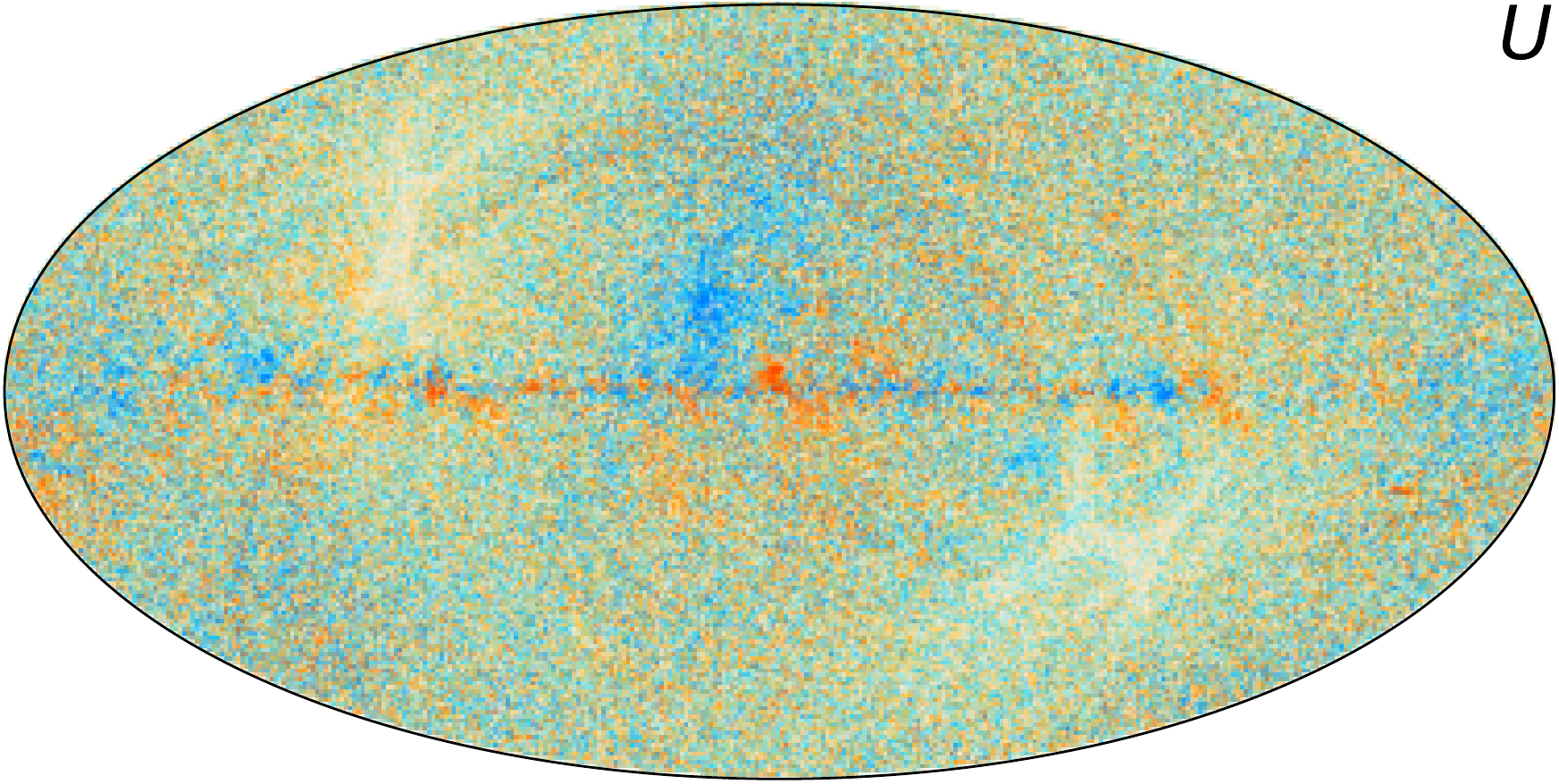}
  \includegraphics[width=0.33\linewidth]{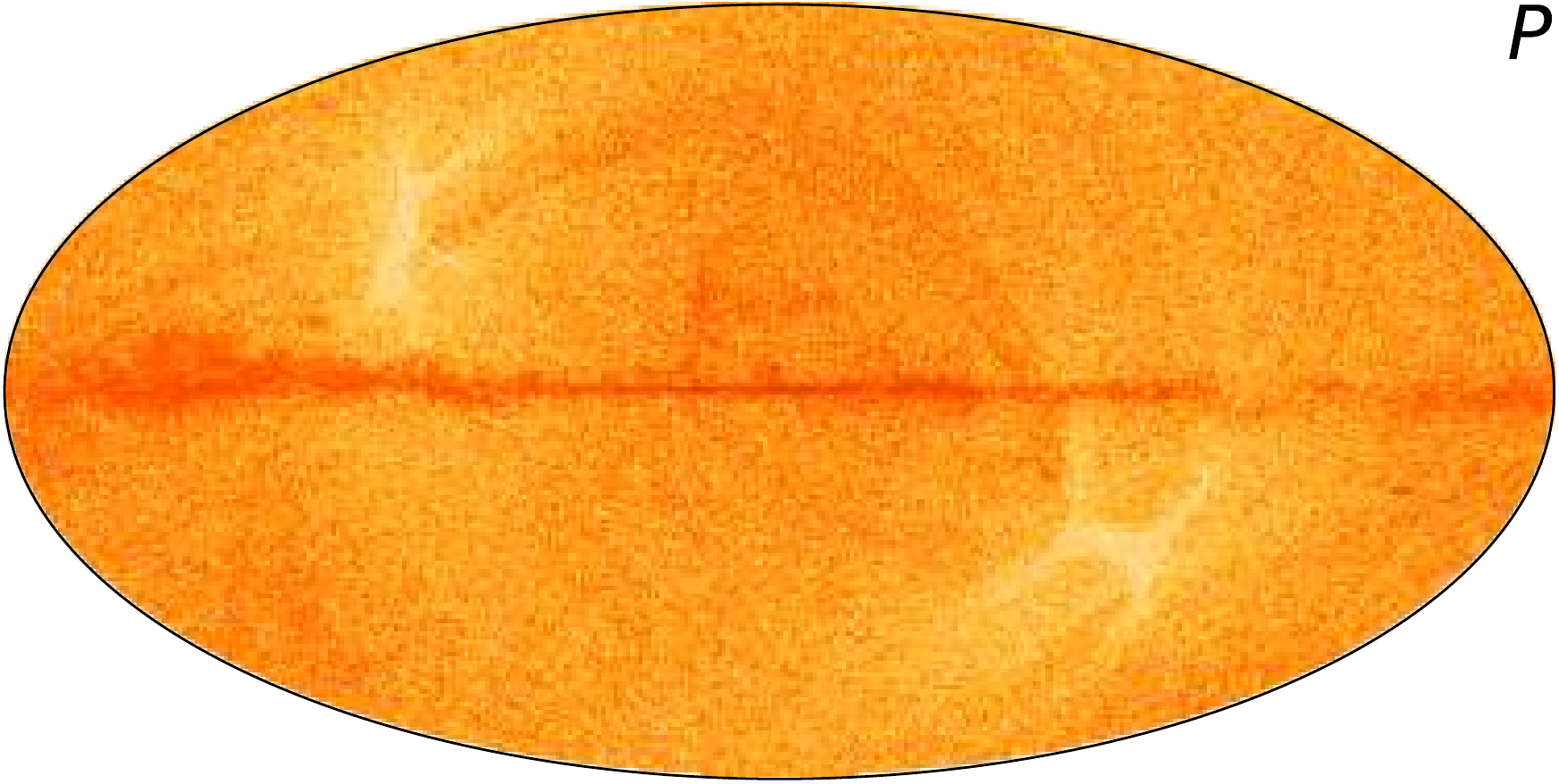}
  \\
  \includegraphics[width=0.33\linewidth]{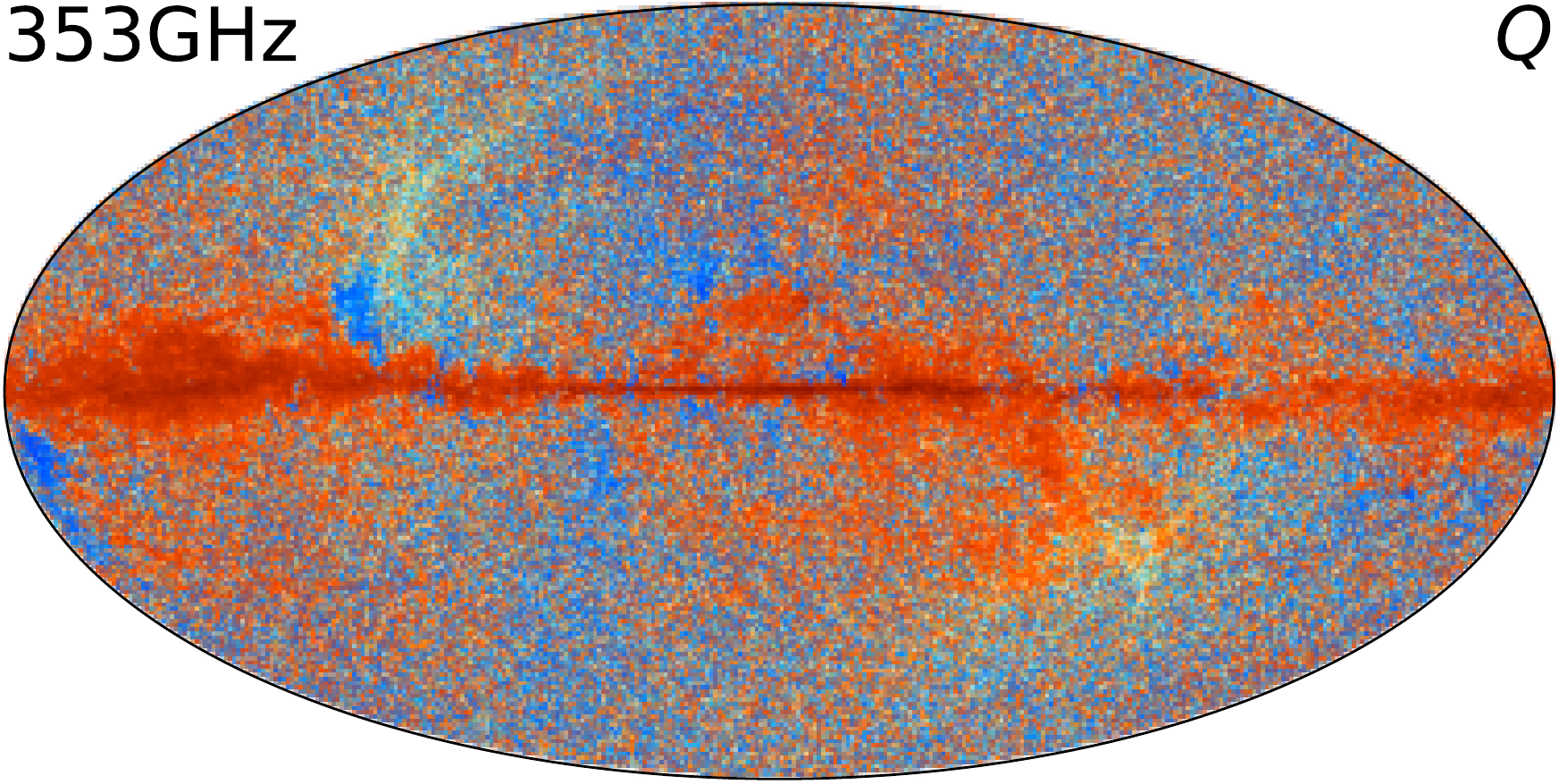}
  \includegraphics[width=0.33\linewidth]{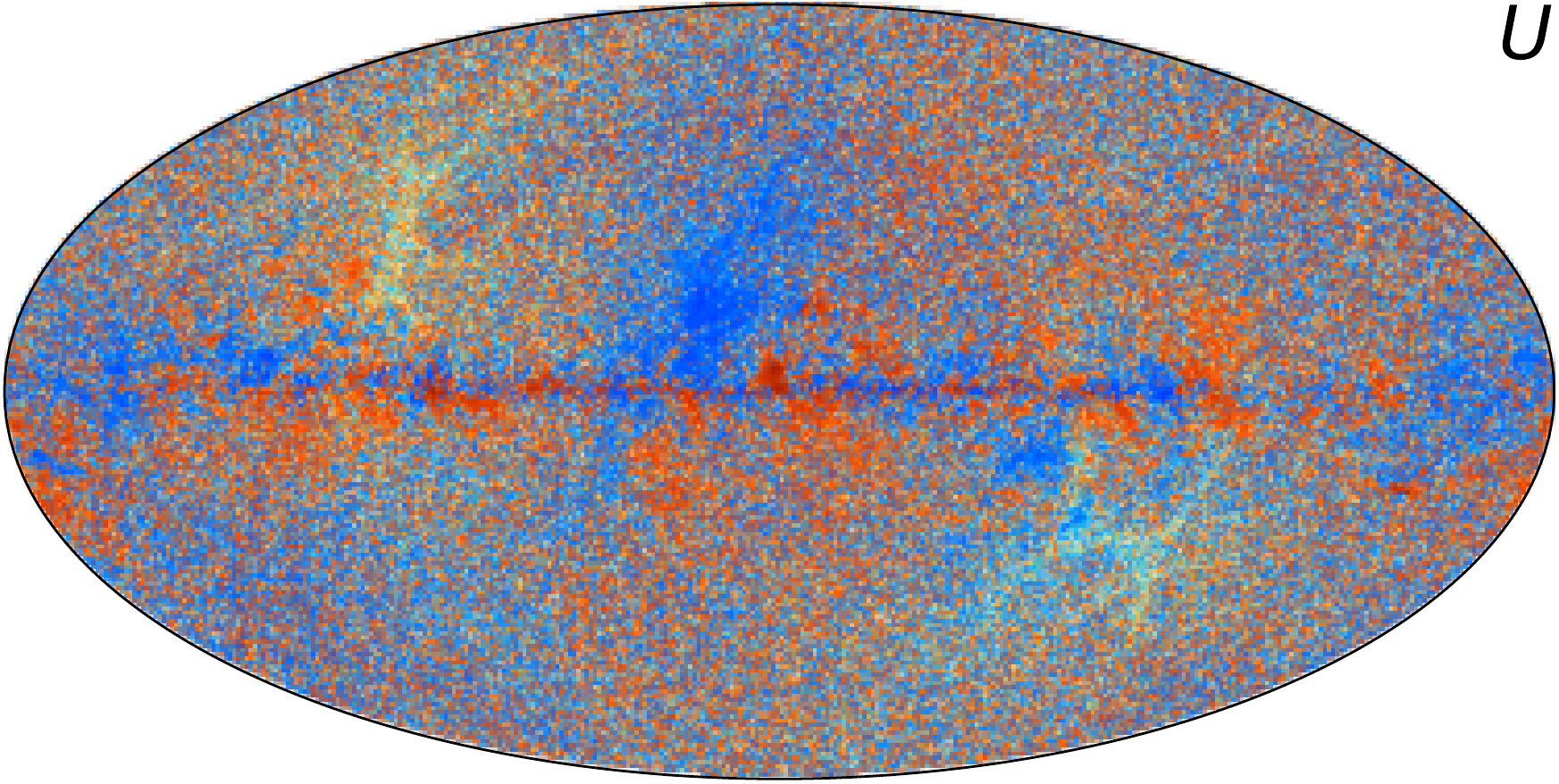}
  \includegraphics[width=0.33\linewidth]{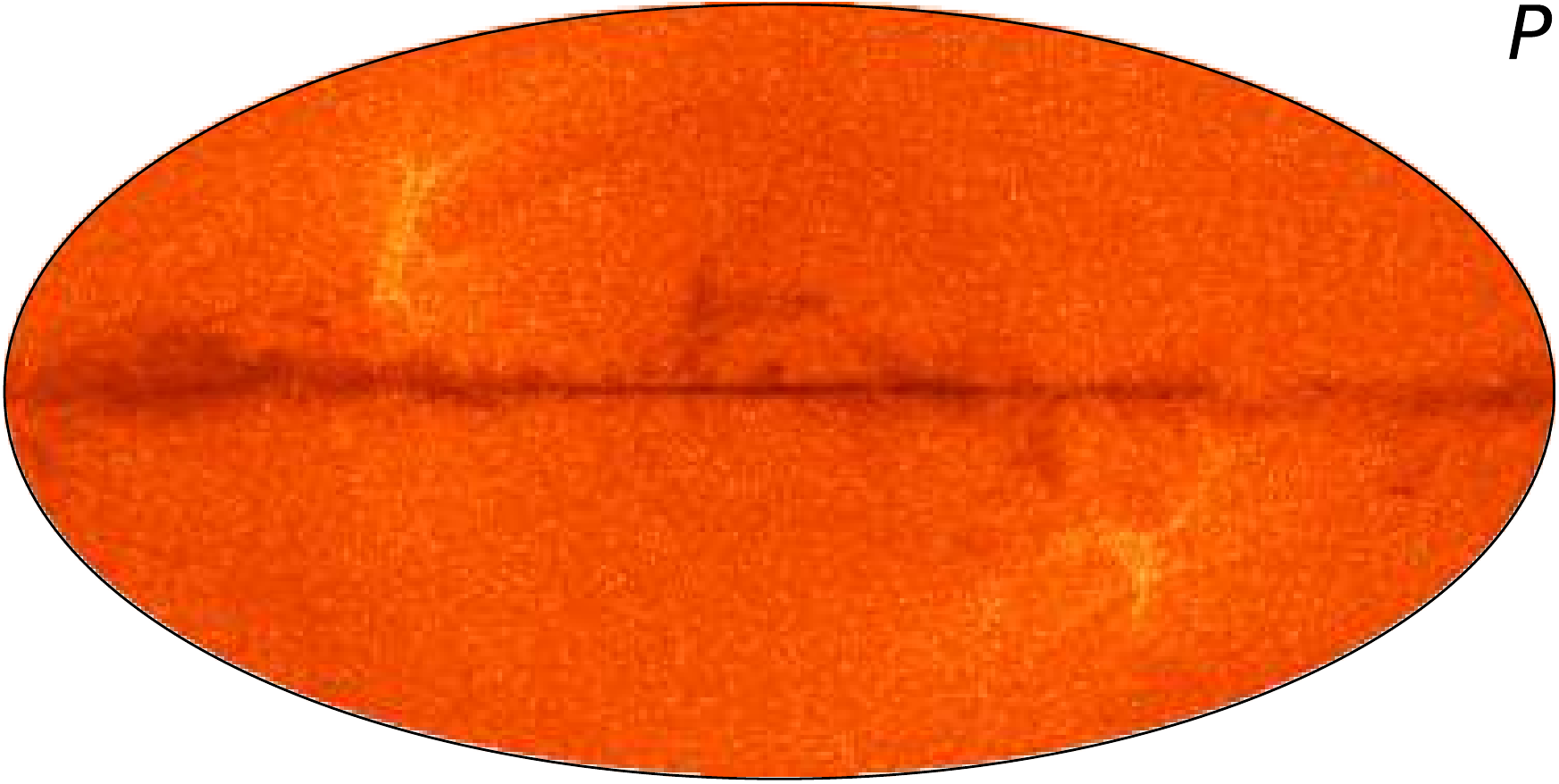}
  \\
  \includegraphics[width=1.0\linewidth]{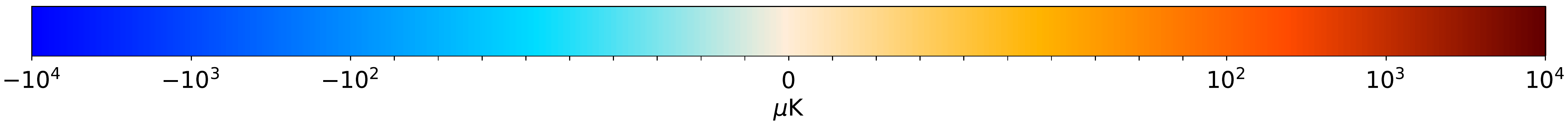}
  \caption{\npipe\ polarization maps.  The scaling is linear between $-100$ and $100\,\mu$K.  Here $P=(Q^2+U^2)^{1/2}$.
  }
  \label{fig:freqmaps_P}
\end{figure*}

\begin{figure*}[htpb!]
  \includegraphics[width=0.33\linewidth]{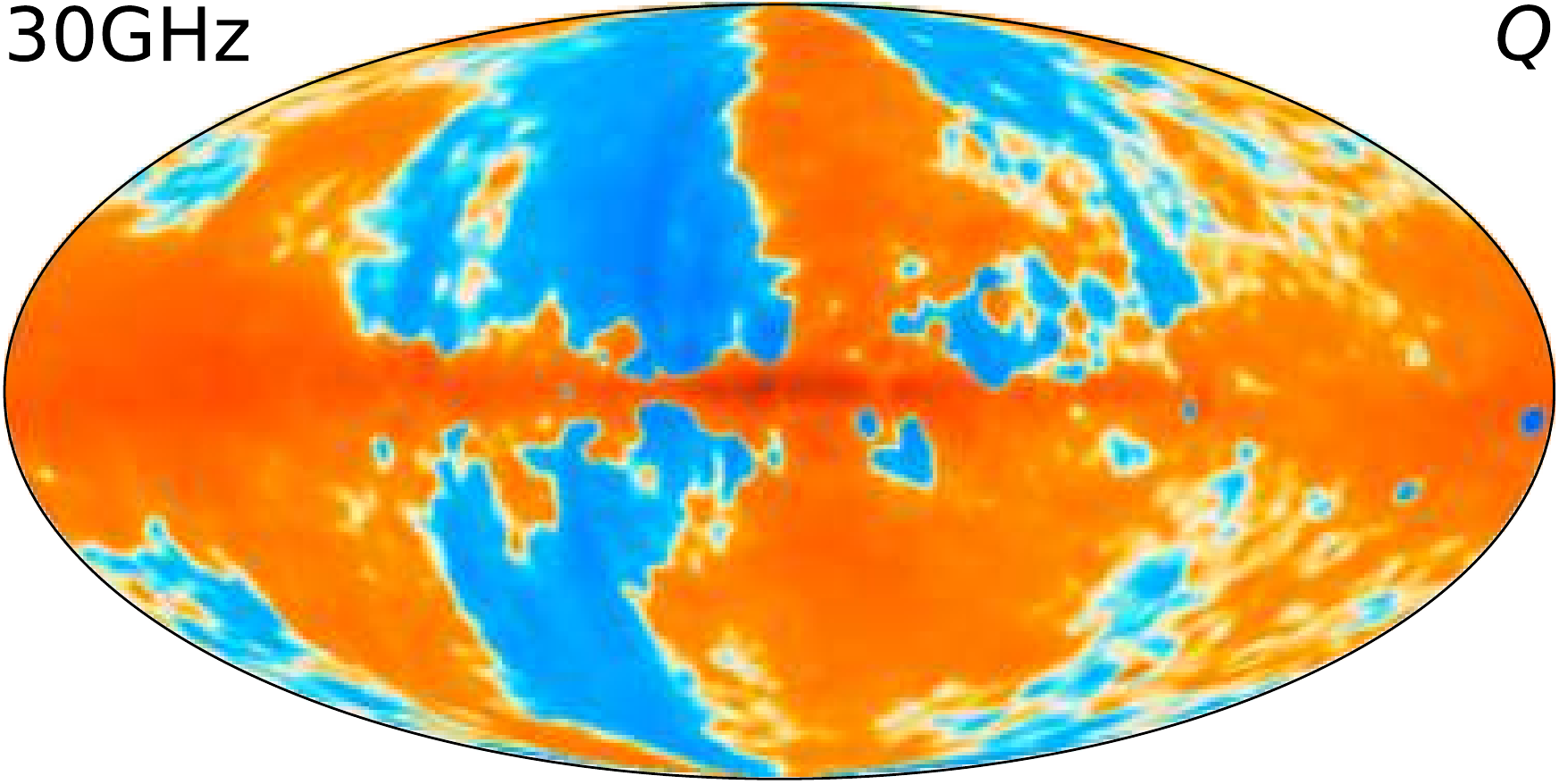}
  \includegraphics[width=0.33\linewidth]{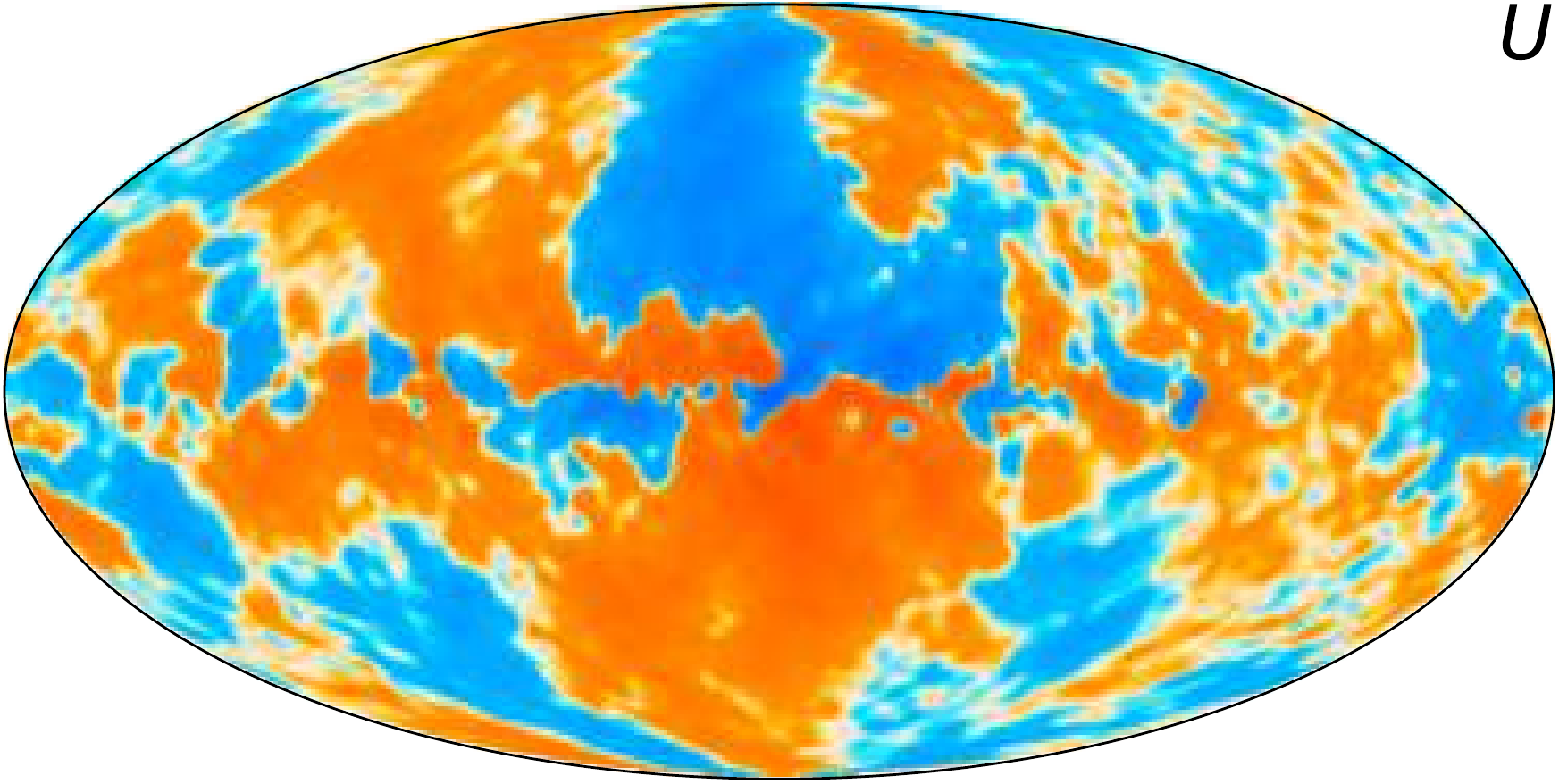}
  \includegraphics[width=0.33\linewidth]{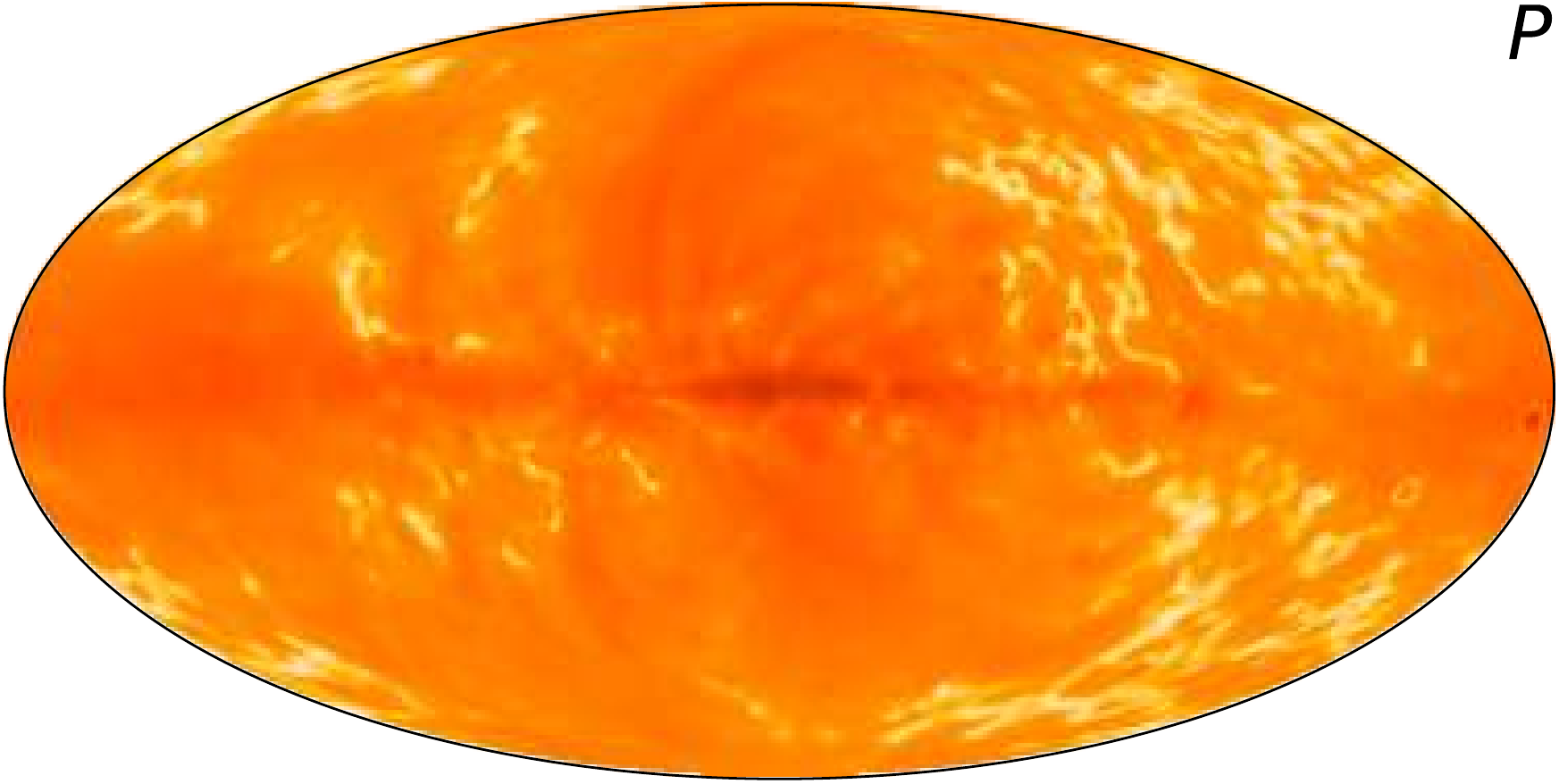}
  \\
  \includegraphics[width=0.33\linewidth]{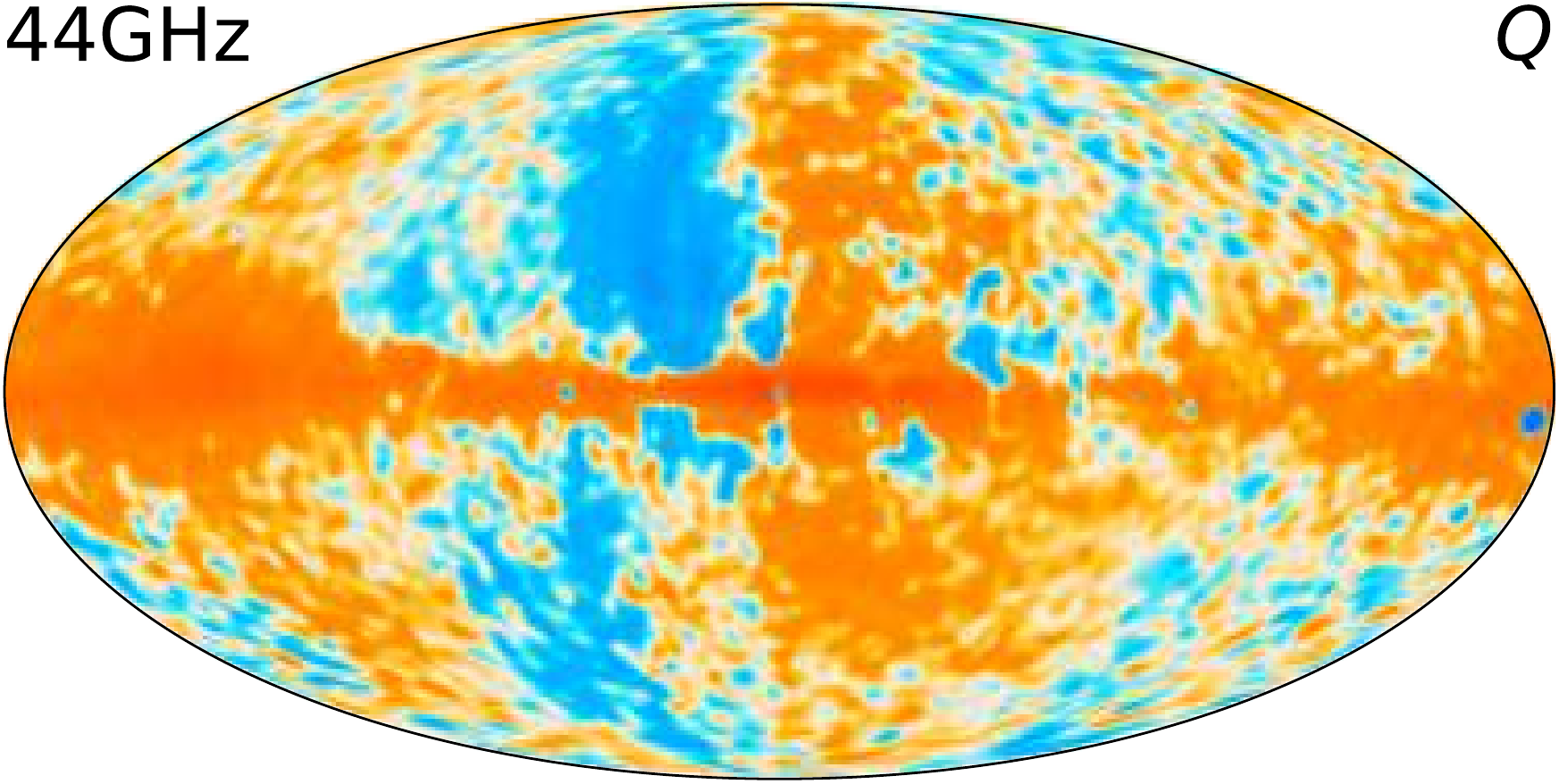}
  \includegraphics[width=0.33\linewidth]{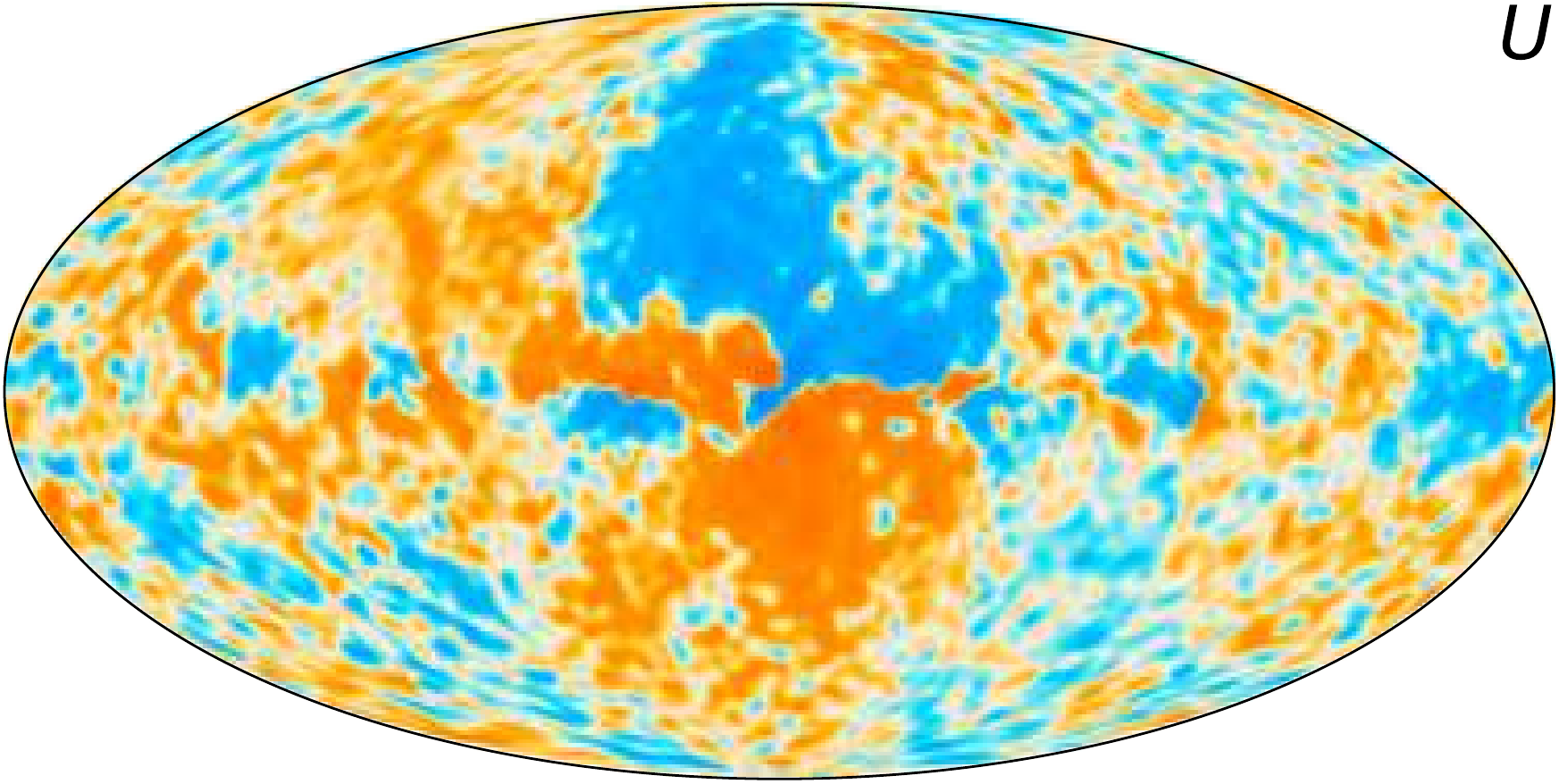}
  \includegraphics[width=0.33\linewidth]{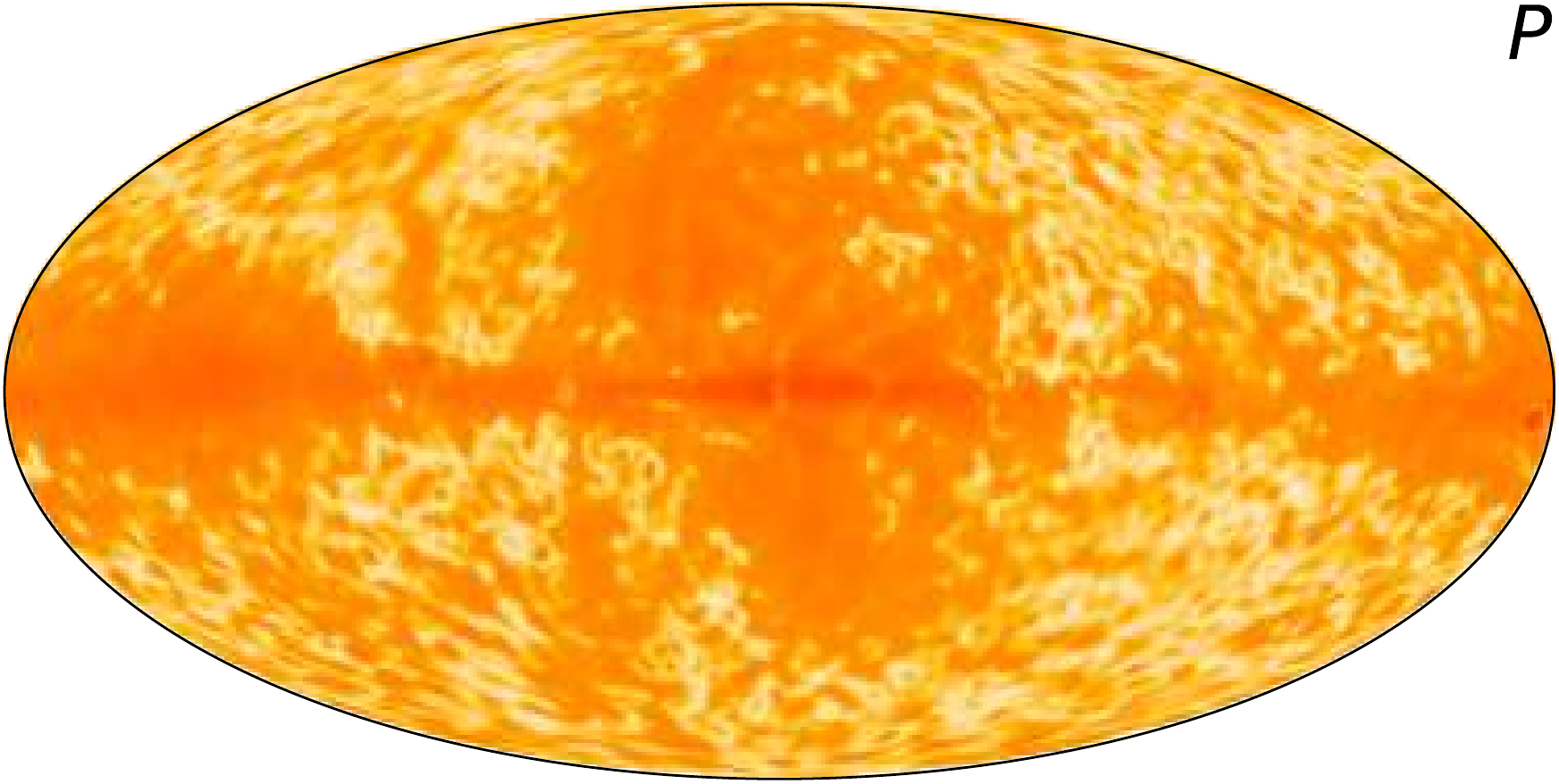}
  \\
  \includegraphics[width=0.33\linewidth]{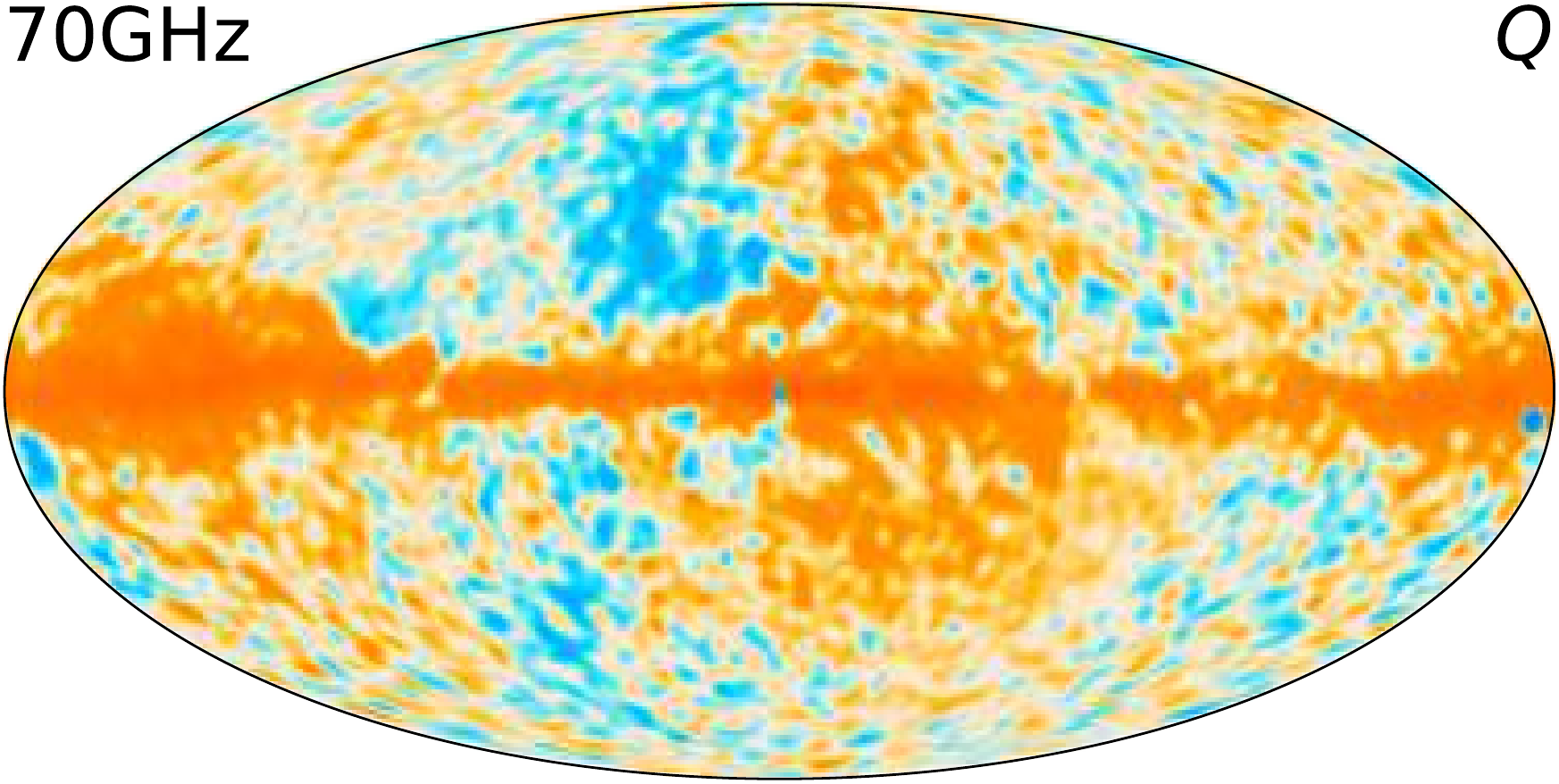}
  \includegraphics[width=0.33\linewidth]{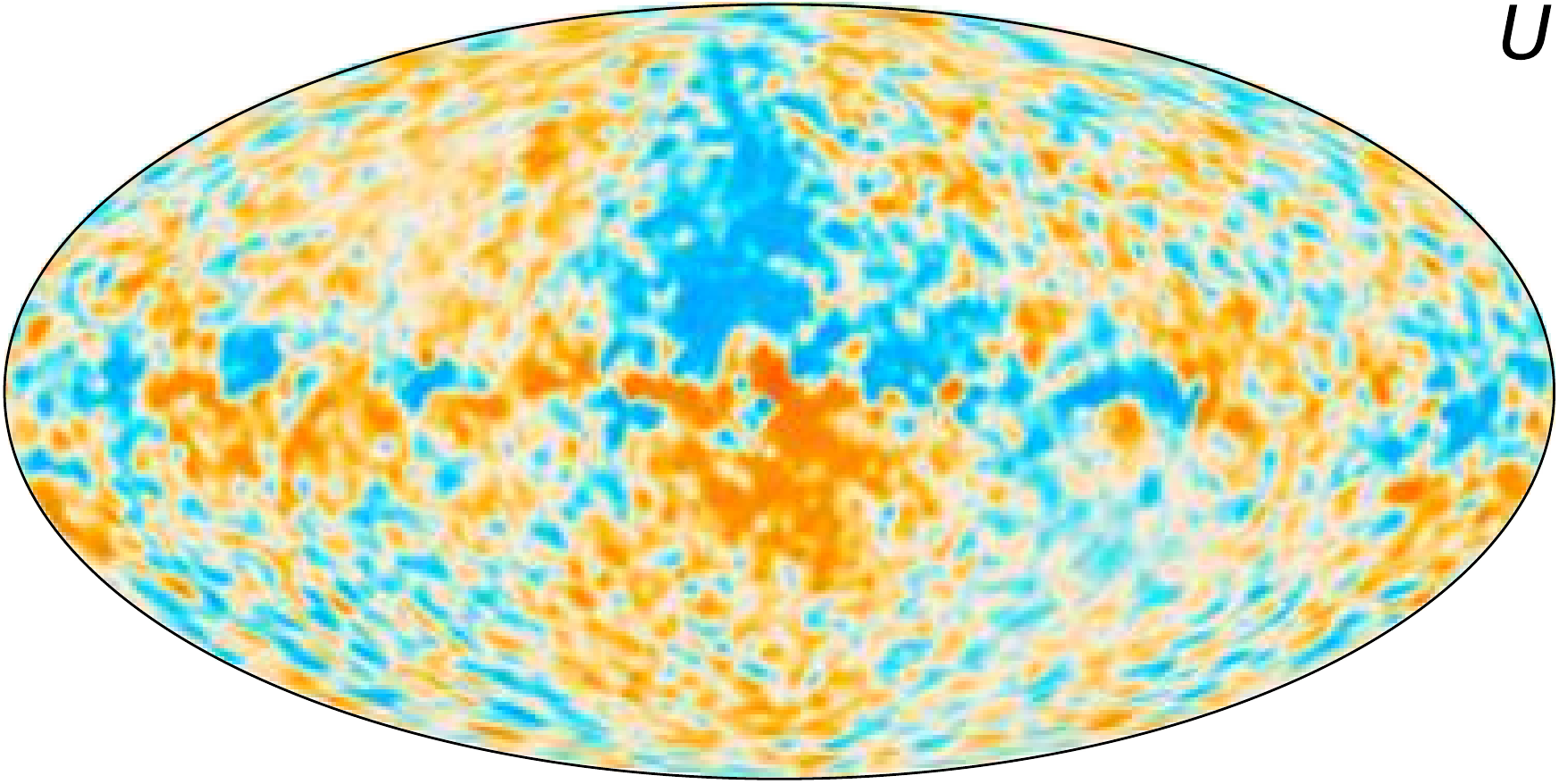}
  \includegraphics[width=0.33\linewidth]{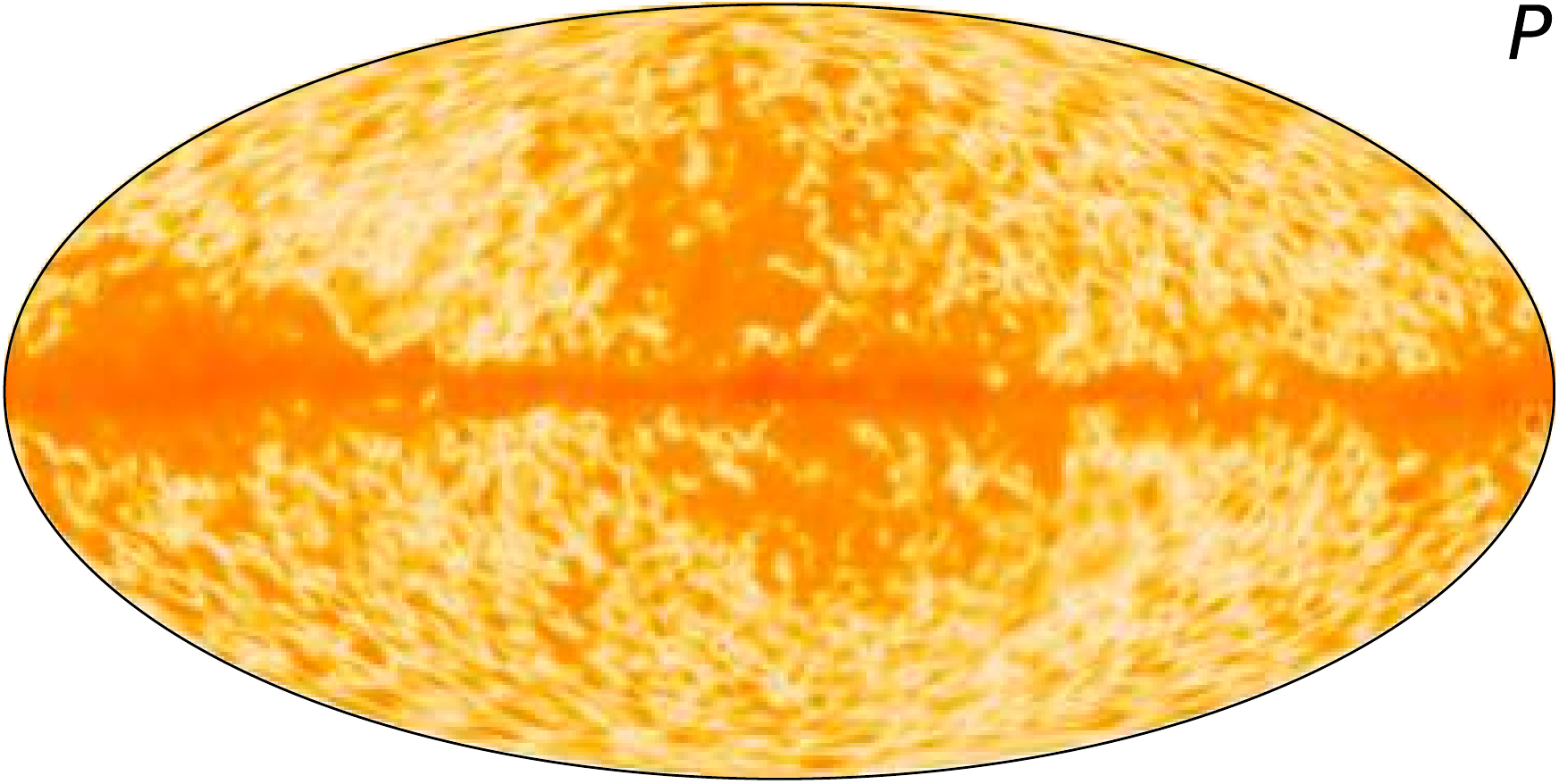}
  \\
  \includegraphics[width=0.33\linewidth]{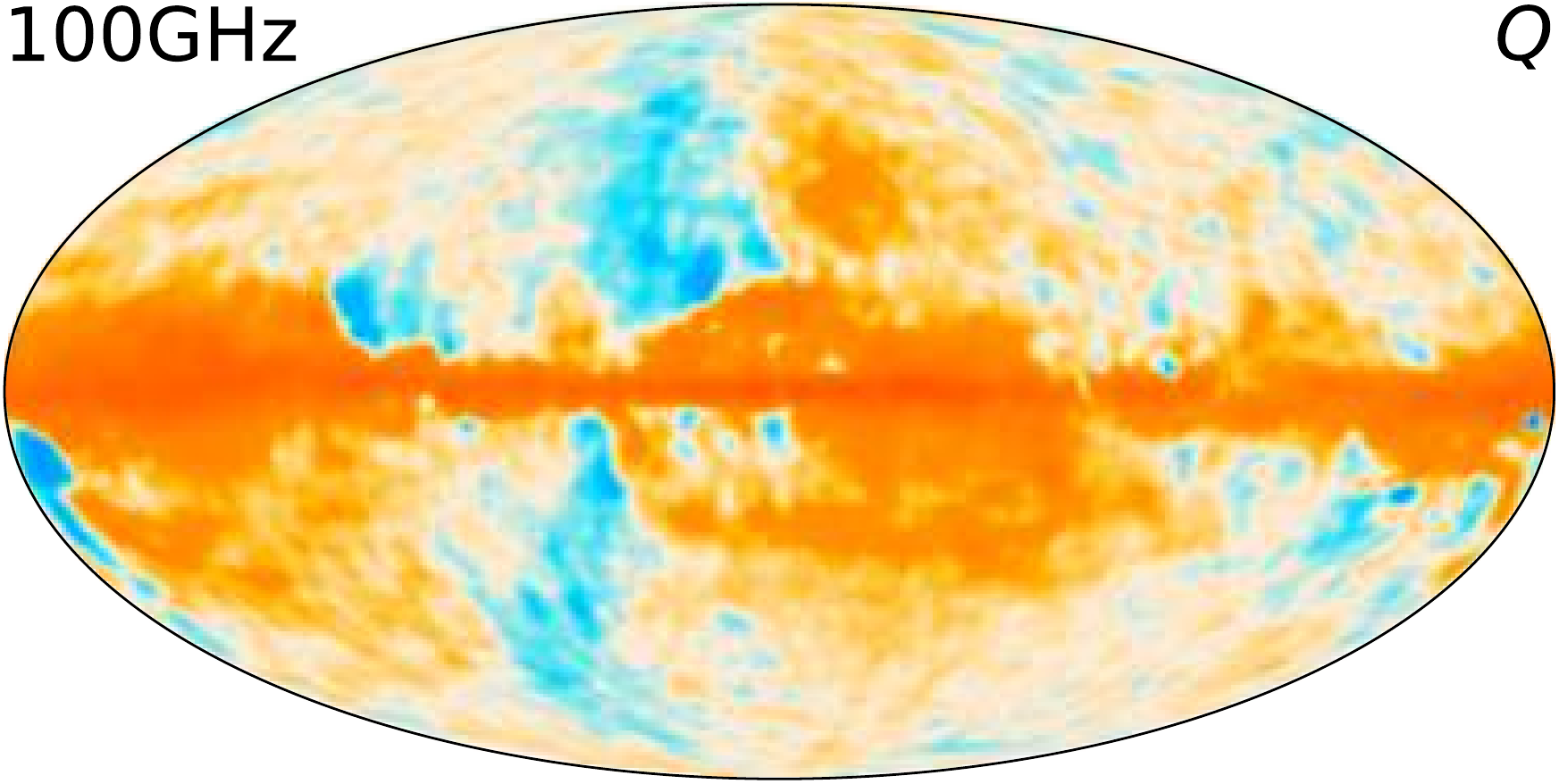}
  \includegraphics[width=0.33\linewidth]{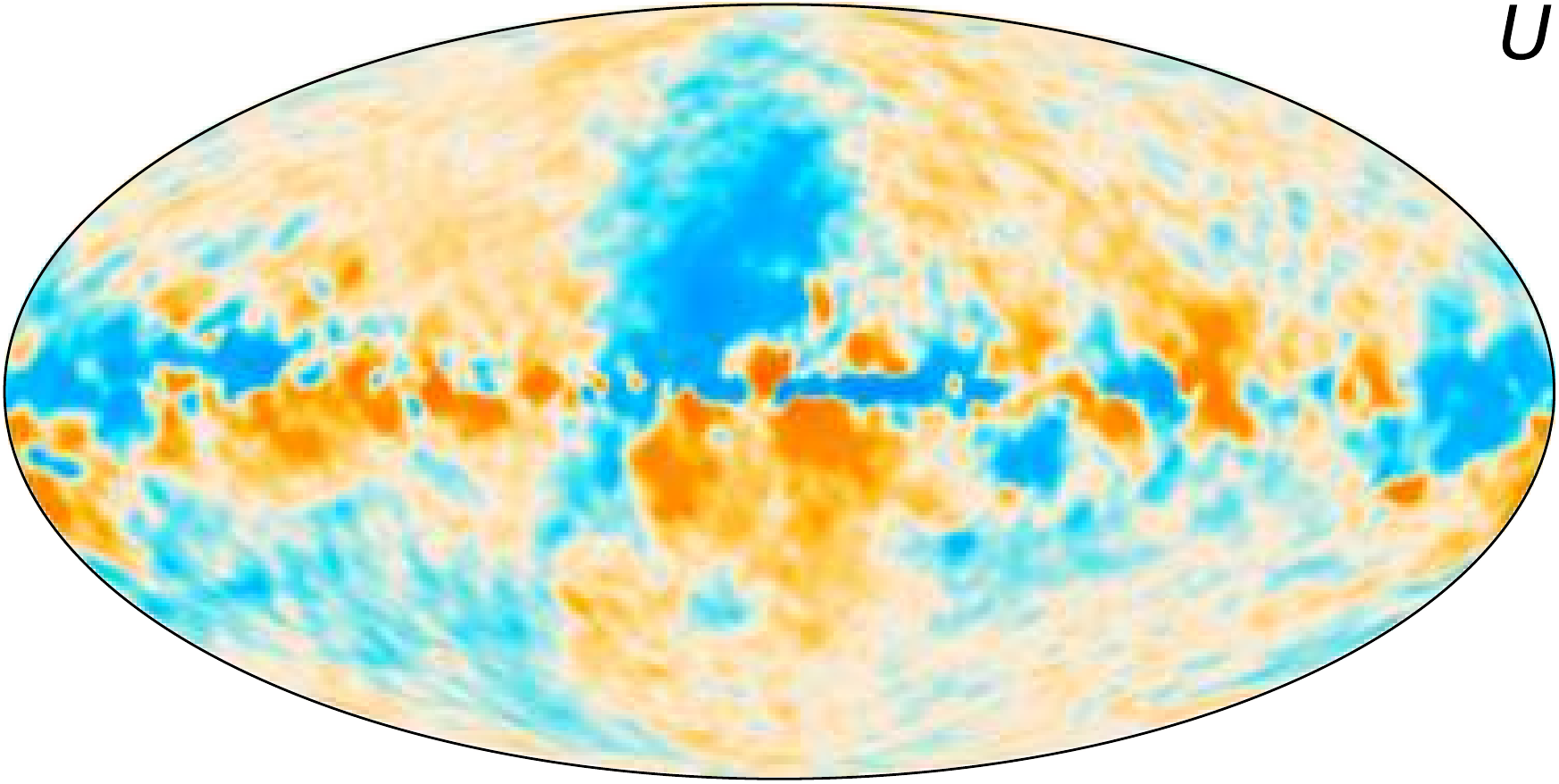}
  \includegraphics[width=0.33\linewidth]{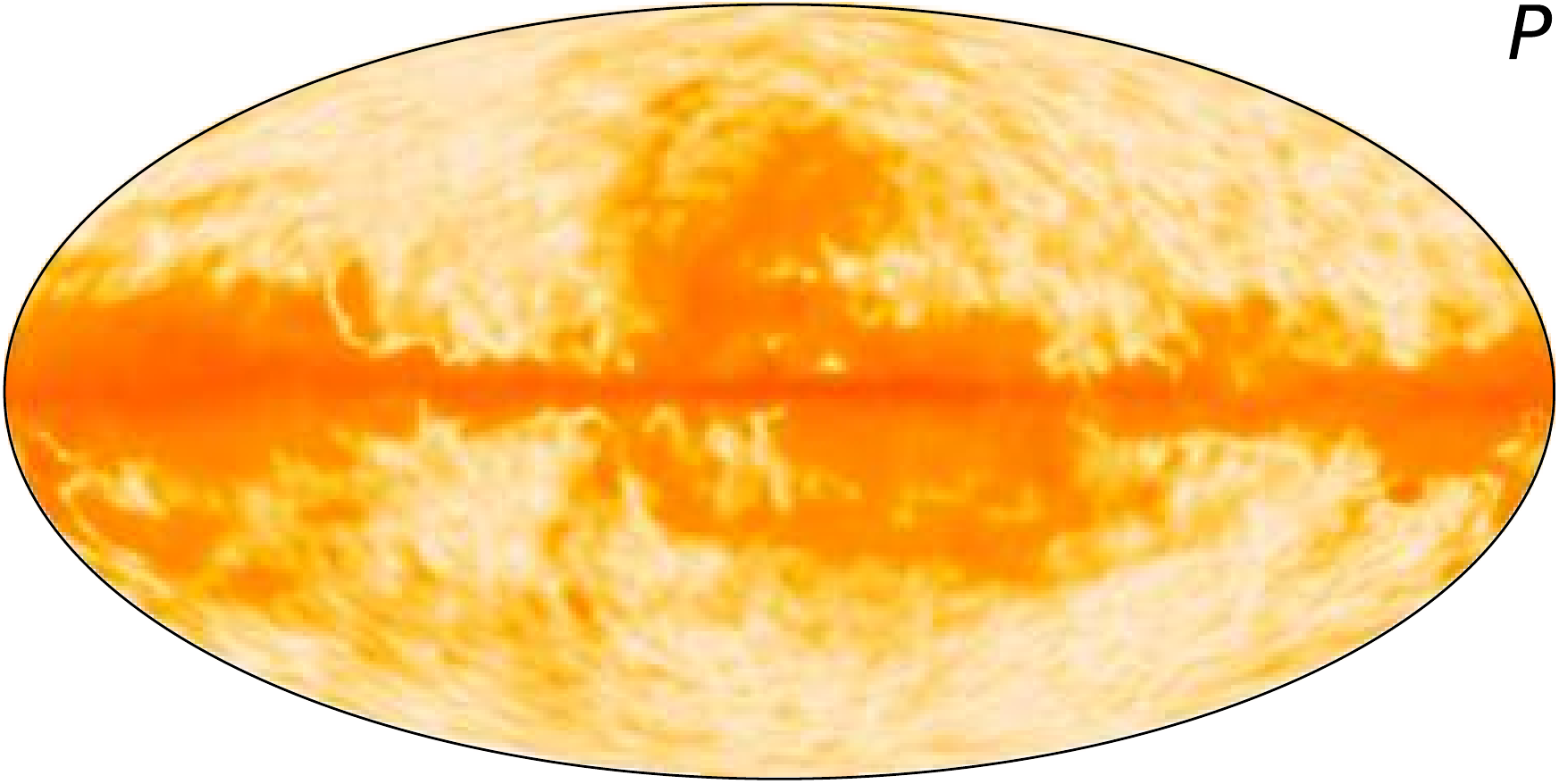}
  \\
  \includegraphics[width=0.33\linewidth]{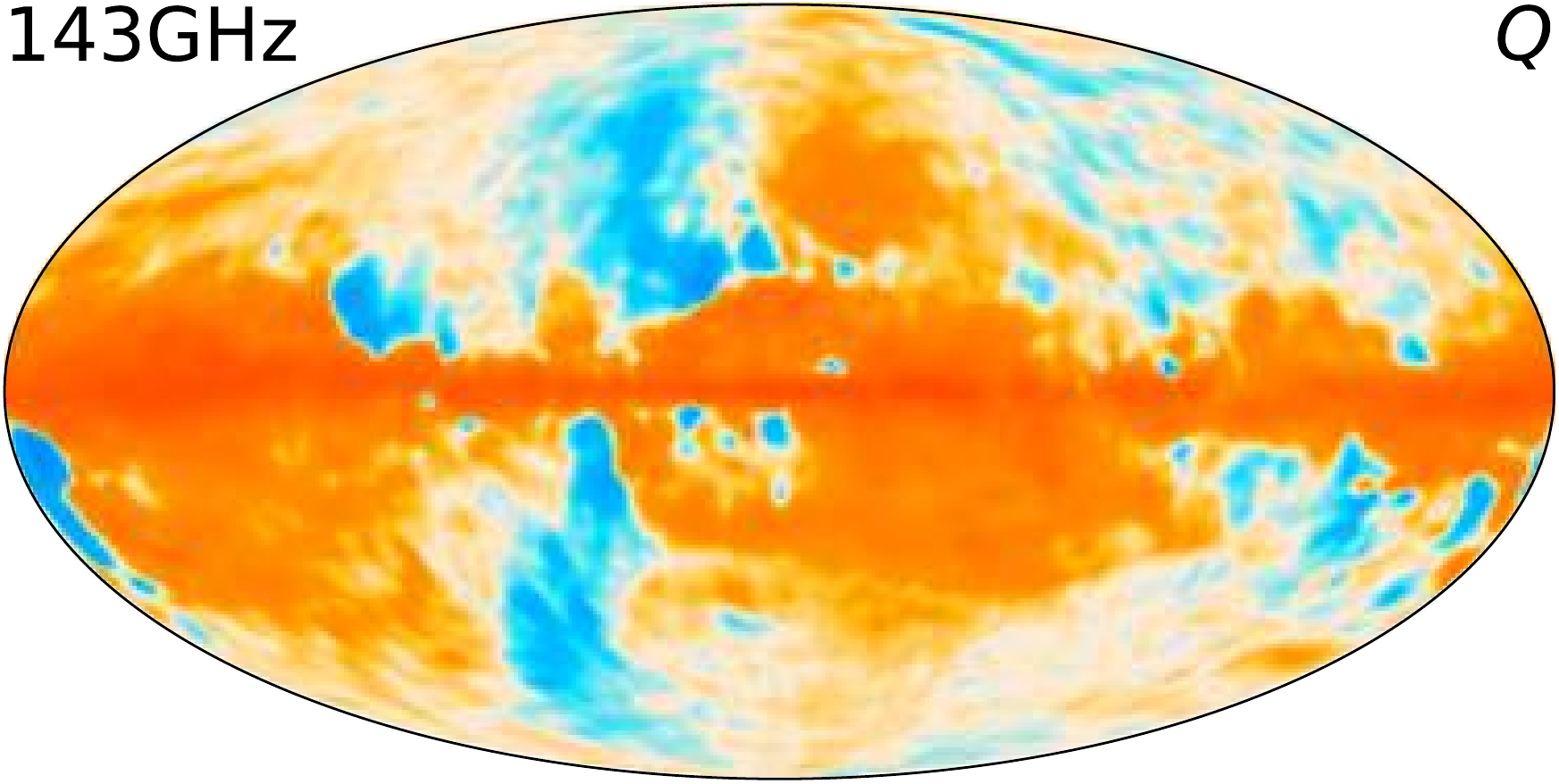}
  \includegraphics[width=0.33\linewidth]{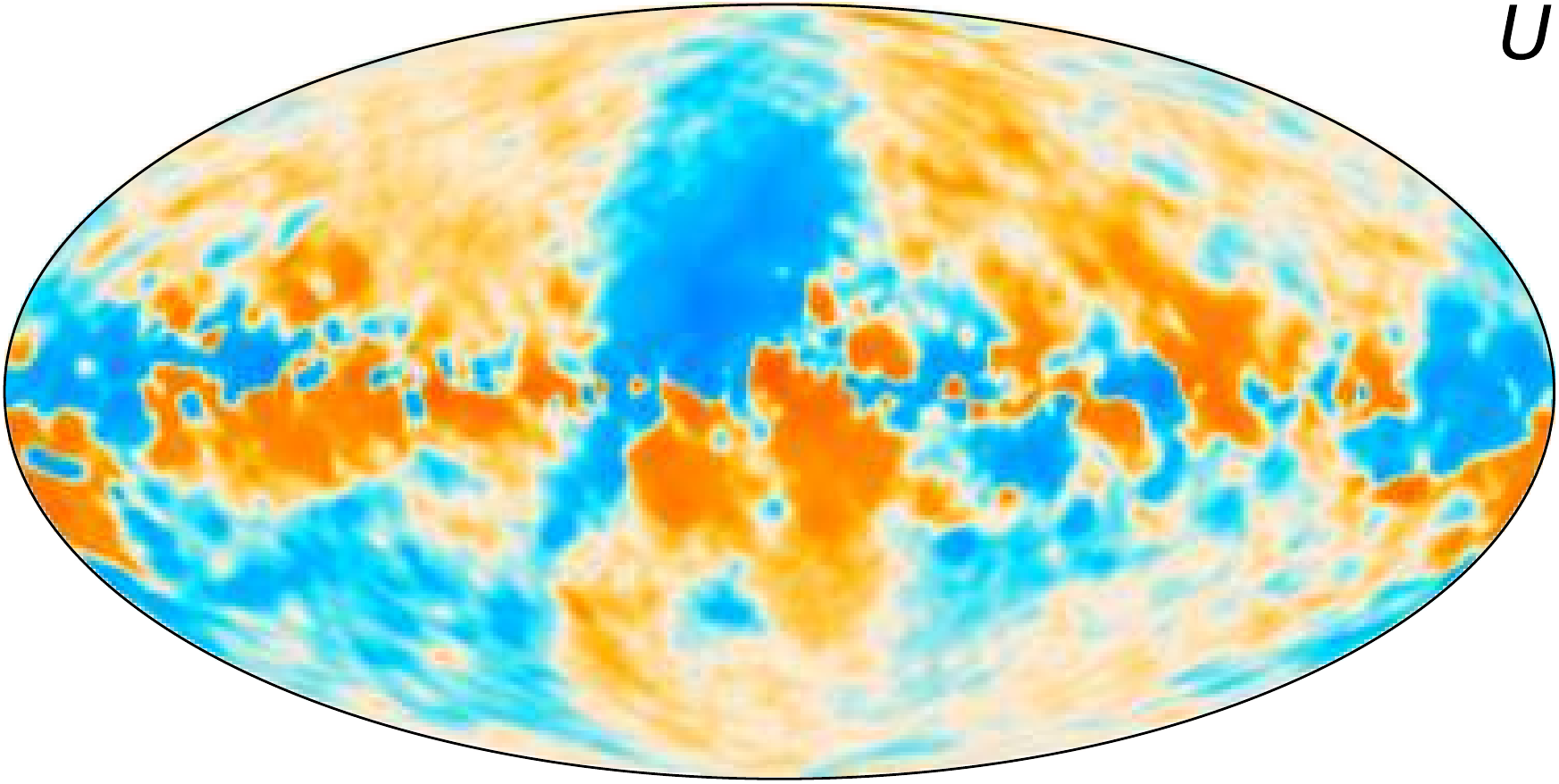}
  \includegraphics[width=0.33\linewidth]{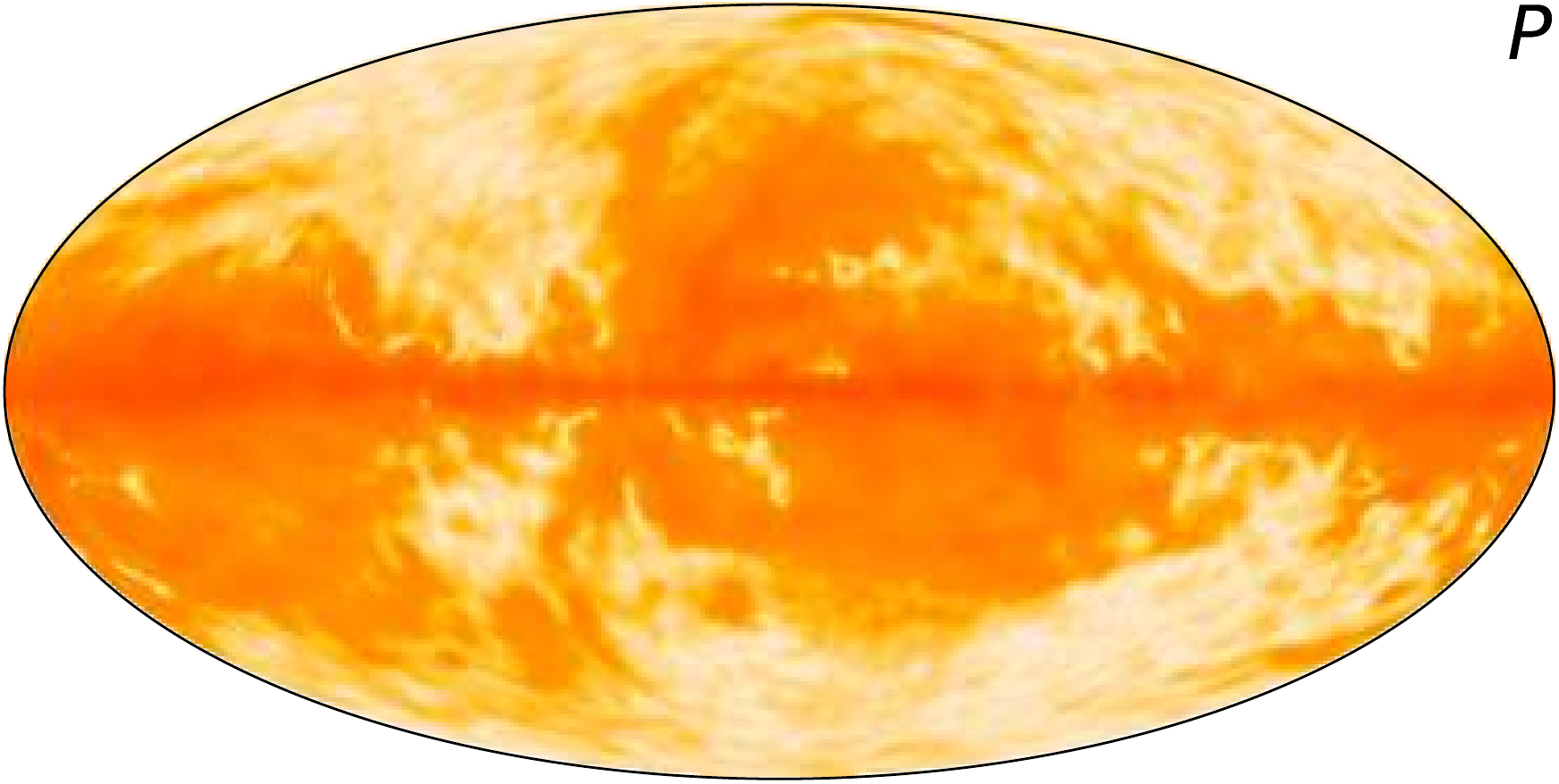}
  \\
  \includegraphics[width=0.33\linewidth]{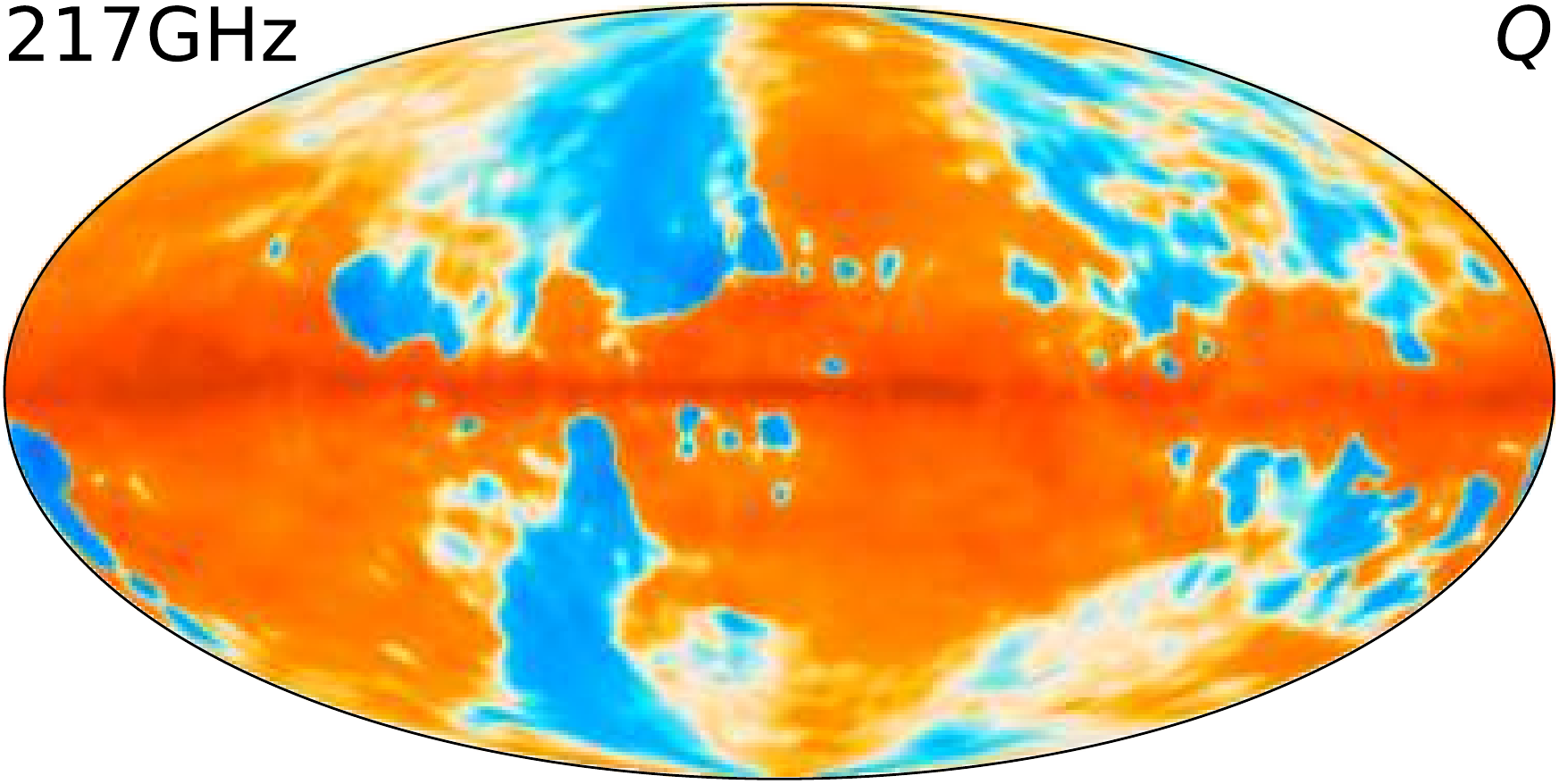}
  \includegraphics[width=0.33\linewidth]{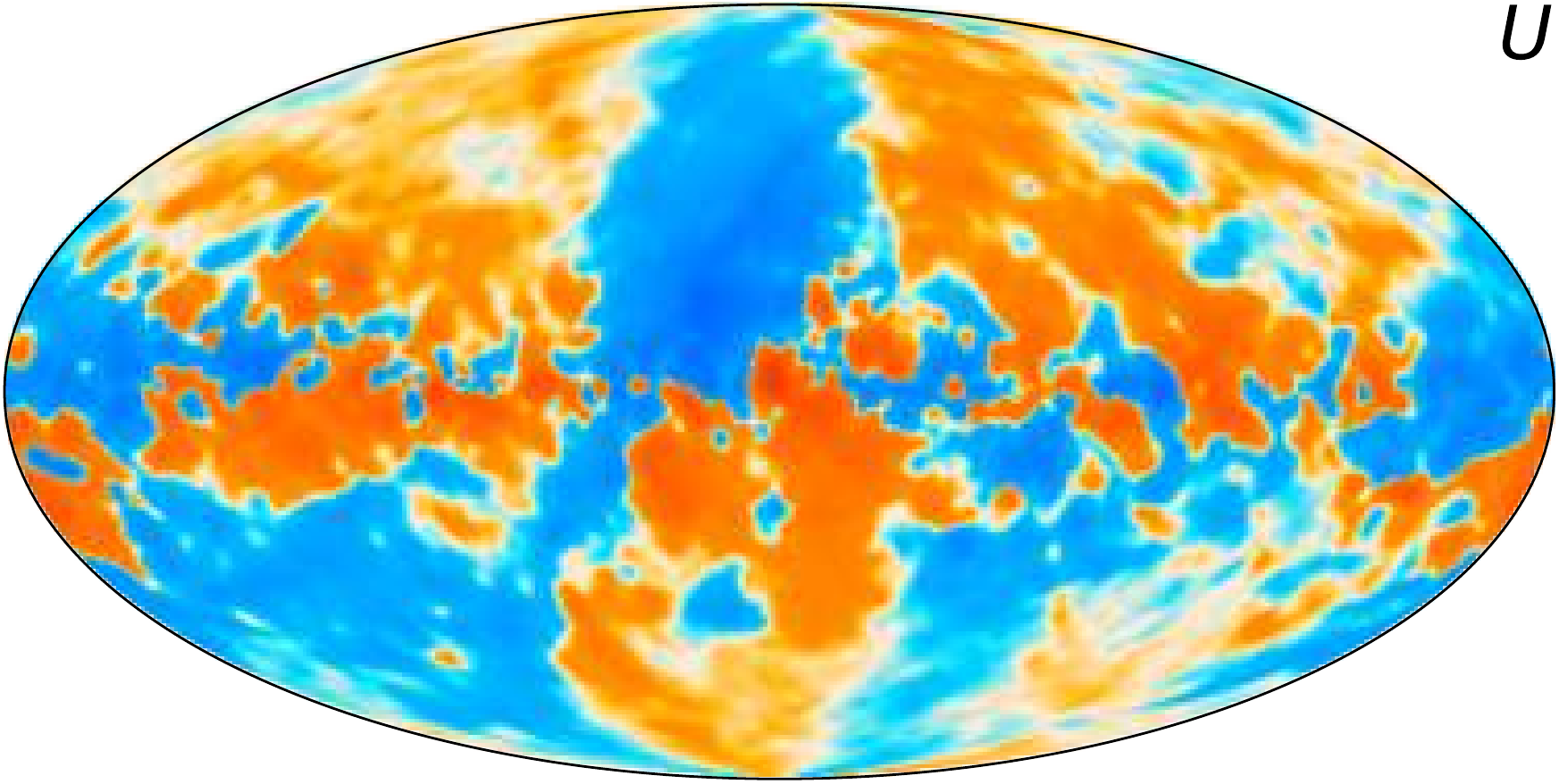}
  \includegraphics[width=0.33\linewidth]{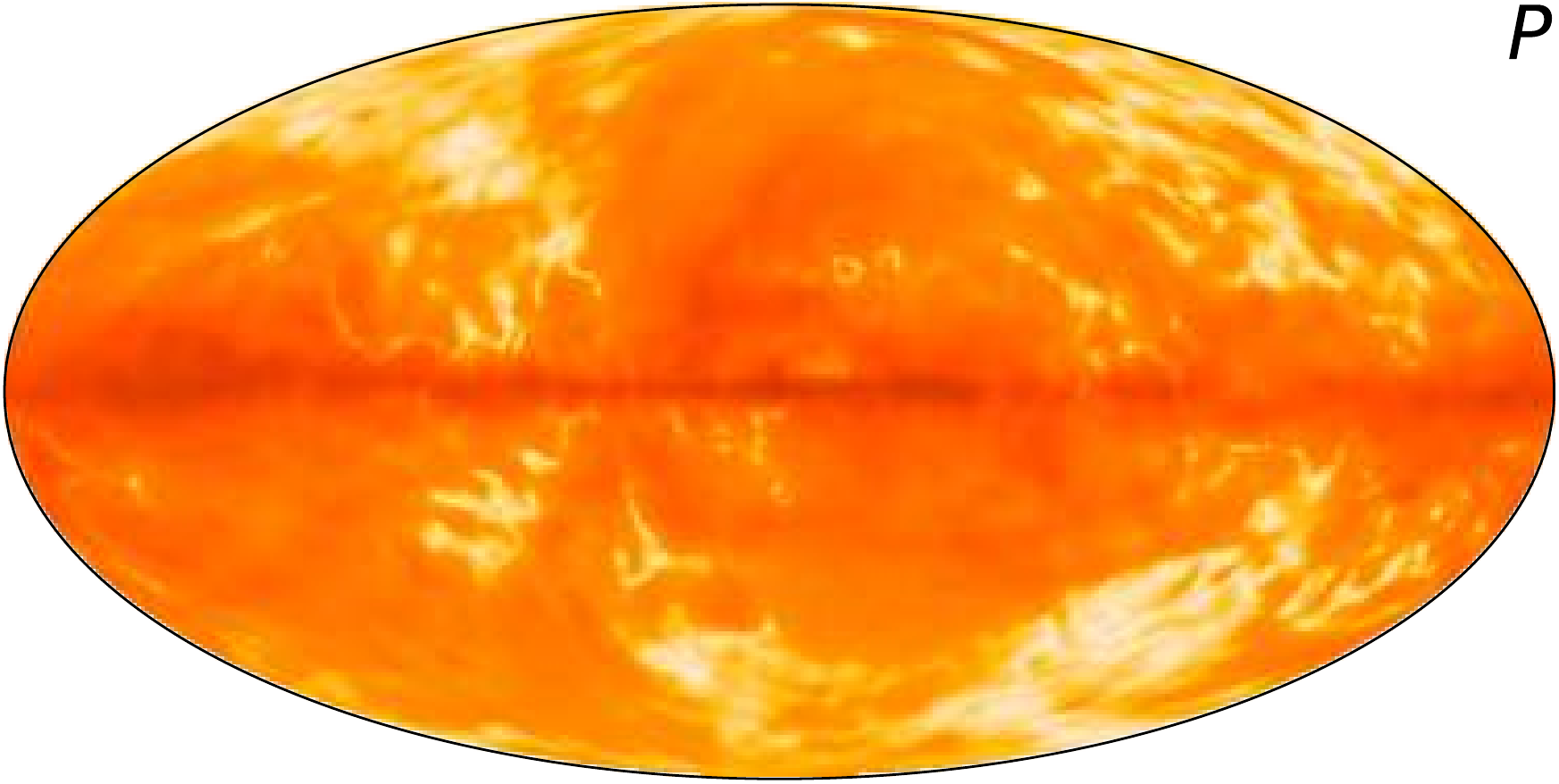}
  \\
  \includegraphics[width=0.33\linewidth]{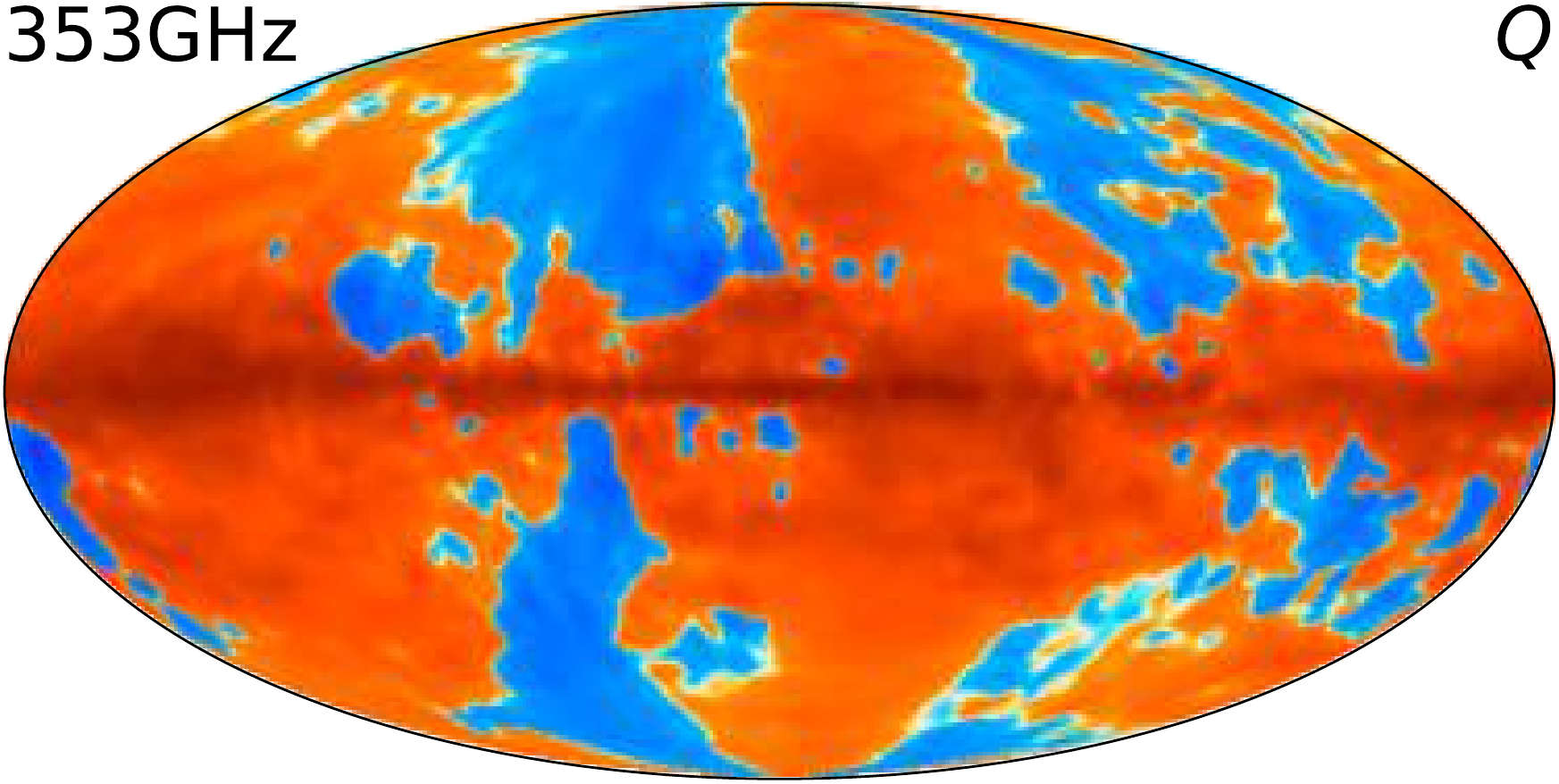}
  \includegraphics[width=0.33\linewidth]{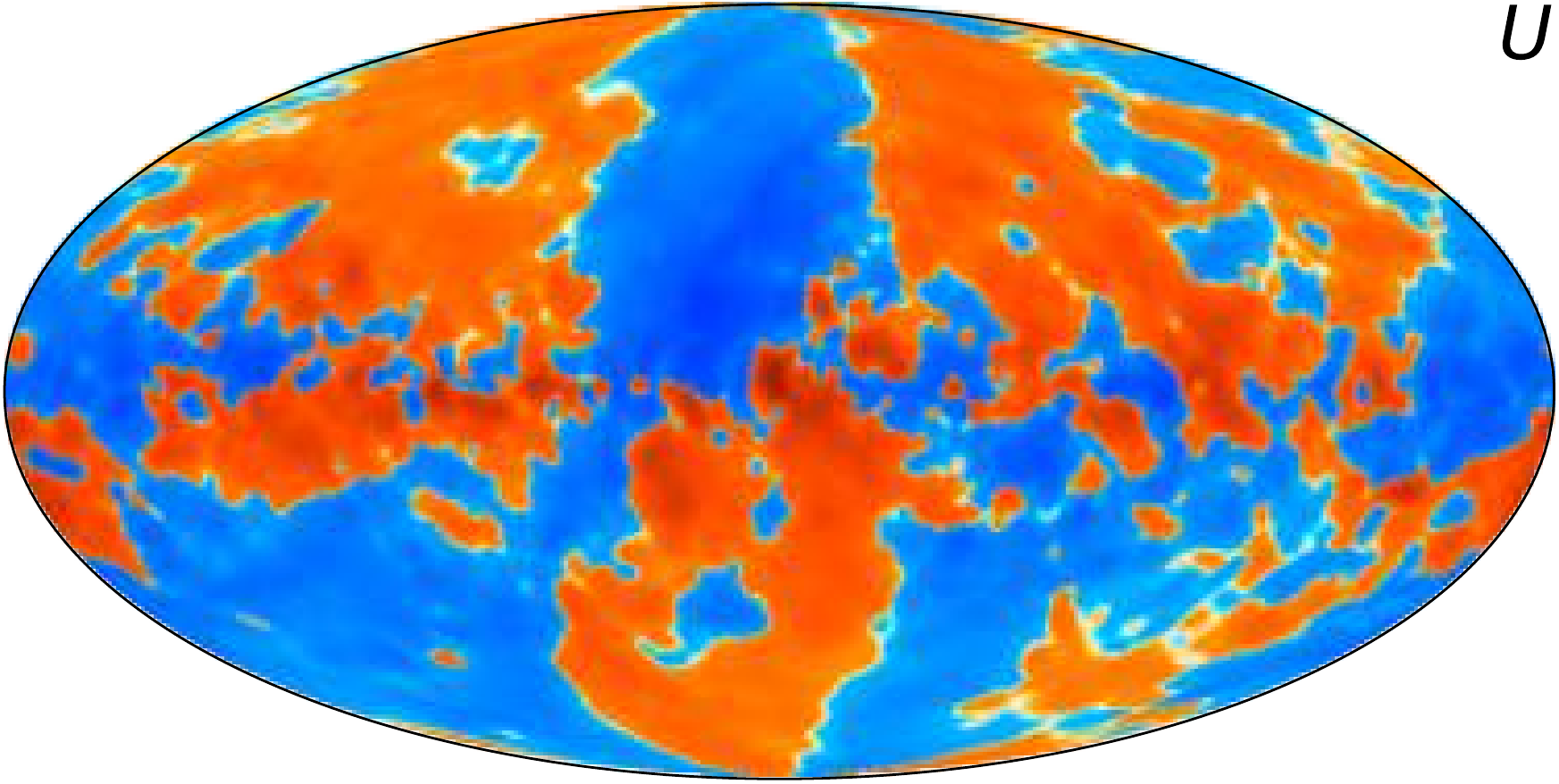}
  \includegraphics[width=0.33\linewidth]{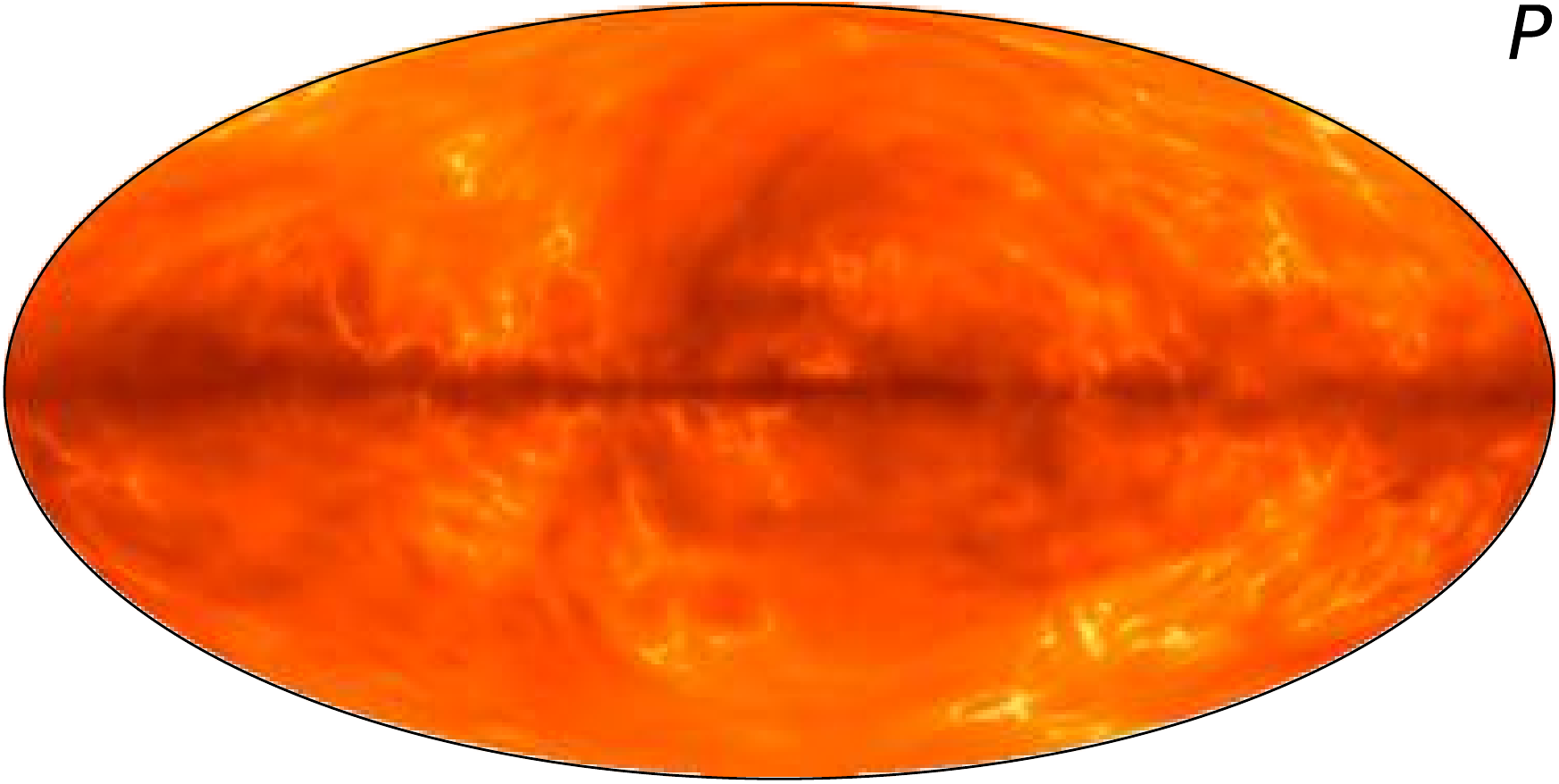}
  \\
  \includegraphics[width=1.0\linewidth]{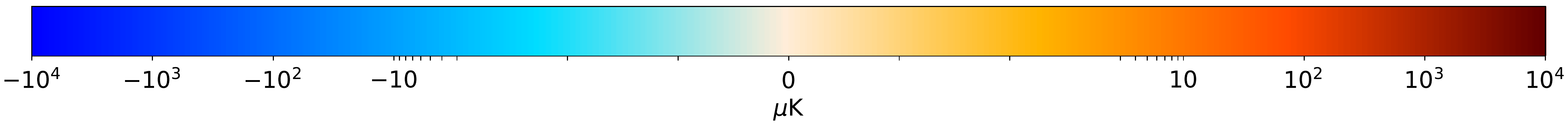}
  \caption{\npipe\ polarization maps smoothed with a Gaussian beam with FWHM 3\deg.  The scaling is linear between $-3$ and
    $3\,\mu$K.
  }
  \label{fig:freqmaps_sP}
\end{figure*}

We present temperature maps at all nine \Planck\ frequencies in Fig.~\ref{fig:freqmaps_Tw}, and with the Solar dipole removed in Fig.~\ref{fig:freqmaps_T}.  The zero levels of the maps are adjusted for plotting by evaluating the monopole outside a Galactic mask.  The mapmaking procedure leaves the true monopole undetermined.  Stokes $Q$ and $U$ polarization maps and the polarization amplitude at the seven polarized frequencies are shown in Fig.~\ref{fig:freqmaps_P}.  Polarization maps smoothed to 3\deg\ are shown in Fig.~\ref{fig:freqmaps_sP}.

\subsubsection{Half-ring maps} \label{sec:hrmaps}

The repetitive \Planck\ scanning strategy (each scanning circle is repeated 30--75 times) allows us to build subset maps that have effectively identical systematics but independent instrument noise.  We split each pointing period into two half-rings, and assign the halves to separate subsets, which we call ``HR1'' and ``HR2.''  The half-ring difference (i.e., HR1 $-$ HR2) is a useful measure of the instrumental noise, but should not be assumed to be unbiased for large-scale cross-spectral analysis.  The half-ring maps share all the calibration, bandpass mismatch, and other template residuals.

\begin{figure*}[htpb!]
  \includegraphics[width=1.0\linewidth]{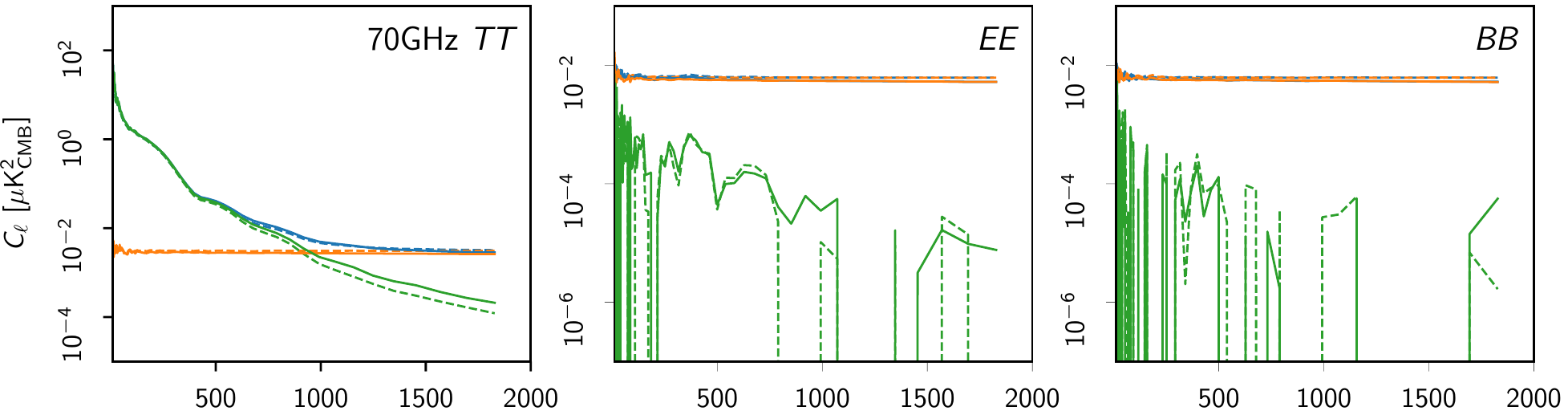}\\
  \includegraphics[width=1.0\linewidth]{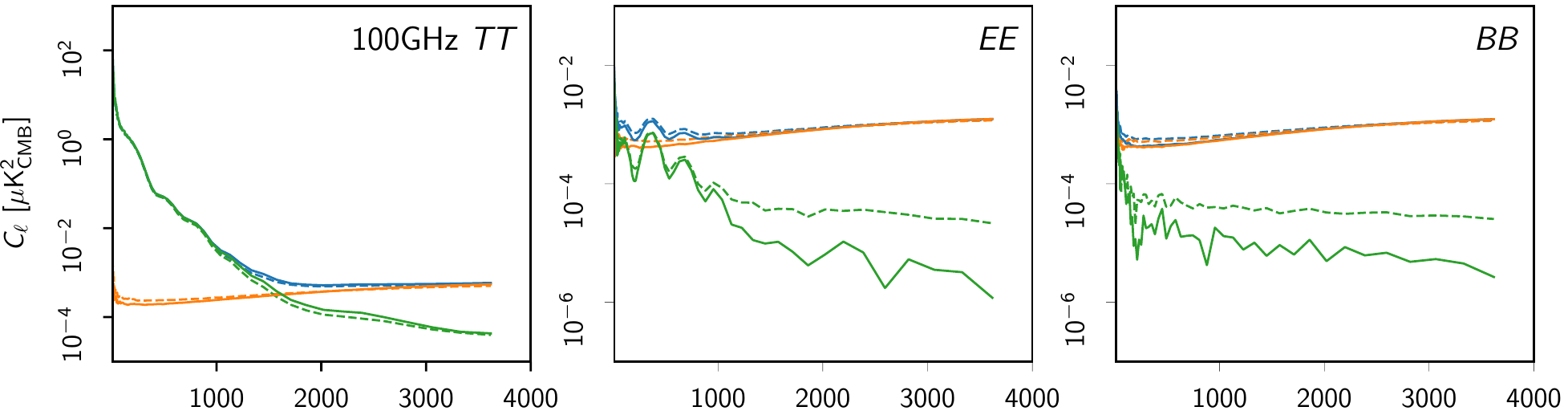}\\
  \includegraphics[width=1.0\linewidth]{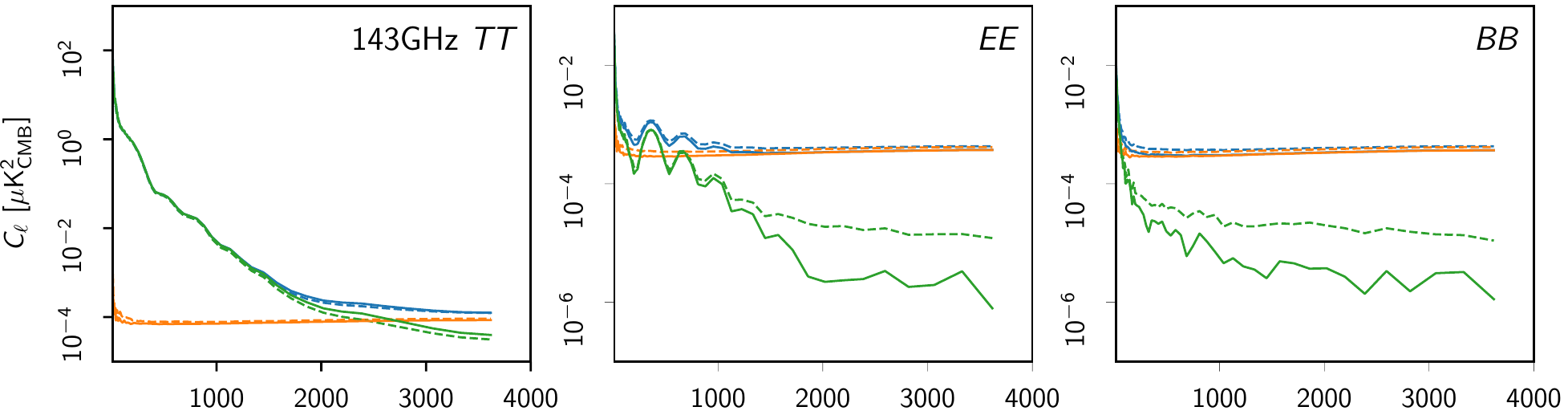}\\
  \includegraphics[width=1.0\linewidth]{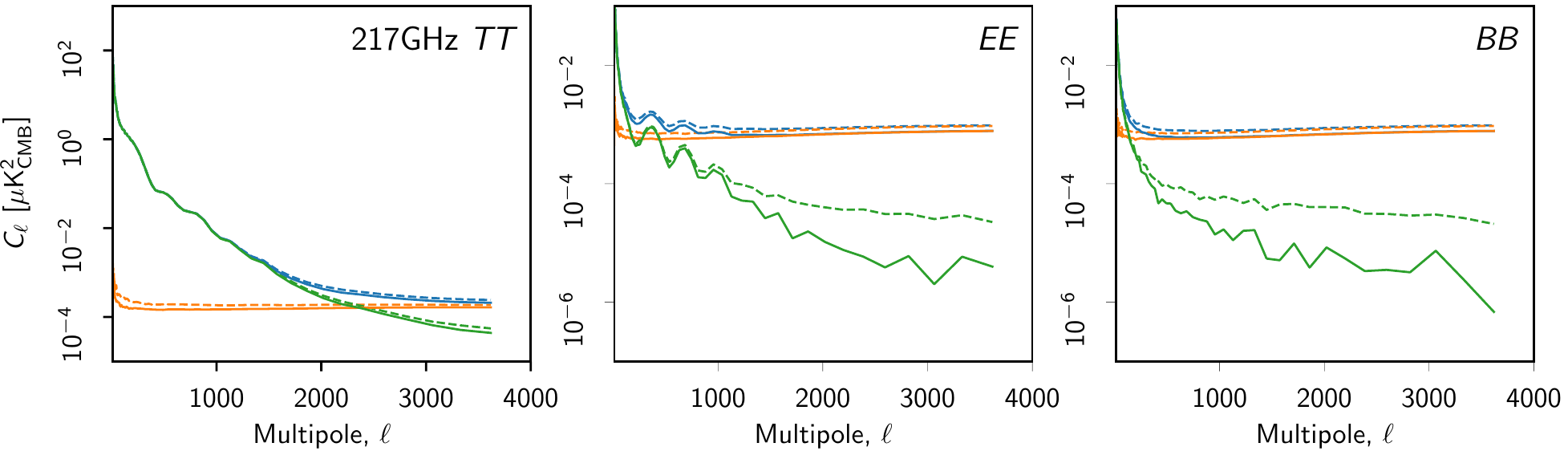}\\
  \caption{Half-ring sum (blue), difference (orange), and cross (green) power spectra from \npipe\ (solid lines) and \prthree\ (dashed lines).  Improvements in gap-filling and transfer-function deconvolution reduce the small-scale half-ring correlations in \npipe\ data from \prthree\ levels.  The power spectra are binned into 100 logarithmically-spaced bins, and corrected for sky fraction.  The 70-GHz results are shown as a reference case without half-ring correlations.
  }
\label{fig:cl_hrcross}
\end{figure*}

The noise in the \hfi\ half-ring maps contains a small amount of correlated error between the two ring halves (cf.\ Sect.~\ref{sec:glitch_removal} and Fig.~\ref{fig:cl_hrcross}).  This error comes from a noisy signal estimate used in the removal of
the glitches.

\subsubsection{Single-detector maps}  \label{sec:dmaps}

\npipe\ builds unpolarized single-detector maps from the reprocessed TOD by subtracting an estimate of the polarization response from the full-frequency map.  The single-detector maps are destriped with \madam\ independently, using the same destriping parameters as for the full-frequency data.  The maps are provided with and without bandpass-mismatch corrections.  Mismatch-corrected maps provide a useful diagnostic for reprocessing performance, while the uncorrected maps are used as inputs to temperature component separation.  This provides component-separation algorithms with samples of the sky for each detector bandpass, which is necessary, for example, to measure the CO line emission.  Sampling and subtracting the polarization signal from the full-frequency map injects a small amount of correlated noise into the single-detector maps.  This noise is attenuated by the fact that we downgrade the frequency map to the same resolution as our bandpass-mismatch templates, i.e., \nside{512} for 30--100\GHz\ and \nside{1024} for 143--353\GHz.  Most of this correlated noise is cancelled in the single-horn maps (Sect.~\ref{sec:hmaps}), so applications that are particularly sensitive to small-scale noise correlations are best served using the horn maps instead.

\subsubsection{Single-horn maps} \label{sec:hmaps}

In addition to polarization-corrected single-detector maps, we also provide unpolarized single-horn maps that naturally cancel most if not all of the polarization.  These maps are linear combinations of the polarization-orthogonal detector data in the polarized \Planck\ horns, with the weights chosen to maximally cancel polarization in the averaged map.  The process also eliminates
noise present in the resampled polarization estimate.

\subsubsection{A/B maps} \label{sec:abmaps}

\npipe\ provides only one data split where the systematics between the splits are expected to be uncorrelated (unlike the case for the half-ring splits discussed in Sect.~\ref{sec:hrmaps}).  We have split the horns in the focal plane into two independent sets, A and B, and performed reprocessing independently on each set.

The \Planck\ scanning strategy requires at least two polarized horns to solve for a full-sky polarized map.  The detector-set splitting was not possible at either 30 or 44\GHz\ because of the lack of redundant polarized horns.  Instead, for these two frequencies, the split is done time-wise: set A comprises operational years 1 and 3; while set B has years 2, 4, and the month of integration time from the final, fifth observing year.  (A half-mission split is not feasible because the second half mission does not have full sky coverage.)  The time-wise split is not expected to have independent systematics, since the two subsets will share initial beam and bandpass mismatch; however, the instrument noise and gain fluctuations will be uncorrelated between the two subsets, making the time-wise split at these frequencies a reasonable compromise to provide two consistent and maximally disjoint sets of \Planck\ data.
Full descriptions of the A/B splits are provided in Table~\ref{tab:absplit}.

\begin{table}[htpb!]
  \begingroup
  \newdimen\tblskip \tblskip=5pt
  \caption{Details of the A/B subset split.
  }
  \label{tab:absplit}
  \nointerlineskip
  \vskip -3mm
  \footnotesize
  \setbox\tablebox=\vbox{
    \newdimen\digitwidth
    \setbox0=\hbox{\rm 0}
    \digitwidth=\wd0
    \catcode`*=\active
    \def*{\kern\digitwidth}
    \newdimen\signwidth
    \setbox0=\hbox{$-$}
    \signwidth=\wd0
    \catcode`!=\active
    \def!{\kern\signwidth}
    \halign{
      \hbox to 2.2cm{#\leaderfil}\tabskip 1.5em&
      #\hfil\tabskip 1.5em&
      #\hfil\tabskip 0pt\cr
      \noalign{\doubleline}
      \omit\hfil Frequency\hfil\cr
     \omit\hfil [GHz]\hfil&
      \omit\hfil Set A\hfil&
      \omit\hfil Set B\hfil\cr
      \noalign{\vskip 3pt\hrule\vskip 4pt}
      *30& Years 1 and 3& Years 2, 4, and start of 5\cr
      *44& Years 1 and 3& Years 2, 4, and start of 5\cr
      *70& Horns 18, 20, and 23& Horns 19, 21, and 22\cr
      100& Horns 1 and 4& Horns 2 and 3\cr
      143& Horns 1, 3, 5, and 7& Horns 2, 4, and 6\cr
      217& Horns 1, 3, 5, and 7& Horns 2, 4, 6, and 8\cr
      353& Horns 1, 3, 5, and 7& Horns 2, 4, 6, and 8\cr
      545& Horn 1& Horns 2 and 4\cr
      857& Horns 1 and 3& Horns 2 and 4\cr
      \noalign{\vskip 3pt\hrule\vskip 5pt}
    }
  }
\endPlancktable 
\endgroup
\end{table}

The \npipe\ approach of processing maximally-independent subsets of data is orthogonal to the approach adopted by \hfi\ in  \prthree.  There the systematic templates were fitted on the entire frequency data set, and the cleaned data were split only for the purpose of projecting the TOD into subset maps.  Such processing introduces correlations between systematic residuals in the subset maps, making it necessary to estimate and correct for noise bias even in cross-spectra between the subsets.  This is part of the reason \citet{planck2016-l03} emphasizes the importance of using inter-frequency cross-spectra in extracting scientific information from \prthree\ maps.

\subsection{Low-resolution data set} 
\label{sec:lowres}

We provide low-resolution versions of the \npipe\ maps for maximum-likelihood pixel-domain analysis of the large scales.  The maps are downgraded to \healpix\ resolution \nside{16} and convolved with a cosine apodizing kernel:
\begin{linenomath*}
\begin{equation}
  \label{eq:cosine_window}
  b_\ell = \left\{
    \begin{array}{ll}
      1,& \ell \leq \ell_1, \\
     {1\over2} \left(1 + \cos\left[ \pi {(\ell-\ell_1)\over(\ell_2-\ell_1)}\right]\right),
   & \ell_1 < \ell \leq \ell_2, \\
      0,& \ell > \ell_2,
    \end{array} \right.
\end{equation}
\end{linenomath*}
with the choice of $\ell_1\,{=}\,1$, $\ell_2\,{=}\,3N_\mathrm{side}$.  The lower threshold has been decreased from the usual $\ell_1\,{=}\,N_\mathrm{side}$ to reduce ringing in the 857-GHz maps.  In addition to the smoothing kernel, the maps include the standard \nside{16} pixel window function.  Our downgrading tool noise-weights each high resolution pixel with the estimated noise level and parallel-transports the polarization parameters between pixel centres.  Initial high resolution pixel window function was not deconvolved as it is negligible at the angular scales that are represented in the low resolution maps.

\begin{figure}[hbtp]
  \centering
  \includegraphics[width=0.48\textwidth]{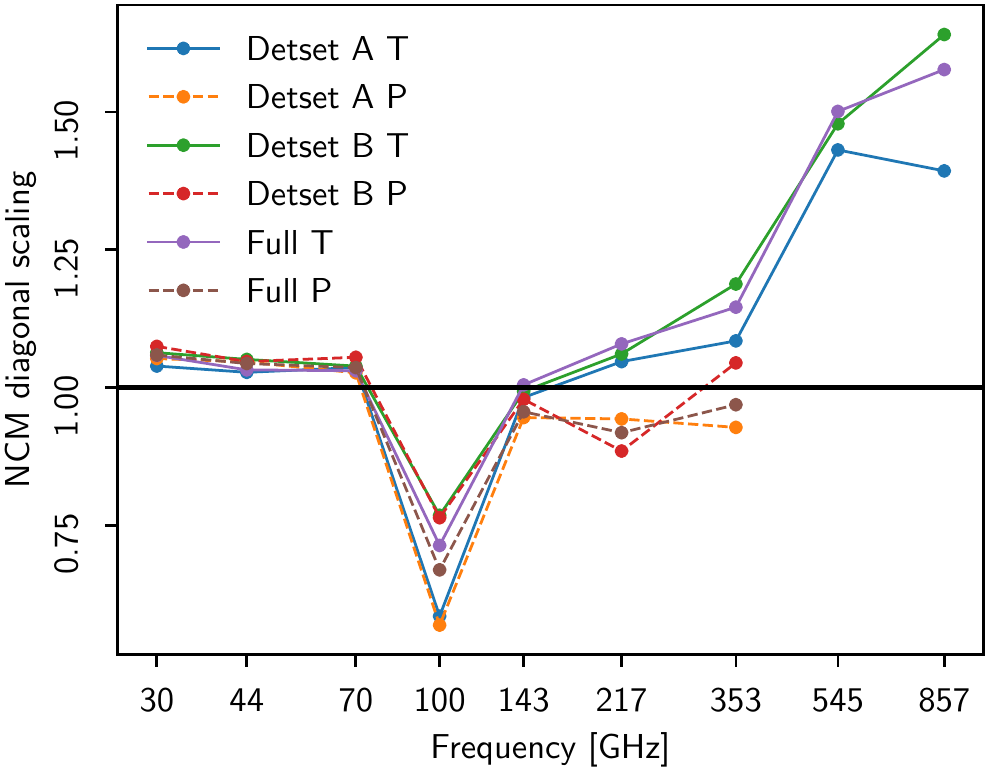}
  \caption{Scaling factors applied to the \madam\ noise matrices to match the variance in half-ring-difference pixels.  We apply different scalings to the temperature and polarization parts of the noise matrices, and scale the detector-set matrices and the full-frequency matrices independently.  The 100\GHz\ channel stands out because of the prominent bump at the high-frequency end of the noise PSD (cf.\ Fig.~\ref{fig:psds}).
  }
  \label{fig:ncm_scaling}
\end{figure}

The low-resolution maps are accompanied by pixel-pixel noise-covariance matrices that reflect the baseline uncertainties and the high-frequency instrumental noise approximated as uniform white noise.  Since the high-frequency noise in \npipe\ is not uniform for any of the \Planck\ detectors, the approximation is inexact and, as a result, requires an extra scaling step performed on the diagonal of the matrix to match the half-ring difference maps.  The off-diagonal elements correspond to low frequency noise and need not be scaled.  We show the scaling factors in Fig.~\ref{fig:ncm_scaling}.  For the CMB channels, these factors are close to unity except at 100\GHz.

\subsection{Large-scale polarization transfer function} 
\label{sec:ee_tf}

\npipe\ calibration at the CMB frequencies ($44$--$217\GHz$) acts like a constrained filter.  The polarized sky model that we derive from the extreme frequencies contains, for all practical purposes, only the foregrounds.  This leaves \npipe\ actively trying to suppress the CMB polarization with the available templates.  While obviously undesirable, the effect of this filtering is kept small by the number and structure of the templates.  It is also straightforward to measure the amount of suppression using simulations and
incorporate the effect into the transfer functions.

Allowing for the non-trivial transfer function in \npipe\ calibration is a compromise between measuring very noisy but unbiased large-scale polarization from all large-scale modes, and filtering out the modes that are most compromised by the \Planck\ calibration uncertainties left in the data by the \Planck\ scan strategy (see Fig.~\ref{fig:detsetdiff} and Fig.~\ref{fig:detsetdiff_polcal}).  Based on similarities between \lfi\ 2018 processing and \npipe\ it is possible that the \lfi\ 2018 results were also subject to filtering.  Unfortunately the accompanying simulations were not sufficient to determine this as only one CMB realization was processed with the calibration pipeline.  Algebraically there is no reason to anticipate the \hfi\ 2018 results to contain a similar effect but, again, the 2018 simulations cannot be used to demonstrate this.

\begin{figure*}[htpb]
  \includegraphics[width=0.48\linewidth]{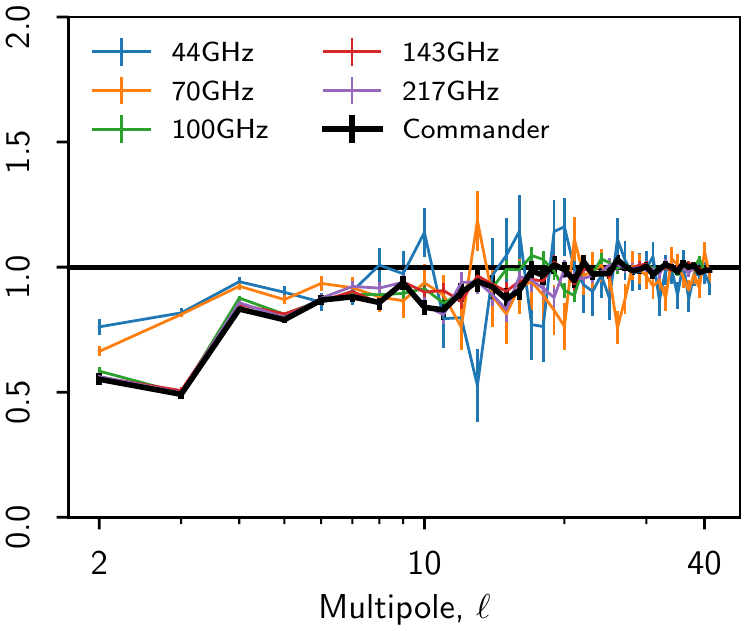}
  \includegraphics[width=0.48\linewidth]{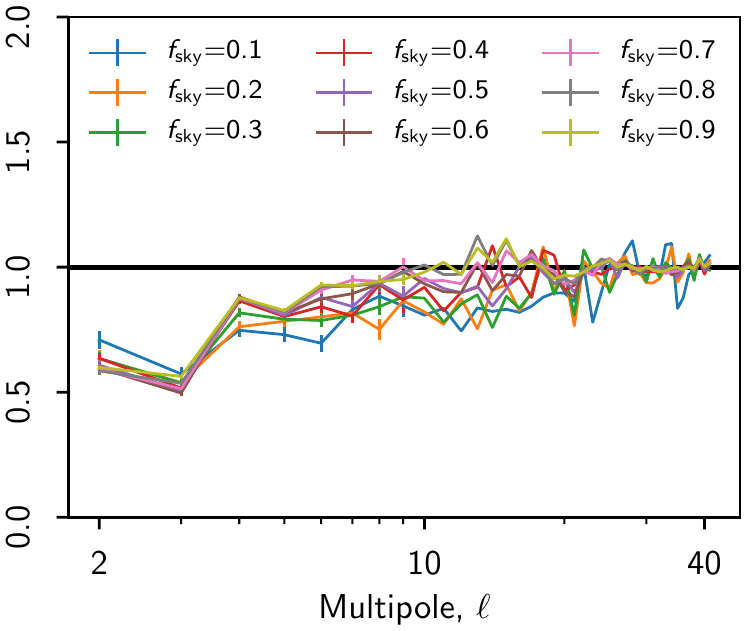}
  \caption{\npipe\ $E$-mode transfer functions measured by comparing simulated CMB input and foreground-cleaned output maps.
    \emph{Left:} CMB frequencies and component-separated \commander\ maps (Sect.~\ref{sec:compsep}) over 60\,\% of the sky.  The apparent mismatch between the \lfi\ and \hfi\ transfer functions results from the quantity and structure of the template corrections; templates that are specific to \hfi, especially the ADC distortion, provide more degrees of freedom to suppress the CMB power.  The 44-GHz transfer function is closer to unity because the 30-GHz template shields about 22\,\% of the CMB polarization.  The error bars reflect the statistical uncertainty of the measured transfer function, not the total Monte Carlo scatter.  Tabulated values of the transfer functions are listed in Table~\ref{tab:tf}.
    \emph{Right:} $E$-mode transfer function for 100\GHz\ over multiple sky fractions.  The error bars at $\ell\,{\geq}\,10$ were suppressed to show more structure.
  }
  \label{fig:ee_bias}
\end{figure*}

The filtering approach is typically adopted by ground and balloon-borne experiments to suppress atmospheric noise in their data.  The only requirement for the filtering to be workable is that the resulting $\ell$-dependent transfer function be measurable.  We show in Fig.~\ref{fig:ee_bias} that it is well characterized in the case of \npipe\ analysis of \Planck\ data.  Tabulated values of the transfer function can be found in Appendix~\ref{app:tf}.

\begin{figure}[htpb]
  \center{
    \includegraphics[width=0.48\textwidth]{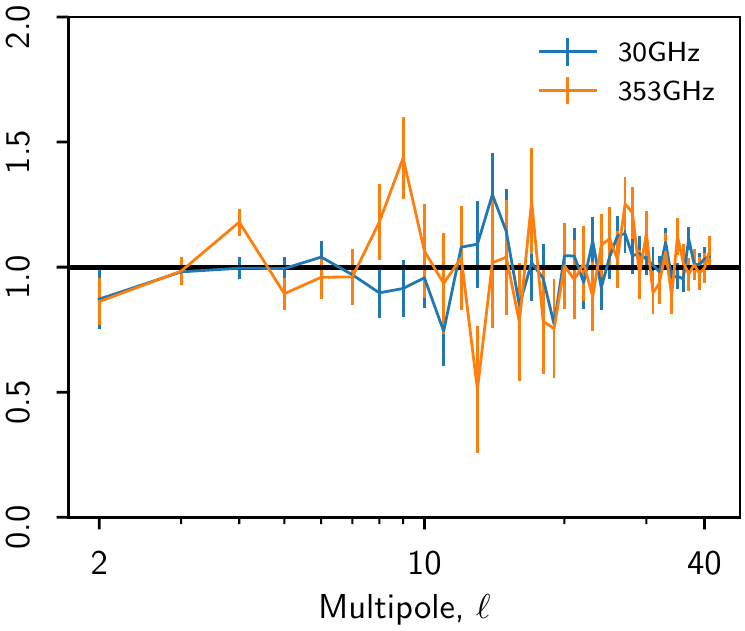}
  }
  \caption{\npipe\ $E$-mode transfer functions, measured by comparing simulated CMB input and foreground-cleaned output maps over 60\,\% of the sky.  The 30 and 353\GHz\ frequencies are not expected to have a measurable transfer function because they are not calibrated with a polarization prior.  These two transfer functions are merely of diagnostic value (to demonstrate the absence of signal suppression) and are not applied in any analysis.  The error bars reflect the statistical uncertainty of the measured transfer function, not the total Monte Carlo scatter.   Tabulated values of the transfer functions are listed in Table~\ref{tab:tf}.
  }
  \label{fig:ee_bias_fg}
\end{figure}

\begin{figure*}[htpb]
  \includegraphics[width=\textwidth]{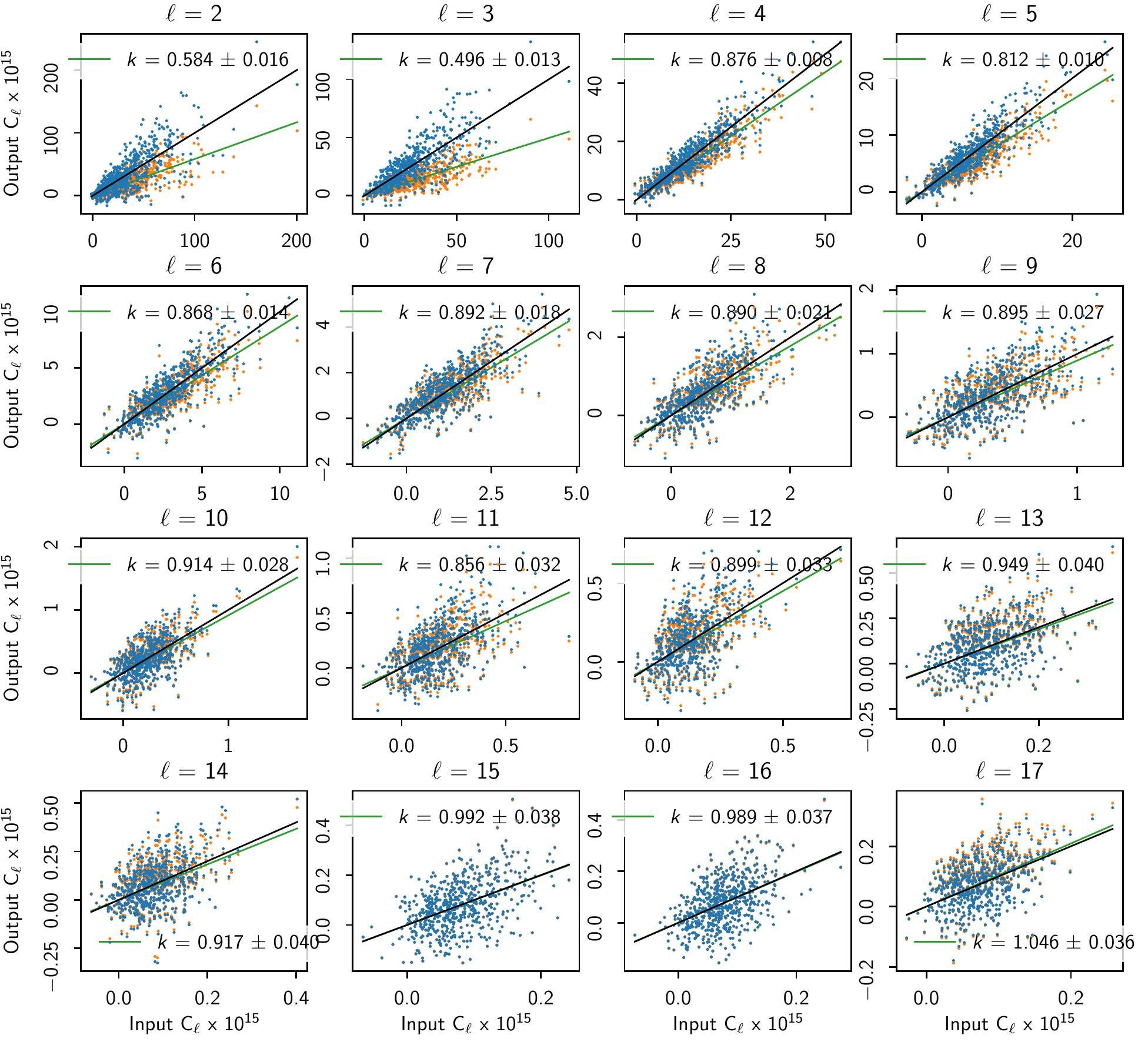}
  \caption{Measuring the \npipe\ large-scale polarization transfer function by comparing the simulated input and output CMB power multipole by multipole at 100\GHz\ and with 60\,\% sky fraction.  The output power spectrum is measured by removing the foreground from the simulated frequency map and then measuring the cross-spectrum with the input CMB map.  The orange data points show input versus output before the transfer function correction.  Their slope is shown in green.  The blue data points show the corrected input versus output.  Their (unity) slope is shown in black.
  }
  \label{fig:ee_bias_100}
\end{figure*}

We measure the transfer function from our simulations by comparing the input CMB map to the output foreground-cleaned map.  We smooth and downgrade both maps and compare the input ($\hbox{CMB} \times \hbox{CMB}$) cross-spectrum to the $\hbox{CMB} \times \hbox{cleaned map}$.  If the effect is multiplicative, as is evident from Fig.~\ref{fig:ee_bias_100}, we expect to find
\begin{linenomath*}
\begin{equation}
  \label{eq:bias}
  C_\ell^{\mathrm{CMB}\times\mathrm{output}}
  = k_\ell\,C_\ell^{\mathrm{CMB}\times\mathrm{CMB}},
\end{equation}
\end{linenomath*}
where $k_\ell$ is a potentially scale-, frequency-, and mask-dependent suppression factor.  The left panel of Fig.~\ref{fig:ee_bias} shows the $k_\ell$ for each CMB frequency.  The right panel of Fig.~\ref{fig:ee_bias} compares the 100-GHz transfer functions evaluated over sky fractions ranging from 0.1 to 0.9 and shows to what degree the shape of the transfer function depends on the mask.  These transfer functions, like the beam window functions, must be squared to obtain the full impact on a power spectrum.  We find that the $EE$ quadrupole and octupole are significantly suppressed by our calibration, with about 64\,\% of the quadrupole and 73\,\% of the octupole missing in the \npipe\ power spectra.  Multipoles 4 and 5 are missing about 30\,\% of the power, and $\ell\,{=}\,6$ is missing about 20\,\%.  The simulations do not include enough CMB $B$ modes to measure the $B$-mode transfer function and changing the signal content would compromise our interpretation due to the nonlinear calibration process.

The suppression of $EE$ power is limited to frequencies that are calibrated with the polarization prior.  There is no suppression at 30\GHz\ or 353\GHz\ (Fig.~\ref{fig:ee_bias_fg}).  Measuring the transfer function using the CMB polarization at these frequencies is much harder because of the foregrounds and noise.  This is evidenced by the large upwards fluctuations at 353\GHz, particularly at $\ell = 4$ and 9.  These fluctuations are associated with excess power at the level of the CMB $E$-mode fluctuations, rather than suppression of power.  Their importance diminishes when these frequency maps are scaled to the CMB frequencies to estimate the foregrounds.

Figure~\ref{fig:ee_bias_100} shows individual input-output $EE$ multipole pairs for $100\GHz$.  The data points show a great deal of scatter owing to the limited S/N.  Nevertheless, fitting for a linear model between the input and output values robustly identifies the degree to which the input $EE$ power is suppressed by the \npipe\ calibration.  It is also apparent that a simple multiplicative correction will remedy the bias for all realizations.

\begin{figure*}[htpb]
  \includegraphics[width=0.48\linewidth]{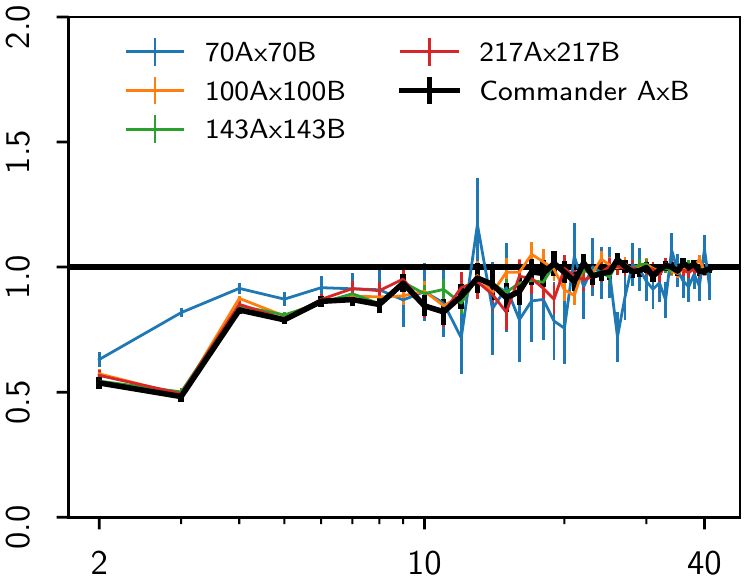}
  \includegraphics[width=0.48\linewidth]{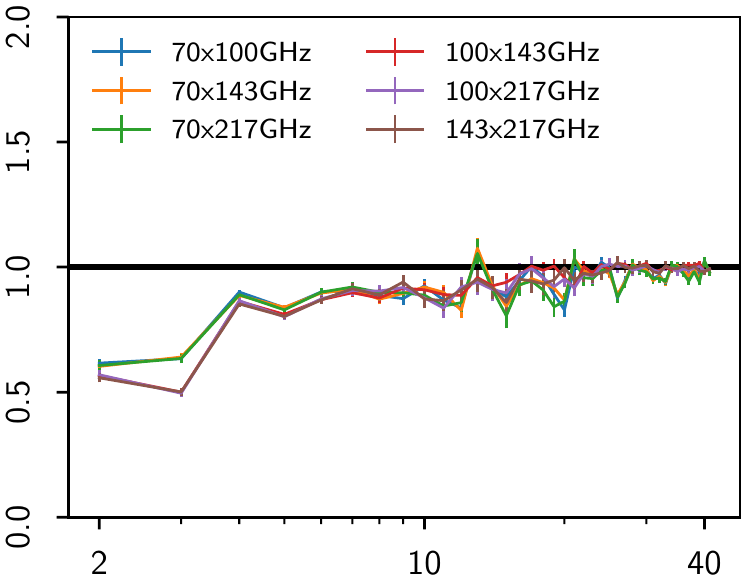}
  \caption{\npipe\ $E$-mode cross-spectrum transfer functions measured by comparing simulated CMB input and foreground-cleaned output maps over 60\,\% of the sky.   \emph{Left:} Detector-set cross-spectra.  \emph{Right:} Frequency cross-spectra.
  }
  \label{fig:ee_bias2}
\end{figure*}

Frequency and detector-set cross-spectra may be impacted by a different transfer function.  The measurement also potentially differs, since we must compare the input spectrum to the foreground-cleaned cross-spectrum:
\begin{linenomath*}
\begin{equation}
  \label{eq:bias2}
  C_\ell^{\mathrm{output}1\times\mathrm{output}2}
  = k_\ell^2\,C_\ell^{\mathrm{CMB}\times\mathrm{CMB}}.
\end{equation}
\end{linenomath*}
We find that the use of the simulated cross-spectra to be problematic for two reasons: they are much noisier than the $\hbox{CMB} \times \hbox{output spectra}$; and they may contain persistent bandpass-mismatch residuals.  For these reasons, we instead derive the cross-spectral transfer functions as the geometric mean of the individual transfer functions.  The geometric mean was found to be in agreement with the Eq.~(\ref{eq:bias2}) and offered superior noise performance.  We show the measured $E$-mode transfer functions for detector-set and frequency cross-spectra in Fig.~\ref{fig:ee_bias2}.

\begin{figure*}[htpb]
  \includegraphics[width=\textwidth]{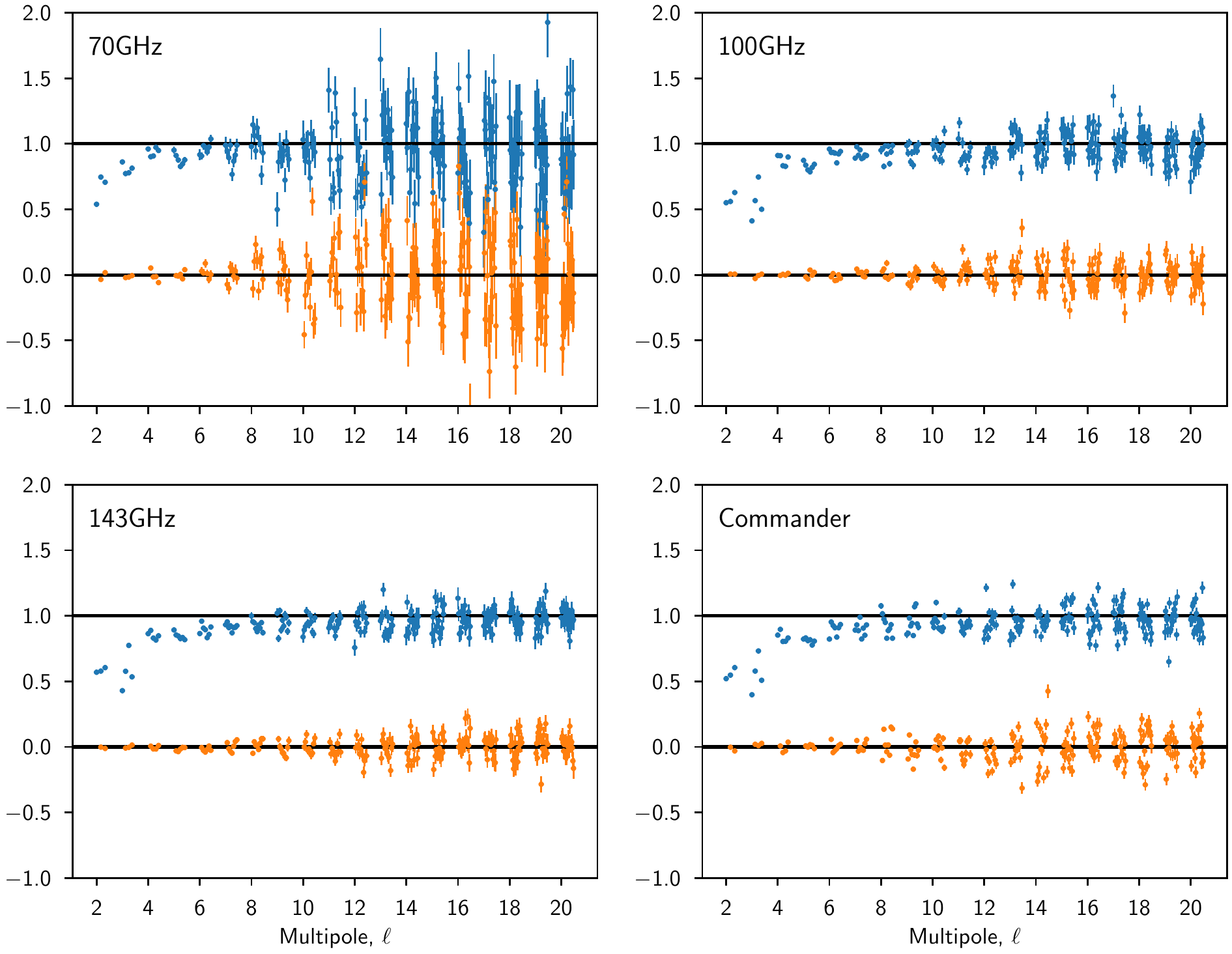}
  \caption{Full anisotropic transfer functions for 70, 100, and 143\GHz\ and \commander.  These demonstrate that the signal suppression is not completely uniform for the quadrupole and the octupole.  The real part of the transfer function is shown in blue.  The imaginary part (orange) of the transfer function is consistent with zero.  These transfer functions correspond to minimizing Eq.~(\ref{eq:anisotropic_chisq}) over 80\,\% of the sky.  The different $m$ modes are slightly staggered here, with $m=0...\ell$ plotted left to right, to aid visualization.
  }
  \label{fig:anisotropic_tf}
\end{figure*}

We also considered measuring the full, anisotropic transfer function, a complex multiplier applied to every $a_{\ell m}$ mode of the input CMB sky.  The results are shown in Fig.~\ref{fig:anisotropic_tf}.  We cannot reliably expand the foreground-cleaned CMB map from the simulations without masking out the strongest Galactic residuals.  This is problematic for a study of anisotropy, because the mask already includes a preferred orientation.  Instead, we write a least-squares minimization problem in terms of a transfer function acting on the input CMB expansion:
\begin{equation}
  \label{eq:anisotropic_chisq}
  \chi^2 = \vec r\trans \vec r
  \quad\mathrm{with}\quad
  \vec r = \vec m - \sum_{X\ell m} f^X_{\ell m}a^X_{\ell m}Y^X_{\ell m},
  \quad X \in [T,E,B]
\end{equation}
where the residual, $\vec r$, is estimated by subtracting a CMB expansion ($a^X_{\ell m}Y^X_{\ell m}$) convolved with a transfer
function($f^X_{\ell m}$).  In this shorthand notation each basis function, $Y^X_{\ell m}$, is a full $IQU$ map with either the $I$ or the $QU$ part identically zero.  The sum in Eq.~(\ref{eq:anisotropic_chisq}) is conveniently carried out using the \healpix\ \texttt{alm2map} facility.  This formulation allows us to use full-sky input expansions and still evaluate the residual over a masked sky.

The anisotropic transfer functions in Fig.~\ref{fig:anisotropic_tf} show a statistically significant anisotropy, especially at $\ell = 3$.  This suggests that fully isotropic methods for power spectrum estimation and transfer function correction are not optimal in analysing the large-scale polarization in the \npipe\ maps.  Certain modes could be down-weighted to minimize uncertainty.  Nevertheless, specializing the methods to utilize this information is complicated and not attempted in this paper.

\begin{figure*}[htpb]
  \begin{center}
    \includegraphics[width=0.9\linewidth]{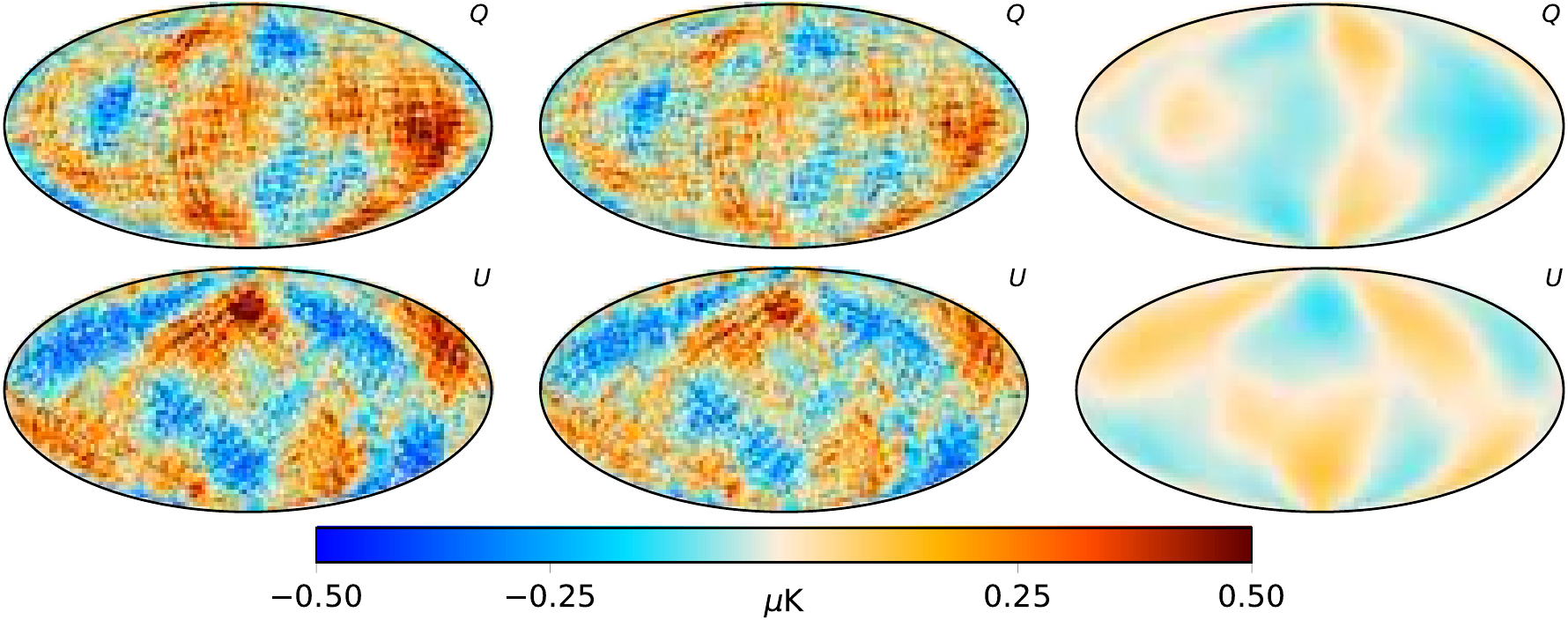}
  \end{center}
  \caption{Impact of the anisotropic \npipe\ transfer function (Fig.~\ref{fig:anisotropic_tf}) on a simulated CMB sky smoothed to 3\deg.  The columns (left to right) are the input CMB, the transfer-function-convolved CMB, and the difference.
  }
  \label{fig:cmb_and_tf}
\end{figure*}

To gain insight into the magnitude of the effect, we take a simulated CMB sky with a realistic amount of large-scale polarization power, and apply the 143-GHz anisotropic transfer function from Fig.~\ref{fig:anisotropic_tf} to it.  The input, output, and difference maps are shown in Fig.~\ref{fig:cmb_and_tf}.

\begin{figure*}[htpb]
  \begin{center}
    \includegraphics[width=1.0\linewidth]{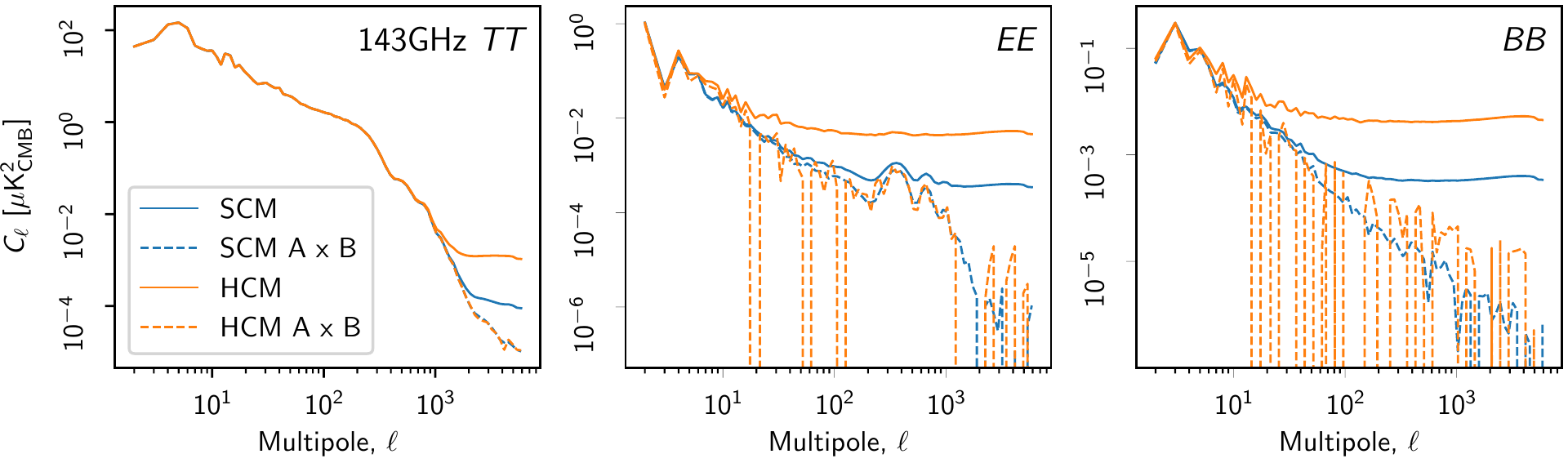}
  \end{center}
  \caption{Comparing the stable science scan (SCM) and the repointing manoeuvre (HCM) data.  We find excellent consistency between the two disjoint data sets.  The solid lines are auto spectra computed over 50\,\% of the sky.  The dashed lines are corresponding A/B cross-spectra.  The spectra are binned into 100 logarithmically-spaced bins.  The dashed orange spectrum in the $EE$ panel demonstrates that even the 143\GHz\ repointing-manoeuvre data alone (which were excluded from all previous releases) are sensitive enough to probe at least the first three acoustic $EE$ spectrum peaks.  The agreement in small-scale $TT$ cross-spectra indicates that the HCM pointing reconstruction is accurate enough not to widen the effective beam.  The agreement in large-scale $EE$ and $BB$ power suggests that potential thermal effects from the thruster burns do not leak into large-scale polarization.  The small scale noise in the HCM data set is higher than in the SCM data set from having only $8\%$ of the integration time.
  }
  \label{fig:repointing_comparison}
\end{figure*}

We provide the measured $E$-mode transfer functions for all frequency auto-spectra, frequency cross-spectra, and detector-set cross-spectra.  The format of the files is exactly the same as the \quickpol\ beam-window-function files used in \cite{planck2016-l05}.  In these files, the transfer function is unity everywhere except for the $\ell\,{<}\,42$ $E$ modes.  These transfer functions were measured with the 60\,\% sky mask.  It is impossible to anticipate all sky masks and map combinations that a user may require, so we also provide software that enables measurement of the transfer function for arbitrary sky masks.  To ensure fidelity of the science results, any statistic that employs large-scale ($\ell\,{<}\,10$) CMB polarization should be corrected for the transfer function or otherwise calibrated with the simulations.  This is the baseline for all \npipe\ products.

\subsubsection{Non-CMB transfer function}
\label{sec:noncmbtransfer}

There is no algebraic reason to expect that the suppression of large-scale CMB polarization would extend to Galactic or extragalactic foregrounds.  Indeed, measurement of these signals is driven by the 30- and 353-GHz channels, neither of which is subject to the polarization prior in the calibration.  Furthermore, the presence of polarized foregrounds is accounted for in the calibration process by marginalizing over the polarization templates.

Foreground experts may object to the use of foreground templates to model the polarized foregrounds in the calibration process.  Such templates are inherently incapable of supporting spatial variation of the spectral index of the foregrounds.  When these templates make up the polarization prior, the calibration process does have the potential to suppress true spatial variation of the spectral index.  We are confident that the limitations of our polarization prior have not compromised the foreground polarization for three reasons.
  \begin{enumerate}
  \item The calibration is performed using a Galactic mask, excluding the more intense Galactic plane and the source of more probable spatial variation of the spectral index.
  \item Spatial variation in the dust spectral index has limited support by fitting both 217- and 353-GHz templates.
  \item Our simulations that are based on the spatially-varying \commander\ sky model do not indicate any suppression of the spatially varying spectral indices (see Fig.~\ref{fig:dust_transfunc}).
  \end{enumerate}

\subsection{Data taken during repointing manoeuvres} \label{sec:repointings}

It is not immediately obvious that including data taken during the 4-minute repointing manoeuvres is going to improve the resulting frequency maps and hence it is important to check for consistency.  The three thruster burns of a given manoeuvre lead to a faint but measurable impact on the \hfi\ thermal baseline, and the attitude reconstruction during the manoeuvre is admittedly less robust due to the dynamic nature of the data.  We explored the consistency between ``stable science mode'' (labelled ``SCM'' in the \Planck\ attitude history files) and the repointing mode (``HCM'') by building 143-GHz full-frequency and detector-set maps exclusively from either SCM or HCM data.  We show the power spectra of such maps in Fig.~\ref{fig:repointing_comparison}.  These power spectra demonstrate excellent consistency between the two data subsets.  Accordingly, data taken during pointing manoeuvres are used in all the \npipe\ data products.   {\it Only\/} SCM data were used in previous \Planck\ products and results.  This additional integration time reduces the small-scale noise uncertainty in \npipe\ by approximately $9\,\%$.

\clearpage

\section{Simulations}
\label{sec:simulations}

\npipe\ is released with 600 high fidelity Monte Carlo simulations that include CMB, foreground, noise and systematics.  These simulations allow testing and debiasing analysis tools with life-like frequency and detector-set maps with known inputs.  For further details of the release, see Appendix~\ref{app:release}.

Calibrating against the same sky signal we are trying to measure is an inherently nonlinear problem and calibration and other template uncertainties in the \npipe\ maps are not negligible.  For these reasons we found it necessary to simulate the entire reprocessing part (Sect.~\ref{sec:reprocessing}) of the \npipe\ processing.  Each Monte Carlo iteration begins with a simulated CMB sky, represented as an expansion in spherical harmonics.  We convolve the CMB with high-order expansions of the final \Planck\ scanning beams ($\ell_\mathrm{max}=2048$ for \lfi\ and 4096 for \hfi).  The convolution is carried out at the sample level \citep{Prezeau:2010mx}, respecting the actual beam orientation as a function of time.  Foregrounds are simulated by evaluating the \commander\ sky model at the target frequencies.  Static zodiacal emission is included by adding the same nuisance templates that \commander\ marginalized over.  When components of the model are measured to much higher resolution than the target frequency (e.g., dust at 30\GHz), we smooth the foreground component with an azimuthally-symmetric beam expansion \citep{Hivon:2016qyw} specifically calculated for the \npipe\ data set.  Bandpass mismatch is included by adding appropriately-weighted maps of foreground frequency derivatives and CO, customizing the foreground for each detector.  The foreground map is then sampled into TOD, using the appropriate detector pointing weights, and co-added with the CMB TOD.  Finally, we add the expected dipole and far-sidelobe signals.  Noise is simulated from the measured noise PSDs as in \cite{planck2014-a14}, including a correlated noise component in each polarized horn.

The simulated CMB sky includes lensing and frequency-dependent frame-boosting effects, as described in \cite{planck2014-a14}.  The CMB realizations are the FFP10 simulations used in \prthree, and the cosmological parameters are listed in Table~\ref{tab:simparams}.

\begin{table}[htpb!]
  \begingroup
  \newdimen\tblskip \tblskip=5pt
  \caption{Cosmological parameters of the FFP10 simulations.
  }
  \label{tab:simparams}
  \nointerlineskip
  \vskip -3mm
  \setbox\tablebox=\vbox{
    \newdimen\digitwidth
    \setbox0=\hbox{\rm 0}
    \digitwidth=\wd0
    \catcode`*=\active
    \def*{\kern\digitwidth}
    \newdimen\signwidth
    \setbox0=\hbox{$-$}
    \signwidth=\wd0
    \catcode`!=\active
    \def!{\kern\signwidth}
    \halign{
\hbox to 2.0cm{#\leaderfil}\tabskip 4em&
\hfil#\hfil \tabskip 0pt\cr
      \noalign{\doubleline}
      \omit\hfil Parameter \hfil&
      \omit\hfil Value \hfil\cr
      \noalign{\vskip 3pt\hrule\vskip 4pt}
      $h^\star$  & $0.6701904$\cr
      $\omega_\mathrm b^\dagger$& $0.02216571$\cr
      $\omega_\mathrm c$& $0.1202944$\cr
      $\tau$& $0.06018107$\cr
      $A_\mathrm s$& $2.119631 \times 10^{-9}$\cr
      $n_\mathrm s$& $0.9636852$\cr
      $r$& $0.01$\cr
      \noalign{\vskip 3pt\hrule\vskip 5pt}
    }
  }
  \endPlancktable
  \tablenote {{\star}} $h$ is the dimensionless Hubble parameter:
  $H_0 = h \times 100\kmsMpc$.\par
  \tablenote {{\dagger}} $\omega\equiv\Omega h^2$.\par
  \endgroup
\end{table}

The gain fluctuations of several percent seen in \lfi\ detectors are applied to the simulated TOD once all the components are added.  Rather than trying to develop a statistical model that captures all the relevant features of the measured fluctuations, we apply the same smoothed version of the measured gain fluctuation to all Monte Carlo realizations.

\hfi\ ADC nonlinearity is included into the simulations using the same tools as for \lfi\ gain fluctuations, i.e., we apply a time-dependent gain fluctuation derived by smoothing the measured apparent gain fluctuations in the flight data.  Similarly, we apply the measured transfer-function residuals to the simulated bolometer TOD.

\subsection{Instrument noise}
\label{sec:instrument_noise}

The $1/f$ instrument noise is simulated using the inverse fast Fourier transform (FFT) technique.  A vector of complex Gaussian random numbers is generated, multiplied by a measured power spectral density (PSD), and then Fourier transformed into the time domain.  We treat each pointing period as independent, enforcing no continuity across the pointing period boundaries.

Our noise estimation uses the same technique as described in \cite{planck2014-a14}, i.e., we subtract an estimate of the signal by resampling the full-frequency map into the time domain, evaluate the correlation function over unflagged and unmasked samples, and Fourier transform to obtain the PSD.  The low-frequency part of the PSD is measured from a downsampled version of the timestream to lessen the cost of estimating the covariance function.  The traditional method of estimating the PSD directly from the TOD by FFT scales as $\mathcal O(N\log N)$, but is forced to use all of the timestream samples, even the ones that contain a constrained realization of the data (e.g., to fill gaps) or are subject to steep gradients in the sky signal.  The covariance function estimation used in \npipe scales as $\mathcal O(N^2)$, but allows for omission of the flagged samples and use of only the part of the sky where the signal estimate is robust.

The noise estimates are derived from preprocessed TOD to best match the noise seen during reprocessing.  We apply the measured \lfi\ gains during the estimation to remove the several-percent gain fluctuations from the noise estimates.

\begin{figure*}[htpb]
  \includegraphics[width=1.0\linewidth]{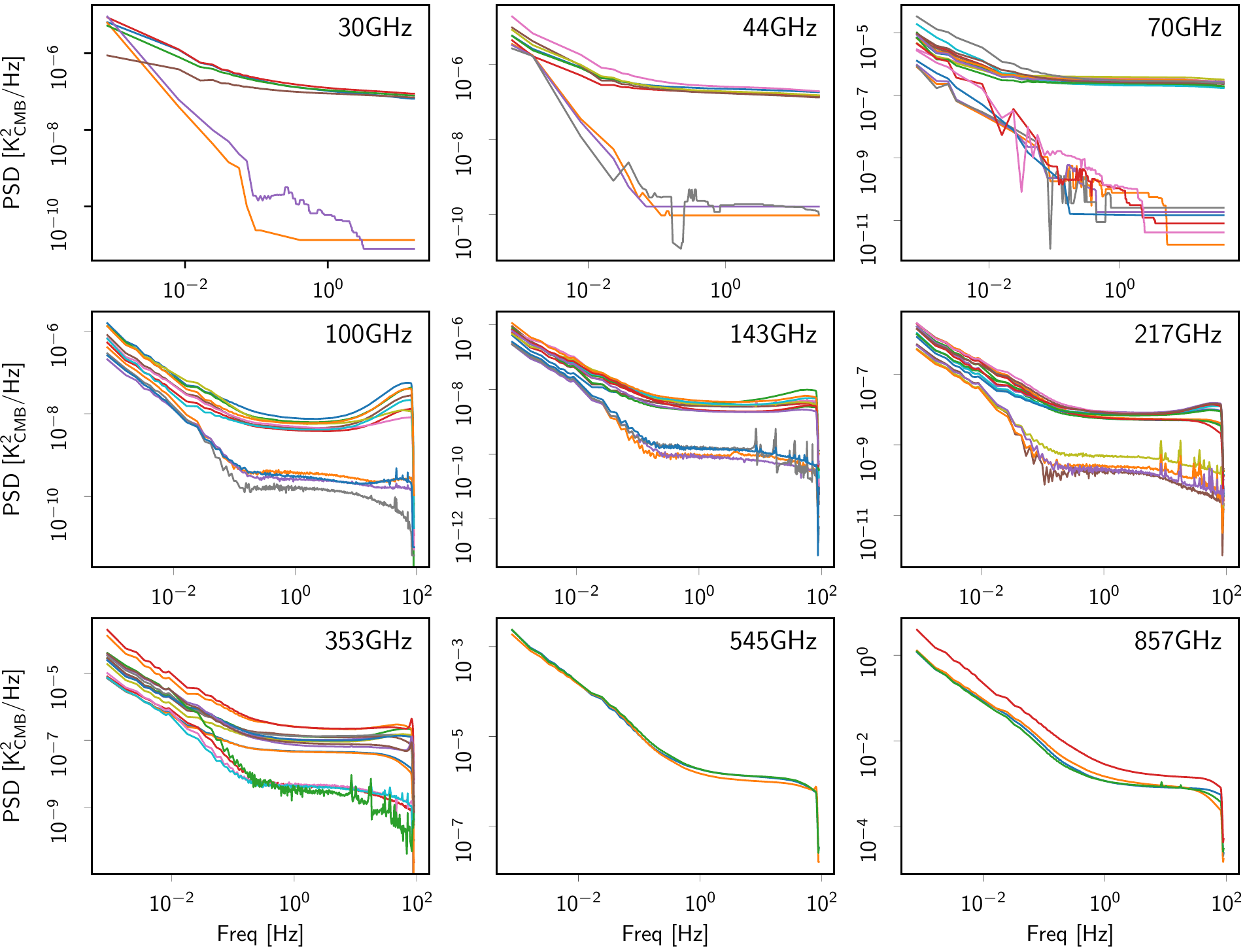}
  \caption{Averaged noise PSDs for each detector (upper curves) and correlated-noise modes for each polarized horn (lower curves).  The total noise power is the sum of the correlated and uncorrelated modes.  These noise PSDs are measured from the data by subtracting a signal estimate and then evaluating the sample-sample covariance function.  The \hfi\ noise is suppressed near the Nyquist frequency ($\approx 90\Hz$) by the bolometric transfer function filtering.  The PSDs are used for simulating the $1/f$ noise fluctuations, and as inputs to the \madam\ noise filter for destriping.
  }
  \label{fig:psds}
\end{figure*}

We have further developed the noise estimation code to use the covariance function method for measuring the correlated noise between detectors.  Of particular interest is the correlated noise within each polarized horn (between the two polarization-orthogonal receivers).  The correlated noise affects the sum of the two receivers, and thus the estimation of sky temperature; however, it cancels in the difference of the two receivers, leaving the polarization estimate unaffected.  The noise estimates in Fig.~\ref{fig:psds} demonstrate that all polarized horns show significant internal noise correlations at low frequency, but only the \hfi\ horns have significant noise correlations at high frequency.  The \hfi\ noise correlations are thought to result from coincident cosmic-ray glitches.  The correlated \hfi\ modes also exhibit several 4-K line residuals, which is understandable, given that the changes in the measured line-template amplitudes are highly correlated.

We simulate the correlated modes using the same inverse FFT technique as used for the uncorrelated modes, and then co-add them to the uncorrelated timestreams.  Only the intra-horn component is simulated.  We ignore the smaller and less relevant correlation between horns.

\subsection{What is \emph{not} included in the simulations}
\label{sec:notincluded}

\begin{figure*}[htpb]
  \includegraphics[width=1.0\linewidth]{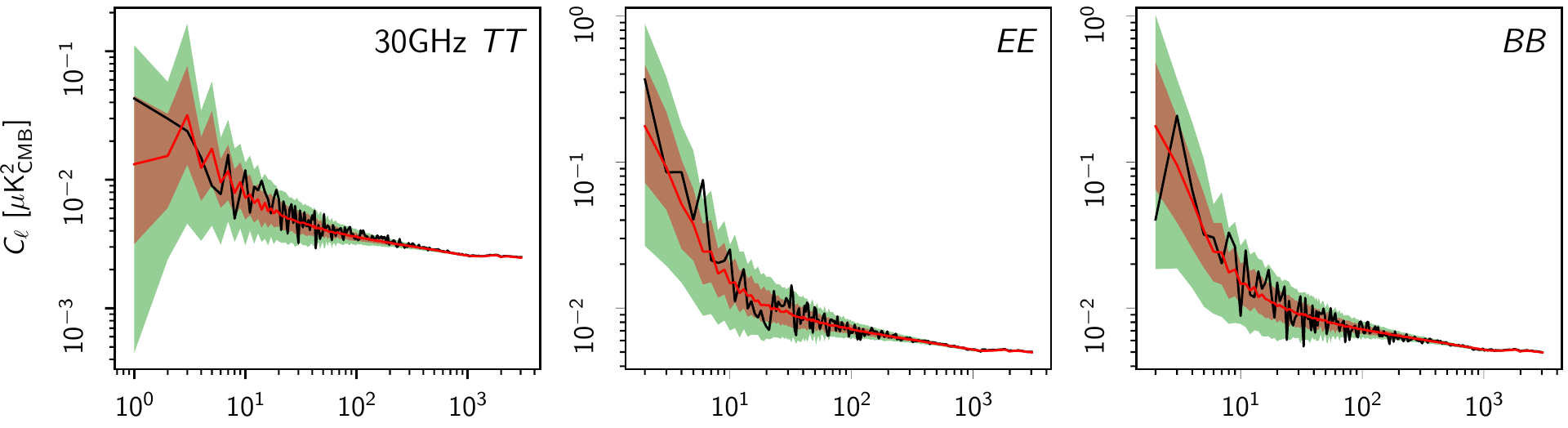}\\
  \includegraphics[width=1.0\linewidth]{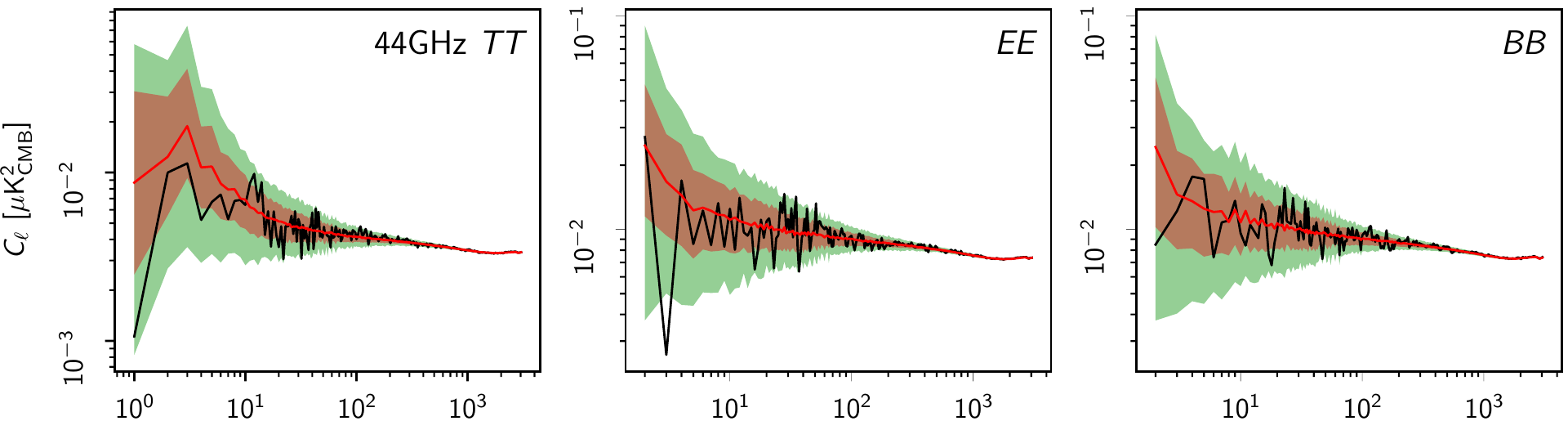}\\
  \includegraphics[width=1.0\linewidth]{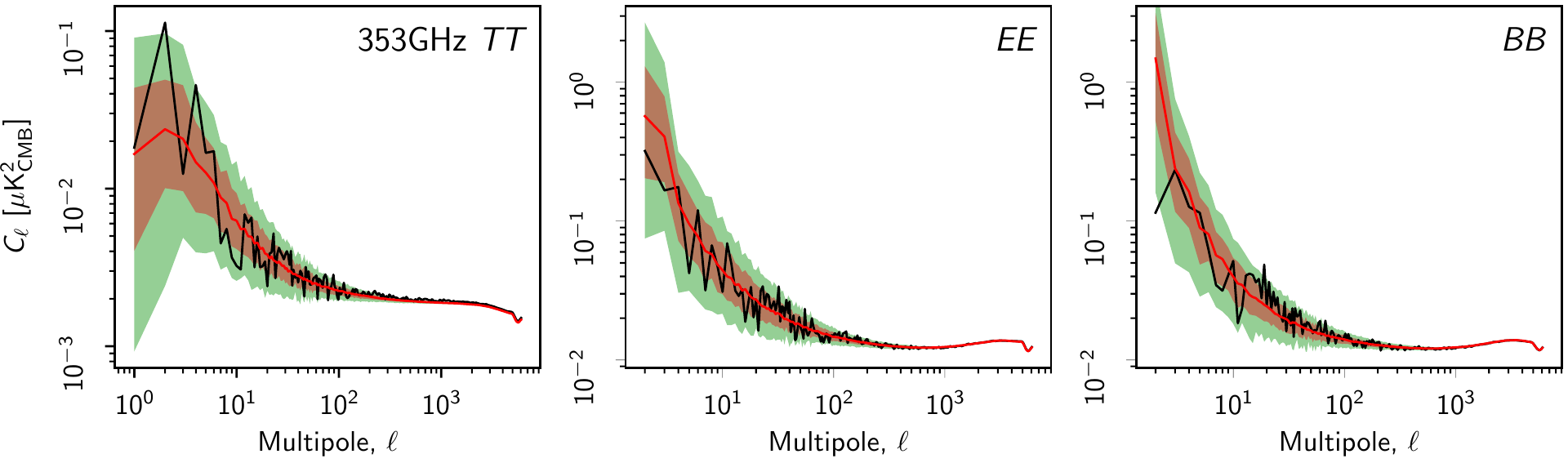}\\
  \centering
  \includegraphics[width=0.66\linewidth]{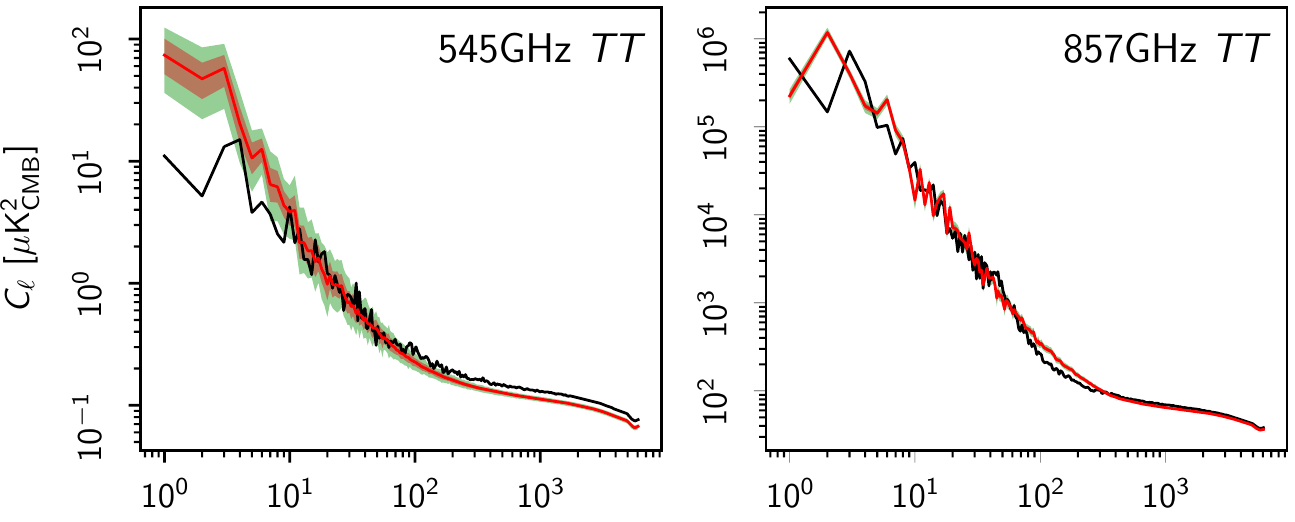}
  \caption{Simulated A/B difference versus flight data A/B difference at 30, 44, 353, 545, and 857\GHz.  The flight data are shown in black, and the median simulated power in each bin is plotted in red.  The coloured bands represent the asymmetric 68\,\% and 95\,\% confidence regions.  The power spectra are binned into 300 logarithmically spaced bins.  These spectra are shown again in Fig.~\ref{fig:abdiff_relative_fg}, but divided by the median simulated spectrum.
  }
  \label{fig:abdiff_fg}
\end{figure*}

\begin{figure*}[htpb]
  \includegraphics[width=1.0\linewidth]{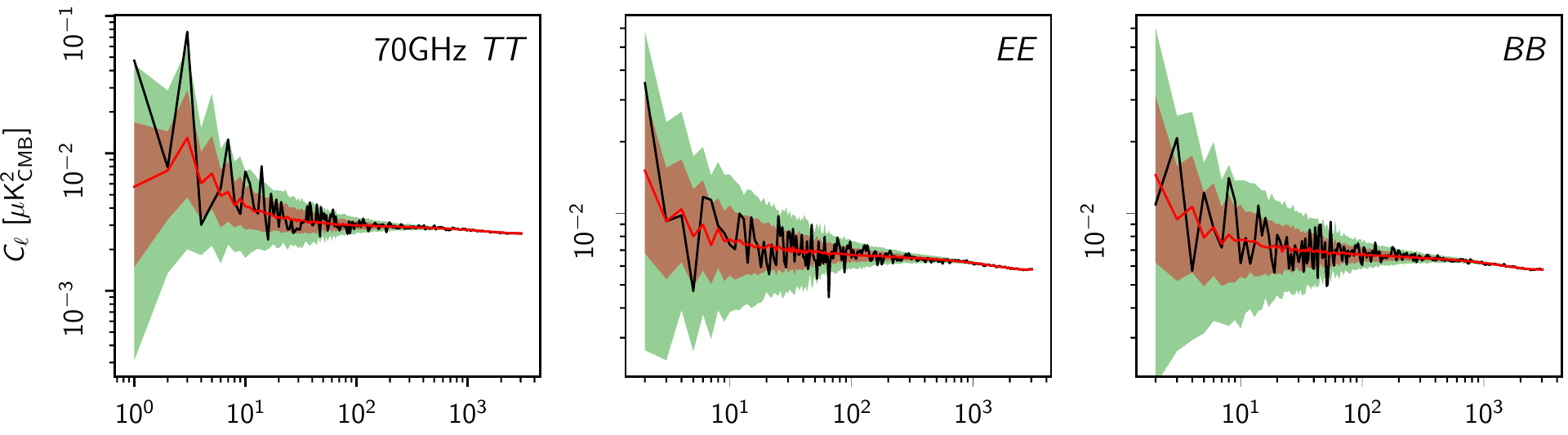}\\
  \includegraphics[width=1.0\linewidth]{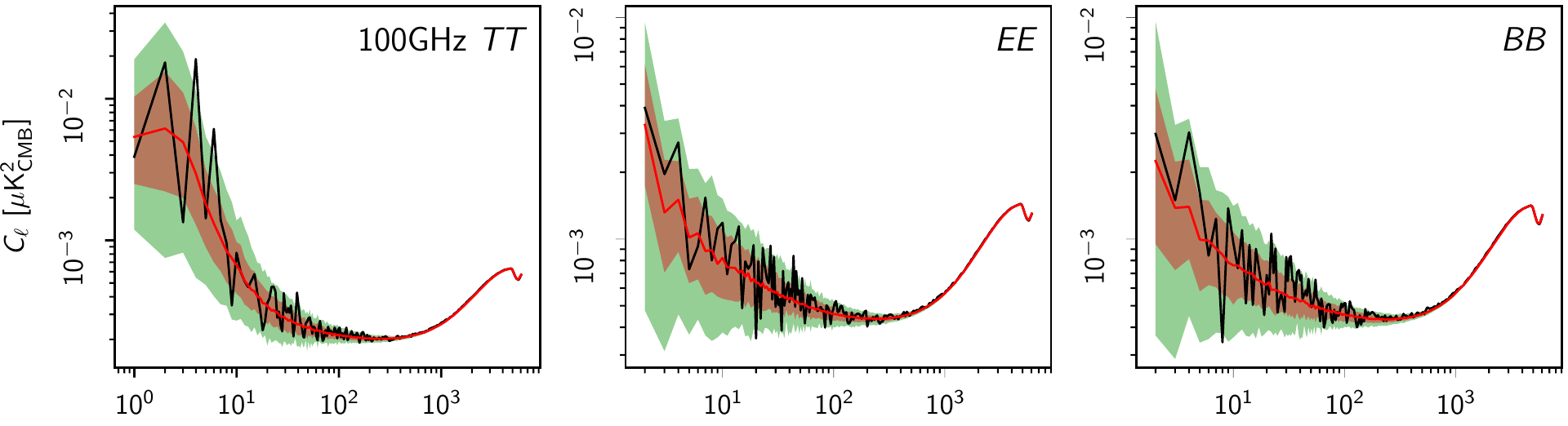}\\
  \includegraphics[width=1.0\linewidth]{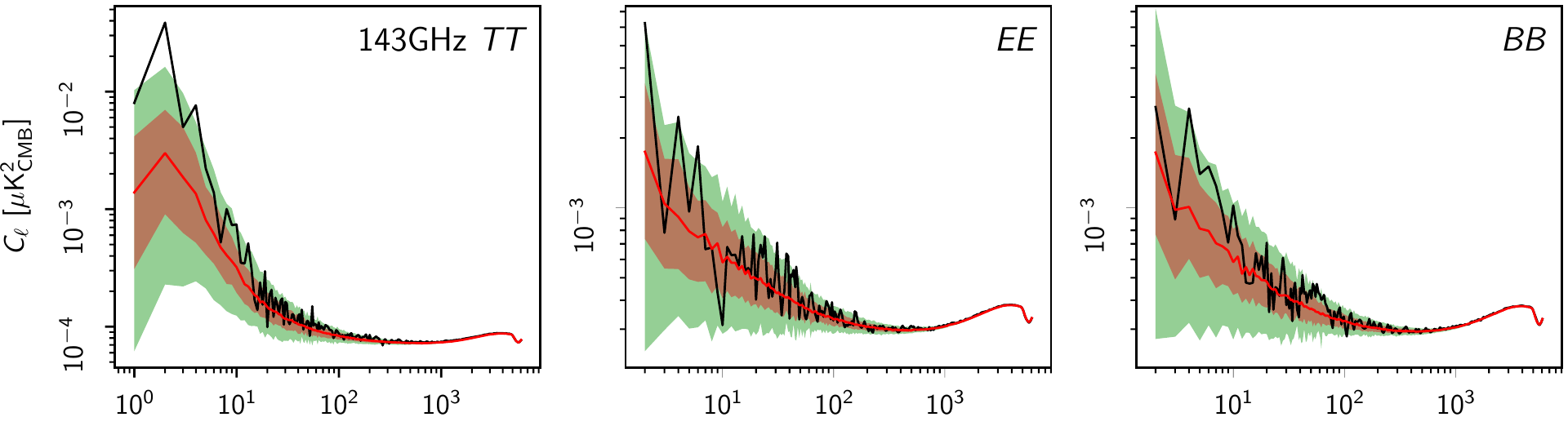}\\
  \includegraphics[width=1.0\linewidth]{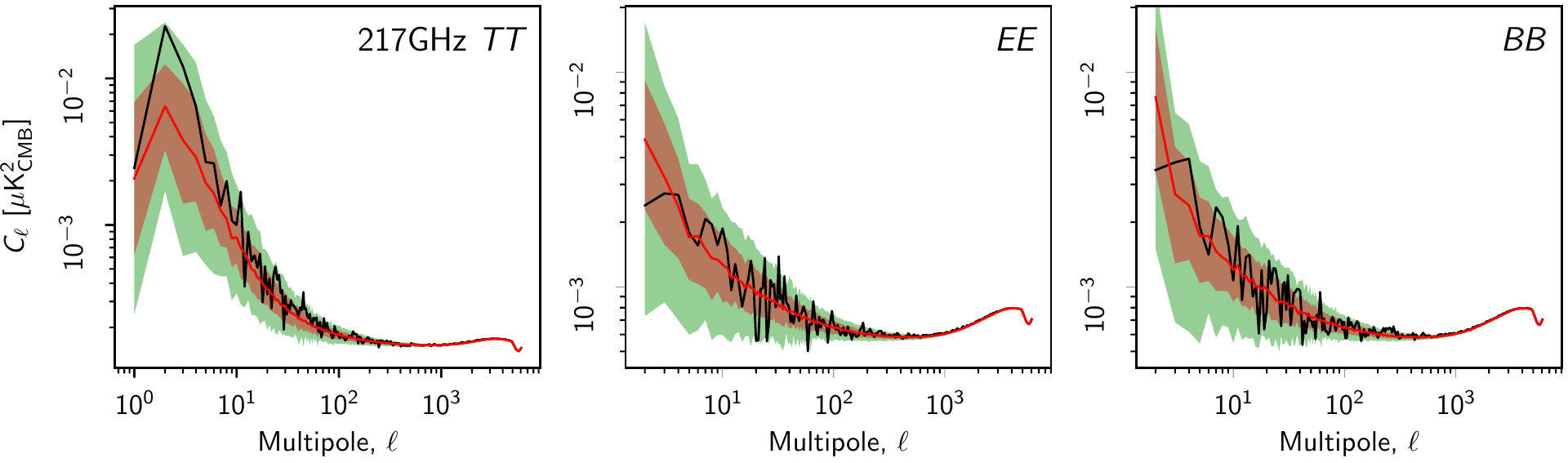}
  \caption{Same as Fig.~\ref{fig:abdiff_fg}, at 70, 100, 143, and 217\GHz.  These spectra are shown again in Fig.~\ref{fig:abdiff_relative_cmb} but divided by the median simulated spectrum.
  }
  \label{fig:abdiff_cmb}
\end{figure*}

\begin{figure*}[htpb]
  \centering
  \includegraphics[width=1.0\linewidth]{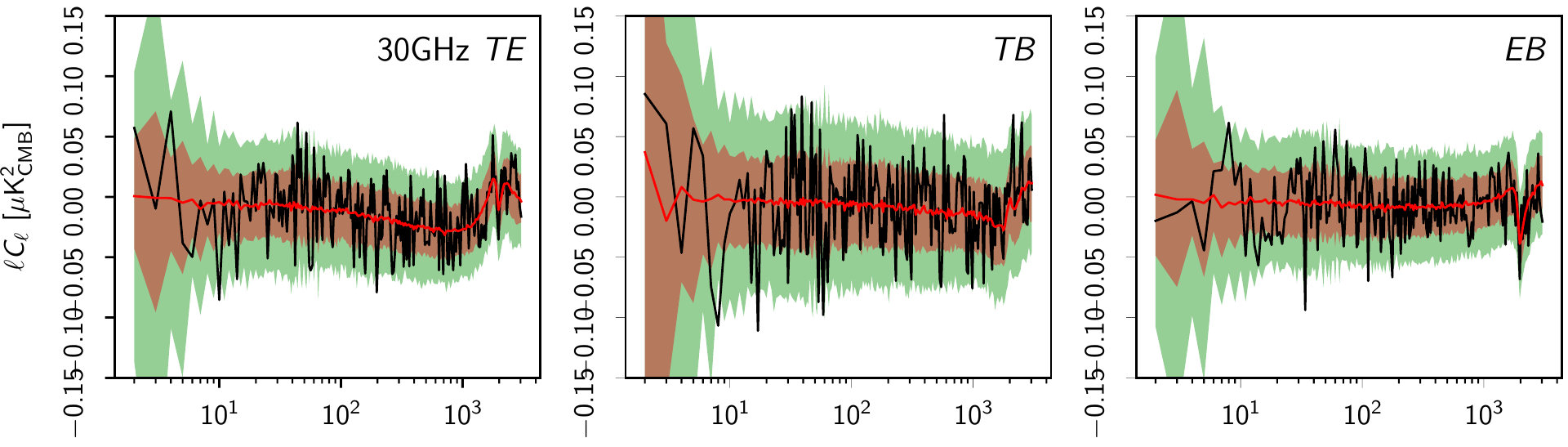}\\
  \includegraphics[width=1.0\linewidth]{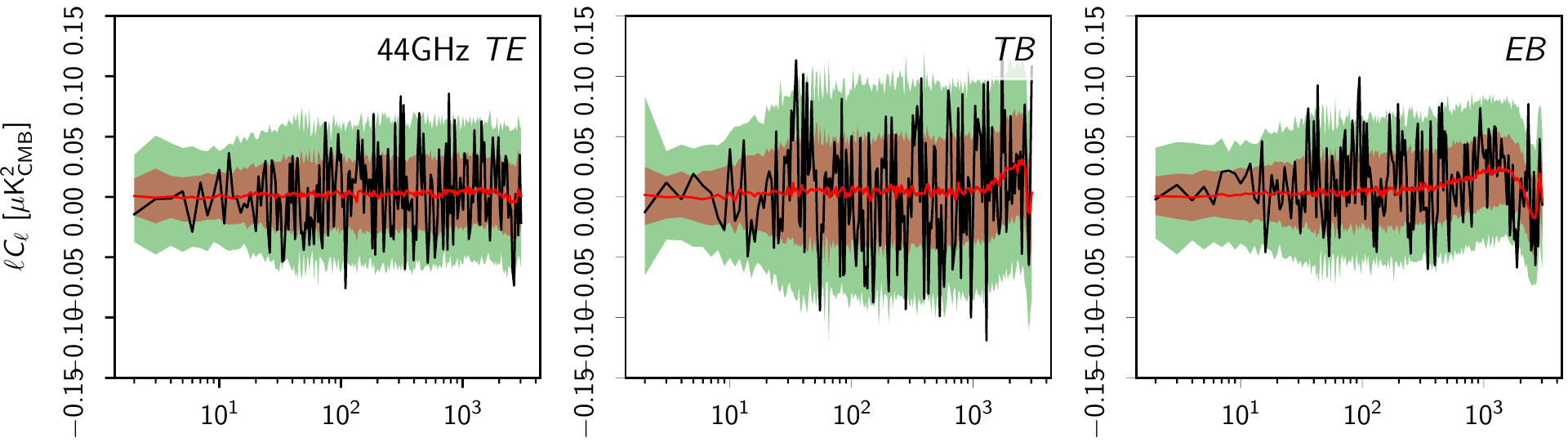}\\
  \includegraphics[width=1.0\linewidth]{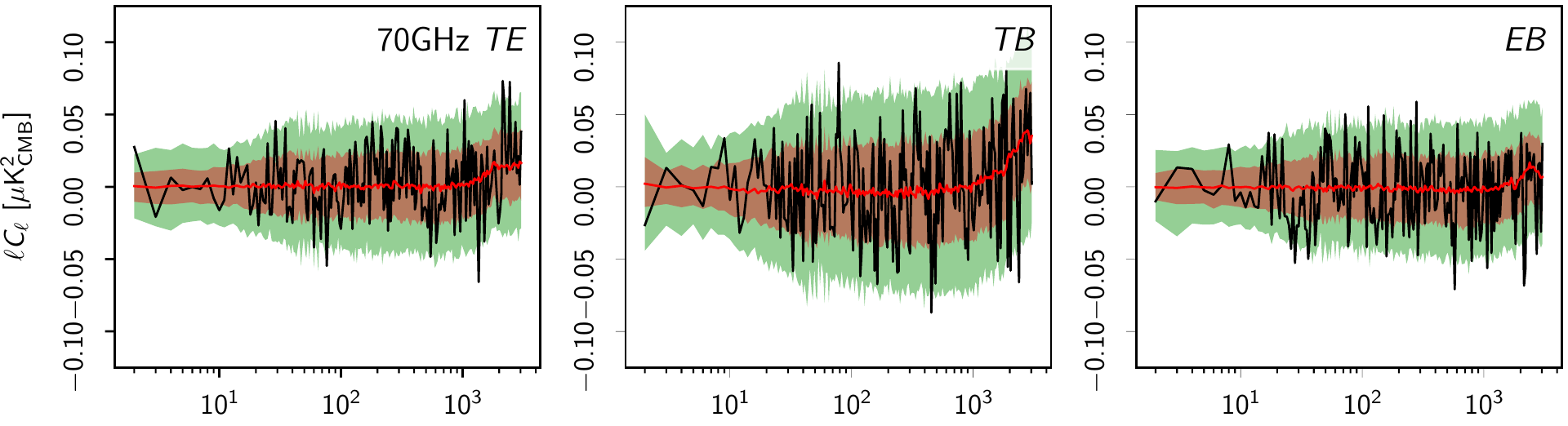}
  \caption{Simulated A/B difference versus flight data A/B difference for $TE$, $TB$, and $EB$ at 30, 44, and 70\GHz.  Similar to 
   Fig.~\ref{fig:abdiff_fg}, except for the $\ell$-scaling.
  }
  \label{fig:abdiff_cross_lfi}
\end{figure*}

\begin{figure*}[htpb]
  \centering
  \includegraphics[width=1.0\linewidth]{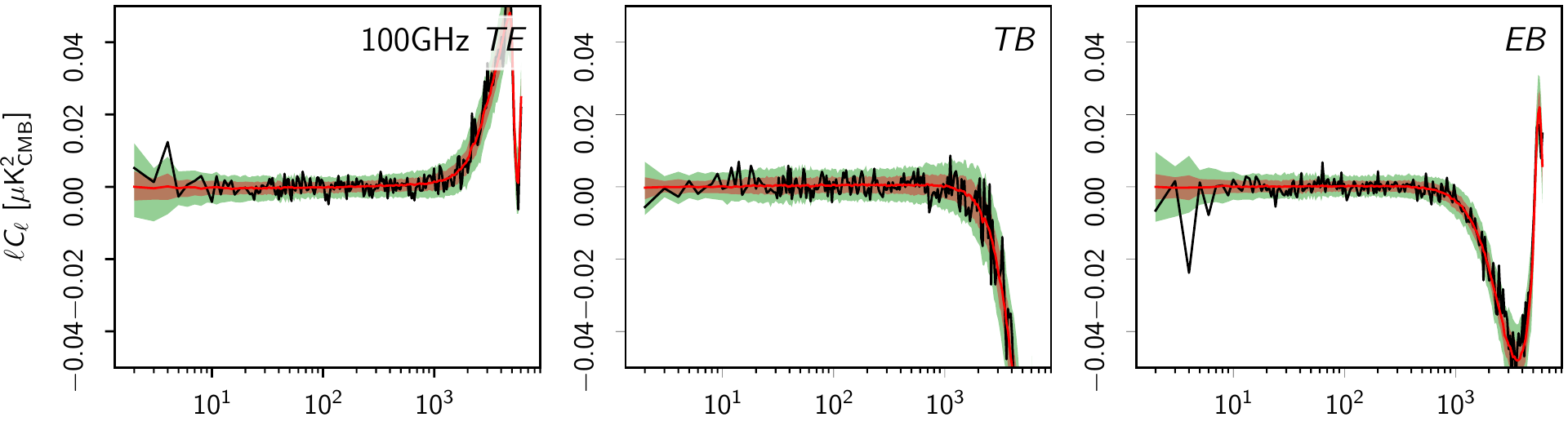}\\
  \includegraphics[width=1.0\linewidth]{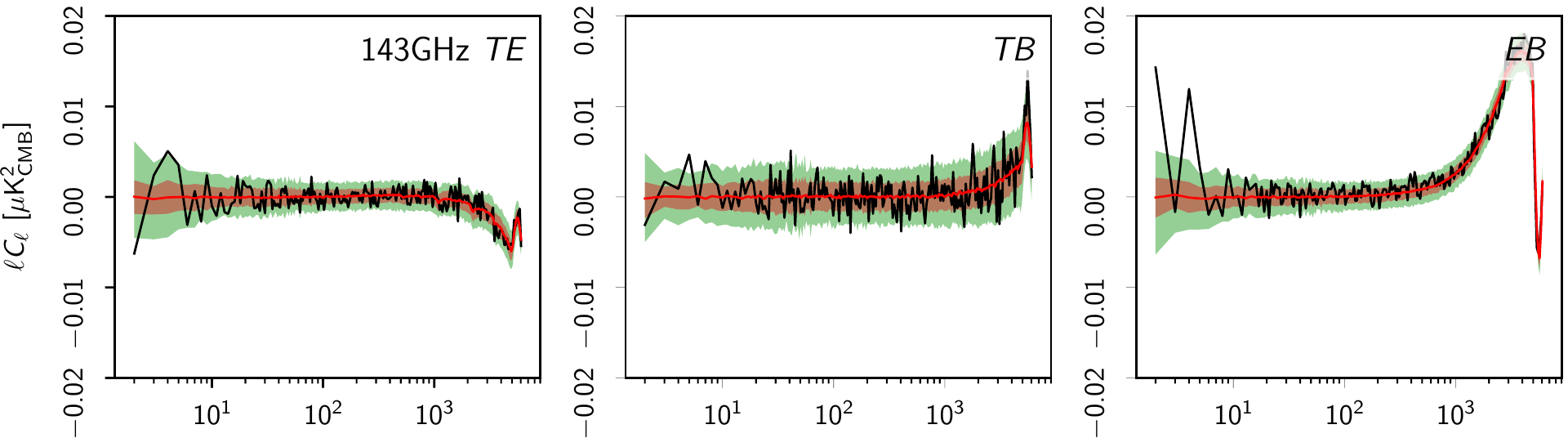}\\
  \includegraphics[width=1.0\linewidth]{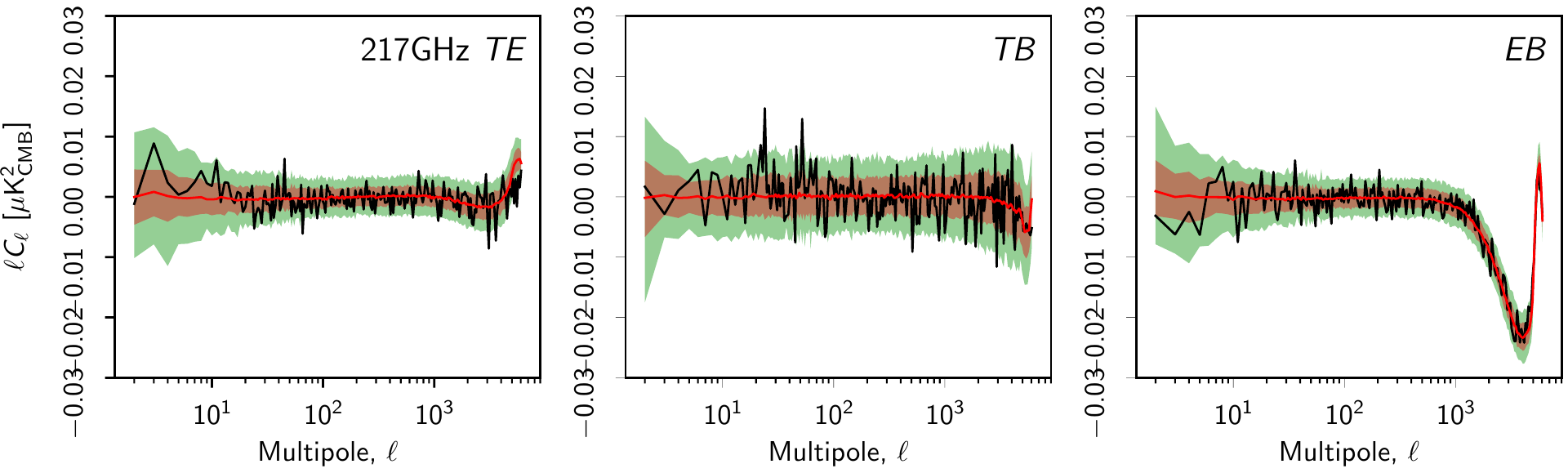}\\
  \includegraphics[width=1.0\linewidth]{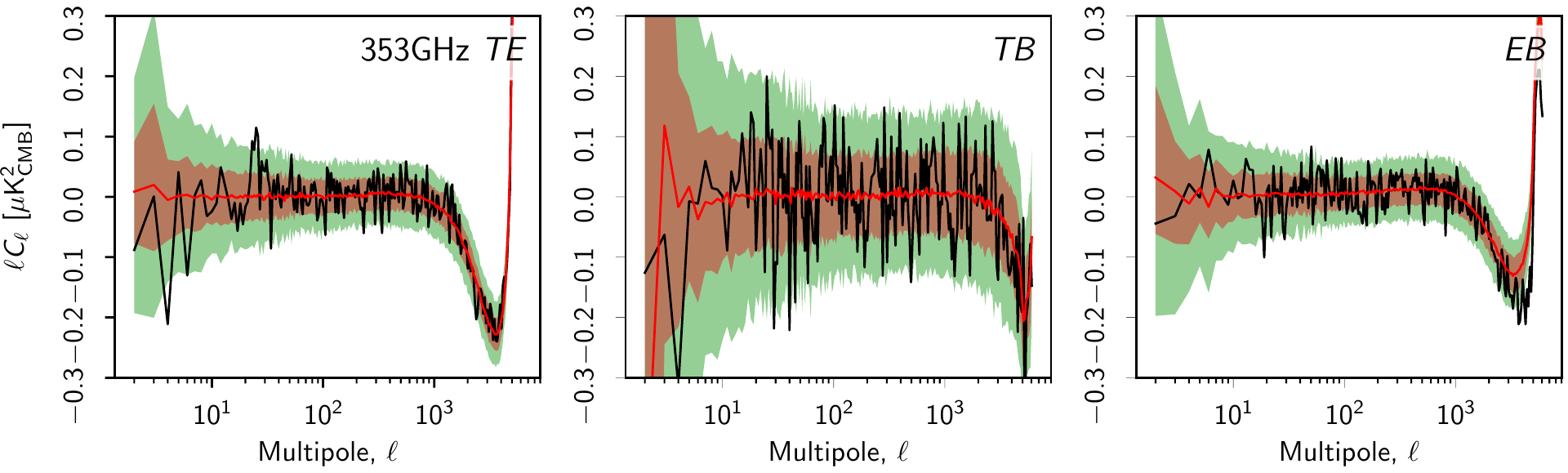}
  \caption{Same as Fig.~\ref{fig:abdiff_cross_lfi}, for 100, 143, 217, and 353\GHz.
  }
  \label{fig:abdiff_cross_hfi}
\end{figure*}

\npipe\ simulations make an effort to capture all of the relevant systematics and their coupling in the final maps; however, the prohibitive cost of preprocessing precludes running the preprocessing module on simulated timelines.  This leads to some notable omissions:
\begin{itemize}
\item ADC nonlinearity residuals beyond the linear gain fluctuation model (the higher-order distortion fluctuations are shown to be small in Fig.~\ref{fig:distortions});
\item 1-Hz and 4-K line residuals, except for what is captured by noise estimation;
\item cosmic-ray glitch residuals, except for what is captured by noise estimation and modelled as Gaussian noise;
\item measuring and updating the detector polarization parameters; and
\item errors in the bandpass-mismatch templates, i.e., the sky model and the CO maps that are solved with \commander\ and fed back in future iterations of \npipe.
\end{itemize}
Despite these omissions, we take the general agreement between flight data and simulated residuals in Figs.~\ref{fig:abdiff_fg}, \ref{fig:abdiff_cmb}, \ref{fig:abdiff_cross_lfi}, and \ref{fig:abdiff_cross_hfi} as a validation of our approach.

\hfi\ ADC nonlinearities are simulated in \npipe\ using the linear gain model.  This notably excludes the higher-order distortion modes discussed in Sect.~\ref{sec:adc}.  These modes are very faint and the fits include a fair amount of statistical uncertainty.  Without these higher modes of distortion, the simulated maps are affected only by the statistical uncertainty associated with the distortion correction.  The excellent agreement between the large-scale polarization residuals in flight data and simulations suggests that the statistical uncertainties are sufficient to describe the large-scale polarization uncertainties in the \npipe\ maps.

The line and glitch residuals would manifest themselves as small-scale excess power in the \npipe\ maps.  We only model these residuals to the extent that they can be measured in the detector noise PSD.  The binned PSD cannot fully capture the Poissonian nature of the glitch residuals, nor the finely localized features associated with the lines.  However, since the small-scale power of our simulated maps agrees very well with the real maps, we deduce that the mismatch is negligible.

Updating the instrument model using the single-detector residual maps and polarization estimates has not been formally included in the \npipe\ processing plan, but has been done incrementally as post-processing over multiple revisions of the \npipe\ maps.  Simulating this process was deemed unfeasible given the time constraints.  \npipe\ simulations are therefore based on the final instrument model, including the adjustments to polarization efficiency and angle determined from flight data processing and shown in Table~\ref{table:polparams} .

\npipe\ \hfi\ simulations differ from the FFP10 simulations used in \cite{planck2016-l03} in some important ways.  FFP10 implemented a light-weight version of the low-level data processing (without glitch or 4-K line removal) allowing for application of the ADCNL directly in the fast, modulated samples.  It also allowed simulating the information loss from compression and decompression.  A noteworthy complication of this approach was that the instrumental noise also needed to be simulated in the raw data, necessitating an extra noise-alignment step later in the pipeline.  Analysis of the FFP10 results showed that the ADCNL was by far the most important systematic included in the simulation.  \npipe\ uses a more approximate approach, simulating the data directly in their preprocessed form based on noise estimates from the same stage data and adding ADCNL consistent with the linear gain model.  Despite their differences, most effects do get included in both pipelines but using very different ways to express them.  Our simulations are the first \Planck\ simulations to apply the asymmetric scanning beam to each CMB realization separately in the time domain.  In FFP10 only one beam-convolved CMB realization was passed through the full processing pipeline.

\subsection{\npipe\ simulation results}
\label{sec:sim_results}

\begin{figure*}[htpb]
  \includegraphics[width=1.0\linewidth]{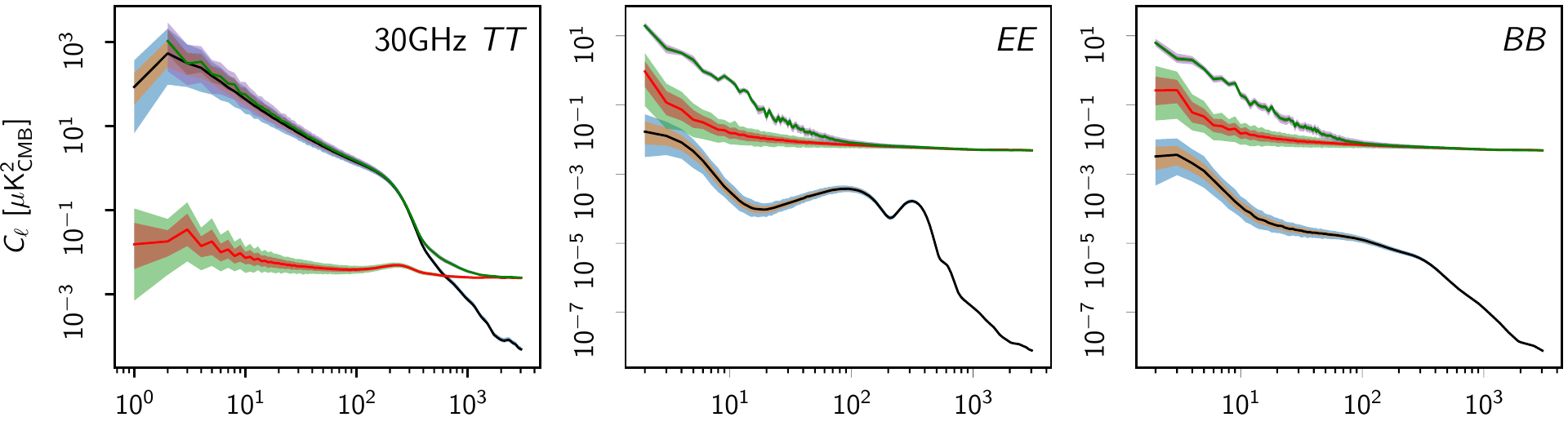}\\
  \includegraphics[width=1.0\linewidth]{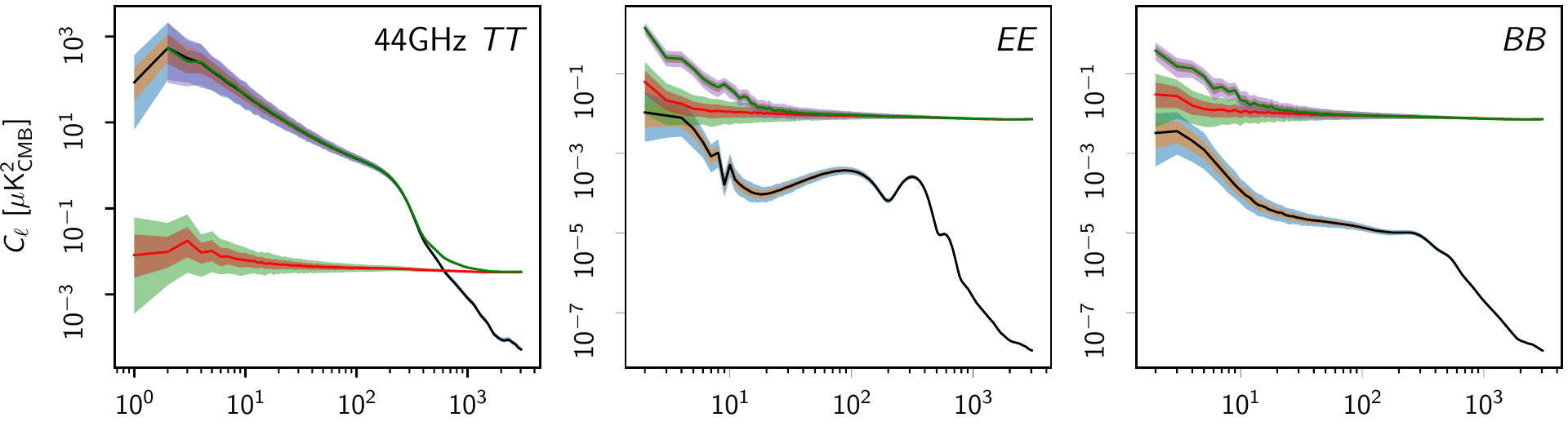}\\
  \includegraphics[width=1.0\linewidth]{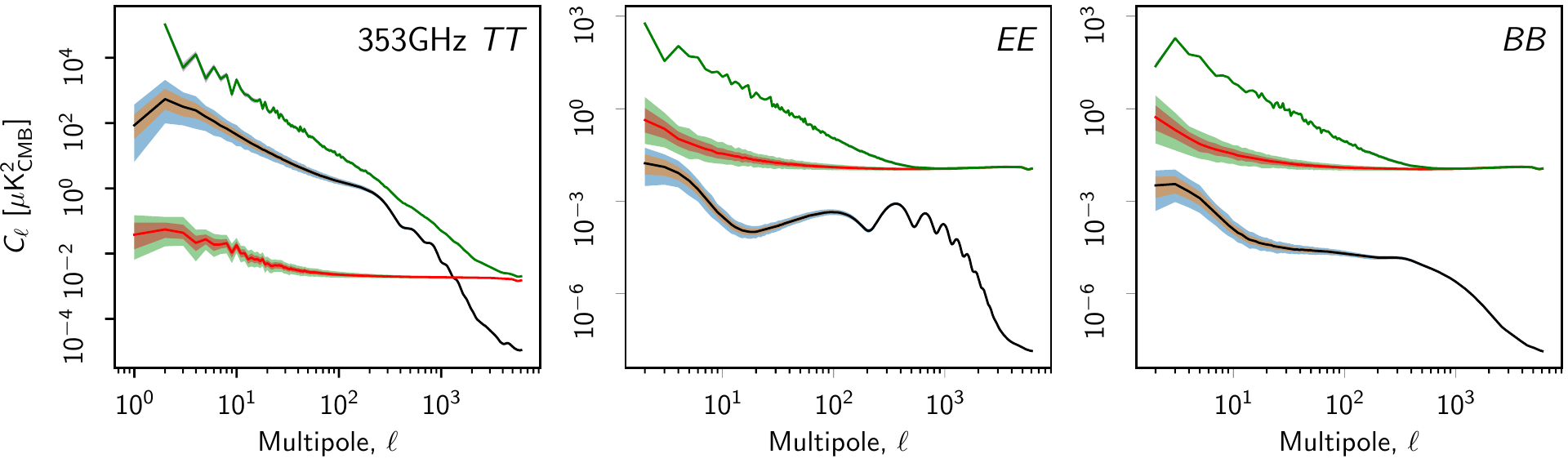}\\
  \centering
  \includegraphics[width=0.66\linewidth]{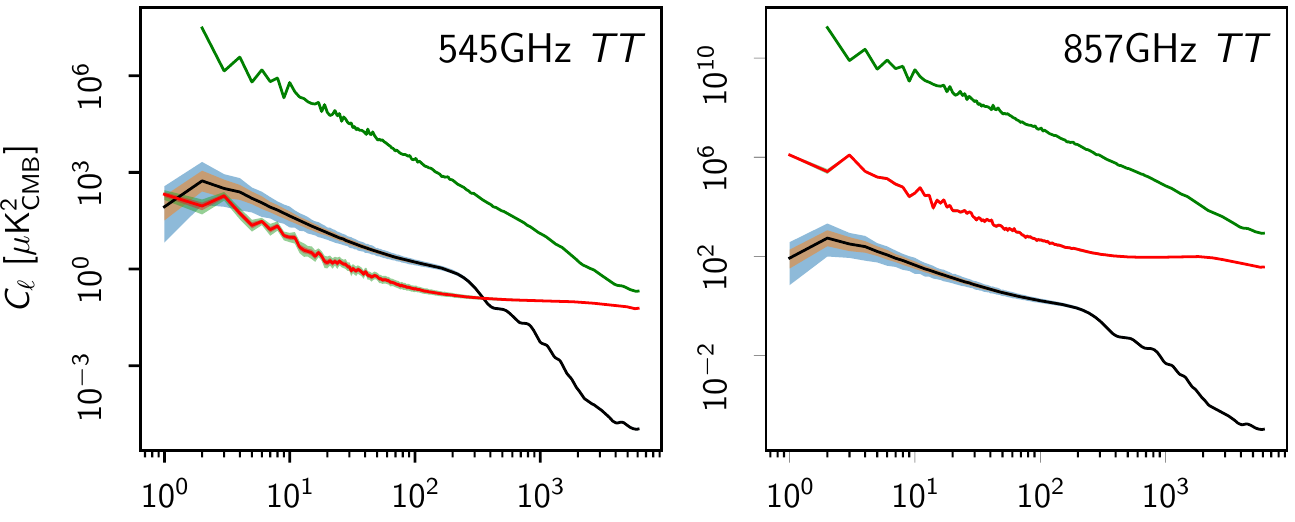}
  \caption{Simulated CMB, noise, and systematics pseudo-spectra at 30, 44, 353, 545, and 857\GHz.  The median CMB power in each bin is plotted in black, and the median noise and systematics power (difference between the input and output maps) in red.  The coloured bands represent the asymmetric 68\,\% and 95\,\% confidence regions.  The power spectra are binned into 300~logarithmically-spaced bins.  The CMB is convolved with the beam and $E$-mode transfer function (Sect.~\ref{sec:ee_tf}).  The large-scale CMB $B$-mode power in these plots follows from mode coupling and is not intrinsic to the simulations.
  }
  \label{fig:residual_cl_fg}
\end{figure*}

\begin{figure*}[htpb]
  \includegraphics[width=1.0\linewidth]{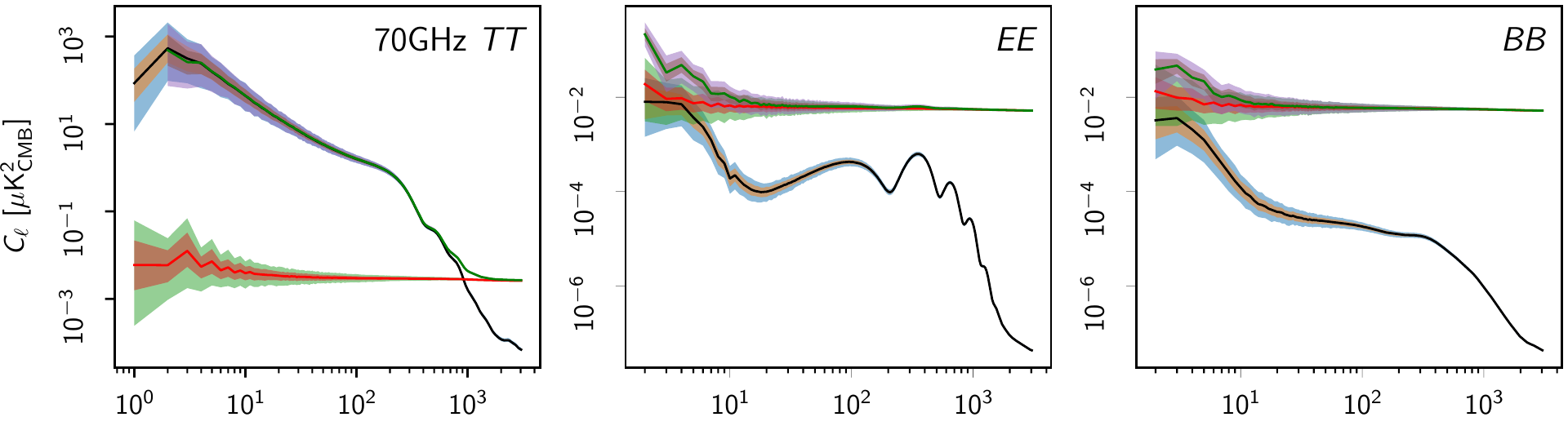}\\
  \includegraphics[width=1.0\linewidth]{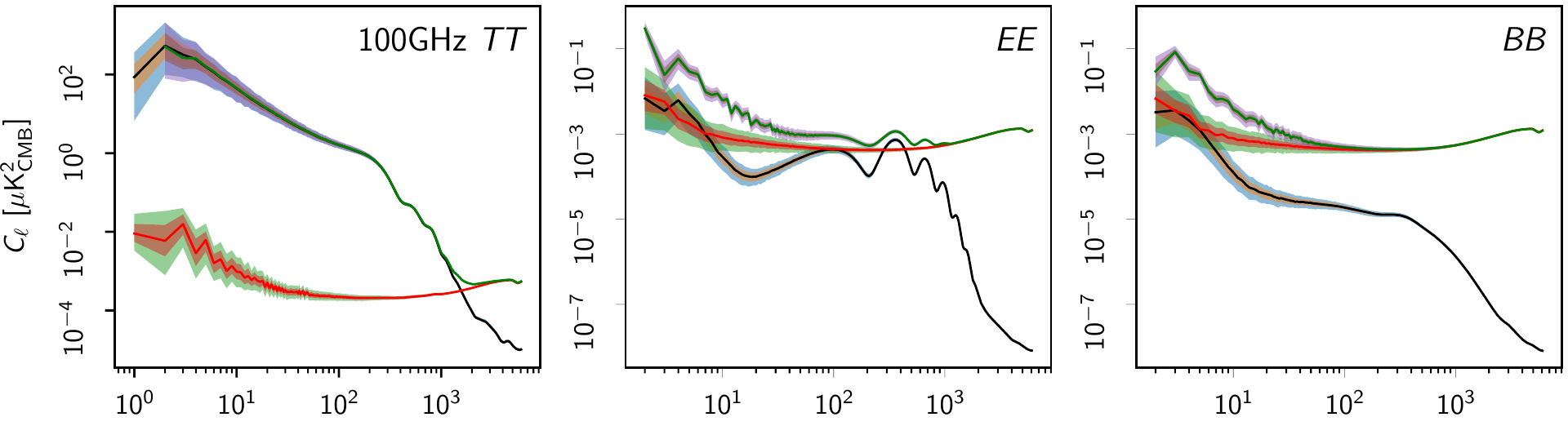}\\
  \includegraphics[width=1.0\linewidth]{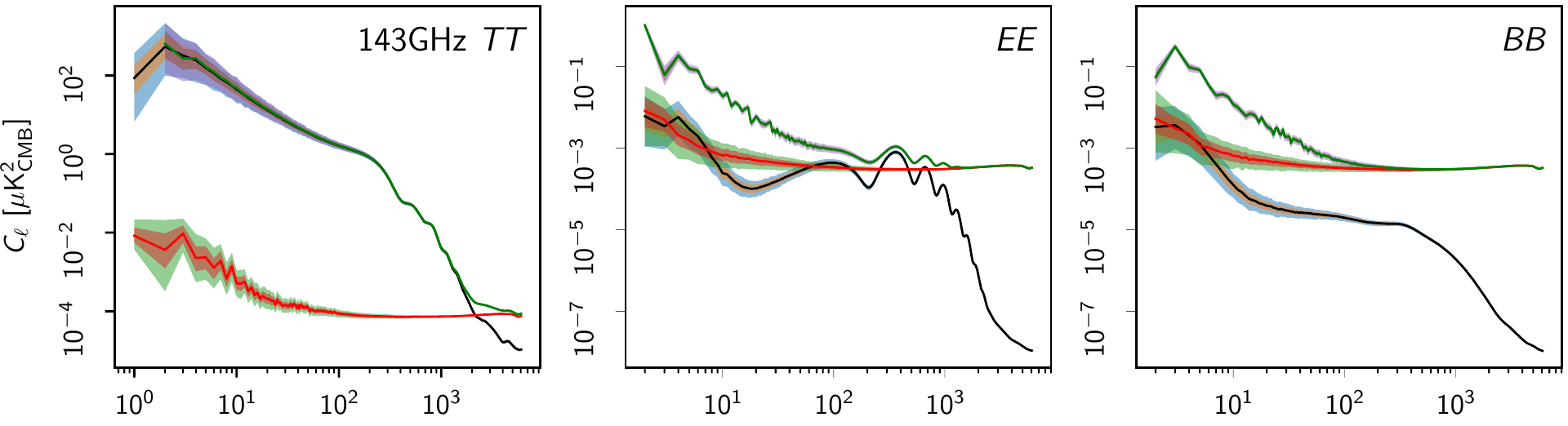}\\
  \includegraphics[width=1.0\linewidth]{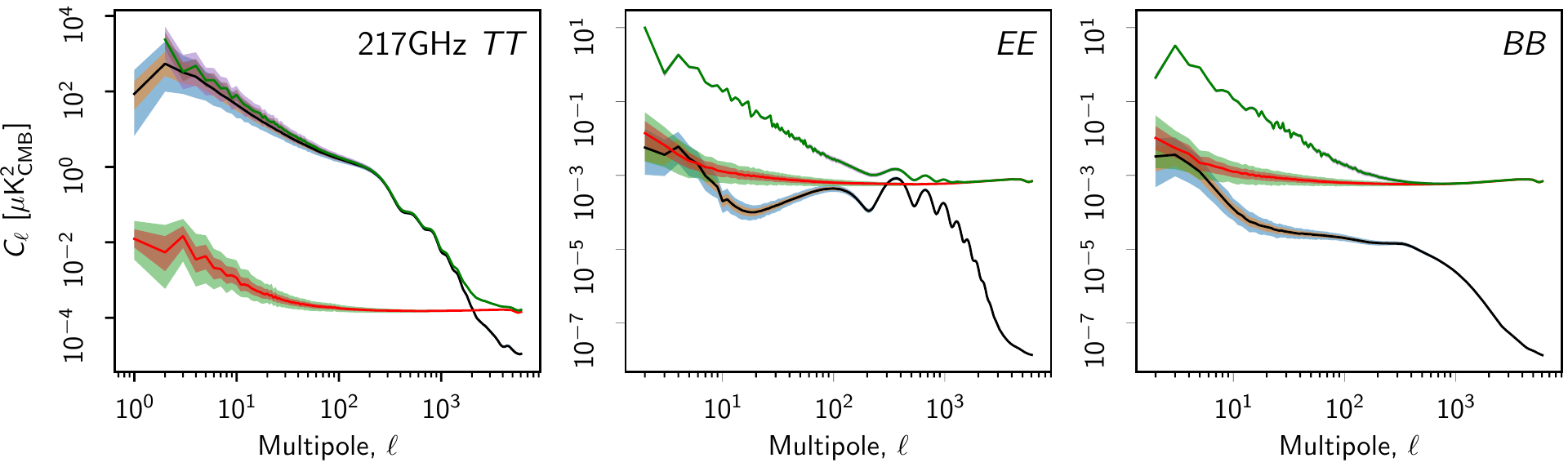}
  \caption{Same as Fig.~\ref{fig:residual_cl_fg}, for 70, 100, 143
    and 217\GHz.
  }
  \label{fig:residual_cl_cmb}
\end{figure*}

We compare the simulated CMB power to the simulated residuals in Figs.~\ref{fig:residual_cl_fg} and \ref{fig:residual_cl_cmb}.  They demonstrate that we are able to suppress large-scale noise and systematic uncertainties at or below the faint large-scale $EE$ power at the \Planck\ CMB frequencies between 70 and 217\GHz.

\subsubsection{Null maps and null map power spectra}

\begin{figure*}[htpb]
  \includegraphics[width=1.0\linewidth]{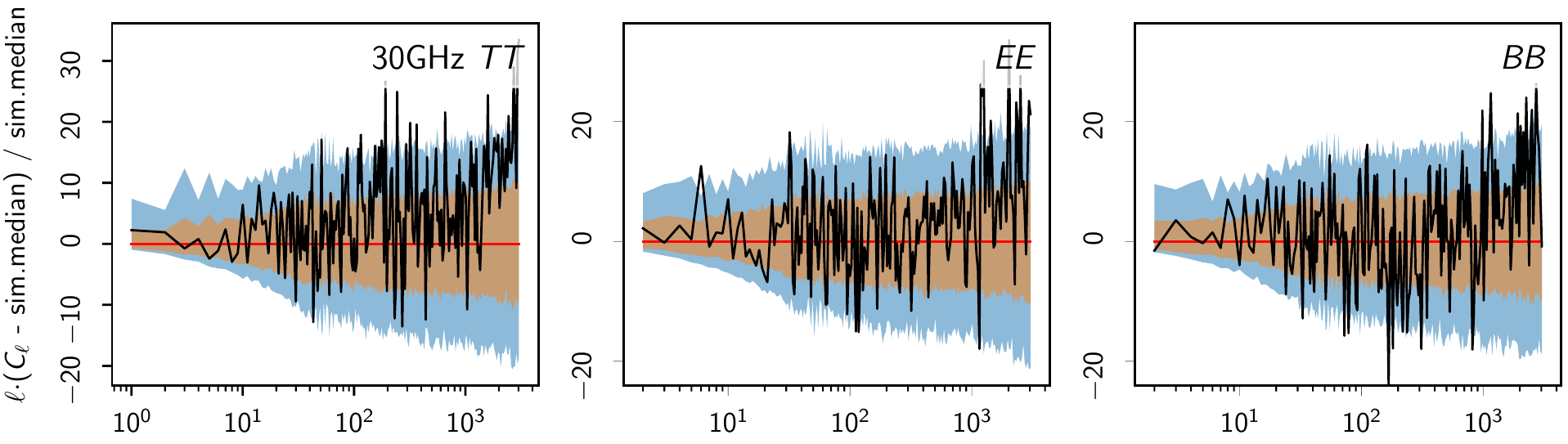}\\
  \includegraphics[width=1.0\linewidth]{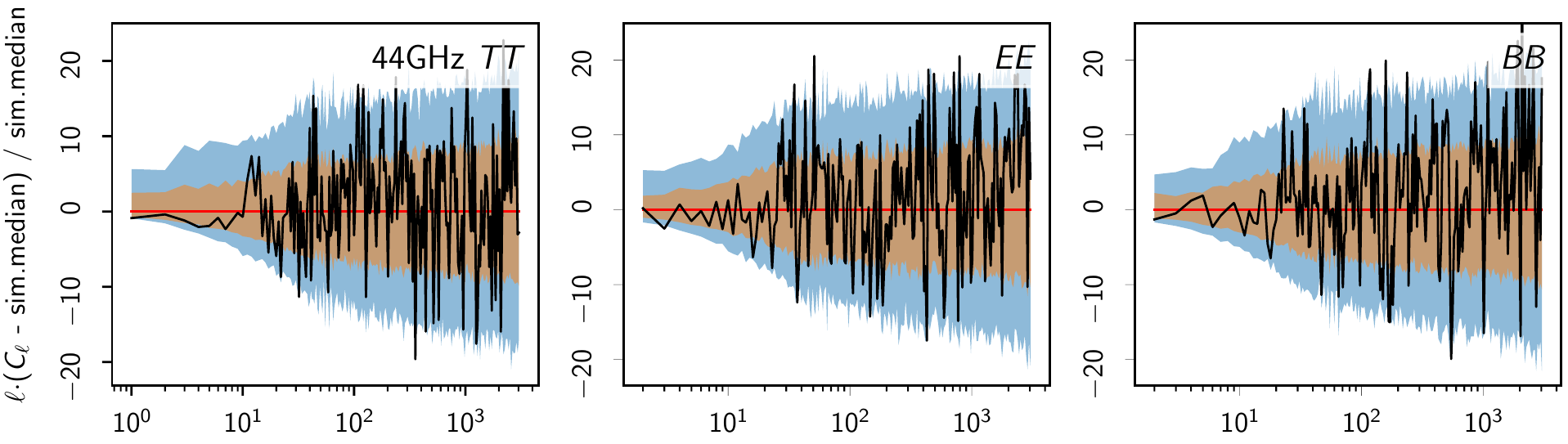}\\
  \includegraphics[width=1.0\linewidth]{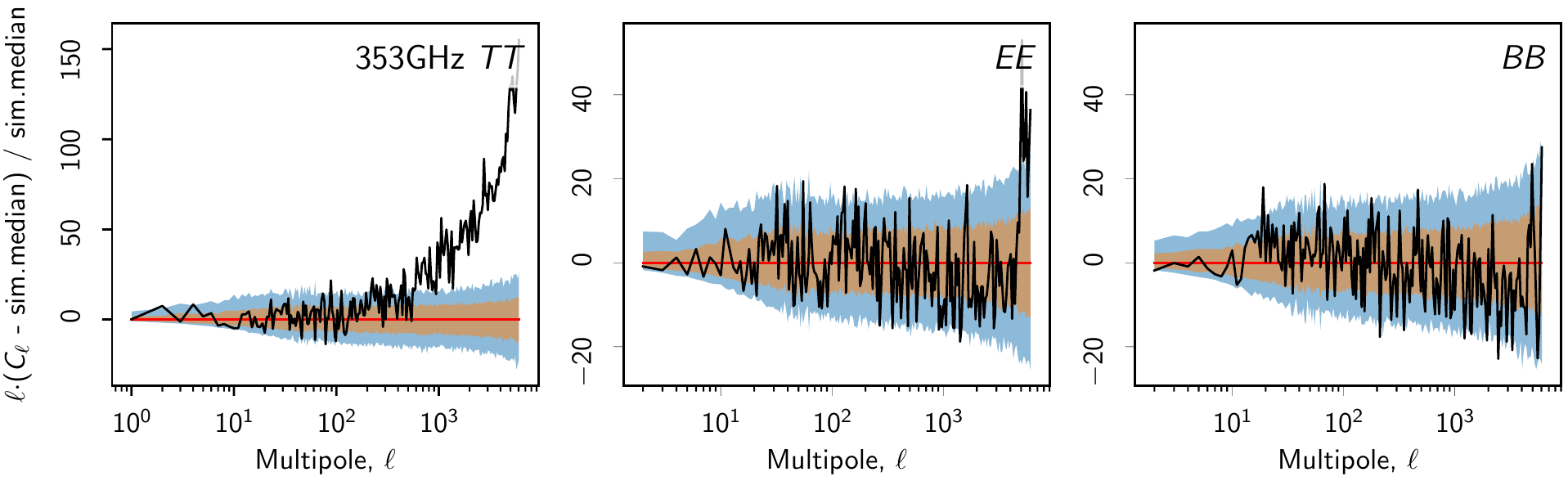}\\
  \centering
  \includegraphics[width=0.66\linewidth]{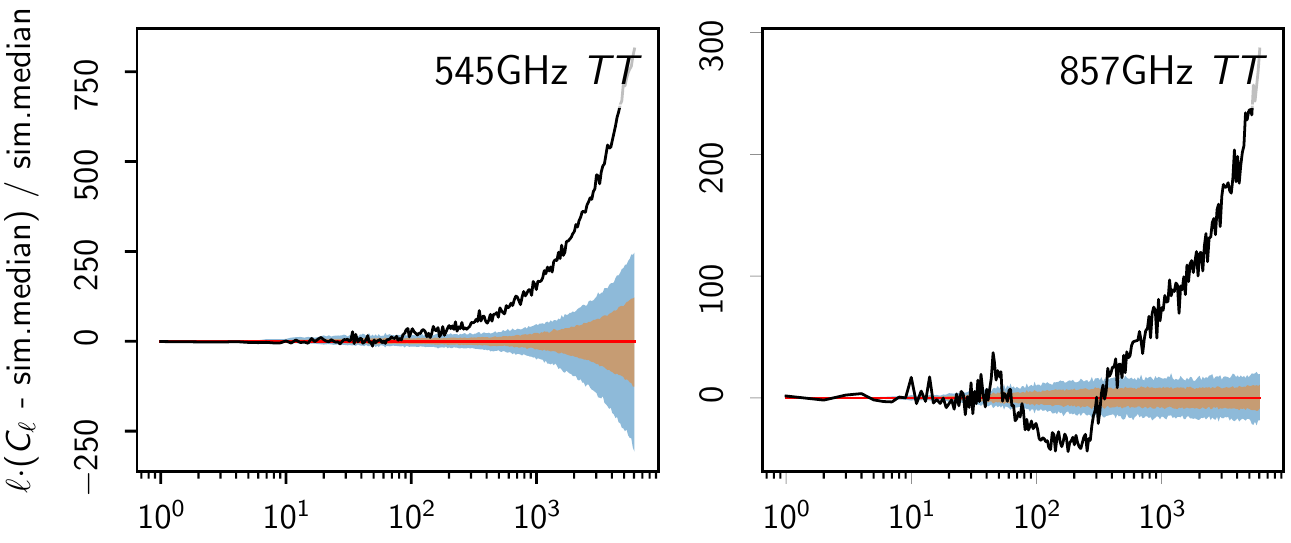}
  \caption{Simulated A/B difference versus flight data at 30, 44, 353, 545, and 857\GHz.  All spectra are differenced and divided
by the median simulated spectrum.  The flight data are shown in black, and the median simulated power in each bin is plotted in red.  The coloured bands represent the asymmetric 68\,\% and 95\,\% confidence regions.  The power spectra are binned into 300~logarithmically-spaced bins.  The spectra shown here are the same as in Fig.~\ref{fig:abdiff_fg}.  Notice the $\ell$-scaling.
  }
  \label{fig:abdiff_relative_fg}
\end{figure*}

\begin{figure*}[htpb]
  \includegraphics[width=1.0\linewidth]{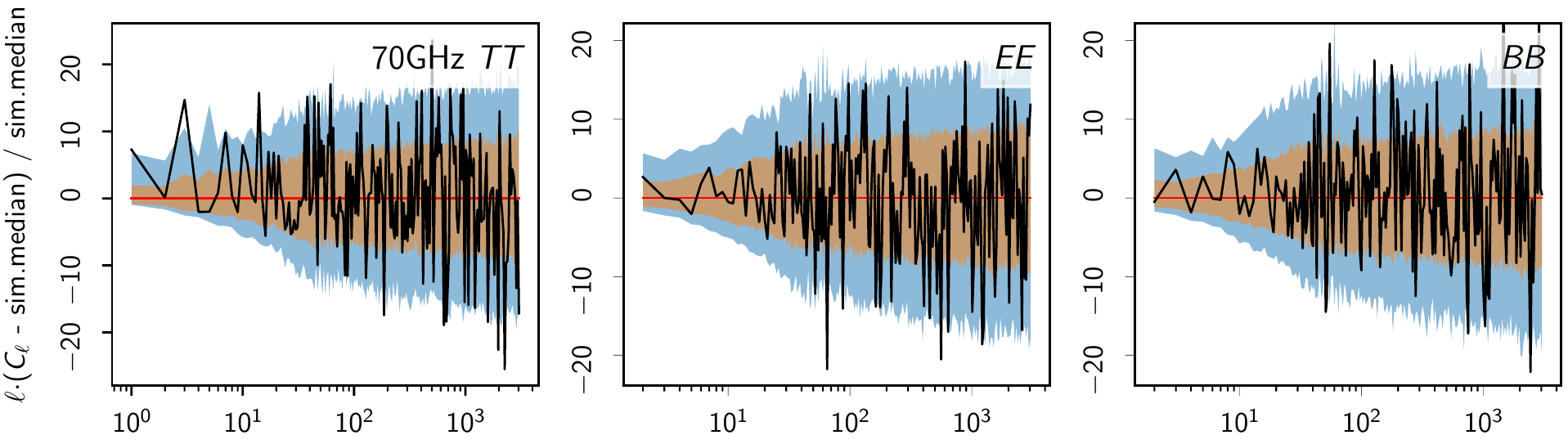}\\
  \includegraphics[width=1.0\linewidth]{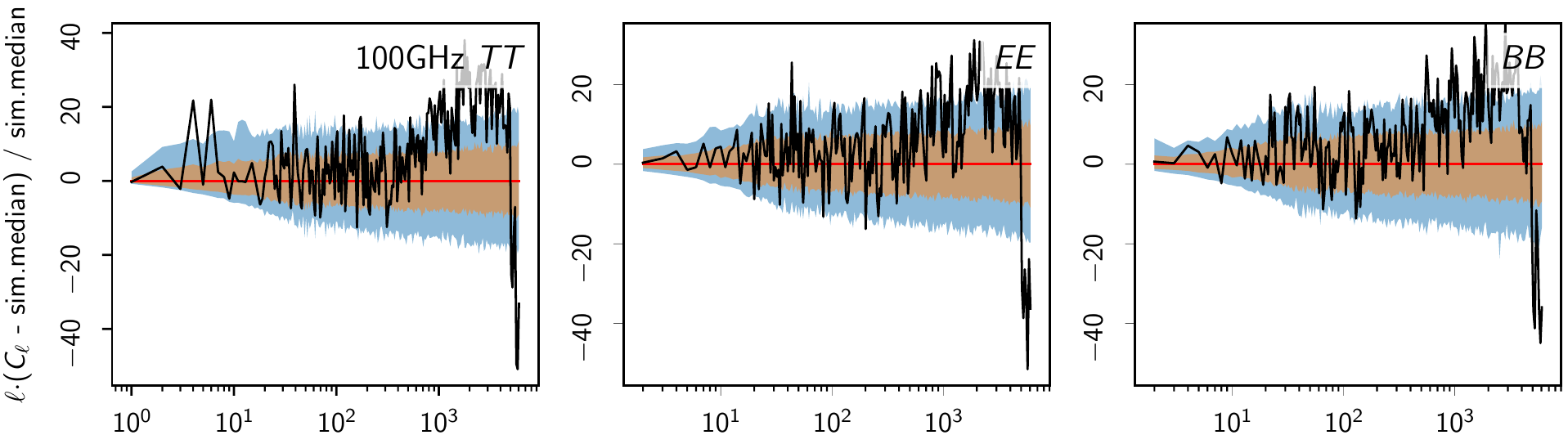}\\
  \includegraphics[width=1.0\linewidth]{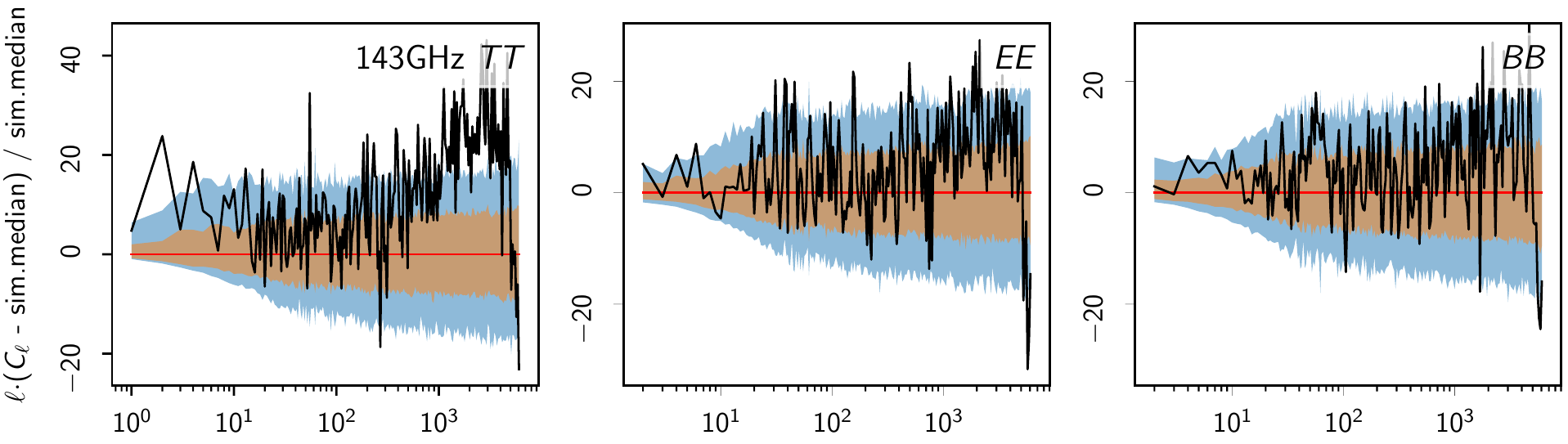}\\
  \includegraphics[width=1.0\linewidth]{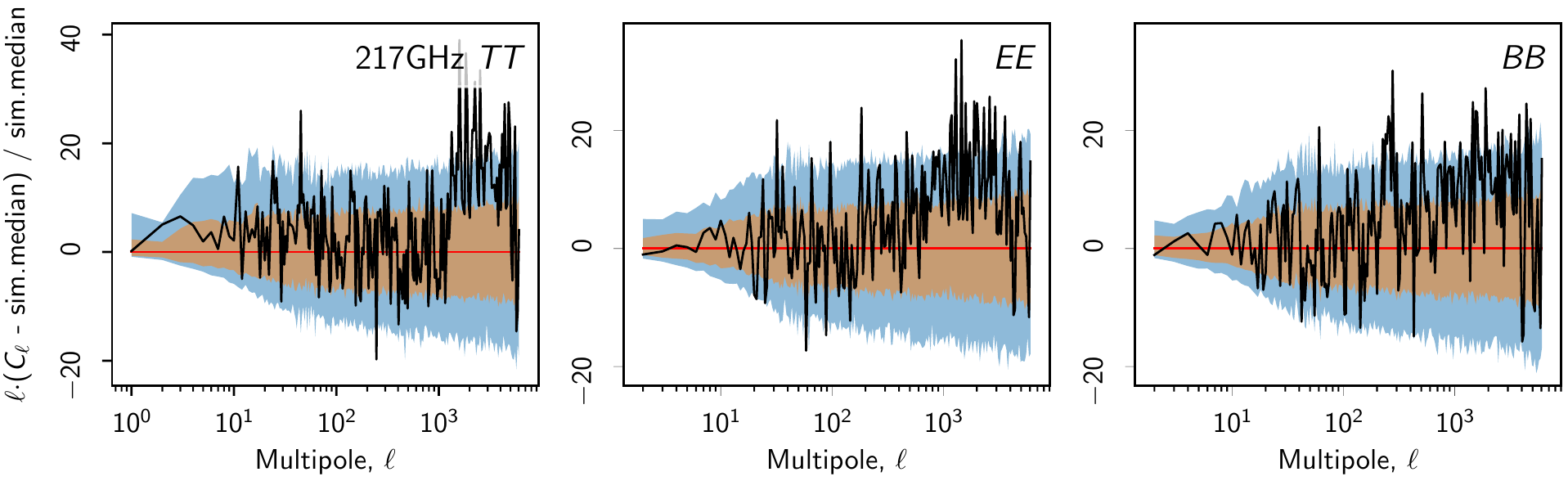}
  \caption{Same as Fig.~\ref{fig:abdiff_relative_fg}, but at 70, 100, 143, and 217\GHz.  The spectra shown here are the same as in Fig.~\ref{fig:abdiff_cmb}.  
    }
  \label{fig:abdiff_relative_cmb}
\end{figure*}

We can demonstrate that the simulations succeed in capturing the relevant residuals by comparing real and simulated detector-set-difference power spectra in Figs.~\ref{fig:abdiff_fg}, \ref{fig:abdiff_cmb}, \ref{fig:abdiff_relative_fg}, and \ref{fig:abdiff_relative_cmb} (for $TE$, $TB$, and $EB$ spectra, which are expected to be null, see Figs.~\ref{fig:abdiff_cross_lfi} and \ref{fig:abdiff_cross_hfi}).  The detector-set differences exhibit excellent agreement between the flight data and the simulated detector-set differences in polarization.  The temperature difference is not as well matched at the largest scales, owing to three factors.  First, at 143 and 217\GHz\ the solved gain fluctuations seem to converge to a slightly different large-scale temperature sky for the two detector sets.  The resulting roughly 1\muKCMB\ mismatch is not fully captured in the simulations, suggesting that it is driven by a systematic that is not simulated.  This residual and the associated mismatch are roughly $100$ times fainter than the simulated CMB anisotropy at the same angular scales.

Second, at 545\GHz\ the simulations suggest almost an order of magnitude more   large-scale temperature mismatch than is found in the flight data.  This is explained by the crude bandpass-mismatch model (different delta-function bandpasses), coupled with the lack of redundancy at 545\GHz, where the ``A'' detector set comprises only a single bolometer.  Even the amplified detector-set mismatch is $1000$ times fainter than the dust in the 545-GHz band.

Third, at 857\GHz\ the dominant failure mode is the mismatch at intermediate scales between $\ell\,{=}\,100$ and $\ell\,{=}\,1000$.  This mismatch is not seen in the half-ring differences, so it is not associated with the detector noise, but rather with systematic residuals related to the bandpass mismatch and bolometer transfer function.  In the power-spectrum domain, this mismatch is 100--10\,000 times fainter than the sky signal over the 50\,\% of sky with the lowest dust emission.  The 857-GHz detector sets only consist of two SWBs.

\subsubsection{Noise alignment}

The detector-set difference power spectra in Figs.~\ref{fig:abdiff_relative_fg} and \ref{fig:abdiff_relative_cmb} testify to a good overall agreement between the simulations and the flight data.  Upon closer inspection it is possible to identify a percent level, scale-dependent deficit in simulated noise power at $\ell\,{>}\,100$ affecting the polarized \hfi\ frequencies ($100$--$353\GHz$).  The deficit is exclusive to the \hfi\ simulations, suggesting that its origins lie in \hfi\ specific phenomena, such as Poissonian glitch residuals that are not included in the simulations and could not be reproduced by a Gaussian noise simulation.

\begin{figure*}[htpb]
  \includegraphics[width=1.0\linewidth]{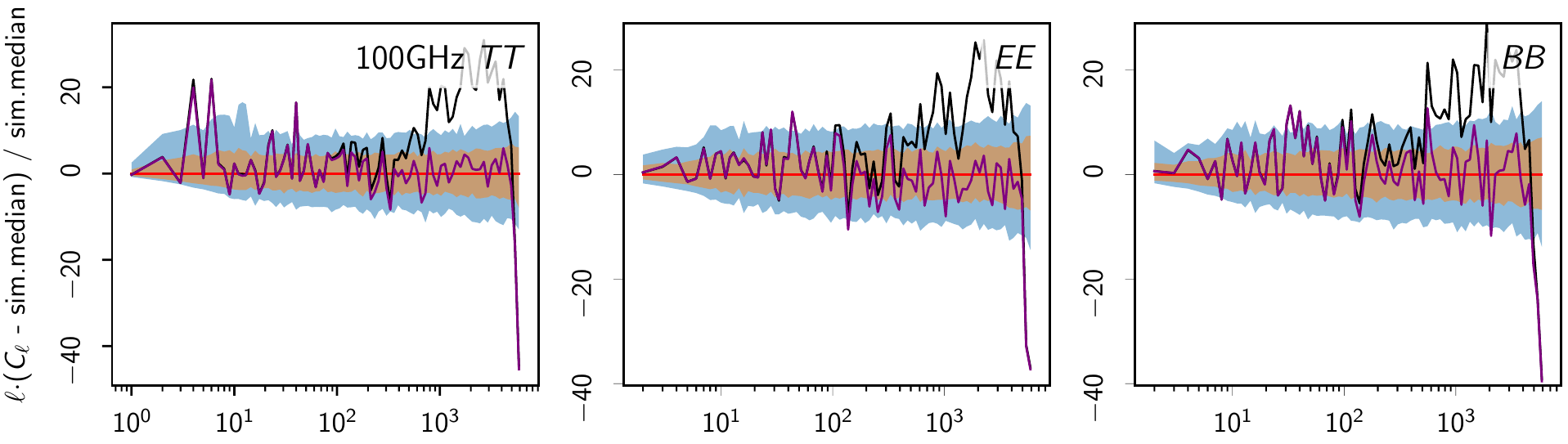}\\
  \includegraphics[width=1.0\linewidth]{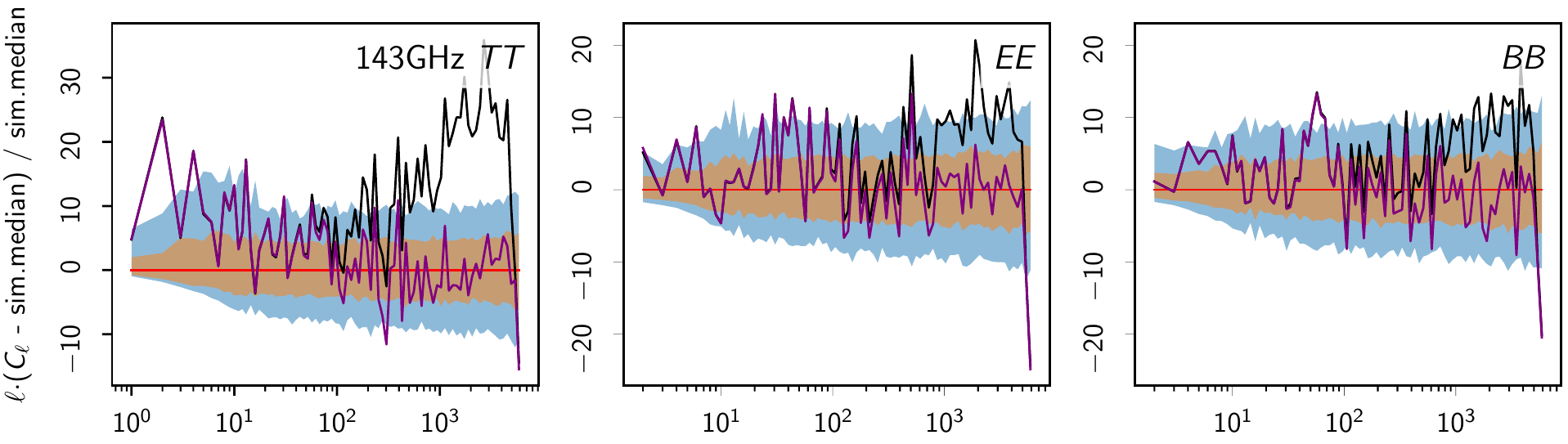}\\
  \includegraphics[width=1.0\linewidth]{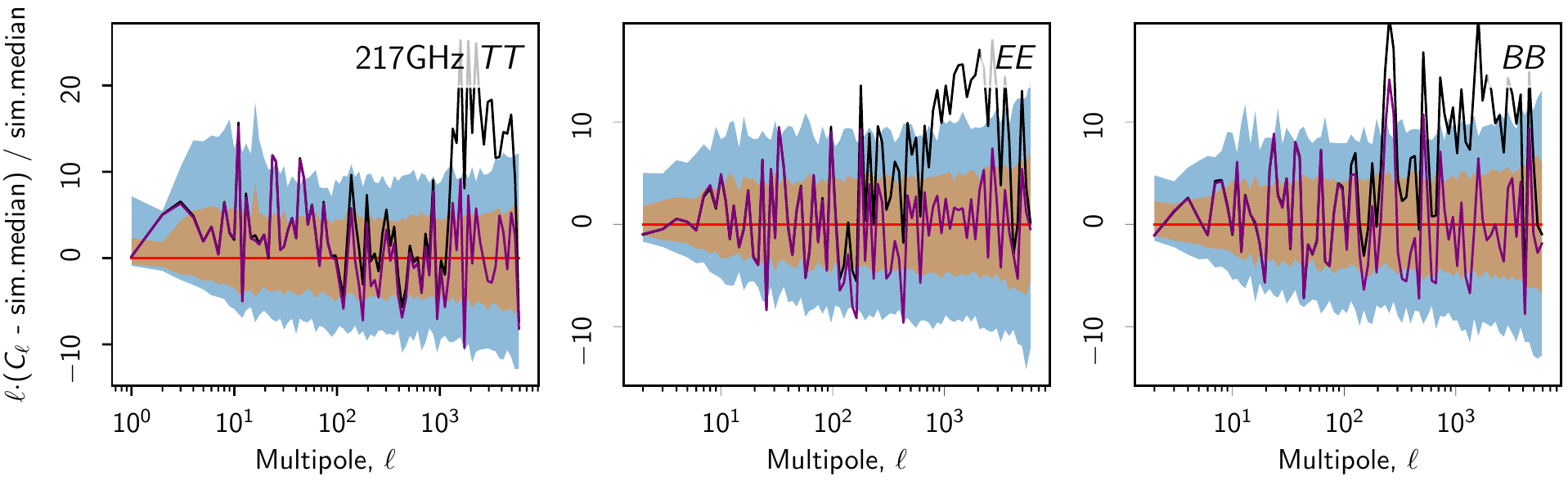}\\
  \includegraphics[width=1.0\linewidth]{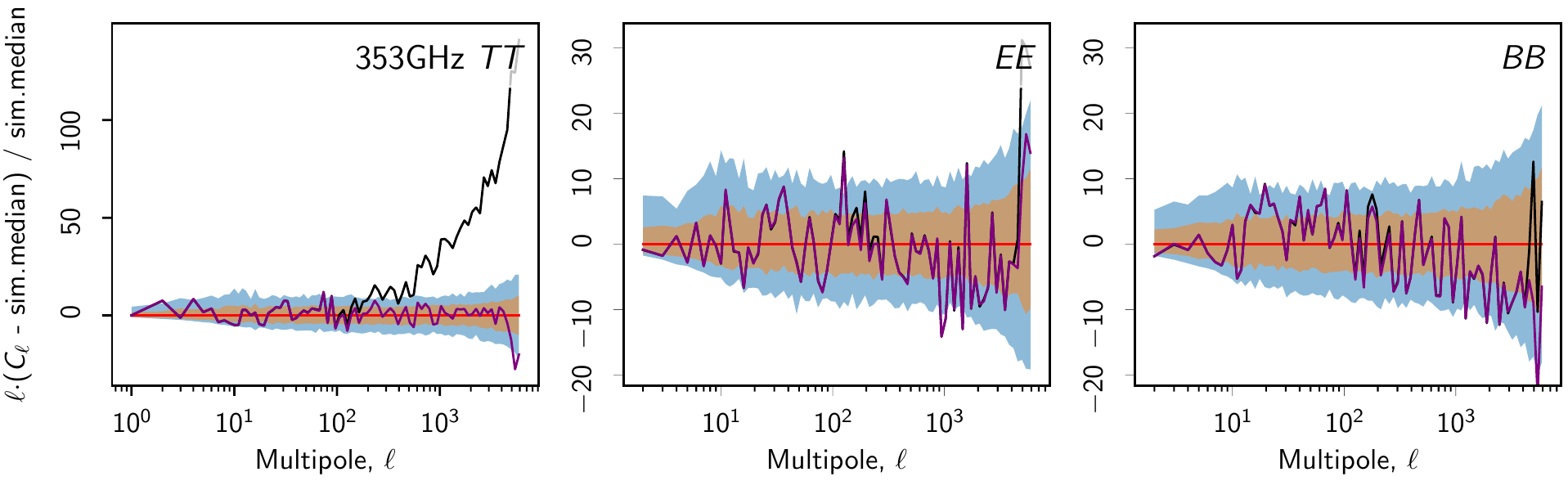}
  \caption{Simulated A/B difference versus flight data at the \hfi\ frequencies where we identify a percent-level, scale-dependent deficit in simulated noise power.  The flight data power spectrum is differenced and divided by the median of the uncorrected simulations (black) and, alternatively, by the simulated maps that include the noise correction (purple).  The coloured bands represent the asymmetric 68\,\% and 95\,\% confidence regions.  The power spectra are binned into 100 logarithmically spaced bins.  Apart from the wider binning, the black curves and the confidence limits are the same as those shown in Figs.~\ref{fig:abdiff_fg} and \ref{fig:abdiff_cmb}.
  }
  \label{fig:abdiff_relative_fix}
\end{figure*}

Overlooking the noise mismatch implies up to a percent level bias in statistical error measures that are derived from the simulations.  For some analyses this degree of uncertainty is acceptable.  For estimators that are sensitive to the mismatch in total power between the flight data and simulations, we offer a set of additional, $100$--$353\GHz$, small-scale noise maps that can be added to the simulated \npipe\ maps to align the total noise power.  We stress that the additional noise maps only adjust the overall noise power, without any effort to match the spatial structure of these missing noise modes.  We refer to this as ``noise alignment.''  We demonstrate the effect of the noise-alignment maps in Fig.~\ref{fig:abdiff_relative_fix}.

\section{\npipe\ and the earlier \Planck\ releases} 
\label{sec:comparison}

\npipe\ differs from the earlier \Planck\ releases in several ways.  We compare \npipe\ to both \prtwo\ and \prthree, allowing the reader to assess the magnitude of the changes against the 2015/2018 differences, which are themselves substantial.  In the  2018 release, the \lfi\ team made a critical change in calibration procedure, taking into account foreground polarization and thus removing a source of bias that generated spurious polarization.  Also in 2018, \hfi\ adopted a new mapmaking and TOD cleaning method called \sroll, which included time-domain corrections for bandpass mismatch, far sidelobes, and bolometer transfer-function residuals.

\subsection{Calibration}

\npipe\ photometric calibration is measured by fitting the orbital dipole template (Sect.~\ref{sec:diporb}) to the data.  The overall calibration is based solely on matching the far-sidelobe-corrected orbital dipole template against the data, rather than fitting a combination of the orbital dipole and some prior estimate of the total dipole.  Our orbital dipole template includes the faint,
frequency-dependent, dipole-induced quadrupole.

We follow the \lfi\ approach, correcting the orbital dipole template for subtle suppression and phase shifts induced by the far sidelobes, and extend this treatment to the \hfi\ frequencies.  Details of the convolution are presented in Appendix~\ref{app:fsl_dipole}.

\begin{figure}[htpb]
  \center{
    \includegraphics[width=0.48\textwidth]{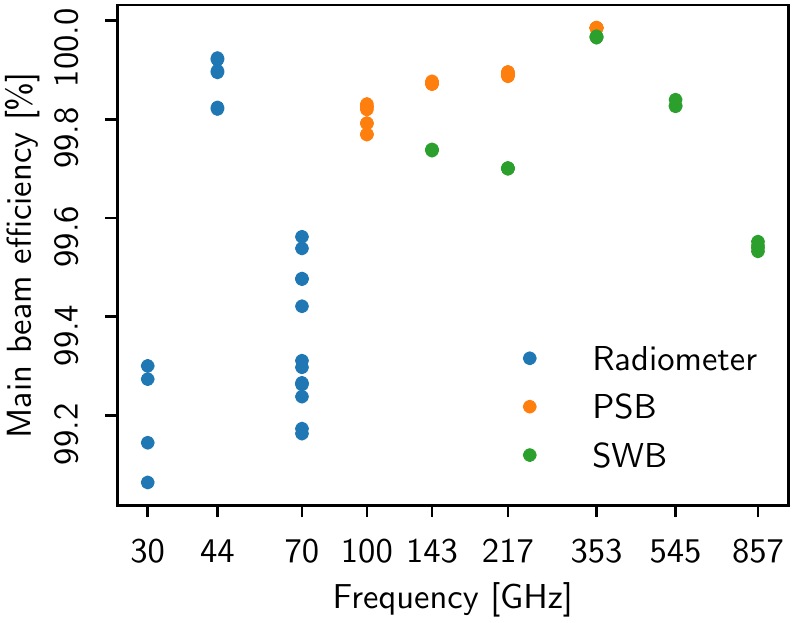}
  }
  \caption{Uncorrected main-beam efficiencies for all \Planck\ detectors.  The \npipe\ calibration procedure corrects each frequency map to have full main-beam efficiency.  The 545- and 857-GHz efficiencies are less certain due to uncertainties in main beam and sidelobe estimates, and are based on fitting the 353-GHz sidelobe templates to the data.  The 100--353-GHz sidelobe estimates are based on first-order \grasp\ simulations, and overestimate the main-beam efficiency by a small but unknown amount.
  }
  \label{fig:mb_efficiencies}
\end{figure}

\npipe\ processing effectively deconvolves the far sidelobes from the data, which are scaled to have unit response to the main beam rather than the $4\pi$ beam.  This is reflected in the effective beam window functions, which have unit response to the monopole and near unit response to the dipole.  The uncorrected main-beam efficiencies are shown in Fig.~\ref{fig:mb_efficiencies}.

We measure the calibration uncertainty and potential bias from template degeneracy, using the full-frequency simulated maps.  To measure the calibration of each simulated map, we subtract the input CMB and foreground map, mask out 50\,\% of the sky most affected by foreground residuals, and regress the input dipole template against the cleaned map.  The resulting overall gain distributions are shown and compared to \prthree\ calibration in Table~\ref{tab:mcgains}.  The simulations indicate that we recover the input dipole at better than $1\muK$ (combined statistical and systematic uncertainty) in each of the CMB frequencies.  The
table also shows that there are small but statistically significant biases in the procedure, which cause the \hfi\ CMB frequencies to be calibrated high by less than 0.03\,\%, amplifying the Solar dipole amplitude by less than $1\muK$.

\begin{table}[htpb!]
  \begingroup
  \newdimen\tblskip \tblskip=5pt
  \caption{Relative calibration between \npipe\ and \prthree\ maps and the \npipe\ gain uncertainty estimated from recovery of the injected dipole in simulated full-frequency maps.
  }
  \label{tab:mcgains}
  \nointerlineskip
  \vskip -3mm
  \footnotesize
  \setbox\tablebox=\vbox{
    \newdimen\digitwidth
    \setbox0=\hbox{\rm 0}
    \digitwidth=\wd0
    \catcode`*=\active
    \def*{\kern\digitwidth}
    \newdimen\signwidth
    \setbox0=\hbox{$-$}
    \signwidth=\wd0
    \catcode`!=\active
    \def!{\kern\signwidth}
    \halign{
      \hbox to 2.2cm{#\leaderfil}\tabskip 1.5em&
      \hfil#\hfil\tabskip 1.25em&
      \hfil#\hfil\tabskip 1.5em&
      \hfil#\hfil\tabskip 0pt\cr
      \noalign{\doubleline}
      \omit\hfil Frequency\hfil&
      \omit\hfil 2018 gain$^\mathrm a$\hfil&
      \omit\hfil MC gain$^\mathrm b$\hfil&
      \omit\hfil MC bias$^\mathrm c$\hfil\cr
      \omit\hfil [GHz]\hfil&
      \omit\hfil [\%]\hfil&
      \omit\hfil [\%]\hfil&
      \omit\hfil [\%]\hfil\cr
      \noalign{\vskip 3pt\hrule\vskip 4pt}
      *$30$& *$99.793$& *$99.975 \pm 0.045$& $-0.025 \pm 0.005$\cr
      *$44$& *$99.763$& *$99.990 \pm 0.040$& $-0.010 \pm 0.004$\cr
      *$70$& *$99.959$& *$99.997 \pm 0.030$& $-0.003 \pm 0.003$\cr
      $100$& $100.061$& $100.016 \pm 0.023$& !$0.016 \pm 0.002$\cr
      $143$& $100.082$& $100.016 \pm 0.018$& !$0.016 \pm 0.002$\cr
      $217$& $100.005$& $100.024 \pm 0.024$& !$0.024 \pm 0.002$\cr
      $353$& $100.463$& $100.052 \pm 0.053$& !$0.052 \pm 0.005$\cr
      $545^\mathrm d$& $115.123$& *$99.586 \pm 0.571$& !$0.414 \pm 0.057$\cr
      \noalign{\vskip 3pt\hrule\vskip 5pt}
    }
  }
  \endPlancktable
\tablenote {{a}} Relative calibration, \npipe\ / 2018, measured over 50\,\% of the sky after adding the Solar dipole back into 2018 maps and smoothing the maps with a 1\deg\ Gaussian beam.\par
\tablenote {{b}} Distribution centre and width, measured relative to the input signal.\par
\tablenote {{c}} This is the average calibration error and the associated uncertainty.  A 0.03\,\% bias in calibration corresponds to about $1\muKCMB$ in Solar dipole amplitude.\par
\tablenote {{d}} The \npipe\ 545-GHz calibration is based on the orbital dipole, and does not suffer from the 10\,\% modelling uncertainty affecting the planet-based calibration in the 2018 release.  We also calibrate directly into \KCMB, without needing a unit conversion factor with additional uncertainty.\par
\endgroup
\end{table}

\subsection{Noise and systematics}

We measure the level of residual noise and systematics in the \Planck\ maps by splitting the available detectors into disjoint sets and performing the same processing on each detector set.  The independent processing guarantees that the residuals between the detector sets are not correlated.  Unlike making a split in the time domain (the so-called half-mission split), the detector-set difference does not cancel systematics that vary detector by detector, such as bandpass mismatch, far sidelobes, or the \hfi\ transfer function
residuals.

The definition of the detector-set null test has been somewhat ambiguous in past \Planck\ publications.  In an independent, local processing, the template corrections (calibration, transfer function residuals, etc.) and noise offsets are fitted to each detector set separately.  Alternatively, it is possible to fit the templates globally to all of the data and project subsets of the cleaned data into detector-set and half mission maps.  The global fit (sometimes referred to as ``turbo destriping'' because it only requires one run of the destriper) is suitable for measuring the achieved internal consistency in the cleaned TOD.  However, residual systematics in turbo-destriped subsets are highly correlated, making it impossible to infer the total level of residuals from their differences.  In this paper, all turbo-destriped subset maps are presented with an explicit mention of the global fit being used.

\begin{figure*}[htpb]
  \center{
    \includegraphics[width=0.75\linewidth]{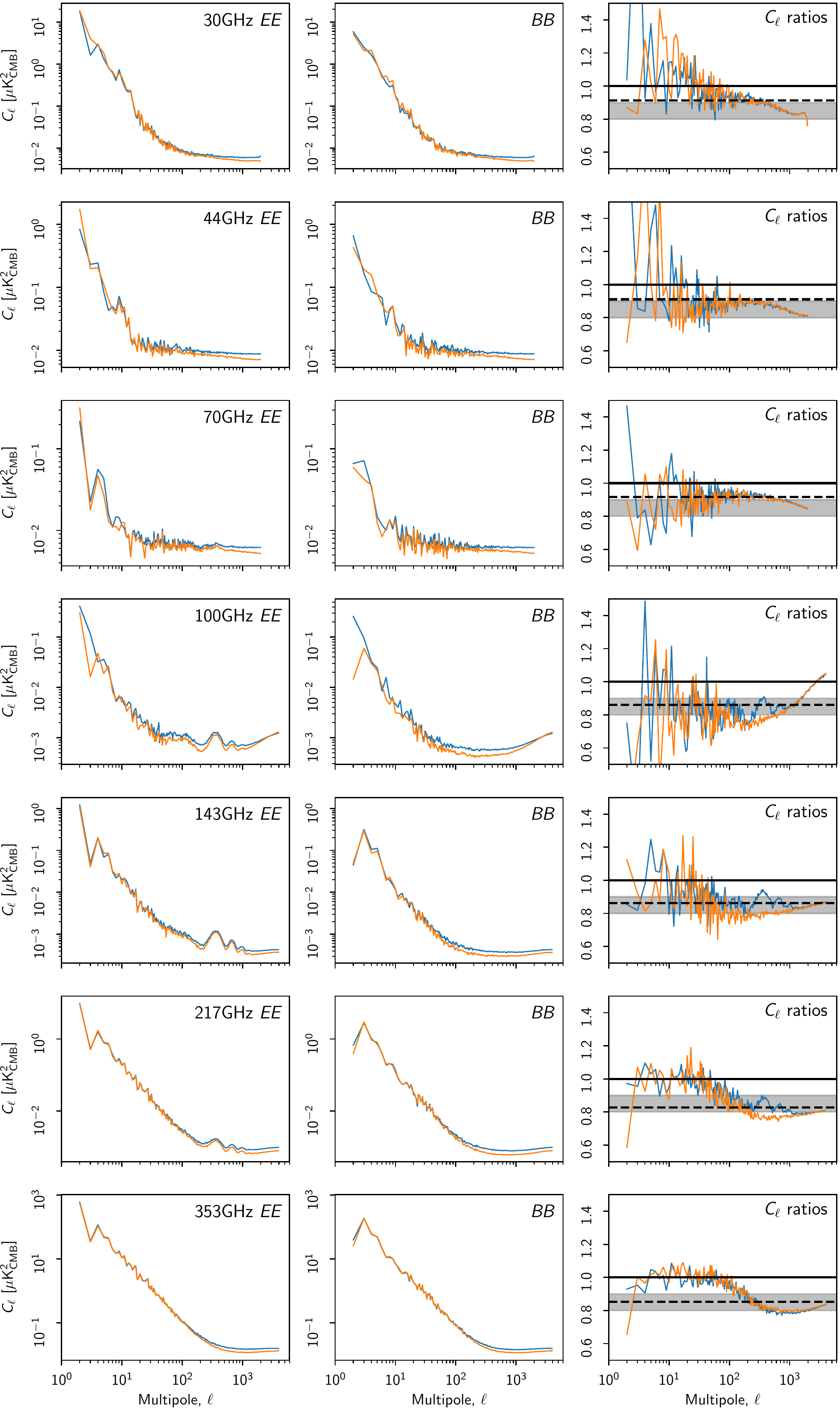}
  }
  \caption{$EE$ and $BB$ auto-spectra of the polarized frequency maps.  The first two columns show \prthree\ (blue) and \npipe\ (orange) auto-spectra, while the third column shows the ratios (\npipe/2018) with $EE$ in blue and $BB$ in orange.  For noise-dominated angular scales, \npipe\ maps have 10--20\,\% lower noise variance, indicated by the grey band in the ratio plot.  We show a naive estimate of the ratios, based on the ratio of masked pixel hits in Table~\ref{tab:discarded}, as a dashed black line.  These spectra are computed over 50.4\,\% of the sky, corrected for the sky fraction and binned into 300 logarithmically-spaced bins.
  }
  \label{fig:autospectra}
\end{figure*}

\begin{figure*}[htpb]
  \center{
    \includegraphics[width=1.0\linewidth]{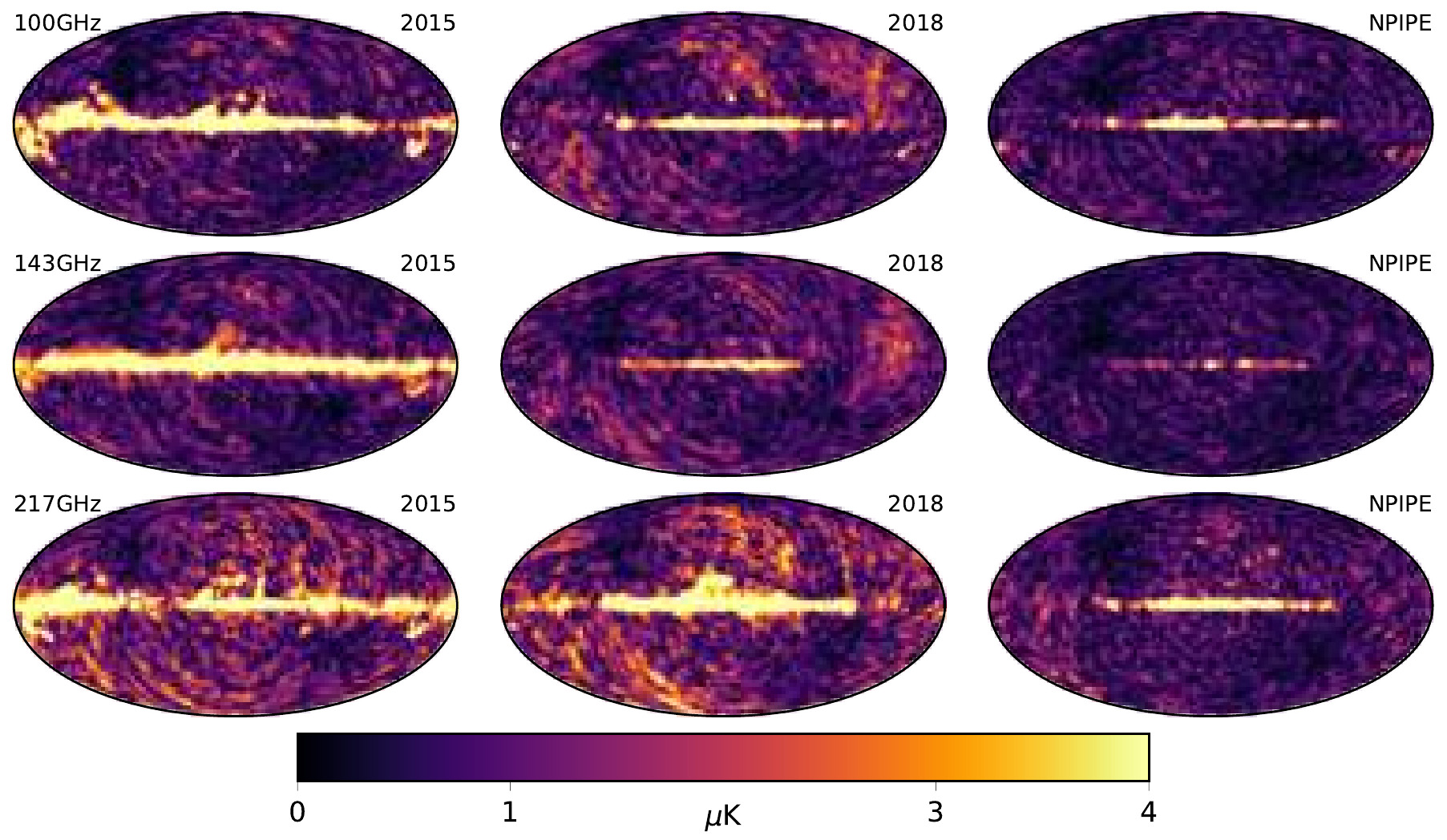}\\
    \includegraphics[width=1.0\linewidth]{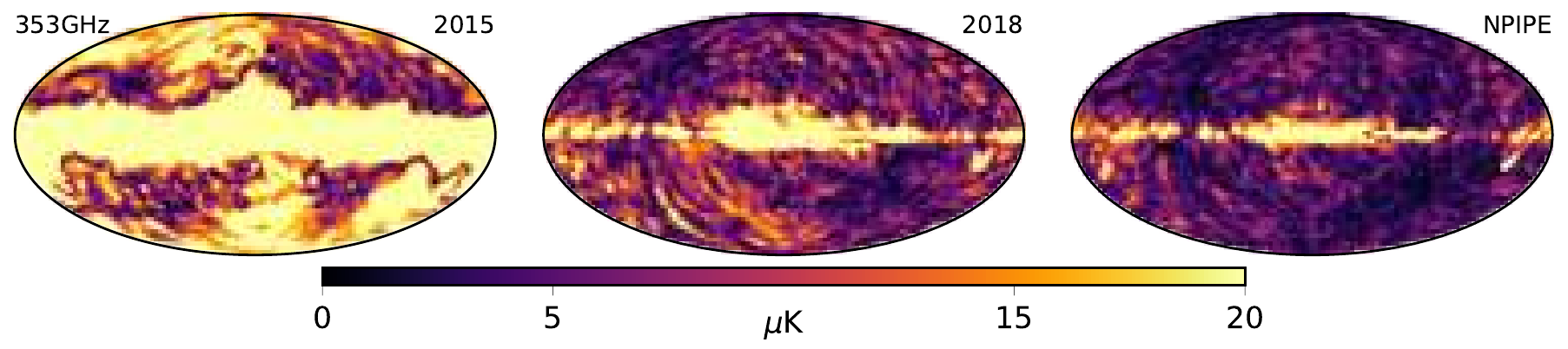}\\
  }
  \caption{Polarization amplitudes of the detector-set difference null maps.  The angular power spectra of \prthree\ and \npipe\ maps are shown in Fig.~\ref{fig:detsetdiff}.  Independent processing of the two detector-sets means that these maps reflect the level of total residuals in the frequency maps.
  }
  \label{fig:detsetdiff_map}
\end{figure*}

\begin{figure*}[htpb!]
  \center{
    \includegraphics[width=1.0\linewidth]{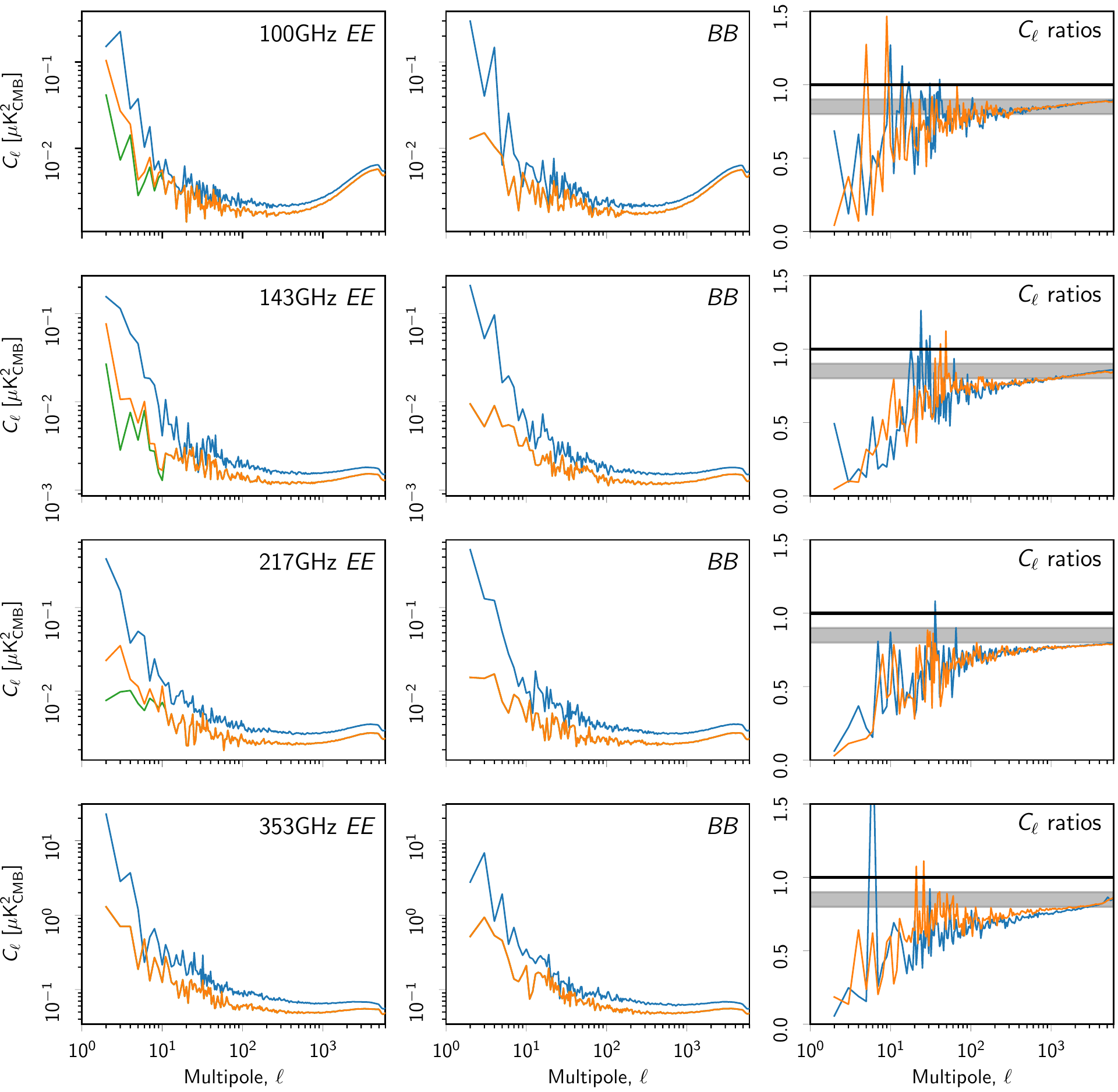}
  }
  \caption{$EE$ and $BB$ detector-set difference power spectra.  The first two columns show \prthree\ (blue), raw \npipe\ (green), and transfer-function-corrected \npipe\ (orange) null-map power spectra.  Note that \prthree\ detector sets are \emph{not} the same as were differenced for figure~14 in \cite{planck2016-l03}, but rather ones that were destriped independently.  The third column of panels shows the transfer-function-corrected \npipe/2018 $EE$ and $BB$ ratios in blue and orange, respectively.  \npipe\ has notably less power at all angular scales.  The grey band in the third column indicates a 10--20\,\% improvement in power.  These spectra are computed over 50.4\,\% of the sky, corrected for the sky fraction and binned into 300 logarithmically-spaced bins.  The polarization amplitudes of 2015, 2018, and \npipe\ detector-set difference maps are shown in Fig.~\ref{fig:detsetdiff_map}.
  }
  \label{fig:detsetdiff}
\end{figure*}

\begin{figure*}[htpb!]
  \center{
    \includegraphics[width=1.0\linewidth]{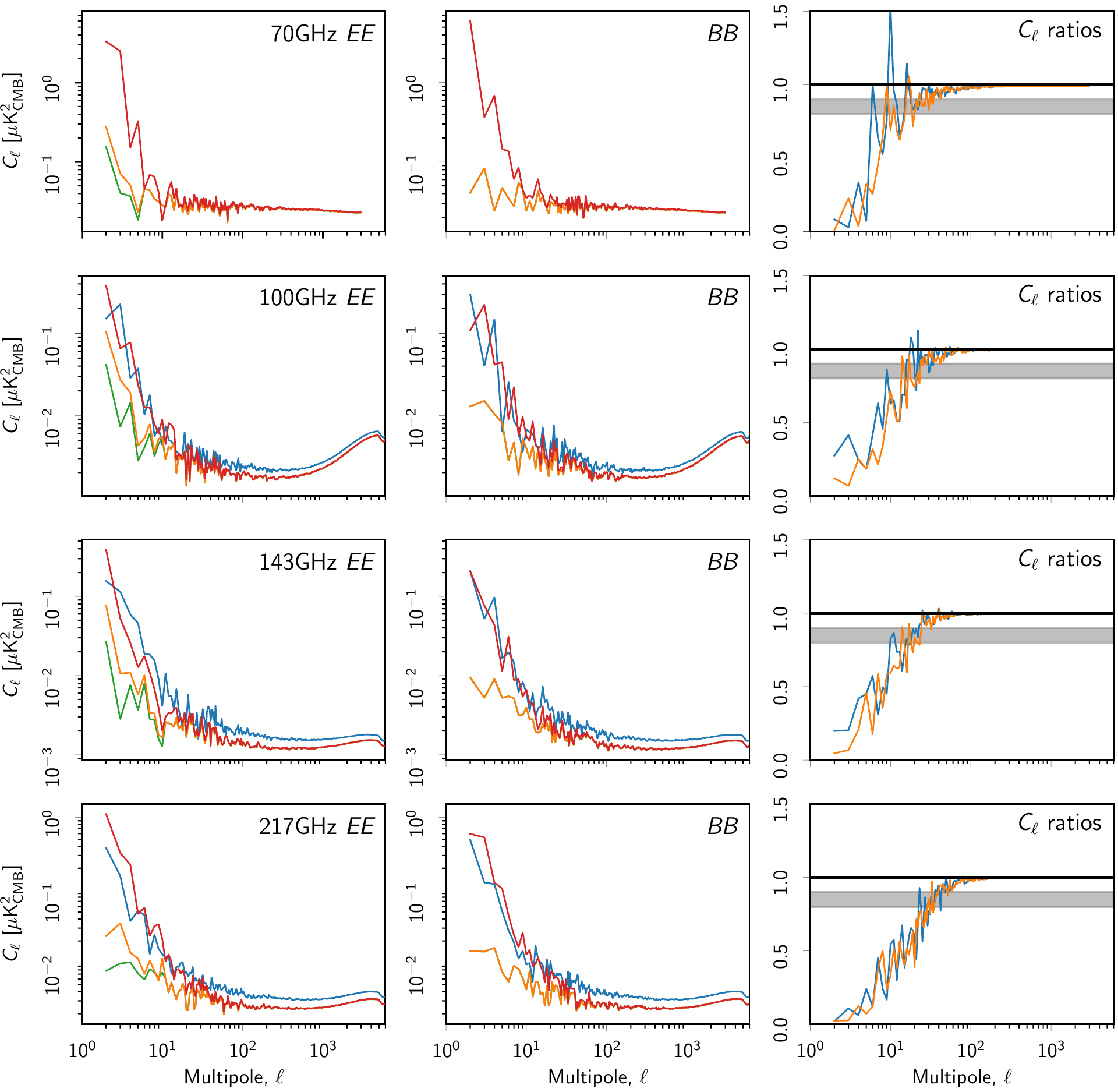}
  }
  \caption{Effect of the polarization prior on $EE$ and $BB$ detector-set difference power spectra.  For 100--217\GHz, the green, orange, and blue lines on the $EE$ and $BB$ plots are the same as in Fig.~\ref{fig:detsetdiff} but here we add a red line, showing the power spectra for an alternative version of the \npipe\ detector-set maps that are computed {\it without\/} the polarization prior.  There is no blue line at 70\GHz\ because there is no comparable detector-set split in \prthree, and 353\GHz\ is not shown because it is always calibrated without the polarization prior.
  }
  \label{fig:detsetdiff_polcal}
\end{figure*}

We compare the $EE$ and $BB$ total power in each polarized \Planck\ frequency in Fig.~\ref{fig:autospectra}.  In  Figs.~\ref{fig:detsetdiff_map} and \ref{fig:detsetdiff} we study the detector-set null maps for the polarized \hfi\ frequencies. 
These maps and spectra indicate that \npipe\ products have less noise than \prthree\ maps. There are several factors that contribute.  The \npipe\ maps include data taken during repointing manoeuvres, which account for roughly 8\,\% of
the \Planck\ mission.  \lfi\ noise levels are reduced further by the adaptive sky-load differencing (Sect.~\ref{sec:diff}).
\hfi\ noise levels are improved by the extended flagging kernel during transfer-function deconvolution (Sect.~\ref{sec:deconv}).  
Large-scale \hfi\ residuals, particularly in polarization, are significantly suppressed by our use of the polarized-foreground prior (Sect.~\ref{sec:pestimates}) during calibration, as is demonstrated by the increase in \npipe\ large-scale polarization uncertainty when the prior is disabled (Fig.~\ref{fig:detsetdiff_polcal}).  These improvements hold even after deconvolution of the $E$-mode transfer function (Sect.~\ref{sec:ee_tf}).  Finally, we fit the $1/f$ noise fluctuations with 167-ms baseline steps, compared to the 0.25-s (30\GHz), 1-s (44 and 70\GHz), and full-pointing-period (35--75\,min at \hfi\ frequencies) baseline steps used in \prthree.

\subsection{Signal}

\begin{figure*}[htpb!]
  \center{
    \includegraphics[width=0.9\linewidth]{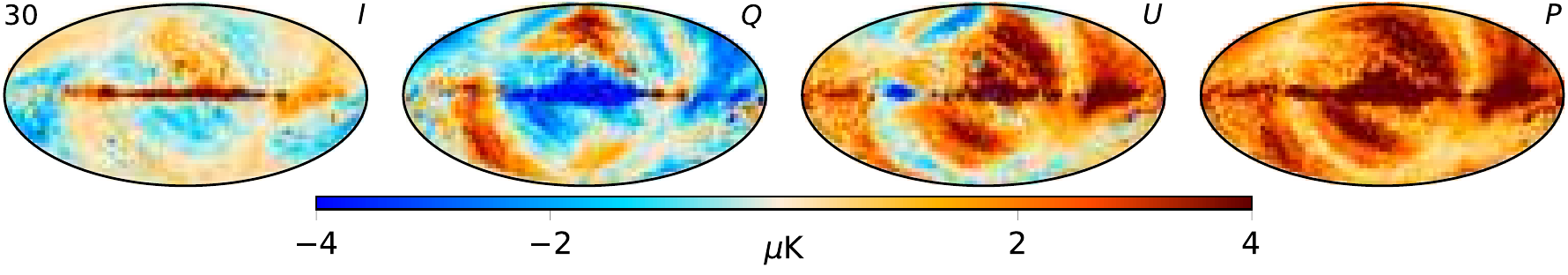}\\
    \includegraphics[width=0.9\linewidth]{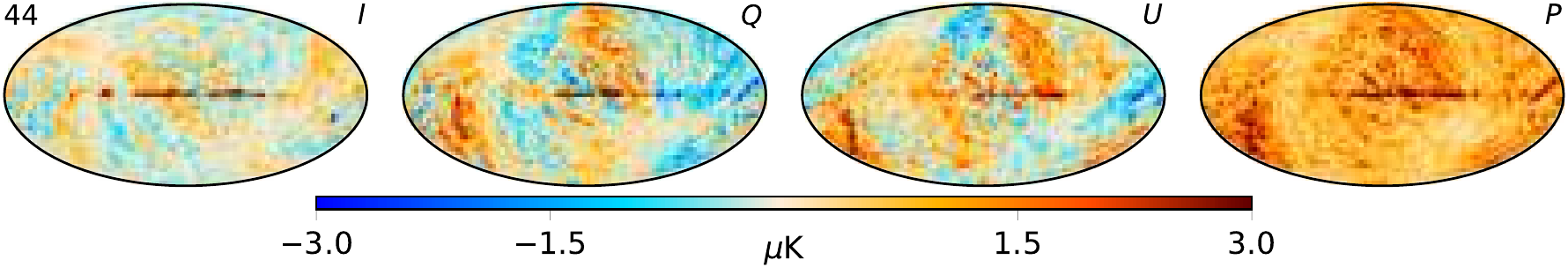}\\
    \includegraphics[width=0.9\linewidth]{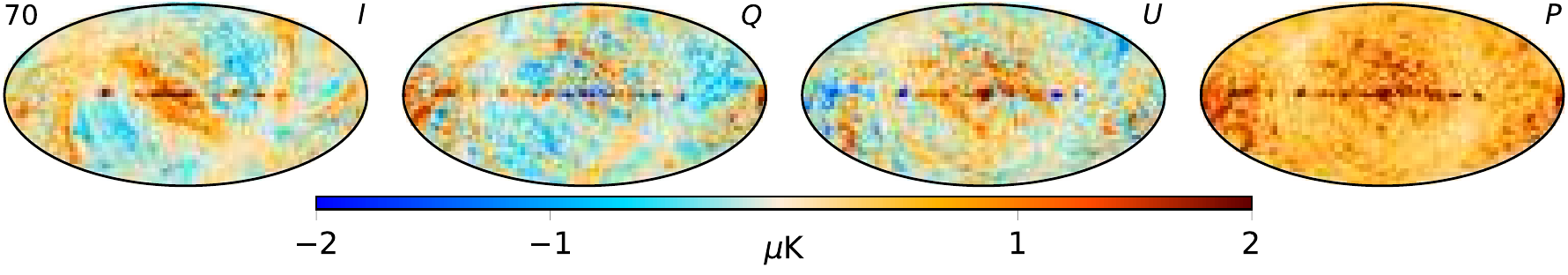}\\
    \includegraphics[width=0.9\linewidth]{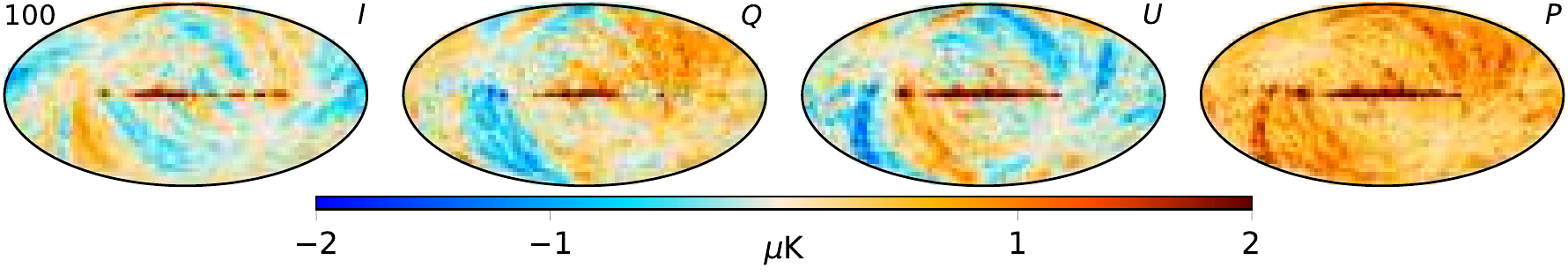}\\
    \includegraphics[width=0.9\linewidth]{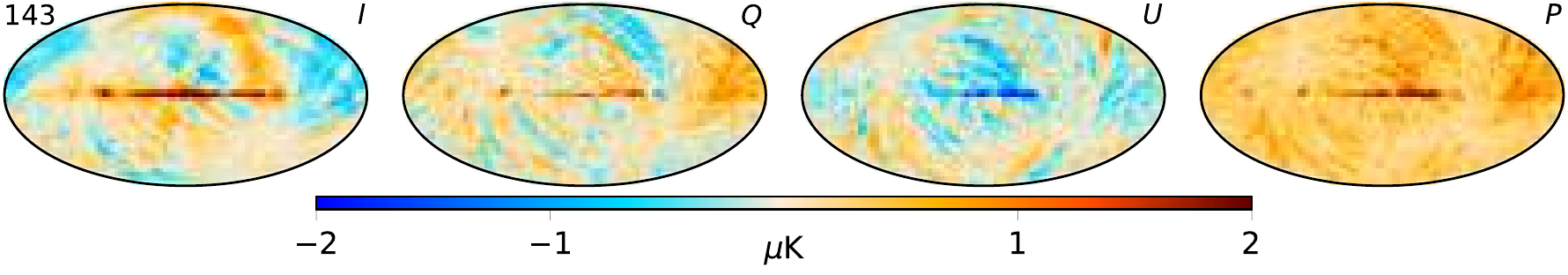}\\
    \includegraphics[width=0.9\linewidth]{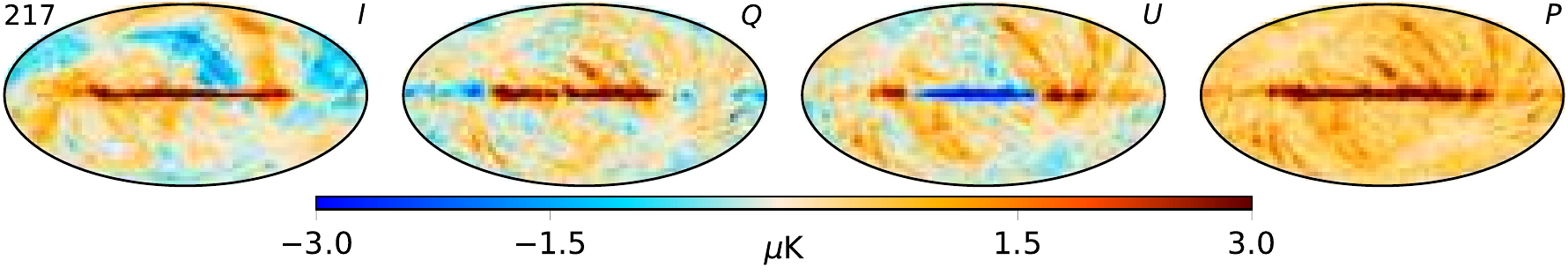}\\
    \includegraphics[width=0.9\linewidth]{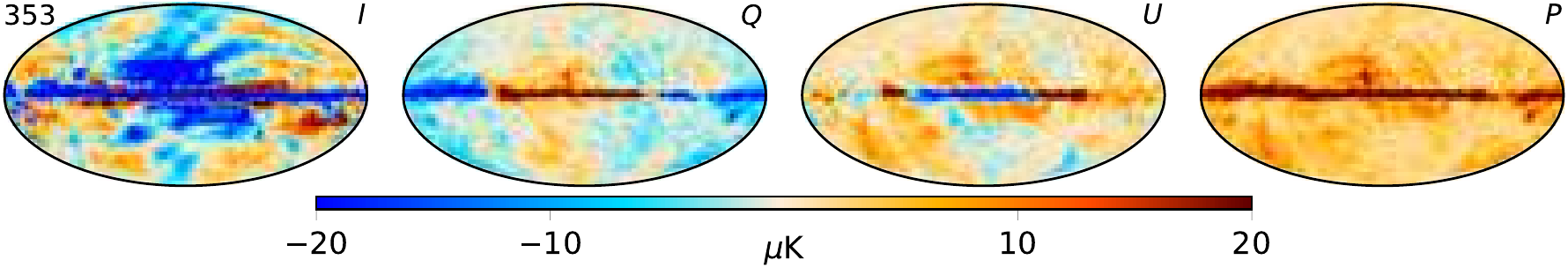}\\
    \includegraphics[width=0.22\linewidth]{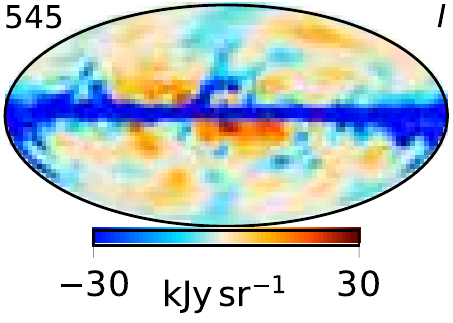}
    \includegraphics[width=0.22\linewidth]{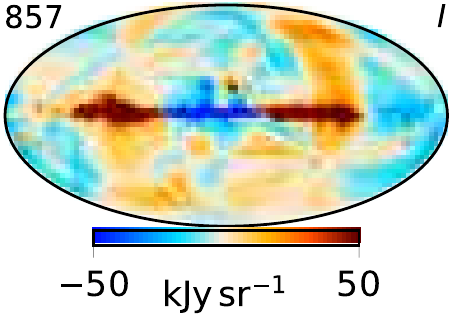}\\
  }
  \caption{\npipe$-$2018 release difference maps in temperature and polarization. We have projected out the Solar dipole and zodiacal emission templates from the temperature differences, and performed a relative calibration using 50\,\% of the sky to highlight differences beyond these trivial mismatch modes.  All maps are smoothed with a 3\deg\ Gaussian beam to suppress small-scale noise.
  }
  \label{fig:diffmaps}
\end{figure*}

\begin{figure*}[htpb!]
  \center{
    \includegraphics[width=1.0\linewidth]{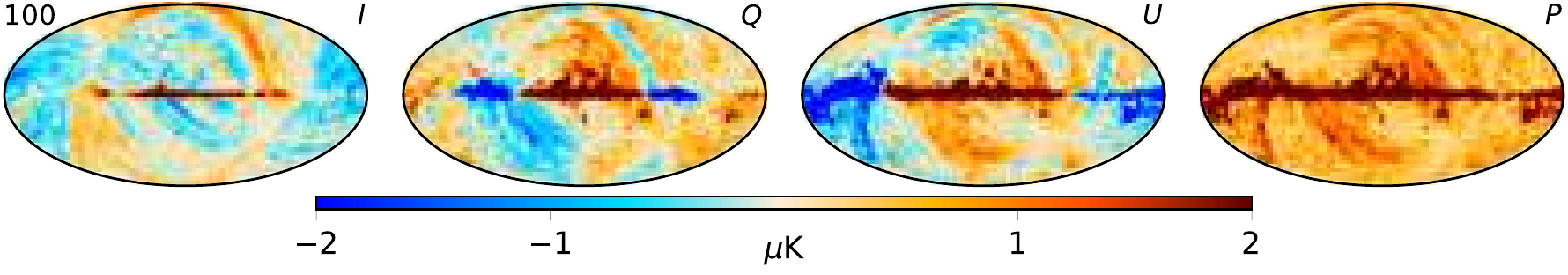}\\
    \includegraphics[width=1.0\linewidth]{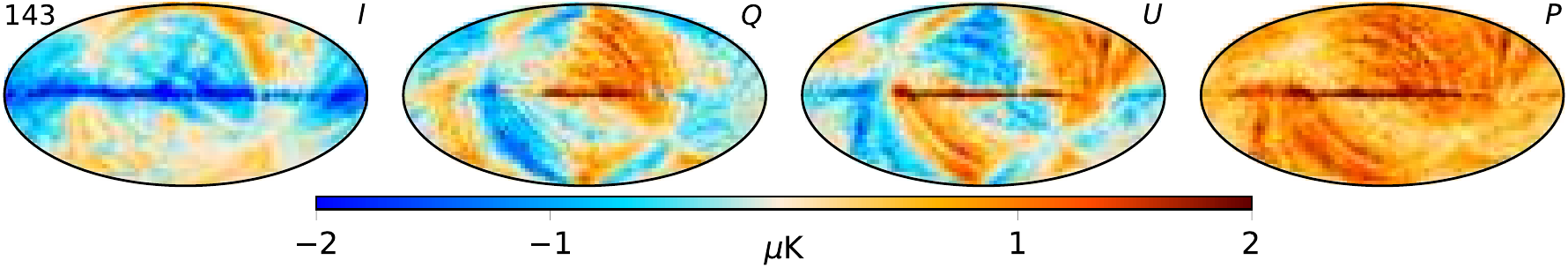}\\
    \includegraphics[width=1.0\linewidth]{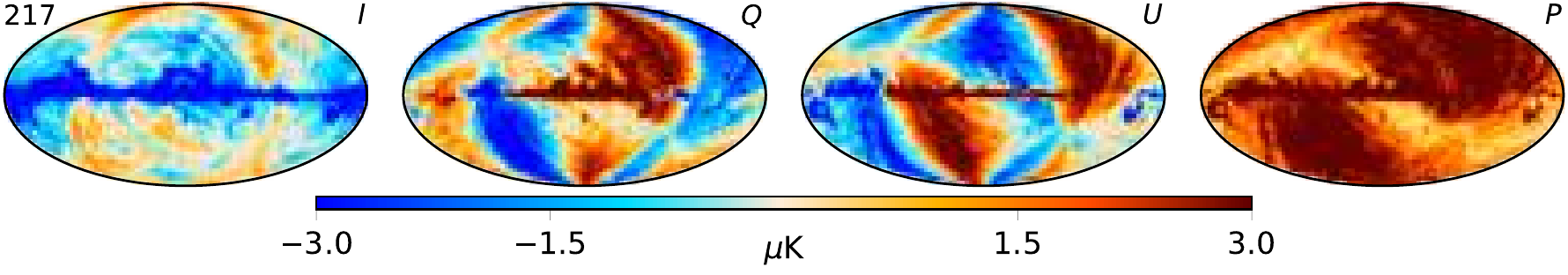}\\
    \includegraphics[width=1.0\linewidth]{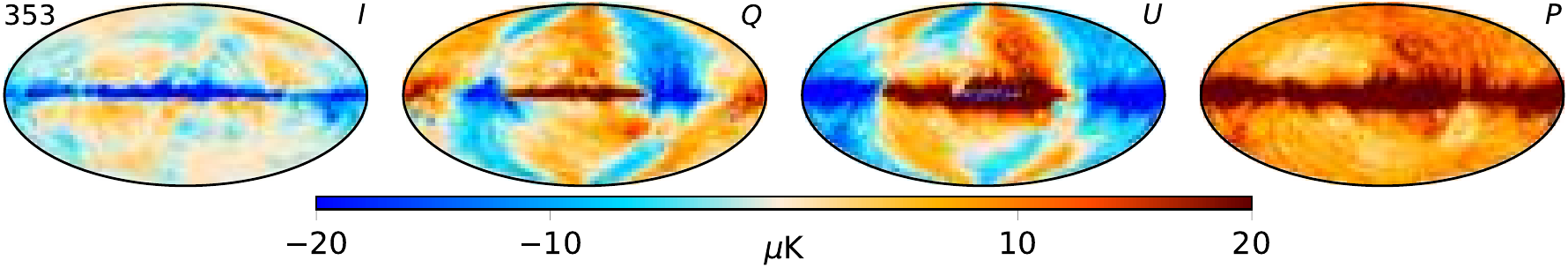}\\
  }
  \caption{
    \npipe$-$2015 release difference maps in temperature and polarization to compare to Fig.~\ref{fig:diffmaps}.  Note that the Solar dipole model for \prtwo\ did not include the frequency-dependent part of the quadrupole term (Table~\ref{tab:qfactors}), so we also omit that correction from the dipole template here.  The 353-GHz difference shows that the ``zebra'' pattern is exclusive to the 2018 temperature map.  The polarization differences are indicative of an overall calibration mismatch between polarization-sensitive bolometers, causing substantial temperature-to-polarization leakage from the Solar dipole in the 2015 maps.
  }
  \label{fig:diffmaps_pr2}
\end{figure*}

We present the temperature and polarization difference maps between \npipe\ and \prthree\ in Fig.~\ref{fig:diffmaps}.  To make the temperature map differences more informative, we have projected out the relativistic dipole, zodiacal emission, and an overall relative calibration mismatch.  The 353-GHz temperature difference has a peculiar ringing pattern, understood to originate from the coarse-grained transfer function model used in the 2018 processing.  For comparison, we show select \npipe--2015 differences in Fig.~\ref{fig:diffmaps_pr2}.

\begin{figure*}[htpb!]
  \center{
    \includegraphics[width=1.0\linewidth]{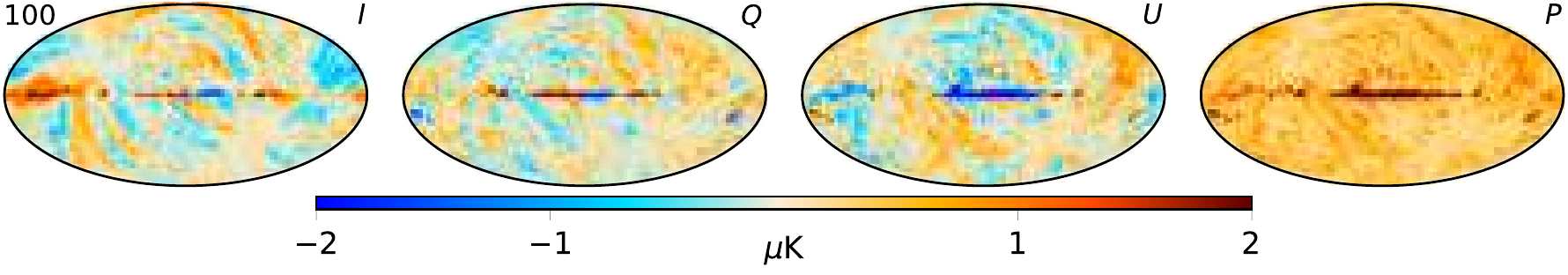}\\
    \includegraphics[width=1.0\linewidth]{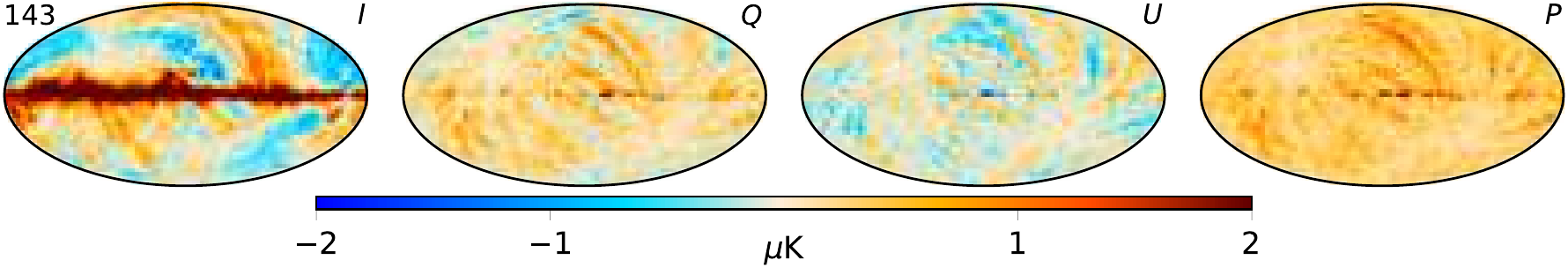}\\
    \includegraphics[width=1.0\linewidth]{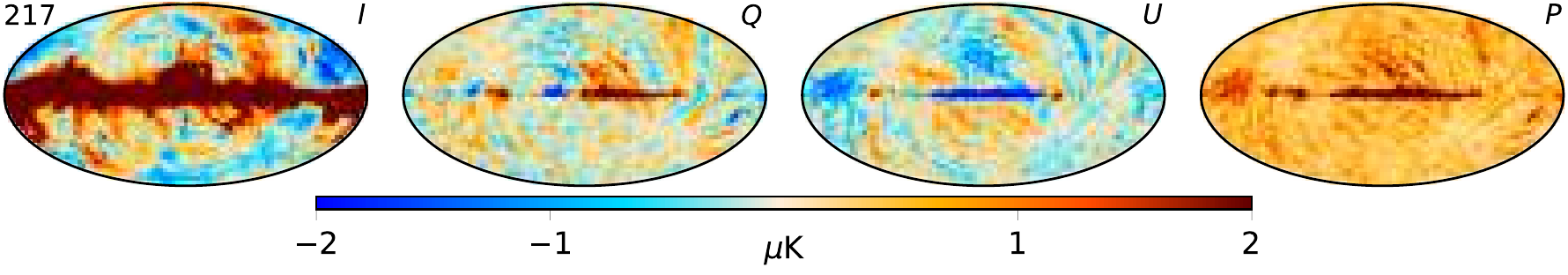}\\
    \includegraphics[width=1.0\linewidth]{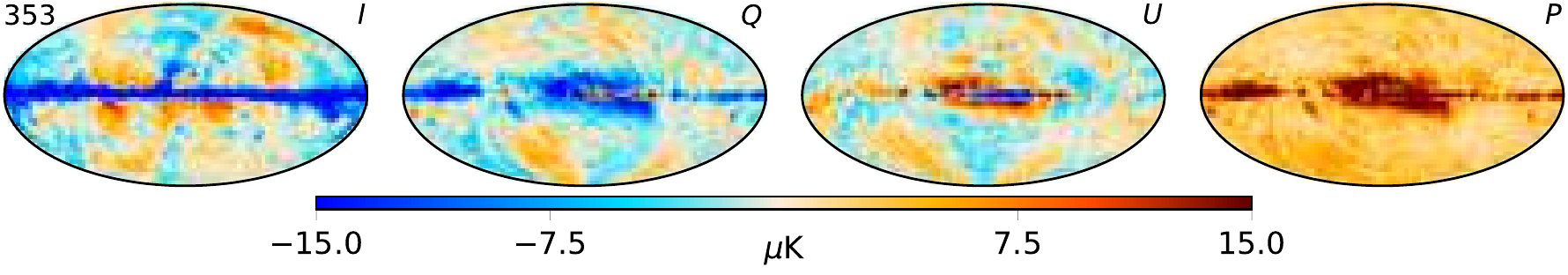}\\
  }
  \caption{\npipe$-$\srolltwo\ difference maps in temperature and polarization to compare to Fig.~\ref{fig:diffmaps}.
  }
  \label{fig:diffmaps_sroll2}
\end{figure*}

While our work was being prepared for publication, a new processing of \Planck-\hfi\ polarization data, known as \srolltwo, was published in \cite{Delouis:2019bub}.  It features a new model of the ADCNL that does not rely on the linear gain fluctuation approximation.  The new ADCNL model is able to capture the bulk of the residual ADCNL with considerable economy, thus limiting the degeneracy between the $IQU$ map and the ADCNL correction.  We show in Fig.~\ref{fig:diffmaps_sroll2} the \npipe$-$\srolltwo\ difference maps, and observe that for polarization, the \npipe\ and \srolltwo\ maps are considerably closer to each other than \npipe\ and \prthree, except at 143\GHz.  But even the \npipe$-$\prthree\ differences are very small.  In temperature, maps from 100 to 217\GHz\ have diverged in the Galactic plane, owing to the exclusion of all SWBs from \srolltwo\ processing.  The 353-GHz zebra stripe pattern is gone for both \npipe\ and \srolltwo.

\subsubsection{Internal consistency}
\label{sss:internal_consistency}

Each \Planck\ operational year divides into two sky surveys.  The two surveys are complementary in the sense that most sky regions are scanned in approximately opposite directions between the two surveys.  This makes the difference between odd and even surveys sensitive to a number of systematic effects, such as pointing error, far sidelobes, and transfer-function residuals. Processing the odd and even surveys into separate maps would be possible, but not representative of the full-map uncertainty, since the opposite scanning directions are essential in fitting the systematic templates.  Instead, we can project the calibrated and template-corrected data into odd and even survey maps, and use their difference to gauge the level of internal consistency achieved.  This test reveals if the considered templates have enough freedom to model the systematics that are responsible for the odd--even survey differences.

\begin{figure*}[htpb!]
  \center{
    \includegraphics[width=0.75\linewidth]{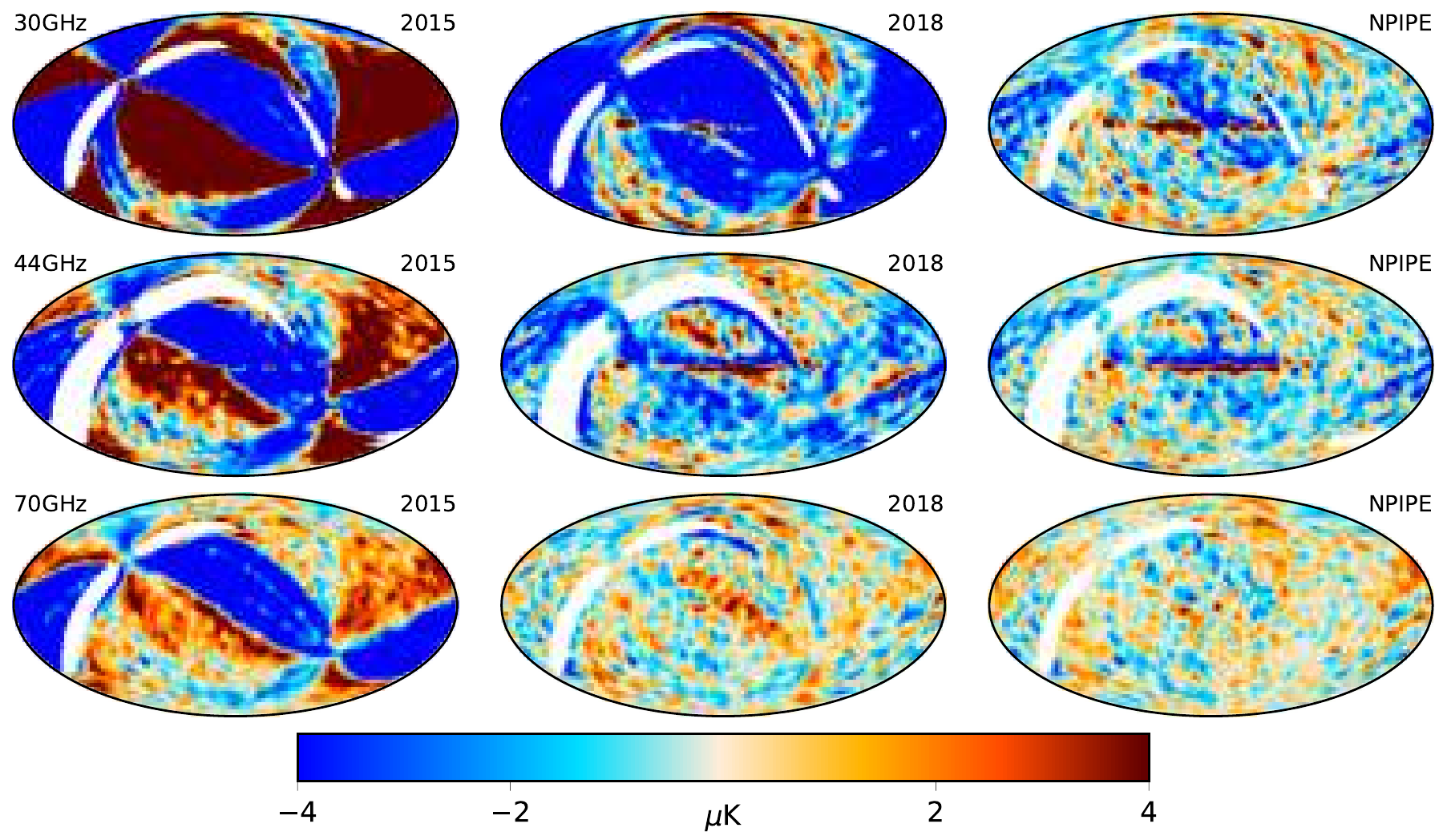}\\
  }
  \caption{Odd--even survey intensity differences for LFI smoothed to 5\deg.  These maps reflect the internal consistency achieved, not the total residuals, since the calibration errors are correlated between the surveys.  To match the 2015 and 2018 processing, the \npipe\ 167\-ms baseline offsets for this plot are solved using individual survey data.
  }
  \label{fig:oddeven1}
\end{figure*}

We compare \lfi\ odd$-$even survey differences between \prtwo, \prthree, and \npipe\ in Fig.~\ref{fig:oddeven1}.  This figure shows a clear improvement in internal consistency between the 2015 and 2018 maps, owing to the improvements in calibration.  The \npipe\ survey differences show further improvement, likely due to the full-frequency calibration scheme that corrects for integrated bandpass mismatches.

\begin{figure*}[htpb!]
  \center{
    \includegraphics[width=1.0\linewidth]{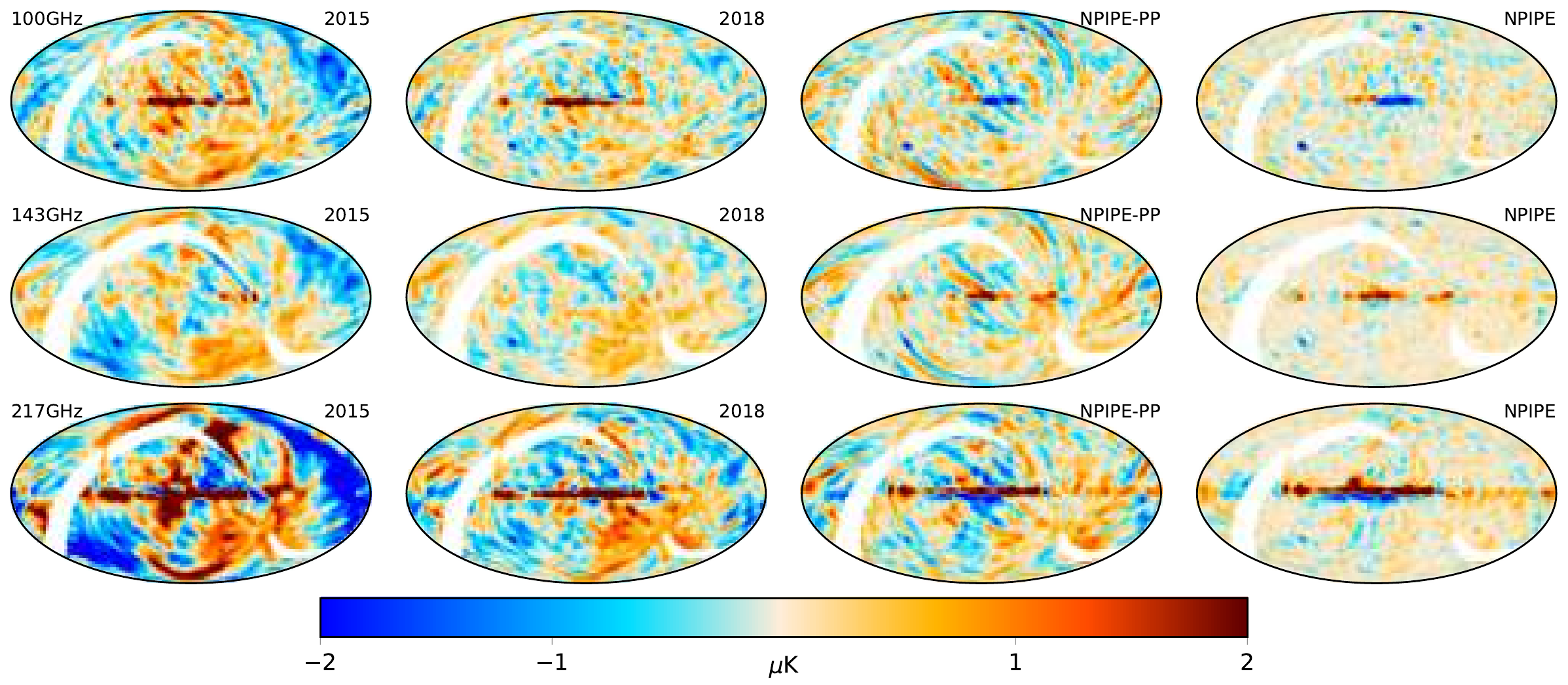}\\
  }
  \caption{Odd$-$even survey intensity differences for 100, 143, and 217\GHz, smoothed to 5\deg.  The 2015 and 2018 maps are the same as in Fig.~12 of \cite{planck2016-l03}.  The NPIPE-PP column shows the difference obtained if \npipe\ is solved only for pointing-period offsets (like \prthree), rather than for the 167-ms baseline offsets.  The stripes visible in the \mbox{NPIPE-PP} maps are glitch and ADC nonlinearity-correction residuals that are well captured by the short-baseline solution. The comparison is not perfect, because \prtwo\ (2015) baseline offsets were solved using the survey TOD, while the other versions use full-mission baselines.  Three variable radio sources can be identified across the frequencies in the \npipe\ maps.  These maps reflect the internal consistency achieved, not the total residuals, as the calibration errors are correlated between the surveys.
  }
  \label{fig:oddeven2}
\end{figure*}

We compare the $100$--$217\GHz$ survey differences in Fig.~\ref{fig:oddeven2}.  These figures contrast the apparent lack of large-scale structure in the \npipe\ null map with the visible residuals in the earlier releases.  The improvement is due to the polarization prior, short-baseline destriping, and additional bins in the transfer function correction.  It is worth pointing out that the \npipe\ 143-GHz map has a noticeable residual in the Galactic plane, possibly from the transfer-function correction that
down-weights these pixels when fitting the templates.  The level of the residual is negligible given the strength of the foreground signal in these pixels.  The 217-GHz 2018 map outperforms the 2015 results due to improvements in FSL and zodiacal-light removal.

\begin{figure*}[htpb!]
  \center{
    \includegraphics[width=1.0\linewidth]{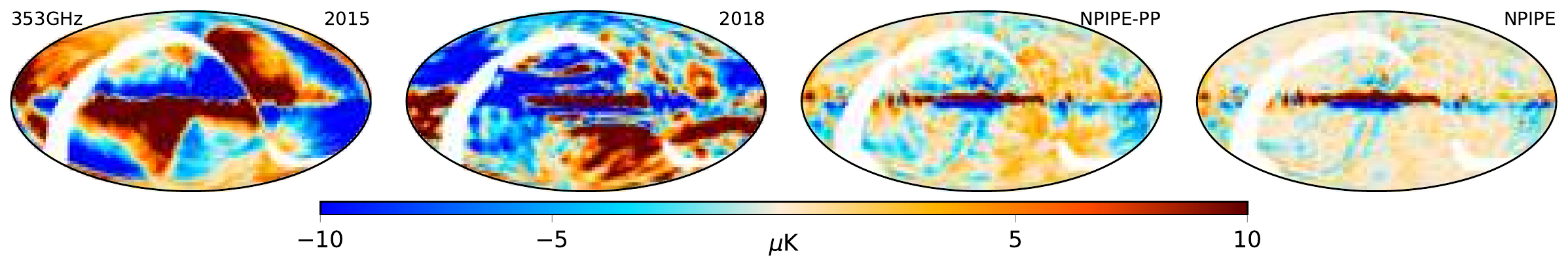}\\
    \includegraphics[width=1.0\linewidth]{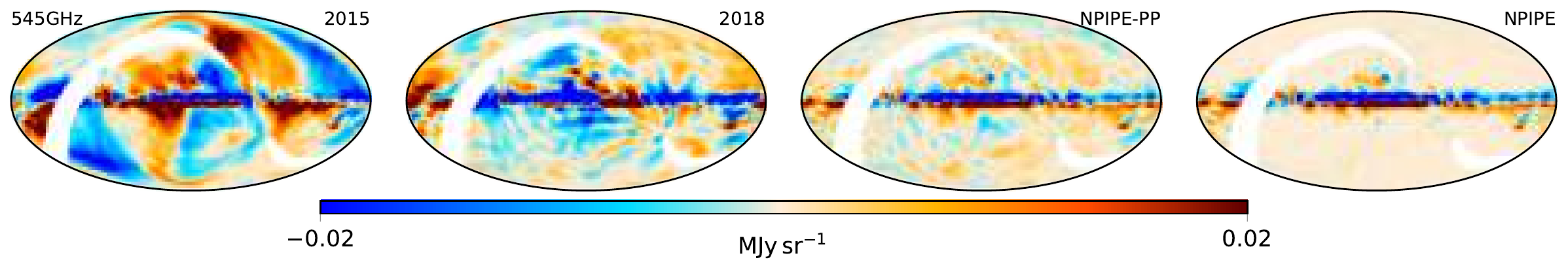}\\
    \includegraphics[width=1.0\linewidth]{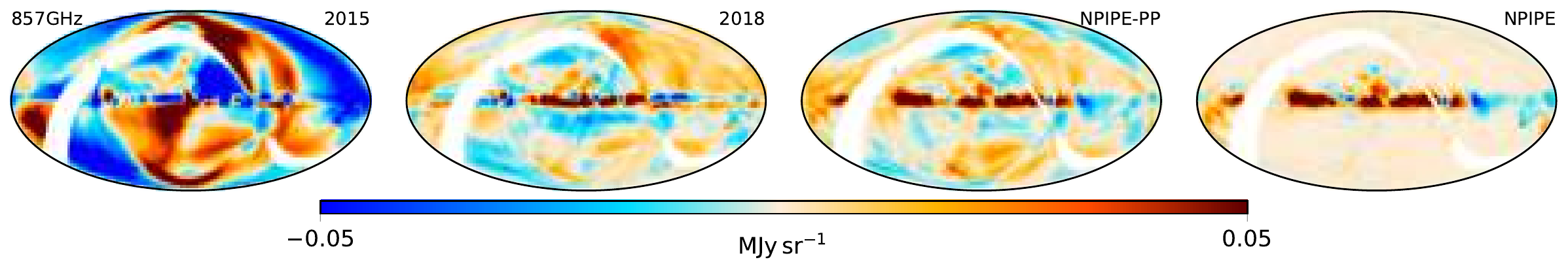}\\
  }
  \caption{Same as Fig.~\ref{fig:oddeven2}, but for 353, 545, and 857\GHz.  The S-shaped residual along the Ecliptic equator, especially in the 353\GHz\ \npipe\ results, is residual zodiacal emission.
  }
  \label{fig:oddeven3}
\end{figure*}

In Fig.~\ref{fig:oddeven3} we extend the comparison from 353 to 857\GHz.  The 353-GHz results are particularly notable because they show an apparent degradation in the null map between the 2015 and 2018 releases, visible in a pattern referred to as ``zebra stripes'' in \cite{planck2016-l03}.  This is a result of insufficient granularity in the measured and applied transfer-function correction.  The regression is corrected in \npipe\ by increasing the number of spectral bins.

\begin{figure}[hbtp!]
   \centering
   \includegraphics[width=0.48\linewidth]{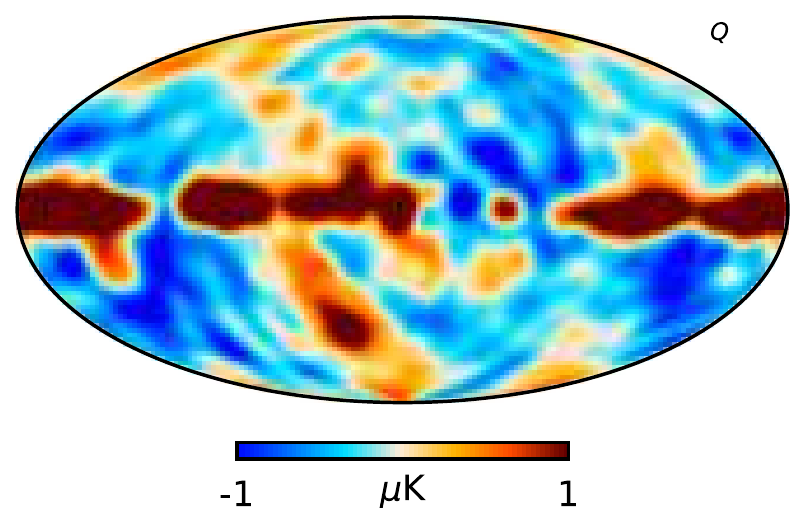}
   \includegraphics[width=0.48\linewidth]{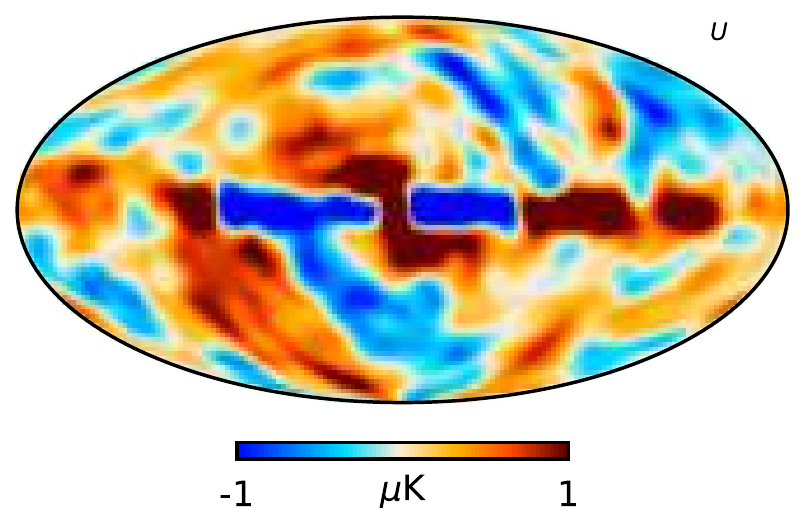}\\
   \includegraphics[width=0.48\linewidth]{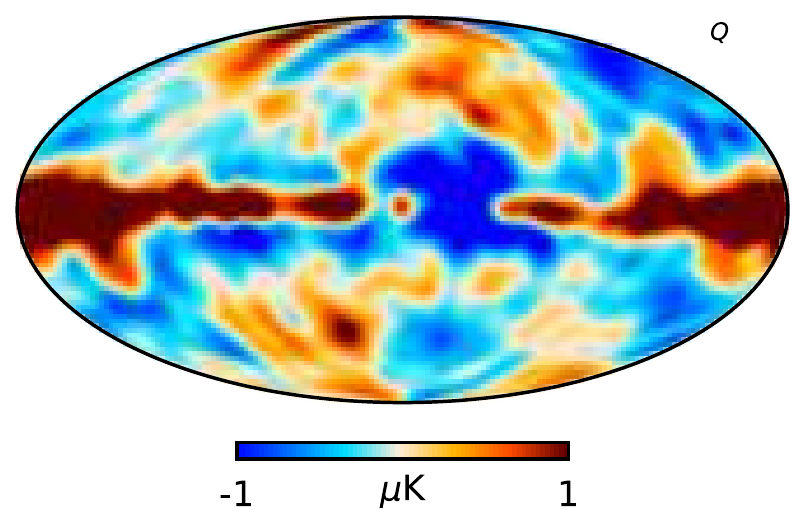}
   \includegraphics[width=0.48\linewidth]{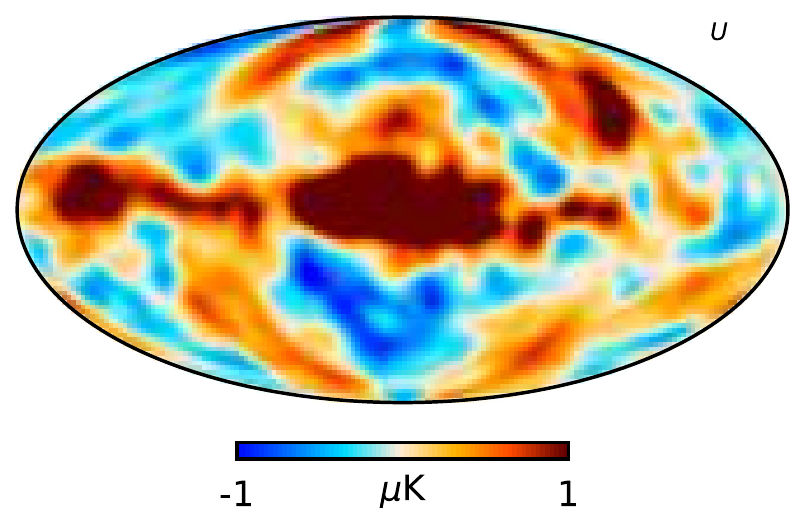}\\
   \includegraphics[width=0.48\linewidth]{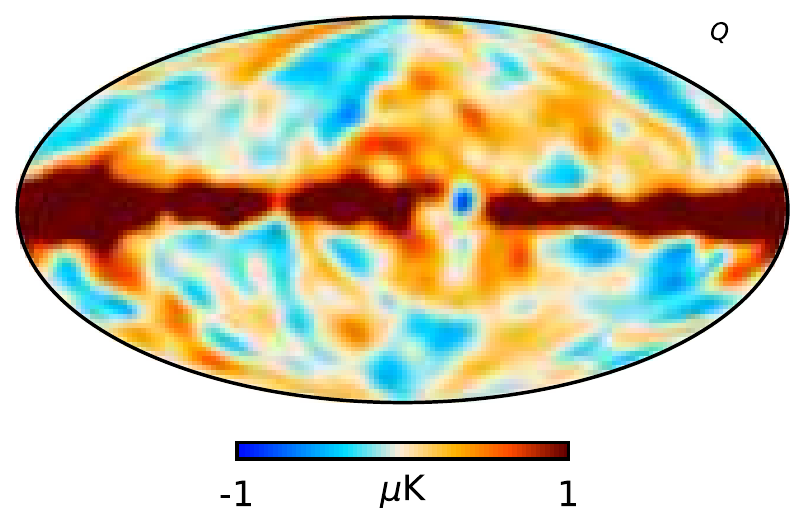}
   \includegraphics[width=0.48\linewidth]{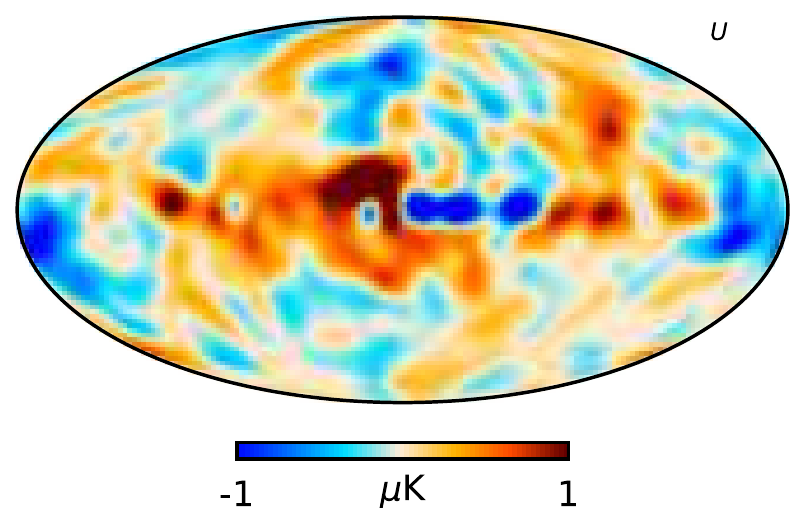}
   \caption{Internal frequency-difference polarization maps of the form $\m_{217}-0.128\,\m_{353}$, where the 353-GHz scaling factor is designed to suppress thermal dust emission at 217\GHz. From top to bottom, the three rows show difference maps based on \sroll, \srolltwo, and \npipe.  The left and right columns show Stokes $Q$ and $U$ parameters, respectively.  All maps have been smoothed to a common angular resolution of 10\deg.}
   \label{fig:comp_217v353}
\end{figure}

In Fig.~\ref{fig:comp_217v353} we consider polarization difference maps of the form $\m_{217}-0.128\,\m_{353}$ as derived by \sroll, \srolltwo, and \npipe, where the scaling factor corresponds to a modified-blackbody spectral energy distribution (SED) ratio between 217 and 353\GHz\ evaluated for $\beta_{\mathrm{d}}=1.6$ and $T_{\mathrm{d}}=19.6\,\mathrm{K}$.  These maps are thus designed to suppress thermal dust emission without having to resort to component-separation methods, and highlight potential residual systematic effects in the high-frequency channels. Focusing first on high Galactic latitudes, we first note that all three map solutions exhibit notable large-scale fluctuations. Some fluctuations are expected simply from true CMB fluctuations (which are only suppressed by 12.8\,\% in these difference maps), and some fluctuations are expected from random statistical noise. However, some fluctuations are also likely due to residual instrumental systematics. For comparison, we note that a standard $\Lambda$CDM CMB sky with $\tau\approx0.06$ exhibits peak-to-peak fluctuations of about $0.5\muK$, with an overall standard deviation of $0.15\muK$. At low Galactic latitudes, the residuals are dominated by possible spatial variations in the spectral index not captured by the constant scaling factor of 0.128 and temperature-to-polarization leakage. We note that \npipe\ exhibits notably lower fluctuations at high latitudes than both \sroll\ and \srolltwo, and Galactic-plane residuals appear more strongly correlated with the morphology of thermal dust.

\begin{figure}[hbtp!]
   \centering
   \includegraphics[width=0.48\textwidth]{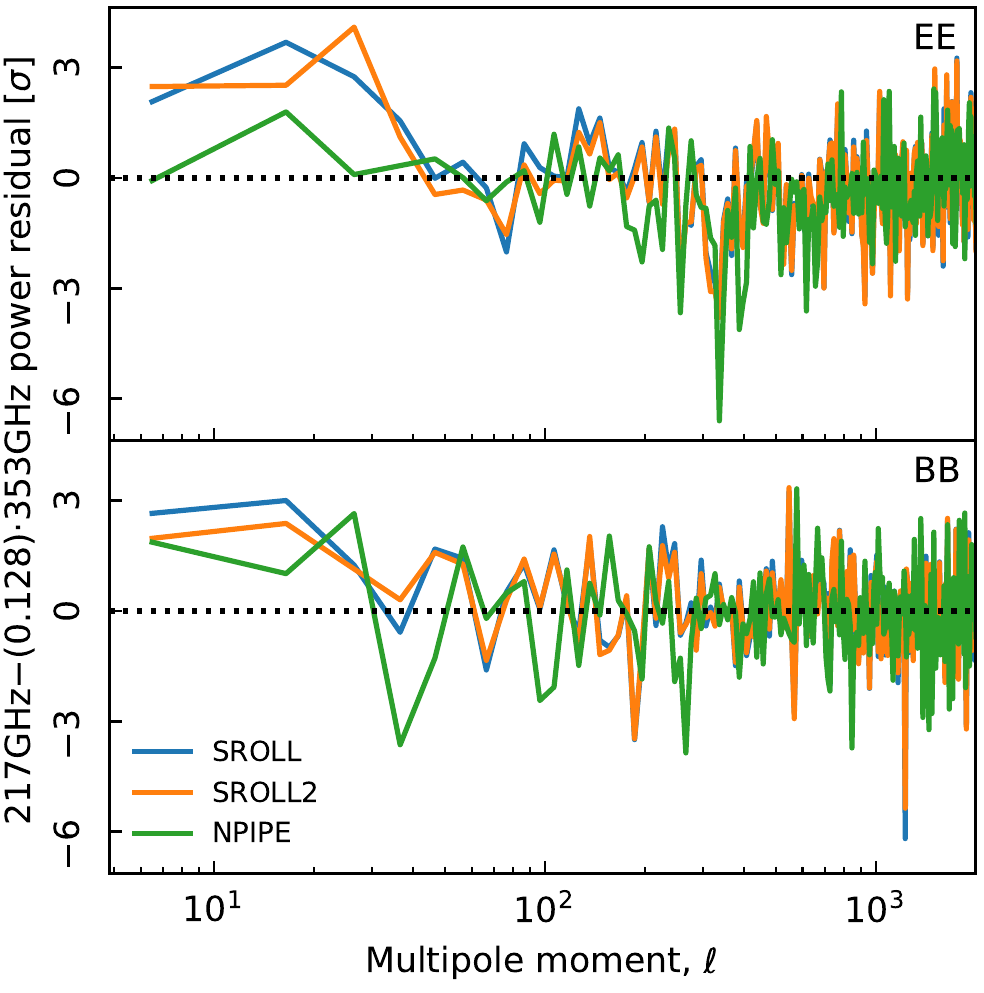}
   \caption{Angular cross-spectra evaluated from $\m_{217}-0.128\,\m_{353}$ difference maps. For \npipe, the cross-spectra are evaluated from detector-split difference maps, while for \sroll\ and \srolltwo\ they are evaluated from half-mission split difference maps. }
   \label{fig:comp_217v353_cl}
\end{figure}

In Fig.~\ref{fig:comp_217v353_cl} we show corresponding angular power spectra, evaluated as cross-spectra between detector-split difference maps (for \npipe) and half-mission split maps (for \sroll\ and \srolltwo) outside the \Planck\ 2018 polarization analysis mask \citep{planck2016-l04}. Consistent with the above visual impression, we see that \npipe\ exhibits significantly lower $EE$ power at large angular scales, while at small scales all three maps appear broadly consistent.

The observed differences between \Planck\ polarization maps illustrated in Figs.~48 and 49 may reflect elements specific to the \npipe\ data processing.  As discussed in Sect.~\ref{sec:noncmbtransfer}, polarized Galactic emission is not affected by the same transfer function at low multipoles as the CMB, because polarization templates are part of the data model in Eq.~9.  However, this approach introduces an interdependence between the \npipe\ frequency maps, which may impact studies of Galactic polarization (See Appendix~H).  Notably, this is an issue to have in mind when using the \npipe\ frequency maps to characterize the frequency correlation of dust polarization \citep{planck2016-l11A}, an essential question in the search for primordial CMB $B$-modes.

The bandpass mismatch template used in \npipe\ is based on the same sky model as the simulated sky signal, and (as discussed in Sect.~\ref{sec:notincluded}) we do not simulate errors in the template itself.  The net effect is that the uncertainty in the polarized emission by foreground Galactic dust is underestimated in the simulations, by the amount that errors in the sky model affect the bandpass mismatch correction.  Since the magnitude of this uncertainty is unknown, it may be misleading to rely on simulations alone to assess uncertainties in polarized Galactic emission on large scales, or to investigate how Galactic polarization decorrelates with frequency. \npipe\ is not unique in its approach of not fully sampling the space of template errors.  Issues related to bandpass mismatch are a generic feature in \Planck\ polarized mapmaking, and caution should be exercised when analysing other Planck releases as well.

\subsubsection{External consistency}

\Planck\ is calibrated without reference to \wmap\ \citep{bennett2012} polarization, although in \npipe, \wmap\ temperature data do contribute to the sky model used to derive bandpass-mismatch templates.  It is informative to compare the two experiments for agreement in synchrotron polarization.  The comparison is particularly interesting due to differences in the two experiments.  \wmap's differencing-assembly design allows for gain and bandpass mismatch to be separated into special spurious maps, while \Planck\ requires an estimate of the foreground intensity to correct for bandpass mismatch.

\begin{figure*}[htpb!]
  \center{
    \includegraphics[width=1.0\linewidth]{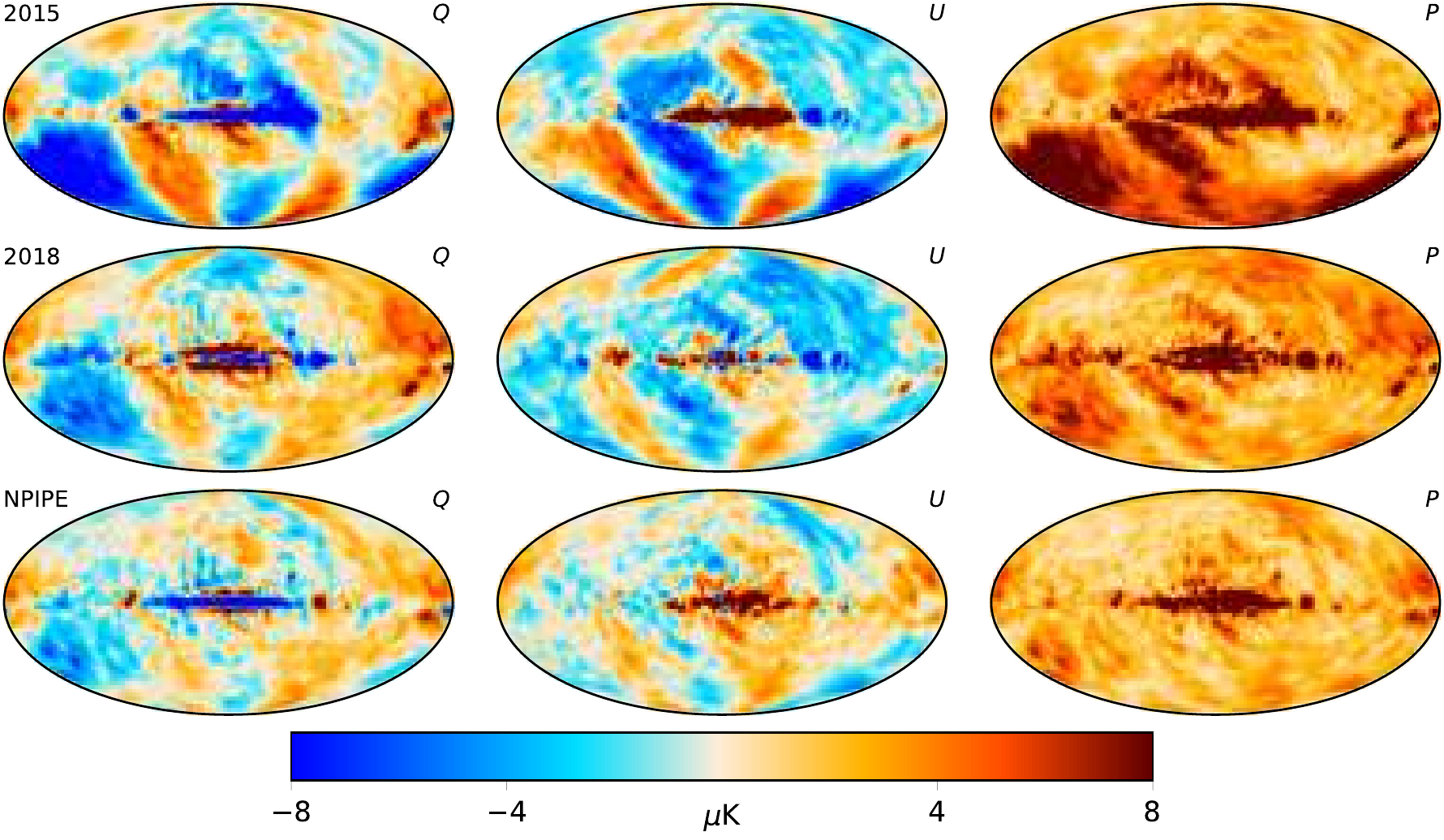}
  }
  \caption{\Planck\ $30\GHz$ -- \wmap\ K-band difference.  The K-band regression coefficients by row (top to bottom) are 0.406, 0.451, and 0.462.
  }
  \label{fig:wmap_comparison_30}
\end{figure*}

\begin{figure*}[htpb!]
  \center{
    \includegraphics[width=1.0\linewidth]{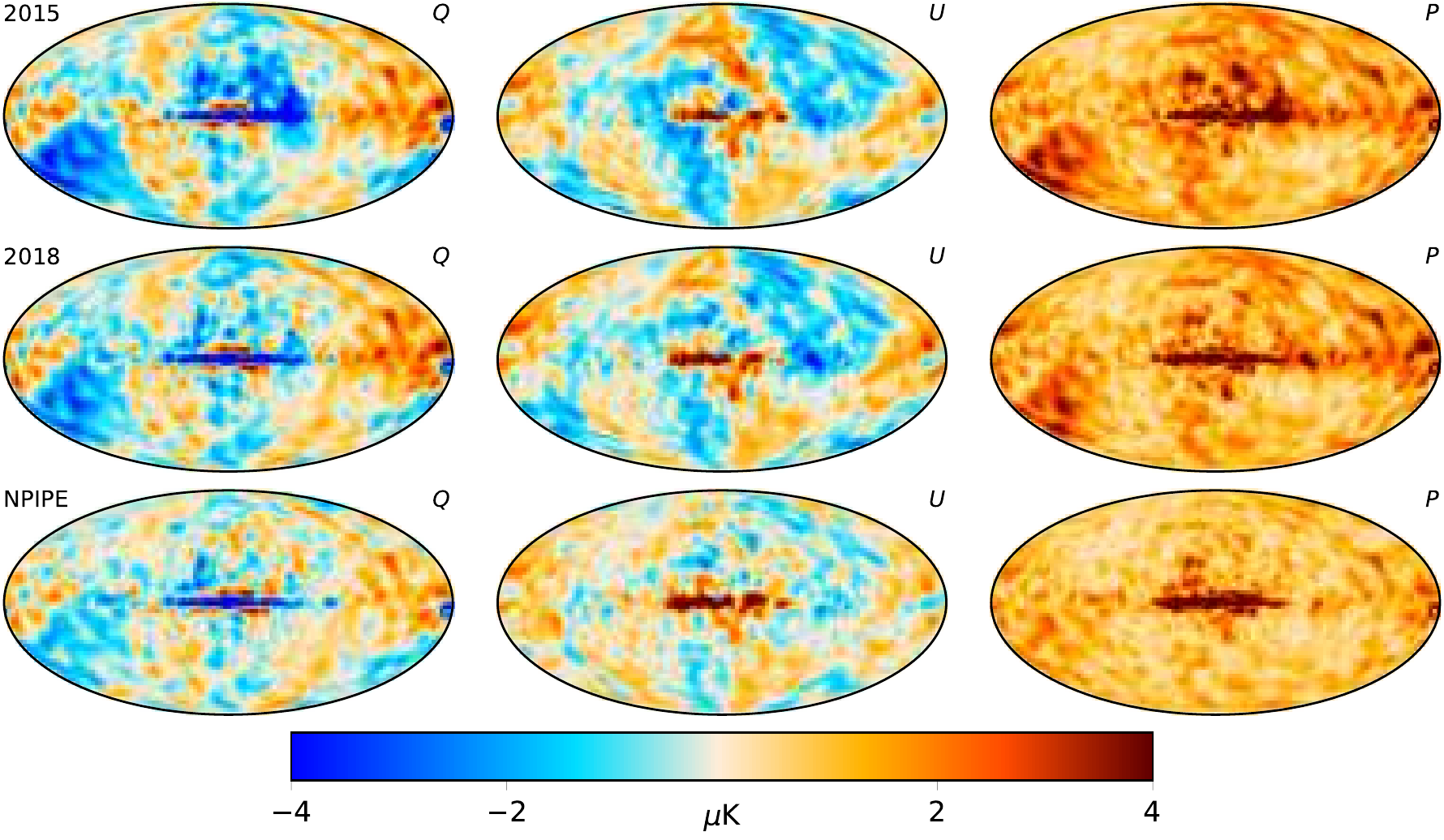}
  }
  \caption{\Planck\ 44\GHz$-$\wmap\ K-band difference.  The K-band regression coefficients by row are 0.104, 0.117, and 0.125.
  }
  \label{fig:wmap_comparison_44}
\end{figure*}

We measure the agreement by smoothing the \Planck\ and \wmap\ K-band (23\GHz) maps with a 5\deg\ Gaussian beam and then regressing the \wmap\ K-band and \Planck\ 353-GHz maps from the \Planck-\lfi\ maps.  The residuals for 30\GHz\ are shown in Fig.~\ref{fig:wmap_comparison_30}, and for 44\GHz\ in Fig.~\ref{fig:wmap_comparison_44}.  These figures show that improvements in the \lfi\ calibration procedure between the 2015 and 2018 releases significantly improved the agreement between \Planck\ and \wmap.  They also show that the \npipe\ large-scale polarization is more compatible with \wmap\ than is that of \prthree.  The fitting was carried out on the 30\,\% of the sky that has the highest polarization amplitude in the smoothed K-band map, and we masked out a further 5\,\% of the sky with the highest foreground intensity to reduce bandpass-mismatch-leakage effects, leaving 25.7\,\% of the sky for fitting.

\clearpage

\section{Component separation}
\label{sec:compsep}

We now derive new astrophysical component maps from the \npipe\ data.
For CMB extraction, we employ two different algorithms,
\commander\ \citep{eriksen:2004,eriksen2008,seljebotn:2017} and
\sevem\ \citep{leach2008,fernandez2012}; for foreground estimation, we
only use \commander.  Both methods have been used extensively in
previous \Planck\ publications; for full details see
\citet{planck2016-l04} and references therein. The main motivation for
the present analysis is to characterize the internal consistency of
the \npipe\ data themselves, rather than to derive an ultimate
\npipe\ sky model.  In particular, external data sets are not
considered in this paper.  Instead, combined analyses are deferred to
future studies.

\subsection{\commander\ methodology}
\label{sec:methodology}

As described by \citet{eriksen2008}, \commander\ adopts a parametric approach to component separation.  The first step in the process is to specify an explicit parametric model for the data.  Since the current analysis considers only \Planck\ data, we adopt a model similar to the one employed for the \Planck\ 2018 analysis \citep{planck2016-l04}.  In temperature, this includes CMB, a power-law low-frequency component, and a single modified-blackbody thermal-dust component.  Since \npipe\ produces single-detector temperature maps, it can also support individual CO components at all relevant frequencies (i.e., 100, 217, and 353\GHz).  This is in contrast to \prthree, which includes only integrated full-frequency maps.  We therefore model CO emission using three independent components, with single-detector line ratios taking into account the relative bandpass differences between detectors.

One important variation with respect to previous \Planck\ releases is the fact that the \npipe\ maps retain the CMB Solar dipole. This allows for joint component separation and high-precision relative calibration, eliminates the need to model additive dipole components in the frequency maps \citep{wehus:2014,planck2014-a12,planck2016-l04}, and removes important large-scale degeneracies from the model.  As a consequence, only three global parameters are fitted per sky map in the following analysis, namely a monopole/zero-level, an absolute calibration, and an overall bandpass correction \citep{planck2014-a12}, compared to six parameters in previous analyses.
  
The temperature model employed in this paper may be written as a vector in pixel space:
\begin{align}
  \s_{\nu} =\   u_{\nu}\,g_{\nu}\,\B_{\nu}\biggl[& \a_{\mathrm{cmb}} \gamma(\nu) \nonumber\\
     &+ \a_{\mathrm{lf}}
     \left(\frac{\nu}{\nu_{\mathrm{lf}}}\right)^{\beta_{\mathrm{lf}}(p)} \nonumber\\
     &+ \a_{\mathrm{d}}
     \left(\frac{\nu}{\nu_{\rm d}}\right)^{\beta_{\mathrm{d}}(p)+1}
     \left(\frac{e^{h\nu_{\mathrm{d}}/kT_{\mathrm{d}}(p)}-1}{e^{h\nu/kT_{\mathrm{d}}(p)}-1}
     \right) \nonumber\\
     &+ \a^{100}_{\mathrm{co}} h^{100}_{\nu} + \a^{217}_{\mathrm{co}}
     h^{217}_{\nu} + \a^{353}_{\mathrm{co}} h^{353}_{\nu}\biggr] \nonumber\\
  &+ m_{\nu},
  \label{eq:comm_model_T}
\end{align}
where $\nu$ denotes channel and $p$ denotes pixel number. The terms correspond, from top to bottom, to the cosmological CMB signal, a combined power-law low-frequency component with reference frequency $\nu_{\rm lf}$, a single modified-blackbody thermal-dust component with a references frequency $\nu_{\rm d}$, three CO components ($J\,{=}\,1\,{\rightarrow}\,0$, 
$J\,{=}\,2\,{\rightarrow}\,1$, and $J\,{=}\,3\,{\rightarrow}\,2$), and, finally, a monopole ``offset.'' The pre-factors before the astrophysical components are, from right to left: a per-channel beam-convolution operator $\B_{\nu}$, accounting for the azimuthally symmetric component of the true beam; an overall relative calibration factor $g_{\nu}$ per channel, with the 143-GHz calibration fixed to unity; and a unit conversion factor $u_{\nu}$, scaling from brightness to thermodynamic temperature. 

Each component is defined in terms of an amplitude map $\a_{i}$ at some reference frequency, multiplied by an SED that translates this amplitude map to any other frequency.  All SEDs are defined in brightness temperature units.  For the CMB component, $\gamma(\nu) = x^2e^{x}/(e^{x}-1)^2$ is therefore simply the conversion factor from thermodynamic to brightness temperature, where $x=h\nu/k_{\mathrm{B}}T_{0}$, $h$ is Planck's constant, $k_{\mathrm{B}}$ is Boltzmann's constant, and $T_0$ is the mean CMB temperature.  For the low-frequency component, the SED is given as a power law with a free spectral index, $\beta_{\textrm{lf}}$.  For thermal dust emission, the SED has two free parameters, a spectral index $\beta_{\textrm{d}}$ and a temperature $T_{\textrm{d}}$.  Finally, the multiplicative CO line ratios $h^{i}_{j}$ are set to zero for sky maps outside the bandpass range of a given CO component, and set to unity for one arbitrarily-chosen reference sky map inside the bandpass range of the same component (i.e., for the map from one detector at the relevant frequency); all other line ratios for the same component are fitted freely. For instance, $h^{217}_j$ is set to unity for the 217-2 channel, fitted freely for all other 217-GHz channels, and set to zero for all non-217-GHz channels.

For notational simplicity, all unit-conversion and bandpass-integration effects are suppressed in the above model. However, all
SEDs are integrated over the full bandpass $\tau(\nu)$ of each detector, using the formulae described in \citet{planck2013-p02b} and \citet{planck2013-p03d}, ensuring that the amplitude maps correspond directly to the signal that would be observed at a sharp reference frequency, not as bandpass-integrated quantities.  To account (at least partially) for the uncertainties in the bandpasses measured on the ground \citet{planck2014-a12}, we allow for an overall additive shift, $\Delta_{\nu}$, in each bandpass, such that $\tau(\nu) \rightarrow \tau(\nu+\Delta_{\nu})$.

For polarization, we adopt a similar, but simpler, model:
\begin{align}
  \s_{\nu} =\   u_{\nu}\,g_{\nu}\,\B_{\nu}\biggl[& \a_{\mathrm{cmb}} \gamma(\nu) \nonumber\\
     &+ \a_{\mathrm{s}}
     \left(\frac{\nu}{\nu_{\mathrm{s}}}\right)^{\beta_{\mathrm{s}}(p)} \nonumber\\
     &+ \a_{\mathrm{d}}
     \left(\frac{\nu}{\nu_{\rm d}}\right)^{\beta_{\mathrm{d}}(p)+1}
     \left(\frac{e^{h\nu_{\mathrm{d}}/kT_{\mathrm{d}}(p)}-1}{e^{h\nu/kT_{\mathrm{d}}(p)}-1}
     \right)\biggr].
    \label{eq:comm_model_P}
\end{align}
No CO components or monopoles are fitted for polarization, and the general low-frequency component has been renamed to just ``synchrotron,'' since neither free-free nor spinning dust emission are expected to be significantly polarized \citep{planck2014-a31}.  Additionally, the thermal dust temperature $T_{\rm d}$ is fixed to the values derived from the temperature analysis, since \Planck\ is not sensitive to polarization above 353\GHz.  Note, however, that $T_{\rm d}$ is spatially varying even though it is fixed throughout the analysis.  In contrast, the two spectral indices, $\beta_{\mathrm{s}}$ and $\beta_{\mathrm{d}}$, are fitted directly with polarization data.

All free parameters in these models are fitted jointly using the \commander\ software suite, following the same procedures as
detailed in \citet{planck2014-a12} and \citet{planck2016-l04}.  The starting point of these fits is (as usual) Bayes theorem,
\begin{linenomath*}
\begin{equation}
P(\theta|\d) = \frac{P(\d|\theta)P(\theta)}{P(\d)},
\end{equation}
\end{linenomath*}
where $P(\theta|\d)$ is the posterior distribution, $P(\d|\theta)=\mathcal{L}(\theta)$ is the likelihood, $P(\theta)$ is a set of priors, and $P(\d)$ is, for our purposes, an irrelevant normalization constant.  In most of the following analysis, we report the maximum-posterior solution as our best-fit estimate, and we use corresponding simulations to quantity uncertainties.
 
We adopt the same set of priors as in \citet{planck2016-l04}. Most importantly, for SED parameters such as $\beta_i$ or $T_{\textrm{d}}$, we adopt a product of informative Gaussian priors, with central values informed by the high signal-to-noise parts of the data set, and a non-informative Jeffreys prior to account for posterior volume effects due to nonlinear parameterizations. For spatial priors, we adopt angular-power-law-spectrum priors on all foreground amplitudes, but no spatial prior on the CMB amplitudes (for further details, see Appendix~A in \citealt{planck2016-l04}). For global parameters, we impose no active priors on either calibration factors or offsets ($g_\nu$ and $m_\nu$, respectively, in Eqs.~19--23), except that we fix one overall calibration factor (at 143\GHz), and one offset per free diffuse component (30\GHz\ for the low-frequency-component; 100-1a for CO $J$=1$\rightarrow$0; 143\GHz\ for CMB; 217-2 for CO
$J$=2$\rightarrow$1; 353-3 for $J$=3$\rightarrow$2; and 545-2 for thermal dust emission.) The reference calibration factor is fixed to unity, while the reference offsets are determined such that the resulting component maps obtain physically-meaningful
zero-levels. Specifically, for 100-1a, 143, 217-2, and 353-3, the reference offsets are set such that the CO and CMB maps have vanishing mean offsets at high Galactic latitudes. For 545-2, the offset is set such that a $T$--$T$ scatter plot between the derived thermal dust map and an \ion{H}{i} survey \citep{Lenz_et_al:2017} has vanishing intercept. For 30\GHz, it is set such that the derived low-frequency spectral index map does not correlate strongly with the corresponding amplitude map, which is a typical artefact resulting from incorrectly-set offsets \citep{wehus:2014}.  The 30-GHz offset determination is the most uncertain among these offsets, since the spectral index information in the map has relatively low S/N.

We approximate the instrumental noise of each detector as Gaussian, and the likelihood therefore is
\begin{linenomath*}
\begin{equation}
P(\d|\theta) \propto e^{-\frac{1}{2}(\d-\s(\theta))\trans
\N^{-1} (\d-\s(\theta))},
\end{equation}
\end{linenomath*}
where $\N$ is the noise covariance matrix. Since the following analyses are performed at the full angular resolution of the \Planck\
instrument, we approximate this matrix by its diagonal, and thereby only take into account the scan-modulated white noise component, not correlated noise features or instrumental systematics.  However, since $\N$ affects only the relative weighting between channels, the derived marginal map products remain unbiased, albeit slightly sub-optimal in terms of variance.  Final uncertainties are assessed with simulations, for which the same effects are present.

\subsection{\sevem\ methodology}

\sevem\  \citep{leach2008, fernandez2012} is based on internal template cleaning in real space. In previous \Planck\ releases, it was one of four approaches used to obtain the CMB signal from frequency maps \citep{planck2016-l04}.  The internal templates trace foreground emission at the corresponding frequency range, and are constructed as difference maps between two neighbouring \Planck\ channels, convolved to the same resolution to facilitate removal of the CMB contribution.  A linear combination of these templates is then subtracted from a CMB-dominated frequency map, in such a way that the coefficients of the combination minimize the variance of the resulting map outside a given mask.  Different single-frequency-cleaned maps are produced in the range 70 to 217\GHz, then a set of them is co-added, in harmonic space, into a single map to produce a final CMB map with higher S/N.

The single-frequency, cleaned maps produced by \sevem\ are useful in testing the robustness of results versus the presence of foreground residuals and systematics, for instance for isotropy and statistics estimators \citep{planck2016-l07} or the integrated Sachs-Wolfe stacking analysis \citep{planck2014-a26}. They are also valuable in constructing cross-frequency estimators, which allow one to minimize the impact of certain types of systematic effects, such as possible correlated noise in the data splits.

We used the same pipeline as in \citet{planck2016-l04} to extract the CMB signal from the \npipe\ frequency maps (unlike \commander, the \sevem\ pipeline does not use single-bolometer maps).  However, since the \npipe\ frequency maps retain the full stationary signal at each frequency, the first step is to subtract the Solar dipole, the frequency-dependent second-order quadrupole contributions, as well as the stationary component of the zodiacal light; the time-dependent component has already been removed during mapmaking.  To reduce contamination from point sources, we use specific catalogues derived from the \npipe\ maps.  As in previous releases, point sources are detected in each frequency map using the {\tt Mexican-Hat-Wavelet 2} code in intensity \citep{lopezcaniego2006, planck2014-a35} and the {\tt Filtered Fusion} code in polarization \citep{argueso2009}.  In general, the number of point sources detected in temperature is similar to that in \Planck\ 2018, with some changes close to the sensitivity limit due to the lower noise in the \npipe\ maps.  However, the number of polarized sources detected in the \npipe\ maps is smaller than in the 2018 maps (except at 44\GHz), because better treatment of leakage from temperature to polarization reduces the number of spurious sources.

In addition to the full-survey maps, we also consider the \npipe\ A/B detector split maps described in Sect.~\ref{sec:abmaps}. These are constructed to be as independent as possible, and the \sevem\ coefficients are independently fitted for each case. This is different from earlier \Planck\ releases, in which data splits were propagated through the pipeline using the same coefficients as derived from the full-mission data.  Because the A/B splits do not have the same effective central frequency, foregrounds do not completely cancel in the AB half-difference map, and independent propagation is as such even more important for this data split.\footnote{In contrast to the case in previous \Planck\ releases, the beams for the full and split data sets are slightly different, and this is also the case in the cleaned single-frequency maps. However, by construction, the final combined CMB map has the same resolution (Gaussian beam with a FWHM of 5\arcm) for the full-frequency and split-data maps.}

The linear coefficients of the templates used to extract the single-frequency, cleaned maps with \sevem\ are shown in Tables~\ref{table:sevem_coef_T} (for intensity) and \ref{table:sevem_coef_P} (for polarization).  For intensity, the coefficients are quite similar to those from \Planck\ 2018, with some differences in the contribution of the low-frequency templates to the cleaned HFI maps, especially 143\GHz.  For polarization, the impact in the value of the coefficients is somewhat larger. Given the lower S/N of the polarization data than the temperature data, these values are more affected by the noise in the maps. Therefore, the different processing of the data using \npipe\ introduces larger differences in the coefficients. For completeness, the linear coefficients used for the \npipe\ splits are also given. Some differences between the coefficients for the \npipe\ full-mission and \npipe\ split data are seen, mainly due to the larger noise contribution in the A/B splits with respect to the full mission. In most cases, the template fitting tries to minimize the effect of this larger noise in the cleaned CMB map by reducing the absolute value of the corresponding coefficients. In particular, the most pronounced deviations are seen in the coefficients for the LFI templates, specially when cleaning the 100-GHz map. For intensity, we checked that the major effect is actually due to the different noise level in the 44-70 template, which affects to the coefficient of both LFI templates. For polarization, the larger noise levels of the 30-44 templates constructed with the splits also affect the coefficients, although in this case some differences are also introduced by the HFI templates.

\begin{table*}[htbp!]
\begingroup \newdimen\tblskip \tblskip=5pt
  \caption{Linear coefficients of the templates used to clean individual frequency maps with \sevem\ for temperature. The  353$-$143 template has been produced at the same resolution as the 70\GHz\ frequency channel, while the 857\GHz\ map has been convolved with the 545-GHz beam. The rest of the templates are constructed such that the first map in the subtraction is convolved with the beam of the second map, and vice versa.}
\label{table:sevem_coef_T}
\nointerlineskip
\vskip -3mm
\footnotesize
\setbox\tablebox=\vbox{
\newdimen\digitwidth
\setbox0=\hbox{\rm 0}
\digitwidth=\wd0
\catcode`*=\active
\def*{\kern\digitwidth}
\newdimen\signwidth
\setbox0=\hbox{+}
\signwidth=\wd0
\catcode`!=\active
\def!{\kern\signwidth}
\newdimen\expsignwidth
\setbox0=\hbox{$^{-}$}
\expsignwidth=\wd0
\catcode`@=\active
\def@{\kern\expsignwidth}
\halign{\hbox to 1.3in{#\leaderfil}\tabskip=2em&
     \hfil#\hfil\tabskip=1em&
     \hfil#\hfil&
     \hfil#\hfil&
     \hfil#\hfil\tabskip=0pt\cr
\noalign{\doubleline}
\omit&\multispan4\hfil \sevem\ coefficients \hfil\cr
\noalign{\vskip -3pt}
\omit&\multispan4\hrulefill\cr
\noalign{\vskip 3pt}
\omit\hfil Template\hfil& 70\GHz&100\GHz& 143\GHz& 217\GHz\cr
\noalign{\vskip 3pt\hrule\vskip 5pt}
\omit\parbox{3cm}{\textbf{Full mission}}\cr
   *30$-$44*& $1.64 \times 10^{-1}$&$-8.34 \times 10^{-2}$& !$1.70\times 10^{-2}$&$-1.49\times 10^{-1}$\cr  
   *44$-$70*&\ldots&!$3.90 \times 10^{-1}$& !$1.56\times 10^{-1}$&!$3.95 \times 10^{-1}$\cr 
   353$-$143& $6.85 \times 10^{-3}$&\ldots& \ldots&\ldots\cr  
   545$-$353&\ldots&!$4.12\times 10^{-3}$&$!6.28\times 10^{-3}$&!$1.71\times 10^{-2}$\cr 
      **857**&\ldots&$-3.18\times 10^{-5}$&$-5.09\times 10^{-5}$&$-1.05\times 10^{-4}$\cr 
\noalign{\vskip 5pt\hrule\vskip 5pt}
\omit\parbox{3cm}{\textbf{A split}}\cr
   *30$-$44*& $ 1.66\times 10^{-1}$&$ -4.58\times 10^{-2}$& !$2.30\times 10^{-2}$&$-1.22\times 10^{-1}$\cr  
   *44$-$70*&\ldots&!$ 2.92\times 10^{-1}$& !$1.40\times 10^{-1}$&!$ 3.14\times 10^{-1}$\cr 
   353$-$143& $ 6.82\times 10^{-3}$&\ldots& \ldots&\ldots\cr  
   545$-$353&\ldots&!$4.00\times 10^{-3}$&$!6.15\times 10^{-3}$&!$1.67\times 10^{-2}$\cr 
         **857**&\ldots&$-3.21\times 10^{-5}$&$-5.14\times 10^{-5}$&$-1.05\times 10^{-4}$\cr 
\noalign{\vskip 5pt\hrule\vskip 5pt}
\omit\parbox{3cm}{\textbf{B split}}\cr
   *30$-$44*& $ 1.62\times 10^{-1}$&$ -4.80\times 10^{-2}$& !$1.91\times 10^{-2}$&$-1.05\times 10^{-1}$\cr  
   *44$-$70*&\ldots&!$ 2.92\times 10^{-1}$& !$1.43\times 10^{-1}$&!$2.84\times 10^{-1}$\cr 
   353$-$143& $ 6.96\times 10^{-3}$&\ldots& \ldots&\ldots\cr  
   545$-$353&\ldots&!$4.15\times 10^{-3}$&$!6.33\times 10^{-3}$&!$1.73\times 10^{-2}$\cr 
         **857**&\ldots&$-3.20\times 10^{-5}$&$-5.06\times 10^{-5}$&$-1.06\times 10^{-4}$\cr 
\noalign{\vskip 5pt\hrule\vskip 3pt}}}
\endPlancktablewide
\endgroup
\end{table*} 

\begin{table*}[htbp!]
\begingroup
\newdimen\tblskip \tblskip=5pt
  \caption{Linear coefficients for each of the templates used to
    clean individual frequency maps with \sevem\ for polarization. Here, the 30$-$44 template is constructed such that the 30-GHz map is smoothed with the 44-GHz beam and vice versa. The 217$-$143, 217$-$100 and 143$-$100 templates are produced with $1^{\circ}$ resolution, and 353$-$217 and 353$-$143 with 10\arcm. In addition, this last template is also constructed at the resolution of the 70-GHz beam, in order to clean that channel.}
\label{table:sevem_coef_P}
\nointerlineskip
\vskip -3mm
\footnotesize
\setbox\tablebox=\vbox{
\newdimen\digitwidth
\setbox0=\hbox{\rm 0}
\digitwidth=\wd0
\catcode`*=\active
\def*{\kern\digitwidth}
\newdimen\signwidth
\setbox0=\hbox{+}
\signwidth=\wd0
\catcode`!=\active
\def!{\kern\signwidth}
\newdimen\expsignwidth
\setbox0=\hbox{$^{-}$}
\expsignwidth=\wd0
\catcode`@=\active
\def@{\kern\expsignwidth}
\halign{\hbox to 1.2in{#\leaderfil}\tabskip=1.2em&
     \hfil#\hfil\tabskip=0.7em&
     \hfil#\hfil&
     \hfil#\hfil&
     \hfil#\hfil&
     \hfil#\hfil&
     \hfil#\hfil&
     \hfil#\hfil&
     \hfil#\hfil\tabskip=0pt\cr
\noalign{\doubleline}
\omit&\multispan8\hfil \sevem\ coefficients \hfil\cr
\noalign{\vskip -3pt}
\omit&\multispan8\hrulefill\cr
\noalign{\vskip 3pt}
\omit\hfil Template\hfil& 70\GHz\ $Q$& 70\GHz\ $U$& 100\GHz\ $Q$& 100\GHz\ $U$
& 143\GHz\ $Q$& 143\GHz\ $U$& 217\GHz\ $Q$& 217\GHz\ $U$\cr
\noalign{\vskip 3pt\hrule\vskip 5pt}
\omit\parbox{3cm}{\textbf{Full mission}}\cr
   *30$-$44*&$3.07\times 10^{-2}$&$3.60\times 10^{-2}$&$1.10\times
   10^{-2}$& $1.32 \times 10^{-2}$& $4.74 \times 10^{-3}$& $8.43
   \times 10^{-3}$& $1.11 \times 10^{-2}$& $1.37 \times 10^{-2}$\cr  
143$-$100& \dots& \ldots& \ldots& \ldots& \ldots& \ldots& $0.79$& $0.62$\cr
217$-$100& \ldots& \ldots& \ldots& \ldots& $1.29 \times 10^{-1}$& $1.20
\times 10^{-1}$& \ldots& \ldots\cr
217$-$143& \ldots& \ldots&  $0.87 \times 10^{-1}$& $0.84
\times 10^{-1}$& \ldots& \ldots& \ldots& \ldots\cr
353$-$143& $1.18 \times 10^{-2}$& $0.98 \times 10^{-2}$& \ldots& \ldots& \ldots& \ldots& $1.18 \times 10^{-1}$& 1.17 $\times 10^{-1}$\cr
353$-$217& \ldots& \ldots&  $1.38 \times 10^{-2}$& $1.32
\times 10^{-2}$& $3.15 \times 10^{-2}$& 3.07 $\times 10^{-2}$& \ldots& \ldots\cr
\noalign{\vskip 5pt\hrule\vskip 5pt}
\omit\parbox{3cm}{\textbf{A split}}\cr
   *30$-$44*&$2.03\times 10^{-2}$&$1.77\times 10^{-2}$&$0.67\times
   10^{-2}$& $0.68 \times 10^{-2}$& $1.48 \times 10^{-3}$& $3.65
   \times 10^{-3}$& $1.45 \times 10^{-2}$& $1.31 \times 10^{-2}$\cr  
143$-$100& \dots& \ldots& \ldots& \ldots& \ldots& \ldots& $0.90$& $0.68$\cr
217$-$100& \ldots& \ldots& \ldots& \ldots& $1.75 \times 10^{-1}$& 
$1.64\times 10^{-1}$& \ldots& \ldots\cr
217$-$143& \ldots& \ldots&  $1.12\times 10^{-1}$& 
$1.04\times 10^{-1}$& \ldots& \ldots& \ldots& \ldots\cr
353$-$143& $1.18 \times 10^{-2}$& $0.95 \times 10^{-2}$& \ldots& \ldots& \ldots& \ldots& $1.17 \times 10^{-1}$& $1.13 \times 10^{-1}$\cr
353$-$217& \ldots& \ldots&  $1.13 \times 10^{-2}$& $1.12
\times 10^{-2}$& $2.58 \times 10^{-2}$& $2.55 \times 10^{-2}$& \ldots& \ldots\cr
\noalign{\vskip 5pt\hrule\vskip 5pt}
\omit\parbox{3cm}{\textbf{B split}}\cr
   *30$-$44*&$2.00\times 10^{-2}$&$2.48\times 10^{-2}$&$0.64\times
   10^{-2}$& $0.88 \times 10^{-2}$& $3.62 \times 10^{-3}$& $6.48
   \times 10^{-3}$& $1.61 \times 10^{-2}$& $1.62 \times 10^{-2}$\cr  
143$-$100& \dots& \ldots& \ldots& \ldots& \ldots& \ldots& $1.31$ & $1.12$\cr
217$-$100& \ldots& \ldots& \ldots& \ldots& $2.09 \times 10^{-1}$& 
$2.13\times 10^{-1}$& \ldots& \ldots\cr
217$-$143& \ldots& \ldots&  $1.31 \times 10^{-1}$& $1.29
\times 10^{-1}$& \ldots& \ldots& \ldots& \ldots\cr
353$-$143& $1.19 \times 10^{-2}$& $0.94 \times 10^{-2}$& \ldots& \ldots& \ldots& \ldots& $1.03 \times 10^{-1}$& $0.96 \times 10^{-1}$\cr
353$-$217& \ldots& \ldots&  $0.94 \times 10^{-2}$& $0.86
\times 10^{-2}$& $2.11 \times 10^{-2}$& $1.90 \times 10^{-2}$& \ldots& \ldots\cr
\noalign{\vskip 5pt\hrule\vskip 3pt}}}
\endPlancktablewide
\endgroup
\end{table*}

\subsection{Data selection and goodness of fit}
\label{sec:data_selection}

An important feature of the Bayesian parametric analysis framework, as implemented in \commander, is its ability to provide robust and intuitive goodness-of-fit statistics in the form of residual maps, $\r=\d-\s$, where $\d$ denotes the observed channel maps and $\s$ is the fitted sky model defined in Eqs.~\ref{eq:comm_model_T} and \ref{eq:comm_model_P}, and $\chi^2$ statistics. These statistics are powerful probes of residual systematics in the data, since they highlight any discrepancies between the raw data and the assumed model. By inspecting the morphology of these maps, it is often possible to understand the physical origin of a given model failure, which in turn can suggest improvements either to the data processing or to the fitted model. Indeed, many of the issues discussed in Sect.~\ref{sec:processing} were discovered precisely through these statistics, which in turn allowed us to improve the overall quality of the \npipe\ data processing.

The first detailed analysis of the \Planck\ data based on this framework was presented in \citet{planck2014-a12}. These data 
included single-detector and detector-set temperature sky maps, in addition to co-added full-frequency maps, allowing a much more fine-grained data-inspection and data-selection process than did either the 2013 or the 2018 releases, both of which included only full-frequency maps. This is only relevant for temperature, not polarization, since the \Planck\ scan strategy does not allow us to solve for polarized maps from single-detector data.  In the following, therefore, we compare the new \npipe\ data products to the \Planck\ 2015 release in temperature, and to the 2018 data release in polarization.

\begin{figure*}[htpb!]
   \centering
      \hspace*{\fill}%
      \begin{minipage}[b]{0.45\textwidth}
      \centering
      \vspace{0pt}
      \includegraphics[width=0.48\textwidth]{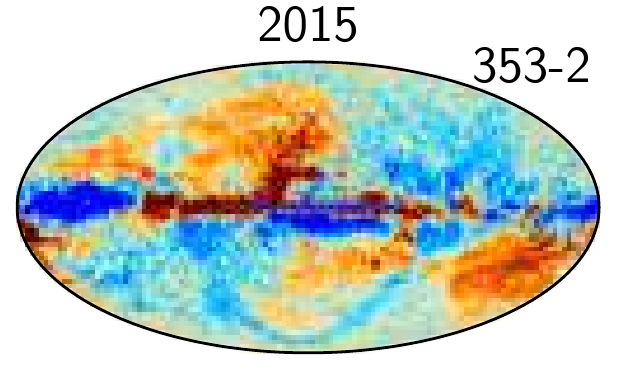}
      \includegraphics[width=0.48\textwidth]{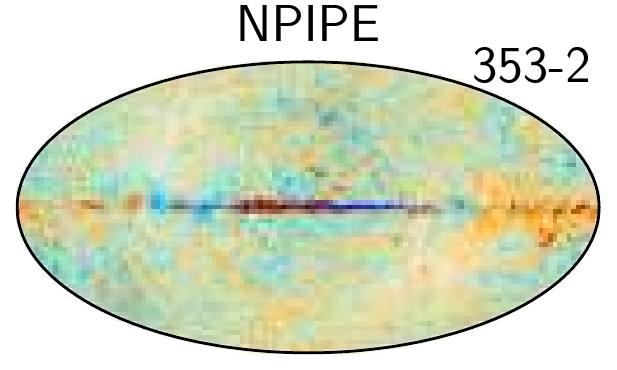}\\
      \includegraphics[width=0.48\textwidth]{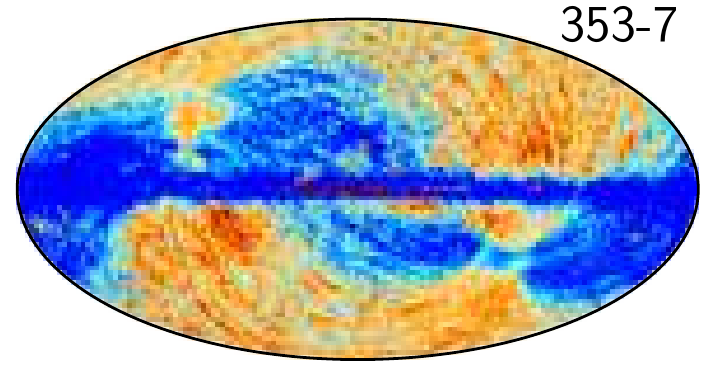}
      \includegraphics[width=0.48\textwidth]{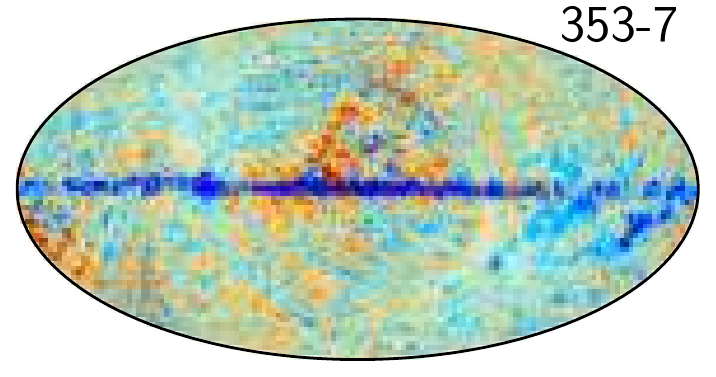}\\
      \includegraphics[width=0.48\textwidth]{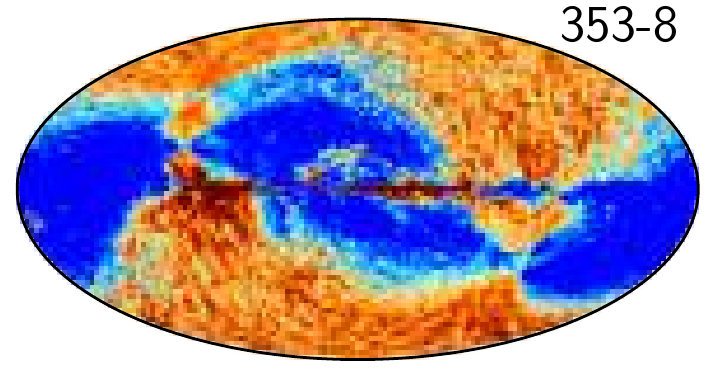}
      \includegraphics[width=0.48\textwidth]{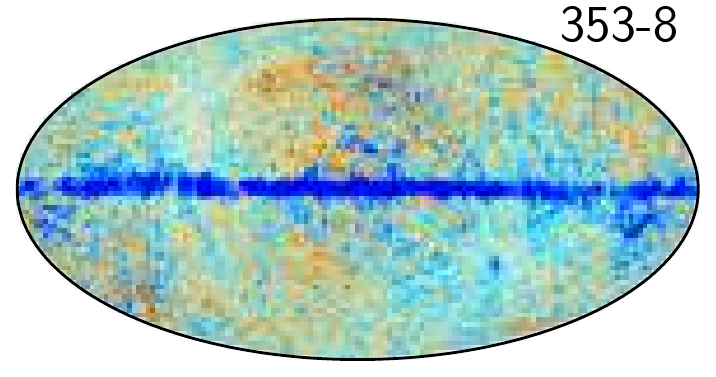}\\
      \includegraphics[width=0.8\textwidth]{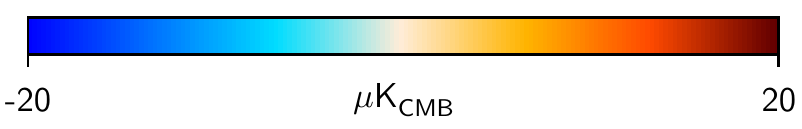}
      \end{minipage}%
      \hspace{0.5cm}
      \begin{minipage}[b]{0.45\textwidth}
      \centering
      \vspace{0pt}
      \includegraphics[width=0.48\textwidth]{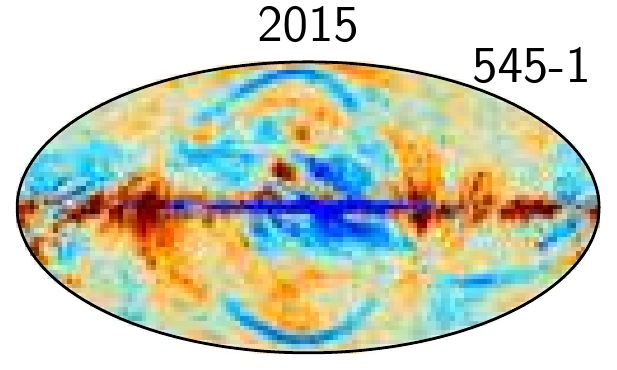}
      \includegraphics[width=0.48\textwidth]{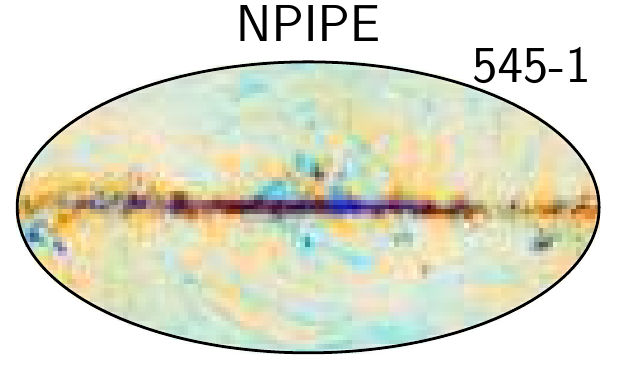}
      \includegraphics[width=0.48\textwidth]{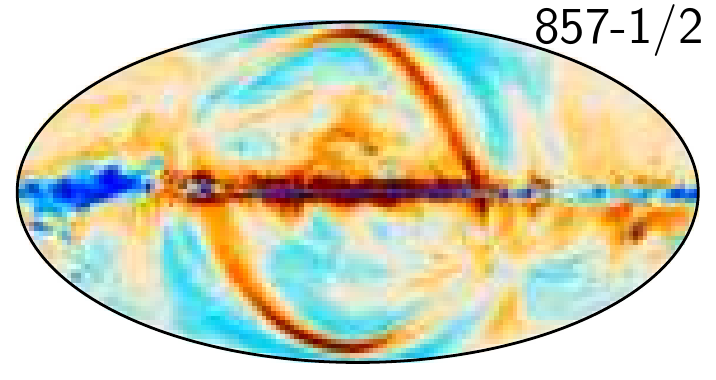}
      \includegraphics[width=0.48\textwidth]{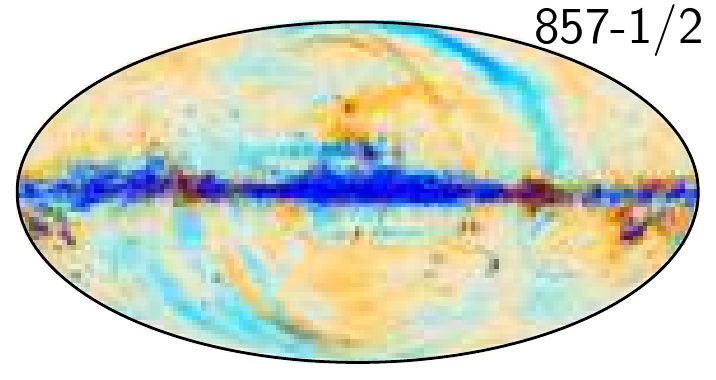}
      \includegraphics[width=0.48\textwidth]{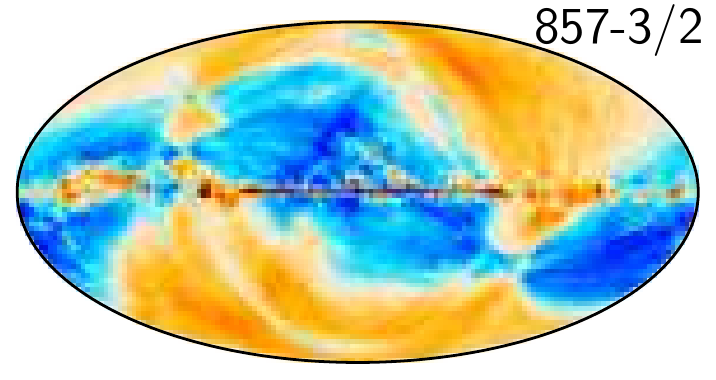}
      \includegraphics[width=0.48\textwidth]{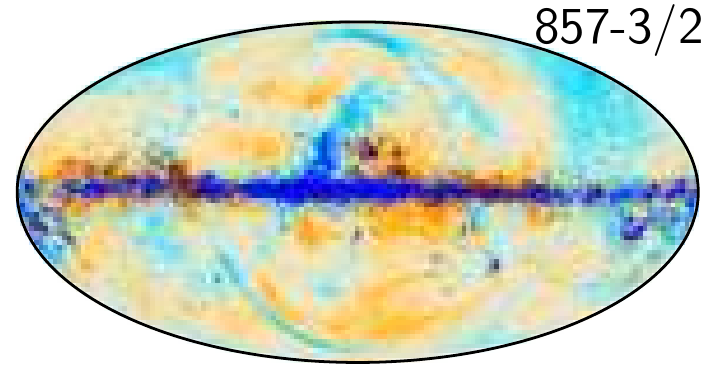}
      \includegraphics[width=0.48\textwidth]{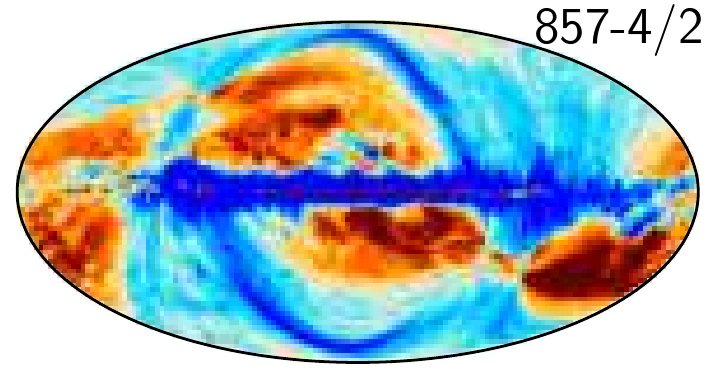}
      \includegraphics[width=0.48\textwidth]{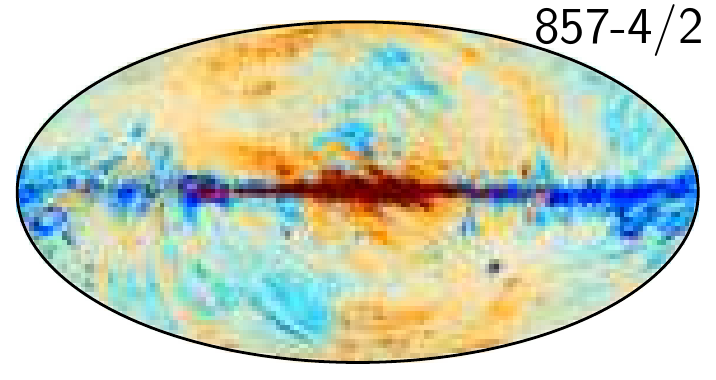}\\
      \includegraphics[width=0.8\textwidth]{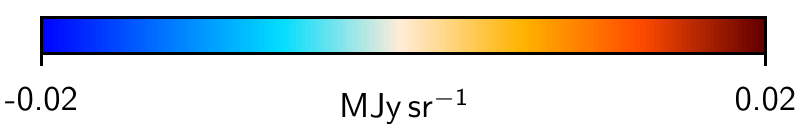}
      \end{minipage}%
      \hspace*{\fill}
      \caption{\Planck\ 2015 residual maps of rejected single bolometers \citep[see][]{planck2014-a12} and the corresponding \npipe\ residual maps. All maps have a smoothing of 60\arcm\ FWHM. The fraction in the label of the 857-GHz detectors indicates that the maps are divided by 2 with respect to the colour bar.  The overall level of coherent systematic residuals in these maps is dramatically reduced in \npipe, to the level that all except 857-4 (see text) can now be included in the final analysis.}
   \label{fig:res_compsep_npipe_T_data_select}
\end{figure*}

In the 2015 \commander\ temperature analysis, 21~single-detector, detector-set, and full-frequency \Planck\ maps were
considered sufficiently clean from residual instrumental systematic effects to be included in the final analysis, while 10~maps (three detector-set and seven single-detector maps) were excluded due to large unexplained systematic effects. Figure~\ref{fig:res_compsep_npipe_T_data_select} compares \Planck\ 2015 and \npipe\ residual maps for the seven excluded single-detector sky maps for which a direct head-to-head comparison is possible (the \npipe\ data release does not provide detector-set maps in the same form as \Planck\ 2015).

Four important effects can be seen in Fig.~\ref{fig:res_compsep_npipe_T_data_select}.  First and foremost, we see that the overall level of coherent systematic residuals in these channels is greatly reduced in \npipe, to the level that all except one can be included in the final analysis.  The only exception is 857-4, which we find would contribute as much to $\chi^2$ as all other \Planck\ channels combined if included in the analysis.  Accounting for the fact that \npipe\ provides single-detector maps also for polarized channels, a total of 36~\Planck\ maps are included in the following \commander\ temperature analysis.  To understand how these improvements are achieved, it is useful to inspect the residual maps in Fig.~\ref{fig:res_compsep_npipe_T_data_select}. Starting with 353-8, the dominant feature is a large-scale red and blue pattern at high Galactic latitudes, extending between the ecliptic poles. This pattern arises when multiple detectors with slightly different bandpasses are destriped simultaneously. Specifically, the destriping (or mapmaking) process intrinsically assumes that there is only one true value within a single sky pixel, and that any deviation from this must therefore be due to noise in the detectors. However, because different detectors have different bandpasses, they also see a slight difference in foreground sky signal. The destriper therefore attempts to suppress this residual sky signal in the same way as actual correlated noise, and effectively ``drags'' the signal along the scan path of the instrument, resulting in the large-scale features seen in Fig.~\ref{fig:res_compsep_npipe_T_data_select}. To solve this problem, one can either destripe each detector independently (after removing a polarization template in the time domain), or fit a set of residual foreground templates in the time domain jointly with the destriping offsets. \npipe\ implements both solutions, the former for the single-detector temperature maps, and the latter for the polarized full-frequency and detector-set maps.
  
A second, and visually striking, effect seen in Fig.~\ref{fig:res_compsep_npipe_T_data_select} is the result of sub-optimal instrumental polarization parameters, in the form of the assumed polarization angle and efficiency of each detector. This is most clearly seen in 353-2, in the form of an alternating blue and red Galactic plane, with amplitudes following the overall polarization amplitude. When either polarization parameter is incorrectly set prior to mapmaking, some of the polarization signal will be interpreted as a temperature signal and vice versa. These residuals are precisely what allows us to perform in-flight determination of the \Planck\
polarization parameters, as described in Sect.~\ref{sec:processing}.

Third, for 857-1 in Fig.~\ref{fig:res_compsep_npipe_T_data_select} we clearly see the imprint of residual sidelobe contamination in the form of two red arcs at high Galactic latitudes. These are greatly suppressed (albeit still visually clear) in \npipe, due to better sidelobe fitting.

Fourth, the effect of improved transfer-function estimation is seen for 857-4 in Fig.~\ref{fig:res_compsep_npipe_T_data_select}. The thick blue region around the Galactic plane is caused by sub-optimal detector transfer functions, which in effect smear out the very bright Galactic signal. Overall, transfer-function estimation is greatly improved in the \npipe\ processing, even though this channel has proved particularly challenging, and is still not deemed sufficiently clean for detailed analysis.

\begin{figure*}[htpb!]
   \centering
      \includegraphics[width=0.24\textwidth]{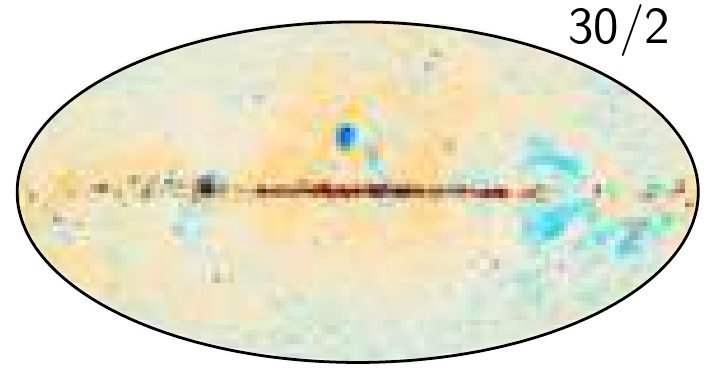}
      \includegraphics[width=0.24\textwidth]{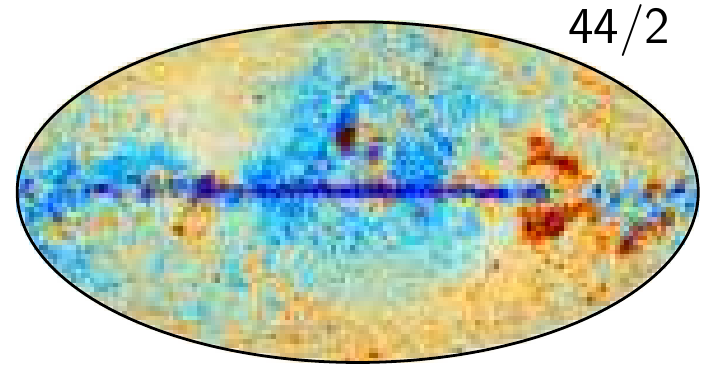}
      \includegraphics[width=0.24\textwidth]{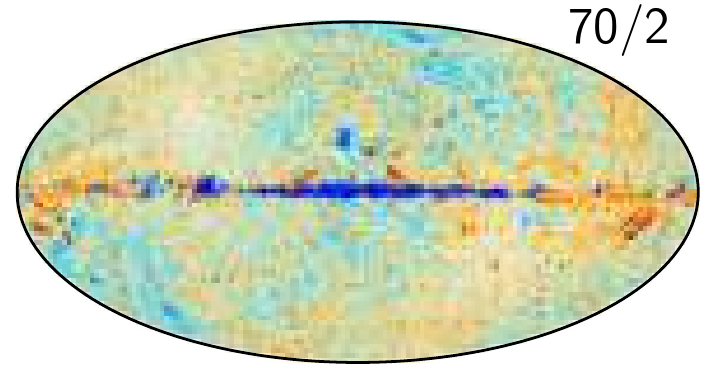}
      \includegraphics[width=0.24\textwidth]{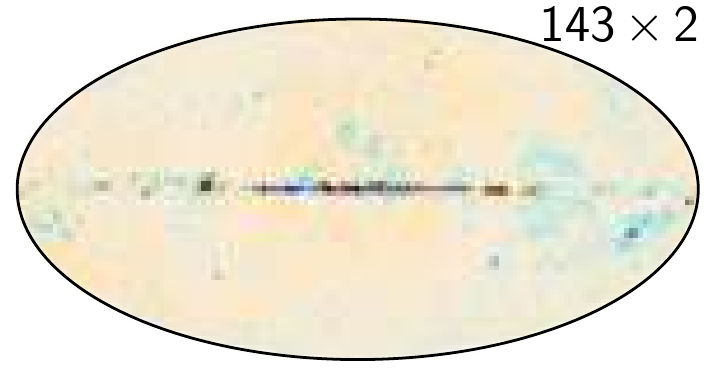}\\
      \includegraphics[width=0.24\textwidth]{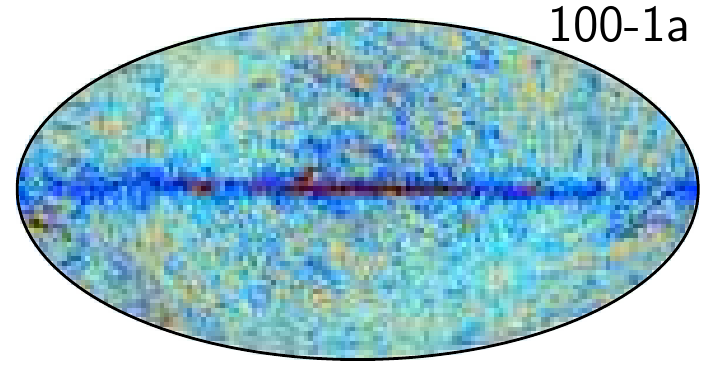}
      \includegraphics[width=0.24\textwidth]{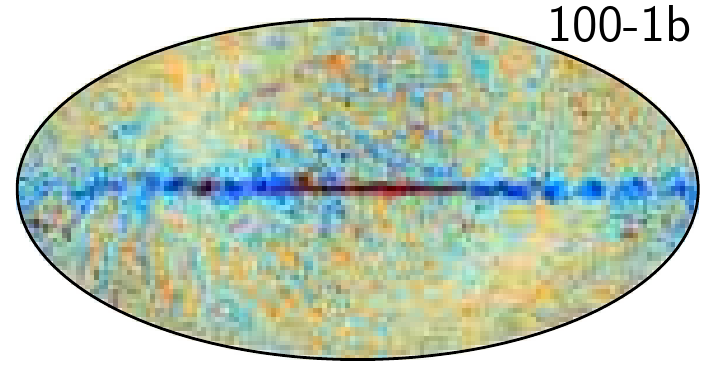}
      \includegraphics[width=0.24\textwidth]{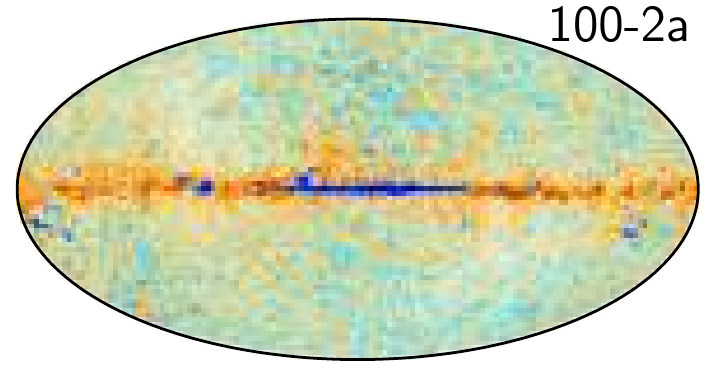}
      \includegraphics[width=0.24\textwidth]{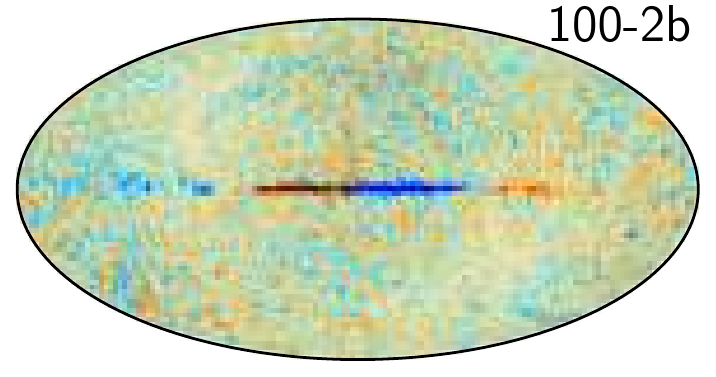}\\
      \includegraphics[width=0.24\textwidth]{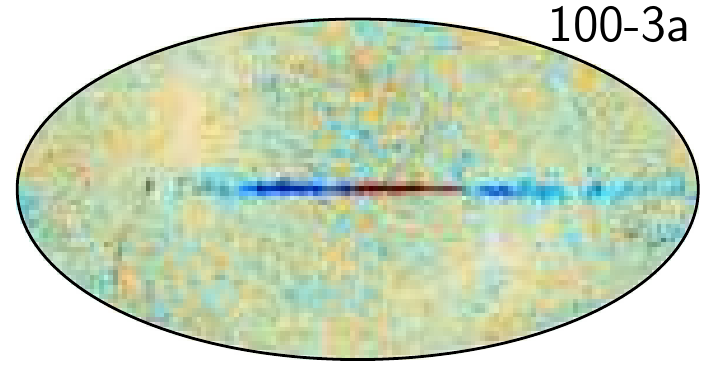}
      \includegraphics[width=0.24\textwidth]{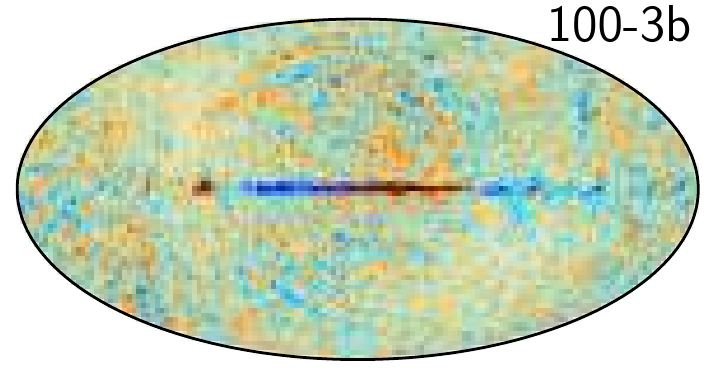}
      \includegraphics[width=0.24\textwidth]{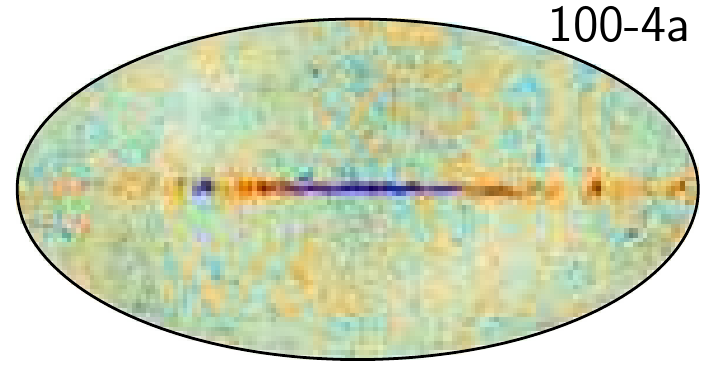}
      \includegraphics[width=0.24\textwidth]{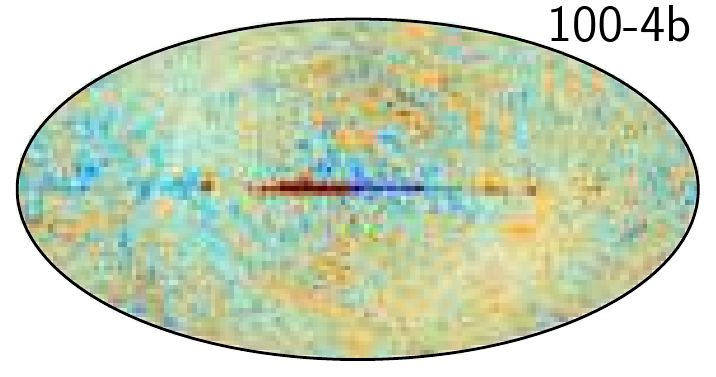}\\
      \includegraphics[width=0.24\textwidth]{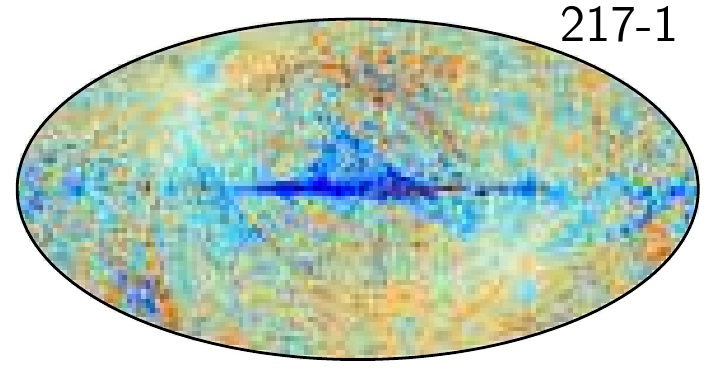}
      \includegraphics[width=0.24\textwidth]{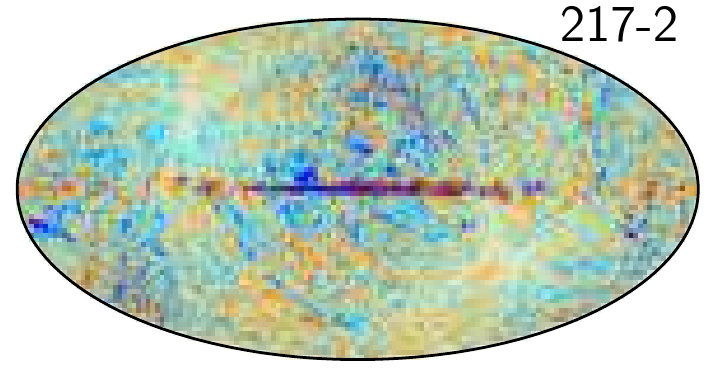}
      \includegraphics[width=0.24\textwidth]{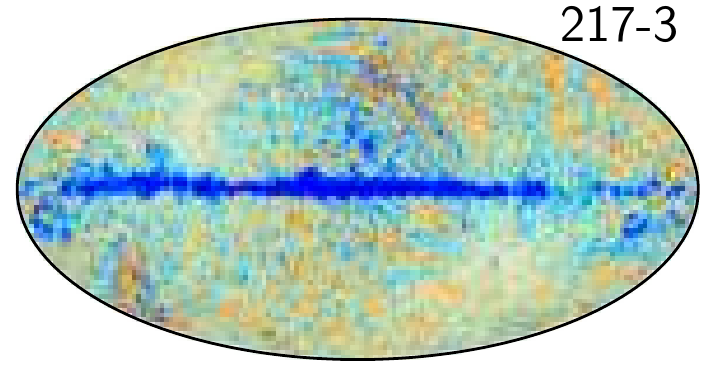}
      \includegraphics[width=0.24\textwidth]{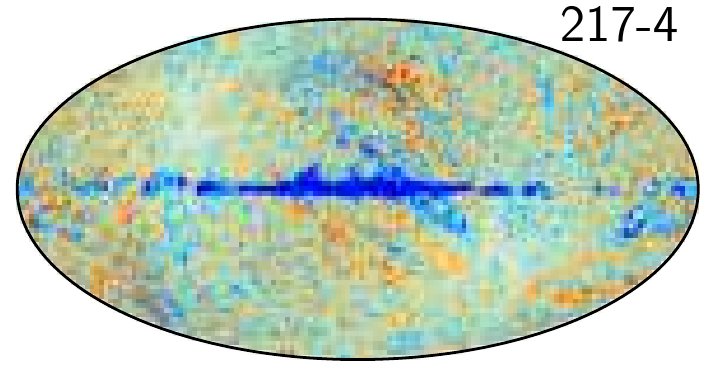}\\
      \includegraphics[width=0.24\textwidth]{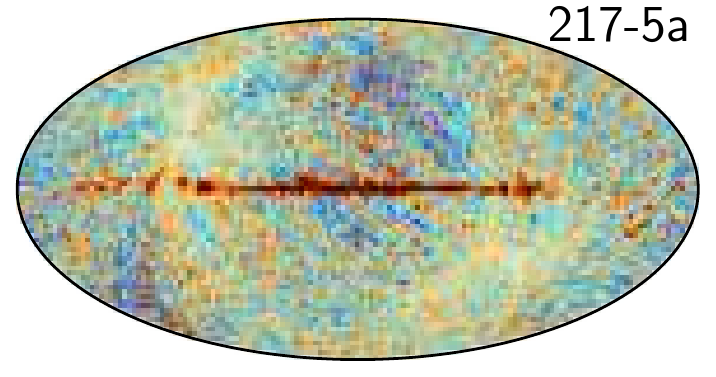}
      \includegraphics[width=0.24\textwidth]{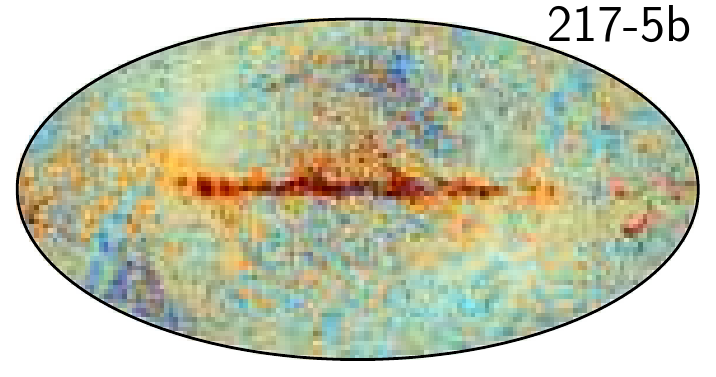}
      \includegraphics[width=0.24\textwidth]{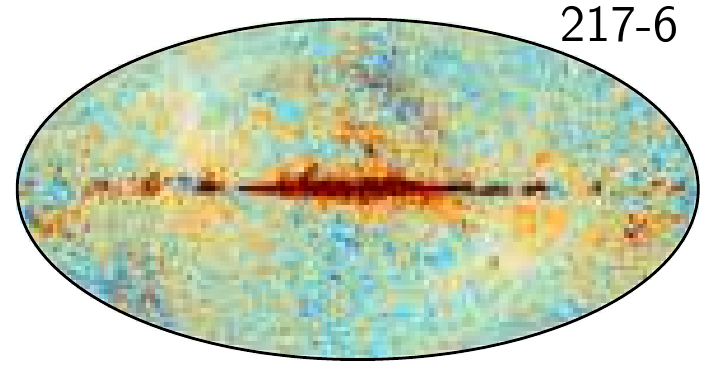}
      \includegraphics[width=0.24\textwidth]{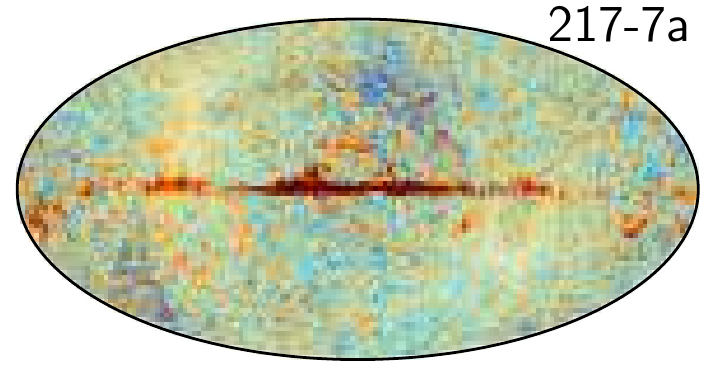}\\
      \includegraphics[width=0.24\textwidth]{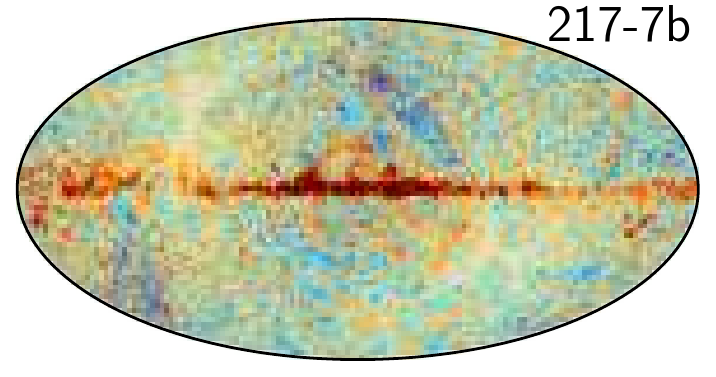}
      \includegraphics[width=0.24\textwidth]{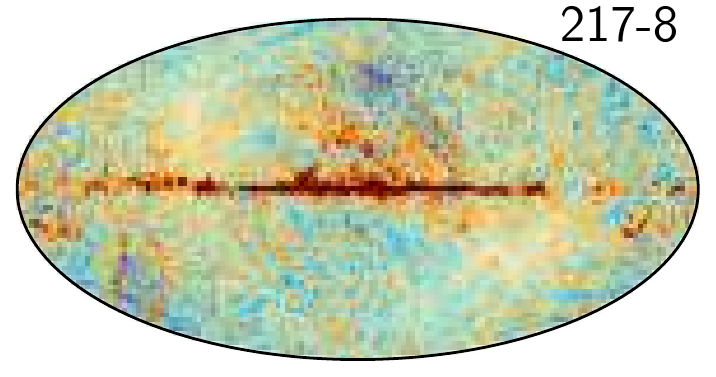}   
      \includegraphics[width=0.24\textwidth]{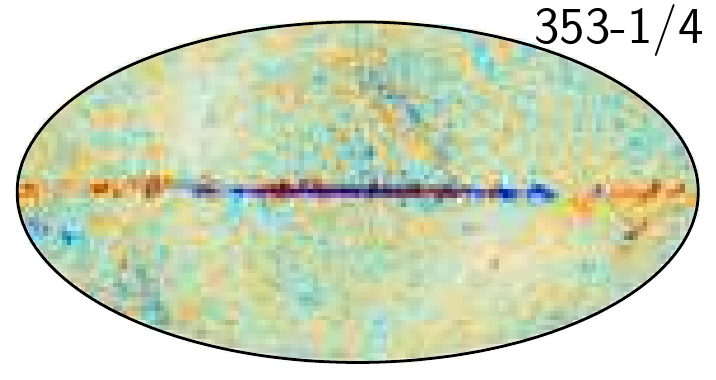}
      \includegraphics[width=0.24\textwidth]{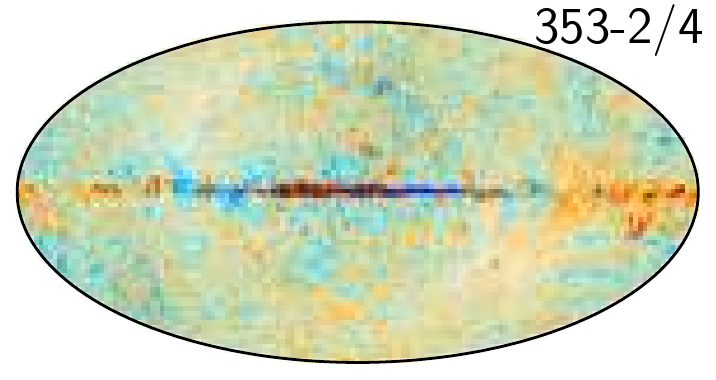}\\
      \includegraphics[width=0.24\textwidth]{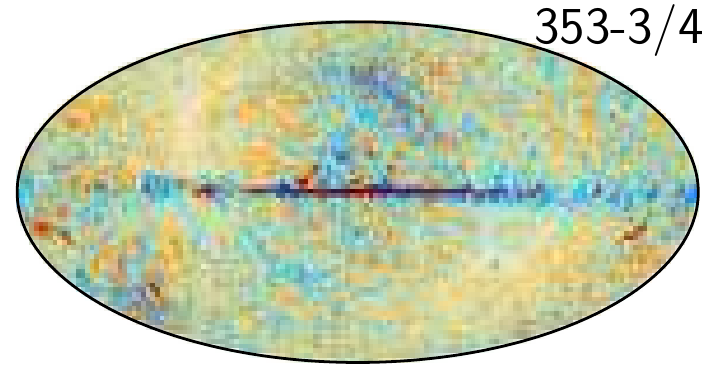}
      \includegraphics[width=0.24\textwidth]{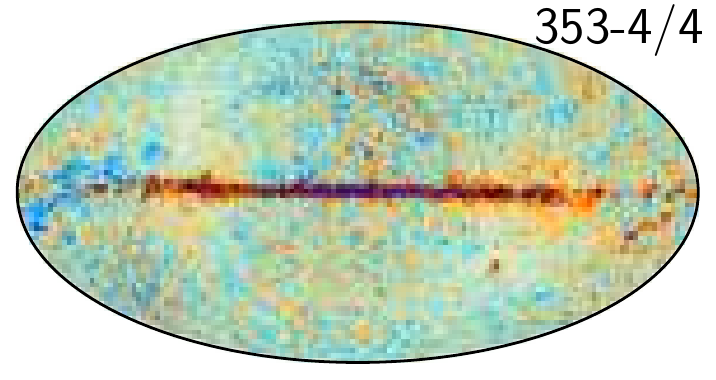}
      \includegraphics[width=0.24\textwidth]{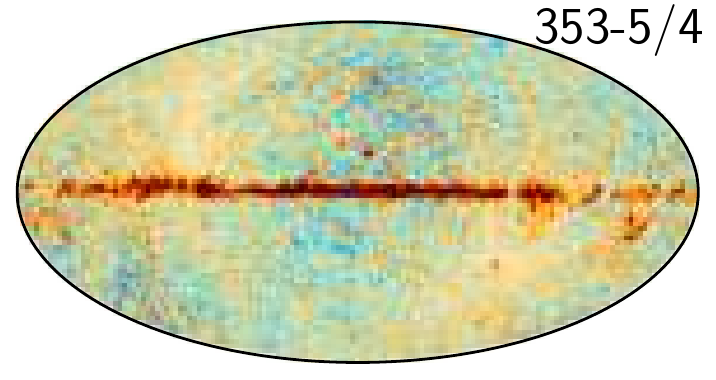}
      \includegraphics[width=0.24\textwidth]{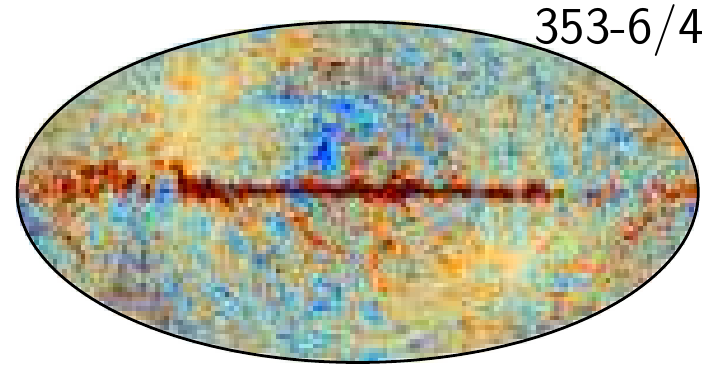}\\
      \includegraphics[width=0.24\textwidth]{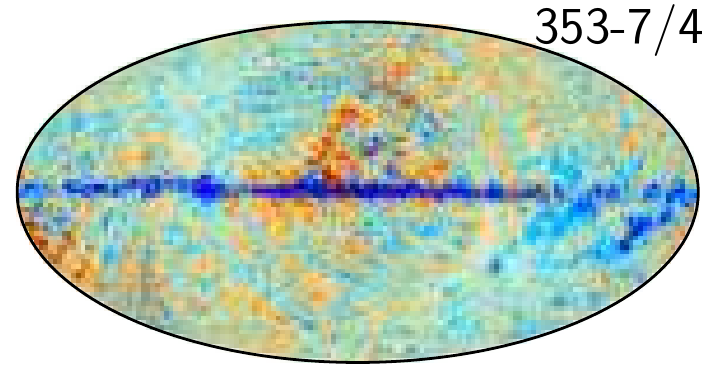}
      \includegraphics[width=0.24\textwidth]{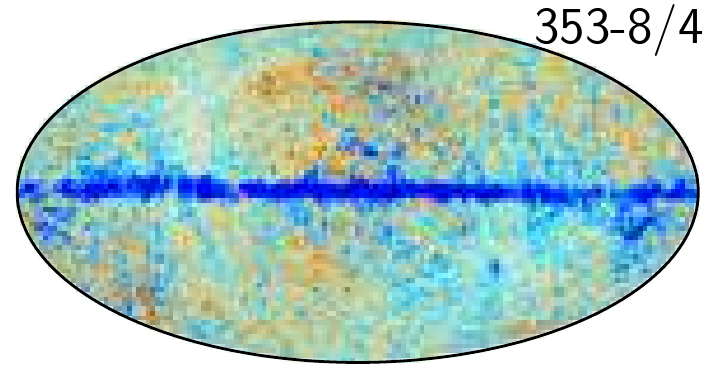}
      \includegraphics[width=0.24\textwidth]{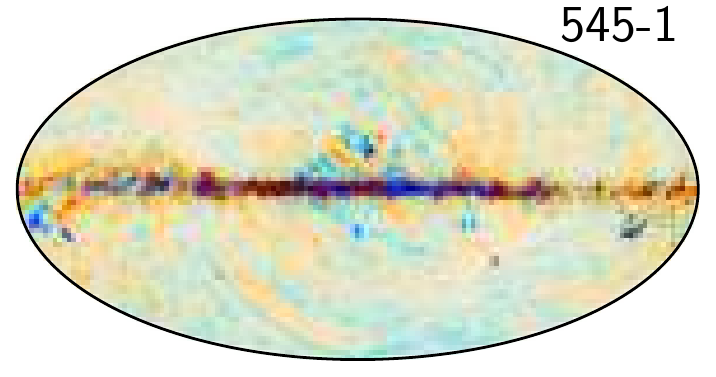}
      \includegraphics[width=0.24\textwidth]{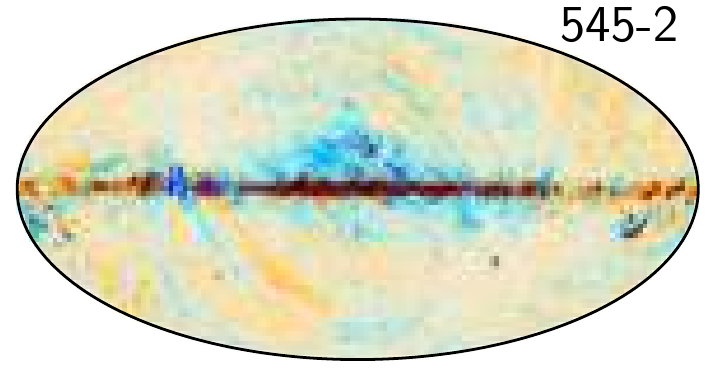}\\
      \includegraphics[width=0.24\textwidth]{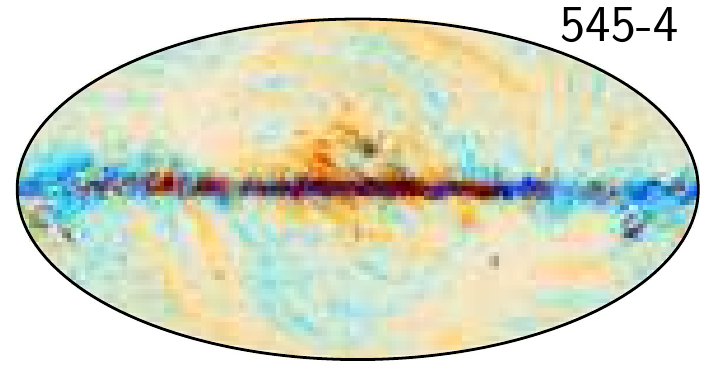}
      \includegraphics[width=0.24\textwidth]{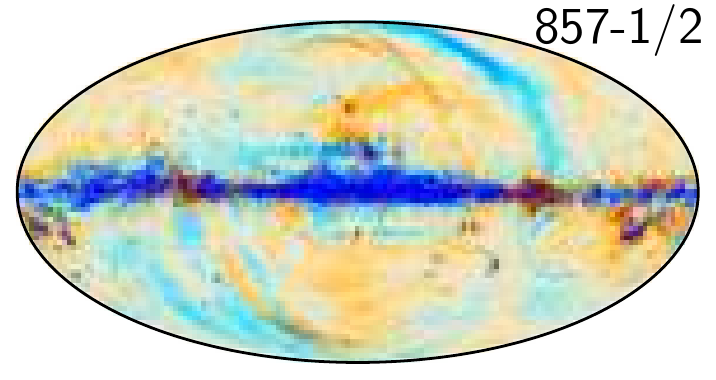}
      \includegraphics[width=0.24\textwidth]{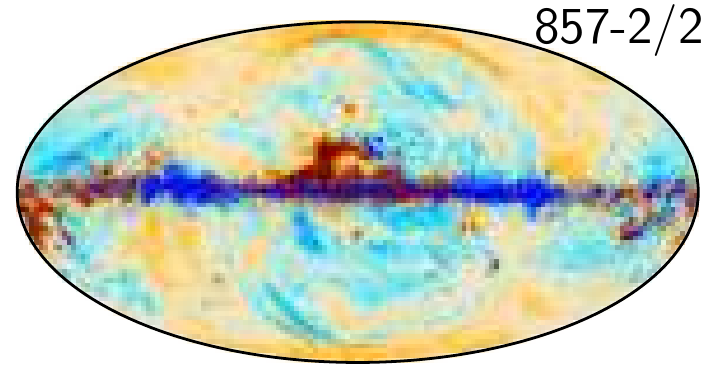}
      \includegraphics[width=0.24\textwidth]{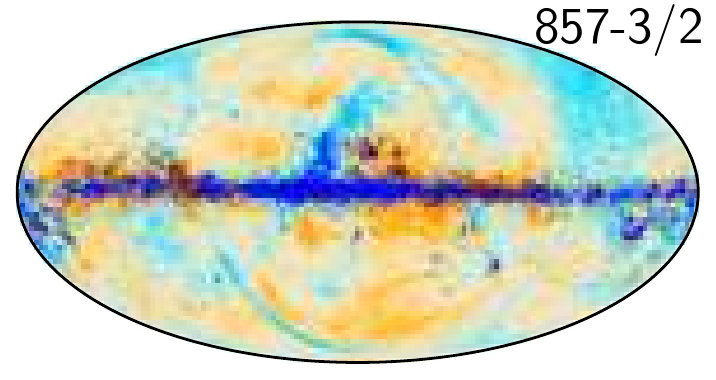}\\
      \includegraphics[width=0.55\textwidth]{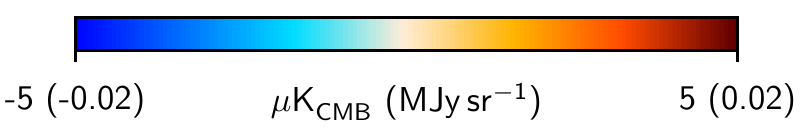}

\caption{\Planck-only \npipe\ residual maps for frequency bands and individual detectors.  Maps up to 353\GHz\ are plotted in $\mathrm{\mu K_{\small{CMB}}}$, while maps at 545 and 857\GHz\ are plotted in $\mathrm{MJy\,sr}^{-1}$.  Maps with a fraction in the label have been divided with respect to the colour bar, while the 143\GHz\ band has been multiplied by a factor of 2.  All maps are smoothed to 60\arcm\ FWHM.}
\label{fig:res_compsep_npipe_T_LFI_217}
\end{figure*}

Figure~\ref{fig:res_compsep_npipe_T_LFI_217} shows data-minus-signal residual maps for the 36 \npipe\ maps included in the \commander\ temperature analysis.  Even though some coherent structures are visible, for frequencies below 545\GHz\ these are visually dominated by white noise at high Galactic latitudes. Starting with the 44-GHz channel, the most striking features are patterns consistent with Galactic residuals, some bright red with free-free-like morphology and some faint blue with the morphology of dust emission. These are due to the fact that only a single low-frequency component is fitted to these \Planck-only data, and a single component cannot account separately for synchrotron, free-free, and spinning dust emission. For 100-1a, the dominant feature is a roughly 1-$\mu$K negative monopole, resulting from the fact that the offset for this particular channel is not fitted freely in the analysis, but rather determined by enforcing vanishing CMB and CO zero-level at high Galactic latitudes. 

For 100-1b, the most notable features are a slight red CO residual in the central Galactic plane, countered by a faint blue residual with thermal-dust-like morphology extending into the Galactic wings. These structures suggest that CO is not perfectly modelled in the current setup.  One possible explanation may be the presence of multiple CO isotopes (i.e., $^{12}$CO and $^{13}$CO) in this band, while we only fit for one overall component in our analysis.  Extensions of this model will be considered in the future.  A second possibility is residual thermal dust temperature uncertainties.  In any case, this model failure appears to bias the bandpass corrections very slightly, resulting in a slightly negative thermal-dust imprint outside the Galactic centre.  This particular case illustrates very directly the importance of joint global component separation and instrumental characterization, as implemented in \npipe, as well as the importance of employing a complete physical model for this task. 

A third model failure of similar type is seen in the 100-2b map, in the form of sharp blue and red residuals on either side of the
Galactic centre. These are due to the rotation of the Milky Way, resulting in a relative blue- or red-shift of the CO lines. Since the
derivative of the detector bandpasses at the CO centre frequency is not zero, an effective shift in the apparent line ratio results. This effect was already noted in the \Planck\ 2015 data set \citep{planck2014-a12}, but with the improved \npipe\ processing, it is now the dominant signal for several channels. 

For channels below 217\GHz, sky-model failures dominate over instrumental effects, whether they come in the form of residual
free-free, spinning dust, or CO features. Resolving these will require both a more fine-grained astrophysical model and combination with external data sets. This is beyond the scope of the current paper, and will be addressed in future publications.

At frequencies above 143\GHz, the picture is more complicated.  Here we see many features that have a clear instrumental signature, most of which have already been discussed qualitatively.  Thus, even though the \npipe\ data set exhibits lower systematic uncertainties than previous \Planck\ releases, the data are clearly still not consistent with white noise.  It is important to bear these residuals in mind in subsequent analyses, and analysing realistic simulations is important to quantify the resulting uncertainties.  On the other hand, in terms of their potential effect on cosmology, one should bear in mind that these residuals are at a low level compared to the CMB signal, especially since they will average to an even lower level in the combination of maps

\begin{figure*}[htpb!]
   \centering
   \includegraphics[width=0.22\linewidth]{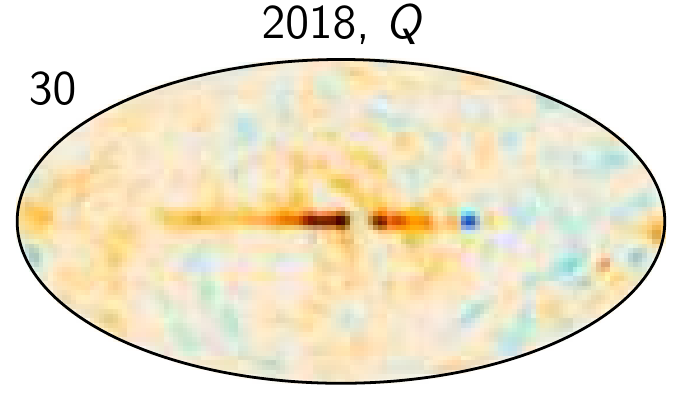}
   \includegraphics[width=0.22\linewidth]{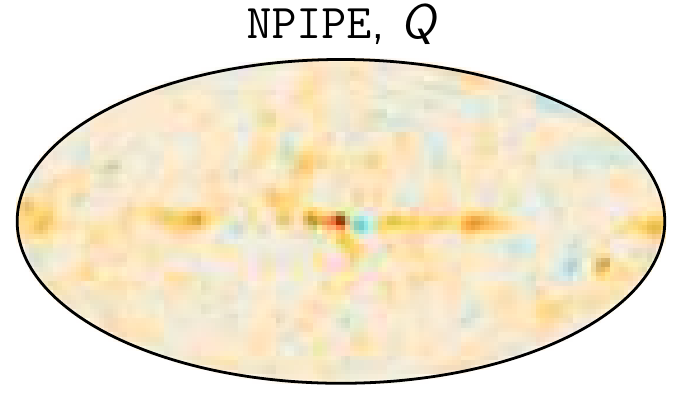}\hspace*{7mm}
   \includegraphics[width=0.22\linewidth]{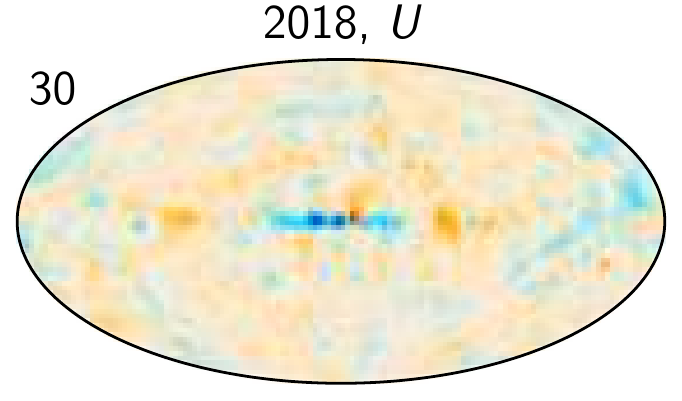}
   \includegraphics[width=0.22\linewidth]{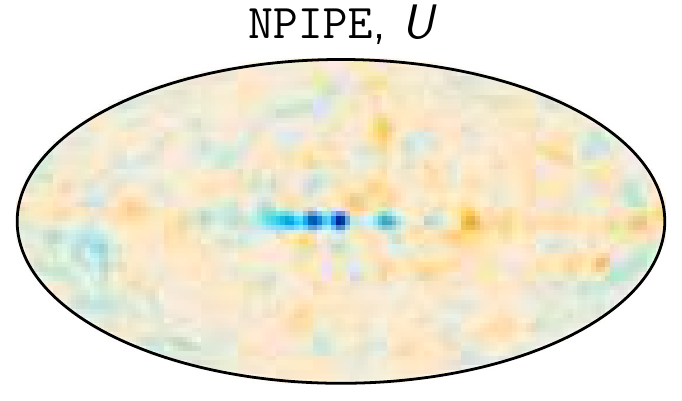}\\
   \includegraphics[width=0.22\linewidth]{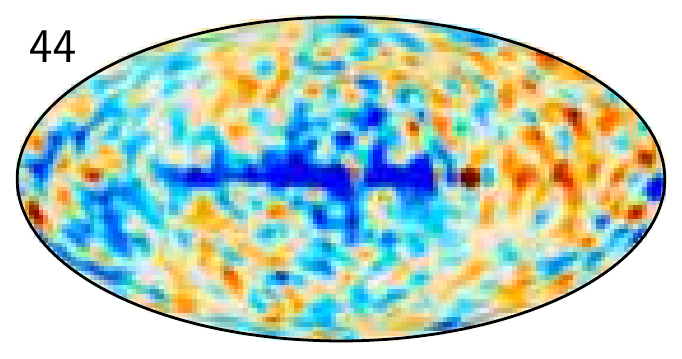}
   \includegraphics[width=0.22\linewidth]{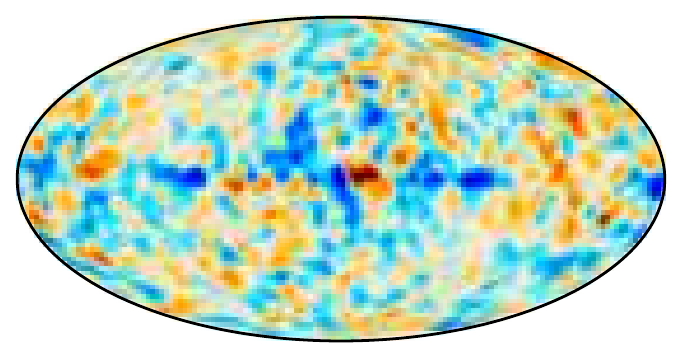}\hspace*{7mm}
   \includegraphics[width=0.22\linewidth]{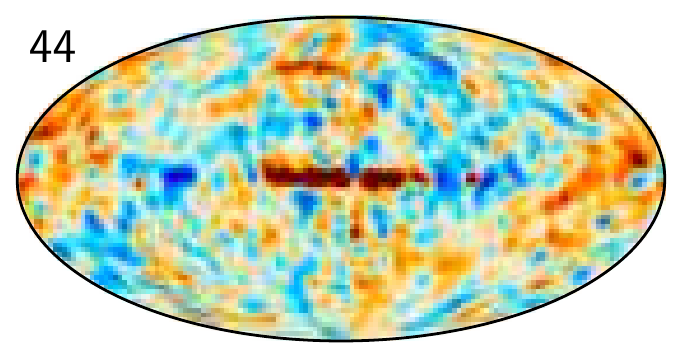}
   \includegraphics[width=0.22\linewidth]{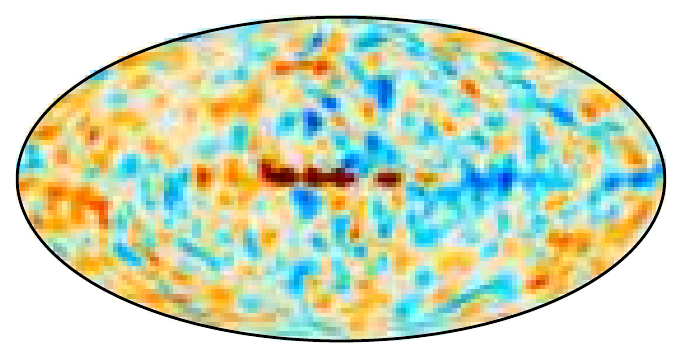}\\
   \includegraphics[width=0.22\linewidth]{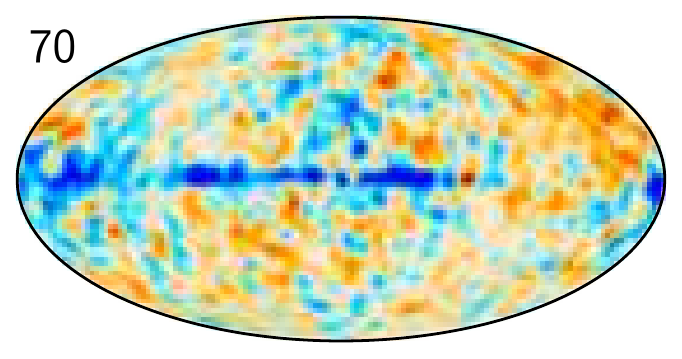}
   \includegraphics[width=0.22\linewidth]{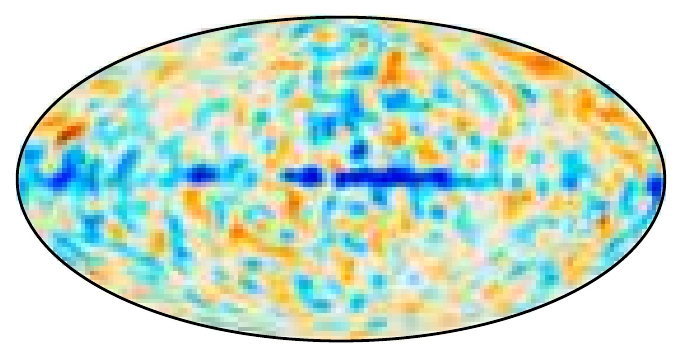}\hspace*{7mm}
   \includegraphics[width=0.22\linewidth]{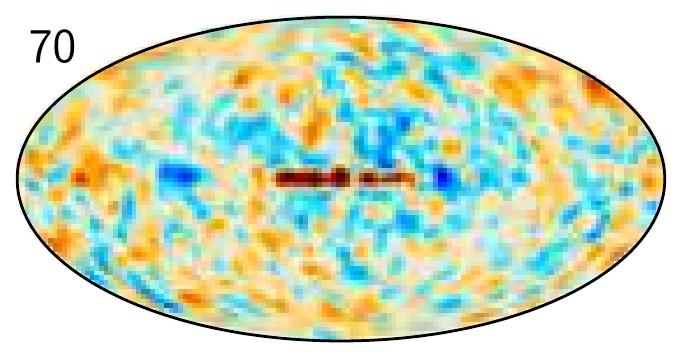}
   \includegraphics[width=0.22\linewidth]{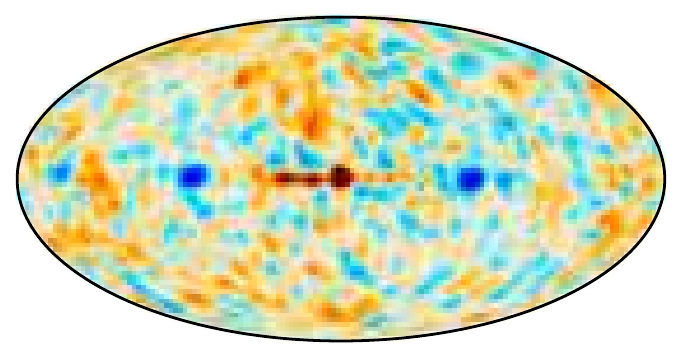}\\
   \includegraphics[width=0.22\linewidth]{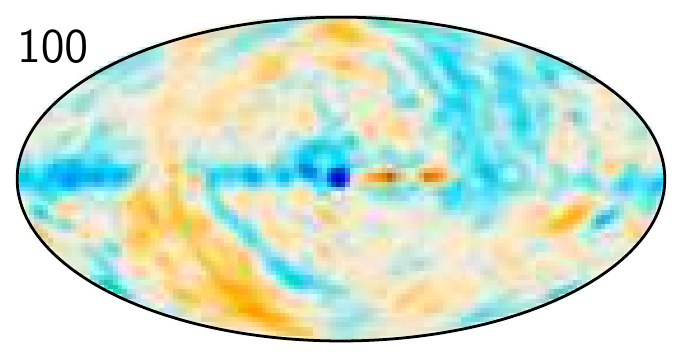}
   \includegraphics[width=0.22\linewidth]{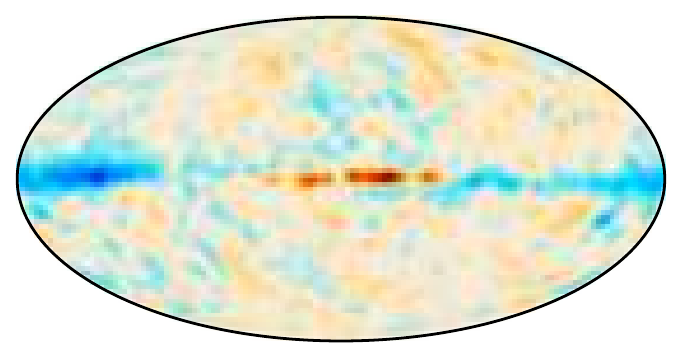}\hspace*{7mm}
   \includegraphics[width=0.22\linewidth]{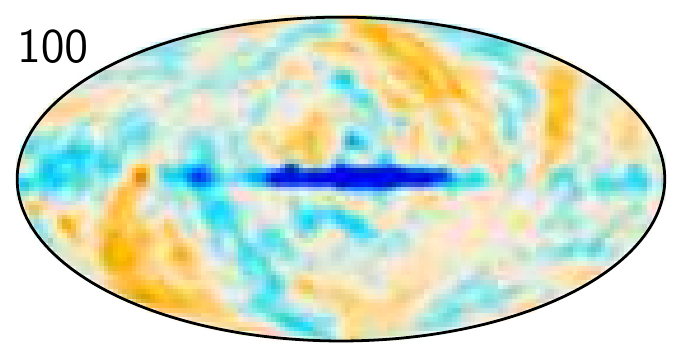}
   \includegraphics[width=0.22\linewidth]{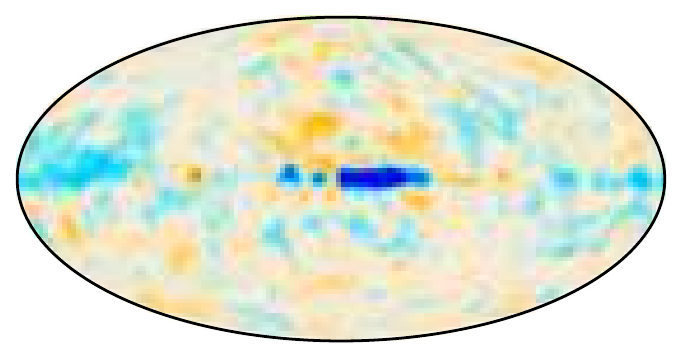}\\
   \includegraphics[width=0.22\linewidth]{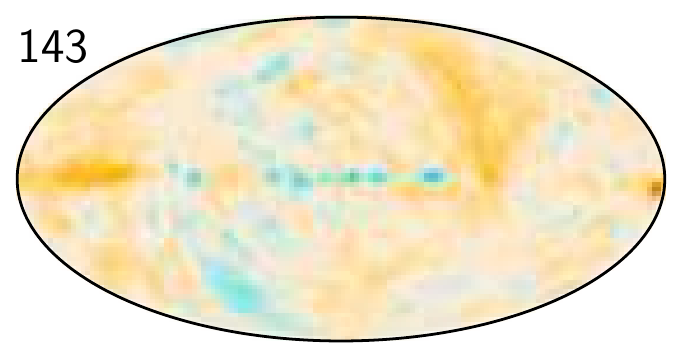}
   \includegraphics[width=0.22\linewidth]{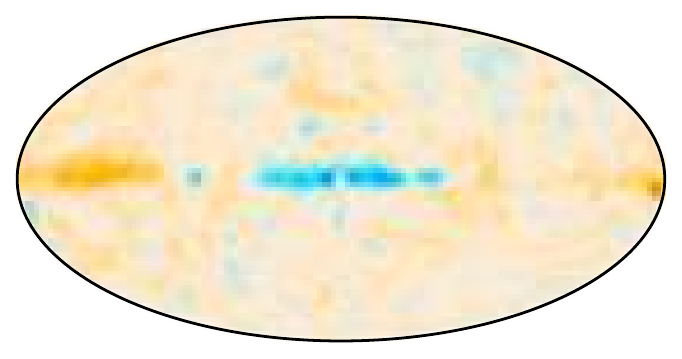}\hspace*{7mm}
   \includegraphics[width=0.22\linewidth]{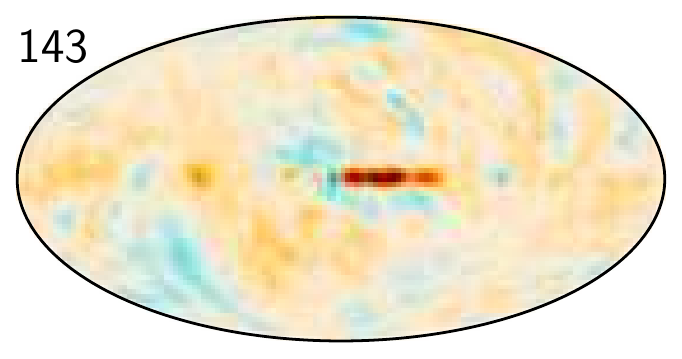}
   \includegraphics[width=0.22\linewidth]{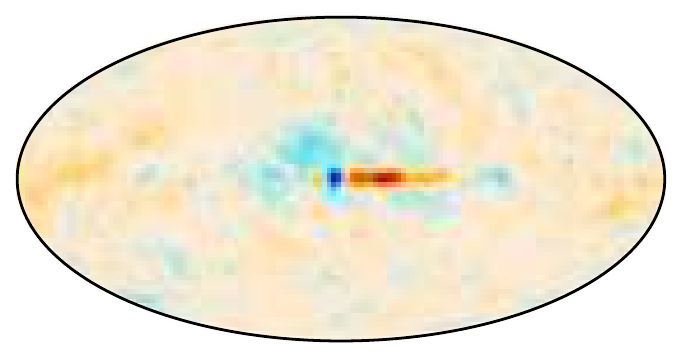}\\
   \includegraphics[width=0.22\linewidth]{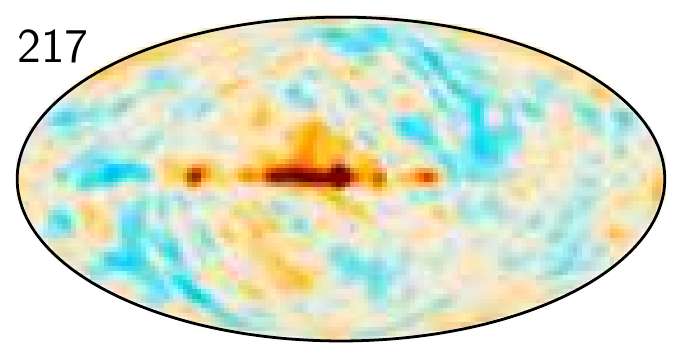}
   \includegraphics[width=0.22\linewidth]{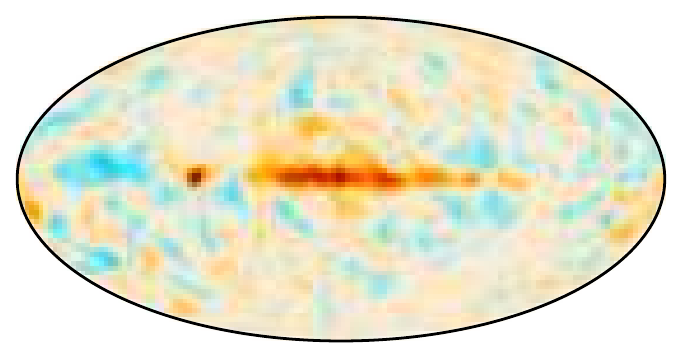}\hspace*{7mm}
   \includegraphics[width=0.22\linewidth]{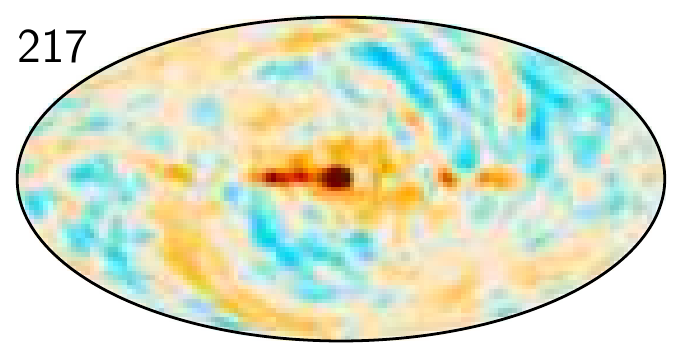}
   \includegraphics[width=0.22\linewidth]{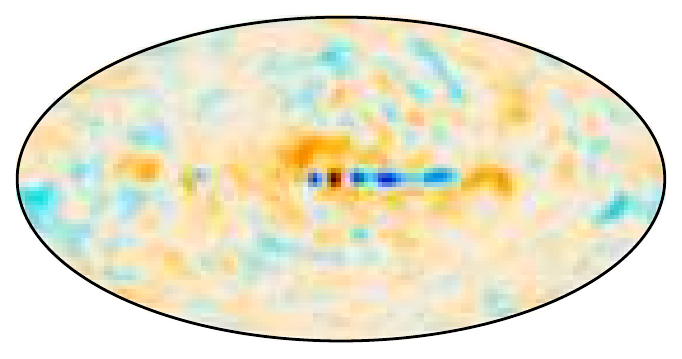}\\
   \includegraphics[width=0.22\linewidth]{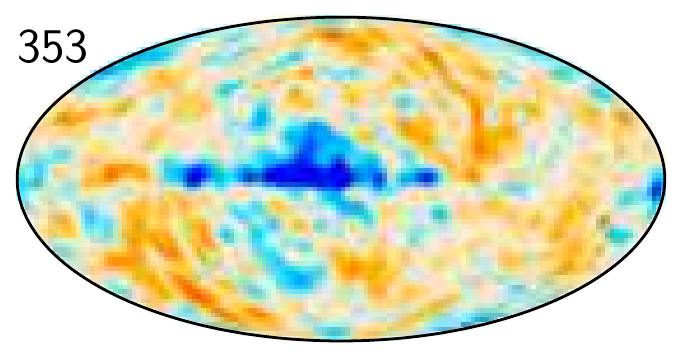}
   \includegraphics[width=0.22\linewidth]{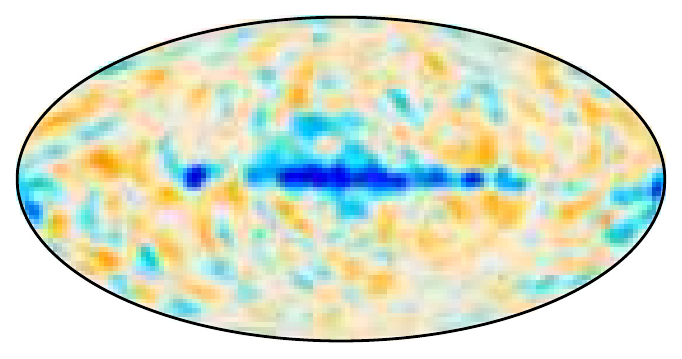}\hspace*{7mm}
   \includegraphics[width=0.22\linewidth]{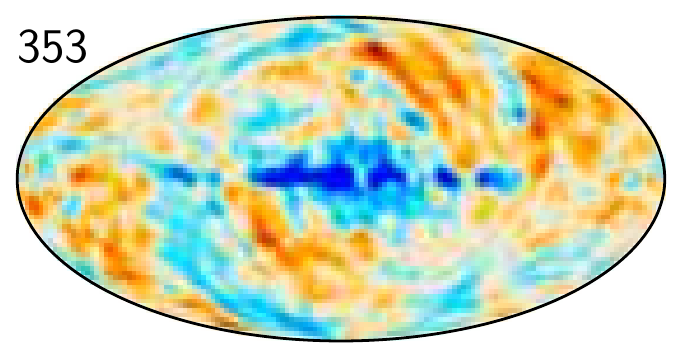}
   \includegraphics[width=0.22\linewidth]{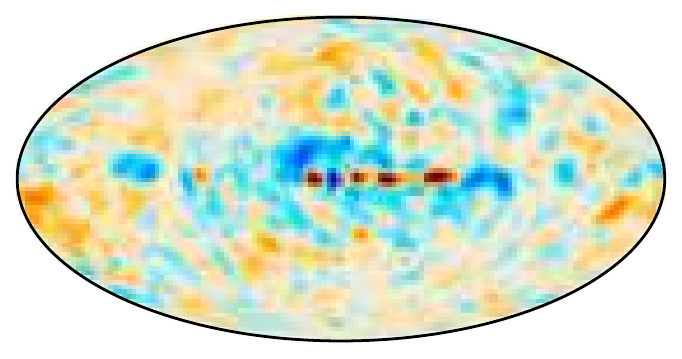}\\
   \includegraphics[width=0.48\linewidth]{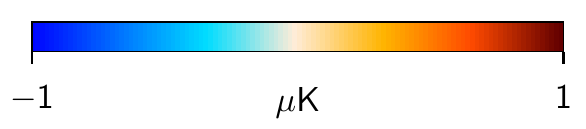}
   \caption{Comparison between \commander\ residual maps, $\d_{\nu}-\s_{\nu}$, for each polarized \Planck\ frequency map between (from top to bottom) 30 and 353\GHz. The two leftmost columns show residual maps for Stokes $Q$ for \Planck\ 2018 and \npipe, while the two rightmost columns show the same for Stokes $U$.  All maps have been smoothed to a common angular resolution of 3\deg\ FWHM. }
   \label{fig:pol_residuals}
\end{figure*}

Figure~\ref{fig:pol_residuals} shows similar residual maps from the polarization analysis, compared with the corresponding \Planck\ 2018 residuals. The single most striking difference between these two data sets is the lower level of coherent and scan-aligned large-scale features at high Galactic latitudes. This is primarily due to improved calibration in \npipe, resulting in lower levels of dipole leakage into the polarization sector.

\subsection{Global parameters}
\label{sec:global_parameters}

We start our discussion of the derived sky model by considering global parameters computed by \commander\ and listed in
Table~\ref{tab:monopole}.  For each sky map included in the analysis, three values are listed: monopole (or offset); calibration factor; and bandpass shift.  Values that are kept fixed during a \commander\ run are marked with a superscript ``a.''   Note, however, that the fixed monopoles are adjusted between preliminary and final runs, as described in Sect.~\ref{sec:methodology}.

\begin{table*}[htbp!] 
\begingroup 
\newdimen\tblskip \tblskip=5pt
\caption{Monopoles, calibration factors, and bandpass corrections derived within the baseline temperature model
\label{tab:monopole}.}
\nointerlineskip
\vskip -2mm
\footnotesize 
\setbox\tablebox=\vbox{
\newdimen\digitwidth 
\setbox0=\hbox{\rm 0}
\digitwidth=\wd0
\catcode`*=\active
\def*{\kern\digitwidth}
\newdimen\signwidth
\setbox0=\hbox{+}
\signwidth=\wd0
\catcode`!=\active
\def!{\kern\signwidth}
\newdimen\decimalwidth
\setbox0=\hbox{.}
\decimalwidth=\wd0
\catcode`@=\active
\def@{\kern\decimalwidth}
\def\s#1{\ifmmode $\rlap{$^{\rm #1}$}$ \else \rlap{$^{\rm #1}$}\fi}
\halign{\hbox to 0.6in{#\leaderfil}\tabskip=0.1em&
    \hfil#\hfil\tabskip=2em&
    \hfil$#$\hfil\tabskip=2em&
    \hfil$#$\hfil\tabskip=2em&
    \hfil$#$\hfil\tabskip=0pt\cr
\noalign{\doubleline}
 Frequency& Detector&\omit\hfil Monopole\hfil&\omit\hfil Calibration\hfil&\omit\hfil Bandpass shift\hfil\cr
 **[GHz]**& label& [\muK]& [\%]&\omit\hfil[GHz]\hfil\cr                                                                                                                                                                                               
 \noalign{\vskip 5pt\hrule\vskip 5pt}
          *30&\dots&      !*773@*\pm 5\s{a}@*&         -0.75\pm0.07& ! 0.4\pm0.1\cr
          *44&\dots&     !1187.3\pm0.3&     -0.40\pm0.08&  -0.6\pm0.1\cr
          *70&\dots&      !**70.8\pm0.3&     -0.09\pm0.01&  -1.1\pm1.0\cr
          100&    1a&     **-68.9\pm1\s{a}@*&     -0.08\pm0.03&    !0.1\pm0.8\cr
        \omit&    1b&     **-70.0\pm0.1&     -0.07\pm0.02&    -0.4\pm0.8\cr
        \omit&    2a&     **-69.4\pm0.1&     -0.07\pm0.02&    -2.1\pm0.8\cr
        \omit&    2b&     **-69.5\pm0.1&     -0.08\pm0.02&    -1.2\pm0.3\cr
        \omit&    3a&     **-69.7\pm0.1&     -0.07\pm0.02&    -1.7\pm0.4\cr
        \omit&    3b&     **-69.9\pm0.1&     -0.06\pm0.02&    -0.7\pm0.2\cr
        \omit&    4a&     **-69.3\pm0.1&     -0.08\pm0.02&    -2.2\pm0.4\cr
        \omit&    4b&     **-69.6\pm0.1&     -0.08\pm0.02&    -1.4\pm0.3\cr
          143&\dots&     **-81.1\pm2\s{a}@*&  0\s{a}&        !0.4\pm0.1\cr
          217&      1&    *-182.6\pm0.2&     -0.01\pm0.01&    0.0\s{a}\cr
        \omit&      2&     *-182.4\pm 2\s{a}@*&     -0.02\pm0.01&    -0.3\pm0.1\cr
        \omit&      3&     *-182.8\pm0.2&     -0.03\pm0.01&    !0.4\pm0.1\cr
        \omit&      4&     *-182.6\pm0.2&     -0.02\pm0.01&    !0.1\pm0.1\cr
        \omit&    5a&     *-182.2\pm0.2&     -0.03\pm0.01&    !0.3\pm0.1\cr
        \omit&    5b&     *-182.2\pm0.2&     -0.04\pm0.01&    !0.4\pm0.1\cr
        \omit&      6&     *-182.1\pm0.2&     -0.01\pm0.01&    !0.6\pm0.1\cr
        \omit&    7a&     *-182.1\pm0.2&     -0.02\pm0.01&    -0.2\pm0.1\cr
        \omit&    7b&     *-182.2\pm0.2&     -0.03\pm0.01&    !0.2\pm0.1\cr
        \omit&      8&     *-182.1\pm0.2&     -0.02\pm0.01&    !0.0\pm0.1\cr
          353&      1&     !*395.2\pm1.0&    ! 0.27\pm0.08&    !0.2\pm0.1\cr
        \omit&      2&     !*394.6\pm1.0&    ! 0.27\pm0.08&    -0.4\pm0.1\cr
        \omit&      3&      !*394@*\pm 4\s{a}@*&    ! 0.23\pm0.08&    !0.0\pm0.1\cr
        \omit&      4&  !*   394.7\pm1.0&    ! 0.24\pm0.08&    !0.3\pm0.1\cr
        \omit&      5&  !*   397.8\pm1.0&    ! 0.24\pm0.08&    -0.2\pm0.1\cr
        \omit&      6&  !*   398.2\pm1.0&    ! 0.20\pm0.08&    -0.1\pm0.1\cr
        \omit&      7&  !*   386.1\pm1.0&    ! 0.26\pm0.08&    -0.5\pm0.1\cr
        \omit&      8&  !*   383.8\pm1.0&    ! 0.26\pm0.08&    -0.2\pm0.1\cr
          545&      1&  -5003@*\pm30@&       *1.5\pm0.3&     -0.4\pm0.1\cr 
        \omit&      2&  @-4891@*\pm852\s{a}&  *1.5\pm0.3&     !0.4\pm0.1\cr
        \omit&      4&   -4951@*\pm30@&        *2.4\pm0.3&     !0.0\pm0.1\cr
          857&      1& !** -0.72\pm0.01 \s{b}&      *4.7\pm0.5&     !0.0\s{c}\cr
        \omit&      2& ! **-0.72\pm0.01 \s{b}&     16.1\pm0.6&     !0.0\s{c}\cr
        \omit&      3& ! **-0.72\pm0.01 \s{b}&     *7.9\pm0.5&     !0.0\s{c}\cr
\noalign{\vskip 5pt\hrule\vskip 5pt}
}}
\endPlancktablewide

\tablenote {{\rm a}} Fixed at reference value. \par
\tablenote {{\rm b}} The units for 857\GHz\ are kelvin. \par
\tablenote {{\rm c}} Fixed to the input value due to strong degeneracies with calibration. \par
\endgroup
\end{table*}

Starting with the monopoles, the reported values are numerically large, and do not correspond directly to typical values reported for previous data sets. In this respect, we note that \Planck\ is not meaningfully sensitive to the zero-level of any frequency channel. Accordingly, the \npipe\ analysis pipeline makes no attempt to adjust the raw outputs from the
\madam\ mapmaker (from which the monopoles are spurious), but rather leaves this task to the component-separation stage, which is far better equipped to solve this problem, exploiting both physical priors and correlations between frequency channels. We therefore recommend subtraction of the values reported in Table~\ref{tab:monopole} prior to any application that is sensitive to the absolute zero-level of the maps.

Regarding the calibration coefficients, the CMB Solar dipole is, for the first time in \Planck\ data, retained in all sky maps. This provides a very bright relative calibration target that allows for high-precision calibration and component separation.  The 143-GHz calibration factor is fixed to unity in these analyses, and all other values are therefore in effect defined relative to this channel.  For the 30- and 44-GHz data, we find calibration corrections of $-0.75$ and $-0.40$\,\%, respectively.  These channels have relatively bright foregrounds combined with lower signal-to-noise ratios, and are therefore more susceptible to potential foreground residuals during calibration.  In particular, we repeat that the foreground model used in this \Planck-only analysis relies on a single joint low-frequency component, as opposed to separate synchrotron, free-free, and spinning dust components.

We observe excellent relative calibration among the 70- and 100-GHz sky maps, with an overall shift of about $-0.08$\,\% relative to 143\GHz, and internal variations smaller than 0.02\,\%.  The same holds true at 217\GHz, for which the overall shift is 0.02\,\%, and the scatter is about the same.  For the 353-GHz channels, where the thermal dust foregrounds become significantly brighter, we find an overall shift of about 0.25\,\%, with a scatter of about 0.05\,\%.

\npipe\ represents the first \Planck\ processing for which the 545-GHz channel is calibrated with the CMB orbital
dipole along with the lower-frequency channels.  After component separation, these estimates appear accurate with a precision of about 2\,\%.

We obtain very large correction factors for the 857-GHz channels, with values up to 16\,\%. The reason for this is that
the \npipe\ processing natively adopts the same planet-based calibration procedure as the 2018 DPC processing, defined in
units of MJy\,sr$^{-1}$, but subsequently converts these to thermodynamic units of $K_{\textrm{CMB}}$.  The primary advantage of this conversion is that all maps then have the same units.  The main disadvantage is that the conversion factor between flux density and thermodynamic units is highly sensitive to small variations and uncertainties in the shape of the bandpass. This is particularly striking for the 857-2 channel, with its total correction factor of 16.1\,\%. 

The bandpass corrections listed in the rightmost column of Table~\ref{tab:monopole} are generally small. The largest numerical values are observed within the 100-GHz channel, for which a mean negative shift of about 1\GHz\ is seen. This corresponds to a reduction in the effective thermal-dust-emission amplitude of about 1.5\,\% in this channel, and an increase in the free-free emission of about 2\,\%, compared to the nominal bandpasses. Both effects are relatively small in an absolute sense, but still statistically significant given the very high signal-to-noise ratio of the \Planck\ observations.

For the 70-GHz channel, we observe an average bandpass shift of about ($-1.1\pm1.0$)\GHz. For comparison, a value of ($0.4\pm1.0$)\GHz\ was found in the corresponding \Planck+WMAP analysis presented in \citet{planck2014-a10}.  The \Planck\
70-GHz bandpass shifts are associated with a large uncertainty for three main reasons.  First, the 70-GHz bandpasses were measured with lower accuracy on the ground than other detectors, and significant power is missing from the tails of the bandpass profiles (see figure~18 of \citealt{villa2010}).  Second, the foreground minimum occurs close to 70\GHz\ \citep{planck2014-a10}, leaving a relatively faint calibration target for the bandpass shift parameters. Third, the strongest foreground components -- thermal dust, synchrotron, and free-free -- are all about equally strong around 70\GHz.  Modelling
errors and internal degeneracies are therefore significant for this channel. The net result of these three effects is a large overall uncertainty, in which the modelling aspects dominate the error budget.

We conclude this section with a brief discussion of uncertainties.  Because of \Planck's high S/N, these are dominated by systematic contributions.  For example, the conditional statistical uncertainty on the 100-1a monopole from instrumental noise alone is about 10\,nK.  This contribution is completely negligible compared to foreground modelling uncertainties and instrumental systematics.  Deriving a statistically rigorous estimate for these uncertainties is therefore complicated. However, as a useful approximation, we exploit the fact that the \npipe\ data set has undergone hundreds of larger or smaller variations during its development phase, where each variation has included different processing features and foreground modelling.  The uncertainties quoted in Table~\ref{tab:monopole} represent the typical variations observed among the two most mature \npipe\ versions.  The uncertainty in the 545-2 monopole is a special case however, being derived from a linear correlation with H{\scriptsize I} observations.

\subsection{CMB maps and power spectra}

\begin{figure}[htpb!]
   \centering
   \includegraphics[width=0.48\textwidth]{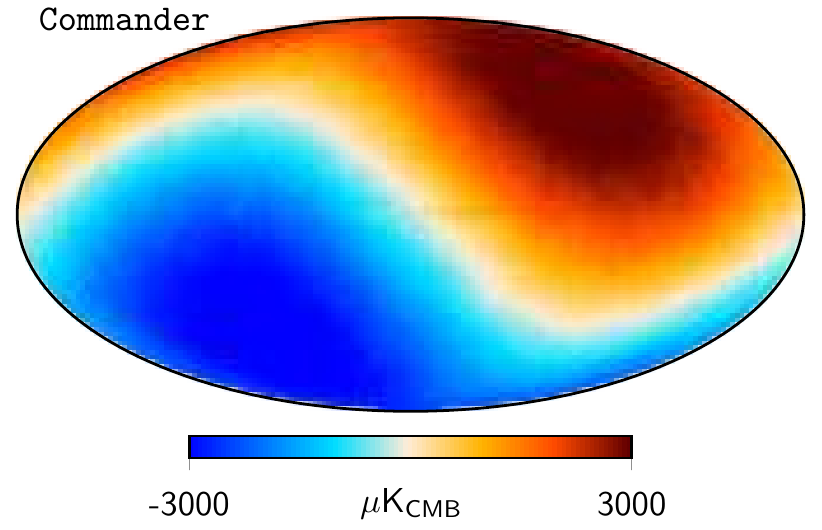}\\
   \includegraphics[width=0.48\textwidth]{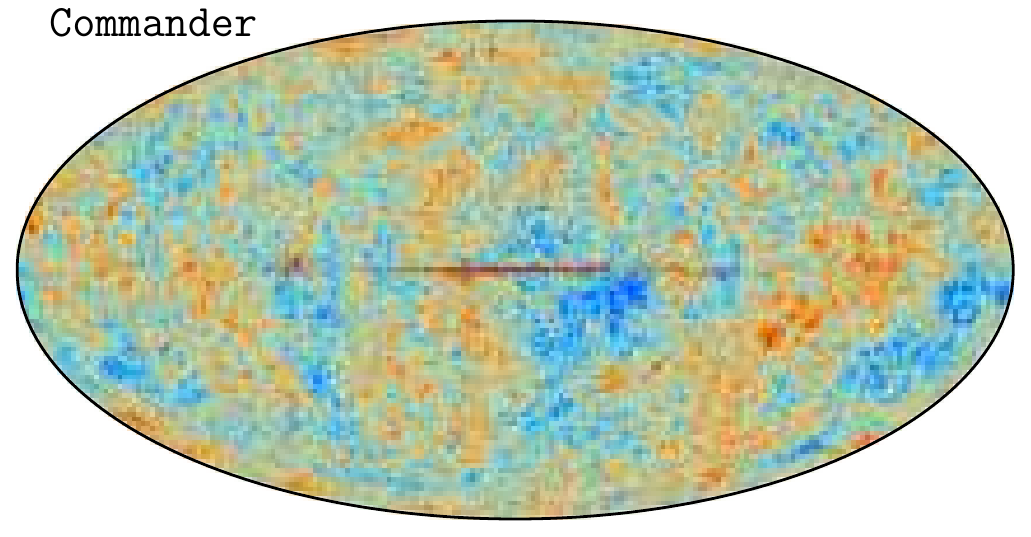}\\
   \includegraphics[width=0.48\textwidth]{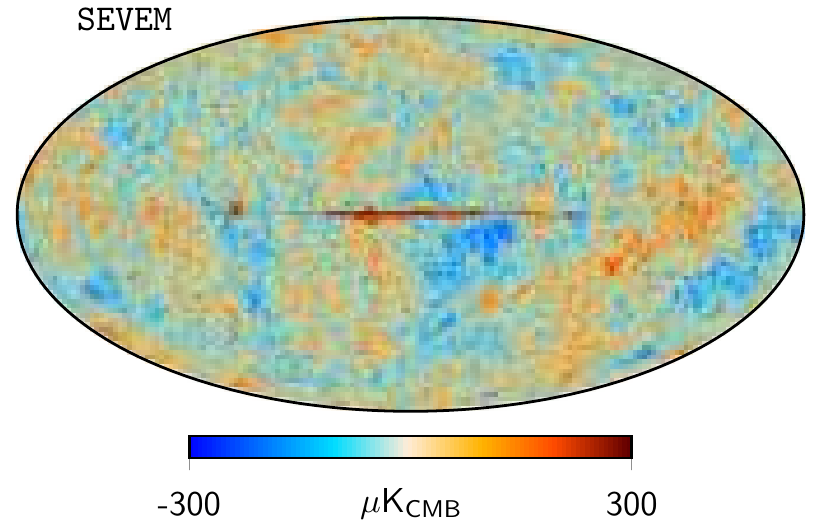}
   \caption{\emph{Top:} CMB $I$ map derived from the \npipe\ data set with \commander, plotted with an angular resolution of 5\arcm\ FWHM. The Solar dipole is retained in the \npipe\ data set, and in this map.  \emph{Middle:} Same as above, but after subtracting the best-fit CMB Solar dipole described in Sect.~\ref{sec:dipole}.  \emph{Bottom:} Dipole-subtracted CMB temperature map derived with \sevem.}
   \label{fig:CMB_T}
\end{figure}

\begin{figure*}[htpb!]
   \centering
   \includegraphics[width=0.48\textwidth]{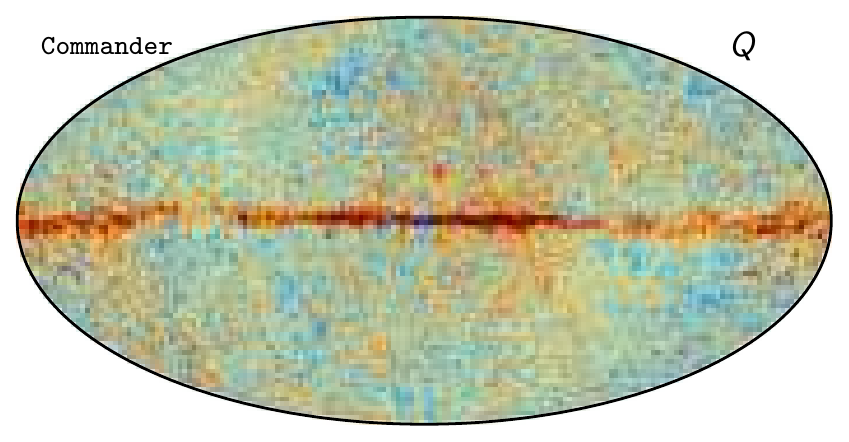}
   \includegraphics[width=0.48\textwidth]{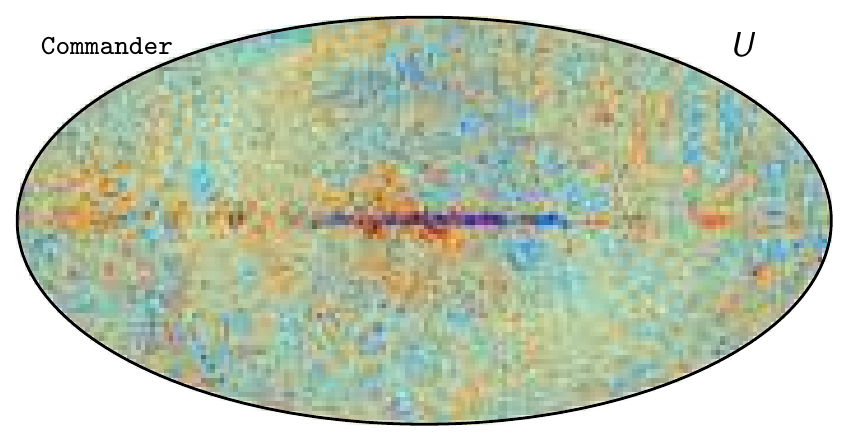}\\
   \includegraphics[width=0.48\textwidth]{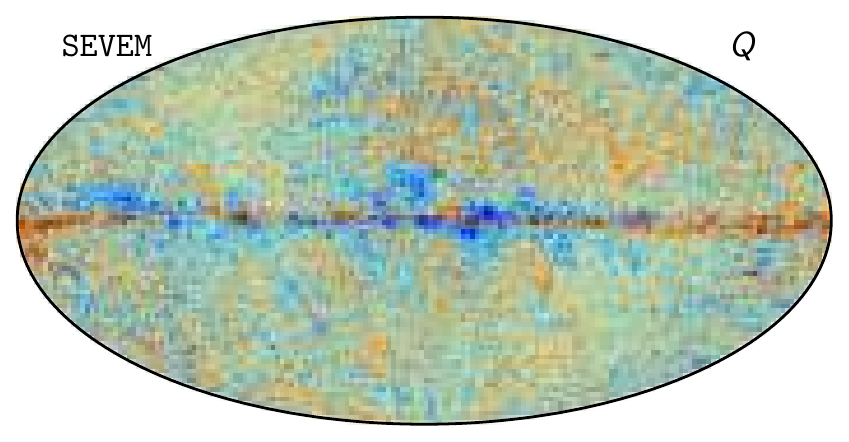}
   \includegraphics[width=0.48\textwidth]{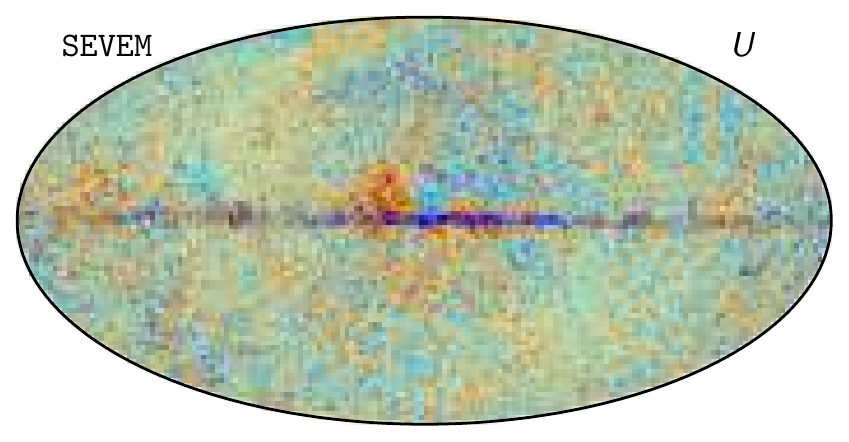}\\
   \includegraphics[scale=1.]{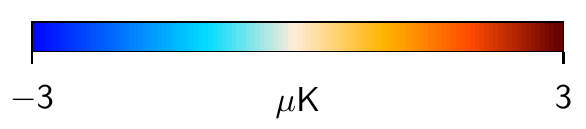}
   \caption{CMB $Q$ and $U$ maps derived from the \npipe\ data set with \commander\ (\emph{top row}) and \sevem\ (\emph{bottom row}).  All maps are plotted with an angular resolution of 40\arcm\ FWHM.}
   \label{fig:CMB_P}
\end{figure*}

Next we consider the cosmological CMB signal as extracted with both \commander\ and \sevem.  Figure~\ref{fig:CMB_T} shows three versions of the \npipe\ temperature CMB map.  The top panel shows the full CMB temperature map as estimated with \commander.  This map is the first component-separation-based product that allows direct estimation of the Solar dipole across the entire \Planck\ frequency range, as discussed in detail in Sect.~\ref{sec:dipole}.  The middle panel of Fig.~\ref{fig:CMB_T} shows the same CMB temperature map after subtracting the best-fit \npipe\ dipole, while the bottom panel shows the same, but estimated with \sevem.  Figure~\ref{fig:CMB_P} shows the corresponding maps of Stokes $Q$ and $U$ from both \commander\ (top row) and \sevem\ (bottom row).

\begin{figure}[htpb!]
   \centering
   \includegraphics[width=0.45\textwidth]{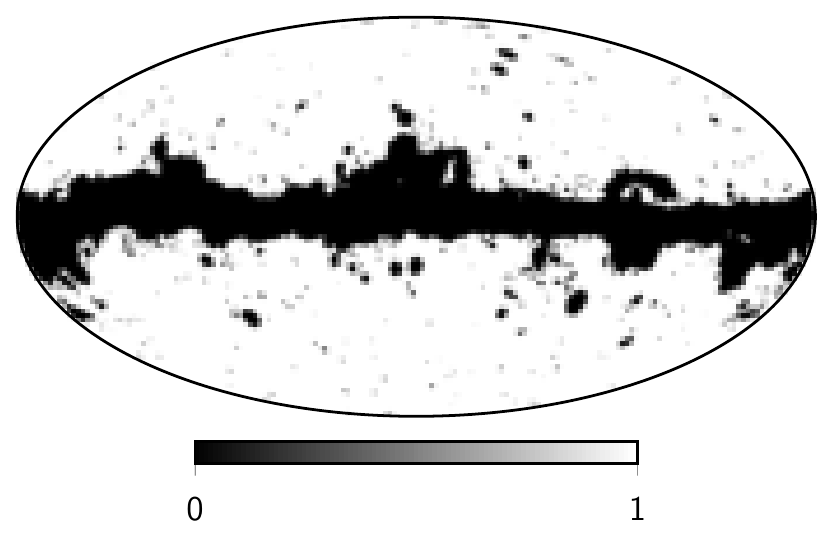}\\
   \caption{\npipe\ CMB temperature analysis mask. This mask is derived by thresholding the standard-deviation map evaluated among the \commander\ CMB temperature maps resulting from three independent \Planck\ processings (\Planck\ 2015, \Planck\ 2018, and \npipe), multiplied by another mask, which is thresholded on an overall $\chi^2$ cut corresponding to the \commander\ \npipe\ analysis. }
\label{fig:cmb_mask_T}
\end{figure}

While the overall morphology of the CMB $I$, $Q$, and $U$ maps at high Galactic latitudes looks generally consistent with expectations based on a Gaussian and isotropic $\Lambda$CDM universe, there are clear visual indications of significant foreground-induced residuals at low Galactic latitudes.  Consequently, proper masking is required before subjecting these maps to detailed scientific analysis.  For temperature, we construct an analysis mask in a manner analogous to the approach taken
in \citet{planck2016-l04}, where a smoothed standard-deviation map evaluated between four different CMB component-separation estimates was thresholded at a given value.  In this paper, we instead threshold the standard-deviation map evaluated from the \commander\ CMB temperature maps derived from the three latest \Planck\ data releases, namely the \Planck\ 2015 and 2018 data sets and the new \npipe\ data set.  This is a conservative approach, since it only admits the parts of the sky that all three of the latest \Planck\ processing pipelines agree on, and that are robust with respect to the very different foreground models adopted for the three data sets. Specifically, for \Planck\ 2015 we combined the \Planck\ data with Haslam 408-MHz and WMAP maps, and employed a detailed model including synchrotron, free-free, and spinning dust, while for \Planck\ 2018 and \npipe, we included only a single combined low-frequency component.  Similarly, for \Planck\ 2015 and \npipe, we include three independent CO-line components in the model, while for \Planck\ 2018, only one combined CO line was included.  For a pixel to be accepted in the new mask, all three versions must agree to a precision better than 3\muK\footnote{The precise numerical value for the standard-deviation cut-off is somewhat arbitrary, but also of little importance, as the gradient in the map is very sharp near the Galactic plane. Any value between 2 and 5\muK\ results in very similar masks, whereas values below 2\muK\ start picking up noise fluctuations.}.  In addition, all pixels with a reduced $\chi^2$ in \npipe\ greater than 2 are excluded (as it happens, most of these pixels are already excluded by the standard-deviation cut).  The resulting mask is shown in Fig.~\ref{fig:cmb_mask_T}, and retains 76\,\% of the sky.

\begin{figure}[htpb!]
   \centering
   \includegraphics[width=0.45\textwidth]{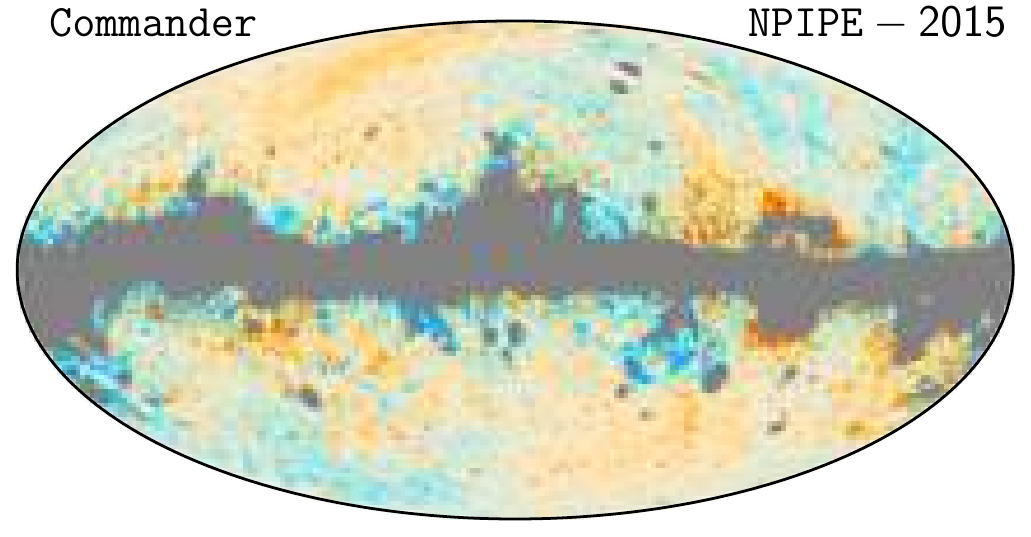}\\
   \includegraphics[width=0.45\textwidth]{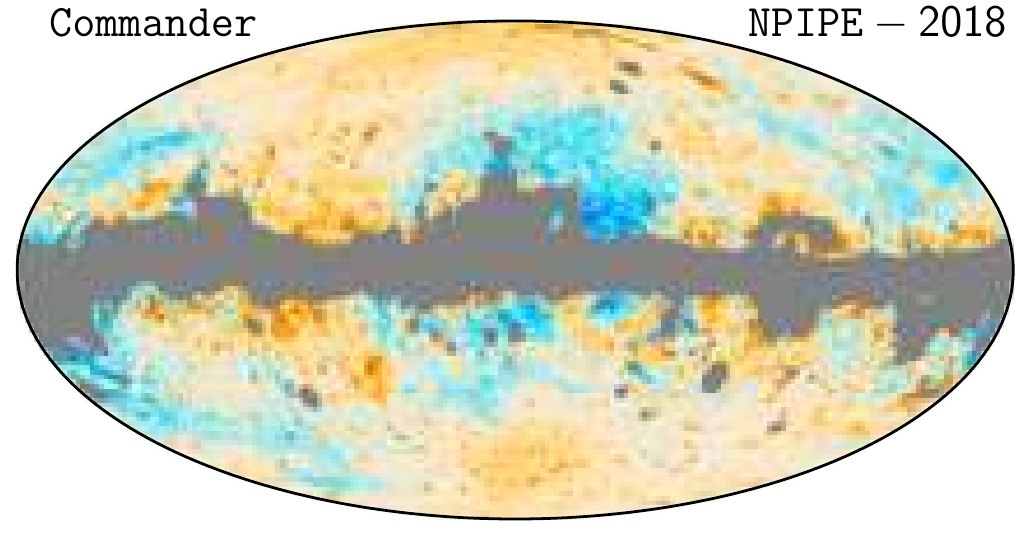}\\
   \includegraphics[width=0.45\textwidth]{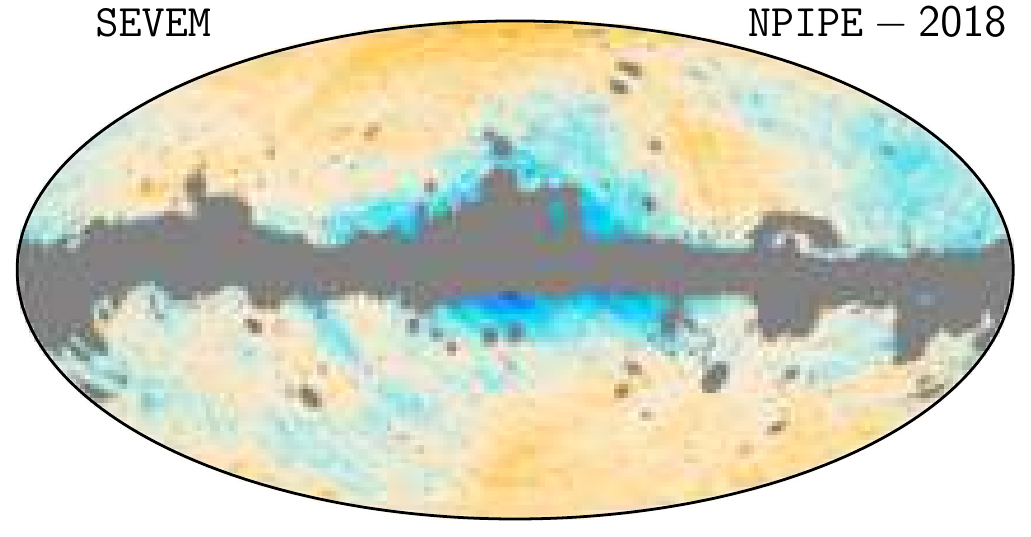}\\
   \includegraphics[width=0.60\linewidth]{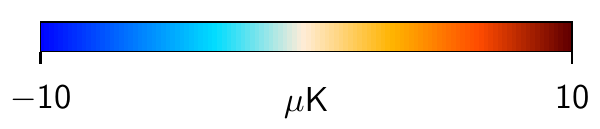}
\caption{\emph{Top:} Stokes $I$ difference map between the \Planck-only \npipe\ and the \Planck\ 2015 \citep{planck2014-a12}
\commander\ CMB maps. Both maps have been smoothed to 1\deg\ FWHM and masked with the mask shown in Fig.~\ref{fig:cmb_mask_T}.  \emph{Middle:} Same, but with the 2018 \commander\ CMB map \citep{planck2016-l04}.  \emph{Bottom:} Same as middle panel, but evaluated for \sevem\ instead of \commander.}
\label{fig:diff_cmb_amplitude_T}
\end{figure}

The top panel of Fig.~\ref{fig:diff_cmb_amplitude_T} shows the difference between the \commander-derived \npipe\ CMB temperature map and the \Planck\ 2015 CMB map smoothed to $1\deg$ FWHM.  The middle panel shows the corresponding difference map with respect to the \Planck\ 2018 map.  The difference between the \npipe\ and \Planck\ 2018 \sevem\ maps is shown in the bottom panel.  The grey regions show the confidence mask described above.  We see that the various CMB estimates outside this mask typically agree to within 1 or 2\muK.  The \commander\ difference map evaluated with respect to the 2015 solution exhibits slightly lower residuals at high latitudes than the corresponding 2018 solution.  In this respect, the \npipe\ and 2015 analyses consider single-detector maps, while 2018 only considers full-frequency maps.  As already noted, a more fine-grained data set allows for better foreground modelling and more selective removal of obviously bad channels.  Comparing the two lower panels, we see that the morphologies are very similar between \commander\ and \sevem.

For polarization, we adopt the same common analysis mask as discussed in \citet{planck2016-l04}.  This choice is motivated by the fact that in the following we will compare the \npipe\ polarization products with corresponding \Planck\ 2018 products, and we note that the \npipe\ maps generally have smaller systematics and foreground residuals than \Planck\ 2018. 

\begin{figure*}[htpb!]
   \centering
   \includegraphics[width=0.44\linewidth]{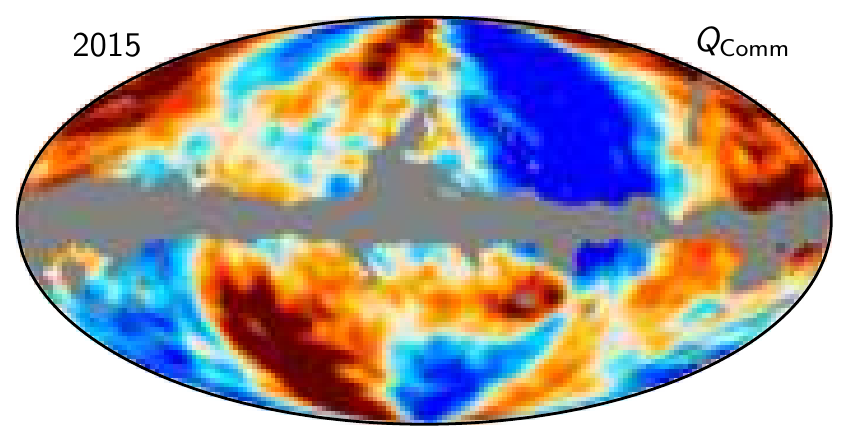}
   \includegraphics[width=0.44\linewidth]{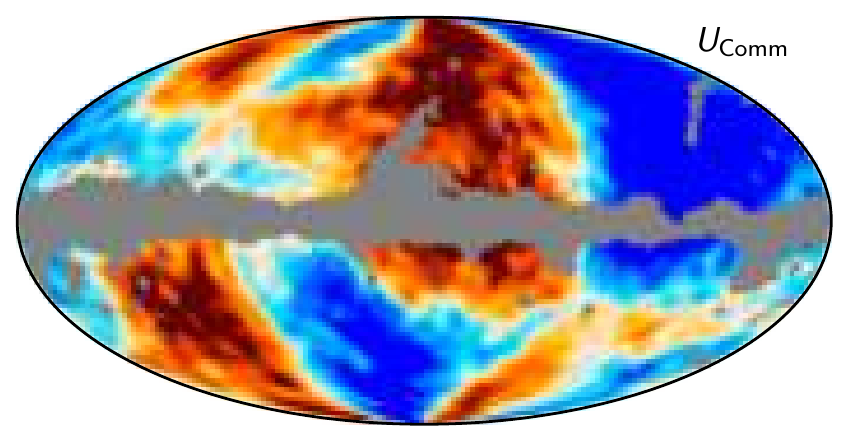}\\
   \includegraphics[width=0.44\linewidth]{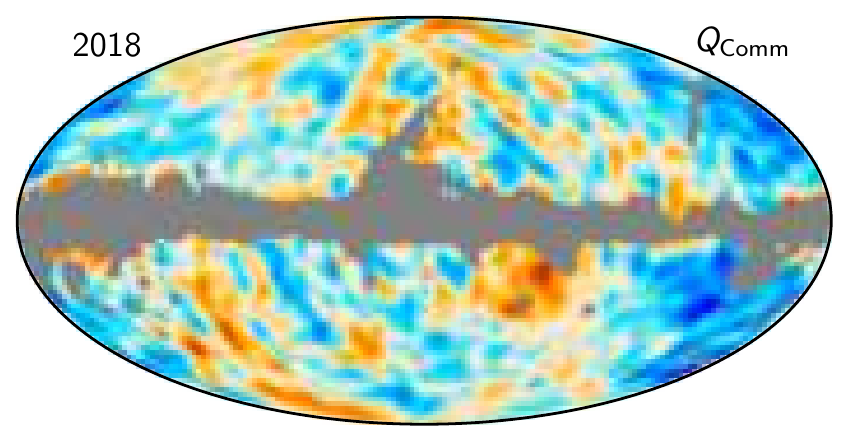}
   \includegraphics[width=0.44\linewidth]{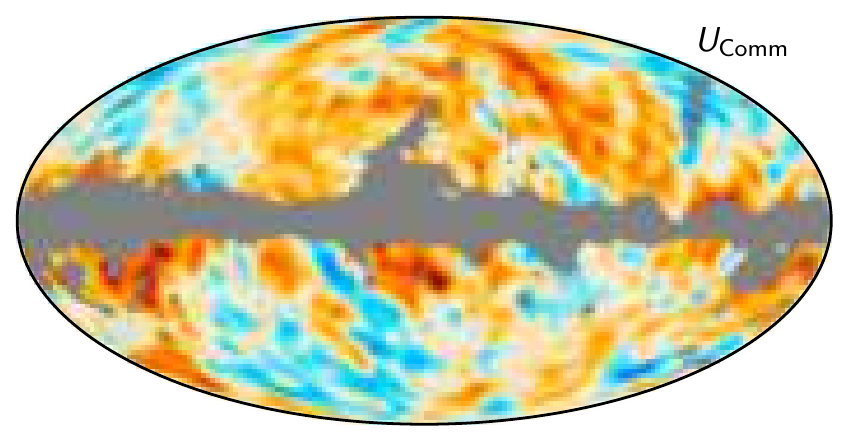}\\
   \includegraphics[width=0.44\linewidth]{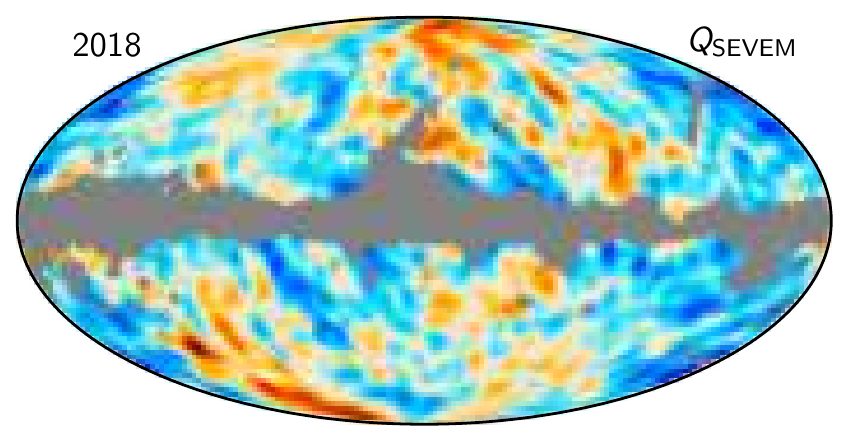}
   \includegraphics[width=0.44\linewidth]{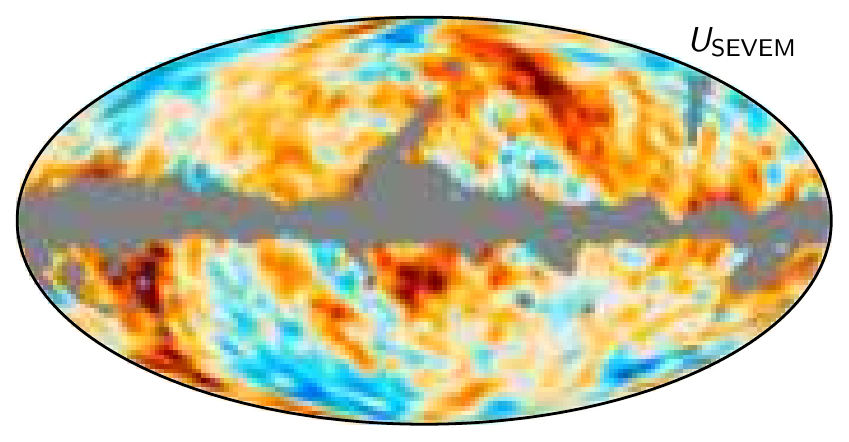}\\
   \includegraphics[width=0.44\linewidth]{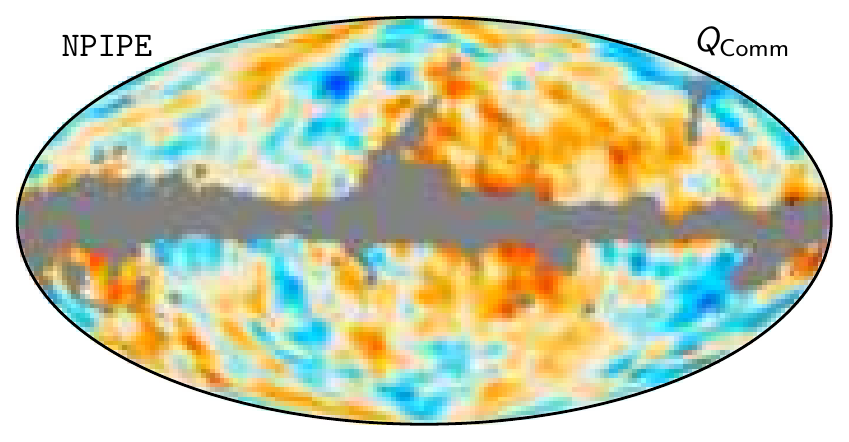}
   \includegraphics[width=0.44\linewidth]{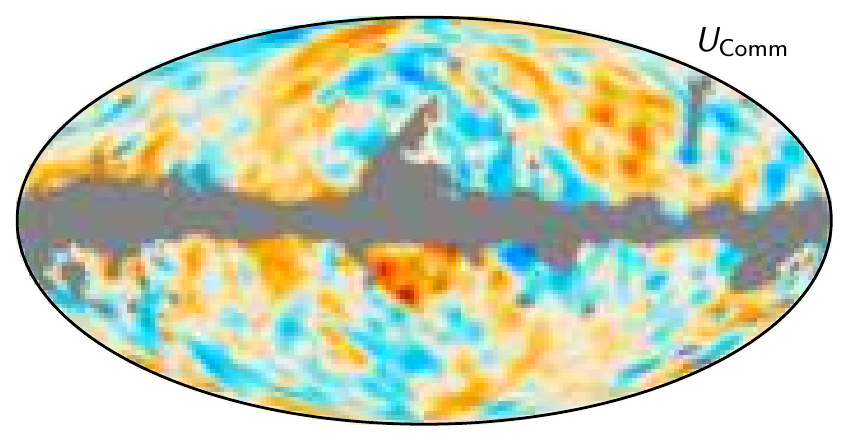}\\
   \includegraphics[width=0.44\linewidth]{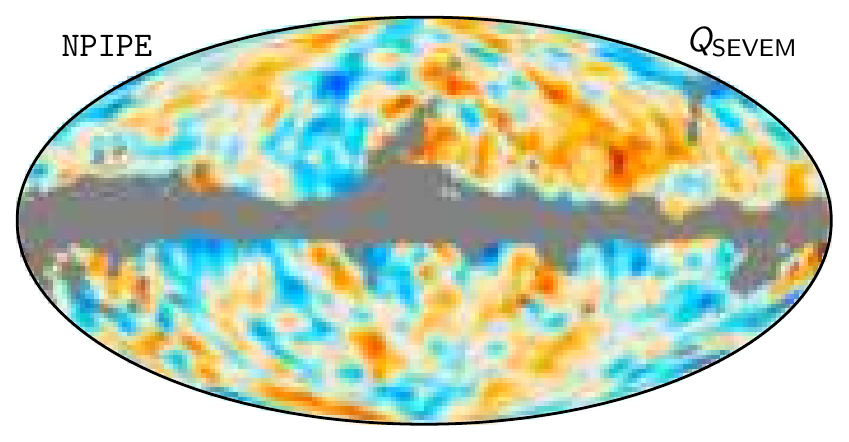}
   \includegraphics[width=0.44\linewidth]{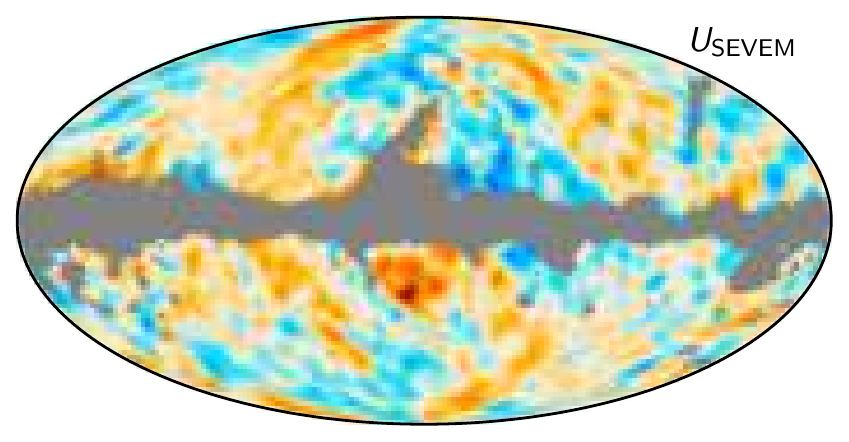}\\
   \includegraphics[width=0.44\linewidth]{colourbar_1uK}
   \caption{Comparison of large-scale CMB $Q$ and $U$ maps from, top to bottom: \commander\ \Planck\ 2015; \commander\ \Planck\ 2018; \sevem\ \Planck\ 2018; \commander\ \npipe; and \sevem\ \npipe.  Note that the large-scale \Planck\ 2015 CMB map in the top row was never publicly released, due to the high level of residual systematic effects.  The grey region corresponds to the \Planck\ 2018 common component-separation mask \citep{planck2016-l04}.  All maps are smoothed to a common angular resolution of 5\deg\ FWHM.}
   \label{fig:cmb_pol_lowres}
\end{figure*}

Figure~\ref{fig:cmb_pol_lowres} compares the \commander\ Stokes $Q$ and $U$ polarization maps from the \Planck\ 2015, \Planck\ 2018, and \npipe\ data sets, as well as the \sevem\ polarization maps from \Planck\ 2018 and \npipe.  All maps are smoothed to a common angular resolution of $3\deg$ FWHM, and the grey regions show the \Planck\ 2018 common confidence mask.  Starting from the top, we see that the \commander\ 2015 polarization map is dominated by large-scale systematic features.  Due to this large systematic contribution, the map was never publicly released in its raw form, but only in the form of a high-pass-filtered map, from which multipoles below $\ell\,{\le}\,20$ had been removed.

Much of the analysis effort of the \Planck\ team between 2015 and 2018 concentrated on understanding and mitigating these residuals.  As seen in the second and third rows of Fig.~\ref{fig:cmb_pol_lowres}, this work was highly successful.  The
large-scale polarization systematics were reduced by an order of magnitude.  Furthermore, this process has continued beyond the final \Planck\ 2018 release within the \npipe\ framework, as shown in the bottom two rows of Fig.~\ref{fig:cmb_pol_lowres}, which exhibits even lower residuals than \Planck\ 2018.

\begin{figure*}[htpb]
   \centering
   \includegraphics[width=0.22\linewidth]{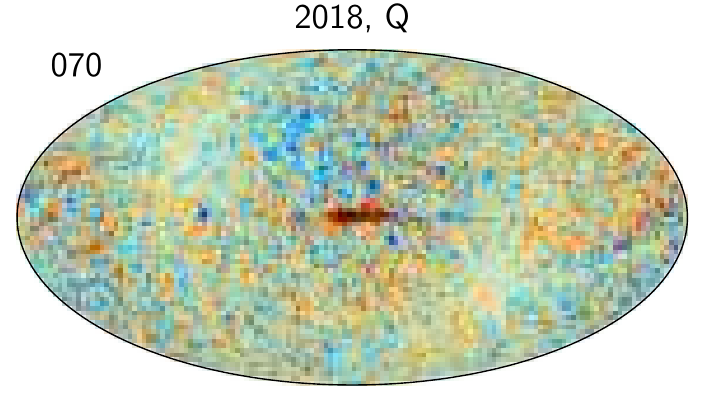}
   \includegraphics[width=0.22\linewidth]{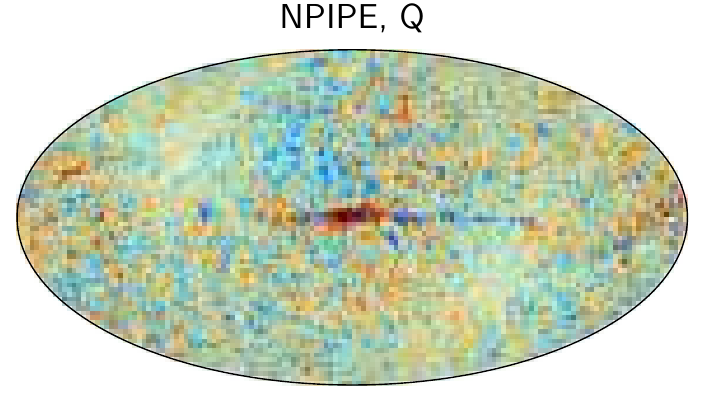}
   \includegraphics[width=0.22\linewidth]{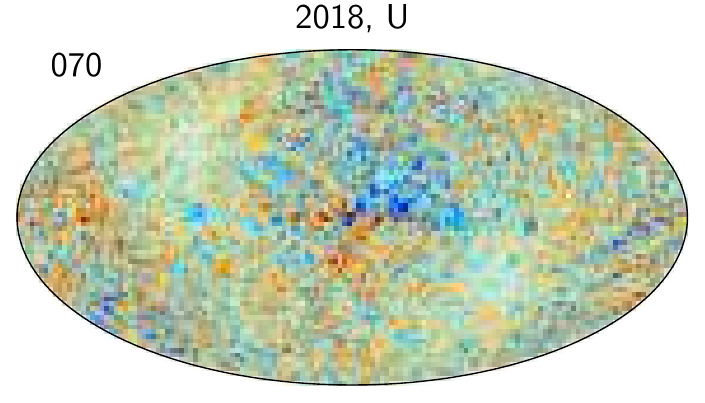}
   \includegraphics[width=0.22\linewidth]{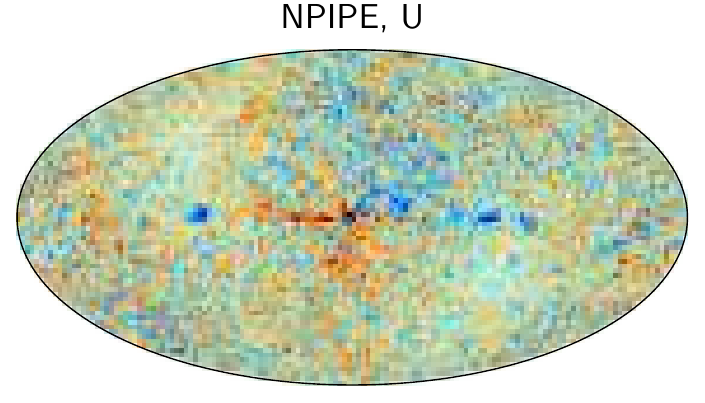} \\
   \includegraphics[scale=1.0]{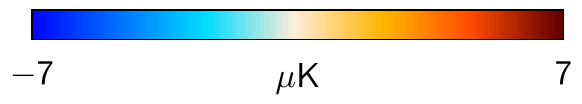} \\
   \includegraphics[width=0.22\linewidth]{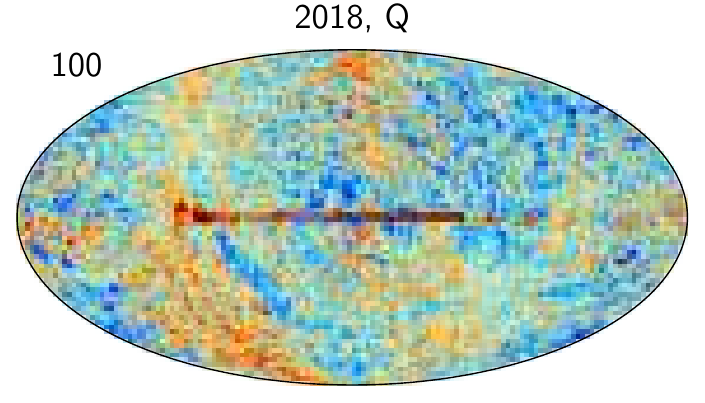}
   \includegraphics[width=0.22\linewidth]{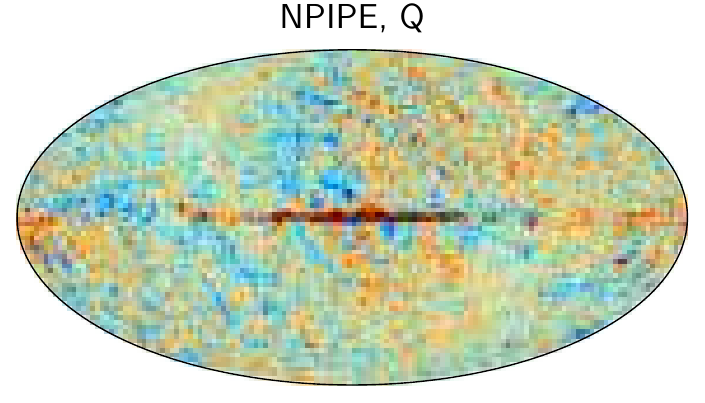}
   \includegraphics[width=0.22\linewidth]{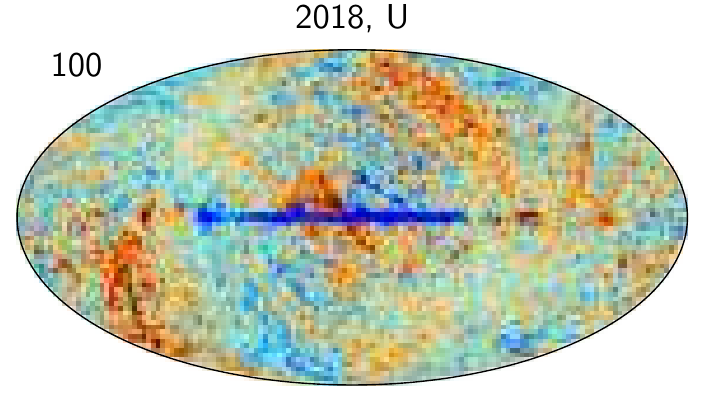}
   \includegraphics[width=0.22\linewidth]{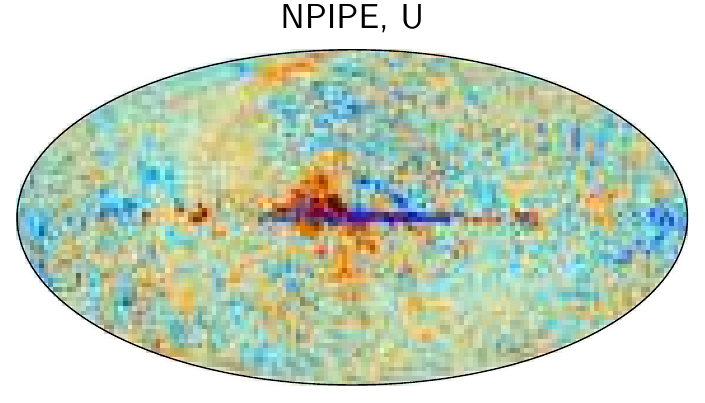} \\
   \includegraphics[width=0.22\linewidth]{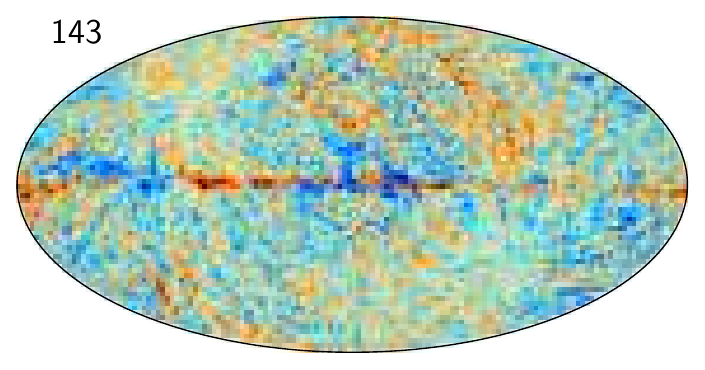}
   \includegraphics[width=0.22\linewidth]{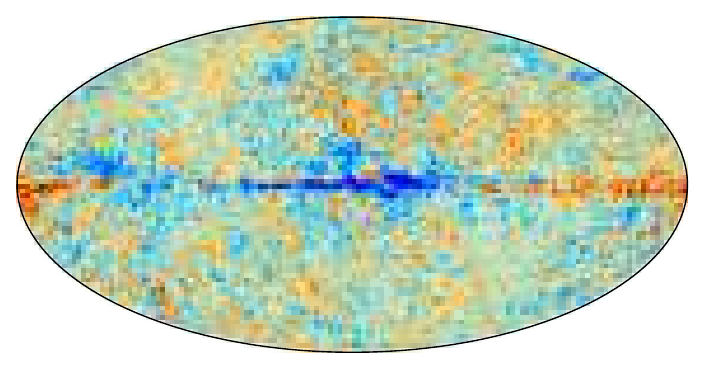}
   \includegraphics[width=0.22\linewidth]{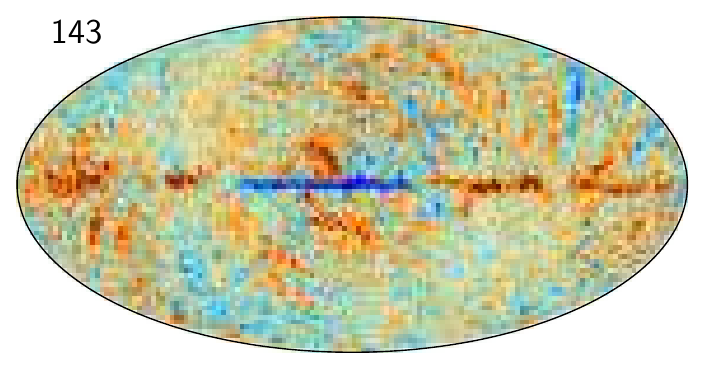}
   \includegraphics[width=0.22\linewidth]{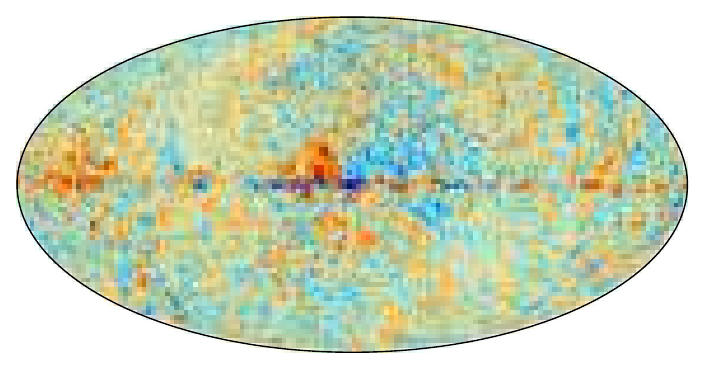} \\
   \includegraphics[scale=1.0]{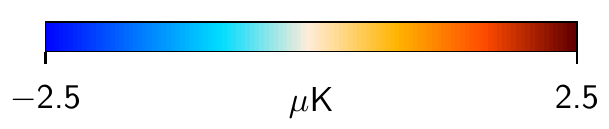} \\
   \includegraphics[width=0.22\linewidth]{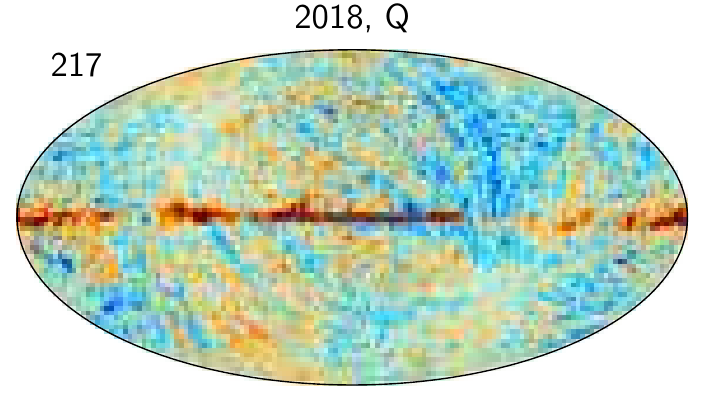}
   \includegraphics[width=0.22\linewidth]{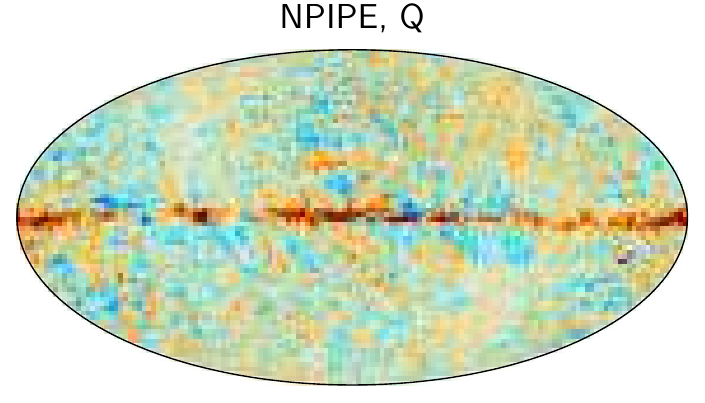}
   \includegraphics[width=0.22\linewidth]{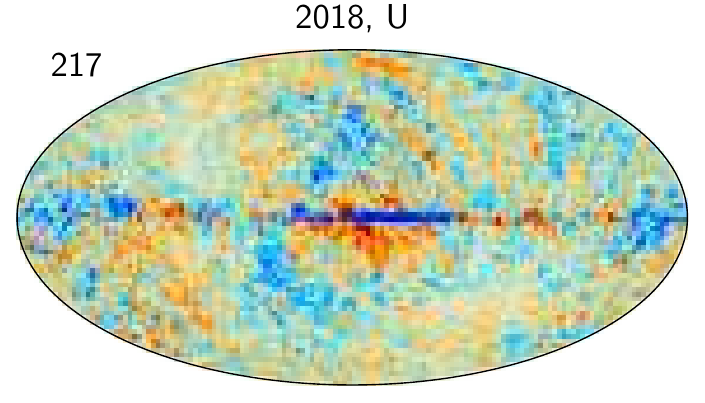}
   \includegraphics[width=0.22\linewidth]{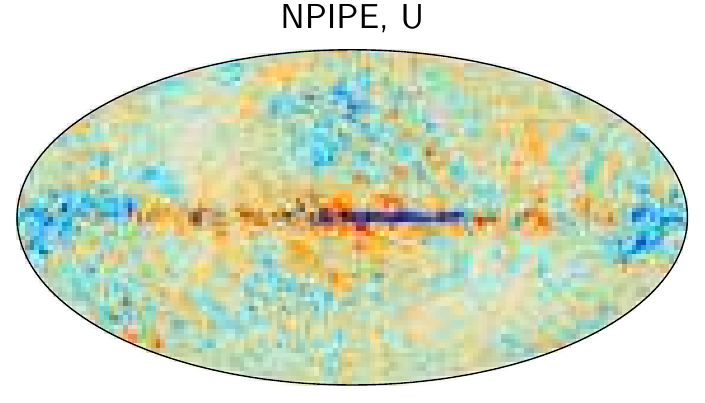} \\
   \includegraphics[scale=1.0]{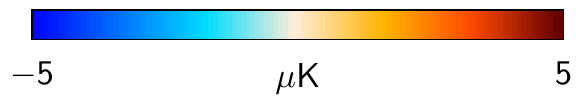} \\
   \caption{Comparison of single-frequency polarization maps derived with \sevem\ from \Planck\ 2018 and from \npipe. All maps are smoothed to a common angular resolution of 80\arcm\ FWHM.}
   \label{fig:cmb_sevem_freq}
\end{figure*}

Figure \ref{fig:cmb_sevem_freq} shows cleaned, single-frequency polarization maps derived with \sevem.  Large-scale systematic features are significantly mitigated in the \npipe\ data.

\begin{figure*}[htpb]
   \centering
   \includegraphics[width=0.48\linewidth]{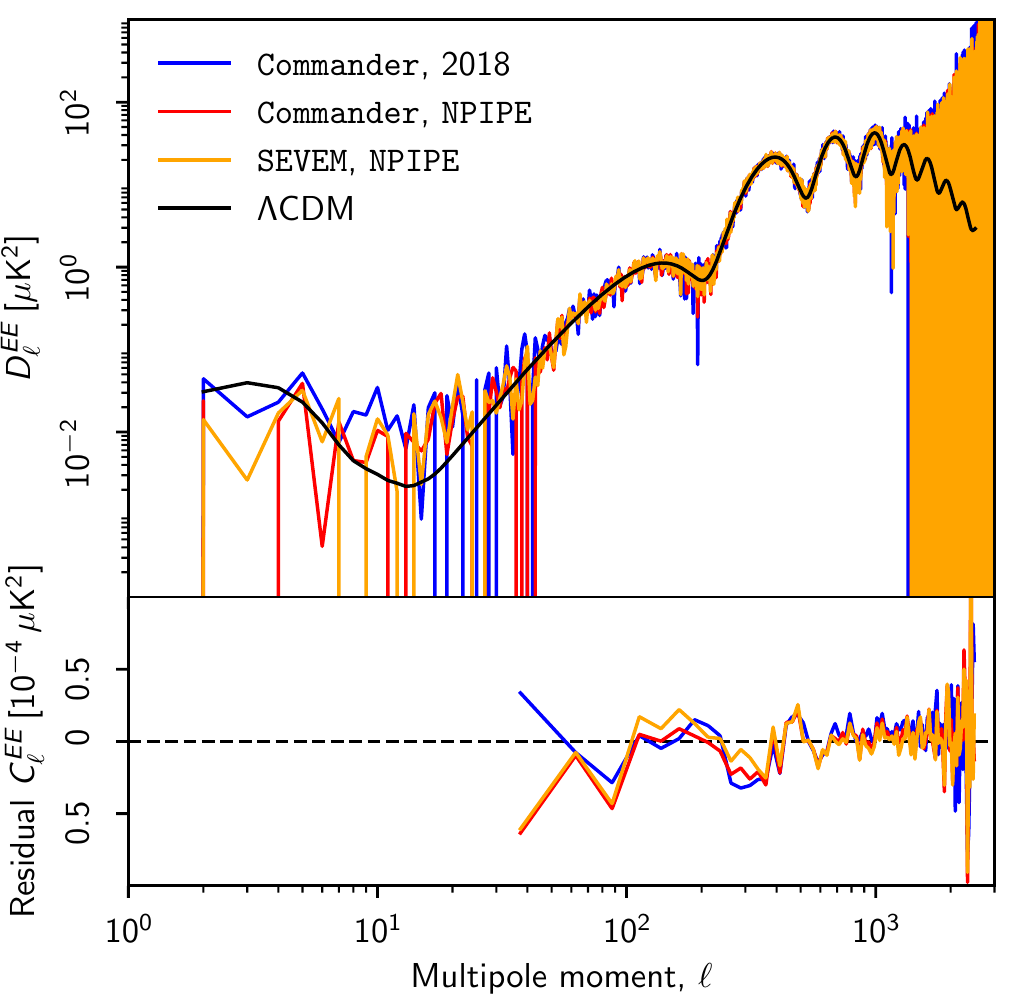}
   \includegraphics[width=0.48\linewidth]{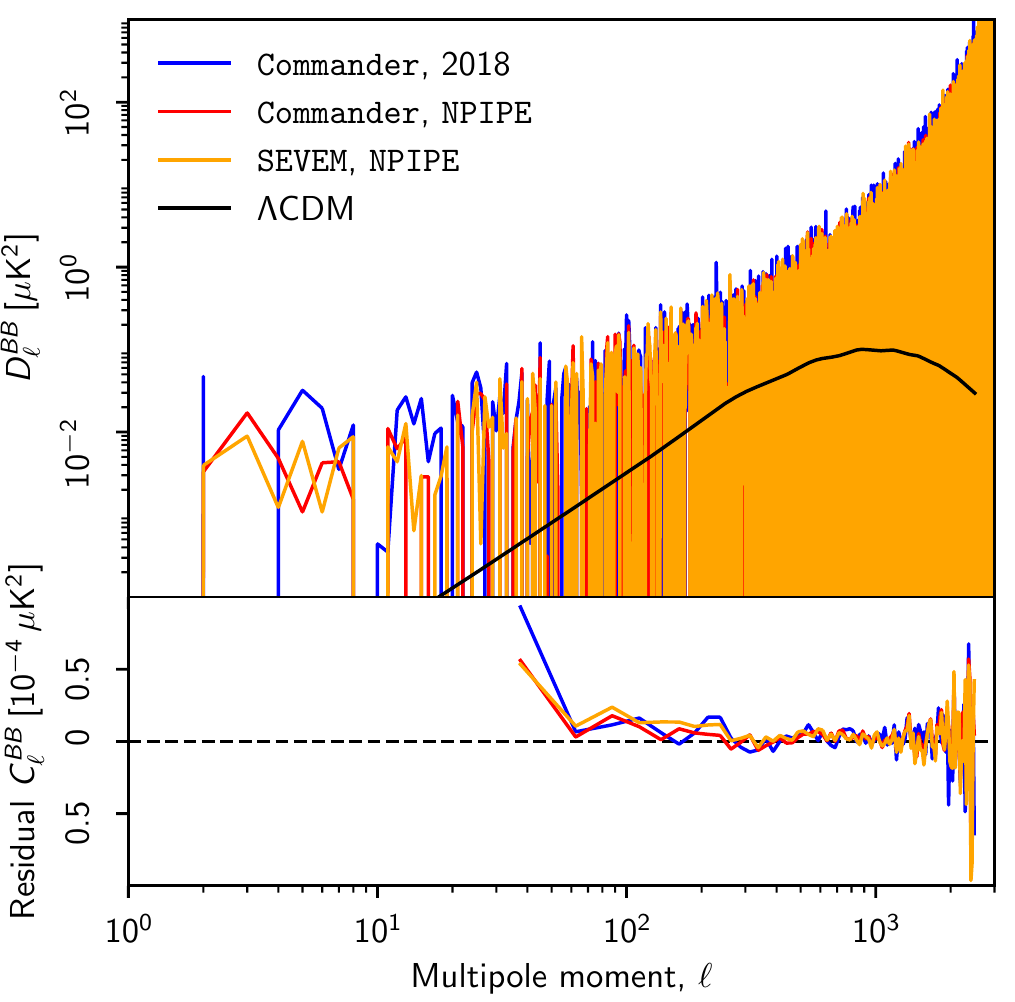}
   \caption{Angular CMB polarization cross-power spectra evaluated from the \Planck\ 2018 (blue curves) and \npipe\ (red curves for \commander; orange curves for \sevem) data sets.  $EE$ and $BB$ spectra are shown in the left and right panels, respectively.  Within each main panel, the full spectrum is shown in the top sub-panel, while the residuals with respect to the \Planck\ 2018 best-fit $\Lambda$CDM spectrum (black curves) are shown in the bottom sub-panels.  The latter have been binned with $\Delta\ell=25$.  The cross-spectra are evaluated from the most independent data split that is available for each data set, corresponding to the A/B detector split for \npipe\ and the half-mission split for \Planck\ 2018.}
   \label{fig:cmb_powspec_highl}
\end{figure*}

To quantify the differences between the \Planck\ 2018 and \npipe\ polarization maps further, Fig.~\ref{fig:cmb_powspec_highl} shows the angular $EE$ and $BB$ cross-spectra evaluated from half-mission (for
\Planck\ 2018) and detector-set (for \npipe) splits, evaluated outside the 2018 common confidence mask.  These two splits represent the two most independent data subsets available within each data set. The most notable feature in these plots is the fact that the blue curves  (corresponding to the \commander\ 2018 spectrum) generally encompass the red and orange curves (corresponding to the \commander\ and \sevem\ \npipe\ spectra). This is a direct reflection of the fact that the overall noise level in the \npipe\ data set is about 15\,\% lower than in \Planck\ 2018, which results in a lower level of $BB$ power at all scales. Other than this noise difference, the two data sets appear statistically quite similar.

\begin{figure*}[htpb]
   \centering
   \includegraphics[width=0.48\linewidth]{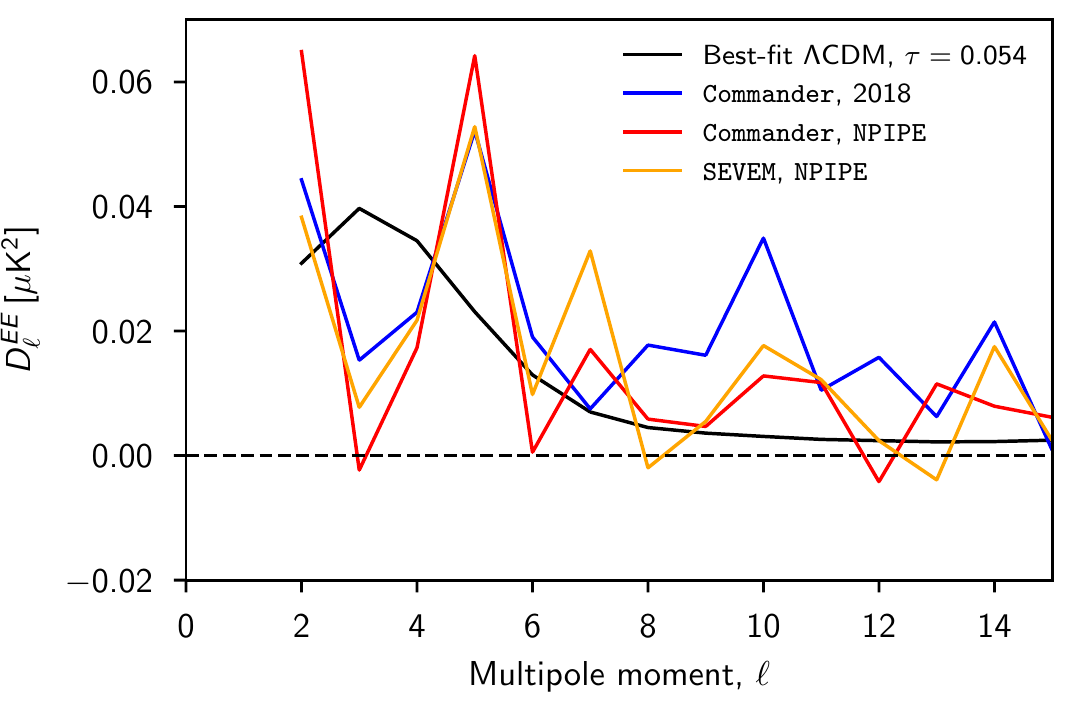}
   \includegraphics[width=0.48\linewidth]{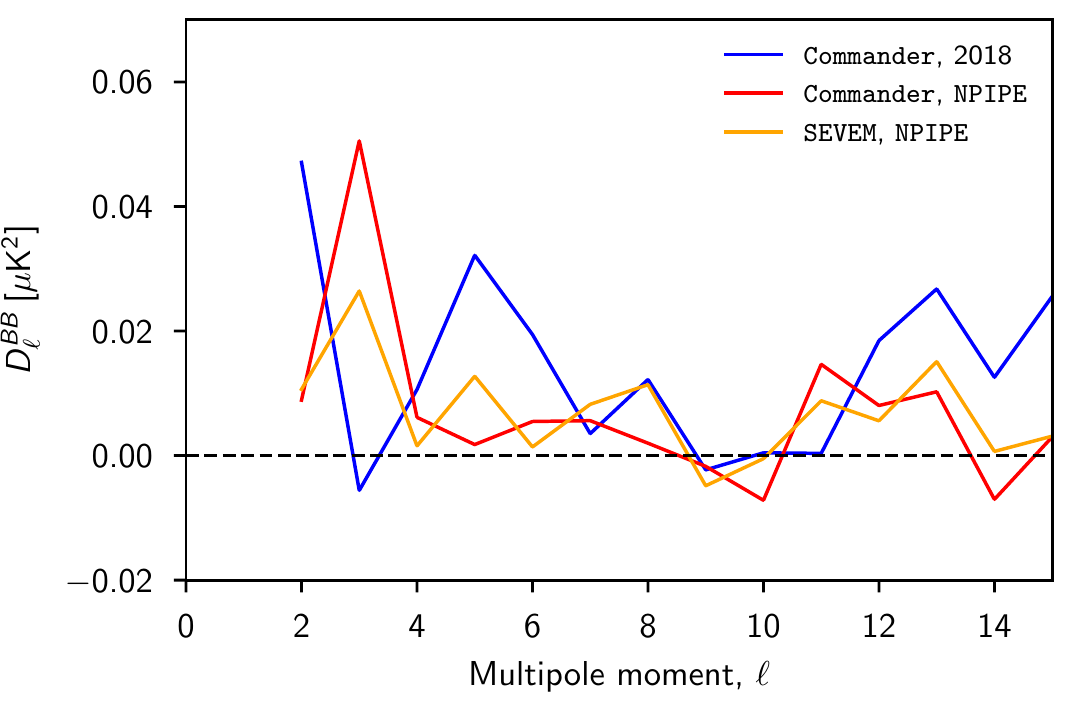}
   \caption{Same as Fig.~\ref{fig:cmb_powspec_highl}, except showing only the lowest multipoles.  Corrections for the low-$\ell$ \npipe\ transfer function have been applied to both sets.  Note that neither of the spectra has been corrected for potential biases from correlations arising due to joint processing of the two data splits.  As shown in \citet{planck2016-l03} and \citet{planck2016-l11A}, these correlations are significant for the 2018 half-mission data split, and reproducible in end-to-end simulations.}
   \label{fig:cmb_powspec_lowl}
\end{figure*}

Figure~\ref{fig:cmb_powspec_lowl} shows the same spectra, but only for $\ell\le15$.  Here we see that the \Planck\ 2018 cross-spectra exhibit a statistically significant power excess on large scales.  This was already noted in \citet{planck2016-l04}, where it was shown that the same excess was present in realistic end-to-end simulations.  The excess was found to be caused by joint processing of the two halves of the data split, which led to cross-split correlations.

In contrast, the two halves of the \npipe\ detector split are processed independently, and, as a result, the cross-spectrum appears consistent with the $\Lambda$CDM prediction, without the need for further simulation-based interpretation.  Indeed, \npipe\ represents the first processing of the \Planck\ data for which the large-scale reionization peak is visually apparent in the raw power spectrum, and not only statistically detected through high-level likelihood analysis.  Likewise, the large-scale \npipe\ $BB$ spectrum is visually consistent with zero.

\subsection{Astrophysical foreground maps}

\begin{figure}[hbtp]
   \centering
\includegraphics[width=0.38\textwidth]{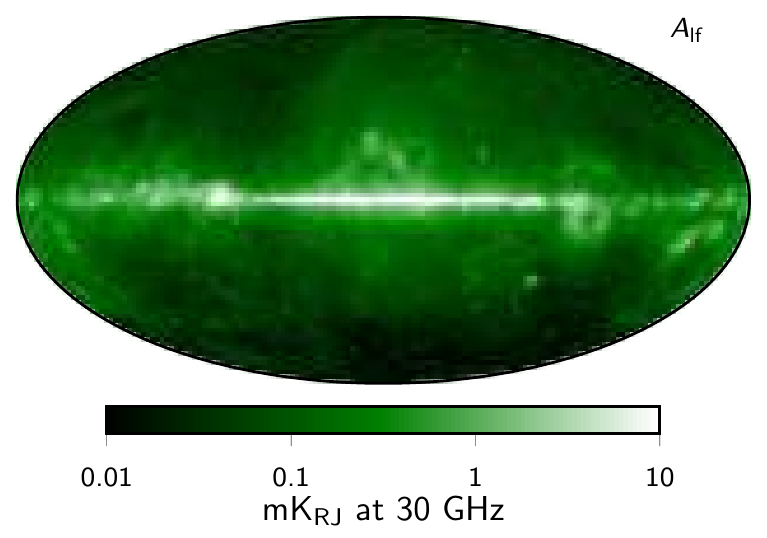}
\includegraphics[width=0.38\textwidth]{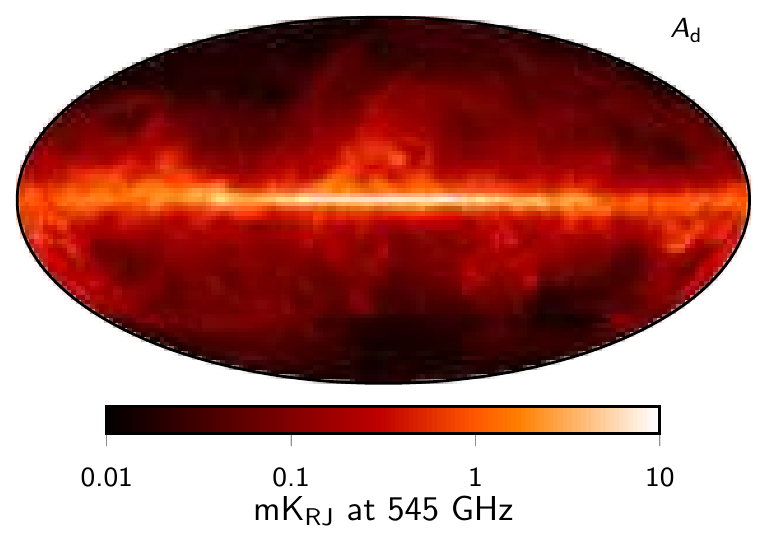}
\includegraphics[width=0.38\textwidth]{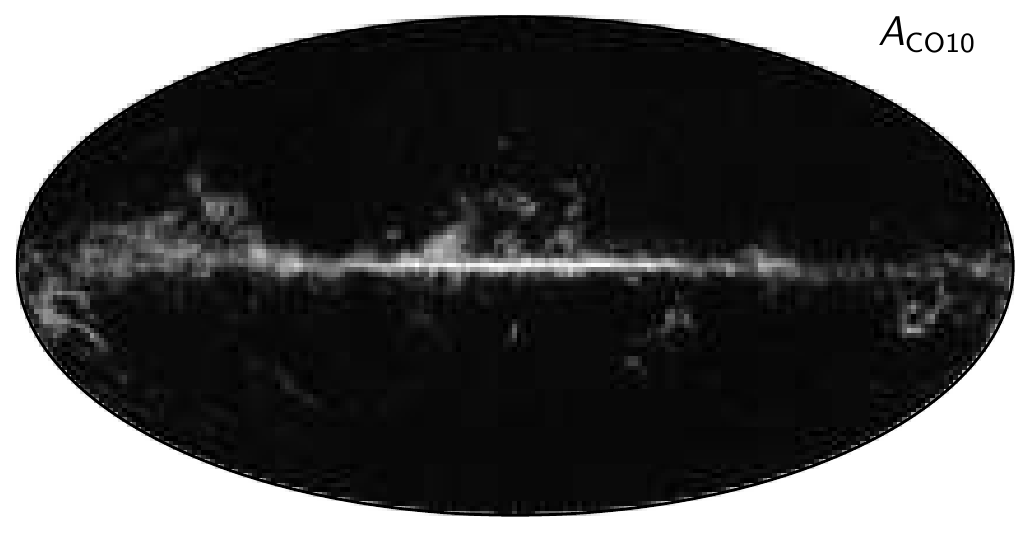}
\includegraphics[width=0.38\textwidth]{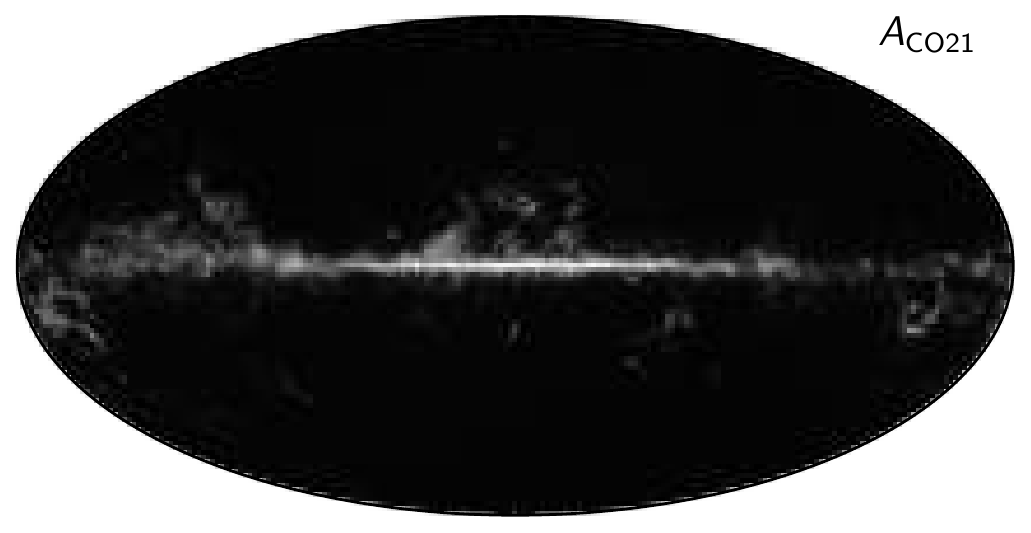}
\includegraphics[width=0.38\textwidth]{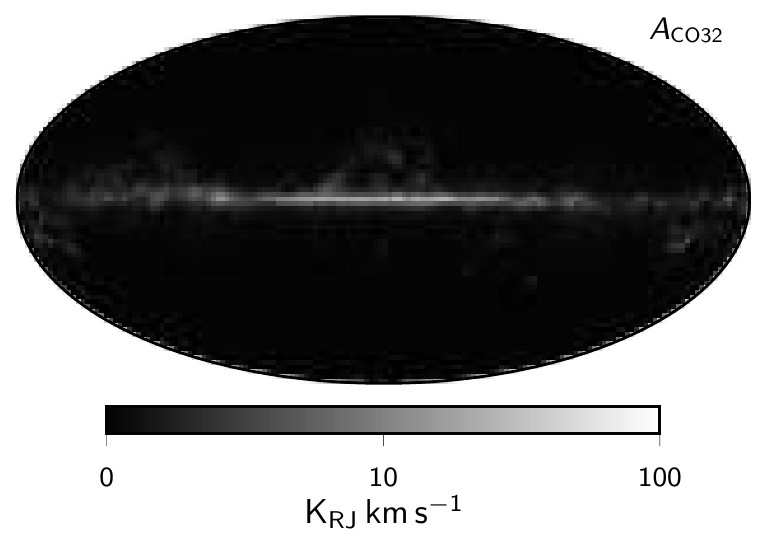}
\caption{{\it Top to bottom:}  Low-frequency (evaluated at 30\GHz, smoothed to 40\arcm\ FWHM), thermal dust (evaluated at 545\GHz, smoothed to 14\arcm\ FWHM), and CO intensity maps.}
\label{fig:fg_amplitude_T}
\end{figure}

\begin{figure}[hbtp]
   \centering
   \includegraphics[width=0.24\textwidth]{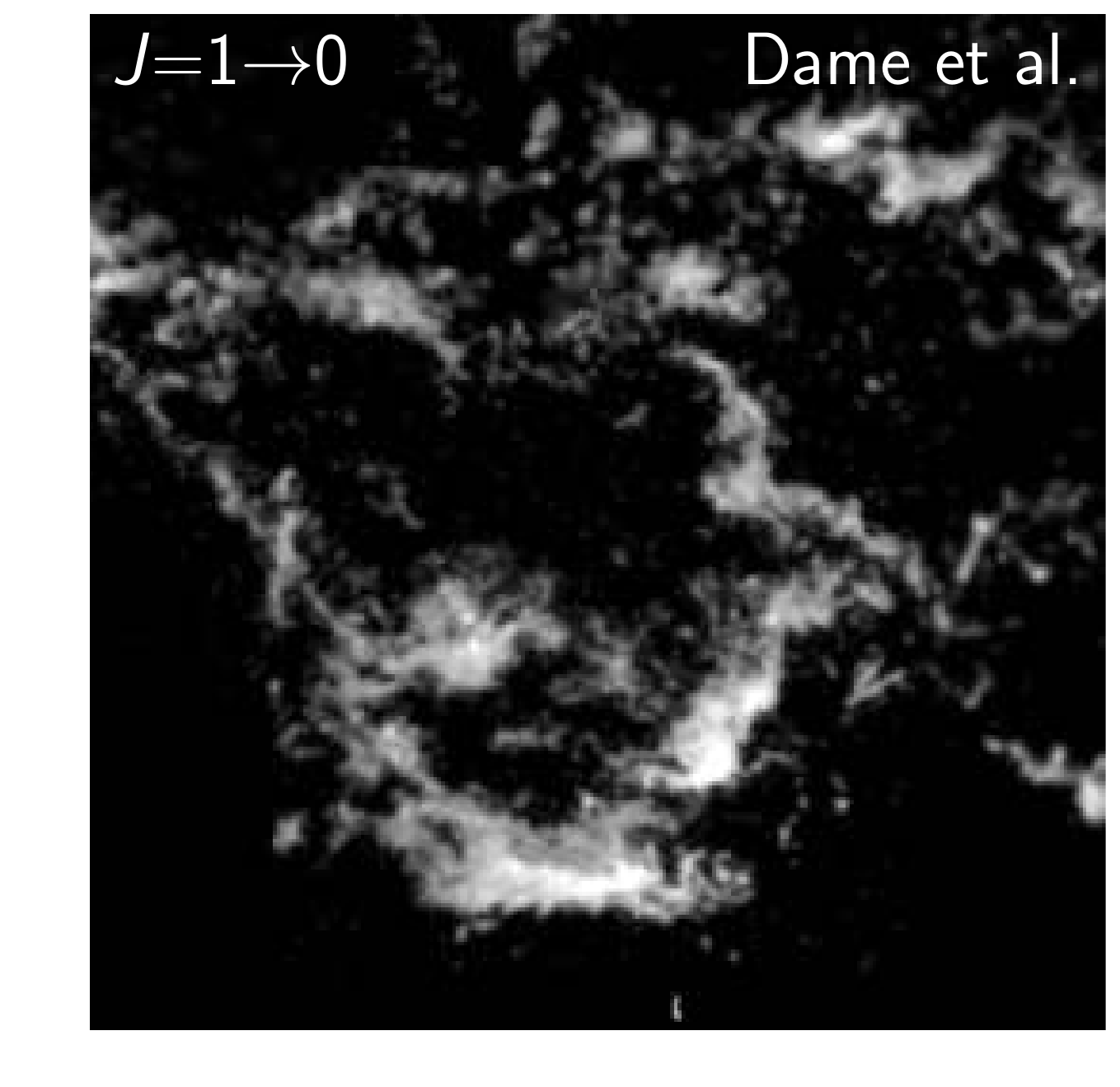}
   \includegraphics[width=0.24\textwidth]{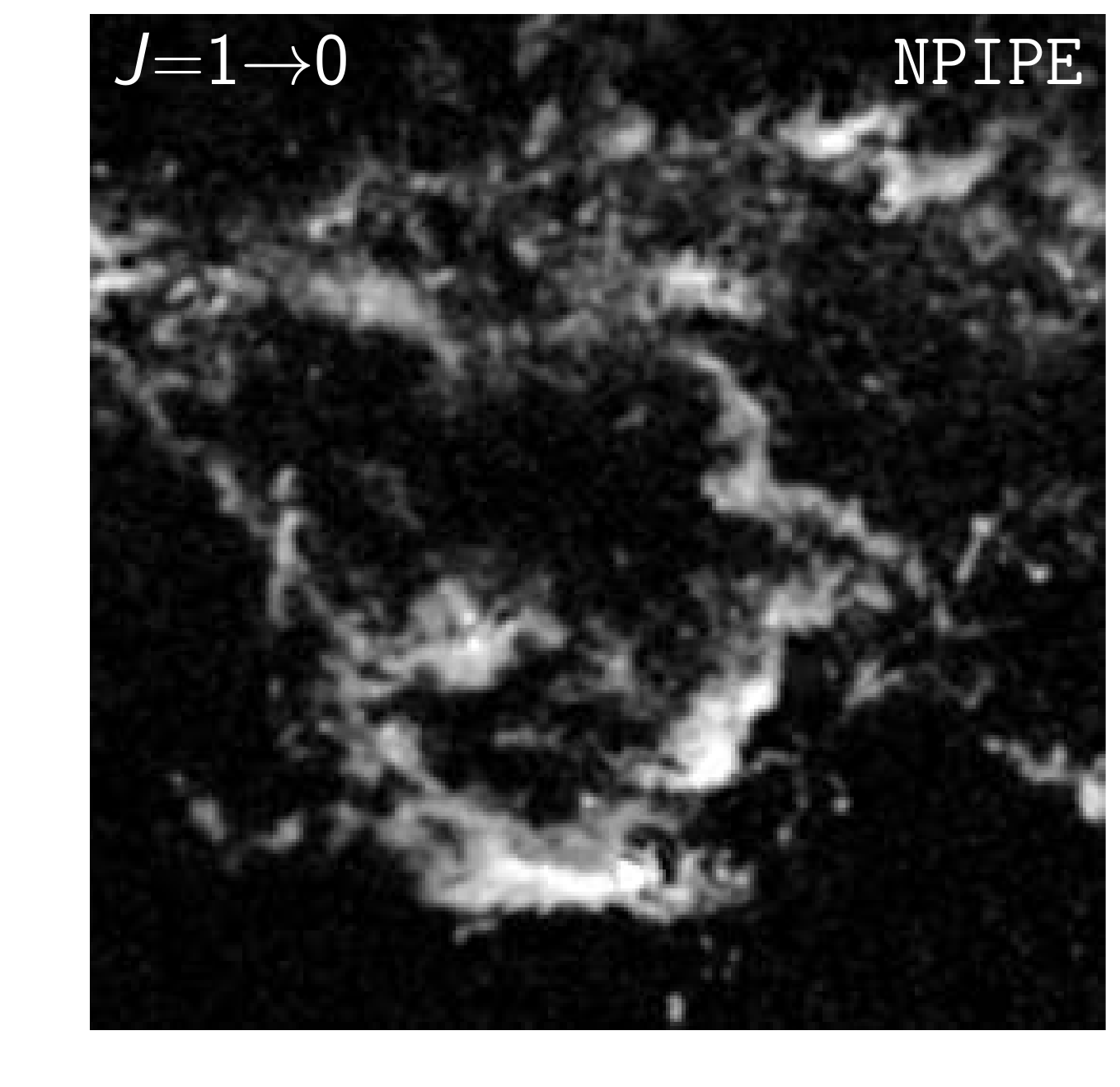}\\
   \includegraphics[width=0.24\textwidth]{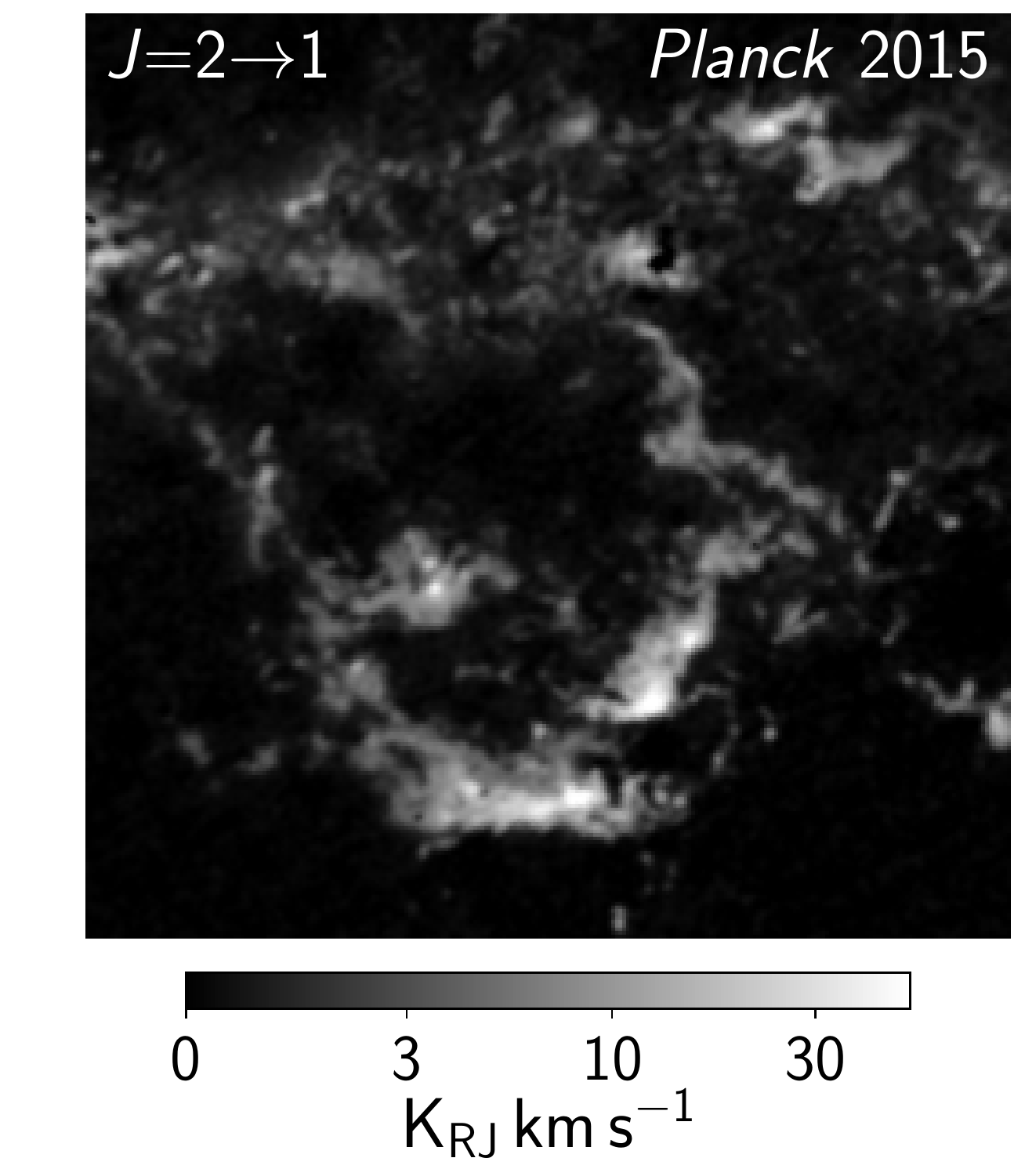}
   \includegraphics[width=0.24\textwidth]{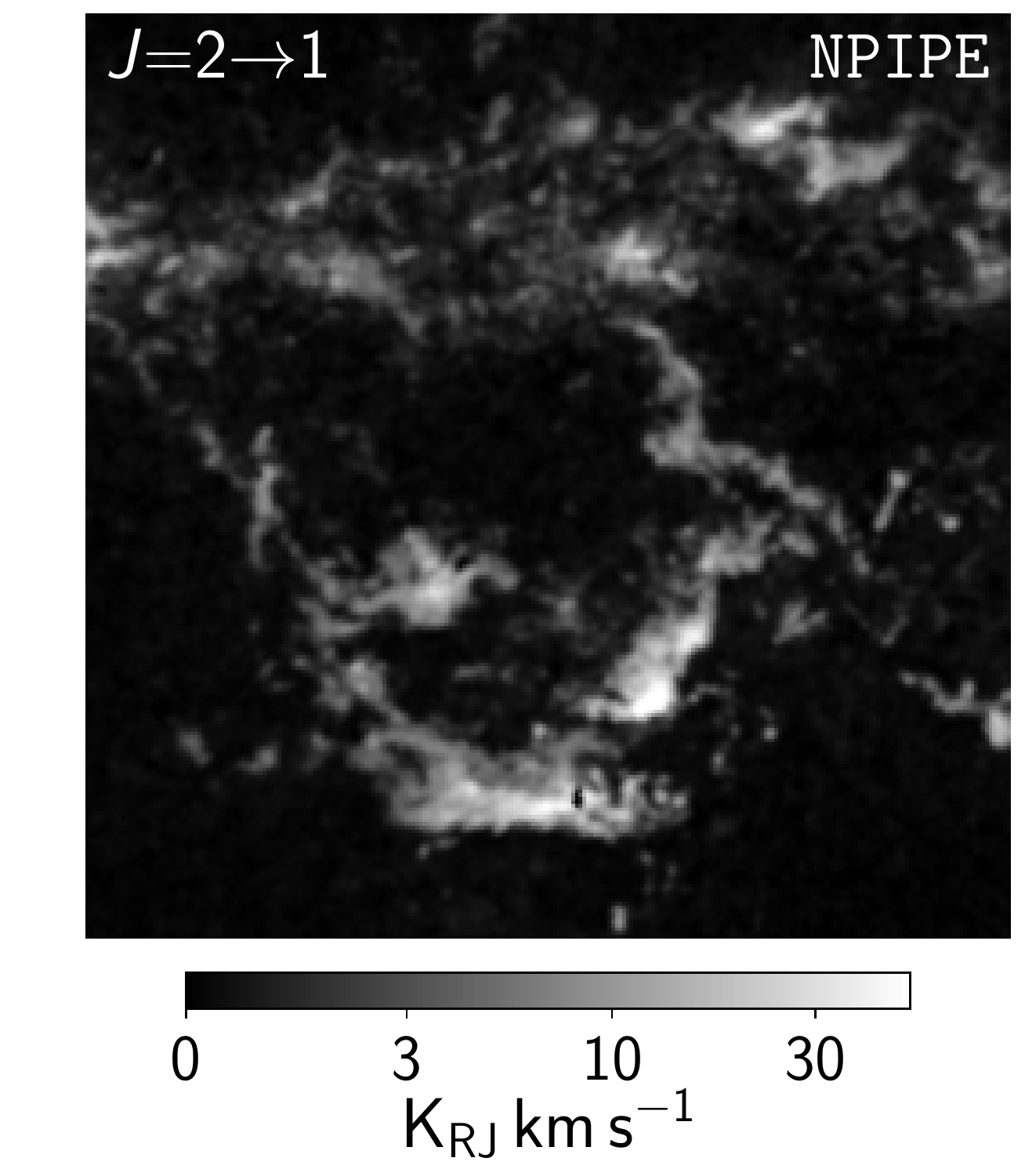}\\
 \caption{$30\deg \times 30\deg$ expansion of various CO emission line maps. All
   maps are smoothed to 14\arcm\ FWHM and are centred on the Orion region, with Galactic coordinates ($l$, $b$) = ($210\deg$, $-9\deg$).}
   \label{fig:orion_co_zoom}
\end{figure}

We now turn our attention to the astrophysical foreground maps resulting from the \commander\ analysis applied to the \npipe\ data, starting with the temperature components.  Figure~\ref{fig:fg_amplitude_T} shows the various temperature amplitude maps included in the analysis. From top to bottom, these are the 1) the combined low-frequency power-law component evaluated at 30\GHz; 2) the thermal dust emission component evaluated at 545\GHz; and 3--5) CO $J$=1$\rightarrow$0,
$J$=2$\rightarrow$1, and $J$=3$\rightarrow$2 line emission. Figure~\ref{fig:orion_co_zoom} compares the CO
$J$=1$\rightarrow$0 and $J$=2$\rightarrow$1 maps with the corresponding Dame et al.\ $J$=1$\rightarrow$0 \citep{dame2001} and the \Planck\ 2015 $J$=2$\rightarrow$1 \citep{planck2014-a12} maps. Overall, \npipe\ maps are in good agreement with previous results.

\begin{figure}[hbtp]
   \centering
   \includegraphics[width=0.48\textwidth]{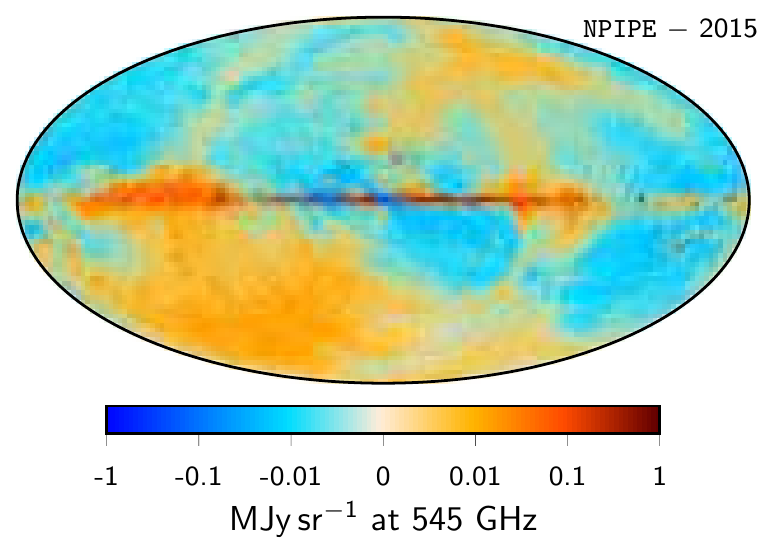}\\
   \includegraphics[width=0.48\textwidth]{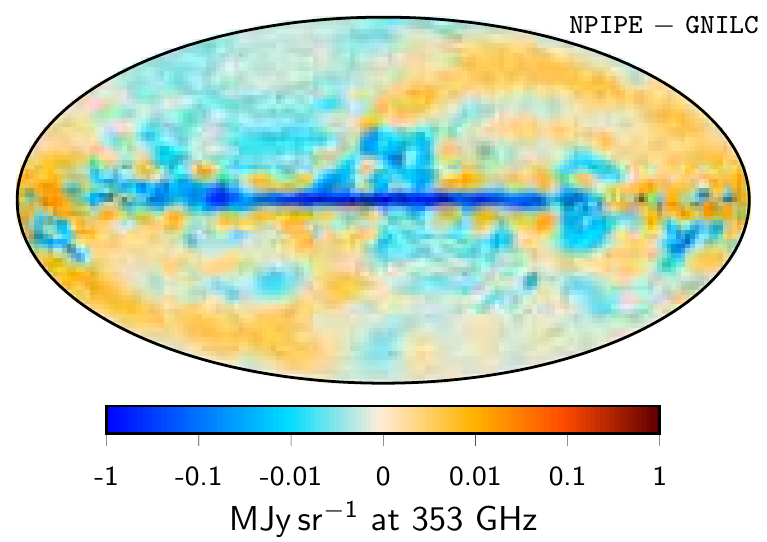}\\
   \caption{\emph{Top:} Stokes $I$ difference map between the \npipe\ and the \Planck\ 2015 \citep{planck2014-a12} dust amplitudes, adjusted for the scaling and offset seen in Fig.~\ref{fig:scatter_dust}. Both dust amplitude maps are smoothed to 1\deg\ FWHM and pixelized with a \healpix\ resolution parameter \nside{256}. (\emph{Bottom:}) Similar difference plot between the \npipe\ dust amplitude map and the \texttt{GNILC} \citep{planck2016-l04} dust amplitudes, adjusted for the scaling and offset seen in Fig. \ref{fig:scatter_dust}. Both dust amplitude maps were smoothed to 80\arcm\ FWHM  and pixelized with a \healpix\ resolution parameter \nside{256}.}
\label{fig:diff_dust_amplitude_T}
\end{figure}

Figure~\ref{fig:diff_dust_amplitude_T} shows difference maps between the \npipe\ and 1) the \commander\ 2015 and 2) the Generalized Internal Linear Combination (\texttt{GNILC}; \citealp{Remazeilles2011b}) 2018 thermal dust amplitude maps. The latter provides an algorithmically independent and recent estimate of thermal dust emission, as discussed in \citet{planck2016-l04}. In both cases, best-fit relative slopes and offsets have been accounted for. We see here that all three maps agree to high precision, with most residuals smaller than 0.01\,MJy\,sr$^{-1}$ at high Galactic latitudes and smaller than 1\,MJy\,sr$^{-1}$ in the Galactic plane.  We further see that the difference between the \npipe\ and \Planck\ 2015 data sets is dominated by the bandpass leakage effect discussed in Sect.~\ref{sec:data_selection}.  For both \Planck\ 2015 and  \texttt{GNILC}, we see a weak imprint of zodiacal light aligned with the ecliptic plane, and for \texttt{GNILC} we additionally note a negative Galactic residual, consistent with the morphology of CO line emission.

\begin{figure}[htpb]
   \centering
   \includegraphics[width=0.35\textwidth]{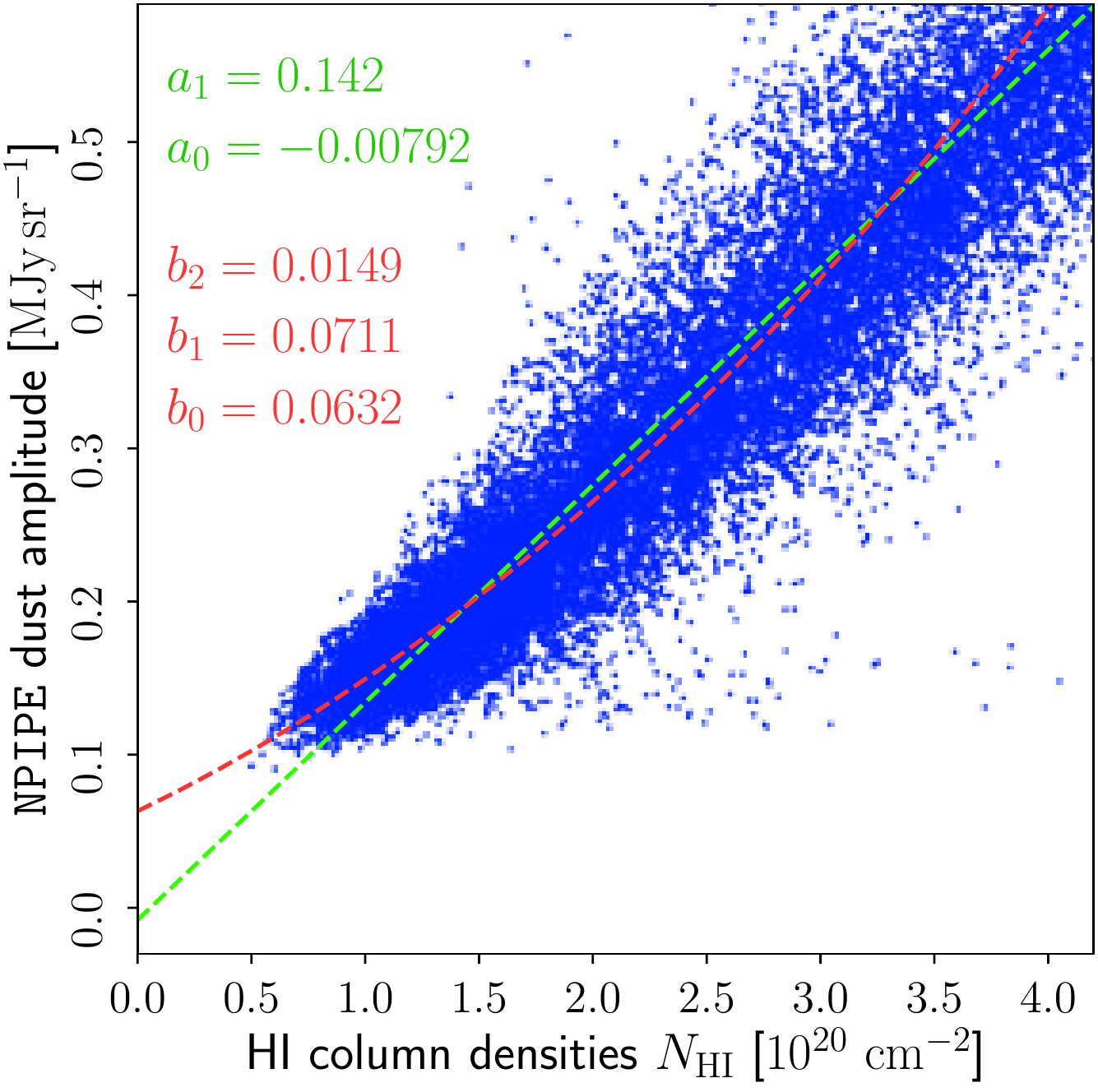}       \\
   \includegraphics[width=0.35\textwidth]{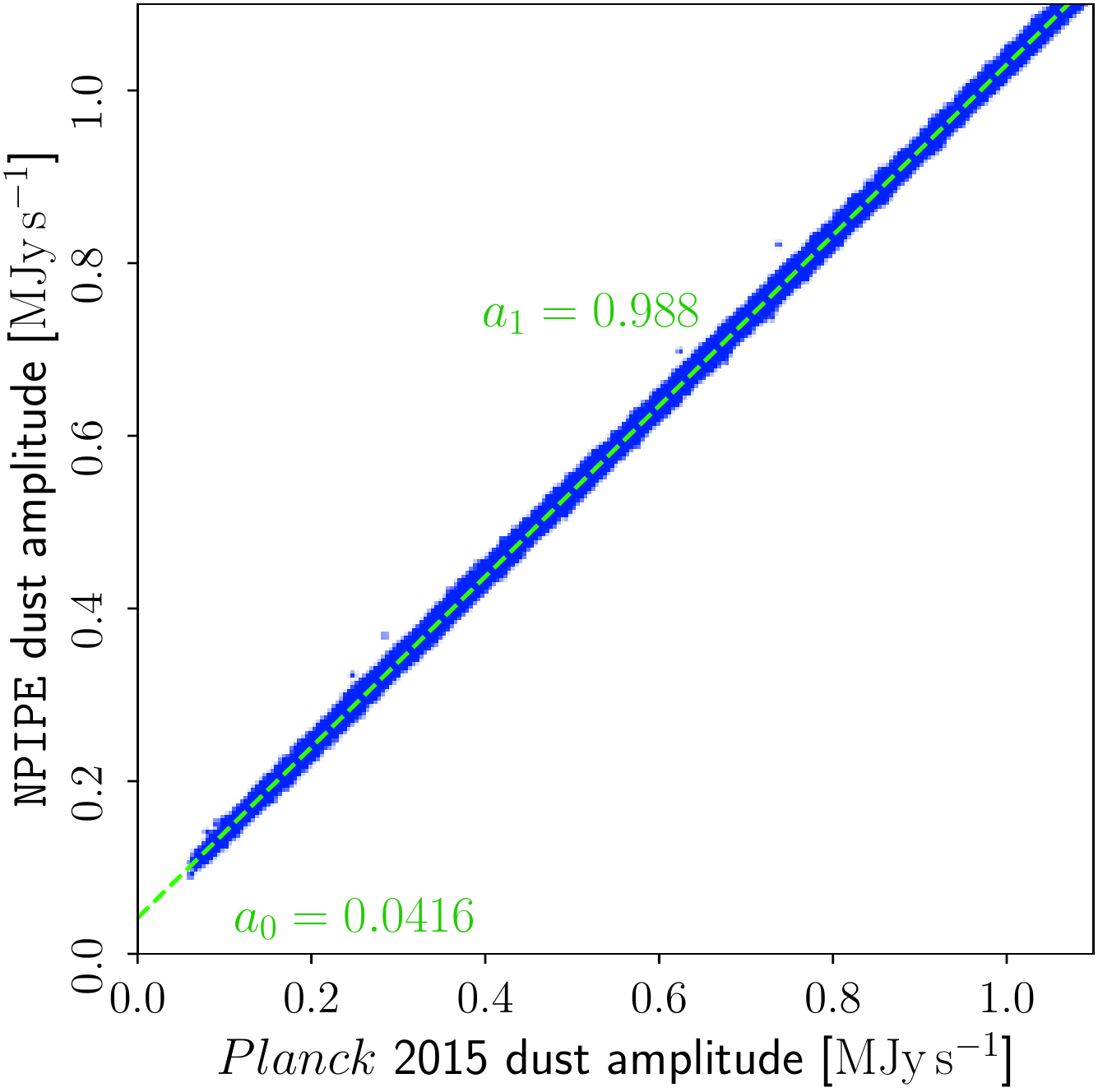}\\
   \includegraphics[width=0.35\textwidth]{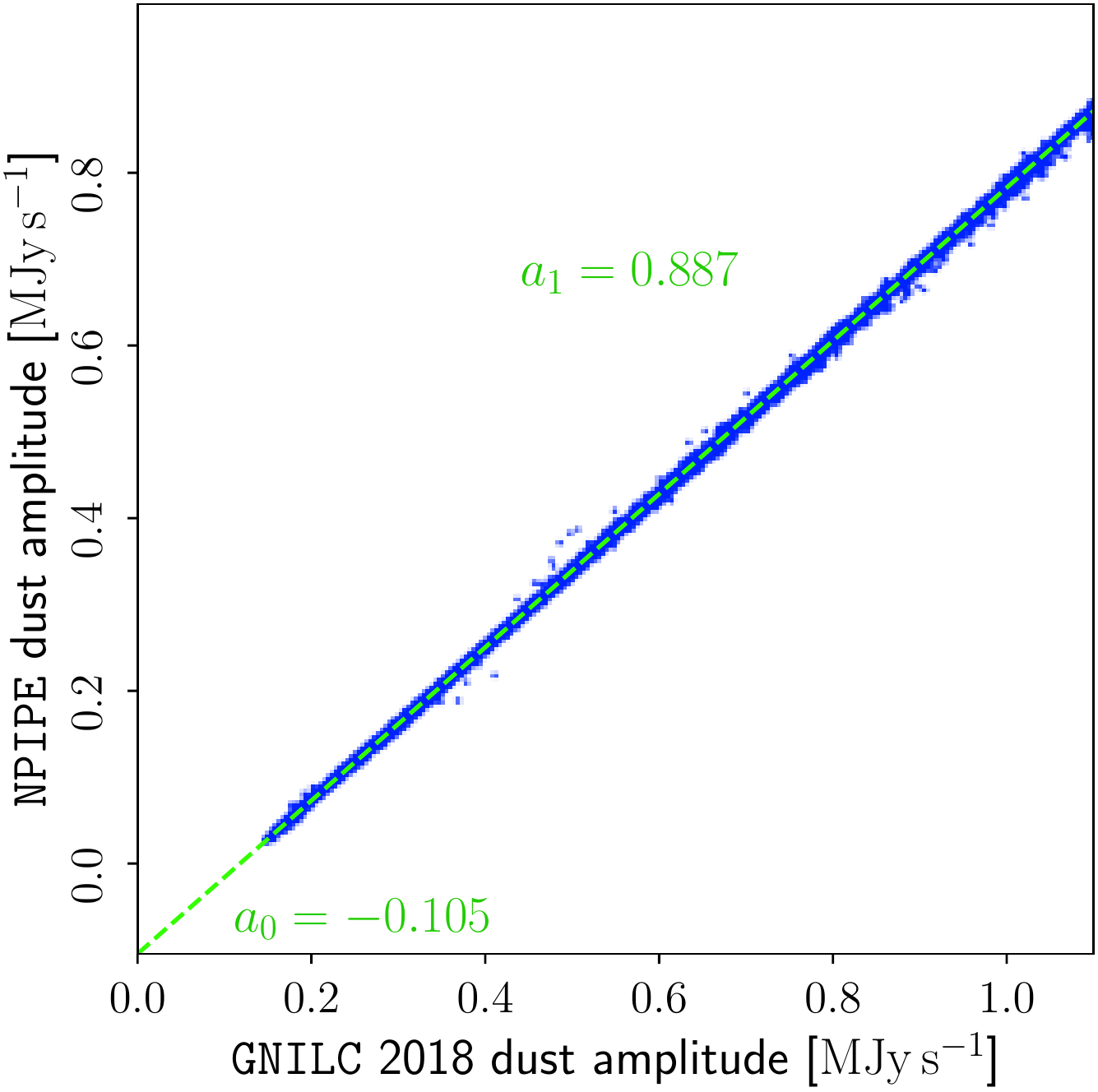}
   \caption{$T$--$T$ scatter plots between the \npipe\ dust amplitude map at 545\GHz\ and three alternative thermal dust estimates.  The panels show, from top to bottom, correlations with respect to: the HI4PI survey \citep{Lenz_et_al:2017}; the \Planck\ 2015 thermal dust amplitude map; and the \texttt{GNILC} 2018 thermal dust amplitude map. The top and middle panels show maps smoothed to a common resolution of 60\arcm\ FWHM, while the bottom panel employs a smoothing scale of 80\arcm\ FWHM, determined by the resolution of the \texttt{GNILC} map.  The numbers marked by $b_2$, $b_1$, and $b_0$ (first plot) correspond to the best-fit polynomial parameters of a quadratic model (red dashed line, $b_0$ being the intercept), while $a_1$ and $a_0$ (all plots) correspond to the slope and intercept of the best fit line (green dashed line) through each distribution of points.}
   \label{fig:scatter_dust}
\end{figure}

Figure~\ref{fig:scatter_dust} shows $T$--$T$ scatter plots between the \npipe\ thermal dust amplitude map and three alternative thermal dust tracers. The top panel shows the correlation with respect to the HI4PI survey \citep{Lenz_et_al:2017} at low column densities, which provides the statistically most independent test of the derived amplitude map.  This correlation is also used to determine the zero-level of the \npipe\ 857-1 detector map (including all pixels with column densities up to
$4\times10^{20}\,\textrm{cm}^{-2}$ \citep{Lenz_et_al:2017}, and thereby also in effect the overall zero-level of the \npipe\ thermal dust component.  As seen in Fig.~\ref{fig:scatter_dust}, the relative residual offset between these two maps after final processing is 0.0079\,MJy\,sr$^{-1}$ according to a linear fit, which is negligible compared to intrinsic systematic uncertainties.  For comparison, a quadratic fit (indicated by a red dashed curve) results in a relative offset of 0.06\,MJy\,sr$^{-1}$.  The middle and bottom panels show similar correlation plots with respect to the \Planck\ 2015 \commander\ and the \Planck\ 2018 \texttt{GNILC} thermal dust amplitude maps.  In both cases we observe very tight correlations. For the \Planck\ 2015 map, the relative slope is 0.988, indicating that the high-frequency calibration of \npipe\ and \Planck\ 2015 agree to about 1\,\%.  This is reassuring, considering that \npipe\ calibrates the 545-GHz channel on the orbital dipole and also adopts thermodynamic units of $K_{\textrm{CMB}}$ at both 545 and 857\,GHz, while \Planck\ 2015 calibrated the 545-GHz channel with planets, and used flux density units of MJy\,sr$^{-1}$ for the two highest frequency channels.  For \texttt{GNILC}, we observe a relative slope of 0.887, which is simply due to the fact that no colour corrections were applied to the \texttt{GNILC} map, whereas the \commander\ maps are all measured relative to a sharp reference frequency of 545\,GHz. \citet{odegard2019} performed a re-calibration analysis of the \Planck\ 2015 HFI maps based on COBE-FIRAS, obtaining results similar to those presented here.  For instance, they derive a slope between \Planck\ 545\,GHz and HI4PI of 0.144 when adopting a threshold of $2.5\times10^{20}\,\textrm{cm}^{-2}$ and (crucially) defining the \Planck\ data in units of MJy\,sr$^{-1}$, while for \npipe\ we find a slope of 0.132.  The resulting relative difference of 5\,\% is likely dominated by uncertainties in the 545\,GHz bandpass profile, given that \npipe\ 545\,GHz is calibrated in thermodynamic units while \citet{odegard2019} calibrate in flux density units.

\begin{figure*}[htpb]
   \centering
   \includegraphics[width=0.32\textwidth]{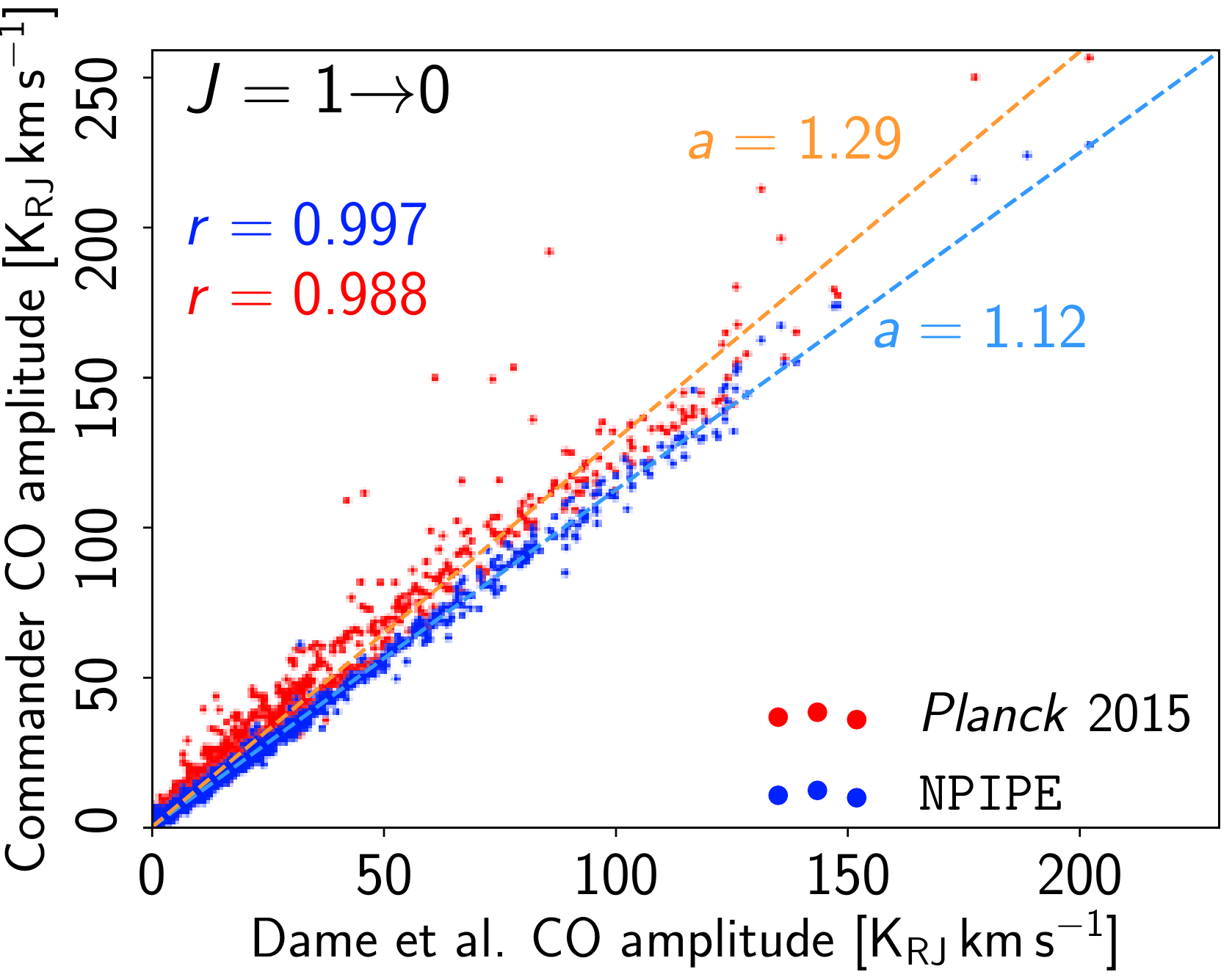}
   \includegraphics[width=0.32\textwidth]{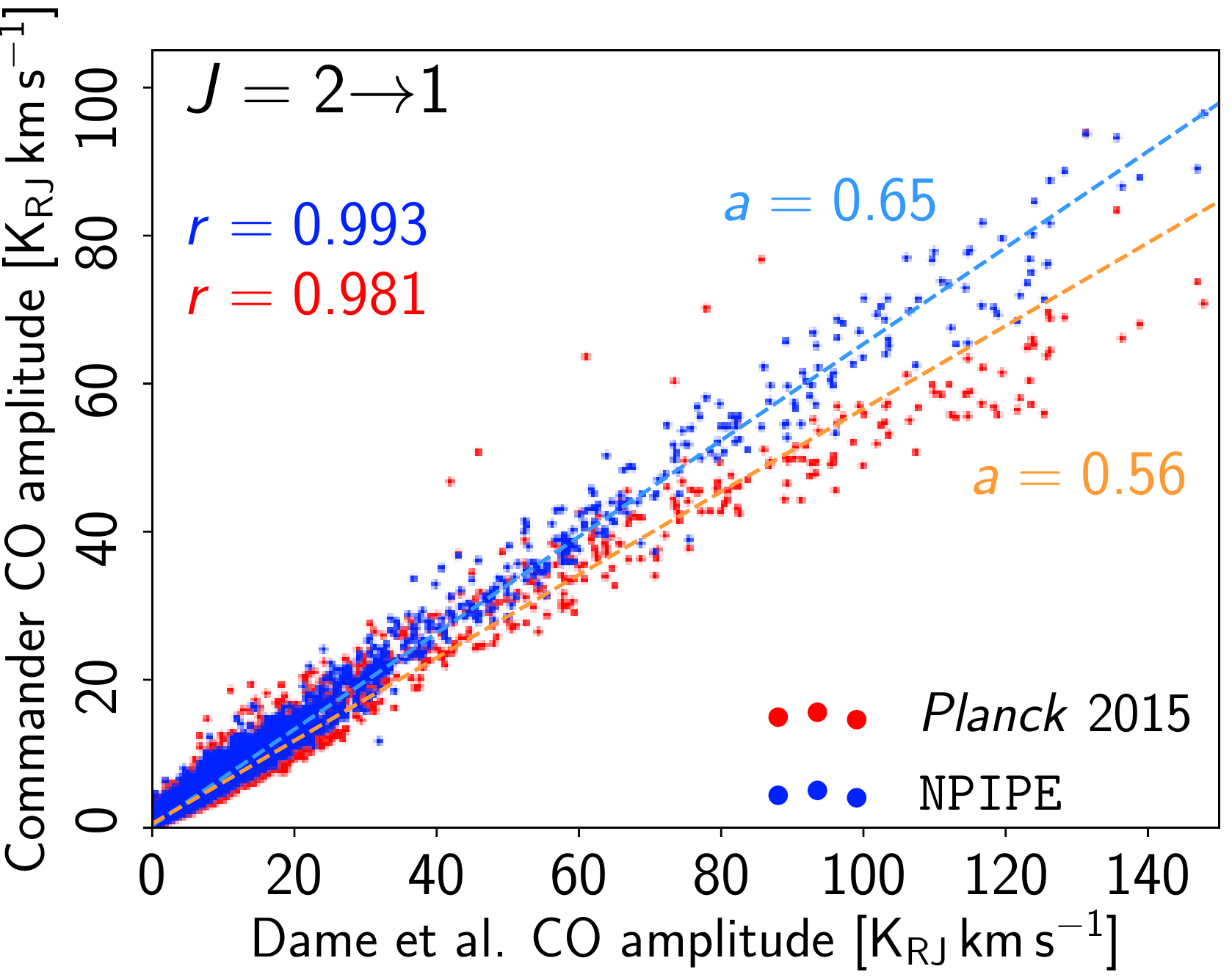}
   \includegraphics[width=0.32\textwidth]{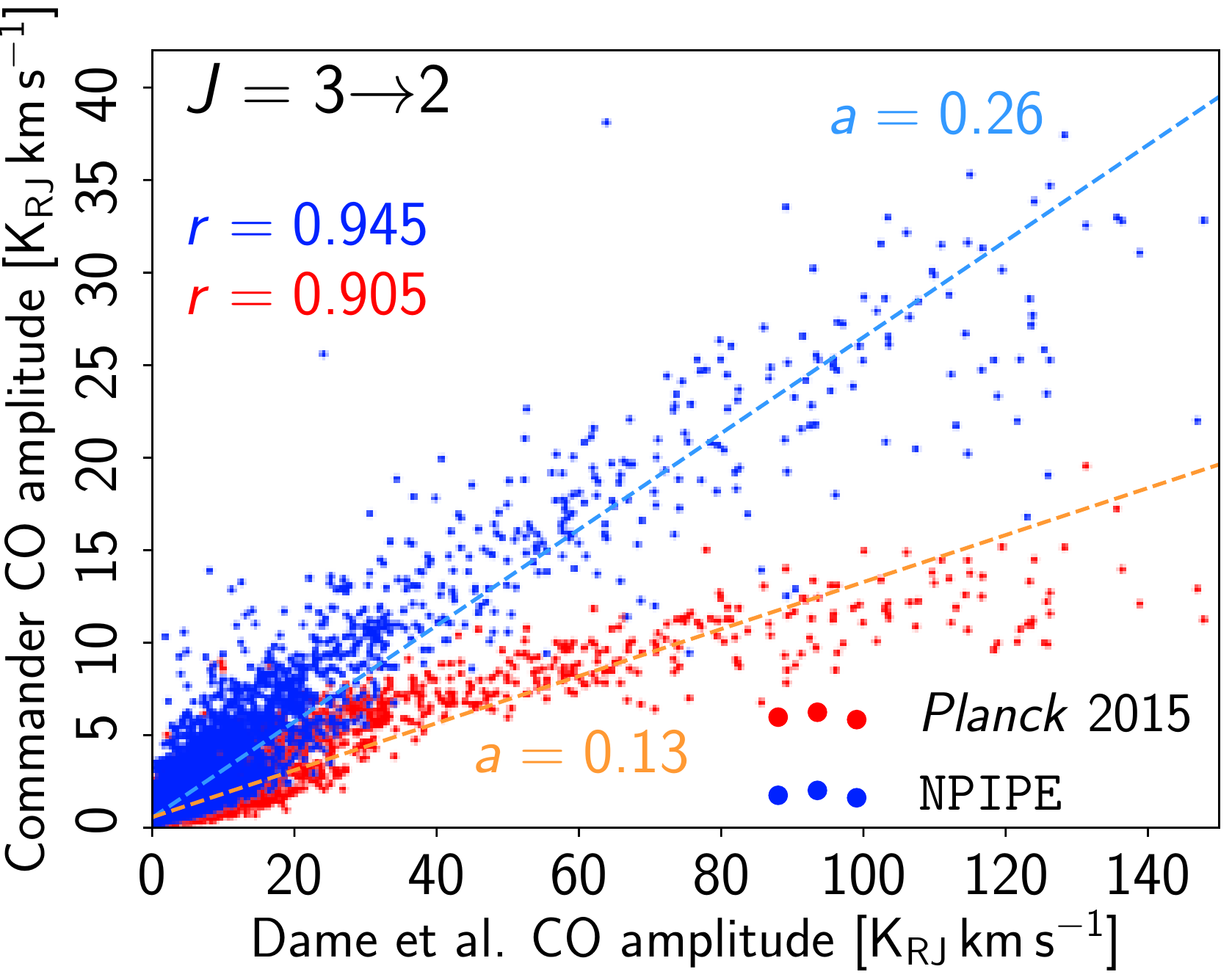}
   \caption{Comparison of $T$--$T$ scatter plots evaluated between the Dame et al. CO map \citep{dame2001} and the \commander\ \npipe\ (blue) and \Planck\ 2015 (red) CO maps. Panels show, from left to right, results for the CO $J$=1$\rightarrow$0, $J$=2$\rightarrow$1, and $J$=3$\rightarrow$2 line maps, respectively.  All maps have been smoothed to $1\deg$ FWHM, and pixelized with $N_\mathrm{side} = 64$.  The parameter marked by ``$a$'' is the best-fit linear slope of the scatter plot including values between 0.01 and 150 $\mathrm{K_{\small{RJ}}\,km\,s^{-1}}$ of the Dame et al. $J$=1$\rightarrow$0 map. The parameter $r$ is the Pearson's correlation coefficient between Dame et al. and the respective CO amplitudes.}
   \label{fig:scatter_co_amplitude_T}
\end{figure*}

Figure~\ref{fig:scatter_co_amplitude_T} compares $T$--$T$ scatter plots between the three \npipe\ and \Planck\ 2015 CO
line maps and the Dame et al.\ maps.  Qualitatively, all estimates agree very well with each other, both visually in the maps and in terms of scatter plots.  However, the \npipe\ maps exhibit generally stronger correlations with respect to the Dame et al.\ survey than do the \Planck\ 2015 maps.  Specifically, the Pearson's correlation coefficients for \Planck\ 2015 (for 1$\rightarrow$0, 2$\rightarrow$1, and 3$\rightarrow$2) are $r=0.988$, 0.981, and 0.905, respectively, while for \npipe\ they are $r=0.997$, 0.993, and 0.945.  These improvements are due to a more fine-grained
set of detector maps available in \npipe\ than in \Planck\ 2015, better control of instrumental systematic effects in high
S/N regions, and better transfer-function and ADC corrections.

\begin{figure*}[htpb]
   \centering \includegraphics[width=0.48\linewidth]{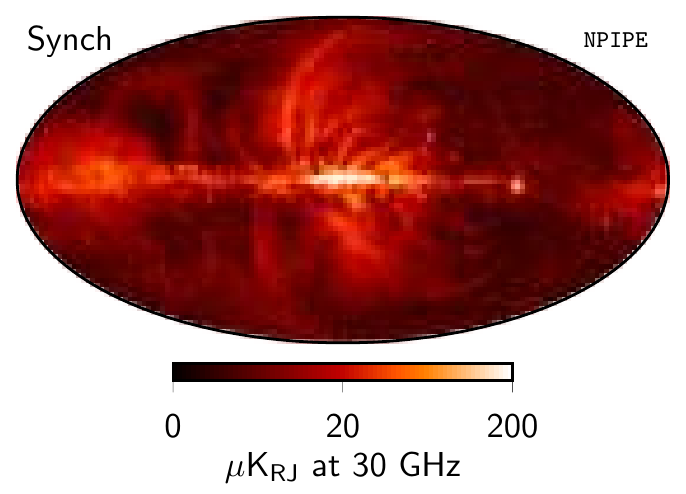} \includegraphics[width=0.48\linewidth]{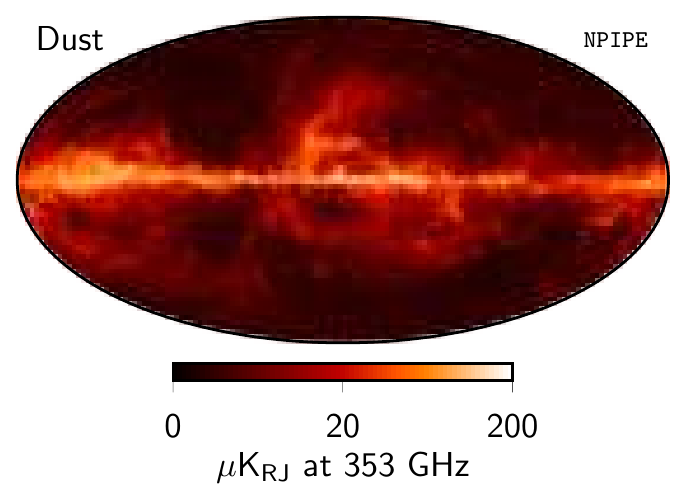}\\ \includegraphics[width=0.48\linewidth]{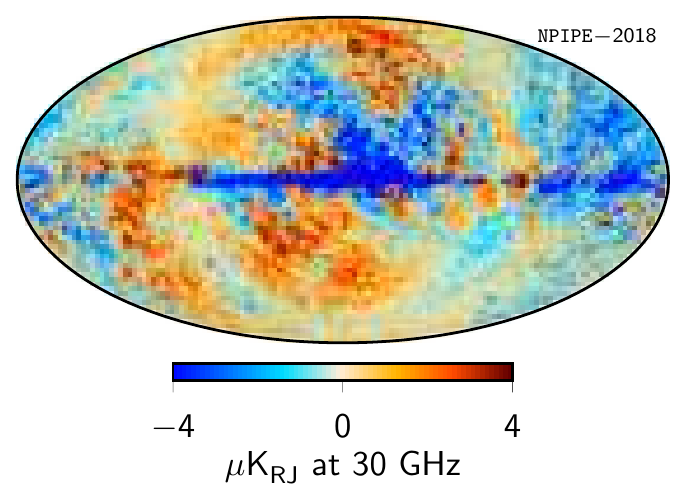} \includegraphics[width=0.48\linewidth]{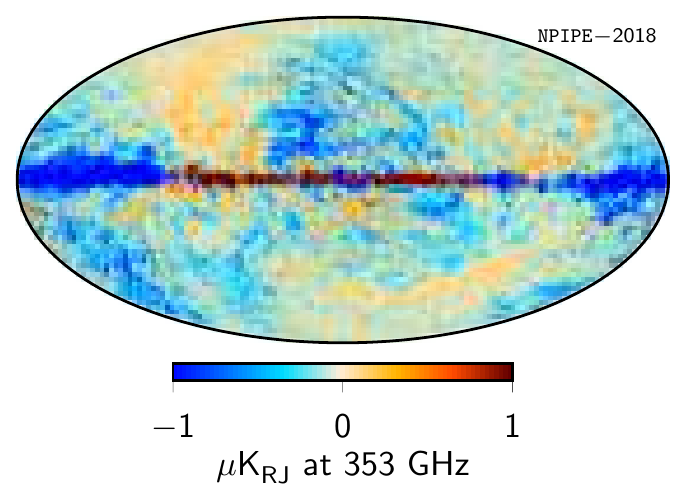} 
\caption{\emph{Top:}  Synchrotron and thermal dust polarization-amplitude ($P=\sqrt{Q^2+U^2}$) maps derived from the \npipe\ data set. The synchrotron map and thermal dust maps are smoothed to angular resolutions of 40\arcm\ and 5\arcm\ FWHM, respectively. \emph{Bottom:} Corresponding polarization-amplitude difference maps taken between the \npipe\ and \Planck\ 2018 component maps.  Both maps are smoothed to a common resolution of 60\arcm\ FWHM.  The top panels use the nonlinear \Planck\ colour scale, while the bottom panels use linear colour scales.  } 
\label{fig:fg_amp}
\end{figure*}

Next we consider the reconstructed polarized foreground emission.  Figure~\ref{fig:fg_amp} shows \npipe\ synchrotron and thermal dust polarization amplitude maps, as well as corresponding difference maps with respect to \Planck\ 2018, all plotted in brightness temperature units.  These two data sets agree very well in terms of thermal dust emission at high Galactic latitudes, with most of the sky exhibiting differences smaller than $0.2\,\mu\textrm{K}_{\mathrm{RJ}}$.  The Galactic plane shows larger differences, with morphology similar to the absolute level of thermal dust emission, but with alternating sign along the plane.  Such features typically arise from spatial variations in the thermal dust temperature or from different instrumental parameters in the form of detector polarization efficiency and angle.  For synchrotron, larger relative differences are observed, both at low and high Galactic latitudes.  \npipe\ uses active polarization priors for the LFI frequencies, as was done in the HFI 2018 DPC processing, but different from what was done in the LFI 2018 processing.  This approach has increased the overall level of polarization in the LFI sky maps, bringing them into better internal agreement among themselves, and also in better agreement with the WMAP observations (see Figs.~\ref{fig:wmap_comparison_30} and \ref{fig:wmap_comparison_44}).

\begin{figure}[htpb]
   \centering
   \includegraphics[width=0.45\textwidth]{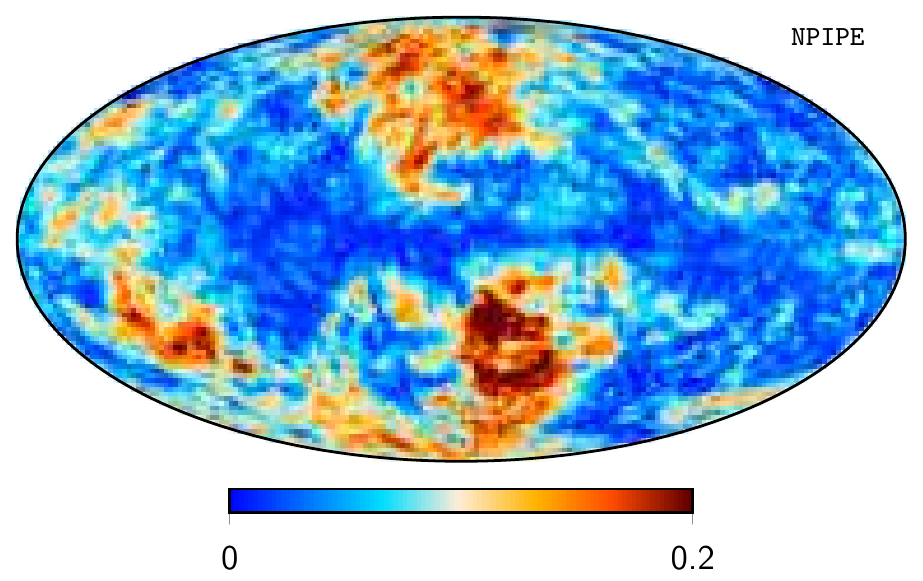}\\
   \includegraphics[width=0.45\textwidth]{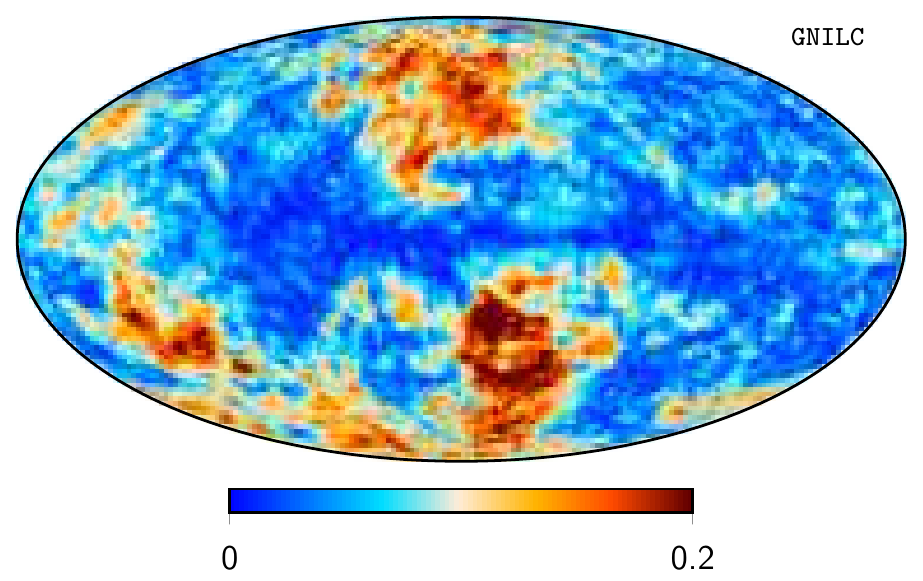}\\
   \includegraphics[width=0.45\textwidth]{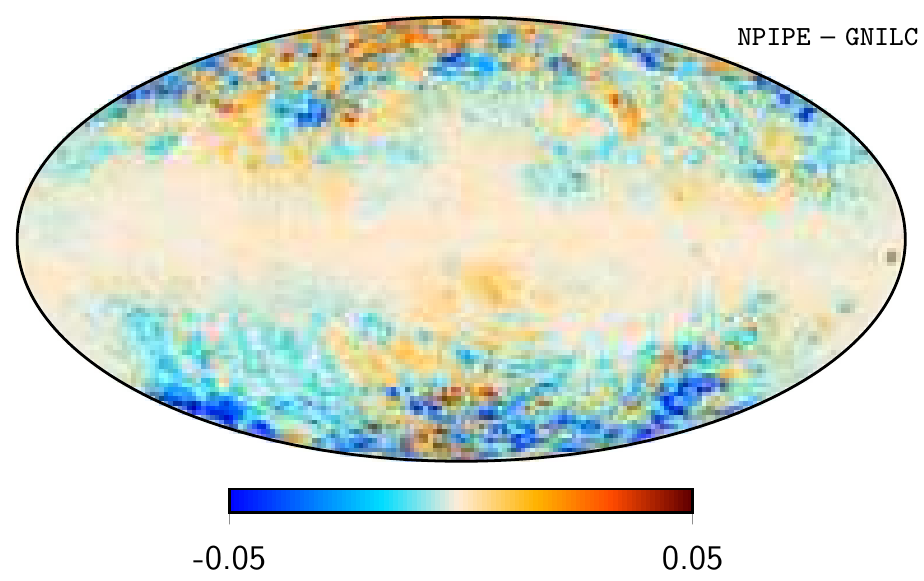}
\caption{Comparison of thermal dust polarization-fraction maps, as evaluated from the \npipe\ (top) and \texttt{GNILC} 2018 (middle) data sets, and a difference map (bottom) between the two.  All maps are smoothed to 80\arcm\ FWHM.  We have subtracted a monopole of 389 $\mathrm{\mu K}$ from the \texttt{GNILC} intensity map \citep{planck2016-l11A}.}
\label{fig:polfrac_compsep}
\end{figure}

Figure~\ref{fig:polfrac_compsep} compares the thermal dust polarization fraction as estimated from the \npipe\ data set
with \commander\ (top panel) and as estimated from the \Planck\ 2018 data set with \texttt{GNILC} (middle panel). The bottom panel shows their difference.  Overall, we see that these data sets agree exceedingly well in terms of polarization fractions, with low latitude differences being much smaller than 1\,\%, and high-latitude differences being dominated by noise-like features.

\subsection{Assessment of uncertainties}

We conclude this section with an assessment of uncertainties in the derived products, for which we adopt two fundamentally different types of estimates. The first type of estimate is based on simulations.  As described in Sect.~\ref{sec:simulations}, the \npipe\ data set is accompanied with a set of 600 complete end-to-end-simulations, and each of these is propagated through the \commander\ analysis described above.  To ensure that the resulting noise levels match the true data set, spectral indices (i.e., mixing matrices) are fixed at the values derived from the real observations, and only amplitudes are fitted freely; this is the same approach as used for the \Planck\ 2013, 2015, and 2018 data releases \citep{planck2013-p06, planck2014-a11, planck2016-l04}.

\begin{figure}[htpb]
   \centering
   \includegraphics[width=0.48\textwidth]{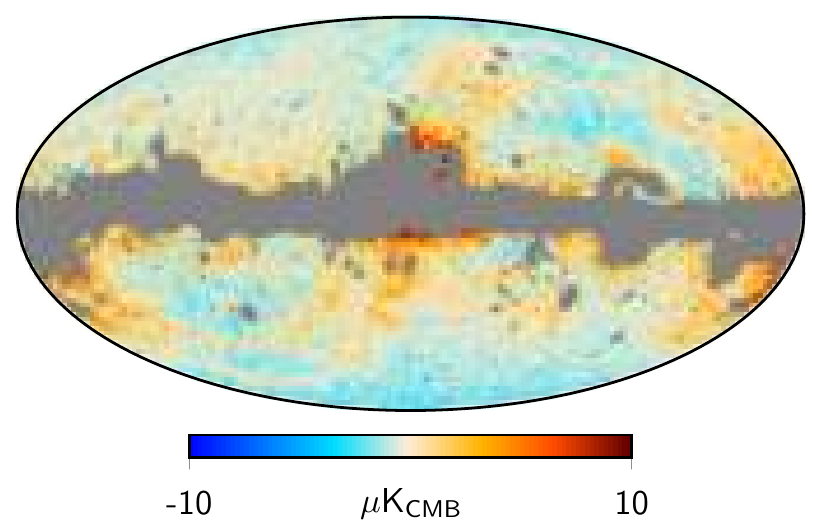}
\caption{Stokes $I$ difference map between the \commander\ \npipe\ CMB map of simulation No.~200 and the corresponding input CMB map of the simulation.  Both maps are pixelized with a \healpix\ resolution parameter \nside{64}.
The map has been masked with the mask shown in Fig.~\ref{fig:cmb_mask_T}. } 
\label{fig:cmb_diff_sim0200}
\end{figure}

Figure~\ref{fig:cmb_diff_sim0200} shows the difference between a single foreground-cleaned \commander\ CMB temperature simulation and the corresponding true input CMB realization, smoothed to a resolution of 1\deg\ FWHM.  The grey region indicates the \npipe\ temperature-analysis mask defined in Fig.~\ref{fig:cmb_mask_T}.  For most of the sky, we see that the reconstruction error is less than about 3\muK, with slightly higher values near the edge of the mask, in particular close to regions with bright free-free emission.  For most cosmological analyses, these errors are very small compared to the CMB fluctuations, which have a standard deviation of about 70\muK\ on these angular scales.  The \commander\ temperature analysis for \npipe\ relies on a fine-grained detector-map analysis, and the computational cost of producing such simulations is very high.  At the time of publication of this paper, 100 such simulations are available.

\begin{figure*}[htpb]
   \centering
   \includegraphics[width=0.46\linewidth]{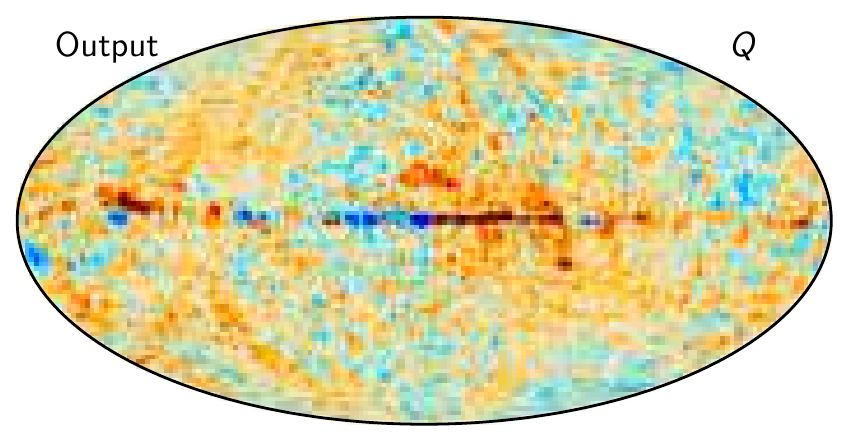}
   \includegraphics[width=0.46\linewidth]{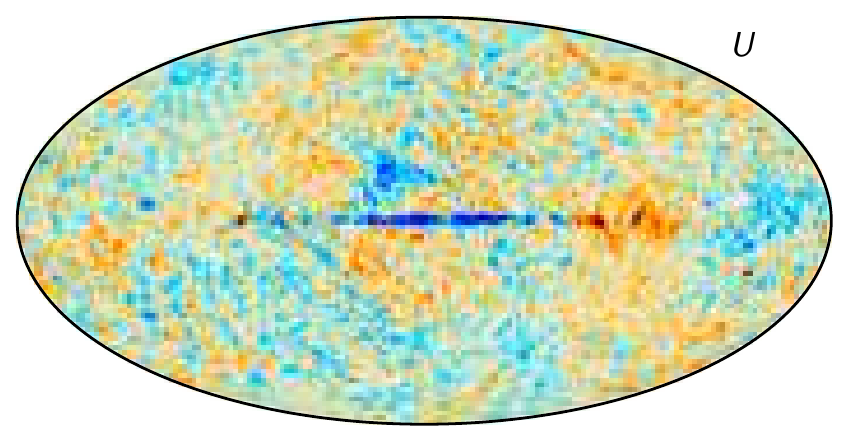}\\
   \includegraphics[width=0.46\linewidth]{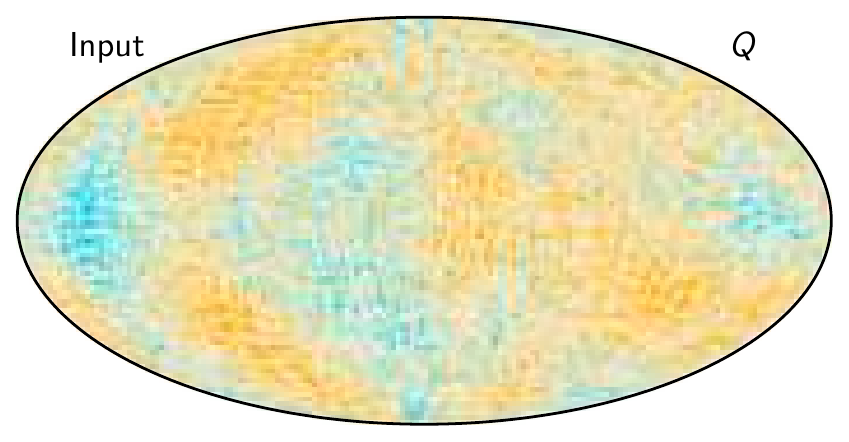}
   \includegraphics[width=0.46\linewidth]{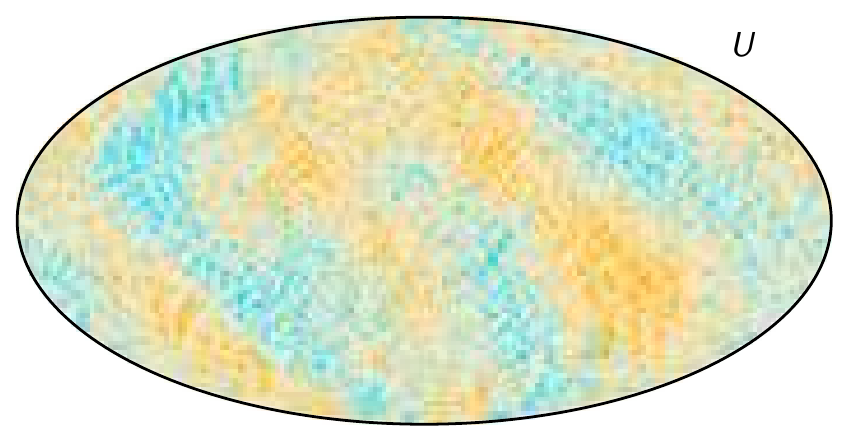}\\
   \includegraphics[width=0.46\linewidth]{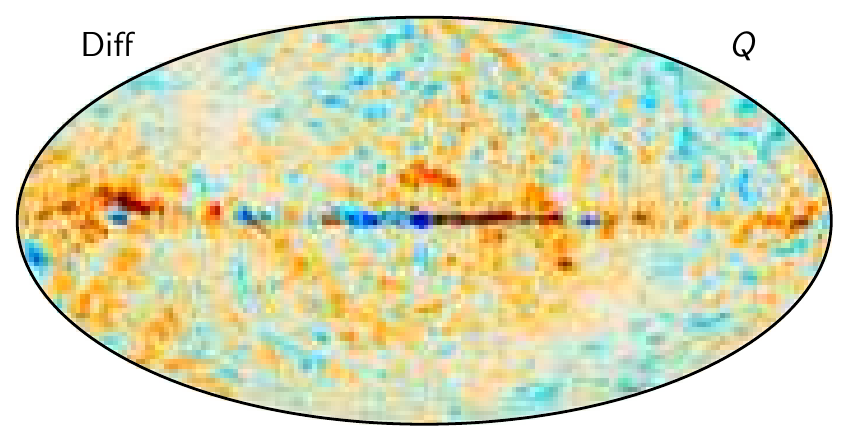}
   \includegraphics[width=0.46\linewidth]{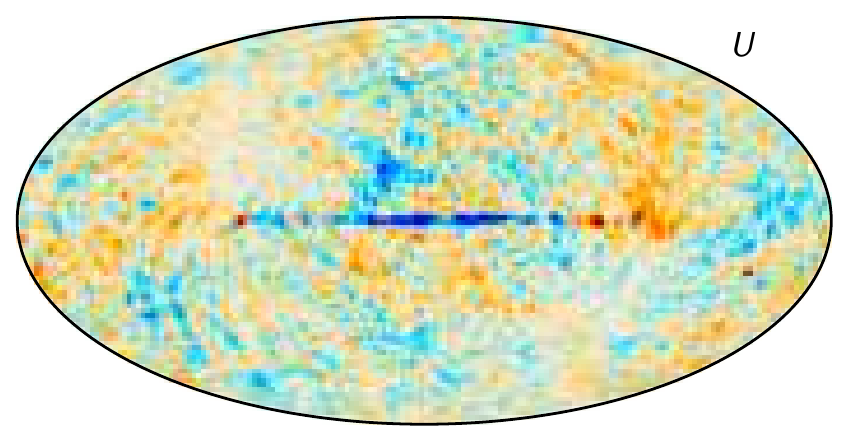}\\
   \includegraphics[width=0.35\linewidth]{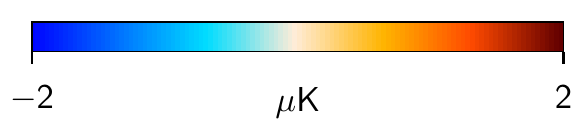}
   \caption{Comparison of end-to-end reconstructed (\emph{top row}) and input (\emph{middle row}) \npipe\ simulations for the Stokes $Q$ and $U$ CMB maps. The bottom row shows the difference between the output and input sky maps. All maps are smoothed to a common angular resolution of $2^{\circ}$ FWHM.}
   \label{fig:cmb_pol_sim}
\end{figure*}

Figure~\ref{fig:cmb_pol_sim} shows a similar comparison for Stokes $Q$ and $U$.  The top panels show the reconstructed, foreground-cleaned, CMB $Q$ and $U$ maps.  The middle panels show the input CMB maps. The bottom panels show the differences.  At a visual level, the simulated process of measuring the sky, processing with \npipe\, and fitting and removing foregrounds, provides a noisy image of the CMB polarization sky, with significant systematic errors present both at low and high Galactic latitudes. However, it is evident, even at a visual level, that the \npipe\ maps provide a clear tracer of true large-scale CMB
features.

\begin{figure}[hbtp]
   \centering
   \includegraphics[width=0.48\textwidth]{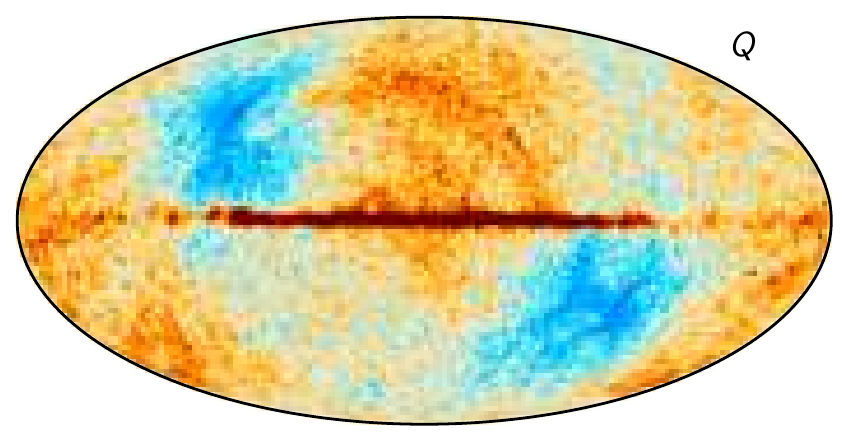}\\
   \includegraphics[width=0.48\textwidth]{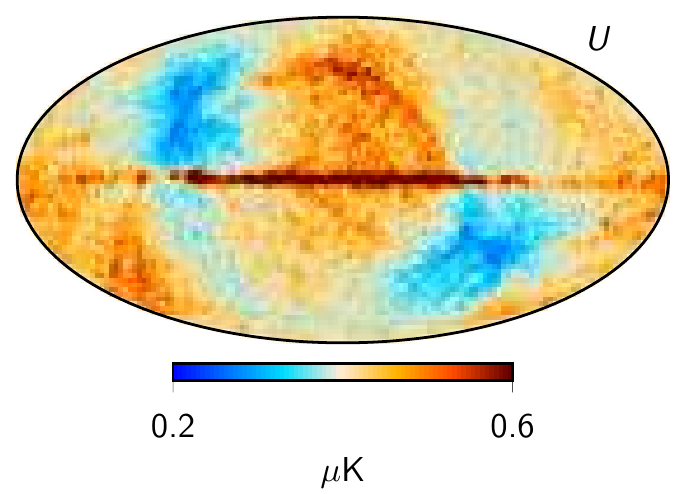}
   \caption{Standard deviation evaluated from 100 end-to-end \npipe\ full-mission simulations of CMB $Q$ and $U$ maps, as derived with \commander.  Both maps are smoothed to 2\deg\ FWHM.}
   \label{fig:cmb_pol_stddev}
\end{figure}

Figure~\ref{fig:cmb_pol_stddev} shows the standard deviation of the foreground-cleaned \commander\ CMB polarization simulations at
1\deg\ FWHM resolution.  We see that the overall noise standard deviation varies between 0.2 and 0.6\muK\ at high Galactic latitudes, with a spatial distribution dominated by the scanning pattern of the \Planck\ telescope.  At low Galactic latitudes, higher values are observed, due to the combination of additional foreground uncertainties and instrumental systematic effects, particularly in the
form of intensity-to-polarization leakage.

\begin{figure*}[htpb]
   \centering
   \includegraphics[width=\textwidth]{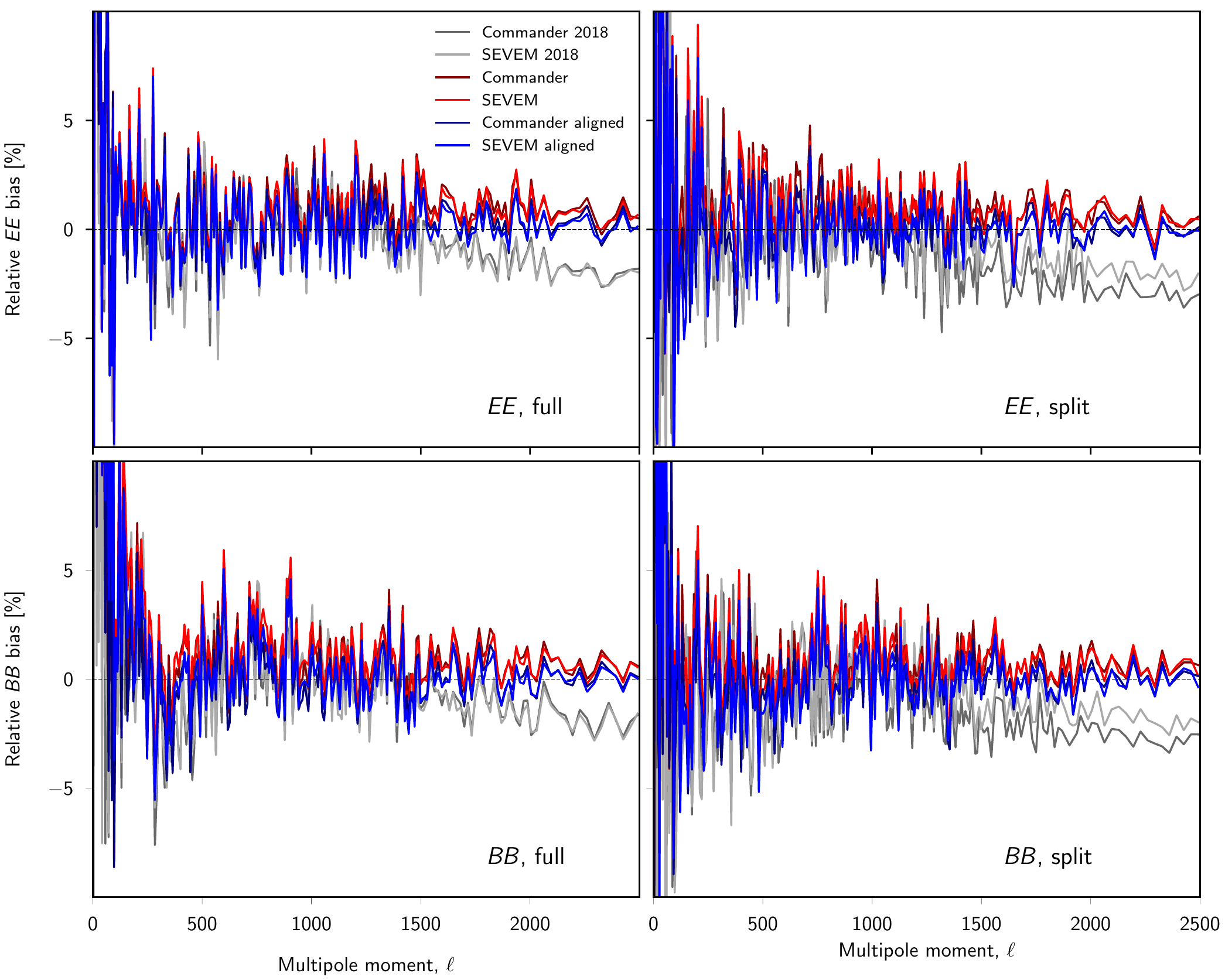}\\
      \caption{Power spectrum consistency between the foreground-cleaned \commander\ (dark curves) and \sevem\ (light curves) CMB polarization map and corresponding
end-to-end-simulations. Each panel shows the fractional difference between the angular power spectrum computed from the observed data and the mean of the simulations. Blue and red curves show results derived for \npipe\ data using simulations \emph{with} and \emph{without} noise alignment, respectively, while grey curves show similar results derived from \Planck\ 2018 data using simulations \emph{with} noise alignment. Rows show results for $EE$ (\emph{top}) and $BB$ (\emph{bottom}) spectra, while columns show results for full-mission (\emph{left}) and split (\emph{right}) data. In the latter case, A-B split results are shown for \npipe, while half-mission splits are shown for \Planck\ 2018.
   } \label{fig:powspec_consistency}
\end{figure*}

Figure~\ref{fig:powspec_consistency} shows the fractional difference between angular power spectra computed from the observed foreground-cleaned \commander\ and \sevem\ CMB polarization maps, and the mean of the simulations\footnote{Note that there are differences between the \commander\ and \sevem\ simulations. While the \commander\ simulations are a foreground-cleaned version of each realization, the \sevem\ ones are generated by propagating CMB plus noise simulations through the pipeline.}, specifically
\begin{linenomath*}
\begin{equation}
  \eta_{\ell} \equiv \frac{D_{\ell}^{\mathrm{data}} - \left<
  D_{\ell}^{\mathrm{sim}}\right>}{\left<
  D_{\ell}^{\mathrm{sim}}\right>},
\end{equation}
\end{linenomath*}
as computed with \texttt{PolSpice} \citep{Chon:2003gx} over the \Planck\ 2018 common polarization mask. Blue and red curves show \npipe\ results with and without noise alignment (i.e., rescaling of the noise), respectively, while grey curves show similar results for \Planck\ 2018, also with noise alignment, as reproduced from figure~12 in \citet{planck2016-l04}. 

Overall, we see that the \npipe\ polarization simulations without noise adjustment (red curves) agree with the observed data to about 1\,\% in power, which is better than the precision of the \Planck\ 2018 simulations {\it after} noise adjustment (grey curves).  However, it is important to note that the raw \npipe\ simulations \emph{under-estimate} the total power in the real data, as opposed to the \Planck\ 2018 simulations, which \emph{over-estimate} the total power level.  This difference is important, because it allows a straightforward path to accurate noise readjustment, simply by adding slightly more noise.  Accordingly, a second set of \npipe\ simulations is also provided, for which the power deficit has been corrected by the addition of scale-dependent noise. The blue curves in Fig.~\ref{fig:powspec_consistency} show the corresponding power consistency after accounting for this extra power.  Following this noise addition, we find excellent agreement between the observed \npipe\ data and simulations.

\begin{figure}[htpb]
   \centering
   \includegraphics[width=0.48\textwidth]{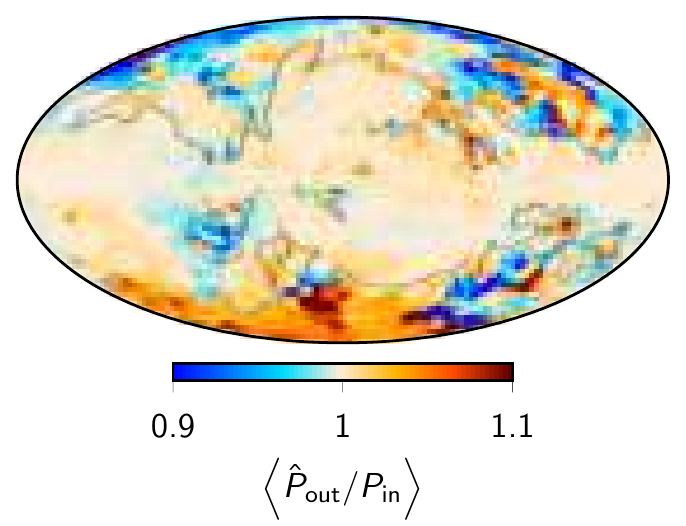}
   \caption{Ratio between simulation output and input thermal dust
     polarization amplitude at 353\GHz, smoothed to an angular
     resolution of $3^{\circ}$ FWHM. Gray lines indicate where the
     polarization amplitude is $3\muK_{\mathrm{RJ}}$, corresponding
     roughly to a signal-to-noise ratio of unity.  }
   \label{fig:dust_transfunc}
\end{figure}

Figure~\ref{fig:dust_transfunc} shows a simulation-based comparison for the thermal dust polarization amplitude. In this case, we estimate the square of the polarization amplitude by cross-correlating the reconstructed A- and B-split thermal dust amplitude maps at 353\GHz, smooth this to an effective angular resolution of $3^{\circ}$ FWHM, and divide by the corresponding square of the simulation input thermal dust amplitude. To reduce noise, we average over 300 simulations, before finally computing the square root to obtain an estimate of the linear polarization amplitude. The grey lines in Fig.~\ref{fig:dust_transfunc} indicate a polarization amplitude of $3\muK_{\mathrm{RJ}}$ at 353\GHz, which corresponds roughly to a signal-to-noise ratio of unity.  Overall, we see that wherever the signal-to-noise ratio is significant, the reconstructed thermal dust amplitude is unbiased to $\lesssim1\,\%$, while where the signal-to-noise ratio is less than unity, biases up to 5--10\,\% may be observed, corresponding to an absolute error of $\lesssim 0.3\muK_{\mathrm{RJ}}$. This value thus defines a systematic uncertainty for the \npipe\ thermal dust polarization map at 353\GHz, and takes into account the full end-to-end processing, including calibration, mapmaking and component separation.

\begin{figure*}[htpb]
   \centering
   \includegraphics[width=0.4\linewidth]{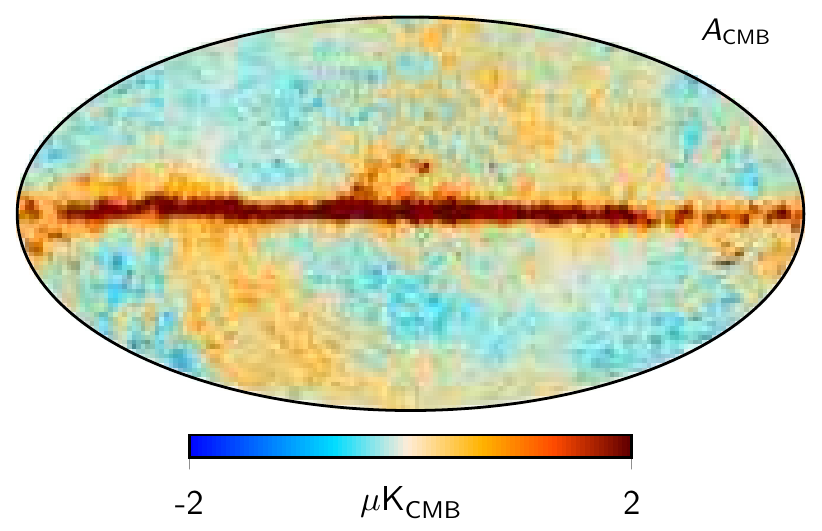}
   \includegraphics[width=0.4\linewidth]{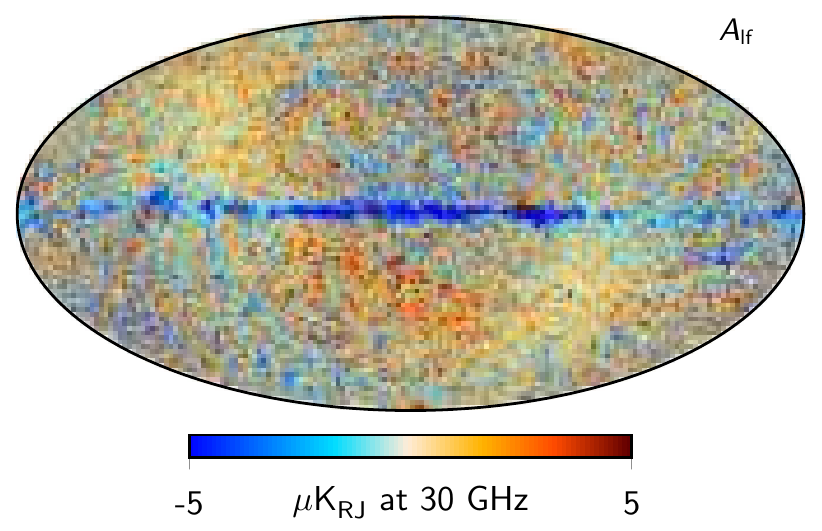}\\
   \includegraphics[width=0.4\linewidth]{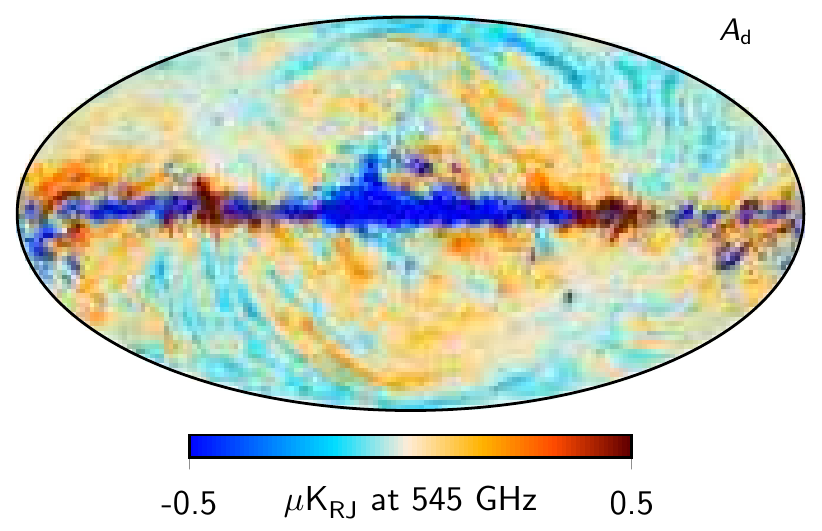}
   \includegraphics[width=0.4\linewidth]{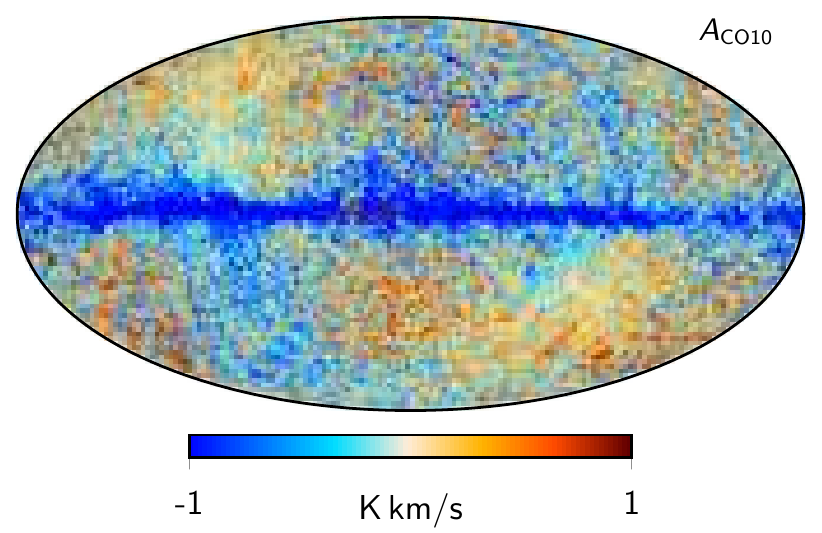}\\
   \includegraphics[width=0.4\linewidth]{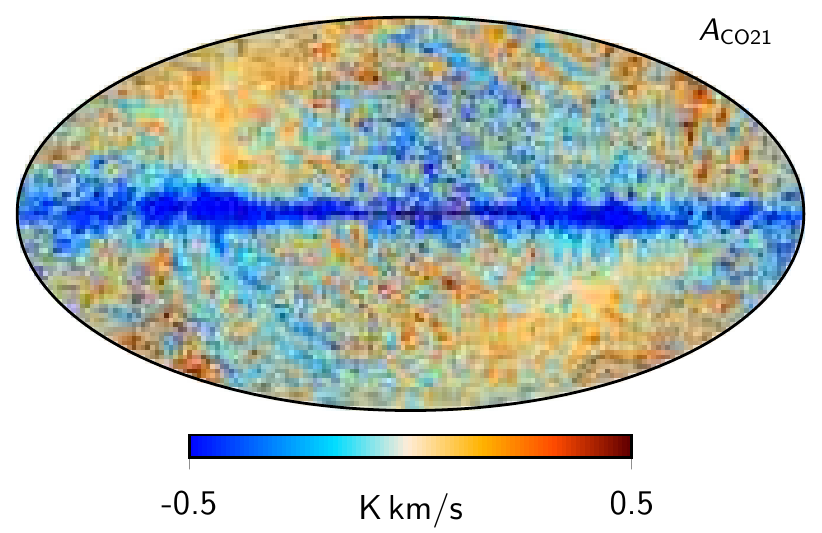}
   \includegraphics[width=0.4\linewidth]{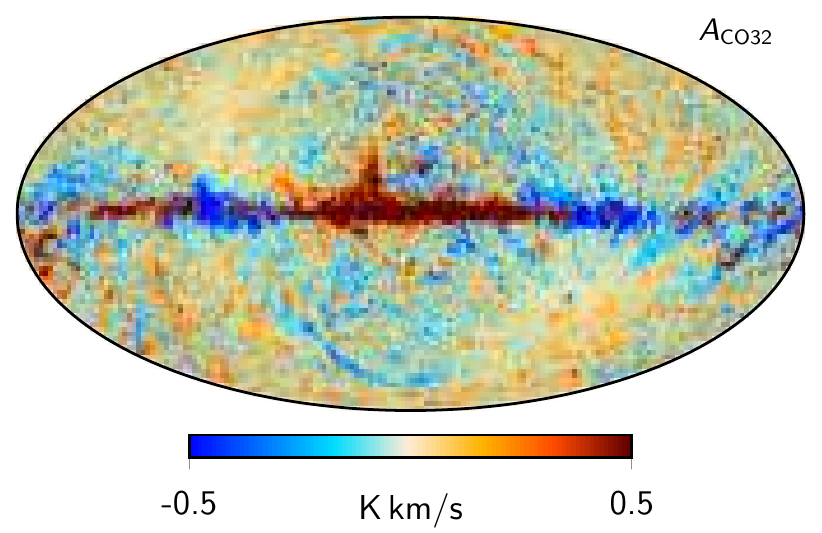}
   \caption{Half-difference plots from the \commander\ high-resolution \npipe\ splits.  All maps are smoothed to a common resolution of 60\arcm\ FWHM beam.}
   \label{fig:halfdiff_T_compsep}
\end{figure*}

The second type of estimate that we use to assess uncertainties is based on detector-split differences. Specifically, each half of the \npipe\ detector-split is processed through the \commander\ analysis framework, establishing two independent estimates of the same quantities.  The resulting half-difference amplitude maps are shown in Fig.~\ref{fig:halfdiff_T_compsep}. These provide a direct view of the overall instrumental noise level in each component map, and residual systematics.  For instance, while the low-frequency component is clearly dominated by instrumental noise at high Galactic latitudes, the thermal dust amplitude is equally clearly dominated by systematic effects, particularly in the form of calibration uncertainties.  Most systematic features seen in these maps can be matched to the instrumental effects discussed in Sect.~\ref{sec:data_selection}.

\clearpage

\section{CMB Solar dipole}
\label{sec:dipole}

\begin{table*}[hbtp!]
\newdimen\tblskip \tblskip=5pt
\caption{Comparison of Solar dipole measurements from COBE, WMAP, and \Planck. }
\label{tab:dipole}
\vskip -4mm
\footnotesize
\setbox\tablebox=\vbox{
 \newdimen\digitwidth
 \setbox0=\hbox{\rm 0}
 \digitwidth=\wd0
 \catcode`*=\active
 \def*{\kern\digitwidth}
  \newdimen\dpwidth
  \setbox0=\hbox{.}
  \dpwidth=\wd0
  \catcode`!=\active
  \def!{\kern\dpwidth}
  \halign{\hbox to 2.5cm{#\leaderfil}\tabskip 2em&
    \hfil$#$\hfil \tabskip 2em&
    \hfil$#$\hfil \tabskip 2em&
    \hfil$#$\hfil \tabskip 2em&
    #\hfil \tabskip 0em\cr
\noalign{\doubleline}
\omit&&\multispan2\hfil\sc Galactic coordinates\hfil\cr
\noalign{\vskip -3pt}
\omit&\omit&\multispan2\hrulefill\cr
\noalign{\vskip 3pt} 
\omit&\omit\hfil\sc Amplitude\hfil&l&b\cr
\omit\hfil\sc Experiment\hfil&[\muK_{\rm
CMB}]&\omit\hfil[deg]\hfil&\omit\hfil[deg]\hfil&\hfil\sc Reference\hfil\cr
\noalign{\vskip 3pt\hrule\vskip 5pt}
COBE--DMR \rlap{$^{\rm a,b}$}&                  3358!**\pm23!**&     264.31*\pm0.16*&
48.05*\pm0.09*&\citet{lineweaver1996}\cr
COBE--FIRAS&                  3343!**\pm16!**&     265.7**\pm0.5**&
48.3**\pm0.5**&\citet{Fixsen:1993rd}\cr
WMAP \rlap{$^{\rm c}$}&                  3355!**\pm*8!**&     263.99*\pm0.14*&
     48.26*\pm0.03*&\citet{hinshaw2009}\cr
\noalign{\vskip 3pt}
LFI 2015 \rlap{$^{\rm b}$}&              3365.5*\pm*3.0*&     264.01*\pm0.05*&
     48.26*\pm0.02*&\citet{planck2014-a03}\cr
HFI 2015 \rlap{$^{\rm d}$}&              3364.29\pm*1.1*&     263.914\pm0.013&
     48.265\pm0.002&\citet{planck2014-a09}\cr
\noalign{\vskip 3pt}
LFI 2018 \rlap{$^{\rm b}$}&              3364.4*\pm*3.1*&     263.998\pm0.051&
     48.265\pm0.015&\citet{planck2016-l02}\cr
HFI 2018 \rlap{$^{\rm d}$}&              3362.08\pm*0.99&     264.021\pm0.011&
     48.253\pm0.005&\citet{planck2016-l03}\cr
\noalign{\vskip 3pt}
\bf\npipe\ \rlap{$^{\rm a,c}$}& \bf3366.6*\pm*2.7*& \bf263.986\pm0.035&
 \bf48.247\pm0.023&Section~\ref{sec:dipole}\cr
\noalign{\vskip 5pt\hrule\vskip 5pt}}}
\endPlancktablewide
\tablenote {{\rm a}} Statistical and systematic uncertainty estimates are added in quadrature.\par
\tablenote {{\rm b}} Computed with a naive dipole estimator that does not account for higher-order CMB fluctuations.\par
\tablenote {{\rm c}} Computed with a Wiener-filter estimator that estimates, and marginalizes over, higher-order CMB fluctuations jointly with the dipole.\par
\tablenote {{\rm d}} Higher-order CMB fluctuations are accounted for by subtracting a dipole-adjusted CMB map from frequency maps prior to dipole estimation. \par
\end{table*}

As discussed in Sects.~\ref{sec:processing} and \ref{sec:compsep}, the \npipe\ maps retain the Solar dipole (only the orbital CMB dipole is removed prior to mapmaking). This makes it possible to include a dipole in the CMB term (Eq.~19) of the temperature model in Sect.~\ref{sec:methodology}, and therefore to fit the Solar dipole simultaneously with the global monopole, calibration, and bandpass shift factors given in Table~\ref{tab:monopole} and the foreground and CMB maps, as discussed in Sect.~\ref{sec:compsep}.  Thus, for the first time, a \Planck\ Solar dipole based on data from both \lfi\ and \hfi\ can be determined, with uniform determination of uncertainties.

The task of determining the CMB Solar dipole parameters is equivalent to fitting a dipole to the CMB map shown in the top panel of Fig.~\ref{fig:CMB_T}.  As usual, regions of brightest Galactic emission must be excluded from the fit.  And as with any such fit on a partial sky, care must be taken to avoid confusion from higher-order CMB moments.  Our approach to this problem was described in general form by \citet{jewell:2004}, \citet{wandelt:2004}, and \citet{eriksen:2004}, and later implemented by \citet{hinshaw2009} for WMAP.  The algorithm has been explored in detail by \citet{thommesen:2019} for application to the \Planck\ observations, and the following description and results are based on the \citet{thommesen:2019} implementation.

The algorithm is a variation of the Gibbs-sampling method described in Sect.~\ref{sec:compsep} and implemented in 
\commander.  In addition to the parametric setup described in Sect.~\ref{sec:compsep}, we assume that the CMB fluctuations constitute an isotropic, Gaussian, random field.  Under this mild assumption, whose validity to high precision is confirmed by the entire body of \Planck\ results, the CMB field is described by an angular power spectrum, $C_{\ell}$.  Combined with the observed phase information of the non-masked CMB fluctuations, this power spectrum then acts as an informative prior for masked regions. Specifically, the large-scale structures can be partially reconstructed from the observed structures outside the mask, given the requirement that the total field must be Gaussian and isotropic. This intuitive idea may be formulated quantitatively in terms of the following equation for Gaussian constrained realizations,
\begin{linenomath*}
\begin{equation}
(\S^{-1} + \N^{-1})\x = \N^{-1}\d + \S^{-1/2}\omega_1 + \N^{-1/2}\omega_2\,,
\end{equation}
\end{linenomath*}
where $\S=\S(C_{\ell})$ is the signal covariance matrix defined by the angular power spectrum, $\N$ is the noise covariance matrix, in which the variance of masked pixels is set to infinity, $\d$ is the (foreground-cleaned, but noisy) CMB map, and $\omega_1$ and $\omega_2$ are two random Gaussian fields, with zero mean and unit variance. The solution $\x$ is known as a ``constrained realization,'' and typically a large number of such samples is produced in order to quantify the uncertainties due to the sky cut and noise. Furthermore, the power spectrum $C_{\ell}$ is in principle unknown, and in practice one therefore computes a Gibbs chain to produce final results, iterating between determining constrained realizations and angular power spectra.  For further details, see  \citet{eriksen:2004} and \citet{thommesen:2019}.

Once the masked region is filled in by this constrained realization, we have a {\it full-sky\/} CMB map that is consistent with the
observed data, from which the dipole can be computed directly without reference to a sky mask.  The statistical uncertainty introduced by noise and this process of dealing with a masked sky is well-represented by the standard deviation measured from the ensemble
of constrained realizations.  In the following, we report values averaged over 100 Gibbs samples per mask.

In addition to the statistical uncertainties just described, there are three main sources of systematic uncertainty that we now estimate.  The first is due to uncertainties in modelling component separation.  We estimate this using the same approach used for other global parameters described in Sect.~\ref{sec:global_parameters}.  Specifically, we quantify the typical scatter seen among various internal \npipe\ versions, analysed with different foreground models.  We estimate the $1\,\sigma$ systematic uncertainties to be $1\muK$ for the CMB dipole amplitude, 1\parcm8 for the Galactic latitude, and 1\parcm3 for the Galactic latitude. 

The second source of systematic uncertainty is due to the uncertainty in absolute calibration between channels. Specifically, the \commander\ analysis presented above adopts the 143-GHz channel as its calibration reference channel, based on the fact that it has the lowest absolute noise of any \Planck\ frequency and no CO contamination; as a result, component separation is generally more robust for this channel.  An equally valid choice, however, might have been the 100-GHz channel, and this channel was indeed adopted for the same purpose by the corresponding HFI DPC analysis.  However, as shown in Table~\ref{tab:monopole}, there is an overall 0.07\,\% relative calibration difference between the 100 and 143-GHz channels.  Given an overall dipole amplitude of about $3360\muK$, this translates directly into a shift of $2.4\muK$ in the overall dipole amplitude, depending on which channels is chosen as the reference.  We do not have any strong reason to prefer one channel over the other, so we adopt this value as another systematic error on the overall dipole amplitude.

The third term is due to the uncertainty in the CMB monopole value, for which we adopt $T_0=2.72548\pm0.00057\,\mathrm{K}$ \citep{fixsen2009}. For a dipole value of 3.3\,mK, this monopole uncertainty translates into a dipole amplitude uncertainty of 0.7\muK. Taking all four terms together, we therefore end up with a total dipole amplitude uncertainty of $(0.25^2 + 1^2 + 2.4^2 + 0.7^2)^{1/2}\muK\,=\,2.7\muK$.  The relative calibration uncertainty between 100 and 143\GHz\ turns out to be the dominant uncertainty in the dipole amplitude.

\begin{figure}[hbtp!]
   \centering
   \includegraphics[width=0.48\textwidth]{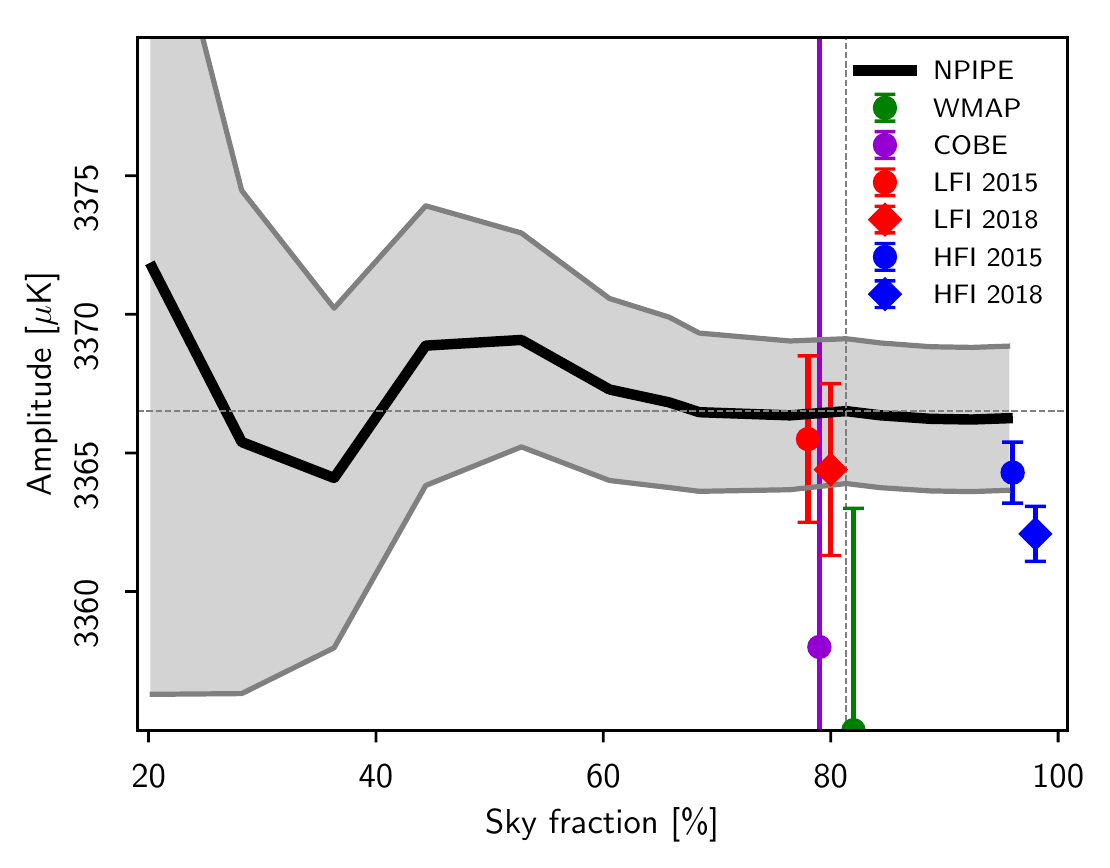}\\
   \includegraphics[width=0.48\textwidth]{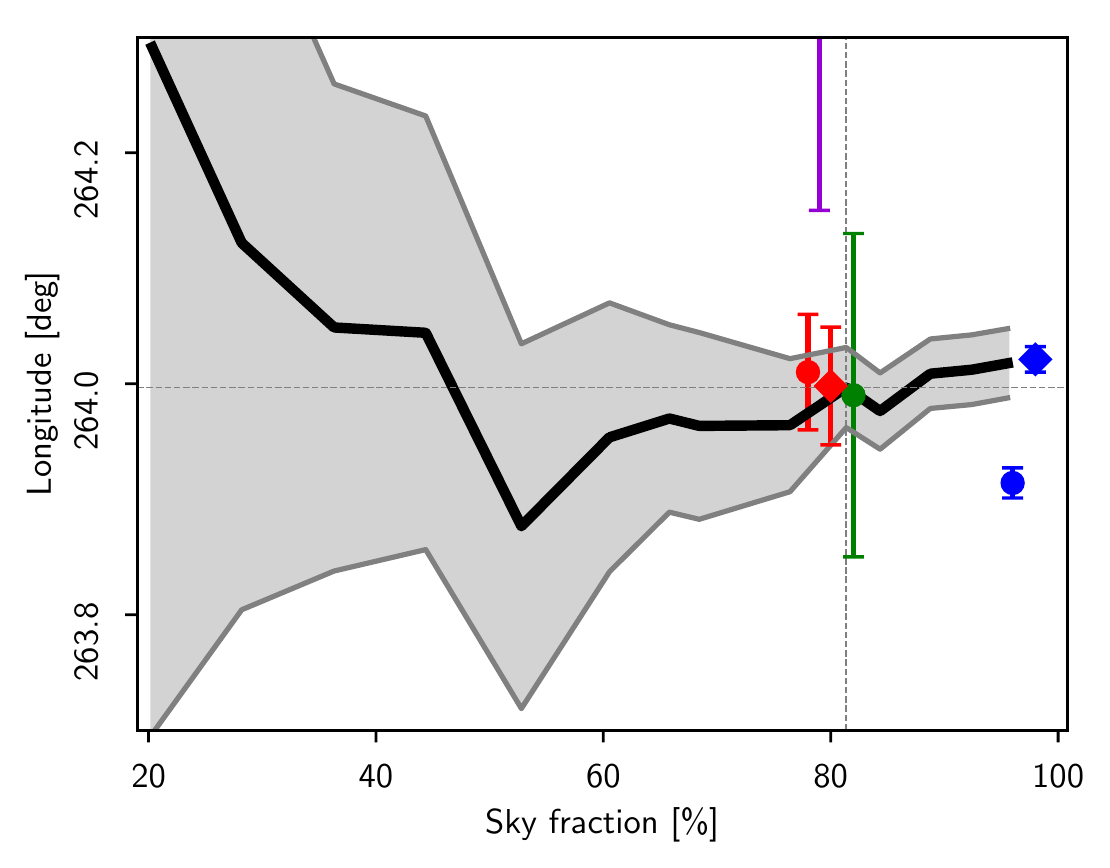}\\
   \includegraphics[width=0.48\textwidth]{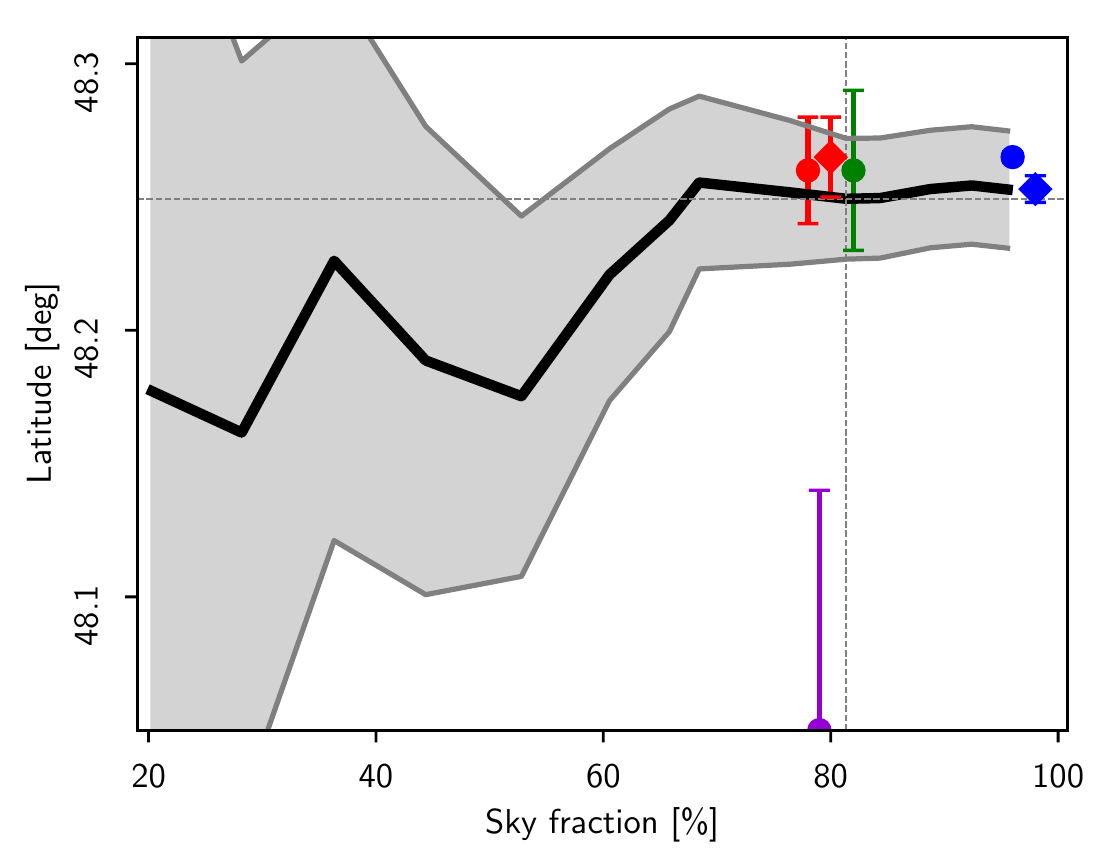}
   \caption{CMB Solar dipole parameters as a function of sky fraction, as estimated from \npipe\ data. The solid black lines show the posterior mean derived with a Wiener-filter estimator, and the grey bands show corresponding $\pm1\,\sigma$ confidence regions including both statistical and systematic uncertainties. From top to bottom, the three panels show amplitude, longitude, and latitude parameters. For comparison, estimates from COBE, WMAP, and \Planck\ LFI and HFI are shown as individual coloured data points. The dotted lines represent the \npipe\ values that are adopted as final optimal estimates, and summarized in Table~\ref{tab:dipole}, defined with a sky fraction of $f_{\textrm{sky}}=0.81$.  }
   \label{fig:dipole_params}
\end{figure}

Figure~\ref{fig:dipole_params} shows results obtained from applying this algorithm to the \commander-based \npipe\ map shown in
Fig.~\ref{fig:CMB_T} as a function of sky fraction.  The series of masks adopted for this analysis is the same as was employed by the \hfi-DPC analyses presented in \citet{planck2016-l03}.  The \npipe\ posterior mean values are shown as solid black lines, while
the $1\,\sigma$ posterior confidence regions are shown as grey regions. For comparison we also plot previously published results as coloured points.\footnote{The \hfi\ measurements are formally assigned a high sky fraction of about 96\,\%, but a precise specification of this sky fraction is difficult to make.  \hfi\ adopted an approach in which foreground-reduced \Planck\ CMB fluctuation maps \citep{planck2016-l04} were subtracted from each frequency map prior to the dipole estimation process.  This procedure is inherently somewhat circular, since the dipole of the CMB maps itself must be adjusted prior to subtraction;  however, the small adjustment to the dipole can be assessed using a smaller mask.  Nevertheless, this approach is fundamentally different from approaches that apply hard masks before the dipole analysis (e.g., \citealt{lineweaver1996}). Due to the use of the strong CMB prior, sky fractions given in Fig.~\ref{fig:dipole_params} are suggestive, not definitive.}

\begin{figure}[hbtp!]
   \centering
   \includegraphics[width=0.48\textwidth]{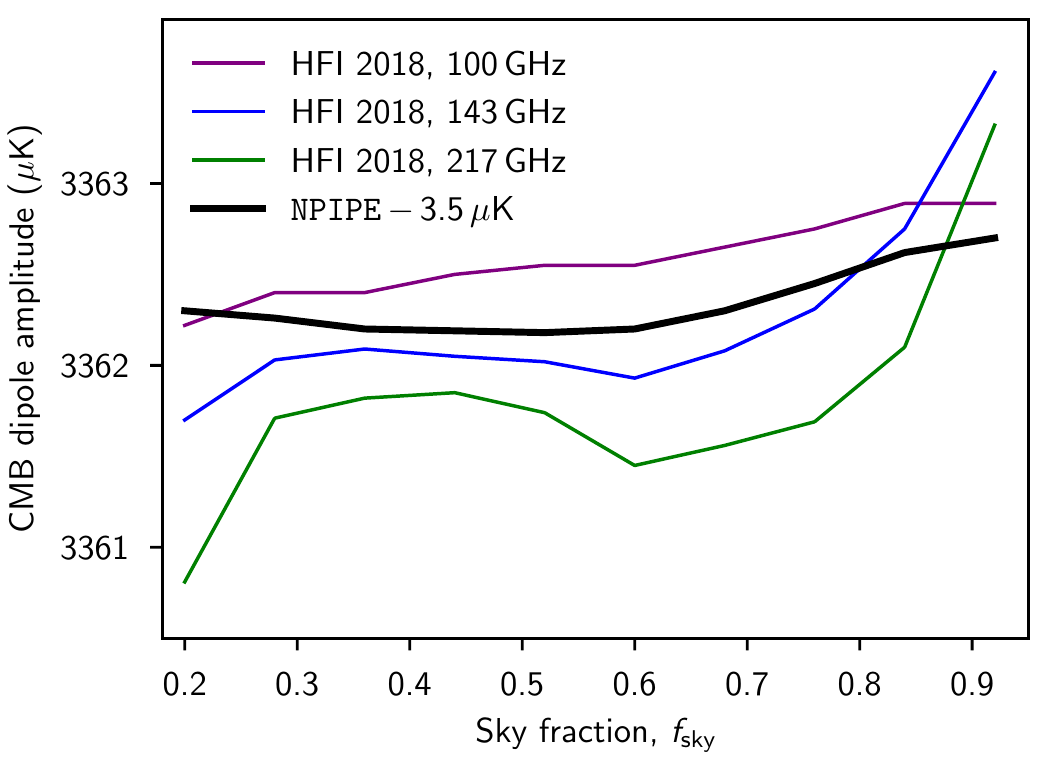}
   \caption{\npipe\ CMB dipole amplitude as a function of sky fraction when estimated using a similar methodology as reported in \citet{planck2016-l03}, i.e., by subtracting a foreground-cleaned \Planck\ CMB temperature map prior to estimating the dipole parameters for each value of $f_{\mathrm{sky}}$. The coloured curves are direct reproductions from Figure~23 in \citet{planck2016-l03}. Note that the \npipe\ results have been offset by 3.5\muK\ for comparison purposes.   }
   \label{fig:dipole_vs_fsky}
\end{figure}

The \npipe\ confidence bands show good stability on sky fractions ranging from $f_{\textrm{sky}}=0.20$ to 0.95, suggesting that the foreground removal process has been successful.  We also observe good qualitative agreement between the estimates derived from different experiments and data sets.  The agreement is particularly striking between the \lfi\ and the \npipe\ results, with less than $1\,\sigma$ shifts in any of the three parameters, despite the fact that the \npipe\ CMB map is strongly dominated by HFI observations. The WMAP data also agree well with both of these, although exhibiting a slightly lower ($1.4\,\sigma$ in units of the WMAP uncertainty) CMB dipole amplitude.

As final \npipe\ dipole estimates, we adopt the values derived for a sky fraction of 81\,\%. As seen in Fig.~\ref{fig:dipole_params}, the
error bars do not decrease further for higher sky fractions, since the fixed systematic uncertainty contribution starts to dominate the error
budget. For convenient reference and comparison, these values are tabulated in Table~\ref{tab:dipole}, together with the previously
published estimates shown in Fig.~\ref{fig:dipole_params}.

Statistically speaking, the most striking feature seen in these plots is the apparent qualitative disagreement between the \npipe\ and HFI DPC uncertainties.  In particular, the HFI directional uncertainties are nominally between 3 and 10 times smaller than the
\npipe\ directional uncertainties, and the HFI 2015 and 2018 longitude estimates disagree at the $8\,\sigma$ level.  Part of this is explained by the fact that the HFI directional uncertainties do not include full estimates of systematic errors.  In addition, as stated above, one of the main algorithmic differences between the HFI DPC and \npipe\ analysis is that in the HFI analysis,   \Planck\ maps of the CMB were subtracted from the sky maps prior to estimation of the CMB dipole. This introduces an uncertainty that is perfectly correlated across the entire sky, and trades a large amount of statistical uncertainty from higher-order CMB fluctuations against a systematic error that is difficult to quantify in terms of the ability of the component separation algorithms to remove foregrounds in the central Galactic plane. To illustrate this point, Fig.~\ref{fig:dipole_vs_fsky} shows the \npipe\ dipole amplitude as a function of sky coverage, where the dipole has now been estimated with a similar methodology as in \citet{planck2016-l03}, i.e., by first subtracting a \Planck\ CMB temperature map prior to estimating the dipole parameters, and not imposing a constrained realization in the masked region. (The coloured curves are direct reproductions from Figure~23 in \citealp{planck2016-l03}.) Here we see that the \npipe\ dipole is in fact nominally stable to a precision smaller than 0.5\,\muK\ when adopting this approach, which is comparable to the 100\GHz\ HFI DPC result. This demonstrates that the variations seen in the top panel of Fig.~\ref{fig:dipole_params} are not due to, say, residual foreground fluctuations in the \npipe\ approach, but rather due to the fact that the CMB dipole is estimated independently over each considered sky fraction. Correspondingly, the apparent stability seen in Fig.~\ref{fig:dipole_vs_fsky} (and Figure 23 in \citealp{planck2016-l03}) is due to the fact that the CMB fluctuations have been fixed based on 95\,\% of the sky for all cases, and not only the nominal $f_{\mathrm{sky}}$ value indicated in the plot. With these important algorithmic points in mind, we conclude that the \npipe\ and HFI DPC dipole amplitudes listed in Table~\ref{tab:dipole} agree well within statistical uncertainties.  The \npipe\ uncertainties are somewhat more conservative than those presented in \citet{planck2016-l03}.

\section{Optical depth to re-ionization}
\label{sec:tau}

In this section we use determination of the optical depth to reionization, $\tau$, as a validation of the quality of the large-scale polarization in the \npipe\ maps.  Considering quadratic maximum-likelihood (QML) approaches, we use both auto-QML and cross-QML methods to extract the large-scale $EE$ power spectrum, and then derive posterior distributions for $\tau$.  Simulations are used to infer the statistical and systematic uncertainties of our methods, and to demonstrate that the large-scale polarization transfer function (Sect.~\ref{sec:ee_tf}) is accounted for properly.  As discussed, e.g., in \cite{planck2016-l05} and \cite{planck2016-l06}, estimated $\tau$ values depend on specific details of the analysis setup, including which part of the data is considered (e.g., $EE$-only or $TT-TE-EE$), whether additional cosmological parameters (e.g., $A_{\rm s}, \, r$) are fixed to a reference value or marginalized over, and whether we model reionization as a sharp or an extended process.  A full assessment of these effects for {\npipe} maps is beyond the scope of this paper.  Rather, here we focus on the consistency of estimated quantities ($\tau$, but also foreground template amplitudes, power spectra, cleaned map goodness-of-fit) between different data cuts (e.g., frequencies, sky masks) as a measure of the cleanliness of the large-scale polarization of \npipe\ maps.

\subsection{Pixel-based analysis}

The approach used for the pixel-based analysis of {\npipe} maps is similar to the one used for the analysis of low-resolution \lfi\ polarization maps described in the 2015 and 2018 releases (see \citealt{planck2016-l05} for methodological details). We summarize here the main differences between the two analyses:
\begin{itemize}
\item{{\bf Temperature Map}. We use the low-resolution 2015 \commander\ map, rather than the 2018 \commander\ map, taking advantadge of the larger sky coverage provided by the former}.
\item{{\bf Map Apodization}. We use cosine apodization to produce a low-resolution data set suitable for pixel-space analysis, as did the 2018 release, but the multipole range is different.  As discussed in Sect.~\ref{sec:lowres}, {\npipe} uses $\ell_1\,{=}\,1$, while the 2018 release uses $\ell_1\,{=}\,16$ ($\ell_2\,{=}\,48$ in both data sets).}
\item{{\bf Frequency channels}. We perform the analysis on 44-, 70-, and 100-GHz channels.}
\item{{\bf Foreground cleaning}. We use the 30- and 353-GHz maps as templates for synchrotron and thermal dust, to produce a cleaned frequency map at the target frequency.  However, no templates modelling instrumental systematics are included in the fit, as was the case for the LFI analysis in the 2018 release}.
\item{{\bf Galaxy masks}. We adopt the same masks as the LFI 2018 analysis.  For an in depth discussion of the derivation and properties of these masks, we refer the reader to section~2.3.1 of \cite{planck2016-l05}; here we just recall that masks are labelled according to the threshold $R$ above which high signal pixels are excluded, so that higher values of $R$ correspond to larger portions of the sky retained in the analysis. For optimum results, a self-consistent sets of masks based on {\npipe} maps should be created following the procedure outlined in \cite{planck2016-l05}. However, to facilitate the comparison between {\npipe} and {\prthree} results, and between different frequencies, we adopt the same masks as the LFI 2018 analysis.}
\item{{\bf Reference foreground scalings}. In the LFI 2018 analysis, the foreground scaling coefficients were fixed to those estimated on the R2.2x polarization mask (retaining a sky fraction $f_{\rm sky} = 0.666$), while the cosmological analysis was performed on less aggressive masks. Here instead we use the same mask for both the template cleaning and the cosmological analysis. In particular, this implies that parameters estimated on different masks have been computed on maps corresponding to different template cleaning solutions.} 
\end{itemize}

\begin{figure}[htpb!]
  \center 
  \includegraphics[width=0.48\textwidth]{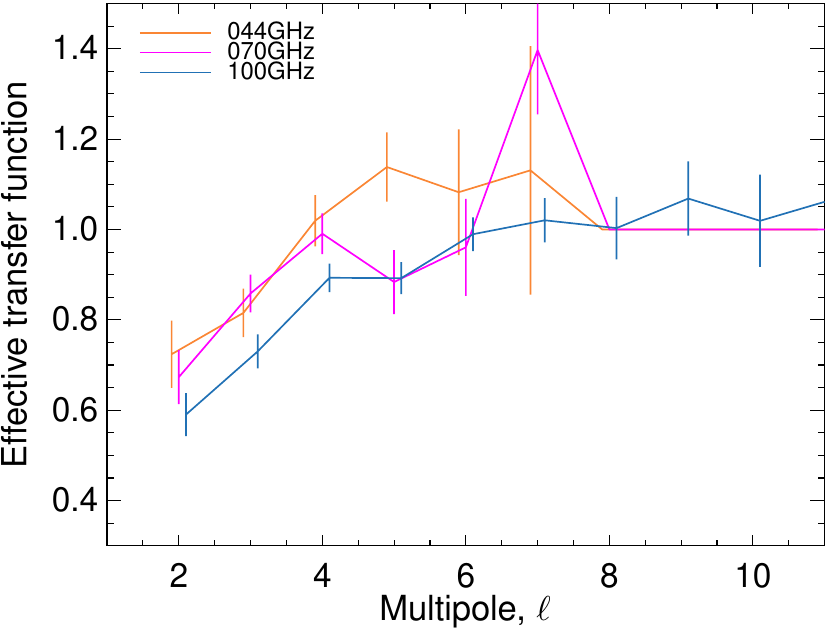}
  \caption{Low-multipole effective transfer functions for the pixel-based analysis.  Due to the higher impact of noise on auto-QML compared to cross-spectra, we are not able to measure the effective transfer function at $\ell\,{>}\,7$ for 44 and 70\GHz, and at $\ell\,{>}\,11$ for 100\GHz.  In the analysis the transfer function is conservatively enforced to be unity at $\ell\,{\ge}\,4$ for 44 and 70\GHz, and at $\ell\,{\ge}\,7$ for 100\GHz.  Note that the transfer functions here are, as expected, different from those shown in Figs.~\ref{fig:ee_bias} and \ref{fig:ee_bias_fg}.  As emphasized in the text, the simulations used to calculate the transfer function must be processed using the same tools and approaches that are applied to the data. The pixel-based analysis here and the spectrum-based analysis represented in Figs.~\ref{fig:ee_bias} and \ref{fig:ee_bias_fg} are different.}
  \label{fig:auto-qml-tf}
\end{figure}

As discussed in Sect.~\ref{sec:ee_tf}, the \npipe\ calibration scheme using a polarized sky model results in suppression of the lowest multipoles, modelled as a spherically-symmetric transfer function (Figs.~\ref{fig:ee_bias}, \ref{fig:ee_bias_fg}, and \ref{fig:ee_bias_100}). The small dependence of the transfer function on the sky mask or on the cross-power spectral estimation approach (QML versus pseudo-$C_\ell$), suggests that this modelling is an effective description of a more complicated structure. It is therefore important that the simulations used to calculate the transfer function be processed using the same tools and approaches that are applied to data. Pixel-based methods are the equivalent of harmonic auto-spectra methods, and we define the corresponding effective transfer function as
\begin{linenomath*}
\begin{equation}
  C_\ell^{\rm output} = k_\ell^2  C_\ell^{\rm CMB}\,,
\end{equation}
\end{linenomath*}
where both $C_\ell^{\rm output}$ and $C_\ell^{\rm CMB}$ are QML estimates on \nside{16} maps, after applying the R1.8x ($f_{\rm sky} = 0.624$) polarization mask. Figure~\ref{fig:auto-qml-tf} shows the resulting $E$-mode transfer functions. Auto-spectra are more impacted by noise bias than cross-spectra, and with the available number of simulations we are not able to reliably measure the transfer function for $\ell\,{>}\,7$ at 44 and 70\GHz, and $\ell\,{>}\,11$ for 100\GHz. In addition, at modes $\ell\,{>}\,4$ for LFI and $\ell\,{>}\,7$ for 100\GHz, the measured transfer function is compatible with unity within 2 standard deviation (with the exception of $\ell\,{=}\,7$ at 70\GHz). Those multipoles also roughly mark the scales above which the $E$-mode $S/N$ ratio falls below unity for values of $\tau$ compatible with current estimates, and have little impact on determination of the optical depth. Therefore, we conservatively enforce the transfer function to be unity at $\ell\,{\ge}\,4$ for 44 and 70\GHz, and at $\ell\,{\ge}\,7$ for 100\GHz.  We nonetheless checked that using the full measured transfer function has only a minimal impact on the estimated values of $\tau$ (assuming a sharp transition to reionization). On the other hand, constraining models with extended reionization history also leverages features in the multipole range $\ell\,{\simeq}\,10$--20, and any such study would need to assess whether pixel-based methods are suitable, given the above limitations on the transfer function. In the following, we do not account for the uncertainty in transfer-function estimates.

\begin{figure}[hbtp!]
  \center
  \includegraphics*[width=0.48\textwidth]{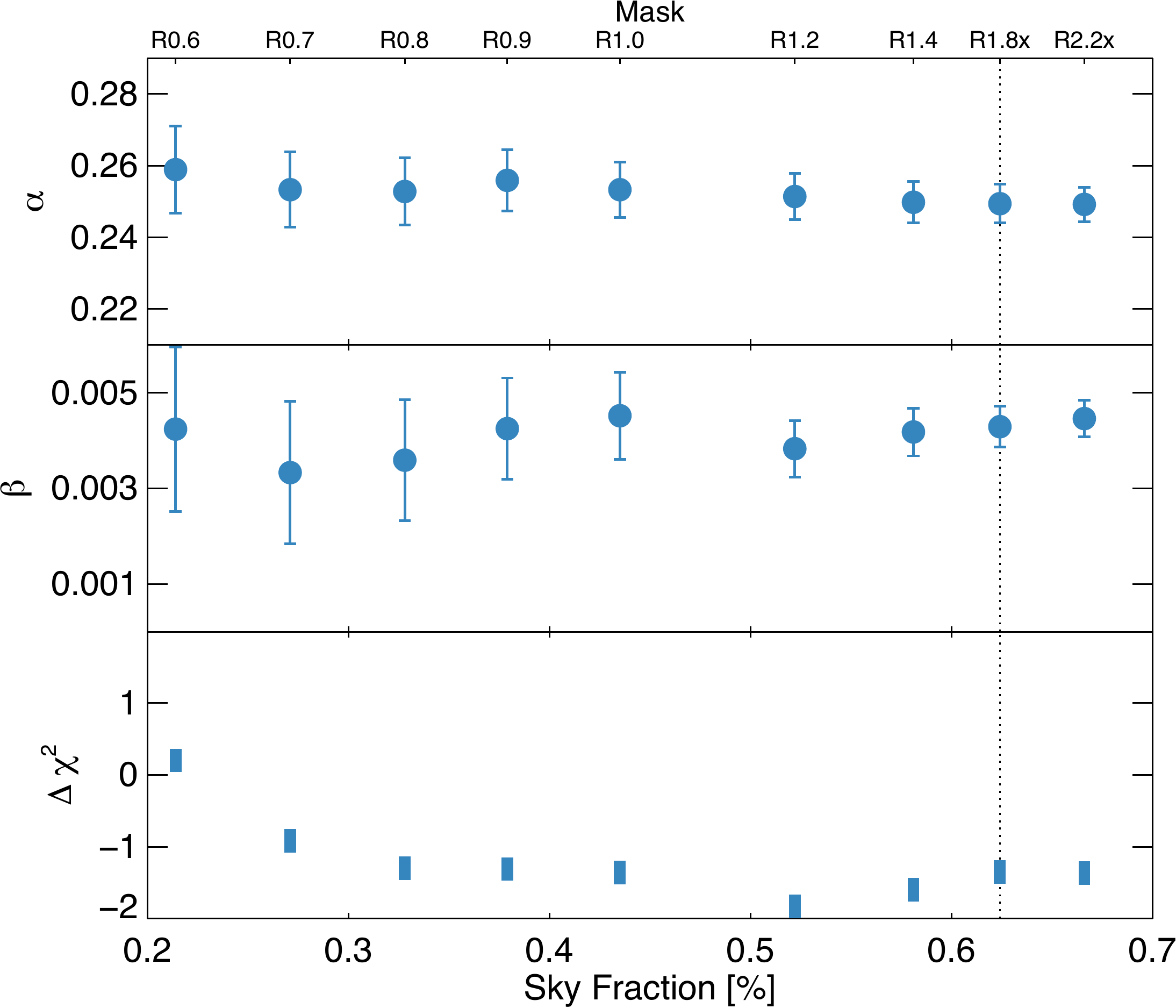}
  \includegraphics*[width=0.48\textwidth]{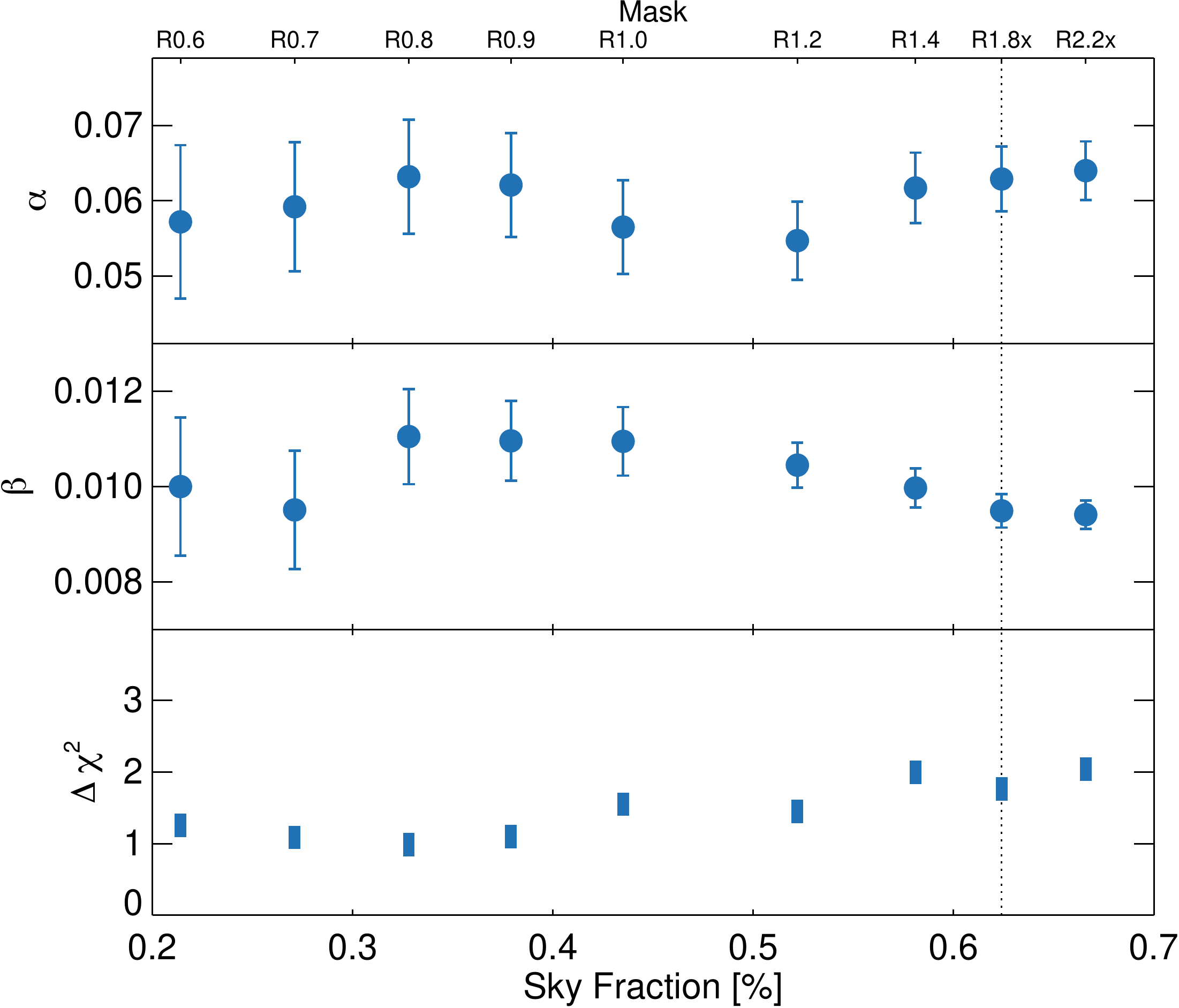}
  \includegraphics*[width=0.48\textwidth]{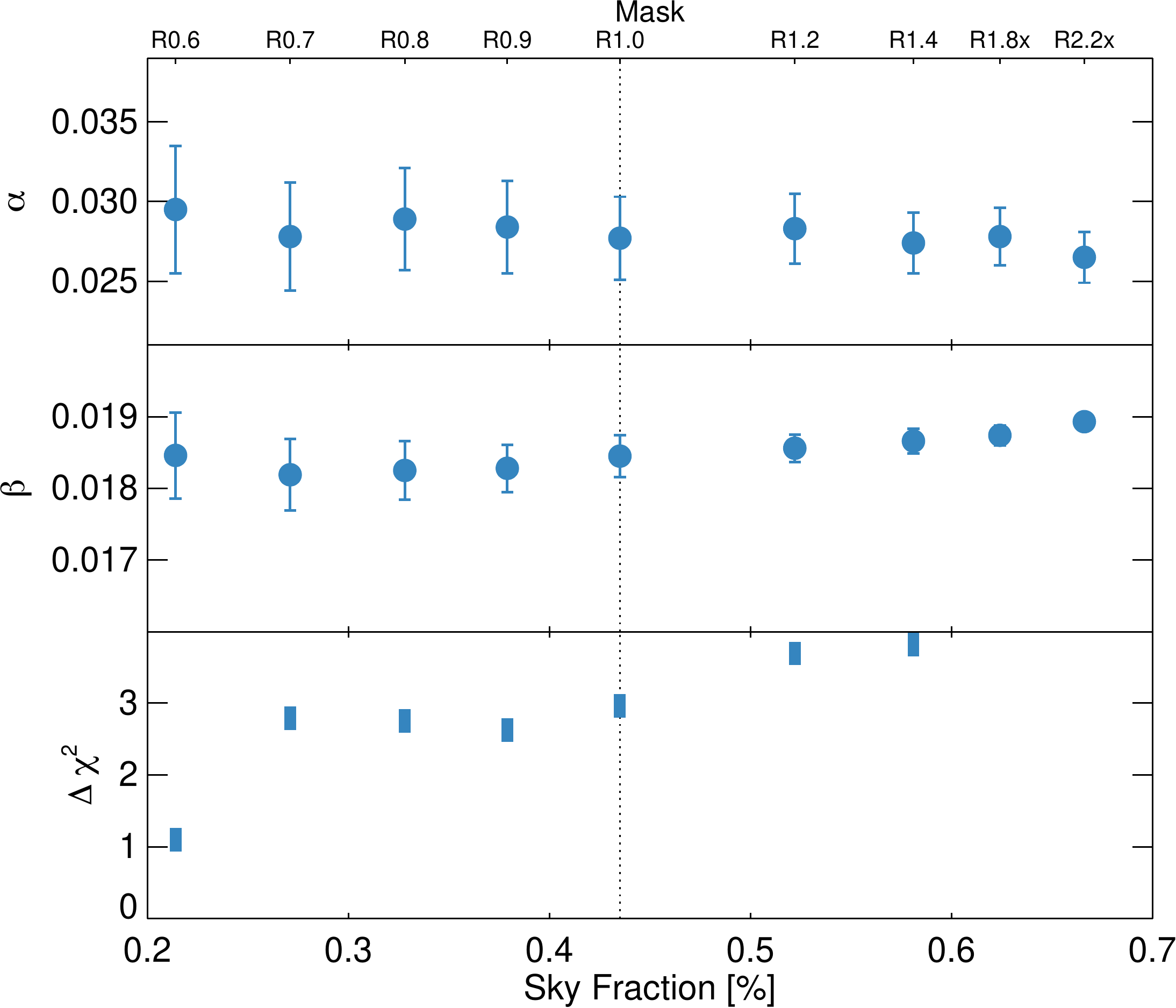}
  \caption{Scaling coefficients for synchrotron spectral index $\alpha$ and thermal dust emissivity $\beta$, and excess $\Delta \chi^2 = (\chi^2 - {\rm N_{dof}})/\sqrt{2{\rm N_{dof}}}$, for 9 different masks.  The panels from top to bottom show results at 44, 70, and 100\GHz, respectively.  For 44- and 70-GHz data, the dotted vertical line shows the mask used for parameter estimation with the low-$\ell$ 2018 LFI likelihood, while for 100-GHz data it shows the largest mask for which $\Delta \chi^2 \le 3$.}
\label{fig:fg_scalings_auto}
\end{figure}

Figure~\ref{fig:fg_scalings_auto} shows the scaling coefficients for synchrotron, $\alpha$, and the thermal dust, $\beta$, and the excess $\chi^2$ (defined as $\Delta \chi^2 = (\chi^2 - {\rm N_{dof}})/\sqrt{2{\rm N_{dof}}}$, where ${\rm N_{dof}}$ is the number of unmasked $Q, U$ pixels as a function of the Galactic mask). In \prthree, the analysis was restricted to masks for which  $|\Delta \chi^2| < 3$, and the R1.8x mask was selected for the final likelihood. Both 44- and 70-GHz \npipe\ maps meet the $|\Delta \chi^2| < 3$ criterion over the range of masks considered, while at 100\GHz\ this criterion is met (marginally) only for the R0.9 ($f_{\rm sky} = 0.379$) or less aggressive masks.  At 70\GHz\, the level of stability over different masks is in line with \cite{planck2016-l05} results, with an overall lower $\chi^2$ excess. However, \npipe maps seem to prefer higher scaling coefficients than what shown there. For the reference R1.8x mask, we find $\alpha = 0.062 \pm 0.004$ and $\beta = 0.0095 \pm 0.0003$, are about $1\,\sigma$ higher than \cite{planck2016-l05} values. On the same R2.2x mask as \cite{planck2016-l05}, we observe only minimal shifts in the estimates ($\alpha = 0.063 \pm 0.004$ and $\beta = 0.0094 \pm 0.0003$). The reason of this discrepancy is currently unclear, but we verified that adopting \cite{planck2016-l05} scalings shifts  $\tau$ estimates (discussed later in this section) by about $0.1\,\sigma$. Modelling the foreground spectral energy distributions as in \cite{planck2016-l04} (see also Sect.~\ref{sec:methodology}), we can convert the measured scalings into estimates of the polarized synchrotron spectral index, $\beta_{\rm s}$, and dust emissivity index $\beta_{\rm d}$. Fixing the thermal dust temperature $T_{\rm d} = 19.5{\rm K}$, we find $\beta_{\rm s} = 3.17 \pm 0.08, \, \beta_{\rm d} = 1.61 \pm 0.02$, in good agreement with \cite{planck2016-l04} results. Results at the other frequencies considered are consistent with 70\GHz\ estimates, with $(\beta_{\rm s} ,\beta_{\rm d}) = (3.23 \pm 0.05, 1.56 \pm 0.05)$ and $(\beta_{\rm s} ,\beta_{\rm d}) = (3.11 \pm 0.08, 1.62\pm 0.01)$ respectively at 44 and 100\GHz.   

\begin{figure}[hbtp!]
  \center
  \includegraphics[width=0.48\textwidth]{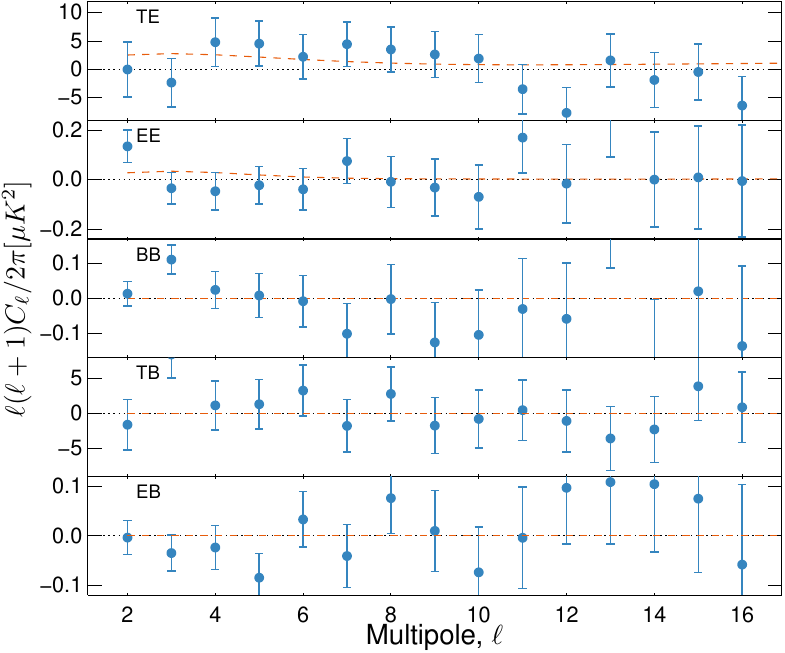}
  \includegraphics[width=0.48\textwidth]{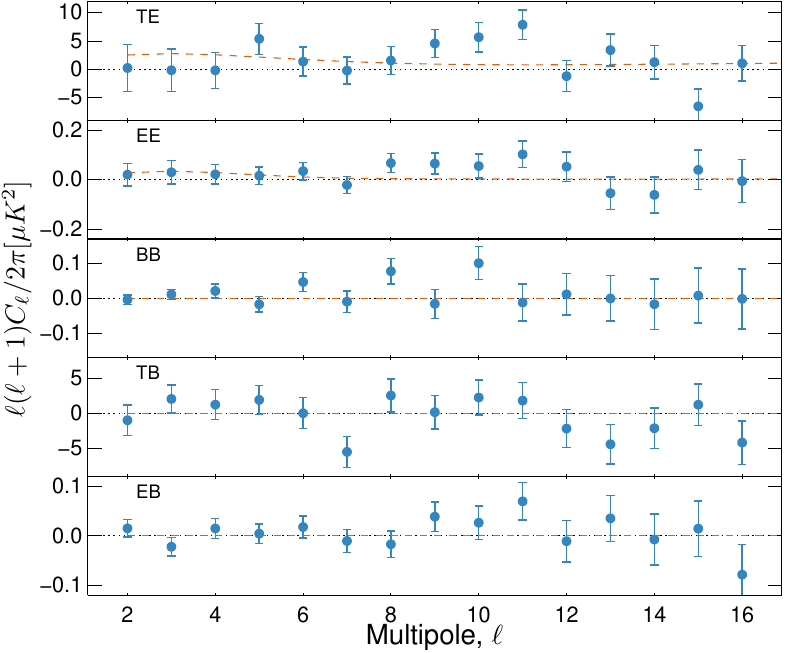}
  \includegraphics[width=0.48\textwidth]{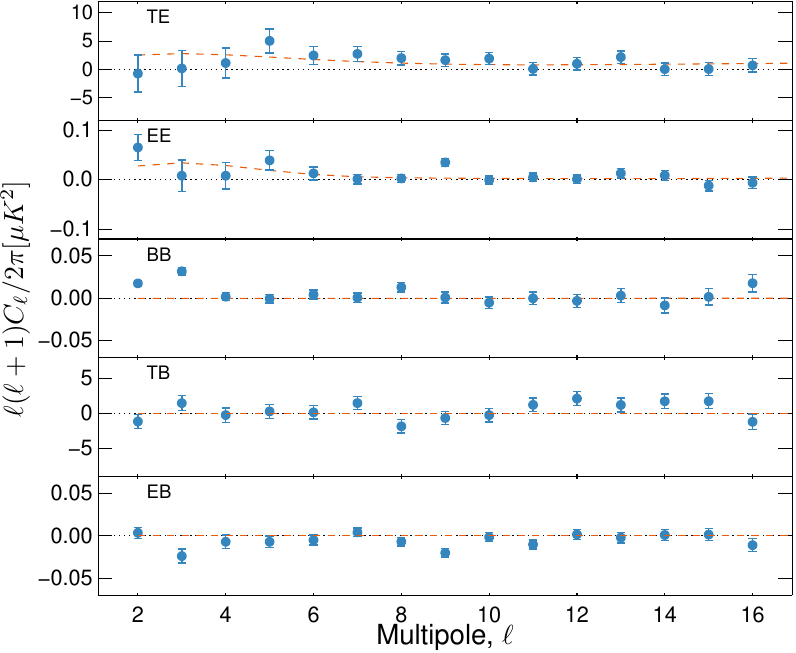}
  \caption{Power spectra of template-cleaned, low-resolution maps.  From top to bottom, the panels show results for 44, 70, and 100\GHz, respectively.  The dashed red lines show a model with $\tau = 0.06$ (and not a fit to the data).  Spectra for $TT$ are not shown, since in all cases the \commander\ 2015 map is used for temperature.  The polarization mask was R1.8x for 44 and 70\GHz, and R0.9 for 100\GHz.  In all cases, we used the \commander\ 2015 mask for temperature.}
\label{fig:spectra-auto-qml}
\end{figure}

Figure~\ref{fig:spectra-auto-qml} shows the power spectra of the corresponding cleaned maps. Some excess power is visible at 44 and 100\GHz, most noticeably for $BB$ at $\ell\,{=}\,2$ and 3 and for the $EE$ at $\ell\,{=}\,2$. A proper estimation of the $BB$ transfer function would require a much larger number of simulations than are presently available, due to the much lower signal-to-noise ratio of $B$ modes compared to $E$ modes.  Similar issues impact the determination of $TB$ and $EB$ transfer function.  It is then not clear whether excesses are driven by residual contamination in the maps, a mismodelling of noise bias, an improper determination of the transfer functions, or a combination of all these effects.

\begin{figure}[!htpb!]
  \center
  \includegraphics[width=0.48\textwidth]{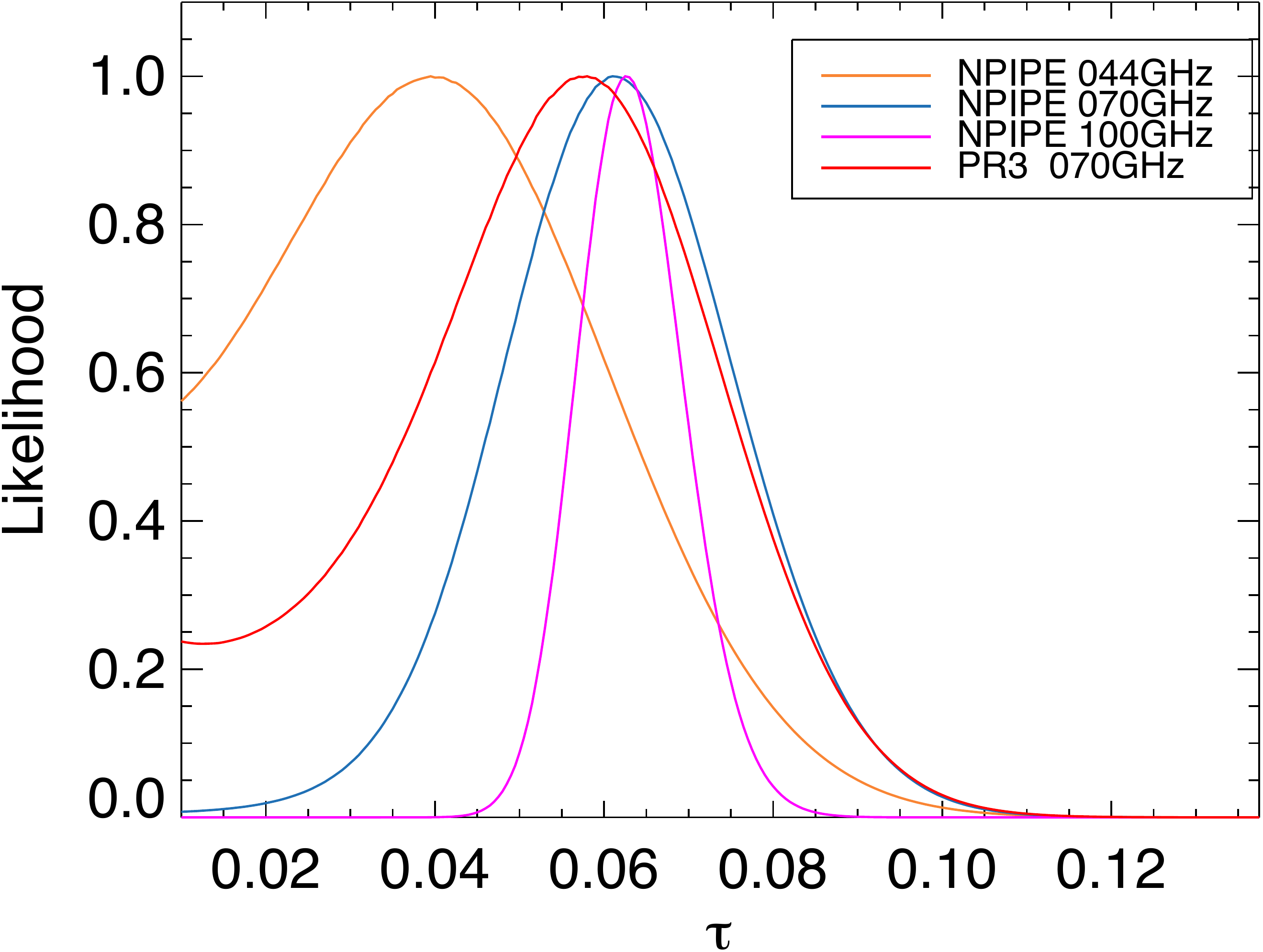}
  \caption{Reionization optical depth $\tau$ likelihood for the \npipe\ 44-, 70-, and 100-GHz maps, compared to that for the {\prthree} 70-GHz map. Results were computed retaining a polarized sky fraction $f_{\rm sky} = 0.624$ (R1.8x mask) at 44 and 70\GHz, and  $f_{\rm sky} = 0.379$ at 100\GHz\ (R0.9 mask).  We fixed $A_{\rm s} e^{-2\tau}$ to 1.884, and the remaining parameters to their $\Lambda$CDM best-fit values.}
  \label{fig:tau_auto_cmp}
\end{figure}

Finally, Fig.~\ref{fig:tau_auto_cmp} shows the $\tau$ likelihood distributions for the three {\npipe} channels considered here, compared to the results for the {\prthree} 70-GHz maps.  The corresponding expectation values are shown in Table~\ref{tab:tau_auto}. Since the focus of this section is mainly on data consistency rather than full exploration of the parameter space, we fix $A_{\rm s} e^{-2\tau}$ to  1.884, and all other parameters to their $\Lambda$CDM best-fit value from \cite{planck2016-l06}. Because of the weak dependence of low-$\ell$ polarization on the other cosmological parameters, this assumption has only a minor impact on the recovered value and uncertainty of $\tau$.  However, in order to make a clearer comparison between different data sets, the value of $\tau$ quoted here for {\prthree} maps has been re-evaluated using the same setup as the {\npipe} analysis.

Estimates from differing data sets are consistent overall, with 44\GHz\ preferring a roughly $1\,\sigma$ lower value than the other channels.  In addition, while the pixel-based 100-GHz value is consistent with those from the other channels, it is about $2\,\sigma$ higher than the estimate based on the $100\times143\GHz$ cross-spectrum (as discussed in the next section). Compared to cross-spectrum methods, a pixel-based approach is more sensitive to inaccuracies in estimation of the effective noise bias.  It is possible that the excess $\chi^2$ and the features seen in the 100-GHz power spectra discussed above are related to a failure of the adopted noise model at that frequency, which would also impact $\tau$ estimates.  A comparison of the noise bias predicted by the 100\GHz\ NCVM with the power spectra of the corresponding half-ring half-difference (HRHD) map supports this idea, even though it is not clear whether this mismatch can completely account for the excess power at $BB$ $\ell\,{=}\,2$ and 3 (given the uncertainty on the $B$-mode transfer function discussed above).

In order to assess the impact of the low-$\ell$ features discussed above on 100-GHz $\tau$ estimates, we repeated the parameter estimation by projecting out $EE$ $\ell\,{=}\,2$, finding $\tau = 0.0598 \pm 0.0064$, decreasing to $\tau = 0.0564 \pm 0.0063$ if we further project out $BB$ $\ell\,{=}\,2$ and 3. Only a marginal shift to $\tau = 0.0560 \pm 0.0064$ is observed if we also exclude $EE$ $\ell\,{=}\,3$. This suggests that the current pixel-based analysis is not yet able to properly model all the sources of uncertainty affecting the largest scales at 100\GHz. Even after projecting out the low-$\ell$ modes, there remains a roughly $1\,\sigma$ discrepancy between the 100-GHz pixel-based results and the $100 \times 143$ cross-spectra results. Figure~\ref{fig:spectra-auto-qml} shows a significant excess also for 100\GHz\ $EE$ $\ell\,{=}\,9$, and we performed a similar test to assess the impact of such a feature on our estimates. Projecting out this multipole leaves $\tau$ estimates virtually unchanged ($\tau = 0.0633 \pm 0.0065$), compared to the baseline estimate, suggesting that feature cannot account for the residual difference between 100\GHz\ pixel-based and 100x143\GHz\ results.  On the other hand, pixel-based estimates are based on ($I$, $Q$, $U$) maps, and therefore include all four power spectra ($TT$, $EE$, $TE$, and $BB$), while the cross-spectrum results are based on $EE$ spectra only. A direct comparison of the two sets of results is therefore not straightforward.

\begin{table}[htpb!]
  \begingroup
  \newdimen\tblskip \tblskip=5pt
  \caption{Reionization optical depth $\tau$ estimates from pixel-based analysis. }
  \label{tab:tau_auto}
  \nointerlineskip
  \vskip -3mm
  \footnotesize
  \setbox\tablebox=\vbox{
    \newdimen\digitwidth
    \setbox0=\hbox{\rm 0}
    \digitwidth=\wd0
    \catcode`*=\active
    \def*{\kern\digitwidth}
    \newdimen\signwidth
    \setbox0=\hbox{$-$}
    \signwidth=\wd0
    \catcode`!=\active
    \def!{\kern\signwidth}    
\halign{
  \hbox to 2.5cm{#\leaderfil}\tabskip 2em&
  \hfil#\hfil\tabskip 0pt\cr
  \noalign{\doubleline}
  \omit\hfil $\nu$ [GHz]\hfil&
  \omit\hfil $\tau$\hfil\cr
  \noalign{\vskip 3pt\hrule\vskip 4pt}
  *$44$& $0.0418 \pm 0.0182$\cr
  *$70$& $0.0617 \pm 0.0142$\cr
  $100$& $0.0634 \pm 0.0063$\cr
  *$70\, (2018)$& $0.0537 \pm 0.0187$\cr
  \noalign{\vskip 3pt\hrule\vskip 5pt}
}
}
\endPlancktable
\endgroup
\end{table}

\newcommand{\lollipop}{{\tt lollipop}}
\newcommand{\lowl}{\mbox{low-$\ell$}}

\subsection{Cross-spectrum analysis}

We use \lollipop, the same \lowl, $EE$ likelihood as was used to constrain the reionization history in \citet{planck2014-a25}.
{\tt Lollipop} is a spectrum-based likelihood following the approach proposed by \citet{hamimeche2008} and extended to cross-spectra in \citet{mangilli2015}.  Uncertainties are propagated using spectral covariance matrices estimated from the \npipe\ end-to-end simulations, which include different realizations for the CMB signal, noise, and systematics (including first-order ADCNL).

\begin{figure}[htpb!]
  \centering
  \includegraphics[width=0.48\textwidth]{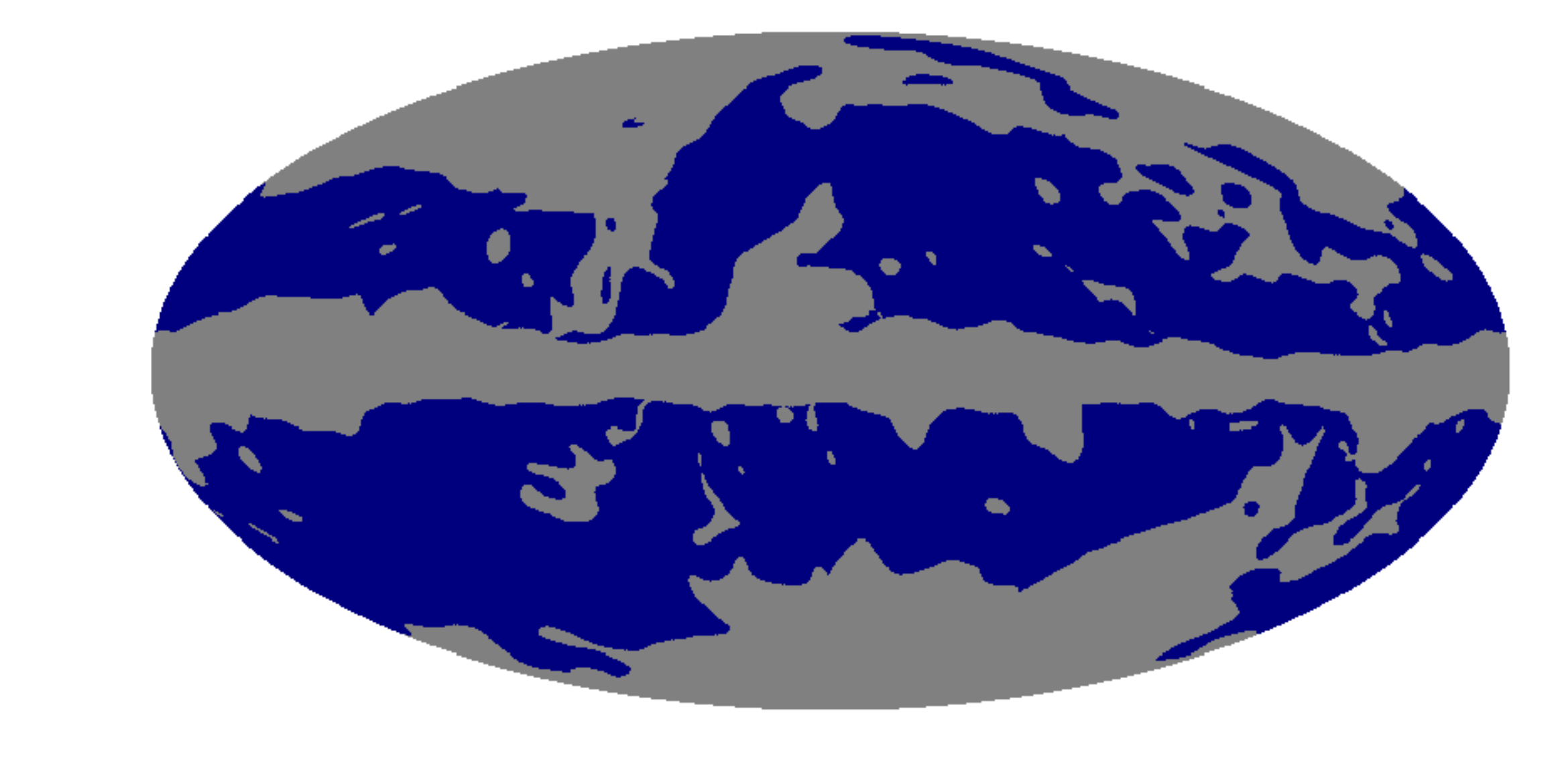}
  \caption{
    Sky region  ($f_{\rm sky} = 0.52$) used for the regression of foreground templates on NPIPE frequency maps (blue pixels are kept for the fit).
  }
  \label{fig:lol:reg_mask}
\end{figure}

We use the cross-correlations of the \Planck\ polarized frequency maps from 70- to 217-GHz.  Polarized foregrounds at those frequencies are dominated by Galactic dust and synchrotron emission.  We use the 353- and 30-GHz \Planck\ maps as templates to subtract dust and synchrotron emission, respectively, using a single coefficient for each component. The fit is performed over 52\,\% of the sky, avoiding the inner Galactic plane, as well as the Galactic poles, where the S/N is too low (Fig.~\ref{fig:lol:reg_mask}). Scaling coefficients (given in Tab.~\ref{tab:reg_coeff}) are estimated using cross-correlation between detector-set maps in order to avoid the bias due to the noise in the template maps during the regression process. Once the coefficients are estimated, we clean each frequency map using detector-set maps at 353 and 30\GHz\ to avoid any noise bias when computing cross-frequency spectra.

\begin{table}[htbp!] 
  \begingroup
  \newdimen\tblskip \tblskip=5pt
  \caption{Template coefficients measured on data. Error bars including statistical noise and systematic uncertainties are estimated from NPIPE simulations. Note that the variation of the coefficient depending on the sky region used for the regression is larger than those error bars.
}
  \label{tab:reg_coeff}
  \nointerlineskip
  \vskip -3mm
  \footnotesize
  \setbox\tablebox=\vbox{
    \newdimen\digitwidth
    \setbox0=\hbox{\rm 0}
    \digitwidth=\wd0
    \catcode`*=\active
    \def*{\kern\digitwidth}
    \newdimen\signwidth
    \setbox0=\hbox{$-$}
    \signwidth=\wd0
    \catcode`!=\active
    \def!{\kern\signwidth}
    \halign{
      \hbox to 2.0cm{#\leaderfil}\tabskip 2em&
    \hfil# \tabskip 10pt&
    \hfil#\hfil\tabskip 0pt\cr
      \noalign{\doubleline}
      \omit\hfil Channel [GHz]\hfil&
      \omit\hfil $c_{30}$\hfil&
      \omit\hfil $c_{353}$\hfil\cr
      \noalign{\vskip 3pt\hrule\vskip 4pt}
*$70$& $0.0671\pm0.0040$& $0.0085\pm0.0005$\cr
$100$& $0.0252\pm0.0023$& $0.0189\pm0.0003$\cr
$143$& $0.0120\pm0.0024$& $0.0399\pm0.0003$\cr
$217$& $0.0101\pm0.0030$& $0.1292\pm0.0004$\cr
      \noalign{\vskip 3pt\hrule\vskip 5pt}
    }
  }
  \endPlancktable 
  \endgroup

\end{table}

Despite the use of these templates, foreground residuals in the cleaned maps are still dominant over the CMB polarized signal near the Galactic plane. For the power-spectrum estimation, we therefore apply a Galactic mask based on the amplitude of the polarized dust emission retaining either 41\,\%, 52\,\%, 63\,\%, or 75\,\% of the sky \citep[similarly as in][]{planck2016-l05}.

\begin{figure*}[htpb!]
  \centering
  \includegraphics[width=0.48\linewidth]{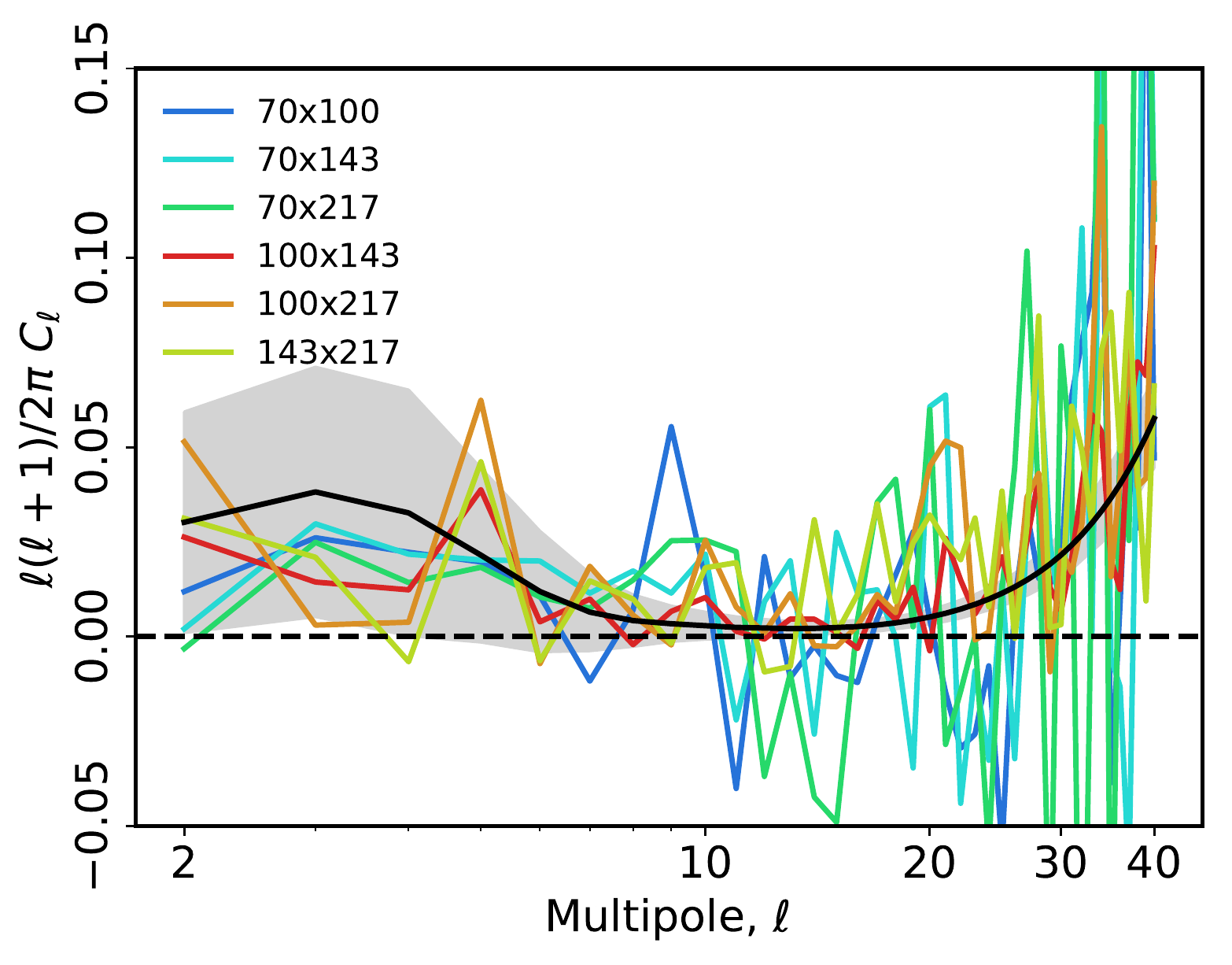}
  \includegraphics[width=0.48\linewidth]{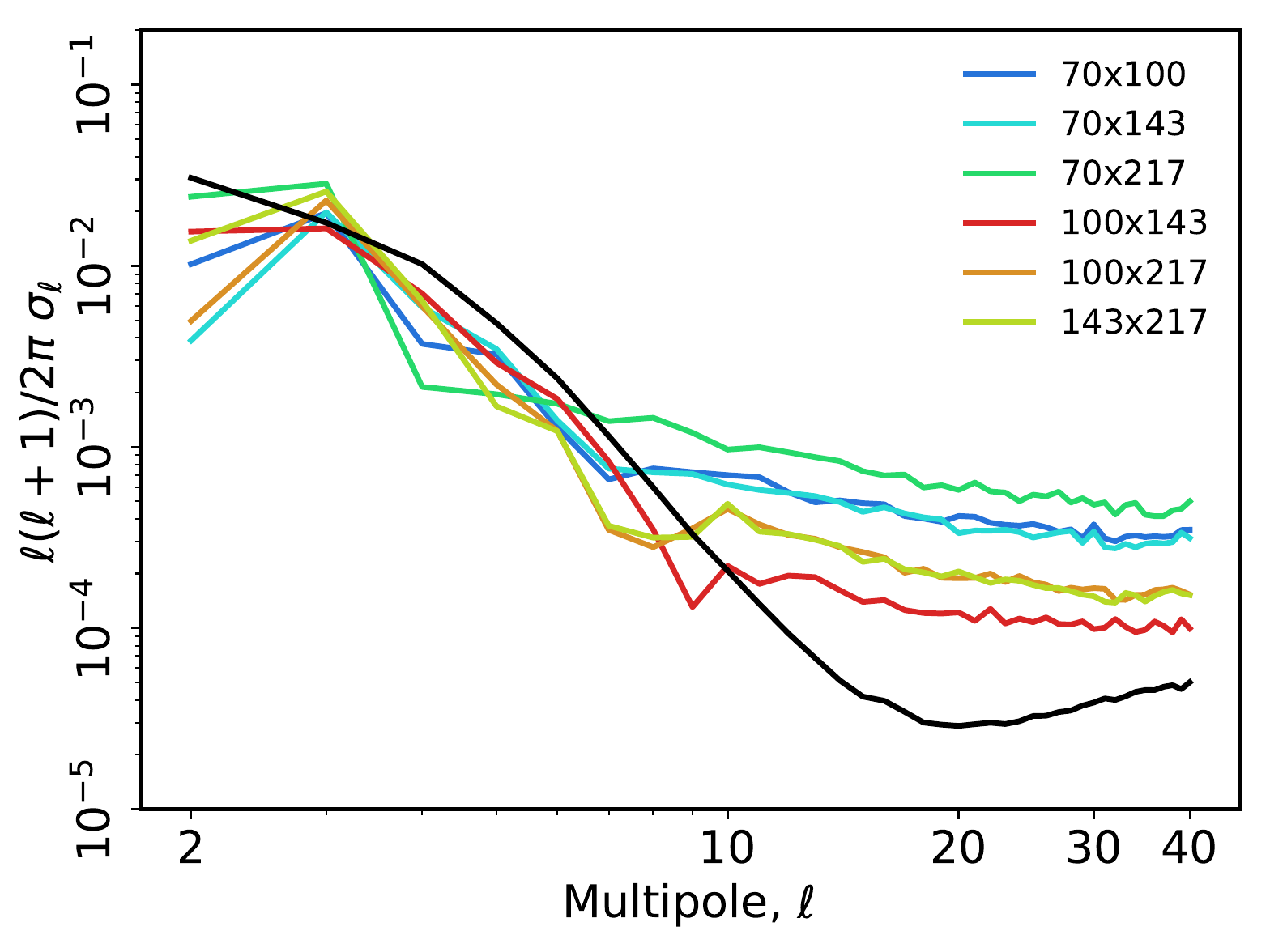}
  \caption{
    {\it Left}: Cross-power spectra for the four \Planck\ frequency maps at 70, 100, 143, and 217\GHz\ compared to the Planck2018 best-fit model for $EE$  (solid black line).  {\it Right}: Standard deviation of the sum of noise, systematics, and uncertainties related to foreground cleaning for the cross-frequency spectra computing using Monte Carlos as compared to the cosmic variance (solid black line).
  }
  \label{fig:lol:cl_freqs}
\end{figure*}

Cross-spectra are estimated using both a pseudo-$C_\ell$ estimator {\tt Xpol} \citep[a generalization to polarization of the algorithm presented in][]{tristram2005} and a so-called cross-QML estimator, which is adapted from {\tt QML} \citep{tegmark2001} for cross-spectra \citep{vanneste2018}, as already used in \prthree\ low-$\ell$ results.  Both estimators lead to compatible results for computing $E$-mode power spectra on \nside{16} maps with slightly less variance for the latter. The final spectra are computed as the average of the cross-QML correlations between the frequency maps cleaned using detector-set~A (or B) and the one cleaned using detector-set~B (or A), thus avoiding any noise bias from the templates.  Figure~\ref{fig:lol:cl_freqs} shows the cross-frequency power spectra corrected for the \npipe\ transfer function (principally affecting multipoles $\ell\,{=}\,2$ and 3, as described in Sect.~\ref{sec:ee_tf}).

We use end-to-end simulations to propagate the uncertainties to the cross-power spectra, and all the way to the estimation of the reionization optical depth $\tau$.  The $C_\ell$ covariance matrix used for the likelihood is directly estimated from these Monte Carlos. Figure~\ref{fig:lol:cl_freqs} shows the $C_\ell$ variance of the Monte Carlos including noise, systematics, and uncertainties related to foreground cleaning compared to cosmic variance. 

\begin{figure}[htpb!]
  \centering
  \includegraphics[width=0.48\textwidth]{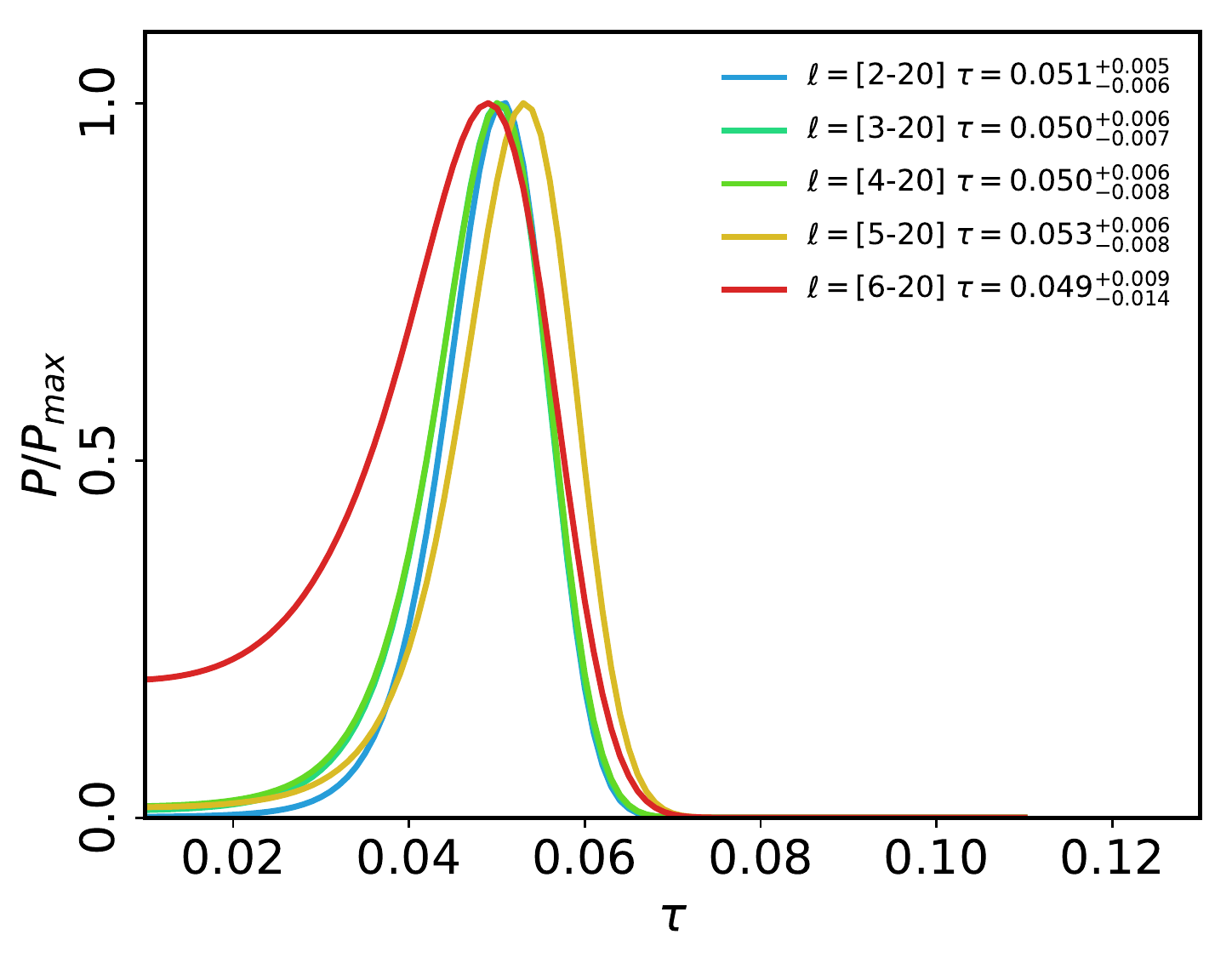}
  \caption{Reionization optical depth $\tau$ posterior for the $100\times143$ $EE$ cross-spectrum, varying the lowest multipole used in the likelihood.}
  \label{fig:lol:tau_lmin}
\end{figure}

\begin{figure}[htpb!]
  \centering
  \includegraphics[width=0.48\textwidth]{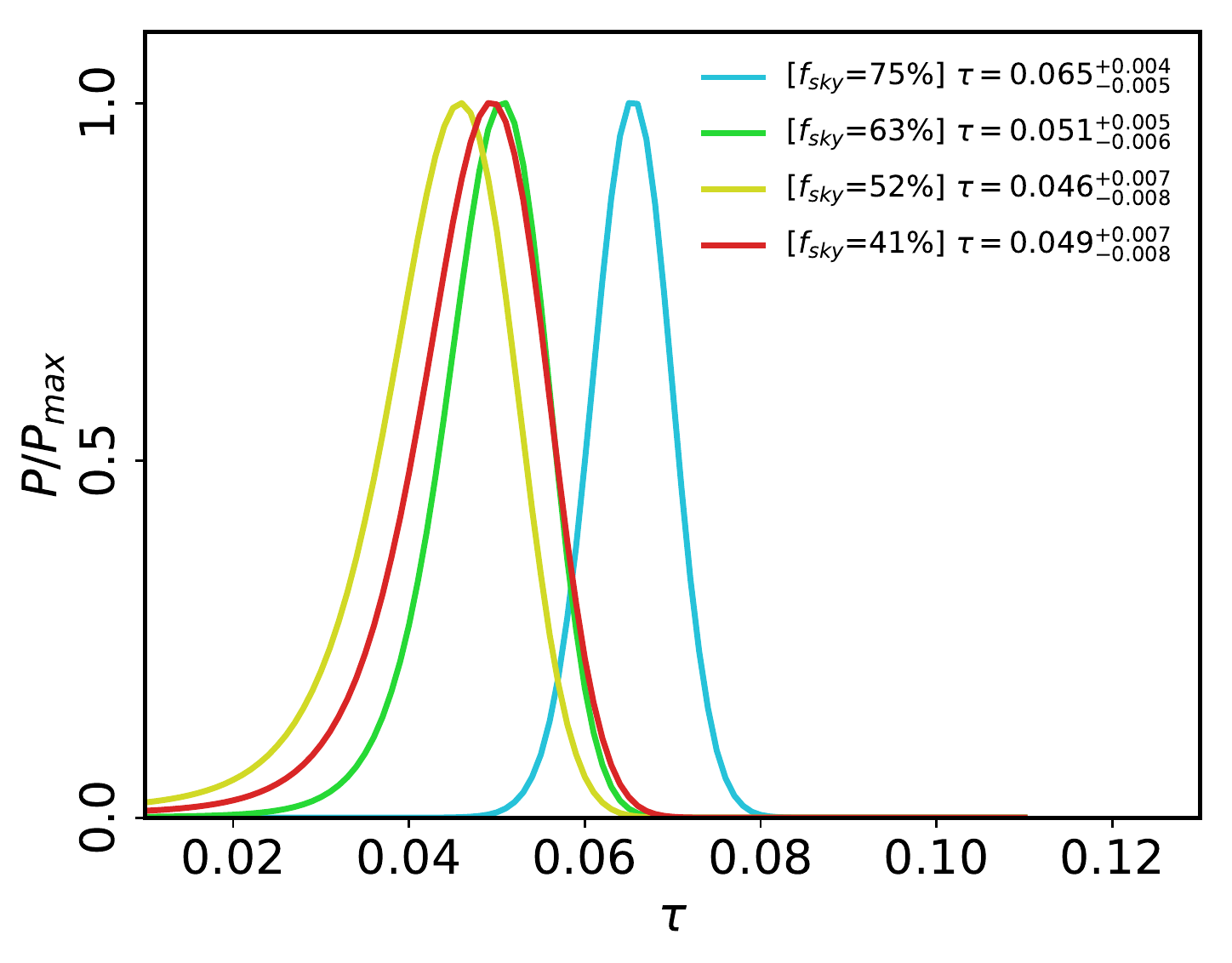}
  \caption{Reionization optical depth $\tau$ posterior for the $100\times143$ $EE$ cross-spectrum, varying the sky fraction used.
    }
  \label{fig:lol:tau_fsky}
\end{figure}

\begin{figure}[htpb!]
  \centering
  \includegraphics[width=0.48\textwidth]{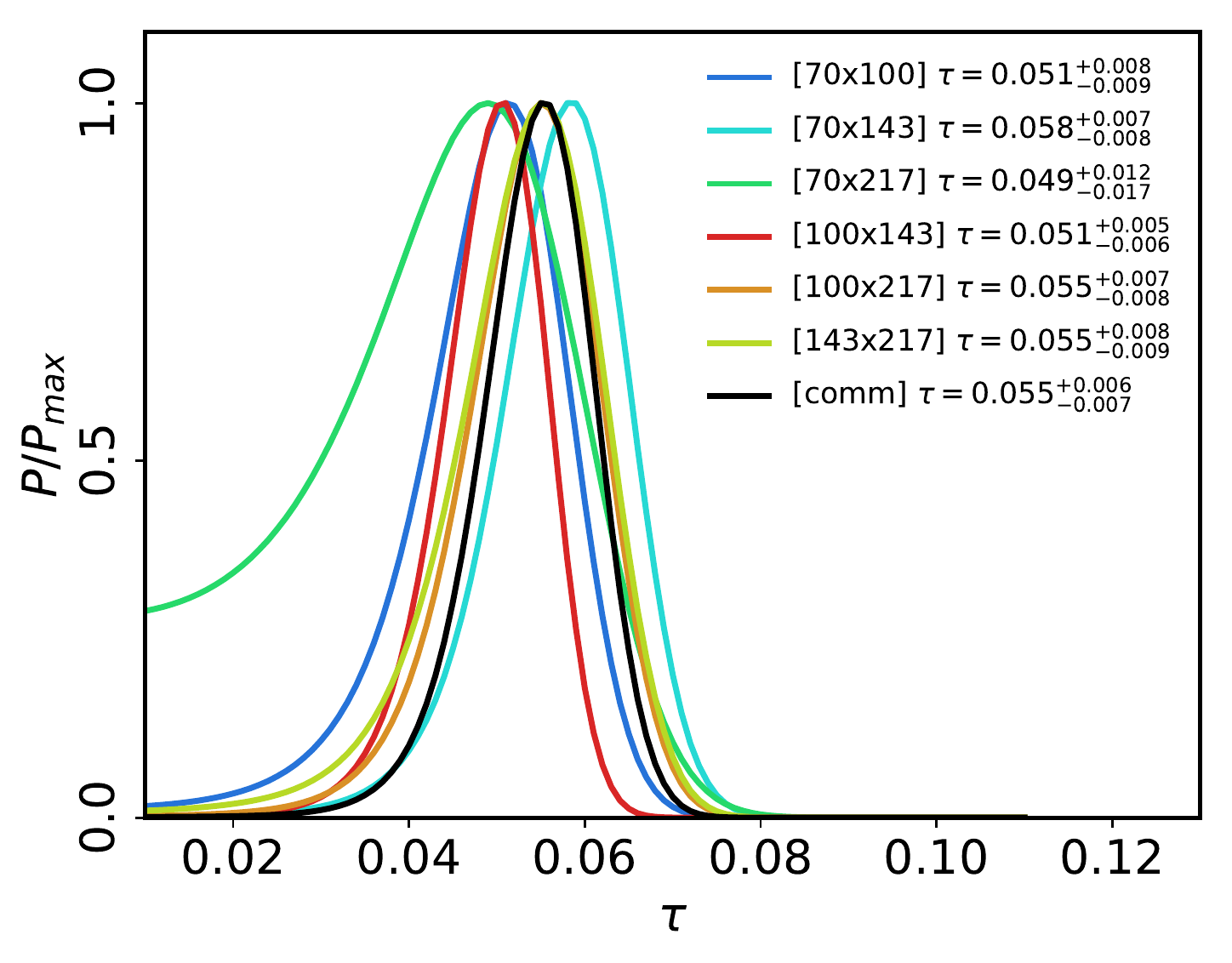}
  \caption{
    Optical depth $\tau$ posterior depending on the cross-spectra used in the likelihood. We use frequency-cleaned maps from 70 to 217\GHz. We also show the posterior for the cross-power spectrum estimated using {\sc commander} detector-set maps.
    }
  \label{fig:lol:tau_freq}
\end{figure}

We construct the likelihood based on the $100\times143$ $EE$ cross-spectrum, and derive the posterior of the reionization optical depth $\tau$ in the $\Lambda$CDM model, fixing $A_{\rm s} e^{-2\tau}$ as well as other parameters to the \Planck\ 2018 best-fit model \citep{planck2016-l06}.  By default, the multipole range used is $\ell\,{=}\,2$--20 containing all the statistical power of the reionization bump in $EE$.  We show the results of the constraints coming from the $100\times143$ cross-spectrum, where we first vary the multipole range used in the likelihood (Fig.~\ref{fig:lol:tau_lmin}) and the sky fraction used for the computation of the power spectra (Fig.~\ref{fig:lol:tau_fsky}).  We also compute the likelihood using the cross-spectrum from \commander\ detector-set maps (using the Monte Carlo simulations accordingly) and compare to the cross-frequency spectra in Fig.~\ref{fig:lol:tau_freq}.

We observe consistent results for all frequencies from 70 to 217\GHz, over the range of multipoles between $\ell\,{=}\,2$ and $\ell\,{=}\,20$, as well as for sky fractions below 75\,\%. Note that, unlike the results in \citet{planck2016-l05}, there is no Monte Carlo correction at the level of the likelihood to take into account residual biases, including foreground residuals depending on sky cuts. 

In particular, the value for the reionization optical depth $\tau$ obtained for the same data set used in \Planck\ 2018 (i.e., the $100\times143$ $EE$ cross-spectrum), using 63\,\% of the sky, is
\begin{equation}
	\tau = 0.051 \pm 0.006,
\end{equation}
which is fully compatible with the value $\tau = 0.0506 \pm 0.0086$ (lowE) derived in \citet{planck2016-l06}, but with a lower uncertainty due to the improvements described in this paper.

\section{Conclusions}
\label{sec:conclusions}

We have presented \npipe, a data-processing pipeline for processing raw \Planck\ timestream data into calibrated frequency maps.  \npipe\ runs outside the \Planck-DPC architecture and can be deployed at almost any supercomputing centre. The software, input data, and configuration files are released (see Appendix~\ref{app:release}) to allow a motivated reader with sufficient computing resources to repeat our analysis and improve upon it.

Compared to \prthree\ results, we have demonstrated significant reduction in the overall noise and systematics across essentially all angular scales, most notably reducing \hfi\ $EE$ and $BB$ noise and systematics variance at $\ell<10$ by 50--90\,\%
and reducing degree-scale statistical noise variance by 10--30\,\% across both instruments.  The improvement in the large-scale polarization uncertainty has been shown to come from the application of a polarized sky model during calibration, and is associated with a measurable and correctable suppression of CMB $E$-mode power at large angular scales.  We have also shown substantial improvements in the internal and external consistency of the frequency maps, providing for a demonstrably-better model of the microwave sky.

\npipe\ processing modules differ, in some cases significantly, from the ones developed, tested, and applied in \Planck\ processing over two decades.   We have tested our results extensively against \prtwo\ and \prthree, and have documented the differences throughout this paper.  Most of these differences are well-understood; they support the notion that \npipe\ maps have lower noise and systematics.   

We summarize here what we consider the benefits of the \npipe\ processing and products, and also give several cautionary comments. 
Advantages of the \npipe\ release:
  \begin{itemize}
  \item reduced levels of noise and systematics at all angular scales;
  \item improved consistency across frequencies, particularly in polarization;
  \item more Monte Carlo realizations of simulated data, and better agreement between the simulated maps and flight data;
  \item availability of single-detector temperature maps from 100 to 857\GHz, with resolution of $N_{\rm side}=4096$ at 217\GHz\ and above;
  \item absence of certain \hfi\ analysis artefacts, in particular ``zebra'' stripes and CO-template pixel boundaries;
  \item one publicly-released pipeline to process \lfi\ and \hfi\ data, accompanied with the release of the raw timestreams for future analysis and improvement.
  \end{itemize}

The following are some cautionary comments about the \npipe\ release.
  \begin{itemize}
  \item Quantitative use of the \npipe\ CMB polarization data on large angular scales ($\ell\,{<}\,20$) requires accounting for the non-negligible transfer function.  This will often necessitate processing the large (36\,TB) body of \npipe\ simulations to determine the degree of signal suppression in the cosmological observables of concern, at considerable computational expense.
  \item At the time of writing, the \npipe\ data products have been extensively tested and validated through to the map level.  With the exception of the demonstration cases of foreground separation and the determination of $\tau$ given in this paper, extraction of the full range of science results represented in the \prtwo\ and  \prthree\ releases remains for the future.  While ongoing tests of cosmological parameter solutions indicate no disparity with the \Planck\ 2018 results, presentation of work on this subject is deferred to a future publication.  Although unlikely, it is possible that future analysis beyond what is presented here will uncover unidentified issues that impact the estimation of cosmological observables from \npipe\ data.
  \item The component separation and sky model described in Sect.~7 use only \Planck\ data, and do not include external data at lower frequencies that help break degeneracies between synchrotron, free-free, and AME, as was done in \prtwo.  We leave this for a future study.  For now, the limitations of the low-frequency foreground model should be recognized.
  \end{itemize}

\npipe\ represents the first comprehensive effort to process all nine \Planck\ frequencies using the same pipeline modules, and to leverage the wide spectral response of the two instruments to derive a consistent, multi-frequency data set.  Our use of the 30-, 217-, and 353-GHz polarization maps as priors in calibrating the CMB frequencies can be considered a first attempt to incorporate component separation \emph{into} the mapmaking pipeline.  On-going and future efforts to expand the treatment into a full-fledged component-separation treatment may yet find significant gains over what is presented here.

\begin{acknowledgements}
  The Planck Collaboration acknowledges the support of: ESA; CNES, and CNRS/INSU-IN2P3-INP (France); ASI, CNR, and INAF (Italy); NASA and DoE (USA); STFC and UKSA (UK); CSIC, MINECO, JA, and RES (Spain); Tekes, AoF, and CSC (Finland); DLR and MPG (Germany); CSA (Canada); DTU Space (Denmark); SER/SSO (Switzerland); RCN (Norway); SFI (Ireland); FCT/MCTES (Portugal); ERC and PRACE (EU). A description of the Planck Collaboration and a list of its members, indicating which technical
  or scientific activities they have been involved in, can be found at
  \href{
    http://www.cosmos.esa.int/web/planck/planck-collaboration
  }{
    \texttt{http://www.cosmos.esa.int/web/planck/planck-collaboration}}.
  This work has received funding from the European Union’s Horizon 2020 research and innovation programme under grant numbers 776282, 772253 and 819478.
  This research would not have been possible without the resources of the National Energy Research Scientific Computing Center (NERSC), a U.S. Department of Energy Office of Science User Facility operated under Contract No. DE-AC02-05CH11231.
\end{acknowledgements}

\bibliographystyle{aat}

\bibliography{Planck_bib,npipe,tau}

\def\eprinttmppp@#1arXiv:@{#1}
\providecommand{\arxivlink[1]}{\href{http://arxiv.org/abs/#1}{arXiv:#1}}
\def\eprinttmp@#1arXiv:#2 [#3]#4@{\ifthenelse{\equal{#3}{x}}{\ifthenelse{
\equal{#1}{}}{\arxivlink{\eprinttmppp@#2@}}{\arxivlink{#1}}}{\arxivlink{#2}
  [#3]}}
\providecommand{\eprintlink}[1]{\eprinttmp@#1arXiv: [x]@}
\providecommand{\eprint}[1]{\eprintlink{#1}}
\providecommand{\adsurl}[1]{\href{#1}{ADS}}
\begin{thebibliography}{75}
\expandafter\ifx\csname natexlab\endcsname\relax\def\natexlab#1{#1}\fi

\bibitem[{{Arg{\"u}eso} {et~al.}(2009){Arg{\"u}eso}, {Sanz}, {Herranz},
  {L{\'o}pez-Caniego}, \& {Gonz{\'a}lez-Nuevo}}]{argueso2009}
{Arg{\"u}eso}, F., {Sanz}, J.~L., {Herranz}, D., {L{\'o}pez-Caniego}, M., \&
  {Gonz{\'a}lez-Nuevo}, J., {Detection/estimation of the modulus of a vector.
  Application to point-source detection in polarization data}. 2009, \mnras,
  395, 649, \eprint{0906.0893}

\bibitem[{{Bennett} {et~al.}(2013){Bennett}, {Larson}, {Weiland}, {Jarosik},
  {Hinshaw}, {Odegard}, {Smith}, {Hill}, {Gold}, {Halpern}, {Komatsu}, {Nolta},
  {Page}, {Spergel}, {Wollack}, {Dunkley}, {Kogut}, {Limon}, {Meyer}, {Tucker},
  \& {Wright}}]{bennett2012}
{Bennett}, C.~L., {Larson}, D., {Weiland}, J.~L., {et~al.}, {Nine-year
  Wilkinson Microwave Anisotropy Probe (WMAP) Observations: Final Maps and
  Results}. 2013, \apjs, 208, 20, \eprint{1212.5225}

\bibitem[{Chon {et~al.}(2004)Chon, Challinor, Prunet, Hivon, \&
  Szapudi}]{Chon:2003gx}
Chon, G., Challinor, A., Prunet, S., Hivon, E., \& Szapudi, I., {Fast
  estimation of polarization power spectra using correlation functions}. 2004,
  Mon. Not. Roy. Astron. Soc., 350, 914, \eprint{astro-ph/0303414}

\bibitem[{{Dame} {et~al.}(2001){Dame}, {Hartmann}, \& {Thaddeus}}]{dame2001}
{Dame}, T.~M., {Hartmann}, D., \& {Thaddeus}, P., {The Milky Way in Molecular
  Clouds: A New Complete CO Survey}. 2001, \apj, 547, 792,
  \eprint{astro-ph/0009217}

\bibitem[{Delouis {et~al.}(2019)Delouis, Pagano, Mottet, Puget, \&
  Vibert}]{Delouis:2019bub}
Delouis, J.~M., Pagano, L., Mottet, S., Puget, J.~L., \& Vibert, L., {SRoll2:
  an improved mapmaking approach to reduce large-scale systematic effects in
  the Planck High Frequency Instrument legacy maps}. 2019, arXiv e-prints,
  \eprint{1901.11386}

\bibitem[{{Eriksen} {et~al.}(2008){Eriksen}, {Jewell}, {Dickinson}, {Banday},
  {G{\'o}rski}, \& {Lawrence}}]{eriksen2008}
{Eriksen}, H.~K., {Jewell}, J.~B., {Dickinson}, C., {et~al.}, {Joint Bayesian
  Component Separation and CMB Power Spectrum Estimation}. 2008, \apj, 676, 10,
  \eprint{0709.1058}

\bibitem[{{Eriksen} {et~al.}(2004){Eriksen}, {O'Dwyer}, {Jewell}, {Wand elt},
  {Larson}, {G{\'o}rski}, {Levin}, {Banday}, \& {Lilje}}]{eriksen:2004}
{Eriksen}, H.~K., {O'Dwyer}, I.~J., {Jewell}, J.~B., {et~al.}, {Power Spectrum
  Estimation from High-Resolution Maps by Gibbs Sampling}. 2004, The
  Astrophysical Journal Supplement Series, 155, 227, \eprint{astro-ph/0407028}

\bibitem[{{Fern{\'a}ndez-Cobos} {et~al.}(2012){Fern{\'a}ndez-Cobos}, {Vielva},
  {Barreiro}, \& {Mart{\'{\i}}nez-Gonz{\'a}lez}}]{fernandez2012}
{Fern{\'a}ndez-Cobos}, R., {Vielva}, P., {Barreiro}, R.~B., \&
  {Mart{\'{\i}}nez-Gonz{\'a}lez}, E., {Multiresolution internal template
  cleaning: an application to the Wilkinson Microwave Anisotropy Probe 7-yr
  polarization data}. 2012, \mnras, 420, 2162, \eprint{1106.2016}

\bibitem[{Fixsen {et~al.}(1994)}]{Fixsen:1993rd}
Fixsen, D. {et~al.}, {Cosmic microwave background dipole spectrum measured by
  the COBE FIRAS}. 1994, Astrophys. J., 420, 445

\bibitem[{{Fixsen}(2009)}]{fixsen2009}
{Fixsen}, D.~J., {The Temperature of the Cosmic Microwave Background}. 2009,
  \apj, 707, 916, \eprint{0911.1955}

\bibitem[{{Godard} {et~al.}(2009){Godard}, {Croon}, {Budnik}, \&
  {Morley}}]{godard2009}
{Godard}, B., {Croon}, M., {Budnik}, F., \& {Morley}, T. 2009, in Proceedings
  of the 21st International Symposium on Space Flight Dynamics (ISSFD),
  Toulouse

\bibitem[{{G{\'o}rski} {et~al.}(2005){G{\'o}rski}, {Hivon}, {Banday},
  {Wandelt}, {Hansen}, {Reinecke}, \& {Bartelmann}}]{gorski2005}
{G{\'o}rski}, K.~M., {Hivon}, E., {Banday}, A.~J., {et~al.}, {HEALPix: A
  Framework for High-Resolution Discretization and Fast Analysis of Data
  Distributed on the Sphere}. 2005, \apj, 622, 759, \eprint{astro-ph/0409513}

\bibitem[{{Hamimeche} \& {Lewis}(2008)}]{hamimeche2008}
{Hamimeche}, S. \& {Lewis}, A., {Likelihood analysis of CMB temperature and
  polarization power spectra}. 2008, \prd, 77, 103013, \eprint{0801.0554}

\bibitem[{{Hinshaw} {et~al.}(2009){Hinshaw}, {Weiland}, {Hill}, {Odegard},
  {Larson}, {Bennett}, {Dunkley}, {Gold}, {Greason}, {Jarosik}, {Komatsu},
  {Nolta}, {Page}, {Spergel}, {Wollack}, {Halpern}, {Kogut}, {Limon}, {Meyer},
  {Tucker}, \& {Wright}}]{hinshaw2009}
{Hinshaw}, G., {Weiland}, J.~L., {Hill}, R.~S., {et~al.}, {Five-Year Wilkinson
  Microwave Anisotropy Probe (WMAP) Observations: Data Processing, Sky Maps,
  and Basic Results}. 2009, \apjs, 180, 225, \eprint{0803.0732}

\bibitem[{{Hivon} {et~al.}(2017){Hivon}, {Mottet}, \&
  {Ponthieu}}]{Hivon:2016qyw}
{Hivon}, E., {Mottet}, S., \& {Ponthieu}, N., {QuickPol: Fast calculation of
  effective beam matrices for CMB polarization}. 2017, \aap, 598, A25,
  \eprint{1608.08833}

\bibitem[{{Jewell} {et~al.}(2004){Jewell}, {Levin}, \&
  {Anderson}}]{jewell:2004}
{Jewell}, J., {Levin}, S., \& {Anderson}, C.~H., {Application of Monte Carlo
  Algorithms to the Bayesian Analysis of the Cosmic Microwave Background}.
  2004, \apj, 609, 1, \eprint{astro-ph/0209560}

\bibitem[{{Keih{\"a}nen} {et~al.}(2010){Keih{\"a}nen}, {Keskitalo},
  {Kurki-Suonio}, {Poutanen}, \& {Sirvi{\"o}}}]{keihanen2010}
{Keih{\"a}nen}, E., {Keskitalo}, R., {Kurki-Suonio}, H., {Poutanen}, T., \&
  {Sirvi{\"o}}, A., {Making cosmic microwave background temperature and
  polarization maps with MADAM}. 2010, \aap, 510, A57, \eprint{0907.0367}

\bibitem[{{Keih{\"a}nen} {et~al.}(2005){Keih{\"a}nen}, {Kurki-Suonio}, \&
  {Poutanen}}]{keihanen2005}
{Keih{\"a}nen}, E., {Kurki-Suonio}, H., \& {Poutanen}, T., {MADAM- a map-making
  method for CMB experiments}. 2005, \mnras, 360, 390,
  \eprint{astro-ph/0412517}

\bibitem[{{Keih{\"a}nen} {et~al.}(2004){Keih{\"a}nen}, {Kurki-Suonio},
  {Poutanen}, {Maino}, \& {Burigana}}]{keihanen2004}
{Keih{\"a}nen}, E., {Kurki-Suonio}, H., {Poutanen}, T., {Maino}, D., \&
  {Burigana}, C., {A maximum likelihood approach to the destriping technique}.
  2004, \aap, 428, 287, \eprint{astro-ph/0304411}

\bibitem[{{Kelsall} {et~al.}(1998){Kelsall}, {Weiland}, {Franz}, {Reach},
  {Arendt}, {Dwek}, {Freudenreich}, {Hauser}, {Moseley}, {Odegard},
  {Silverberg}, \& {Wright}}]{kelsall1998}
{Kelsall}, T., {Weiland}, J.~L., {Franz}, B.~A., {et~al.}, {The COBE Diffuse
  Infrared Background Experiment Search for the Cosmic Infrared Background. II.
  Model of the Interplanetary Dust Cloud}. 1998, \apj, 508, 44,
  \eprint{astro-ph/9806250}

\bibitem[{{Kurki-Suonio} {et~al.}(2009){Kurki-Suonio}, {Keih{\"a}nen},
  {Keskitalo}, {Poutanen}, {Sirvi{\"o}}, {Maino}, \&
  {Burigana}}]{kurki-suonio2009}
{Kurki-Suonio}, H., {Keih{\"a}nen}, E., {Keskitalo}, R., {et~al.}, {Destriping
  CMB temperature and polarization maps}. 2009, \aap, 506, 1511,
  \eprint{0904.3623}

\bibitem[{{Leach} {et~al.}(2008){Leach}, {Cardoso}, {Baccigalupi}, {Barreiro},
  {Betoule}, {Bobin}, {Bonaldi}, {Delabrouille}, {de Zotti}, {Dickinson},
  {Eriksen}, {Gonz{\'a}lez-Nuevo}, {Hansen}, {Herranz}, {Le Jeune},
  {L{\'o}pez-Caniego}, {Mart{\'{\i}}nez-Gonz{\'a}lez}, {Massardi}, {Melin},
  {Miville-Desch{\^e}nes}, {Patanchon}, {Prunet}, {Ricciardi}, {Salerno},
  {Sanz}, {Starck}, {Stivoli}, {Stolyarov}, {Stompor}, \& {Vielva}}]{leach2008}
{Leach}, S.~M., {Cardoso}, J., {Baccigalupi}, C., {et~al.}, {Component
  separation methods for the PLANCK mission}. 2008, \aap, 491, 597,
  \eprint{0805.0269}

\bibitem[{{Lenz} {et~al.}(2017){Lenz}, {Hensley}, \&
  {Dor{\'e}}}]{Lenz_et_al:2017}
{Lenz}, D., {Hensley}, B.~S., \& {Dor{\'e}}, O., {A New, Large-scale Map of
  Interstellar Reddening Derived from H I Emission}. 2017, \apj, 846, 38,
  \eprint{1706.00011}

\bibitem[{{Lineweaver} {et~al.}(1996){Lineweaver}, {Tenorio}, {Smoot},
  {Keegstra}, {Banday}, \& {Lubin}}]{lineweaver1996}
{Lineweaver}, C.~H., {Tenorio}, L., {Smoot}, G.~F., {et~al.}, {The Dipole
  Observed in the COBE DMR 4 Year Data}. 1996, \apj, 470, 38,
  \eprint{astro-ph/9601151}

\bibitem[{{L{\'o}pez-Caniego} {et~al.}(2006){L{\'o}pez-Caniego}, {Herranz},
  {Gonz{\'a}lez-Nuevo}, {Sanz}, {Barreiro}, {Vielva}, {Arg{\"u}eso}, \&
  {Toffolatti}}]{lopezcaniego2006}
{L{\'o}pez-Caniego}, M., {Herranz}, D., {Gonz{\'a}lez-Nuevo}, J., {et~al.},
  {Comparison of filters for the detection of point sources in Planck
  simulations}. 2006, \mnras, 370, 2047, \eprint{astro-ph/0606199}

\bibitem[{{Mangilli} {et~al.}(2019){Mangilli}, {Aumont}, {Bernard}, {Buzzelli},
  {de Gasperis}, {Durrive}, {Ferri{\`e}re}, {Fo{\"e}nard}, {Hughes}, \&
  {Lacourt}}]{2019arXiv190106196M}
{Mangilli}, A., {Aumont}, J., {Bernard}, J.~P., {et~al.}, {The geometry of the
  magnetic field in the Central Molecular Zone measured by PILOT}. 2019, arXiv
  e-prints, arXiv:1901.06196, \eprint{1901.06196}

\bibitem[{{Mangilli} {et~al.}(2015){Mangilli}, {Plaszczynski}, \&
  {Tristram}}]{mangilli2015}
{Mangilli}, A., {Plaszczynski}, S., \& {Tristram}, M., {Large-scale cosmic
  microwave background temperature and polarization cross-spectra likelihoods}.
  2015, \mnras, 453, 3174, \eprint{1503.01347}

\bibitem[{Notari \& Quartin(2015)}]{Notari:2015kla}
Notari, A. \& Quartin, M., {On the proper kinetic quadrupole CMB removal and
  the quadrupole anomalies}. 2015, JCAP, 1506, 047, \eprint{1504.02076}

\bibitem[{{Odegard} {et~al.}(2019){Odegard}, {Weiland}, {Fixsen}, {Chuss},
  {Dwek}, {Kogut}, \& {Switzer}}]{odegard2019}
{Odegard}, N., {Weiland}, J.~L., {Fixsen}, D.~J., {et~al.}, {Determination of
  the Cosmic Infrared Background from COBE/FIRAS and Planck HFI Observations}.
  2019, \apj, 877, 40, \eprint{1904.11556}

\bibitem[{{Planck HFI Core Team}(2011)}]{planck2011-1.7}
{Planck HFI Core Team}, {\textit{Planck} early results. VI. The High Frequency
  Instrument data processing}. 2011, \aap, 536, A6, \eprint{1101.2048}

\bibitem[{{\sorthelp{Planck Collaboration 2014A}}{Planck Collaboration
  I}(2014)}]{planck2013-p01}
{\sorthelp{Planck Collaboration 2014A}}{Planck Collaboration I},
  {\textit{Planck} 2013 results. I. Overview of products and scientific
  results}. 2014, \aap, 571, A1, \eprint{1303.5062}

\bibitem[{{\sorthelp{Planck Collaboration 2014B}}{Planck Collaboration
  II}(2014)}]{planck2013-p02}
{\sorthelp{Planck Collaboration 2014B}}{Planck Collaboration II},
  {\textit{Planck} 2013 results. II. Low Frequency Instrument data processing}.
  2014, \aap, 571, A2, \eprint{1303.5063}

\bibitem[{{\sorthelp{Planck Collaboration 2014E}}{Planck Collaboration
  V}(2014)}]{planck2013-p02b}
{\sorthelp{Planck Collaboration 2014E}}{Planck Collaboration V},
  {\textit{Planck} 2013 results. V. LFI Calibration}. 2014, \aap, 571, A5,
  \eprint{1303.5066}

\bibitem[{{\sorthelp{Planck Collaboration 2014F}}{Planck Collaboration
  VI}(2014)}]{planck2013-p03}
{\sorthelp{Planck Collaboration 2014F}}{Planck Collaboration VI},
  {\textit{Planck} 2013 results. VI. High Frequency Instrument data
  processing}. 2014, \aap, 571, A6, \eprint{1303.5067}

\bibitem[{{\sorthelp{Planck Collaboration 2014I}}{Planck Collaboration
  IX}(2014)}]{planck2013-p03d}
{\sorthelp{Planck Collaboration 2014I}}{Planck Collaboration IX},
  {\textit{Planck} 2013 results. IX. HFI spectral response}. 2014, \aap, 571,
  A9, \eprint{1303.5070}

\bibitem[{{\sorthelp{Planck Collaboration 2014J}}{Planck Collaboration
  X}(2014)}]{planck2013-p03e}
{\sorthelp{Planck Collaboration 2014J}}{Planck Collaboration X},
  {\textit{Planck} 2013 results. X. HFI energetic particle effects:
  characterization, removal, and simulation}. 2014, \aap, 571, A10,
  \eprint{1303.5071}

\bibitem[{{\sorthelp{Planck Collaboration 2014L}}{Planck Collaboration
  XII}(2014)}]{planck2013-p06}
{\sorthelp{Planck Collaboration 2014L}}{Planck Collaboration XII},
  {\textit{Planck} 2013 results. XII. Diffuse component separation}. 2014,
  \aap, 571, A12, \eprint{1303.5072}

\bibitem[{{\sorthelp{Planck Collaboration 2014N}}{Planck Collaboration
  XIV}(2014)}]{planck2013-pip88}
{\sorthelp{Planck Collaboration 2014N}}{Planck Collaboration XIV},
  {\textit{Planck} 2013 results. XIV. Zodiacal emission}. 2014, \aap, 571, A14,
  \eprint{1303.5074}

\bibitem[{{\sorthelp{Planck Collaboration 2014ZB}}{Planck Collaboration
  XXVII}(2014)}]{planck2013-pipaberration}
{\sorthelp{Planck Collaboration 2014ZB}}{Planck Collaboration XXVII},
  {\textit{Planck} 2013 results. XXVII. Doppler boosting of the CMB: Eppur si
  muove}. 2014, \aap, 571, A27, \eprint{1303.5087}

\bibitem[{{\sorthelp{Planck Collaboration 2015A}}{Planck Collaboration
  I}(2016)}]{planck2014-a01}
{\sorthelp{Planck Collaboration 2015A}}{Planck Collaboration I},
  {\textit{Planck} 2015 results. I. Overview of products and results}. 2016,
  \aap, 594, A1, \eprint{1502.01582}

\bibitem[{{\sorthelp{Planck Collaboration 2015B}}{Planck Collaboration
  II}(2016)}]{planck2014-a03}
{\sorthelp{Planck Collaboration 2015B}}{Planck Collaboration II},
  {\textit{Planck} 2015 results. II. Low Frequency Instrument data processing}.
  2016, \aap, 594, A2, \eprint{1502.01583}

\bibitem[{{\sorthelp{Planck Collaboration 2015E}}{Planck Collaboration
  V}(2016)}]{planck2014-a06}
{\sorthelp{Planck Collaboration 2015E}}{Planck Collaboration V},
  {\textit{Planck} 2015 results. V. LFI calibration}. 2016, \aap, 594, A5,
  \eprint{1505.08022}

\bibitem[{{\sorthelp{Planck Collaboration 2015G}}{Planck Collaboration
  VII}(2016)}]{planck2014-a08}
{\sorthelp{Planck Collaboration 2015G}}{Planck Collaboration VII},
  {\textit{Planck} 2015 results. VII. High Frequency Instrument data
  processing: Time-ordered information and beam processing}. 2016, \aap, 594,
  A7, \eprint{1502.01586}

\bibitem[{{\sorthelp{Planck Collaboration 2015H}}{Planck Collaboration
  VIII}(2016)}]{planck2014-a09}
{\sorthelp{Planck Collaboration 2015H}}{Planck Collaboration VIII},
  {\textit{Planck} 2015 results. VIII. High Frequency Instrument data
  processing: Calibration and maps}. 2016, \aap, 594, A8, \eprint{1502.01587}

\bibitem[{{\sorthelp{Planck Collaboration 2015I}}{Planck Collaboration
  IX}(2016)}]{planck2014-a11}
{\sorthelp{Planck Collaboration 2015I}}{Planck Collaboration IX},
  {\textit{Planck} 2015 results. IX. Diffuse component separation: CMB maps}.
  2016, \aap, 594, A9, \eprint{1502.05956}

\bibitem[{{\sorthelp{Planck Collaboration 2015J}}{Planck Collaboration
  X}(2016)}]{planck2014-a12}
{\sorthelp{Planck Collaboration 2015J}}{Planck Collaboration X},
  {\textit{Planck} 2015 results. X. Diffuse component separation: Foreground
  maps}. 2016, \aap, 594, A10, \eprint{1502.01588}

\bibitem[{{\sorthelp{Planck Collaboration 2015L}}{Planck Collaboration
  XII}(2016)}]{planck2014-a14}
{\sorthelp{Planck Collaboration 2015L}}{Planck Collaboration XII},
  {\textit{Planck} 2015 results. XII. Full Focal Plane simulations}. 2016,
  \aap, 594, A12, \eprint{1509.06348}

\bibitem[{{\sorthelp{Planck Collaboration 2015U}}{Planck Collaboration
  XXI}(2016)}]{planck2014-a26}
{\sorthelp{Planck Collaboration 2015U}}{Planck Collaboration XXI},
  {\textit{Planck} 2015 results. XXI. The integrated Sachs-Wolfe effect}. 2016,
  \aap, 594, A21, \eprint{1502.01595}

\bibitem[{{\sorthelp{Planck Collaboration 2015Y}}{Planck Collaboration
  XXV}(2016)}]{planck2014-a31}
{\sorthelp{Planck Collaboration 2015Y}}{Planck Collaboration XXV},
  {\textit{Planck} 2015 results. XXV. Diffuse, low-frequency Galactic
  foregrounds}. 2016, \aap, 594, A25, \eprint{1506.06660}

\bibitem[{{\sorthelp{Planck Collaboration 2015ZA}}{Planck Collaboration
  XXVI}(2016)}]{planck2014-a35}
{\sorthelp{Planck Collaboration 2015ZA}}{Planck Collaboration XXVI},
  {\textit{Planck} 2015 results. XXVI. The Second Planck Catalogue of Compact
  Sources}. 2016, \aap, 594, A26, \eprint{1507.02058}

\bibitem[{{\sorthelp{Planck Collaboration 2018A}}{Planck Collaboration
  I}(2019)}]{planck2016-l01}
{\sorthelp{Planck Collaboration 2018A}}{Planck Collaboration I},
  {\textit{Planck} 2018 results. I. Overview, and the cosmological legacy of
  \textit{Planck}}. 2019, \aap, in press, \eprint{1807.06205}

\bibitem[{{\sorthelp{Planck Collaboration 2018B}}{Planck Collaboration
  II}(2019)}]{planck2016-l02}
{\sorthelp{Planck Collaboration 2018B}}{Planck Collaboration II},
  {\textit{Planck} 2018 results. II. Low Frequency Instrument data processing}.
  2019, \aap, in press, \eprint{1807.06206}

\bibitem[{{\sorthelp{Planck Collaboration 2018C}}{Planck Collaboration
  III}(2019)}]{planck2016-l03}
{\sorthelp{Planck Collaboration 2018C}}{Planck Collaboration III},
  {\textit{Planck} 2018 results. III. High Frequency Instrument data
  processing}. 2019, \aap, in press, \eprint{1807.06207}

\bibitem[{{\sorthelp{Planck Collaboration 2018D}}{Planck Collaboration
  IV}(2019)}]{planck2016-l04}
{\sorthelp{Planck Collaboration 2018D}}{Planck Collaboration IV},
  {\textit{Planck} 2018 results. IV. Diffuse component separation}. 2019, \aap,
  in press, \eprint{1807.06208}

\bibitem[{{\sorthelp{Planck Collaboration 2018E}}{Planck Collaboration
  V}(2019)}]{planck2016-l05}
{\sorthelp{Planck Collaboration 2018E}}{Planck Collaboration V},
  {\textit{Planck} 2018 results. V. Power spectra and likelihoods}. 2019, \aap,
  submitted, \eprint{1907.12875}

\bibitem[{{\sorthelp{Planck Collaboration 2018F}}{Planck Collaboration
  VI}(2019)}]{planck2016-l06}
{\sorthelp{Planck Collaboration 2018F}}{Planck Collaboration VI},
  {\textit{Planck} 2018 results. VI. Cosmological parameters}. 2019, \aap,
  submitted, \eprint{1807.06209}

\bibitem[{{\sorthelp{Planck Collaboration 2018G}}{Planck Collaboration
  VII}(2019)}]{planck2016-l07}
{\sorthelp{Planck Collaboration 2018G}}{Planck Collaboration VII},
  {\textit{Planck} 2018 results. VII. Isotropy and statistics}. 2019, \aap, in
  press, \eprint{1906.02552}

\bibitem[{{\sorthelp{Planck Collaboration 2018K}}{Planck Collaboration
  XI}(2019)}]{planck2016-l11A}
{\sorthelp{Planck Collaboration 2018K}}{Planck Collaboration XI},
  {\textit{Planck} 2018 results. XI. Polarized dust foregrounds}. 2019, \aap,
  in press, \eprint{1801.04945}

\bibitem[{{\sorthelp{Planck Collaboration IntZU}}{Planck Collaboration Int.
  XLVI}(2016)}]{planck2014-a10}
{\sorthelp{Planck Collaboration IntZU}}{Planck Collaboration Int. XLVI},
  {\textit{Planck} intermediate results. XLVI. Reduction of large-scale
  systematic effects in HFI polarization maps and estimation of the
  reionization optical depth}. 2016, \aap, 596, A107, \eprint{1605.02985}

\bibitem[{{\sorthelp{Planck Collaboration IntZV}}{Planck Collaboration Int.
  XLVII}(2016)}]{planck2014-a25}
{\sorthelp{Planck Collaboration IntZV}}{Planck Collaboration Int. XLVII},
  {\textit{Planck} intermediate results. XLVII. Constraints on reionization
  history}. 2016, \aap, 596, A108, \eprint{1605.03507}

\bibitem[{Prezeau \& Reinecke(2010)}]{Prezeau:2010mx}
Prezeau, G. \& Reinecke, M., {Algorithm for the evaluation of reduced Wigner
  matrices}. 2010, Astrophys. J. Suppl., 190, 267, \eprint{1002.1050}

\bibitem[{{Reinecke} \& {Seljebotn}(2013)}]{2013A&A...554A.112R}
{Reinecke}, M. \& {Seljebotn}, D.~S., {Libsharp - spherical harmonic transforms
  revisited}. 2013, \aap, 554, A112, \eprint{1303.4945}

\bibitem[{{Remazeilles} {et~al.}(2011){Remazeilles}, {Delabrouille}, \&
  {Cardoso}}]{Remazeilles2011b}
{Remazeilles}, M., {Delabrouille}, J., \& {Cardoso}, J.-F., {Foreground
  component separation with generalized Internal Linear Combination}. 2011,
  \mnras, 418, 467, \eprint{1103.1166}

\bibitem[{{Rosset} {et~al.}(2010){Rosset}, {Tristram}, {Ponthieu}, {Ade},
  {Aumont}, {Catalano}, {Conversi}, {Couchot}, {Crill}, {D{\'e}sert}, {Ganga},
  {Giard}, {Giraud-H{\'e}raud}, {Ha{\"i}ssinski}, {Henrot-Versill{\'e}},
  {Holmes}, {Jones}, {Lamarre}, {Lange}, {Leroy}, {Mac{\'{\i}}as-P{\'e}rez},
  {Maffei}, {de Marcillac}, {Miville-Desch{\^e}nes}, {Montier}, {Noviello},
  {Pajot}, {Perdereau}, {Piacentini}, {Piat}, {Plaszczynski}, {Pointecouteau},
  {Puget}, {Ristorcelli}, {Savini}, {Sudiwala}, {Veneziani}, \&
  {Yvon}}]{rosset2010}
{Rosset}, C., {Tristram}, M., {Ponthieu}, N., {et~al.}, {\textit{Planck}
  pre-launch status: High Frequency Instrument polarization calibration}. 2010,
  \aap, 520, A13, \eprint{1004.2595}

\bibitem[{{Seljebotn} {et~al.}(2017){Seljebotn}, {B{\ae}rland}, {Eriksen},
  {Mardal}, \& {Wehus}}]{seljebotn:2017}
{Seljebotn}, D.~S., {B{\ae}rland}, T., {Eriksen}, H.~K., {Mardal}, K.~A., \&
  {Wehus}, I.~K., {Multi-resolution Bayesian CMB component separation through
  Wiener-filtering with a pseudo-inverse preconditioner}. 2017, arXiv e-prints,
  arXiv:1710.00621, \eprint{1710.00621}

\bibitem[{{Szapudi} {et~al.}(2001){Szapudi}, {Prunet}, \&
  {Colombi}}]{Szapudi:2000xj}
{Szapudi}, I., {Prunet}, S., \& {Colombi}, S., {Fast Analysis of Inhomogenous
  Megapixel Cosmic Microwave Background Maps}. 2001, \apjl, 561, L11

\bibitem[{{Tegmark} \& {de Oliveira-Costa}(2001)}]{tegmark2001}
{Tegmark}, M. \& {de Oliveira-Costa}, A., {How to measure CMB polarization
  power spectra without losing information}. 2001, \prd, 64, 063001,
  \eprint{astro-ph/0012120}

\bibitem[{Thommesen {et~al.}(2019)}]{thommesen:2019}
Thommesen {et~al.}, {Bayesian CMB dipole estimation on an incomplete sky}.
  2019, in preparation

\bibitem[{{Tristram} {et~al.}(2005){Tristram}, {Mac{\'{\i}}as-P{\'e}rez},
  {Renault}, \& {Santos}}]{tristram2005}
{Tristram}, M., {Mac{\'{\i}}as-P{\'e}rez}, J.~F., {Renault}, C., \& {Santos},
  D., {XSPECT, estimation of the angular power spectrum by computing
  cross-power spectra with analytical error bars}. 2005, \mnras, 358, 833,
  \eprint{astro-ph/0405575}

\bibitem[{Tuttlebee(2013)}]{PT-CMOC-OPS-RP-6435-HSO-GF}
Tuttlebee, M. 2013, Herschel/Planck Star Tracker Performance Assessment and
  Calibration, Tech. Rep. PT-CMOC-OPS-RP-6435-HSO-GF, Herschel/Planck Flight
  Dynamics; SCISYS GmbH

\bibitem[{{Vanneste} {et~al.}(2018){Vanneste}, {Henrot-Versill{\'e}}, {Louis},
  \& {Tristram}}]{vanneste2018}
{Vanneste}, S., {Henrot-Versill{\'e}}, S., {Louis}, T., \& {Tristram}, M.,
  {Quadratic estimator for CMB cross-correlation}. 2018, \prd, 98, 103526,
  \eprint{1807.02484}

\bibitem[{{Villa} {et~al.}(2010){Villa}, {Terenzi}, {Sandri}, {Meinhold},
  {Poutanen}, {Battaglia}, {Franceschet}, {Hughes}, {Laaninen}, {Lapolla},
  {Bersanelli}, {Butler}, {Cuttaia}, {D'Arcangelo}, {Frailis}, {Franceschi},
  {Galeotta}, {Gregorio}, {Leonardi}, {Lowe}, {Mandolesi}, {Maris}, {Mendes},
  {Mennella}, {Morgante}, {Stringhetti}, {Tomasi}, {Valenziano}, {Zacchei},
  {Zonca}, {Aja}, {Artal}, {Balasini}, {Bernardino}, {Blackhurst}, {Boschini},
  {Cappellini}, {Cavaliere}, {Colin}, {Colombo}, {Davis}, {de La Fuente},
  {Edgeley}, {Gaier}, {Galtress}, {Hoyland}, {Jukkala}, {Kettle}, {Kilpia},
  {Lawrence}, {Lawson}, {Leahy}, {Leutenegger}, {Levin}, {Maino}, {Malaspina},
  {Mediavilla}, {Miccolis}, {Pagan}, {Pascual}, {Pasian}, {Pecora},
  {Pospieszalski}, {Roddis}, {Salmon}, {Seiffert}, {Silvestri}, {Simonetto},
  {Sjoman}, {Sozzi}, {Tuovinen}, {Varis}, {Wilkinson}, \& {Winder}}]{villa2010}
{Villa}, F., {Terenzi}, L., {Sandri}, M., {et~al.}, {\textit{Planck} pre-launch
  status: Calibration of the Low Frequency Instrument flight model
  radiometers}. 2010, \aap, 520, A6, \eprint{1005.2541}

\bibitem[{{Wandelt} {et~al.}(2004){Wandelt}, {Larson}, \&
  {Lakshminarayanan}}]{wandelt:2004}
{Wandelt}, B.~D., {Larson}, D.~L., \& {Lakshminarayanan}, A., {Global, exact
  cosmic microwave background data analysis using Gibbs sampling}. 2004, \prd,
  70, 083511, \eprint{astro-ph/0310080}

\bibitem[{{Wehus} {et~al.}(2014){Wehus}, {Fuskeland}, {Eriksen}, {Band ay},
  {Dickinson}, {Ghosh}, {Gorski}, {Lawrence}, {Leahy}, {Maino}, {Reich}, \&
  {Reich}}]{wehus:2014}
{Wehus}, I.~K., {Fuskeland}, U., {Eriksen}, H.~K., {et~al.}, {Monopole and
  dipole estimation for multi-frequency sky maps by linear regression}. 2014,
  arXiv e-prints, arXiv:1411.7616, \eprint{1411.7616}

\bibitem[{Zonca {et~al.}(2019)Zonca, Singer, Lenz, Reinecke, Rosset, Hivon, \&
  Gorski}]{Zonca2019healpy}
Zonca, A., Singer, L., Lenz, D., {et~al.}, healpy: equal area pixelization and
  spherical harmonics transforms for data on the sphere in Python. 2019,
  Journal of Open Source Software, 4, 1298

\end{thebibliography}

\appendix

\section{\npipe\ software and data release} 
\label{app:release}

The \npipe\ software is publicly accessible on Github\footnote{
  \href{https://github.com}{\texttt{github.com}}} at
\href{https://github.com/hpc4cmb/toast-npipe
}{\texttt{github.com/hpc4cmb/toast-npipe}}.
In addition to some general dependencies, \npipe\ is built upon the {\tt TOAST}\footnote{{\tt TOAST} is the time-ordered astrophysics scalable toolset. It is publicly
  available at \href{https://github.com/hpc4cmb/toast}{
    \texttt{github.com/hpc4cmb/toast}}.
} framework for time-ordered data processing.  The final destriping is performed
using \libmadam\footnote{\libmadam\ is a library version of the \madam\ generalized destriper.  It can
  be downloaded from \href{https://github.com/hpc4cmb/toast}{
    \texttt{github.com/hpc4cmb/libmadam}}.
} \citep{keihanen2005,keihanen2010}.  \npipe\ is a Python3 code with Cython and C
extensions. \libmadam\ is a Fortran 2003 code.

The pipeline is parallelized with \texttt{mpi4py}.\footnote{
  \href{https://bitbucket.org/mpi4py/mpi4py}{
    \texttt{bitbucket.org/mpi4py/mpi4py}}
}  The beam-convolved timestreams are produced with
\texttt{libconviqt}\footnote{
  \href{https://github.com/hpc4cmb/libconviqt}{
    \texttt{github.com/hpc4cmb/libconviqt}}
} \citep{Prezeau:2010mx}.  Distributed spherical harmonic operations are performed
using \texttt{libsharp}\footnote{ 
  \href{https://github.com/Libsharp/libsharp}{
    \texttt{github.com/Libsharp/libsharp}}
} \citep{2013A&A...554A.112R}.  HEALPix map operations are from
\texttt{healpy}\footnote{
  HealPy is a Python front-end
  (\href{https://github.com/healpy/healpy}{
    \texttt{github.com/healpy/healpy}})
  to the HEALPix library:
  \href{https://healpix.sourceforge.io/}{
    \texttt{healpix.sourceforge.io}}.
} \citep{Zonca2019healpy}.  Other FITS files are manipulated using
\texttt{PyFITS}\footnote{
  \href{https://pythonhosted.org/pyfits}{
    \texttt{pythonhosted.org/pyfits}}
} from \texttt{Astropy}.\footnote{
  \href{http://www.astropy.org}{
    \texttt{www.astropy.org}}
}  The beam window functions are evaluated using the \quickpol\ \citep{Hivon:2016qyw} code adapted to \npipe\ files and included in the \npipe\ repository.  Power-spectrum estimation with mode decoupling is done using
\texttt{PolSpice}\footnote{
  \href{http://www2.iap.fr/users/hivon/software/PolSpice}{
    \texttt{www2.iap.fr/users/hivon/software/PolSpice/}}
} \citep{Szapudi:2000xj,Chon:2003gx}.

The frequency maps, sky models, simulations, beam window functions, time streams, and auxiliary files are all available at NERSC under \texttt{/global/cfs/cdirs/cmb/data/planck2020}.  Interested parties are invited to apply for an account following the instructions at \href{
  https://crd.lbl.gov/departments/computational-science/c3/c3-research/cosmic-microwave-background/cmb-data-at-nersc
}{
  \texttt{
    crd.lbl.gov/departments/\\computational-science/c3/c3-research/\\cosmic-microwave-background/cmb-data-at-nersc.
  }
}
Flight data products and limited release of the simulations will be made available on the Planck Legacy Archive (PLA) at \href{
  https://www.cosmos.esa.int/web/planck/pla
}{
  \texttt{
    https://www.cosmos.esa.int/web/planck/pla
  }
}

\section{Calibration} \label{app:calibration}

We consider two alternative approaches to correcting gain fluctuations in the detector data.  The first is conceptually simpler and can be described as ``total power calibration.''  The second approach is based on the destriping principle \citep{keihanen2004,keihanen2005,kurki-suonio2009,keihanen2010} and targets the gain fluctuations instead of the total signal strength.

\subsection{Total-power calibration}

A direct approach to calibration is to build a signal model, $\vec s$, and regress it against the detector data, $\vec d$, at some chosen gain steps.  The fitting coefficients, $g$, are directly the detector gains during each gain step:
\begin{linenomath*}
\begin{equation}
  \label{eq:1}
  \vec d = g \cdot \vec s + b \cdot \vec o + \vec n,
\end{equation}
\end{linenomath*}
where $b$ is the noise offset, $\vec o$ is a constant offset template, and $\vec n$ is the instrumental noise without an offset.  Maximum likelihood estimates of the template coefficients $g$ and $b$ follow from the familiar linear regression.  First we collect the templates into columns of a template matrix,
\begin{linenomath*}
\begin{equation}
  \label{eq:3}
  \mat F = \left[ \vec s, \vec o \right],
\end{equation}
\end{linenomath*}
and then solve the template coefficients, $\vec a = [g, b]\trans$, from
\begin{linenomath*}
\begin{equation}
  \label{eq:4}
  \vec d = \mat F\,\vec a + \vec n.
\end{equation}
\end{linenomath*}
Assuming a noise covariance matrix
$\mat N = \langle \vec n \vec n\trans \rangle$ the maximum likelihood solution
is
\begin{linenomath*}
\begin{equation}
  \label{eq:2}
  \vec a = \left(
    \mat F\trans \mat N\inv \mat F
  \right)\inv \mat F\trans \mat N\inv \vec d.
\end{equation}
\end{linenomath*}
Aside from simplicity, the total power calibration has the benefit that it naturally produces gains that both trace gain fluctuations and calibrates the signal against known astrophysical sources such as the Solar dipole.

Multiple gain steps are easily accommodated by adding columns to the template matrix $\mat F$.  Each gain template is zeroed outside the intended gain step or one may even blend the templates to enforce continuity.

Total-power calibration suffers from incomplete knowledge of the sky signal.  The sky estimate, $\vec s$, \emph{has} to be incomplete.  If we knew exactly what the sky was, the measurement would have been unnecessary in the first place.  Any errors (noise and otherwise) in our sky estimate cause a bias towards zero due to a linear regression phenomenon know as ``errors in variables.''  Ignoring the noise offset for a while, it can be shown that the gain is systematically underestimated:
\begin{linenomath*}
\begin{equation}
  \label{eq:5}
  \hat g = \frac{g}{1 + \sigma^2 / \sigma^2_{\rm g}},
\end{equation}
\end{linenomath*}
where $\sigma_g^2$ is the variance of the ``true'' sky signal and $\sigma^2$ is the variance of errors in our estimate.  The error is directly proportional to the overall gain:
\begin{linenomath*}
\begin{equation}
  \label{eq:6}
  \frac{g}{\hat g} = 1 + \sigma^2 / \sigma^2_{\rm g}.
\end{equation}
\end{linenomath*}
This form of the bias $g/\hat g$ is only accurate in the presence of a single template and becomes more complicated when several gain steps or additional templates are considered.

Lack of astrophysical gradients in the signal (e.g., scanning along the dipole equator) makes even this simple system degenerate and causes unchecked transfer of power between the gain and the noise offset.

\subsection{Calibration via destriping}

In this section we describe the calibration approach adopted in \cite{planck2014-a10} for \hfi\ analysis, and is a multi-detector polarized
extension of the \dacapo\ algorithm used for \lfi\ calibration \citep{planck2014-a03,planck2016-l02} with some critical differences.

Destriping assumes that the detector data are a combination of a sky-synchronous signal, a linear combination of noise templates (typically disjoint noise offsets), and uncorrelated white noise:
\begin{linenomath*}
\begin{equation}
  \label{eq:9}
  \vec d = \mat P \vec m + \mat F\,\vec a + \vec n\,.
\end{equation}
\end{linenomath*}
This algorithm works to infer a maximum likelihood estimate of the sky, $\hat{\vec m}$, by first determining the noise template amplitudes in $\vec a$.

It is possible to repurpose the destriping approach to fit for gain \emph{fluctuations} rather than the total gain.  We use the destriping
projection operator: \begin{linenomath*}
\begin{equation}
  \label{eq:7}
  \mat Z = \mat I- \mat P \left(
    \mat P\trans \mat N\inv \mat P
  \right)\inv \mat P\trans \mat N\inv,
\end{equation}
\end{linenomath*}
where $\mat I$ is the identity matrix and $\mat P$ is the pointing matrix.  The operator $\mat Z$ acts on time-domain objects such as $\vec d$ and $g\cdot\vec s$ by:
\begin{enumerate}
\item binning a map of the signal (object);
\item resampling a timeline from the map; and
\item subtracting the resampled signal from the original signal.
\end{enumerate}
The result is a time-domain object with the sky-synchronous part removed.  Such a signal will be void of all stationary sky signal, but the projection will also affect the non-stationary parts, such as noise, orbital dipole, and gain fluctuations.

If we consistently apply this projection operator to every time-domain object in the linear regression equation, Eq.~(\ref{eq:2}), we find
\begin{linenomath*}
\begin{equation}
  \label{eq:8}
  \vec a = \left(
    \mat F\trans \mat N\inv \mat Z \mat F
  \right)\inv \mat F\trans \mat N\inv \mat Z\,\vec d.
\end{equation}
\end{linenomath*}
We have simplified Eq.~(\ref{eq:8}) using the projection matrix properties of $\mat Z$ to write $\mat Z\trans \mat N\inv \mat Z = \mat N\inv \mat Z$.  We note that Eq.~(\ref{eq:8}) is also the maximum likelihood solution of Eq.~(\ref{eq:9}) in the absence of prior knowledge about the template amplitude covariance:
\begin{linenomath*}
\begin{equation}
  \label{eq:10}
  \mat C_\mathrm a = \langle \vec a \vec a\trans \rangle
  \quad \Rightarrow \quad
  \mat C_\mathrm a\inv = 0
\end{equation}
\end{linenomath*}

In Eq.~(\ref{eq:8}) we have translated the total power calibration (Eq.~(\ref{eq:2}) into a subspace that lacks sky-synchronous degrees of freedom.  Recovered gain-template amplitudes no longer reflect the overall gain, but rather deviations about a sky-synchronous average.  The errors-in-variables bias now becomes
\begin{linenomath*}
\begin{equation}
  \label{eq:8b}
  \widehat{\delta g} = \frac{\delta g}{1 + \sigma^2 / \sigma^2_g},
\end{equation}
\end{linenomath*}
making the error proportional to the gain fluctuation, rather than the overall gain.  The bias is now attenuating the magnitude of the fluctuation instead of systematically pulling down the overall gain.  If we iterate over the calibration, the gains rapidly converge to a self-consistent pair of sky and gain estimates $(\hat m, \hat g)$, even with errors in the gain template.

\section{Far-sidelobe corrections to the dipole} 
\label{app:fsl_dipole}

\npipe\ uses the same method as the 2018 \lfi\ DPC processing to convolve the ideal dipole model with the \grasp\ estimates of the \Planck\ sidelobes.  The corrections are applied to the Solar dipole and associated quadrupole.

We begin by noting that the Doppler effect up to the quadrupole term can be written (omitting a monopole term) as
\citep[e.g.,][]{Notari:2015kla}:
\begin{linenomath*}
\begin{equation}
  \label{eq:dipo}
  D(\vec{\hat n}) = T_0\left[
    \vec \beta \cdot \vec{\hat n} \left(
      1 + q\, \vec \beta \cdot \vec{\hat n}
    \right)
  \right],
\end{equation}
\end{linenomath*}
where $T_0$ is the CMB monopole temperature, $\vec{\hat n}$ is the observing direction, $\vec \beta = \vec v / c$ is the total velocity divided by the speed of light and $q$ is the frequency-dependent quadrupole factor:
\begin{linenomath*}
\begin{equation}
  \label{eq:q}
  q = \frac{x}{2}\cdot\frac{e^x + 1}{e^x - 1},
  \quad
  \mathrm{with}
  \quad
  x = \frac{h\nu}{k_\mathrm BT_0},
\end{equation}
\end{linenomath*}
where $\nu$ is the observing frequency \citep[see also][]{planck2013-pipaberration}.  The quadrupole factors for each \Planck\ frequency are shown in Table~\ref{tab:qfactors}.  \lfi-DPC processing used unity in place of $q$.

\begin{table}[tb] 
  \begingroup
  \newdimen\tblskip \tblskip=5pt
  \caption{
    Frequency-dependent, second-order quadrupole factors.
  }
  \label{tab:qfactors}
  \nointerlineskip
  \vskip -3mm
  \footnotesize
  \setbox\tablebox=\vbox{
    \newdimen\digitwidth
    \setbox0=\hbox{\rm 0}
    \digitwidth=\wd0
    \catcode`*=\active
    \def*{\kern\digitwidth}
    \newdimen\signwidth
    \setbox0=\hbox{$-$}
    \signwidth=\wd0
    \catcode`!=\active
    \def!{\kern\signwidth}
    \halign{
      \hbox to 2.0cm{#\leaderfil}\tabskip 2em&
    \hfil#\hfil\tabskip 0pt\cr
      \noalign{\doubleline}
      \omit\hfil $\nu$ [GHz]\hfil&
      \omit\hfil $q$\hfil\cr
      \noalign{\vskip 3pt\hrule\vskip 4pt}
      **$0$& $1.00$\cr
      *$30$& $1.02$\cr
      *$44$& $1.05$\cr
      *$70$& $1.12$\cr
      $100$& $1.25$\cr
      $143$& $1.48$\cr
      $217$& $2.00$\cr
      $353$& $3.12$\cr
      $545$& $4.80$\cr
      $857$& $7.55$\cr
      \noalign{\vskip 3pt\hrule\vskip 5pt}
    }
  }
  \endPlancktable
  \endgroup
\end{table}

A detector observing the sky at direction $\vec{\hat n}_0$ will see the dipole and the quadrupole in Eq.~(\ref{eq:dipo}) convolved with the instrumental beam response, $B(\vec{\hat n})$:
\begin{linenomath*}
\begin{equation}
  \label{eq:dipoconv}
  \widetilde D(\vec{\hat n}_0)
  = \int\mathrm d\Omega\,B(\vec{\hat n}, \vec{\hat n}_0)D(\vec{\hat n}).
\end{equation}
\end{linenomath*}
If we write $D(\vec{\hat n})$ from Eq.~({\ref{eq:dipo}}) in terms of its
Cartesian components,
$\vec \beta = (\beta_\mathrm x, \beta_\mathrm y, \beta_\mathrm z)$ and 
$\vec{\hat n} = (x, y, z)$ we find
\begin{align}
  \label{eq:dipoconv2}
  \widetilde D(\vec{\hat n}_0)
  = T_0\int\mathrm d\Omega\,B(\vec{\hat n}, \vec{\hat n}_0)\bigg[
& \beta_\mathrm x x + \beta_\mathrm y y + \beta_\mathrm z z + 
    \nonumber \\
&
    q\,\bigg(\beta_\mathrm x^2 x^2 +
    \beta_\mathrm x \beta_\mathrm y x y +
    \beta_\mathrm x \beta_\mathrm z x z\, +
    \nonumber \\
&
    \beta_\mathrm y \beta_\mathrm x y x +
    \beta_\mathrm y^2 y^2 +
    \beta_\mathrm y \beta_\mathrm z y z\, +
  \nonumber \\
&
    \beta_\mathrm z \beta_\mathrm x z x +
    \beta_\mathrm z \beta_\mathrm y z y +
    \beta_\mathrm z^2 z^2
    \bigg) \bigg].
\end{align}
If we now assume that we always rotate the velocity, $\vec \beta$, into a constant frame where $\vec{\hat n}_0$ is along the $z$--axis
(or any other frame where the beam, $B$ is constant), then the integrals can be pre-evaluated and we have
\begin{align}
  \label{eq:dipoconv3}
  \widetilde D = T_0\bigg[
&
    S_\mathrm x\, \beta_\mathrm x + 
    S_\mathrm y\, \beta_\mathrm y +
    S_\mathrm z\, \beta_\mathrm z + 
    \nonumber \\
&
    q\,\bigg(S_\mathrm{xx}\, \beta_\mathrm x^2 + 
    S_\mathrm{xy}\, \beta_\mathrm x\, \beta_\mathrm y +
    S_\mathrm{xz}\, \beta_\mathrm x\, \beta_\mathrm z\, +
    \nonumber \\
&
    S_\mathrm{yx}\, \beta_\mathrm y\, \beta_\mathrm x +
    S_\mathrm{yy}\, \beta_\mathrm y^2 + 
    S_\mathrm{yz}\, \beta_\mathrm y\, \beta_\mathrm z\, +
    \nonumber \\
&
    S_\mathrm{zx}\, \beta_\mathrm z\, \beta_\mathrm x +
    S_\mathrm{zy}\, \beta_\mathrm z\, \beta_\mathrm y +
    S_\mathrm{zz}\, \beta_\mathrm z^2
    \bigg) \bigg],
\end{align}
where (for example),
$S_\mathrm{xy} \equiv \int\mathrm d\Omega\,B(\vec{\hat n})\,x\,y$.

We carry out the integrals over entire $4\pi$ beams and store the resulting $S$-parameters for every \Planck\ detector, allowing us to convolve the Doppler field with the full beam response on-the-fly for every detector sample.  We then only need to rotate the total velocity, $\vec \beta$, and carry out the sum in Eq.~(\ref{eq:dipoconv3}).

\section{Anomalies around the Galactic centre}

\begin{figure*}[htpb!]
  \includegraphics[width=1.0\linewidth]{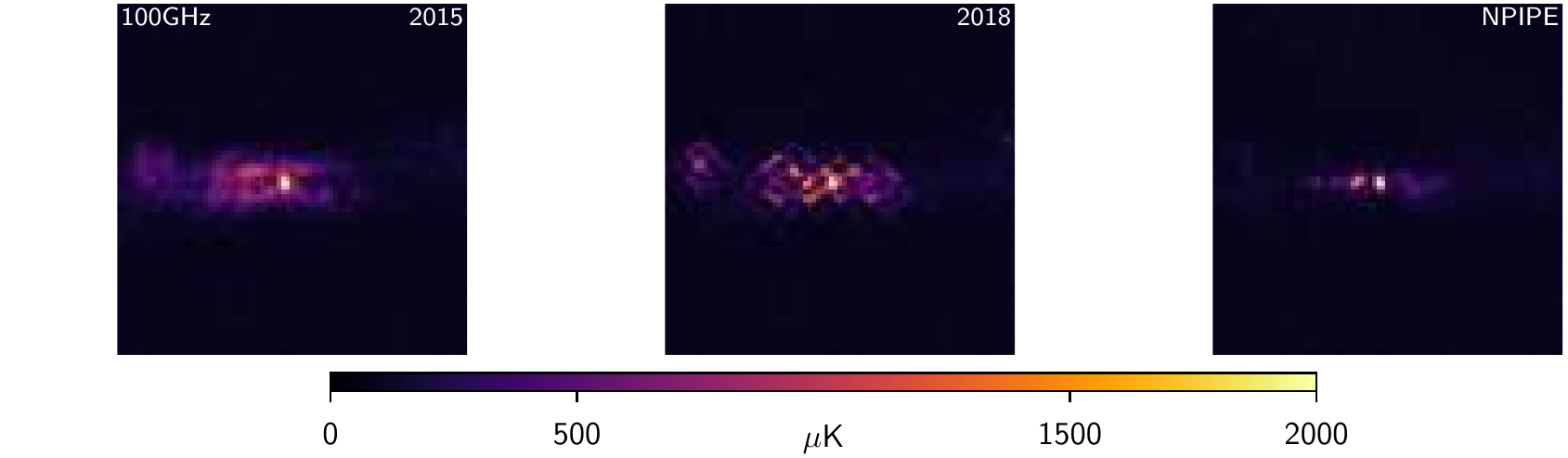}\\
  \includegraphics[width=1.0\linewidth]{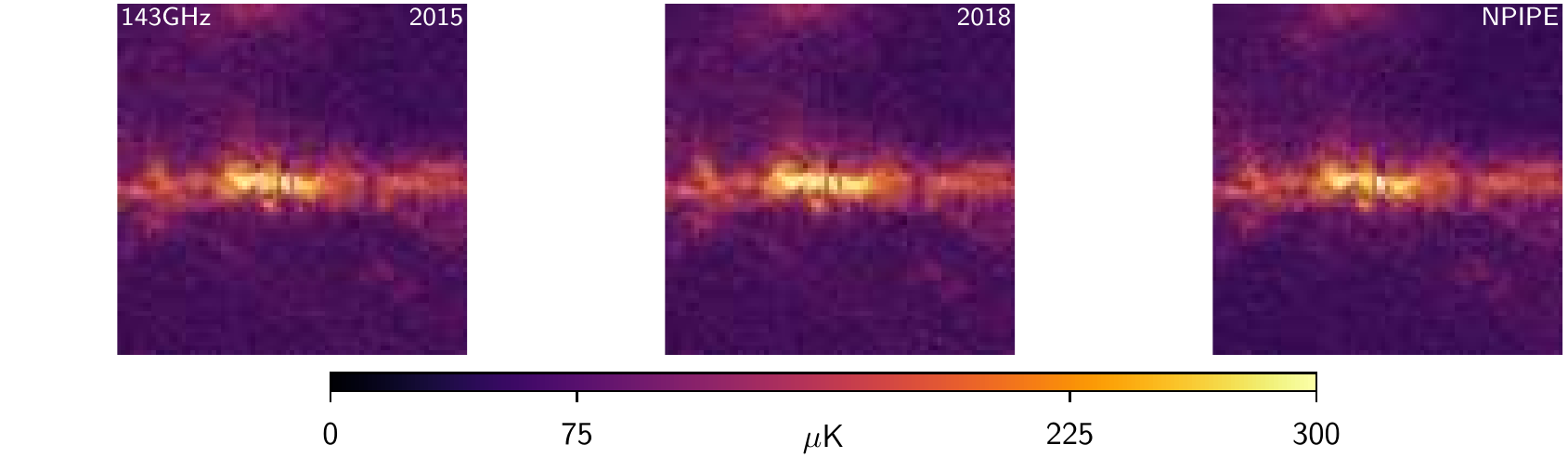}\\
  \includegraphics[width=1.0\linewidth]{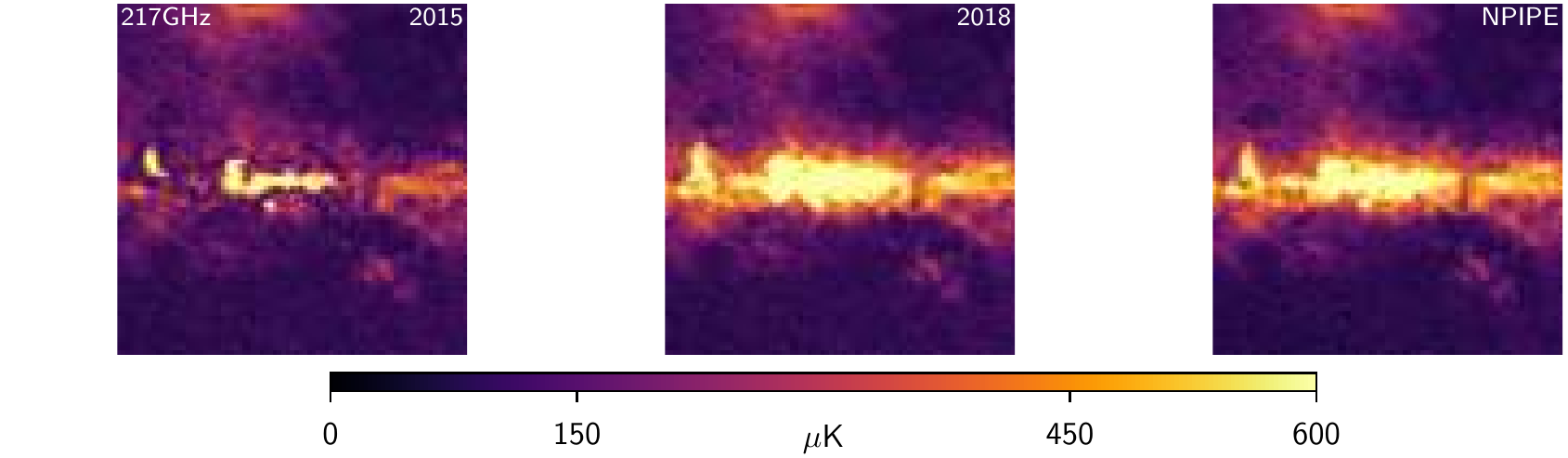}\\
  \includegraphics[width=1.0\linewidth]{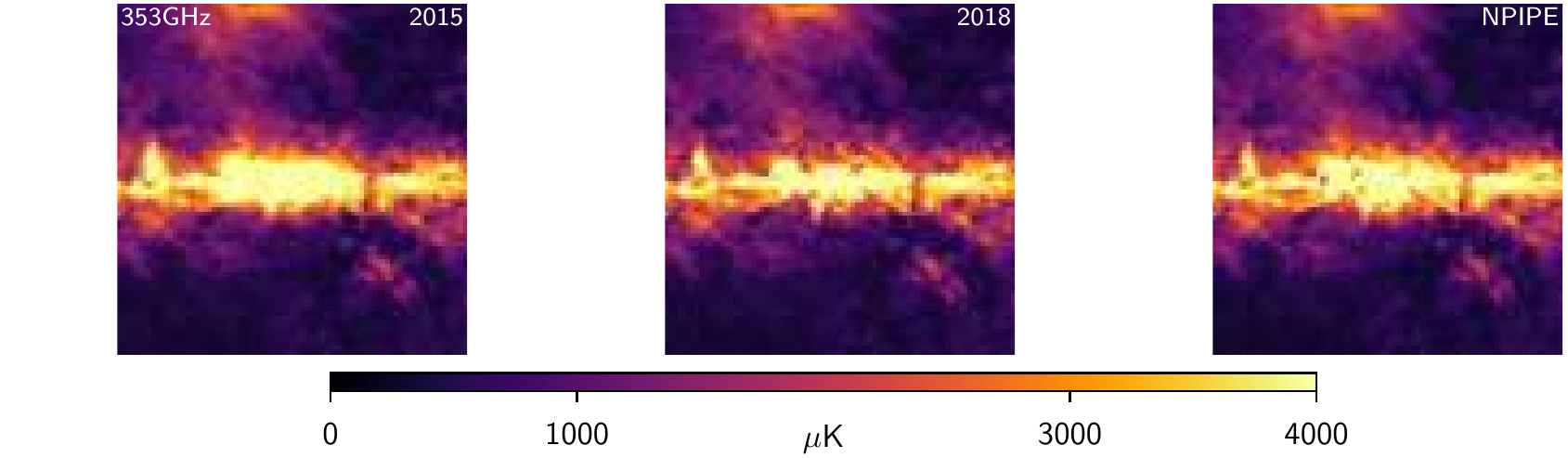}\\
  \caption{Polarization amplitude in an $8\deg\times 8\deg$ patch centred around the Galactic centre.  The linear colour scale was chosen to demonstrate the low-resolution CO template residuals in the 2018 maps.  The residuals are most pronounced at 100\GHz, where the CO corrections are largest, and absent at 143\GHz, where there is no CO correction needed.  Since the 2015 maps are not corrected for bandpass mismatch, they do not display the same artefacts..
  }
  \label{fig:gal_center}
\end{figure*}

\npipe\ fixes a known issue in \prthree\ polarized \hfi\ maps around the Galactic centre \citep[Fig.~9 lower panel]{2019arXiv190106196M}.  The CO templates used in 2018 bandpass-mismatch correction were downgraded to low resolution (\nside{128} or 27\parcm5) and the outlines of these pixels can be detected around the Galactic centre in the polarization maps.  We show images of the Galactic centre in Fig.~\ref{fig:gal_center}.

\section{Visualizations of the destriping templates} \label{app:templates}

\begin{figure*}[htpb!]
  \includegraphics[width=1.0\linewidth]{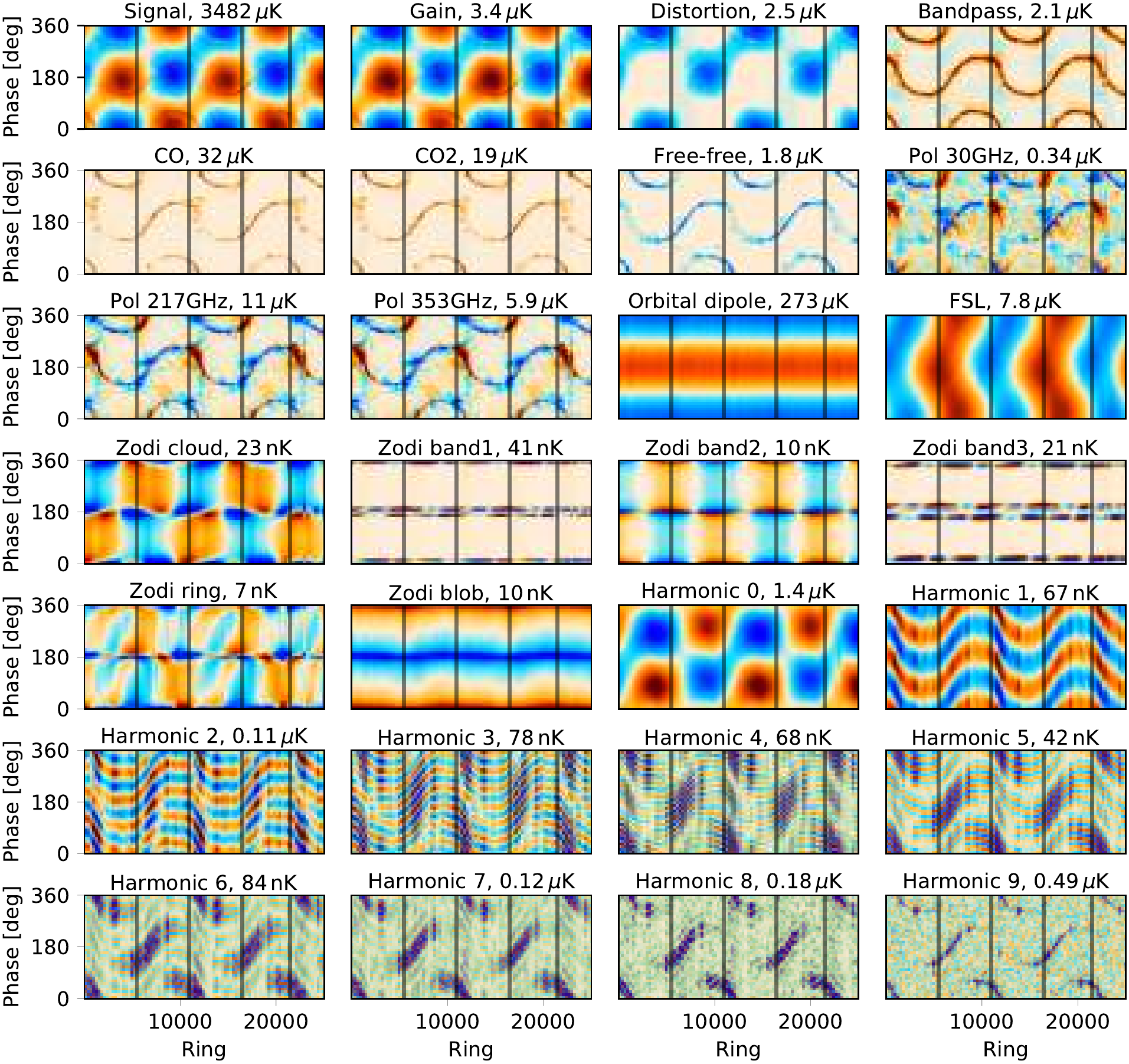}
  \caption{Signal and systematics templates for detector 100-1a, plotted as a function of pointing period (ring) and spacecraft spin phase.  The gain and signal distortion templates are actually split into several disjoint steps that vary in length depending on the S/N.  The templates for 100-1b are otherwise identical, but the 30-, 217-, and 353-GHz polarization templates are multiplied by $-1$.  The far sidelobe (FSL) template is not fitted because of degeneracies, but it is estimated and subtracted.  The polarization templates across all detectors share a single fitting amplitude.  The zodiacal emission-template amplitudes are similarly shared.  For 353\GHz\ and above, the harmonic templates are doubled to include frequency-dependent gain.  At 100--217\GHz, only relative time-shift between frequency bins is modelled.  The last harmonic template includes all frequencies not included in the other harmonic templates.  The templates are scaled to match the rms amplitude of each systematic across the 100-GHz detectors, and the plotting ranges are chosen to match the $2\,\sigma$ range of each panel.  To save space, the amplitude is reported in the title of each panel rather than as a colour bar.  The grey vertical lines indicate the survey boundaries.  Figure~\ref{fig:template_surveymaps} shows \healpix\ maps of these templates that include only the first survey.
  }
  \label{fig:template_phasemaps}
\end{figure*}

\begin{figure*}[htpb!]
  \includegraphics[width=1.0\linewidth]{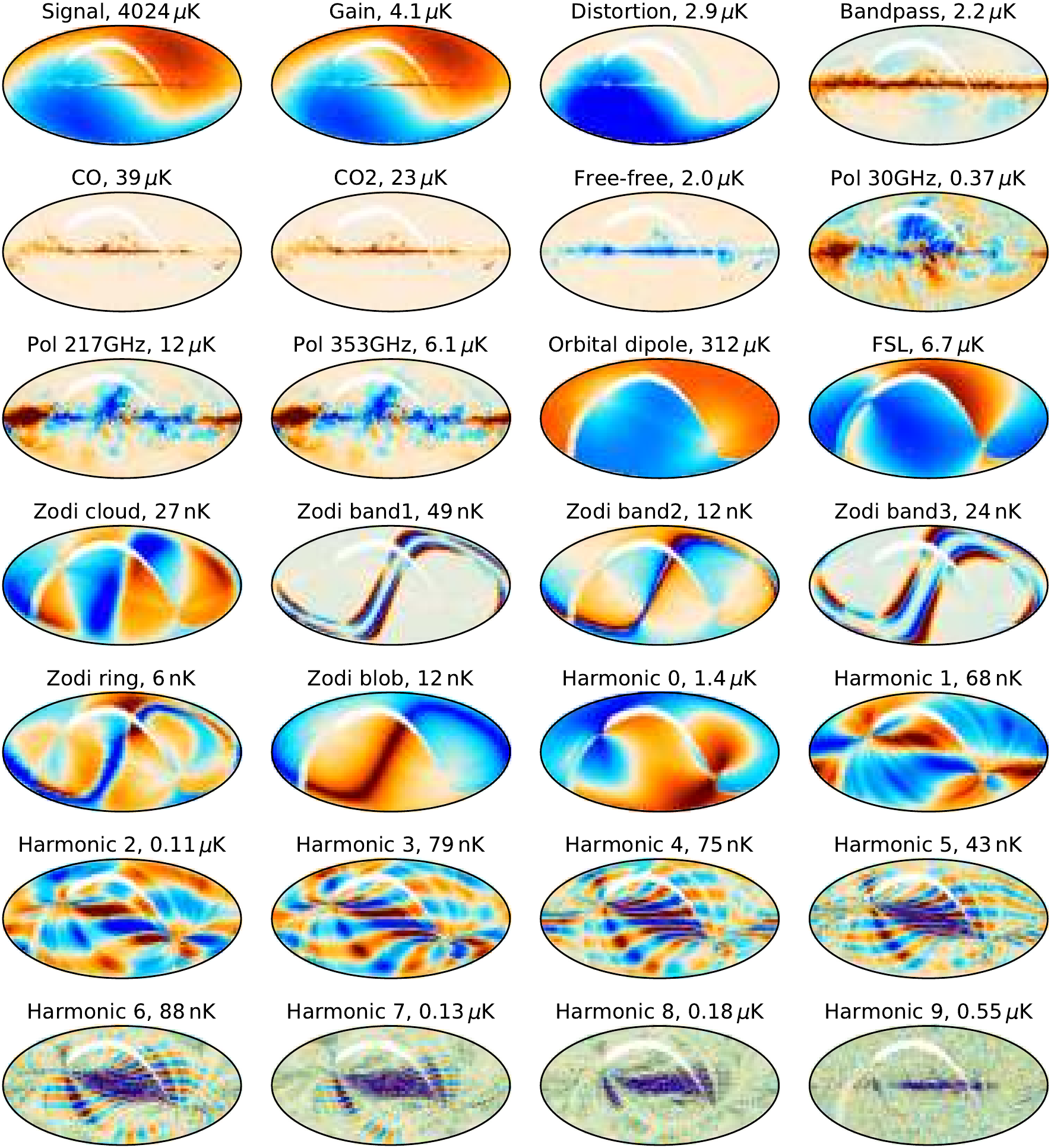}
  \caption{Binned maps of detector 100-1a signal and systematics templates for the first survey.  For details, see the caption for
    Fig.~\ref{fig:template_phasemaps}.
  }
  \label{fig:template_surveymaps}
\end{figure*}

\npipe\ suppresses systematics by fitting and removing time-domain templates (see Sect.~\ref{sec:reprocessing}).  The destriping templates are stored as columns of the template matrix, $F$, in Eq.~(\ref{eq:13a}).  In this Appendix we visualize the templates, first by binning their full time-domain representations as a function of the spacecraft spin phase and pointing period (ring) index (Fig.~\ref{fig:template_phasemaps}).  In Fig.~\ref{fig:template_surveymaps} we bin the first survey (six months and approximately $5\,500$ rings) onto more intuitive \healpix\ maps.  Most of the templates are time-dependent, so the full mission span of the templates
cannot be binned into a single map without loosing the time-dependent features.

The gain and distortion templates are represented as single panels but, in reality, are split into a number of disjoint time steps and columns in the template matrix.  Each of these steps is fitted as a separate template.

\section{Validating the \npipe\ ADCNL formalism}

The \npipe\ model for ADC nonlinearity adds another expansion order over the linear gain model used in \prthree\ \citep{planck2016-l03}.  That puts the complexity of the \npipe\ model between the \sroll\ solution in \prthree\ and the \srolltwo\ solution \citep{Delouis:2019bub}.  It was discussed in Sect.~\ref{sec:simulations} that the \npipe\ simulation set only includes ADCNL compatible with the linear gain model.  This is acceptable, if the applied ADCNL correction is powerful enough to remove the original ADCNL and replace it with the statistical and systematic template uncertainties.

\begin{figure*}[htpb!]
  \includegraphics[width=1.0\linewidth]{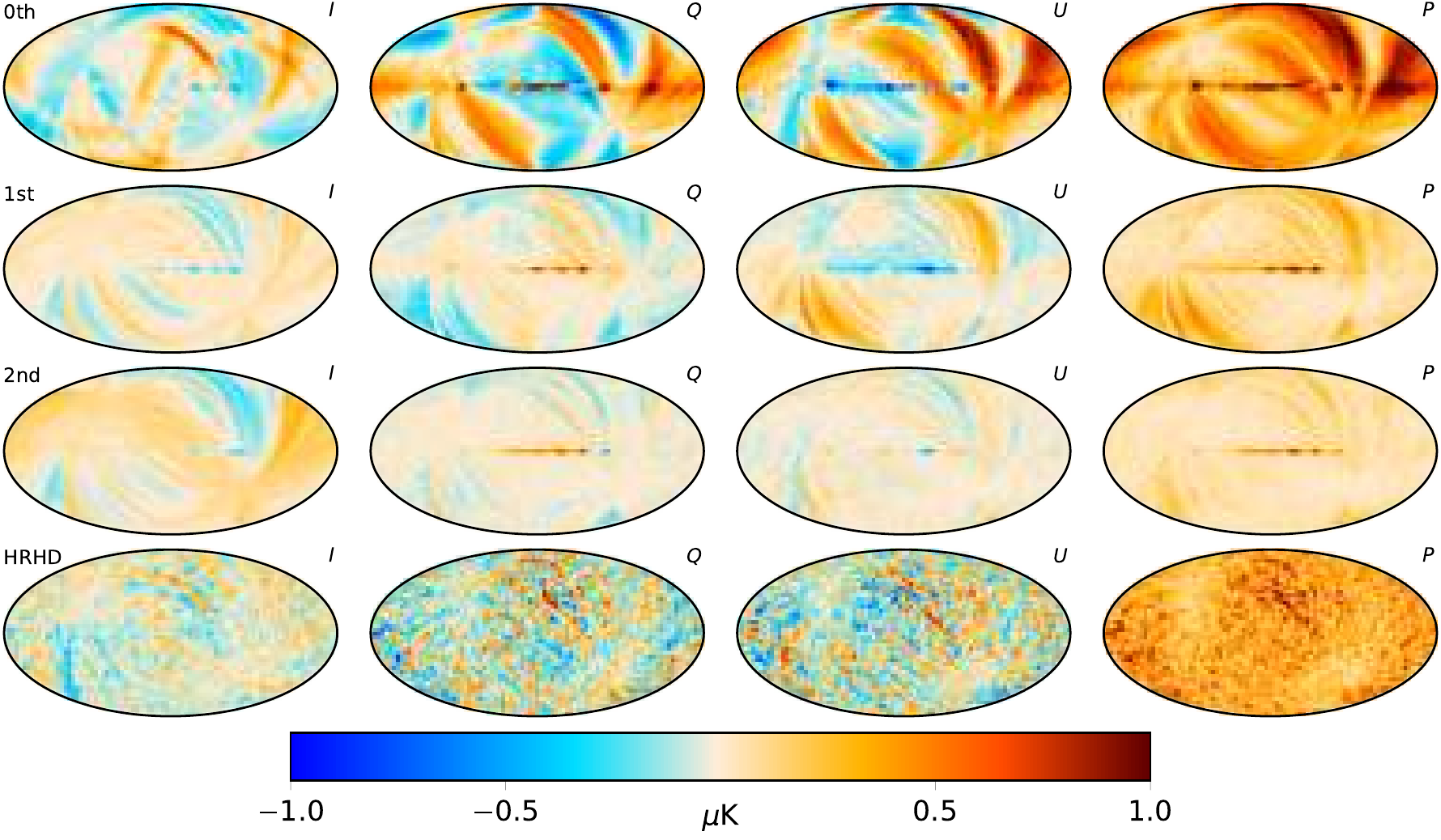}
  \caption{Simulated full-mission maps of \hfi\ ADCNL at 143\GHz.  The top row shows the full, unmitigated effect.  The second row shows the residuals after fitting and correcting ADCNL using the linear gain model (first-order correction), as was done in \prthree.  The third row shows the residual after fitting for gain and distortion terms, as is done in \npipe.  The \npipe\ transfer function (Sect.~\ref{sec:ee_tf}) applies to both the full and ADCNL-free simulations.  The fourth and last row shows the half-ring, half-difference map from the same simulation to compare the magnitude of the effect to instrumental noise.  All maps were smoothed with a 3\deg\ Gaussian beam.
The residual power spectra are shown in Fig.~\ref{fig:ffp10_adcnl_cl}.
  }
  \label{fig:ffp10_adcnl_maps}
\end{figure*}

\begin{figure*}[htpb!]
  \includegraphics[width=1.0\linewidth]{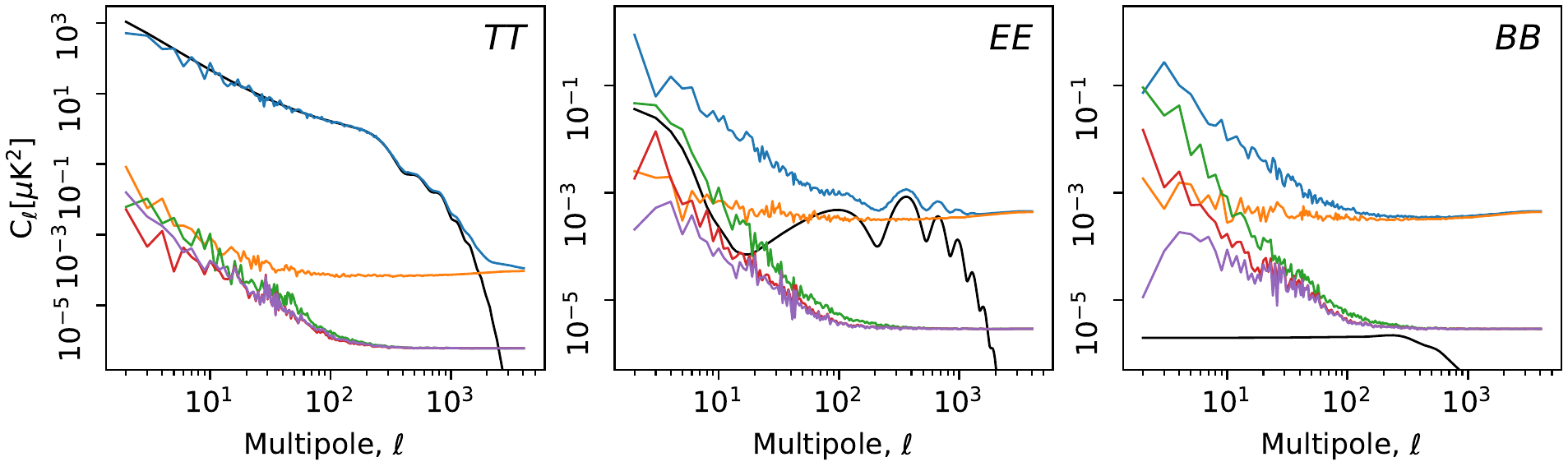}
  \caption{Power spectra of the simulated full-mission ADCNL effect over $50\,\%$ of the sky, corrected for the sky fraction.  The green curve shows the full, unmitigated effect.  Red shows the residual after the linear (first order) gain correction, and magenta shows the residual after applying the gain and distortion templates (second order), as is done in \npipe.  For scale, the power spectrum of the full 143\GHz\ frequency map is shown in blue, and an instrumental noise estimate (from the half-ring, half-difference map) is shown in orange.  The theoretical input power spectrum ($\tau=0.06$) used in the simulation is in black.  The residual maps are shown in Fig.~\ref{fig:ffp10_adcnl_maps}. The $EE$ residuals are deconvolved from the NPIPE transfer function (Sect.~\ref{sec:ee_tf}).
  }
  \label{fig:ffp10_adcnl_cl}
\end{figure*}

To test the efficacy of the \npipe\ ADCNL correction, we produced two sets of simulated $143\GHz$ TOD: one with a full model of ADCNL as was discussed in \cite{planck2016-l03} and one without ADCNL.  Other aspects of the simulated TOD were identical.  Running \npipe\ on these two simulations we may quantify the amount of ADCNL left in the \npipe\ frequency maps by measuring the difference between the full and ADCNL-free maps.  We show the ADCNL residual maps in Fig.~\ref{fig:ffp10_adcnl_maps} and compare them to a noise estimate map derived from the half-ring, half-difference maps.  In Fig.~\ref{fig:ffp10_adcnl_cl} we show the power spectra of the residual maps.  These figures demonstrate that the residual ADCNL is expected at a level lower than the pure instrumental noise.  They also demonstrate that the linear gain correction alone is not enough to meet the instrumental noise threshold.

\section{Tabulated transfer function} \label{app:tf}

In Table~\ref{tab:tf} we show the values of the measured $E$ transfer functions that correspond to Figs.~\ref{fig:ee_bias} and \ref{fig:ee_bias_fg}.

\begin{table*}[hbtb!]
  \begingroup
  \newdimen\tblskip \tblskip=5pt
  \caption{
    NPIPE $E$ transfer functions over 60\,\% of the sky corresponding to Figs.~\ref{fig:ee_bias} and \ref{fig:ee_bias_fg}.  The uncertainties represent the uncertainty of the transfer function fit, not the full dispersion of the simulations.  These transfer function values apply to each element of the spherical harmonic expansion.  Their effect in the power spectrum is squared.
  }
  \label{tab:tf}
  \nointerlineskip
  \vskip -3mm
  \footnotesize
  \setbox\tablebox=\vbox{
    \newdimen\digitwidth
    \setbox0=\hbox{\rm 0}
    \digitwidth=\wd0
    \catcode`*=\active
    \def*{\kern\digitwidth}
    \newdimen\signwidth
    \setbox0=\hbox{$-$}
    \signwidth=\wd0
    \catcode`!=\active
    \def!{\kern\signwidth}
 \halign{
      \hbox to 2.5cm{#\leaderfil}\tabskip 1em&
      \hfil#\hfil \tabskip 1em&
      \hfil#\hfil \tabskip 1em&
      \hfil#\hfil \tabskip 1em&
      \hfil#\hfil \tabskip 1em&
      \hfil#\hfil \tabskip 1em&
      \hfil#\hfil \tabskip 1em&
      \hfil#\hfil \tabskip 0pt\cr
    \noalign{\doubleline}
      \omit\hfil Multipole\hfil&
      \omit\hfil 30\GHz\hfil&
      \omit\hfil 44\GHz\hfil&
      \omit\hfil 70\GHz\hfil&
      \omit\hfil 100\GHz\hfil&
      \omit\hfil 143\GHz\hfil&
      \omit\hfil 217\GHz\hfil&
      \omit\hfil 353\GHz\hfil\cr
      \noalign{\vskip 4pt\hrule\vskip 4pt}
     *2& $1.020 \pm 0.087$& $0.761 \pm 0.031$& $0.663 \pm 0.020$& $0.584 \pm 0.016$& $0.560 \pm 0.015$& $0.562 \pm 0.018$& $0.936 \pm 0.072$\cr
     *3& $0.968 \pm 0.042$& $0.817 \pm 0.018$& $0.811 \pm 0.012$& $0.496 \pm 0.013$& $0.506 \pm 0.013$& $0.498 \pm 0.014$& $1.073 \pm 0.058$\cr
     *4& $1.015 \pm 0.035$& $0.942 \pm 0.019$& $0.924 \pm 0.016$& $0.876 \pm 0.008$& $0.850 \pm 0.008$& $0.858 \pm 0.010$& $1.075 \pm 0.048$\cr
     *5& $1.034 \pm 0.041$& $0.899 \pm 0.025$& $0.870 \pm 0.018$& $0.812 \pm 0.010$& $0.813 \pm 0.010$& $0.798 \pm 0.012$& $0.980 \pm 0.058$\cr
     *6& $1.077 \pm 0.046$& $0.860 \pm 0.035$& $0.935 \pm 0.030$& $0.868 \pm 0.014$& $0.864 \pm 0.013$& $0.876 \pm 0.016$& $0.923 \pm 0.061$\cr
     *7& $1.005 \pm 0.056$& $0.904 \pm 0.054$& $0.918 \pm 0.041$& $0.892 \pm 0.018$& $0.903 \pm 0.016$& $0.923 \pm 0.021$& $0.853 \pm 0.079$\cr
     *8& $0.913 \pm 0.064$& $1.008 \pm 0.071$& $0.881 \pm 0.058$& $0.890 \pm 0.021$& $0.857 \pm 0.020$& $0.916 \pm 0.026$& $0.886 \pm 0.097$\cr
     *9& $0.911 \pm 0.079$& $0.974 \pm 0.092$& $0.866 \pm 0.068$& $0.895 \pm 0.027$& $0.941 \pm 0.024$& $0.940 \pm 0.030$& $0.995 \pm 0.119$\cr
    10& $1.017 \pm 0.088$& $1.139 \pm 0.096$& $0.938 \pm 0.073$& $0.914 \pm 0.028$& $0.900 \pm 0.025$& $0.846 \pm 0.036$& $1.159 \pm 0.133$\cr
    11& $0.831 \pm 0.097$& $0.794 \pm 0.117$& $0.888 \pm 0.083$& $0.856 \pm 0.032$& $0.906 \pm 0.030$& $0.810 \pm 0.039$& $1.274 \pm 0.143$\cr
    12& $0.918 \pm 0.102$& $0.798 \pm 0.127$& $0.760 \pm 0.093$& $0.899 \pm 0.033$& $0.863 \pm 0.030$& $0.939 \pm 0.042$& $0.806 \pm 0.157$\cr
    13& $1.155 \pm 0.115$& $0.526 \pm 0.147$& $1.186 \pm 0.119$& $0.949 \pm 0.040$& $0.965 \pm 0.037$& $0.942 \pm 0.050$& $0.716 \pm 0.165$\cr
    14& $1.163 \pm 0.110$& $0.967 \pm 0.150$& $0.884 \pm 0.122$& $0.917 \pm 0.040$& $0.932 \pm 0.035$& $0.890 \pm 0.046$& $0.908 \pm 0.159$\cr
    15& $0.980 \pm 0.114$& $1.047 \pm 0.148$& $0.812 \pm 0.119$& $0.992 \pm 0.038$& $0.903 \pm 0.036$& $0.830 \pm 0.051$& $1.035 \pm 0.150$\cr
    16& $1.049 \pm 0.098$& $1.143 \pm 0.146$& $0.922 \pm 0.110$& $0.989 \pm 0.037$& $0.944 \pm 0.033$& $0.962 \pm 0.044$& $0.762 \pm 0.157$\cr
    17& $1.012 \pm 0.101$& $0.770 \pm 0.140$& $0.941 \pm 0.111$& $1.046 \pm 0.036$& $0.951 \pm 0.033$& $0.948 \pm 0.047$& $0.919 \pm 0.146$\cr
    18& $0.991 \pm 0.095$& $0.762 \pm 0.141$& $0.888 \pm 0.105$& $1.031 \pm 0.035$& $0.944 \pm 0.033$& $0.915 \pm 0.043$& $1.003 \pm 0.143$\cr
    19& $0.930 \pm 0.104$& $1.142 \pm 0.128$& $0.828 \pm 0.101$& $0.984 \pm 0.034$& $1.021 \pm 0.028$& $0.878 \pm 0.043$& $0.882 \pm 0.143$\cr
    20& $0.950 \pm 0.097$& $1.161 \pm 0.116$& $0.763 \pm 0.093$& $0.907 \pm 0.031$& $0.988 \pm 0.027$& $0.993 \pm 0.036$& $0.603 \pm 0.136$\cr
    21& $0.941 \pm 0.090$& $0.995 \pm 0.110$& $1.112 \pm 0.087$& $0.886 \pm 0.026$& $0.932 \pm 0.024$& $0.955 \pm 0.035$& $0.877 \pm 0.125$\cr
    22& $1.048 \pm 0.084$& $0.929 \pm 0.113$& $0.966 \pm 0.079$& $1.005 \pm 0.025$& $0.994 \pm 0.022$& $0.955 \pm 0.032$& $0.949 \pm 0.118$\cr
    23& $0.958 \pm 0.087$& $0.903 \pm 0.097$& $0.985 \pm 0.076$& $0.974 \pm 0.023$& $0.978 \pm 0.020$& $0.966 \pm 0.029$& $1.142 \pm 0.116$\cr
    24& $0.931 \pm 0.080$& $0.967 \pm 0.093$& $1.006 \pm 0.066$& $1.033 \pm 0.022$& $0.980 \pm 0.019$& $0.981 \pm 0.027$& $1.268 \pm 0.120$\cr
    25& $1.176 \pm 0.081$& $0.875 \pm 0.086$& $0.959 \pm 0.066$& $1.017 \pm 0.021$& $0.961 \pm 0.017$& $1.010 \pm 0.026$& $1.077 \pm 0.115$\cr
    26& $1.043 \pm 0.078$& $1.111 \pm 0.084$& $0.757 \pm 0.066$& $0.994 \pm 0.019$& $1.022 \pm 0.018$& $1.008 \pm 0.025$& $1.068 \pm 0.114$\cr
    27& $1.025 \pm 0.072$& $1.028 \pm 0.078$& $0.873 \pm 0.061$& $1.014 \pm 0.018$& $1.000 \pm 0.017$& $0.997 \pm 0.022$& $1.193 \pm 0.104$\cr
    28& $0.962 \pm 0.078$& $0.976 \pm 0.072$& $1.005 \pm 0.058$& $0.986 \pm 0.018$& $0.995 \pm 0.015$& $0.994 \pm 0.022$& $1.088 \pm 0.108$\cr
    29& $0.999 \pm 0.076$& $0.983 \pm 0.073$& $0.991 \pm 0.060$& $0.994 \pm 0.016$& $1.009 \pm 0.015$& $0.998 \pm 0.020$& $0.855 \pm 0.106$\cr
    30& $1.019 \pm 0.081$& $0.995 \pm 0.067$& $0.964 \pm 0.053$& $1.009 \pm 0.016$& $1.011 \pm 0.015$& $0.999 \pm 0.020$& $1.057 \pm 0.108$\cr
    31& $0.928 \pm 0.078$& $1.042 \pm 0.061$& $0.925 \pm 0.051$& $0.996 \pm 0.015$& $0.969 \pm 0.014$& $0.994 \pm 0.019$& $1.166 \pm 0.106$\cr
    32& $1.012 \pm 0.078$& $0.867 \pm 0.062$& $0.947 \pm 0.048$& $0.989 \pm 0.014$& $0.982 \pm 0.013$& $0.971 \pm 0.019$& $1.150 \pm 0.110$\cr
    33& $1.051 \pm 0.075$& $0.988 \pm 0.061$& $0.875 \pm 0.049$& $0.997 \pm 0.013$& $1.000 \pm 0.013$& $1.017 \pm 0.017$& $1.112 \pm 0.103$\cr
    34& $0.911 \pm 0.073$& $0.992 \pm 0.058$& $1.034 \pm 0.047$& $0.987 \pm 0.013$& $0.986 \pm 0.012$& $1.002 \pm 0.017$& $1.006 \pm 0.103$\cr
    35& $0.934 \pm 0.078$& $0.888 \pm 0.054$& $1.009 \pm 0.044$& $0.975 \pm 0.012$& $0.999 \pm 0.012$& $0.989 \pm 0.017$& $0.968 \pm 0.110$\cr
    36& $0.927 \pm 0.082$& $1.014 \pm 0.053$& $0.961 \pm 0.044$& $0.993 \pm 0.012$& $1.017 \pm 0.011$& $0.994 \pm 0.016$& $0.990 \pm 0.107$\cr
    37& $1.083 \pm 0.080$& $0.871 \pm 0.052$& $0.907 \pm 0.040$& $1.003 \pm 0.012$& $1.013 \pm 0.011$& $0.977 \pm 0.015$& $0.881 \pm 0.106$\cr
    38& $0.932 \pm 0.083$& $0.974 \pm 0.048$& $0.971 \pm 0.039$& $1.000 \pm 0.012$& $1.012 \pm 0.010$& $0.997 \pm 0.015$& $0.960 \pm 0.109$\cr
    39& $1.273 \pm 0.080$& $0.967 \pm 0.049$& $0.914 \pm 0.039$& $1.033 \pm 0.011$& $0.992 \pm 0.010$& $0.976 \pm 0.013$& $0.959 \pm 0.105$\cr
    40& $1.042 \pm 0.080$& $0.976 \pm 0.046$& $1.063 \pm 0.037$& $0.983 \pm 0.011$& $0.981 \pm 0.010$& $0.986 \pm 0.014$& $1.163 \pm 0.110$\cr
      \noalign{\vskip 4pt\hrule\vskip 5pt}
    }
  }
\endPlancktablewide 
 \endgroup
\end{table*}

\section{Degeneracy in bandpass mismatch and polarization templates} \label{app:degeneracy}

During reprocessing (Sect.~\ref{sec:reprocessing}) of the CMB channels, \npipe\ fits the polarized frequency maps from the foreground channels as time-domain templates (Sect.~\ref{sec:pestimates}).  This allows fitting for the other time-domain templates over a temperature-only sky, breaking some significant degeneracies that otherwise render the large-scale polarization in the maps very noisy.  It is tempting to ask, if the polarization templates can be combined into a high S/N estimate of the polarized sky at each of the CMB frequencies.  Unfortunately the answer is negative for two reasons:
\begin{enumerate}
\item the polarization templates are degenerate with the bandpass-mismatch templates and;
\item the polarization template description does not support spectral index variation across the sky.
\end{enumerate}

The fitted polarization template amplitudes are shown in Table~\ref{tab:polamps}.  It is straightforward to demonstrate that a polarization map constructed with these amplitudes is missing some of the total polarization at each of the CMB frequencies.  The remainder is captured in the bandpass-mismatch templates that, through temperature-to-polarization leakage, translate into Galactic polarization resembling the channel-map templates.  We were able to demonstrate this effect with a simplified simulation of the 217-\GHz\ channel processing.  In this test we simulated noise-free TOD and fed them to the \npipe\ reprocessing.  We also replaced the frequency-map-based polarization templates with the actual polarization template used in the simulation.  One would expect this setup to yield unit amplitude for the fitted polarization template but instead we recovered 0.61.  Since the simulation did not include systematics, it was possible to disable templates in the template matrix one by one and repeat the simulation.  We found that the low fitting amplitude persisted despite disabling orbital-dipole fitting, zodiacal emission, and the whole hierarchy of the \hfi\ transfer function residual templates.  Once we disabled the bandpass-mismatch correction, the fitted polarization template amplitude jumped to 1.01.  There is no way to repeat the test with flight data as we cannot disable the true bandpass mismatch.

\begin{table}[htbp!] 
  \begingroup
  \newdimen\tblskip \tblskip=5pt
  \caption{\npipe\ polarization-template amplitudes.
  }
  \label{tab:polamps}
  \nointerlineskip
  \vskip -3mm
  \footnotesize
  \setbox\tablebox=\vbox{
    \newdimen\digitwidth
    \setbox0=\hbox{\rm 0}
    \digitwidth=\wd0
    \catcode`*=\active
    \def*{\kern\digitwidth}
    \newdimen\signwidth
    \setbox0=\hbox{-}
    \signwidth=\wd0
    \catcode`!=\active
    \def!{\kern\signwidth}
    \halign{
      \hbox to 2.5cm{#\leaderfil}\tabskip 1em&
      \hfil#\hfil \tabskip 1em&
      \hfil#\hfil \tabskip 1em&
      \hfil#\hfil \tabskip 0pt\cr
      \noalign{\doubleline}
      \omit\hfil&
      \multispan3\hfil Template frequency\hfil\cr
      \omit\hfil&
      \multispan3\hrulefill\cr
      \omit\hfil Fitting frequency\hfil&
      \omit\hfil $30\GHz$\hfil&
      \omit\hfil $217\GHz$\hfil&
      \omit\hfil $353\GHz$\hfil\cr
      \noalign{\vskip 3pt\hrule\vskip 4pt}
      *$44\GHz$& *$0.221$** & $0.227$ & $-0.0236$\cr
      *$70\GHz$& *$0.0482$* & $0.274$ & $-0.0251$\cr
      $100\GHz$& *$0.0133$* & $0.338$ & $-0.0215$\cr
      $143\GHz$& *$0.00979$ & $0.382$ & $-0.0185$\cr
      $217\GHz$& *$0.0114$* & n/a & $\phantom{-}0.0493$\cr
      \noalign{\vskip 3pt\hrule\vskip 5pt}
    }
  }
  \endPlancktable
  \endgroup
\end{table}

The fact that the polarization templates are degenerate with the bandpass-mismatch correction may seem risky, as it could compromise the vital bandpass-mismatch correction and bias the Galactic polarization in the \npipe\ CMB frequency maps (polarization templates are not fitted at 30 or 353\GHz).  However, the potential bias is avoided by the two-step approach used in reprocessing the CMB frequencies.
\begin{enumerate}
\item During the first $N-1$ iterations, the time-dependent gain or ADCNL correction and other templates are fitted while marginalizing over the bandpass-mismatch and polarization-template amplitudes.  Fitting is done over an intensity-only sky, approximating that the polarized signal is fitted by the TOD templates.  As long as the combination of these templates is a reasonable description of the polarization modulation in the TOD, the other templates are not affected.
\item During the final iteration, the gain solution is held fixed, fitting is done over a polarized $IQU$ sky, and the polarization templates are not involved.
\end{enumerate}

 As the TOD are being calibrated and ADCNL-corrected, it matters only that the \emph{total} polarization model is complete enough not to bias the time-varying gain and ADCNL solution.  The fact that the polarization model is built from degenerate templates does not bias the solution, although it may slow down the convergence of the solver.  The bandpass-mismatch correction is not affected by the degeneracy either, because it is ultimately solved over a polarized sky \emph{without} the polarization prior.

\section{Simulated uncertainty and bias}

\begin{figure*}[htpb]
  \center{
    \includegraphics[width=1.0\linewidth]{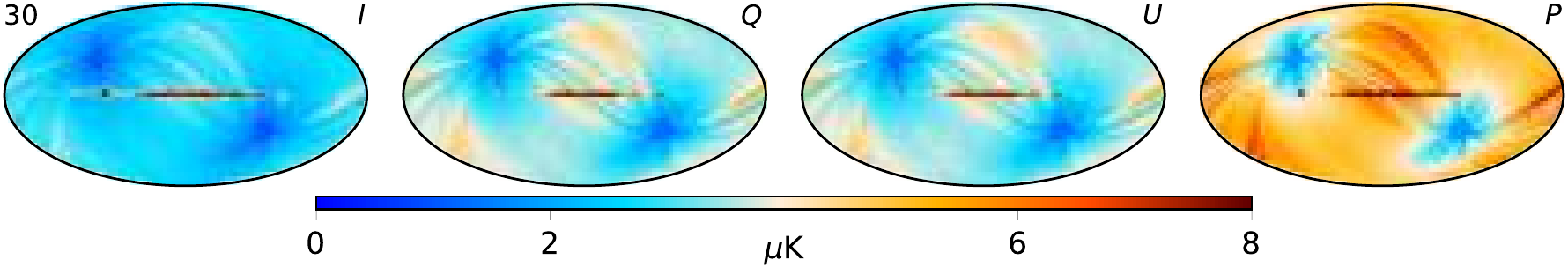}\\
    \includegraphics[width=1.0\linewidth]{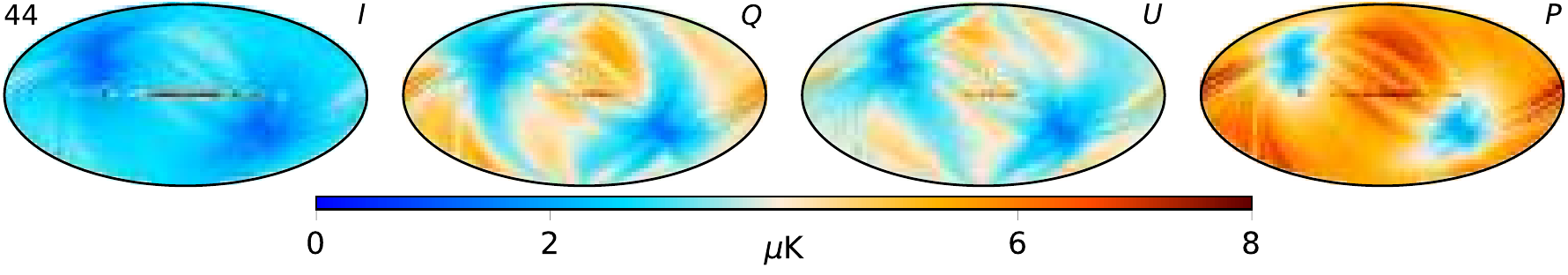}\\
    \includegraphics[width=1.0\linewidth]{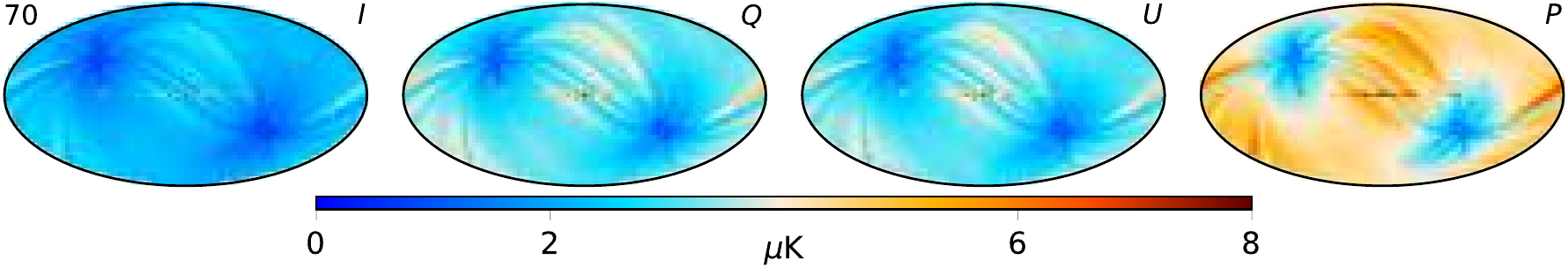}\\
    \includegraphics[width=1.0\linewidth]{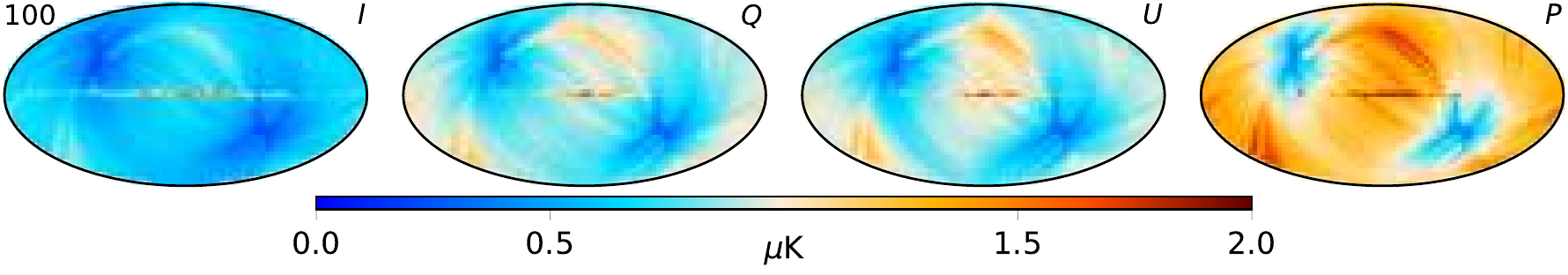}\\
    \includegraphics[width=1.0\linewidth]{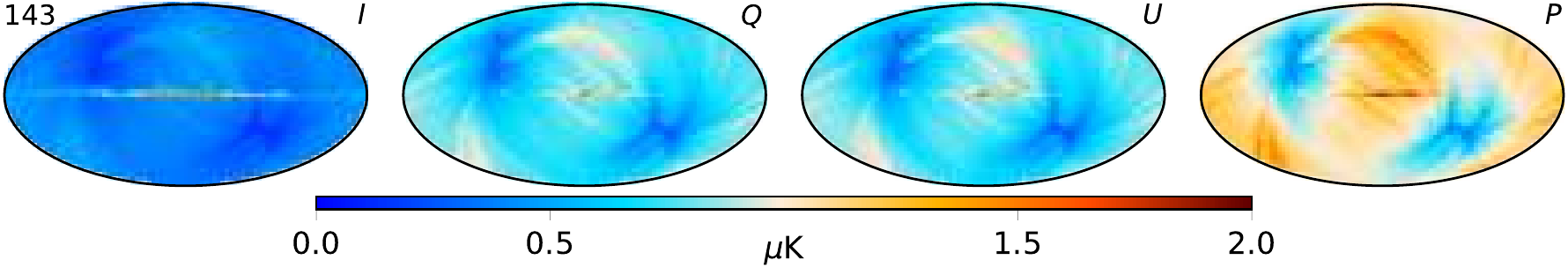}\\
    \includegraphics[width=1.0\linewidth]{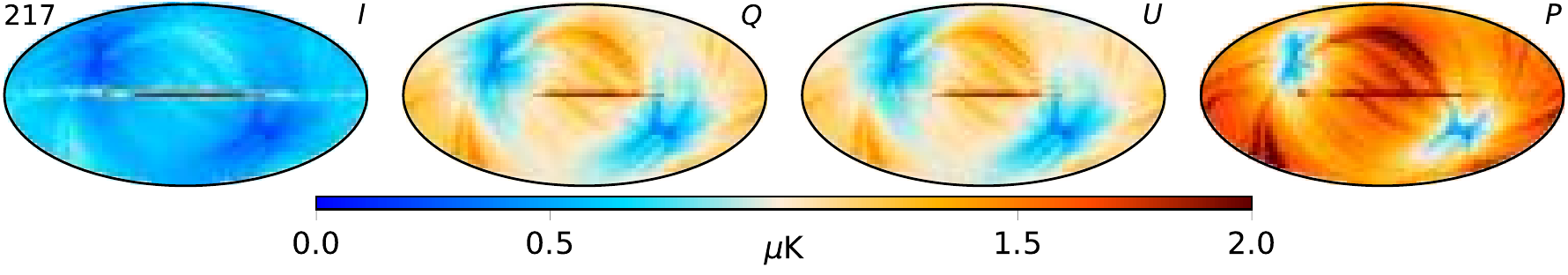}\\
    \includegraphics[width=1.0\linewidth]{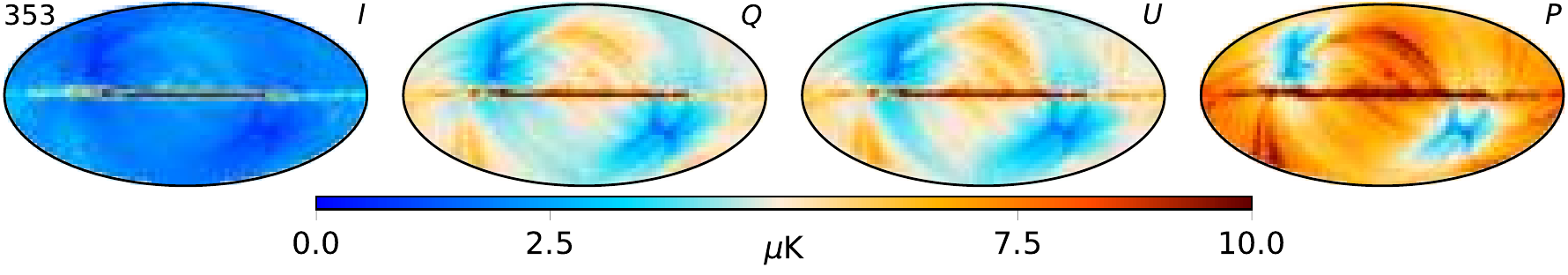}\\
  }
  \caption{Simulation error.  The rms of the residual maps smoothed to 1\deg\ is a measure of the total per-pixel uncertainty at degree scales and larger. 
  }
  \label{fig:total_error_1deg}
\end{figure*}

\begin{figure*}[htpb]
  \center{
    \includegraphics[width=1.0\linewidth]{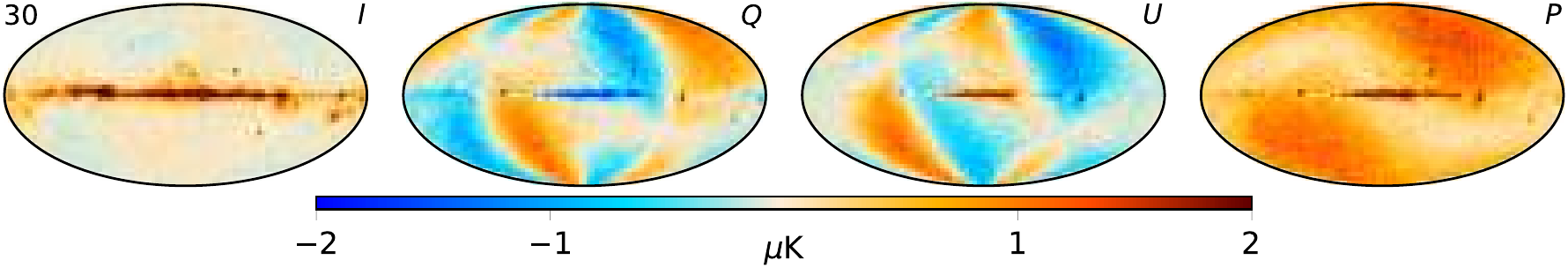}\\
    \includegraphics[width=1.0\linewidth]{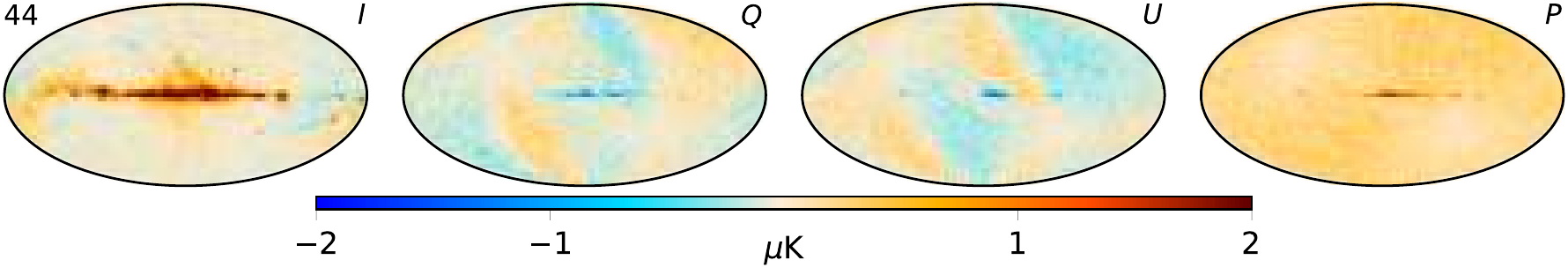}\\
    \includegraphics[width=1.0\linewidth]{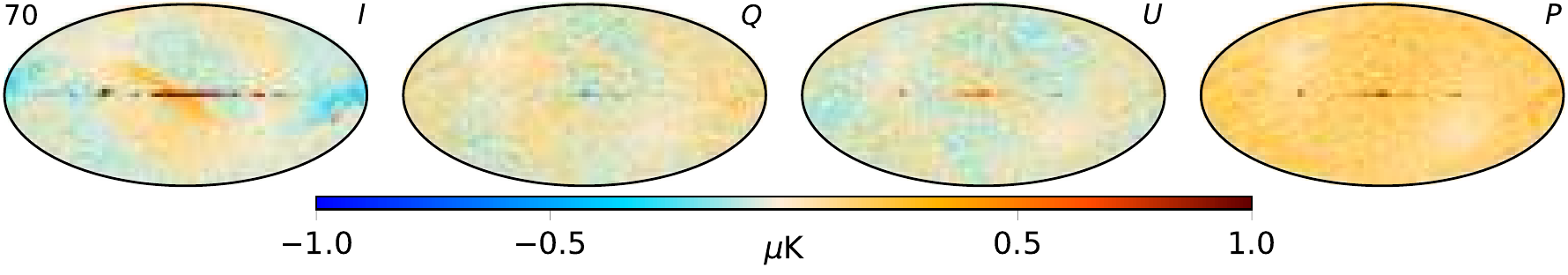}\\
    \includegraphics[width=1.0\linewidth]{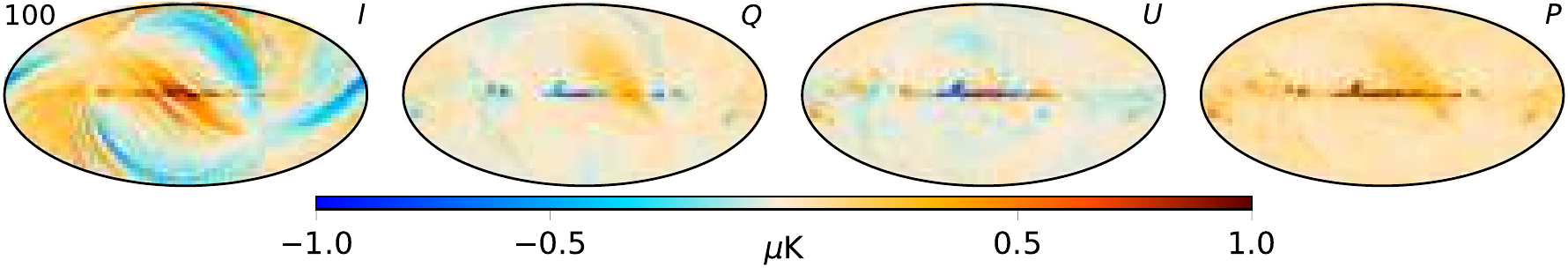}\\
    \includegraphics[width=1.0\linewidth]{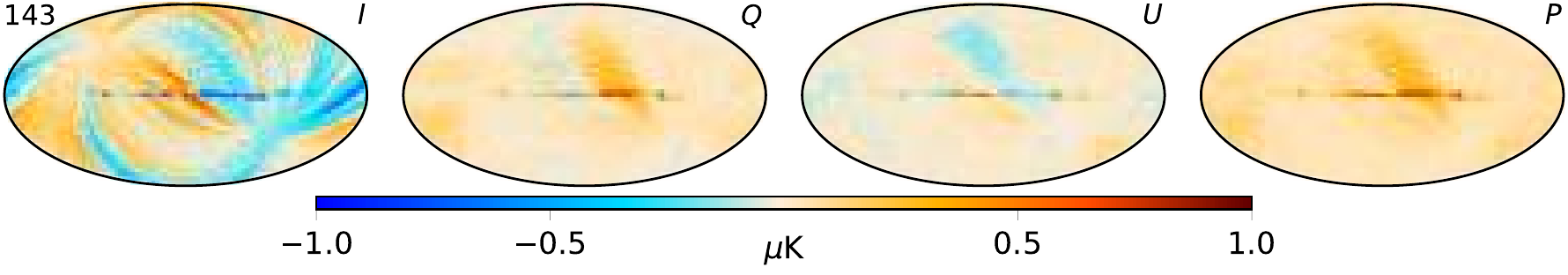}\\
    \includegraphics[width=1.0\linewidth]{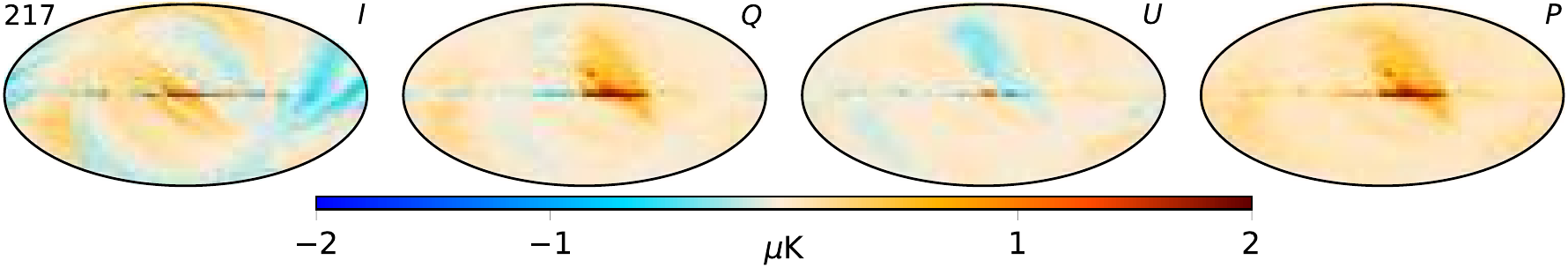}\\
    \includegraphics[width=1.0\linewidth]{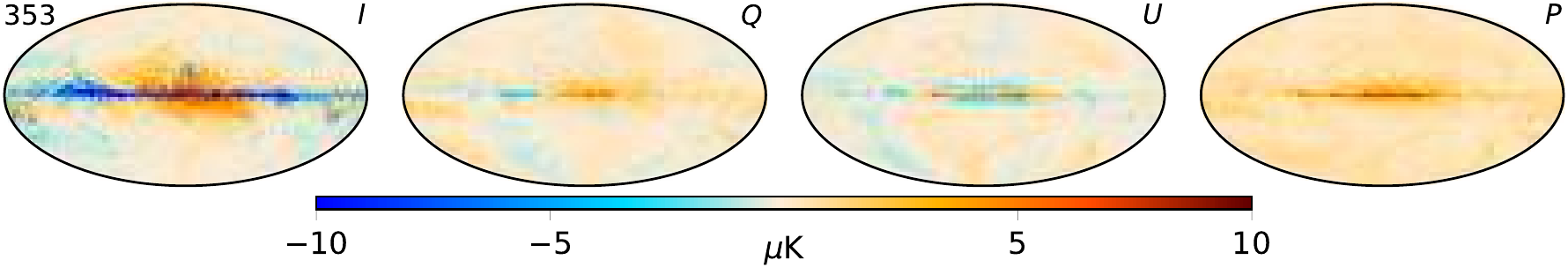}\\
  }
  \caption{Simulation bias. The mean of the residual maps smoothed to 1\deg\ is a measure of persistent error for this particular realization of systematics.  The magnitude of the bias can be compared to the total error shown in Fig.~\ref{fig:total_error_1deg}.  The dipole residual in the 30-GHz polarization maps is consistent with a small relative calibration error between the radiometers, and reflects the degeneracy between calibration and the large scale polarization due to the \Planck\ scan strategy.  The error translates to 44\GHz\ by means of the polarization prior.  \hfi\ maps exhibit a ringing structure coming from the transfer function residuals and a faint ADCNL error above and to the right of the Galactic centre.
  }
  \label{fig:total_bias_1deg}
\end{figure*}

Comparison of simulated and real null (noise) maps in Sect.~\ref{sec:sim_results} shows that the overall uncertainty in the simulations agrees with the flight data.  With 600 Monte Carlo realizations, we may also ask if the processing residuals have zero mean or if there are detectable biases, even much below the overall uncertainty.  Figure~\ref{fig:total_error_1deg} shows the total (noise and systematics) uncertainty of 1-degree smoothed $IQU$ maps;  Fig.~\ref{fig:total_bias_1deg} shows the corresponding bias.  The uncertainty is measured as the per-pixel rms of the smoothed residual (output$-$input) maps, and the bias is the average of those same maps.

We remind the reader that the \npipe\ simulations treat as fixed systematics such as bandpass mismatch, gain fluctuations, ADCNL, bolometric transfer-function residuals, and beam asymmetry.  They are not drawn from a distribution, but rather applied to each Monte Carlo simulation the same way.  This means that the residuals associated with these systematics do not average out across the simulations.  The averaged residual in Fig.~\ref{fig:total_bias_1deg} demonstrates the total level of systematic residuals without the statistical noise but should not be taken as a measurement of the actual residuals in the flight data maps.

\end{document}